\documentclass[aps,prx,reprint,superscriptaddress,twocolumn,amsmath,amssymb,amsfonts, nofootinbib]{revtex4-2}

\AtBeginDocument{\usepackage{booktabs}}               
\makeatletter
\g@addto@macro\bfseries{\boldmath}
\makeatother

\usepackage[varg]{txfonts}
\DeclareMathAlphabet{\mathcal}{OMS}{cmsy}{m}{n}

\usepackage[T1]{fontenc}
\usepackage[utf8]{inputenc}
\usepackage{hyperref}
\usepackage{amsmath}
\usepackage{color}
\usepackage{graphicx}
\usepackage[percent]{overpic}
\usepackage{mathrsfs}
\usepackage{bm}
\usepackage{braket}
\usepackage[nolist,nohyperlinks]{acronym}
\usepackage{comment}
\usepackage{tikz}
\usepackage{dsfont}
\usepackage{mathtools}
\usepackage{enumitem}
\usepackage{stfloats} 

\usepackage{capt-of}

\usepackage{amsthm}
\usepackage{slashed}

\usepackage{float}

\usepackage{tikz-cd}

\usetikzlibrary{decorations.pathmorphing,shapes}
\usetikzlibrary{arrows.meta}
\tikzcdset{arrow style=tikz,
    squigarrow/.style={
        decoration={
        snake, 
        amplitude=.4mm,
        segment length=2mm
        }, 
        rounded corners=.2pt,
        decorate
        }
    }

\usepackage[normalem]{ulem}

\newcommand{\im}{{\rm Im}}
\newcommand{\tr}{{\rm tr}}

\newcommand{\U}{\text{U}}
\newcommand{\SO}{\text{SO}}
\newcommand{\On}{\text{O}}
\newcommand{\cspt}{\text{cSPT}^{d+1}_{G}}
\newcommand{\tcspt}{\widetilde{\text{cSPT}}^{d+1}_{G}}
\newcommand{\csptu}{\mathrm{cSPT}^{d+1}_{\mathrm{univ}}}
\newcommand{\Sq}{\text{Sq}}

\newcommand{\bordd}{{\bf Bord}_{\langle d, d+1\rangle}}
\newcommand{\Bord}{{\bf Bord}_{n}}

\newcommand{\vect}{{\bf Vect}_{\mathbb{C}}}
\newcommand{\Line}{{\bf Line}_{\mathbb{C}}}
\newcommand{\tqft}{\mathcal{TQFT}_{n}}
\newcommand{\tqftd}{\mathcal{TQFT}_{d+1}}

\newcommand{\ofin}{\mathcal{O}_{\text{fin}}}
\newcommand{\qa}{\mathcal{A}_{2}}
\newcommand{\res}{\text{res}}
\newcommand{\tor}{\text{Tor}}
\newcommand{\Z}{\mathbb{Z}}

\newcommand{\catname}[1]{{\textbf{#1}}}
\newcommand{\hoCW}{\catname{hoCW}}
\newcommand{\Ab}{\catname{Ab}}

\usepackage{xfrac}

\definecolor{cred}{RGB}{228,26,28}
\definecolor{cblue}{RGB}{55,126,184}
\definecolor{cdblue}{RGB}{40,96,139}
\definecolor{clblue}{RGB}{205,223,237}
\definecolor{cgreen}{RGB}{77,175,74}
\definecolor{cgray}{RGB}{150,150,150}
\definecolor{clgray}{RGB}{200,200,200}
\definecolor{cpurple}{RGB}{152,78,163}
\definecolor{corange}{RGB}{255,127,0}
\definecolor{cgold}{RGB}{230,171,2}
\definecolor{cL}{RGB}{255,0,0}
\hypersetup{colorlinks=true,linkcolor=cL,citecolor=cL,urlcolor=cL}

\newtheoremstyle{apsstyle}%
  {3pt} 
  {3pt} 
  {\normalfont} 
  {} 
  {\bfseries}
  {.} 
  {.5em} 
  {} 

\theoremstyle{apsstyle}
\newtheorem{theorem}{Theorem}[section]
\newtheorem{proposition}[theorem]{Proposition}
\newtheorem{hypothesis}[theorem]{Hypothesis}
\newtheorem{definition}[theorem]{Definition}
\newtheorem{remark}[theorem]{Remark}
\newtheorem{conjecture}[theorem]{Conjecture}
\newtheorem{corollary}[theorem]{Corollary}
\newtheorem{ansatz}[theorem]{Ansatz}
\makeatletter
\def\thmhead@plain#1#2#3{%
  \thmname{#1}\thmnumber{ #2}%
  \ifx\@empty#3\@empty\else\thmnote{ (#3)}\fi}
\makeatother

\makeatletter
\newcommand{\AppendixTOCSectionsOnly}{%
  \renewcommand{\l@subsection}[2]{}%
  \renewcommand{\l@subsubsection}[2]{}%
}
\makeatother

\newtheorem{lemma}{Lemma}
\newtheorem{claim}{Claim}
\newtheorem{notation}{Notation}
\newtheorem{assumption}{Assumption}

\begin{document}

\title{The Fate of Crystalline Topological Phenomena in the Continuum}
\author{Rajas Chari} 
\author{Taylor L. Hughes} 
\affiliation{Department of Physics and Anthony J. Leggett Institute for Condensed Matter Theory, University of Illinois at Urbana-Champaign, Urbana, IL 61801, USA}

\begin{abstract}
    The continuum limit is a widely used theoretical construct for obtaining continuum field-theories of crystalline systems. We study the formulation of a continuum limit for gapped bosonic phases and ask whether their topological properties survive passage to the continuum. Using methods in algebraic topology and category theory, we give a rigorous formulation of the continuum limit and construct a surjective global map relating crystalline topological phases across all finite point-group symmetries to continuum invertible topological phases. Consequently, we find that some lattice phases admit no continuum limit, while distinct lattice phases that do admit a continuum limit can share the same continuum image, hence implying that some lattice topological data can collapse. Conversely, we find that every continuum invertible phase admits a faithful crystalline realization. In addition to the general framework, we apply it to several examples, including studies of a 2D rotation-symmetric phase, higher-order topological phases, and the mixed spin-lattice anomaly of the deconfined quantum critical point between an antiferromagnet and valence bond solid.
\end{abstract}
\date{\today}

\maketitle 
\makeatletter
\let\oldl@subsubsection\l@subsubsection
\renewcommand{\l@subsubsection}[2]{}
\makeatother

\tableofcontents

\section{Introduction}\label{sec:intro}Models of condensed matter systems are commonly organized into lattice descriptions and continuum field theories. To connect the two, physicists often appeal to a continuum limit that produces a field-theoretic description from a microscopic lattice model. However, mapping lattice models to continuum field theories is often fraught with subtleties. Fortunately, gapped phases of matter can provide an explicit setting to answer precise questions about this mapping as their universal low-energy behavior is captured by relatively simple topological quantum field theories (TQFTs). Providing a mathematically precise continuum-limit construction for TQFTs would therefore yield a rigorous continuum-limit statement about the universal, low-energy data of gapped phases. 

Here we study general questions about the connection between gapped systems in the lattice and continuum. To make the continuum limit precise, we model phases using deformation classes of gapped TQFTs. These are equivalence classes of theories connected by smooth parameter changes that preserve partition functions, anomalies, and quantized responses. Furthermore, to narrow our scope we focus on \emph{invertible, bosonic} phases. Invertible topological phases, abbreviated as invTOs, are short range entangled states whose boundaries exhibit quantum anomalies. InvTOs generalize the notion of topological insulators to strongly interacting systems. As a further distinction, an \emph{intrinsic} invertible phase remains non-trivial even without global symmetries, whereas an invertible \emph{symmetry-protected topological} (SPT) phase becomes trivial once a protecting symmetry is broken. For convenience, and to stay close to experimental settings, we further restrict to systems living on \emph{flat} Euclidean space, $\Sigma^{d}=\mathbb{R}^d$ where $d$ is the spatial dimension. 
Various classification schemes have been proposed to organize and classify bosonic invTOs and SPTs, see e.g.~\cite{Chen2013, ChenScience, Kapustin2014, Kapustin2015, GuXGW2014, Freed2014, Freed2021, WangGu2018, Huang2017}. 
Among these schemes, the group cohomology approach~\cite{Chen2013, ChenScience} classifies bosonic systems on $d$-dimensional  Euclidean space $\mathbb{R}^{d}$ with internal or spatial\cite{ThorngrenElse} symmetry $G$ via the cohomology group:
\begin{equation}
    H^{d+1}(G, \U(1)) \cong H^{d+2}(BG, \mathbb{Z}),
    \label{eq:grpcohclass}
\end{equation}
\noindent where $BG$ is a mathematical object called the classifying space of symmetry $G$, for more details see App.~\ref{app:grpcoh}. However, this classification cannot detect intrinsic invTOs since the classification cohomology group is trivial in the absence of symmetry: if $G=\{1\}$, then $BG=\{\mathrm{pt}\}$ (where pt means a single point) and $H^{d+2}(BG,\mathbb{Z})=H^{d+2}(\{\mathrm{pt}\},\mathbb{Z})=0$.  
In contrast, more complete classifications are given by cobordism theory~\cite{Kapustin2014, Kapustin2015, Freed2014, Freed2021}.  
For a system on $d$-dimensional  Euclidean space $\mathbb{R}^{d}$ with spatial symmetry $G$, the cobordism classification is given by the cobordism group 
\begin{equation}
    \Omega_{O}^{d+1}(BG) = \mathrm{Hom}(\Omega^{O}_{d+1}(BG), \U(1)),
    \label{eq:cobclass}
\end{equation}
see App.~\ref{app:cob} for definitions.  
Unlike group cohomology, the cobordism group can be  nontrivial even in the absence of symmetry  ($G=\{1\}$). For example, when $d=3$ one finds $\Omega^{4}_{O}(B\{1\})\cong\mathbb{Z}_{2}^{2}$, corresponding to two non-trivial generators, and hence four distinct phases. Since much of what we describe below is agnostic to the choice of classification scheme, we package both classification schemes into a single symbol
\begin{equation*}
   \Theta^{d+1}(BG)\;:=\;
   \begin{cases}
     \Omega^{d+1}_{O}(BG) & \text{(cobordism)},\\[2pt]
     H^{d+2}(BG,\mathbb Z) & \text{(group cohomology)}.
   \end{cases}
\end{equation*}
Throughout, $\Theta^{d+1}(BG)$ will refer to the chosen classification model, with the specific choice stated explicitly whenever necessary.

As mentioned above, topological phases of matter often appear in two distinct settings, i.e., lattice models or continuum field theories. For this article we are primarily interested in connections between crystalline systems having point-group or space-group symmetries and their continuum counterparts.  
In Euclidean flat-space, such crystalline symmetry-protected topological phases (cSPTs) in $d$ spatial dimensions are classified by $\Theta^{d+1}(BG)$, where $G$ is the finite symmetry group determined by the lattice structure~\cite{ThorngrenElse, Huang2017}.  
Taking the continuum limit removes all microscopic length scales and yields a genuine infrared (IR) theory.  
In this limit, the original lattice point-group symmetries embed into the emergent continuum symmetry, usually taken to be the orthogonal group $\mathrm{O}(d)$.  
The classification of continuum SPT phases is then given by $\Theta^{d+1}(B\mathrm{O}(d))$~\cite{Fradkin2014, Fradkin2015, Hung2013, Hung2013_2, Rao2023, Gromov2014, Gromov2016}.\footnote{When an internal symmetry is present and decoupled from the spatial data~\cite{note_decouple}, we replace $G\subset \mathrm{O}(d)$ by $H\rtimes_{\rho} G$ throughout; where we allow for spatial operations $G\leq \mathrm{O}(d)$ to act on $H$ via an action $\rho: G \to \mathrm{Aut}(H)$. For readability we suppress $H$ and reinstate it when stating our main results.} Note that continuum theories in which the spatial symmetry is \textit{explicitly} broken down to a subgroup are \textit{derived} theories in the sense that they are obtained from the full continuum classification group $\Theta^{d+1}(B\mathrm{O}(d))$ by the operation called symmetry restriction, which we will define in Sec.~\ref{sec:motivation}; see App.~\ref{app:why_Od_continuum} for a detailed discussion of why $\mathrm{O}(d)$ is taken as the intrinsic continuum spatial symmetry. Hence, within the continuum framework adopted in this work, the IR classification is given by the group $\Theta^{d+1}(B\mathrm{O}(d))$.

\begin{figure}[H]
    \centering
    \hspace*{-0.75cm}%
    \includegraphics[width=1.0\linewidth]{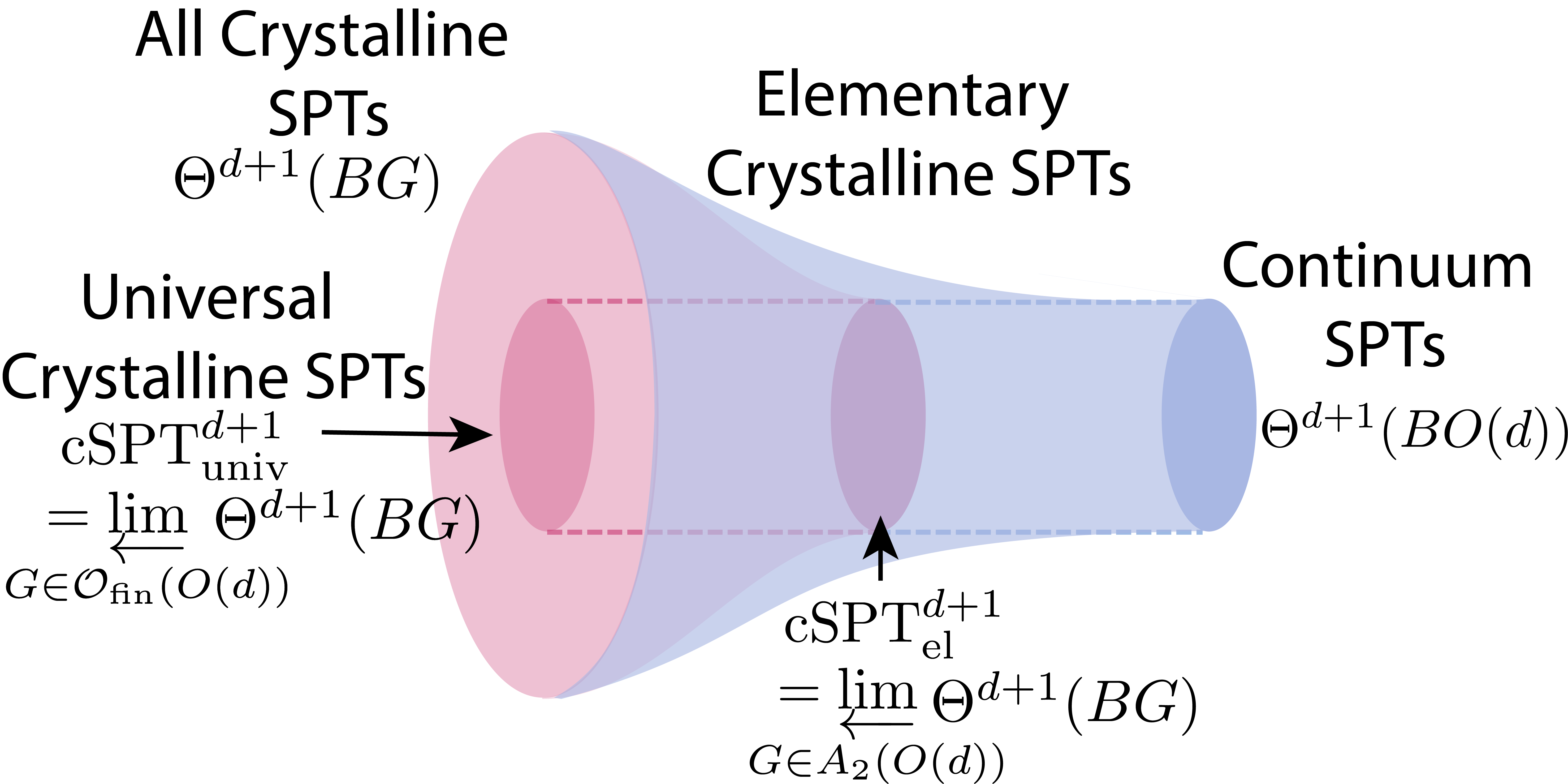}
    \caption{Schematic “funnel” for the stratification theorem.  
    The classification of crystalline SPTs across all finite symmetry groups (left) maps surjectively onto the continuum classification (right) via the stratification map.  
    }
    \label{fig:funnel}
\end{figure}

While the embedding of a point group $G$ into $\mathrm{O}(d)$ is understood at the level of symmetry, the precise relationship between the phases classified by $\Theta^{d+1}(BG)$ and $\Theta^{d+1}(B\mathrm{O}(d))$ remains largely unexplored. This raises two natural questions: does the intrinsic topological data of a crystalline phase always survive in the continuum limit, or can information be lost? And conversely, does every continuum topological phase admit a lattice realization, or do some have no microscopic counterpart at all?  
Put succinctly, our goal is to construct a well-defined map
\begin{equation*}
    \{\text{Crystalline SPTs}\} \longrightarrow \{\text{Continuum SPTs}\},
\end{equation*}
and to understand its structural properties. Our main result demonstrates that such a map always exists and has two key properties: it is \emph{surjective}, so every continuum SPT phase arises from some crystalline phase; but it is generally \emph{not injective}, meaning that distinct crystalline phases can collapse to the same continuum phase, and some microscopic lattice information is inevitably lost. 

\section{Statement of Main Theorem}
\label{sec:maintheorem}
As illustrated in Fig.~\ref{fig:funnel}, our approach to answering these questions will be to organize crystalline classifications into a universal object and then funnel this object to the continuum classification via a stratification map, constructed in Sec.~\ref{sec:rigor}. 
The remainder of this section collects some remarks and definitions that we need to explain the objects in Fig. \ref{fig:funnel}, and to state our main result and some immediate corollaries. The remaining manuscript will then further introduce the necessary framework to explain and prove our main result and provide some important illustrative examples and applications of our results.

First, let us motivate the symmetry context we study. Let $\mathcal{O}_{\mathrm{fin}}(\On(d))$ denote the collection of all finite subgroups $G\subset \On(d)$ which will serve as the point-group symmetries of $(d{+}1)$-dimensional crystalline systems.  Throughout we will restrict to \emph{point groups}: the lattice symmetry collections we consider include finite $G\subset \mathrm{O}(d)$ but exclude lattice translations. This is deliberate because in the continuum, lattice translations $\mathbb{Z}^{d}$ embed into the continuous group $\mathbb{R}^{d}$ whose classifying space is contractible, so
$\Theta^{d+1}\bigl(B(\mathbb{R}^{d}\rtimes \mathrm{O}(d))\bigr)\simeq \Theta^{d+1}(B\mathrm{O}(d))$. Consequently, phases whose protection essentially uses translations (e.g., the family of “weak” SPTs) have trivial image in the continuum classification and will be thus ignored in our analysis.\footnote{Generally speaking, weak SPTs lie outside the scope of our Euclidean space assumption. Considering the spatial manifold to be a torus $\mathbb{T}^d$ instead of $\mathbb{R}^d$ could reveal more features of translation-protected SPT.\cite{kitaev2009}} 
\footnote{There is growing evidence that, in certain regimes, translation can feed nontrivially into the infrared via additional continuum structures (“emanant” symmetries)~\cite{Seiberg_Shao}. Accounting for such effects would require enriching the background beyond $B\mathrm{O}(d)$, which we do not pursue here; our results concern the translation–free, point–group symmetries.} Since we consider only point group symmetries, “crystalline” systems can hence include aperiodic discrete structures such as quasicrystals that have point-group symmetry but not translation symmetry.

In addition to spatial symmetries, we let $H$ denote an \emph{onsite internal} symmetry that is decoupled from the spatial structure (see def.~\ref{def:decoupled_internal} in Sec.~\ref{sec:rigor}). We allow spatial operations $G\le \mathrm{O}(d)$ to act on $H$ via a homomorphism $\rho:G\to\mathrm{Aut}(H)$, so the full symmetry is $\Gamma:=H\rtimes_{\rho}G$. For each crystalline symmetry $G$, cSPT phases protected by $H\rtimes_{\rho} G$ are classified by $\Theta^{d+1}\bigl(B(H\rtimes_{\rho} G)\bigr)$. 
\bigskip 

Turning back to Fig. \ref{fig:funnel}, on the lattice (i.e. left) side of the map we assemble deformation classes of point-group symmetric phases into the \emph{universal} cSPT
\begin{equation}
\mathrm{cSPT}^{d+1}_{\mathrm{univ}, H}
:= \varprojlim_{G\in \mathcal{O}_{\mathrm{fin}}(\On(d))}\Theta^{d+1}\bigl(B(H\rtimes_{\rho} G)\bigr),
\label{eq:universal_cSPTH}
\end{equation}
which can be understood as an abelian group of equivalence classes of crystalline SPTs, with each class representing a set of physically equivalent cSPTs. For now this heuristic understanding is sufficient, and we clarify the notation and precise relations which are used to define the equivalence classes and the concept of an ``inverse limit" (the RHS of Eq. \ref{eq:universal_cSPTH}) in Secs.~\ref{sec:orbit_categories},~\ref{sec:inverselimit}. We will demonstrate that each individual crystalline topological phase that admits a continuum limit naturally belongs to this universal family. For brevity, and without loss of generality, we suppress the internal symmetry factor $H$ in what follows. This is reasonable because we will see that the constructions discussed in this paper depend explicitly on the \emph{spatial} symmetry classification, while a decoupled internal symmetry enters only through the fixed fiber factor $R^{*}:=H^{*}(BH;\mathbb Z_{2})$, see Sec.~\ref{sec:internal_sym}.
\footnote{More precisely, under Hypotheses~\ref{hyp:1}-\ref{hyp:2} of Sec.~\ref{sec:internal_sym}, we have mod-$2$ Leray-Hirsch splittings
\[
\Lambda_{G}: H^{*}(BG;\mathbb Z_{2})\otimes R^{*}
\;\xrightarrow{\;\cong\;}\;
H^{*}\bigl(B(H\rtimes_{\rho}G);\mathbb Z_{2}\bigr).
\]
which makes the $H$-enriched restriction and stratification statements reduce to the purely spatial ones with the fixed auxiliary factor $R^{*}$. See Sec.~\ref{sec:internal_sym} and App.~\ref{app:canon_factor_H}.}

On the continuum side of the map  (i.e., right side of Fig. \ref{fig:funnel}) we focus on TQFT continuum limits. In line with standard usage in the literature~\cite{Atiyah1988, Baez1995, Lurie, Freed2014}, the term TQFT is understood in this paper to mean a diffeomorphism-invariant IR theory whose partition function depends on \textit{only} the standard tangential structure appropriate to the class of systems under study, i.e., $\mathrm{O}$-structure or $\SO$-structure for bosons, and $\mathrm{Spin}$-structure or $\mathrm{Pin}^{\pm}$-structure for fermions, and  no additional geometric choices.  
In contrast, integer-quantized (``free") responses, such as Chern-Simons terms and certain $\eta$-invariant terms, are only \emph{almost} topological\footnote{\label{ft:almost}By almost topological, we mean theories whose low-energy effective theories are \textit{not} topological, but have mild metric dependence (metric tensor proportional to identity).
For a detailed discussion see Sec.5.4~\cite{Freed2021}, and Sec.1.1~\cite{Gaiotto2016}. An example is quantum Chern-Simons theory where it is known that quantization makes the geometric dependence explicit~\cite{Witten1989}. The differential refinement required by free integral response terms in the present framework is discussed in App.~\ref{app:diff_refinement}.}: they do not introduce propagating bulk degrees of freedom, but their partition functions can retain a mild dependence on geometric data through an invertible response term. For example, upon quantization Chern-Simons theory exhibits a gravitational framing anomaly, so its partition function is well-defined only after choosing extra tangential/geometric data, such as a framing~\cite{Witten1989, Atiyah1990, Reshetikhin1990}. Integer-quantized free responses therefore lie outside the notion of TQFT classification adopted here, which does not retain such differential or additional smooth geometric data~\cite{note_free}. 

\begin{figure}[H]
    \centering
    \includegraphics[width=0.65\linewidth]{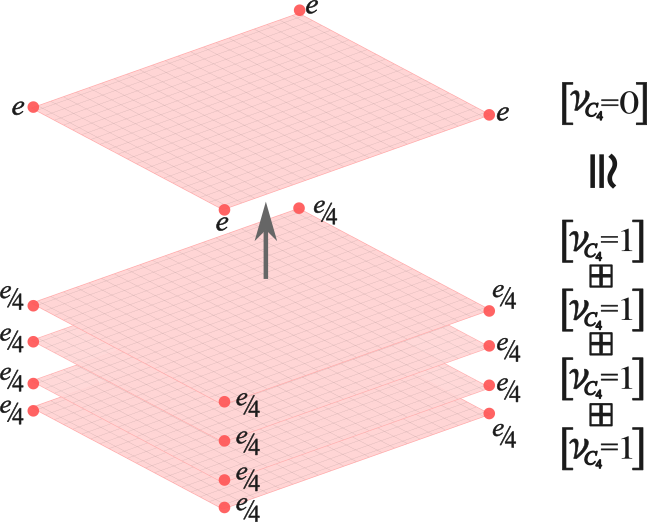}
    \caption{A single $C_{4}$‐symmetric HOTI layer carries fractional corner charge $q_{\mathrm{corner}}=e/4\,(\mathrm{mod} \,e)$ with indicator $\nu_{C_{4}}=1\in \Z_{4}$, and is denoted by $[\nu_{C_{4}}=1]$. Stacking four layers adds the corner charges to an integer $e$, yielding a trivial insulator. This realizes $[\nu_{C_{4}}=1]^{\boxplus 4}=0$, i.e.,\ a $p=2$ torsion class with exponent $k=2$ with $\nu_{C_{4}}=0$.}
    \label{fig:HOTI_stack}
\end{figure}

Accordingly, we focus exclusively on torsion TQFT data—those phases, classified by finite groups, that become trivial after stacking finitely many copies (see Fig. \ref{fig:HOTI_stack}). Explicitly, a cSPT phase $[x]\in\Theta^{d+1}(BG)$ is called $p$–torsion if stacking $p^k$ copies, for some $k\ge 0,$ returns the trivial phase. More explicitly, for $p$-torsion phases there exists $k\ge 0$ such that 
\begin{equation*}
\underbrace{[x]\boxplus \dots \boxplus [x]}_{p^k\; \mathrm{times}}= p^{k}\cdot [x]=[0],
\end{equation*} where $[0]$ denotes the class of trivial insulators, and $\boxplus$ denotes stacking of independent/decoupled physical copies. In Fig.~\ref{fig:HOTI_stack} we show an example of a $2$-torsion phase where stacking four copies of the same phase yields the trivial phase, i.e., $p=k=2$. Note that $p$-torsion does not necessarily imply a phase is of \emph{order}-$p$, that is $p\cdot [x]=0$, and we use the terms order-$p$ phases and $p$-torsion phases to mean different things. We remark that free responses can still enter the torsion classification in reduced form: for example, an orientation-reversing symmetry (e.g., a $\mathbb{Z}_{2}$ reflection) quantizes a Witten–$\theta$ term to $\theta\in\{0,\pi\}$, yielding a $\mathbb{Z}_{2}$ quantized Chern-Simons term on its boundary (see Refs.~\cite{qi2008,hughes2011,turner2012, Vishwanath2013}). Such terms can also appear via decorated-domain-wall constructions, where free responses decorate symmetry defects and, when combined multiplicatively with a torsion phase, produce a torsional response because $\Z\otimes \Z_p = \Z_p$ (see example in Sec.~\ref{sec:example_2_highertorsion} and Refs.~\cite{spatial_symm_3}).

Thus, we consider TQFTs corresponding to the torsion subgroups of the classifications since they are well-defined diffeomorphism-invariant theories whose definition requires no extra tangential structure like framing, metric, or differential refinements of connections (see App.~\ref{app:diff_refinement}). Both the lattice and continuum classification models can contain finite-order torsion classification phases.  With this understanding, we can take any abelian group $X$ and define $\tor_{p}(X)$ to be its $p$–torsion subgroup. In our context we can extract the $p$–torsion universal object on the crystalline side of the classification\footnote{By definition, $\tor_p(\mathrm{cSPT}^{d+1}_{\mathrm{univ}})$ denotes the inverse limit of the $p$-torsion subgroups in Eq.~\eqref{eq:univSPT}; we do not assume $\tor_p$ commutes with the inverse limit.}
\begin{equation}
    \tor_{p}\bigl(\mathrm{cSPT}^{d+1}_{\mathrm{univ}}\bigr)
    :=\;
    \varprojlim_{G\in\mathcal{O}_{\mathrm{fin}}(\mathrm{O}(d))}\tor_{p}\bigl(\Theta^{d+1}(BG)\bigr),
    \label{eq:univSPT}
\end{equation}
\noindent which is the $p$-torsion sector of the universal crystalline classification.
For lattice systems, this universal cSPT classification can be non-trivial, in general, for all primes $p$. 

In contrast, on the continuum side the torsion is entirely $2$-primary (see App.~\ref{app:no_odd_torsion}).
In particular, for bosonic systems one has $\tor_{p}\bigl(H^{d+1}(B\mathrm{O}(d))\bigr)=0$ and $\tor_{p}\bigl(\Omega^{d+1}(B\mathrm{O}(d))\bigr)=0$
for every odd prime $p$.\footnote{For bosons on oriented manifolds the same statement holds with $B\mathrm{O}(d)$ replaced by $B\SO(d)$.
Moreover, the analogous continuum classifications for fermionic systems (built from $\mathrm{Spin}$ or $\mathrm{Pin}^{\pm}$ structures) also seem to have no odd-primary torsion; see App.~\ref{app:no_odd_torsion}.} This gives the following elementary consequence, proved formally in App.~\ref{app:odd_torsion_continuum_limit}.

\begin{proposition}[Odd-primary torsion has no nontrivial continuum image]
\label{prop:main_odd_torsion_collapse}
Let $p$ be an odd prime. In the continuum classification models considered here, odd-primary torsion vanishes:
\begin{equation}
\tor_p\bigl(\Theta^{d+1}(B\mathrm{O}(d))\bigr)=0, \quad \forall \;p\;\text{odd}.
\end{equation}
Therefore any odd $p$-primary torsion lattice class which admits a continuum TQFT limit has trivial continuum image. Equivalently, odd-primary torsion lattice phases either lie outside the continuum-admitting domain, or else collapse to the trivial continuum TQFT.
\end{proposition}

Consequently, odd-primary torsion plays no role in the nontrivial continuum TQFT classification considered in this work. We therefore restrict to the $2$-primary sector from now on. This restriction will help us form the continuum limit from crystalline SPT phases via the construction of the \emph{stratification map},
\begin{equation*}
    F_{2}:\;\tor_{2}\bigl(\mathrm{cSPT}^{d+1}_{\mathrm{univ}}\bigr) \longrightarrow \tor_{2}\bigl(\Theta^{d+1}(B\mathrm{O}(d))\bigr),
\end{equation*}
which maps a compatible family (see Sec.~\ref{sec:inverselimit})
of crystalline phases to their continuum limit. The precise algebraic and categorical details of this map, as well as its construction, are provided in Secs.~\ref{sec:motivation}–\ref{sec:rigor}. Intuitively, $F_{2}$ extracts only the information that survives in the continuum from the entire crystalline classification that includes all finite subgroups of $\mathrm{O}(d)$.

With this heuristic understanding we can state our main result:

\medskip

\begin{theorem}[\textbf{Main theorem ($p=2$)}]
\label{theorem:main}
For every spatial dimension $d\ge 1$,
the stratification map $F_{2}$ is surjective.
Explicitly,
\begin{equation}
   F_{2}:
        \tor_{2}\bigl(\csptu\bigr)
   \;\twoheadrightarrow\;
   \tor_{2}\bigl(\Theta^{d+1}(B\mathrm{O}(d))\bigr).
   \label{eq:maintheorem}
\end{equation}
Moreover, in the presence of a decoupled onsite internal symmetry $H$ (in the sense of Def.~\ref{def:decoupled_internal}) satisfying mild assumptions (Hypotheses~\ref{hyp:1},~\ref{hyp:2} in App.~\ref{app:proof_internal}, see also Sec.~\ref{sec:internal_sym}), on which the spatial symmetry acts by $\rho:G\to \mathrm{Aut}(H)$, the same stratification map extends to a surjection:
\begin{equation}
F_{2,H}:
\tor_{2}\bigl( \mathrm{cSPT}^{d+1}_{\mathrm{univ}, H} \bigr)
\;\twoheadrightarrow\;
\tor_{2}\left(\Theta^{d+1} \bigl(B(H\rtimes_{\rho} \mathrm{O}(d))\bigr)\right).
\label{eq:maintheorem_H}
\end{equation}
\noindent For each lattice symmetry $G$, this universal construction determines a canonical map
\begin{equation*}
\pi_H:\tor_{2}\left(\mathrm{cSPT}^{d+1}_{\mathrm{univ},H}\right)\longrightarrow \tor_{2}\left(\Theta^{d+1}\!\bigl(B(H\rtimes_{\rho}G)\bigr)\right).
\end{equation*}
Its image $\im(\pi_H)\subseteq \Theta^{d+1}\bigl(B(H\rtimes_{\rho}G)\bigr)$ is the subgroup of $G$-lattice TQFT classes which admit a continuum description arising from the universal limit construction.
\end{theorem}
\medskip
\noindent\textbf{Corollaries.}
\begin{enumerate}
\item \emph{Lattice realizability.}
      Every continuum invertible phase contains at least one crystalline
      representative whose boundary anomaly, partition function, and
      cobordism (or cohomology) class agree with those of the continuum
      theory.
\item \emph{Continuum limit admissibility and collapse of lattice physics.}
The converse of lattice realizability is false in general. For a fixed lattice
symmetry $G\leq \mathrm{O}(d)$, only the subgroup
\begin{equation}
\im(\pi_H)
\subseteq
\tor_2\!\left(\Theta^{d+1}\!\bigl(B(H\rtimes_{\rho}G)\bigr)\right)
\end{equation}
admits continuum limit descriptions arising from the universal construction. Lattice classes outside this subgroup are intrinsically crystalline and have no continuum TQFT image. Moreover, the continuum limit need not be faithful: distinct lattice classes in $\mathrm{cSPT}^{d+1}_{\mathrm{univ},H}$ whose difference lies in $\ker(F_{2,H})$ have the same continuum image. Thus the continuum limit both excludes some intrinsically crystalline classes and collapses some lattice distinctions.
\end{enumerate}

\medskip

As noted above, the continuum classification is completely 2-torsion; odd-prime lattice torsion collapses to zero. Moreover, the surviving $2$-primary sector has order-$2$, i.e., $[x]\boxplus [x]=0$, and there are no elements of order $2^{k}$ with $k>1$. Consequently, (i) odd–prime torsion phases do not survive the continuum limit (see App.~\ref{app:no_odd_torsion} for proof), and (ii)  order-$2^{k}$ torsion phases (with $k>1$), which  include some higher order SPTs for example,  collapse to the trivial class in the continuum classification. We give explicit examples demonstrating these consequences in Sec.~\ref{sec:examples}.

Now that we have completed the heuristic discussion of our main result, the remainder of this article is focused on making these concepts precise and providing some illustrative examples. The article is structured as follows: Section~\ref{sec:basics} recalls Atiyah’s axioms for TQFTs, fixes conventions, and motivates the classification approaches to crystalline phases. Section~\ref{sec:lattice_compatible_continuum_limit} introduces the notion of a continuum limit as a map and develops the physical intuition behind it. Section~\ref{sec:RG_flow} provides a microscopic motivation for these maps from RG/EFT flows. Section~\ref{sec:orbit_categories} defines orbit categories of crystalline symmetries which will be used to construct the universal cSPT groups. Section~\ref{sec:inverselimit} defines the universal crystalline SPT, $\mathrm{cSPT}^{d+1}_{\mathrm{univ}}$, as an inverse limit over the orbit category. We then provide a ground-up/first-principle definition of a continuum limit in Sec.~\ref{sec:continuum_limit} and show how it naturally recovers the universal cSPT construction. Section~\ref{sec:rigor} proves the stratification theorem as stated in the main theorem~\ref{theorem:main} for the cohomology model, with the proof for the cobordism based classification model deferred to App.~\ref{app:cob_proof}. Section~\ref{sec:examples} presents three representative case studies illustrating how our framework organizes lattice invertible phases and their continuum limits, including an explicit $(2{+}1)$D crystalline response with boundary inflow, a higher-torsion family that becomes invisible in the continuum, and an application to the Antiferromagnet-Valence bond solid transition and its possible mixed anomaly. Appendix~\ref{app:tqft} goes over Segal's axioms for TQFTs. Appendices~\ref{app:cob}, \ref{app:grpcoh} and \ref{app:grpcoh2} collect background on the two classification frameworks, and Appendices ~\ref{app:groundup_continuumlimit},~\ref{app:proofs1},~\ref{app:proofs2}.~\ref{app:canon_factor_H}.~\ref{app:cob_proof},~\ref{app:so_str} contain a collection of proofs which support the conclusions in the main text.

\section{Preliminaries: Classification of TQFTs and Symmetry Backgrounds}\label{sec:basics} In this section we recall the basic ingredients of TQFTs, extend them to incorporate global symmetries, and set up the classification models.

\bigskip 
\subsection{Axiomatic TQFTs}
\label{sec:axiomaticTQFTs}

\begin{figure}[H]
    \centering
    \includegraphics[width=0.7\linewidth]{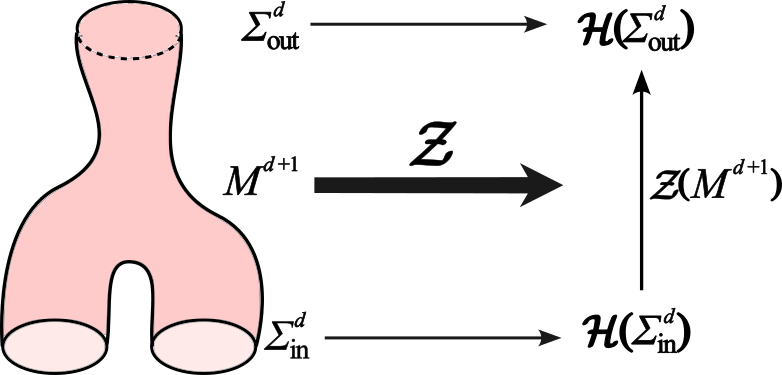}
    \caption{Pictorial description of a TQFT functor.}
    \label{fig:bordism}
\end{figure}

A topological quantum field theory (TQFT) is a local effective theory describing the universal long-distance limit of a gapped system, in which observables are insensitive to smooth changes of the metric and depend on only the topology of spacetime (and any specified background structures). More formally, a TQFT assigns: (i) a finite-dimensional complex vector space of states, denoted $\mathcal H(\Sigma^{d})$, to every spatial slice $\Sigma^{d},$ and (ii) a transition amplitude $\mathcal{Z}(M^{d+1})$ to each spacetime region $M^{d+1}$ that interpolates between two spatial slices. This transition amplitude is realized as a linear evolution map $\mathcal{Z}(M^{d+1}): \mathcal H(\Sigma^{d}_{\mathrm{in}})\to \mathcal H(\Sigma^{d}_{\mathrm{out}})$ which describes how states evolve from the incoming to the outgoing boundary, as illustrated in Fig.~\ref{fig:bordism}.
 
 For a TQFT, transition amplitudes $\mathcal{Z}(M^{d+1})$ depend on only topological information: deforming $M^{d+1}$ smoothly without cutting or gluing leaves $\mathcal Z(M^{d+1})$ unchanged. Mathematically, this structure is summarized by the Atiyah–Segal axioms (see Def.~\ref{def:AtiyahSegal}), according to which a $(d{+}1)$-dimensional TQFT is a symmetric monoidal functor 
\begin{equation*}
    \mathcal{Z}: \bordd \;\longrightarrow\; \mathbf{Vect}^{\mathrm{fd}}_{\mathbb{C}},
\end{equation*}
where $\bordd$ is the category whose objects are closed $d$-dimensional manifolds (the spatial slices) and whose morphisms are $(d{+}1)$-dimensional manifolds interpolating between them (the spacetime regions). The target category of the TQFT functor is that of finite dimensional complex vector spaces, denoted $\mathbf{Vect}^{\mathrm{fd}}_{\mathbb{C}}$, whose morphisms are linear maps. “Symmetric–monoidal” means that $\mathcal Z$ respects the two operations in $\bordd$: disjoint union and gluing.  Disjoint union $\sqcup$ is sent to tensor product $\otimes$ in $\mathbf{Vect}^{\mathrm{fd}}_{\mathbb{C}}$. The identity objects are related accordingly: the point $\{\mathrm{pt}\}$ is the unit for disjoint union in $\bordd$, while $\mathbb{C}$ is the unit for tensor product in $\mathbf{Vect}^{\mathrm{fd}}_{\mathbb{C}}$, and the functor satisfies $\mathcal H(\{\mathrm{pt}\})=\mathbb{C}$. Gluing of bordisms is sent to composition of linear maps.\footnote{One must also fix a \emph{tangential structure} $\xi$ on the bordisms (e.g., $\mathrm{O}$, $\mathrm{SO}$, $\mathrm{Spin}$, $\mathrm{Pin}^{\pm}$, framing), which specifies the allowed manifolds and gluings. The TQFTs and their classification depend on $\xi$ (e.g., in the cobordism model one obtains $\Omega^{d+1}_{\xi}(BG)$). In this article we take $\xi=\mathrm{O}$ (unoriented) for definiteness.} This intuitive picture is all we use in the main text; the formal details appear in Appendix~\ref{app:tqft}.

\subsection{Symmetry background and crystalline equivalence principle}
\label{sec:symBG}

\begin{figure}[H]
    \centering
    \includegraphics[width=0.95\linewidth]{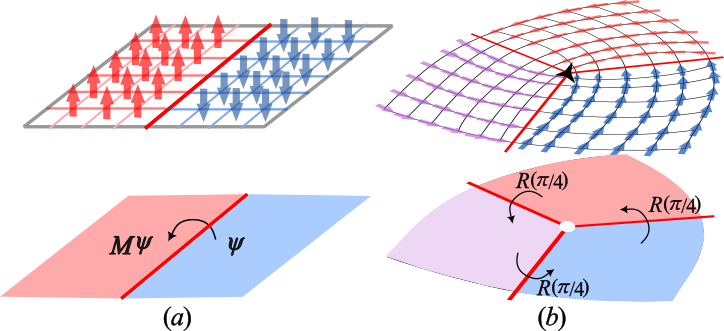}
    \caption{(a) Codimension-1 domain walls for orientation-reversing mirror symmetry $M$. (b) Codimension-2 disclinations defect of discrete rotation symmetry $G=C_{4}$. Non-trivial holonomy around a loop encircling the disclination implies it is a source of curvature. Bottom panels encode the physical defect configurations using triangulations with edges labelled with group elements.}
    \label{fig:defect}
\end{figure}

To describe systems with global symmetries, we begin by formalizing the notion of \emph{symmetry defects}.  The simplest such defects are \emph{domain walls}: codimension-one surfaces across which the system undergoes a symmetry transformation. More generally, several such walls can meet along higher-codimension junctions and produce higher-codimension defects detected by nontrivial holonomy around a small linking cycle.\footnote{A linking cycle is the boundary of a small disk in the directions normal to the defect; for example, a loop encircling a codimension-two defect, a sphere enclosing a codimension-three defect, and so on.} Typical examples include codimension-two vortices or disclinations, or codimension-three monopoles.
Figure~\ref{fig:defect} illustrates two representative examples of symmetry defects for discrete  spatial symmetries.

\begin{figure}[H]
    \centering    \includegraphics[width=0.99\linewidth]{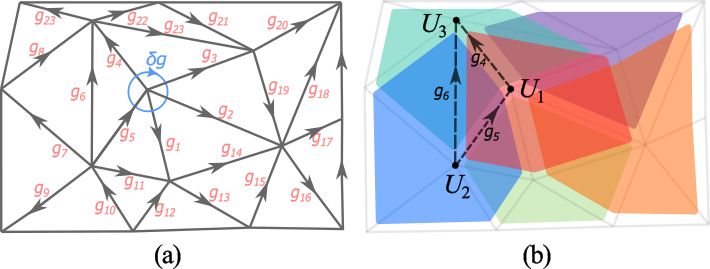}
    \caption{(a) A patch of a triangulated manifold with locally oriented edges labeled by group elements $g_{e}\in G$.  Consistency requires that the ordered product of labels along any contractible loop in the $1$-skeleton equals the identity. 
    (b) Illustrates the dual cellulation of the triangulation. The same consistency condition applied to every face ($2$-simplex above) leads to the Čech cocycle condition on the topology defined by open covers $\{U_{\alpha}\}$ given by polytopes in the dual lattice(or cellulation). For example, in the highlighted example we get $g_{5}\cdot g_{4}=g_{6}$.}
    \label{fig:graph}
\end{figure}

Mathematically, when higher–codimension defect cores are present, we excise (or ``delete") a small tubular neighborhood $D\subset M^{d+1}$ of those defect cores and work on the complement 
\begin{equation*}
    M^{\circ}\;:=\; M^{d+1}\setminus D.
\end{equation*}
For example, a disclination core in $d=2$ would have a small disk excised as marked in Fig.~\ref{fig:defect}(b).
A defect configuration on $M^\circ$ is then described by choosing a triangulation $M^\circ_{\Delta}$ and assigning group elements to branched, oriented edges, recording the symmetry transformation across each domain wall. This defines a simplicial $1$-cochain $g \in C^{1}(M^\circ_{\Delta},G).$ We impose a consistency condition that the ordered product of edge labels around every \emph{contractible} loop along edges in the $1$-skeleton of $M^{\circ}_{\Delta}$ is the identity. Equivalently, the labeling is a simplicial $1$-cocycle, $g\in Z^{1}(M^\circ_{\Delta},G)$.
This guarantees consistency of the domain-wall junctions; see Fig.~\ref{fig:graph}.

The domain wall defect picture serves as a good intuitive tool, but keeping track of triangulations and symmetry labels quickly becomes cumbersome. A convenient way to organize defect configurations is to repackage the edge labels as the gluing data of a single object: a principal $G$-bundle $P\to M^{\circ}$ (see Def.~\cite{note_defPGB}).\footnote{Refining the triangulation or performing a vertex relabeling (“gauge move”) changes only the presentation, not the underlying bundle $P$.} To see that the defect construction using simplicial cochains also defines a principal $G$-bundle, 
choose an open cover $\{U_{\alpha}\}$ of $M^{\circ}$ given by the faces of the dual cellulation shown in Fig.~\ref{fig:graph}(b). Whenever two neighboring patches $U_{\alpha}$ and $U_{\beta}$ overlap there exists an edge in the original simplex connecting the centers of the dual cells, the group label of which determines the transition function
\begin{equation*}
    g_{\alpha\beta}: U_{\alpha}\cap U_{\beta} \to G.
\end{equation*}
For each original $2$-simplex, the corresponding three patches have a non-empty triple overlap $U_{\alpha}\cap U_{\beta}\cap U_{\gamma}\neq 0$, and the local consistency condition is precisely the Čech cocycle relation $g_{\alpha\beta}g_{\beta\gamma}g_{\gamma\alpha}=1$. Thus, the collection $\{g_{\alpha\beta}\}$ defines a Čech $1$-cocycle and glues the local trivializations into a principal $G$-bundle $P\to M^{\circ}$\footnote{More precisely, isomorphism classes of principal $G$-bundles over $M^{\circ}$ are classified by the first Čech cohomology $\check{H}^{1}(M;G)$, see Chapter IV of Ref.~\cite{Bredon1997}.}. 

\bigskip 

For the finite groups considered in the discrete defect picture, the transition functions are locally constant, so the associated principal $G$-bundle is equivalently a topologically flat bundle. In this case one may also phrase the background in terms of its holonomy: the ordered product of transition functions around any contractible loop is trivial. Equivalently, there is no nontrivial $G$-flux through an ordinary domain-wall junction, since a small loop encircling such a junction is contractible. Nontrivial $G$-flux can occur only around noncontractible linking cycles, such as loops encircling excised defect cores. 
This is the flatness condition illustrated by the blue loop $\delta g=1$ in Fig.~\ref{fig:graph}(a). It should be distinguished from the \v{C}ech cocycle condition above: the latter is the local gluing condition on triple overlaps, while flatness is the statement that the resulting discrete holonomy vanishes on contractible loops.

\bigskip 

Importantly, every principal $G$-bundle $P  \to  M^{d+1}$ can be conveniently encoded by a continuous \emph{classifying map}
\begin{equation*}
f_{P} : M^{d+1} \longrightarrow BG,
\end{equation*}
where $BG$ is the classifying space of $G$. By definition, $BG$ carries a universal bundle $EG  \to  BG$ which is defined to be the (up to homotopy) unique principal $G$-bundle whose total space $EG$ is contractible~\cite{note_universalbundle}. The classifying space $BG$ can be thought of as a parameter space for all topological sectors of $G$-gauge fields; and defining a map $f_{P}: M\to BG$ is equivalent to determining (or picking out) the underlying bundle as the pullback of the universal bundle $P \cong f_{P}^{*}EG$. Conversely, two maps $f_P,f_Q : M^{d+1} \to BG$ determine isomorphic bundles $P_{f_P} \cong  Q_{f_Q}$ exactly when they are homotopic, i.e., $f_{P}\simeq f_{Q}$.  Hence isomorphism classes of principal $G$-bundles on $M^{d+1}$ are in one-to-one correspondence with homotopy classes of maps $[M^{d+1}\to BG]$(see standard references~\cite{Hatcher_VB, Nakahara2018, Eguchi1980}). In conclusion, we can be assured that a symmetry defect configuration, also called a \textit{symmetry background}, is uniquely specified by a classifying map. 

\medskip

The bundle–classifying space description above is most directly applicable to \emph{internal} symmetries. 
Spatial (crystalline) symmetries require a subtle refinement, because the symmetry now acts on space itself rather than on only internal degrees of freedom. We review the corresponding notion of a spatial gauge field in detail in App.~\ref{app:bordism_spatial_sym}; here we record only the flat-space specialization needed in the main text.
Concretely, for spatial (crystalline) symmetries one must separately keep track of the defect-free underlying space $\Sigma^{d}$ and the group action on it, $\sigma: (g,\,x)\mapsto g\cdot x,\,\forall \, x\in \Sigma^{d}$\footnote{This is because inserting \emph{spatial} defects changes the space itself. In contrast, inserting internal symmetry defects leaves the space unchanged.}. 
The corresponding generalization of the classifying space is the Borel (homotopy–quotient) construction,
\begin{equation*}
\Sigma^{d}_{G} \;\coloneqq\; (EG \times \Sigma^{d})/G,
\end{equation*}
which packages, in a single object, both the principal $G$-bundle data and the $G$-action on $\Sigma^{d}$. In this work we model our systems on flat space $\Sigma^{d}=\mathbb{R}^{d}$, which is contractible and thus has the special property
\begin{equation*}
\Sigma^{d}_{G}\;\simeq\; BG.
\end{equation*}
Therefore, on flat space the Borel construction reduces to the ordinary classifying space, and spatial symmetries can be treated in the same formalism as internal ones; the same simplification holds for spacetimes of the form $M^{d+1}=\mathbb{R}^{d}\times \mathbb{R}_{t}$. This reduction is often referred to as the \emph{crystalline equivalence principle}~\cite{ThorngrenElse}.\footnote{Note that this means a spatial symmetry background is defined by a classifying map $f:M^{d+1} \to \Sigma^{d}_{G}$. One can prove that this map is equivalent to the datum used in Ref.~\cite{ThorngrenElse}, that is a pair $\{\pi: P \to M^{d+1},\; \hat{f}: P \to \Sigma^{d}\}$, to describe \textit{spatial gauge fields}.}

\subsection{$G$-symmetric TQFTs and classification models}
\label{sec:classification_models}

To define $G$-symmetric TQFTs, we can use the bordism category with generalized $G$-backgrounds introduced in Def.~\ref{def:bordism_spatial_BG_background}. In the flat-space setting relevant for this paper, where $\Sigma^{d}=\mathbb{R}^{d}$, the Borel construction simplifies $\Sigma_{G}\simeq BG$. And thus, using the crystalline equivalence principle the generalized bordism category with spatial $G$-backgrounds reduces to the simpler internal symmetry-background bordism category defined in Def.~\ref{def:bordism_BG_background} (see also remark~\ref{remark:int_simplification}), which we denote simply as $\bordd(BG)$ for notational simplicity. A $G$-symmetric TQFT is then a symmetric monoidal functor
\begin{equation*}
    \mathcal{Z}: \bordd(BG) \to \mathbf{Vect}^{\mathrm{fd}}_{\mathbb{C}}, 
\end{equation*}
\noindent where the objects in the domain category are now pairs $(Y^{d},f)$ of a spatial manifold $Y^{d}$ and a classifying map $f:Y^{d}\to BG$ (defined up to homotopy), and morphisms are spacetime bordisms compatible with the background. At this point we observe that, the only input that went into defining the $G$-symmetric TQFT was $BG$\footnote{This is in addition to the tangential structure which specifies the class of manifolds in the domain category, which we assume to be $O$-structure in this paper, see footnote 8.}. 

\bigskip 

We emphasize that throughout the main text we work with only the ordinary un-extended Atiyah-Segal framework, i.e.,\ with closed spatial slices and bordisms between them. If one instead keeps localized higher-codimension defect cores explicit, then excision introduces additional boundary components around the defect locus, i.e., $\partial M^{\circ} = \partial D$, and a full treatment requires extra boundary conditions or, more generally, an extended TQFT framework.\footnote{For example, in the $(2{+}1)$-dimensional finite-group theory of Freed-Quinn~\cite{FreedQuinn} discussed in App.~\ref{app:grpcoh_functor}, this is precisely the additional structure appearing in their treatment of surfaces with boundary, where one first fixes suitable boundary data and obtains a refined gluing law.} We will not use that boundary-level refinement in the main text. Rather, we restrict to the closed-manifold response sector relevant for bulk quantized responses. This should not be read as excluding symmetry defects altogether: global symmetry twists, including the twisted-trace constructions of Ref.~\cite{ThorngrenElse}, remain included because they are encoded by background classifying maps on closed manifolds. Furthermore, it can be shown that kinematic properties such as charges trapped by symmetry defects can be understood from these twisted trace constructions. For completeness, we note that the same global conclusions are expected in the extended TQFT setting as well; App.~\ref{app:cob} proves the main theorem for the corresponding classification of extended invertible TQFTs.

\bigskip 

\noindent \textit{Invertible TQFTs}: Again, we note that we are considering only invertible TQFTs in this article, so let us clarify the definition of invertible TQFTs in this language. Two decoupled topological phases, denoted by $\mathcal Z_{1}$ and $\mathcal Z_{2}$, placed on the same space-time $M$ with the same symmetry background ($f: M\to BG$) combine by \emph{stacking}. We write stacking at the level of theories as $\boxplus$, and reserve $\otimes$ for the usual tensor product on vector spaces/linear maps:
\begin{equation*}
(\mathcal{Z}_{1}\boxplus \mathcal{Z}_{2})(M,f)= \mathcal{Z}_{1}(M,f)\otimes \mathcal{Z}_{2}(M,f).
\end{equation*}
For a closed $(d{+}1)$-manifold $M$ this simply says the partition functions multiply under stacking. A topological phase is invertible (see App.~\ref{app:tqft} for formal definition)
when there exists a second topological phase such that stacking the two gives a trivial
product phase. A trivial product phase is a topologically trivial phase whose ground state can be continuously deformed to a product state, and has trivial partition function $\mathcal{Z}_{\mathrm{triv}}(M, f)=1$ for all $(M,f)$. It can be shown that for invertible TQFTs the Hilbert spaces assigned to closed manifolds  are always 1-dimensional, see App.~\ref{app:tqft}, and so the product of partition functions is multiplication of complex phases. Thus, two topological phases are inverses if their partition functions are complex conjugates of each other.  

The deformation classes of invertible $(d{+}1)$-dimensional TQFTs with symmetry $G$ form an abelian group which is called the classification of invertible $G$-TQFTs, and denoted $\Theta^{d+1}(BG)$. Intuitively, this group classifies all distinct topological response actions, i.e., all possible ways in which a system can couple nontrivially to $G$-background fields while remaining gapped and invertible.

\bigskip 

\noindent \textit{Classification models}: We are now ready to define what we mean by a \textit{classification model} of TQFTs in the context of this work.  
The input to a classification model is a symmetry group $G$. Rather than working directly with $G$, one passes to its classifying space $BG$, because symmetry backgrounds on a spacetime $M$ are specified by maps $f:M\to BG$.  
In this sense, $BG$ is the natural object on which the classification depends, not $G$ itself.  The output of a classification model is an abelian group describing deformation classes/phases of invertible TQFTs with symmetry $G$. Thus, a classification model $\Theta$ is a rule
\begin{equation*}
\Theta^{d+1}:\; BG \longmapsto \Theta^{d+1}(BG)\in \Ab,
\end{equation*}
that assigns to each classifying space $BG$ an abelian group whose elements represent classes of $(d{+}1)$-dimensional invertible TQFTs.  

\bigskip

More generally, relations between symmetry groups, e.g., equivalence under conjugation, also induce relations among their classifications. Therefore, more formally, classification models are functors $\Theta^{d+1}: \mathbf{hoCW}^{\textbf{op}}\to \mathbf{Ab}$ from the homotopy category of CW-complexes, denoted $\mathbf{hoCW}$ (which includes classifying spaces as objects) to the category of abelian groups $\mathbf{Ab}$. The classification functors take (classifying spaces of) groups $G$ to their classification $\Theta^{d+1}(BG)$. As we will see later, relations among groups, e.g., inclusion $i: G\to P$, result in relations among their classifications, e.g., restriction $\res^{P}_{G}: \Theta^{d+1}(BP)\to \Theta^{d+1}(BG)$. For more details see Sec.~\ref{sec:orbit_categories}, and note~\cite{note_functor}. 

\bigskip 
\begin{remark}[Convention]
\label{rmk:convention}
We define classification models in $(d+1)$-dimensions as contravariant functors
\begin{equation*}
    \Theta^{d+1}: {\bf hoCW}^{\mathrm{op}} \to {\bf Ab}.
\end{equation*}
Here ${\bf hoCW}^{\mathrm{op}}$ denotes the opposite category of ${\bf hoCW}$: it has the same objects as ${\bf hoCW}$, but every morphism is reversed. Thus, a map of classifying spaces $f:X\to Y$ induces a restriction map
\begin{equation*}
    f^*:=\Theta^{d+1}(f):
    \Theta^{d+1}(Y)\longrightarrow\Theta^{d+1}(X).
\end{equation*}
We use the notation $f^*$ for the action of the classification functor
on morphisms, in order to distinguish it from its action on objects.
When discussing symmetry groups, we keep the classifying space explicit and write $\Theta^{d+1}(BG)$. More generally, if $\mathcal{C}$ is a category of groups equipped with the classifying space functor $B:\mathcal{C}\to {\bf hoCW}$, then by abuse of notation we also write
\begin{equation*}
    \Theta^{d+1}: \mathcal{C}^{\mathrm{op}} \to {\bf Ab}\;\;\text{for}\;\; \Theta^{d+1}\circ B^{\mathrm{op}}.
\end{equation*}
Thus for symmetry $P\in\mathcal{C}$ we write $\Theta^{d+1}(BP)$ for its classification group, and for a symmetry morphism $i_g:P\to Q$ in $\mathcal{C}$ we write
\begin{equation*}
    (Bi_g)^*:=
    \Theta^{d+1}(Bi_g):
    \Theta^{d+1}(BQ)\longrightarrow\Theta^{d+1}(BP).
\end{equation*}
\end{remark}

\bigskip 

The classification model framework is consistent with the viewpoint—originating with Kitaev and articulated in Ref.~\cite{Gaiotto2019}—that SPT phases are classified by \emph{generalized cohomology theories} (see note~\cite{note_GCT}). That is, one may write $\Theta^{d+1}(BG)\cong E^{d+1}(BG)$ for a suitable generalized cohomology theory $E$. We will not need the full axiomatic framework of generalized cohomology in what follows. Instead, we fix a classification model $\Theta^{d+1}$ and work with two concrete choices: Borel-equivariant cobordism and group cohomology. In what follows, these two frameworks will serve as the main tools for classifying crystalline topological phases. We record their definitions here.

\bigskip 

\begin{definition}[Cobordism classification model]
\label{def:cob_classmodel}
The model assigns to a system with crystalline symmetry $G$ in $(d+1)$-dimensions the classification group:
\begin{equation*}
    \cspt = \Omega^{d+1}_{O}(BG).
\end{equation*}
A concise derivation is given in App.~\ref{app:cob}. Since we are interested in the 2-torsion subgroup of the classification, we will more specifically use the model: 
\begin{equation}    \mathrm{cSPT}^{d+1}_{G,\mathrm{tor}} = \tor_{2}(\Omega^{d+1}_{O}(BG)),
\label{eq:cob_model}
\end{equation}
where given an abelian group $A$, we denote the subgroup of 2-torsion phases as $\tor_{2}(A)$.
\end{definition}

\bigskip

For explicit calculations we will primarily work in the (coarser) group cohomology model, see Refs.\cite{Chen2013,ChenScience,ThorngrenElse}, defined below. Here, we make an important refinement to our earlier description of the group cohomology classification in Eq.~\ref{eq:grpcohclass}: when $G$ contains orientation-reversing elements, the corresponding topological response must be defined with twisted coefficients. To illustrate what this means, we note that an orientation-reversing background, e.g., a domain wall where a reflection is applied, flips the local choice of spacetime orientation. Hence, a topological response density that would ordinarily be integrated as a volume form acquires a sign change under such a transformation over local patches. Accordingly, the relevant group cohomology is valued not in ordinary $\U(1)$ (or $\Z$), but in the sign local system $\U(1)_{w_{1}}$ (or $\Z_{w_{1}}$) determined by the orientation character $w_{1}:G\to \{\pm1\}$ (see App.~\ref{app:grpcoh_def},~\ref{app:twisted_coefficients} for more details).\footnote{Equivalently, the coefficient system keeps track of the sign picked up when locally oriented patches are glued together. When all symmetry elements preserve orientation, the twist is trivial and one recovers ordinary coefficients.}

\bigskip

\begin{definition}[Classification of crystalline topological phases in the cohomology model]
\label{def:grpcoh_classmodel}
The model assigns to a system with crystalline symmetry $G$ in $(d+1)$-dimensions the classification group:
\begin{equation}
    \tcspt = H^{d+1}(G, \U(1)_{w_1}),
    \label{eq:grpcoh_fullclass}
\end{equation}
\noindent where for finite $G$ this is ordinary group cohomology, while for compact Lie group $G$ it is Borel cohomology (see App.~J4(a) of \cite{Chen2013}). Equivalently, one may use the identity $H^{d+1}(G,\U(1)_{w_{1}})\cong H^{d+2}(BG,\mathbb Z_{w_{1}})$ which works in both cases. The mathematical details of how the group cohomology model specifies a class of invertible TQFTs is given in App.~\ref{app:grpcoh}. 

\bigskip 

Restricting attention to the torsion subgroup allows us to use a convenient substitution
\begin{equation}
    \tor(H^{d+1}(G, \U(1)_{w_{1}})) = \tor(H^{d+1}(BG, \U(1)_{w_{1}})),
    \label{eq:tor_equiv_first}
\end{equation}
so that we can work directly with the singular cohomology of the classifying space $BG$. This gives us the cohomology classification model for 2-torsion phases:
\begin{equation}
\widetilde{\mathrm{cSPT}}^{d+1}_{G,\mathrm{tor}} =  \tor_{2}(H^{d+1}(BG, \U(1)_{w_{1}})).
    \label{eq:grpcohtormodel}
\end{equation}
We note here the identification $\tor_{2}(H^{d+1}(G, \U(1)_{w_{1}})) \cong \tor_{2}(H^{d+2}(BG, \Z_{w_{1}}))$ (see Eq.~\ref{eq:G_finite}-~\ref{eq:G_compactLie} in App.~\ref{sec:grpcohmodel_def}) which is often used to explicitly compute classifications in this model.
\end{definition}

\medskip 

\begin{remark}[Pre-TQFT Classification]
Because our main interest is in $2$-torsion phases, it is natural to first organize the analysis at the mod-$2$ level, where the structure is simpler and the physical $\U(1)$ response can later be recovered. Thus, our computations and proofs in either of the two models above will typically begin with an analysis of an auxiliary cohomology classification model with mod-$2$ coefficients: 
\begin{equation}
  H^{d+1}(BG,\mathbb Z_{2}),
\label{eq:preTQFT}
\end{equation}
which we refer to as the \emph{pre-TQFT} classification. The reason is that mod-$2$ cohomology is technically simpler, while still capturing the basic structure of $2$-torsion phases. The corresponding physical $\U(1)$ response is recovered by the (twisted) Bockstein homomorphism
\begin{equation*}
\beta_{w_{1}}:\ H^{d+1}(BG,\mathbb Z_{2})\to H^{d+2}(BG,\Z_{w_{1}}),
\end{equation*}
which lifts mod-$2$ data to genuine TQFT response phases in the full group cohomology model. 
\end{remark}

Thus, the mod-$2$ theory serves as a convenient starting point for studying $2$-torsion phases, even though it contains some redundancy relative to the full twisted integral classification. We review the relevant constructions in App.~\ref{app:lift}. In the following sections, the cohomology model will be our primary workhorse, however, we prove the main theorem for the cobordism model in App.~\ref{app:cob_proof} and expect the analogous statements in the following sections to hold for the cobordism model as well.

\subsection{From lattice and continuum models with matter to TQFTs}
\label{sec:scales}

In this subsection we describe the emergence of invertible TQFTs from gapped microscopic theories. In a lattice model, integrating out gapped degrees of freedom and coarse-graining to a scale $\ell$ leaves a topological functional of discrete background fields; this ``lattice TQFT" should be understood as a mesoscopic effective theory (made more precise below). In a continuum QFT, on the other hand, the corresponding procedure of integrating out the gapped degrees of freedom produces a true IR TQFT whose partition function depends on only the background tangential data on a smooth manifold, independent of any discretization.

\subsubsection{Lattice systems, mesoscopic limits, and lattice classes.}
Consider a microscopic lattice model defined on a lattice embedded in flat Euclidean space $\Lambda \subset \mathbb{R}^{d}$ with lattice spacing $a,$ and a finite symmetry group $G\le \mathrm{O}(d)$ acting as a crystalline symmetry of the lattice. A schematic imaginary–time partition function can be written as
\begin{equation}
    \mathcal{Z}^{\mathrm{UV}}_{\mathrm{Lat}}\bigl[M^{d+1}_\Lambda, {\mathbf g}_{\Lambda}\bigr]
    =\sum_{\{\psi\}} \exp\left\{
      i \sum_{X \subset \Lambda}
      \Bigl[\mathcal S_{X}\bigl[\psi\vert_{X}\bigr]
      + \mathcal{S}_{X}^{\mathrm{coup}}\bigl[\psi\vert_{X}, {\mathbf g}_{\Lambda}\vert_{X}\bigr] \Bigr]
    \right\},
    \label{eq:PF_lattice_UV}
\end{equation}
where $M^{d+1}_{\Lambda}$ is the spacetime, $X\subset \Lambda$ are local subsets of sites, $\psi\vert_X=\{\psi_i\}_{i\in X}$ are microscopic bosonic matter degrees of freedom living on sites $i\in X$ that can transform under the lattice symmetry, and ${\mathbf g}_{\Lambda}\vert_{X}$ denotes the $G$–symmetry background 
restricted to the oriented edges $\langle ij\rangle\subset X$ (with link variables ${(\mathbf g}_{\Lambda})_{ij}$).

\medskip

To connect microscopic lattice models with the TQFT classification models, we state explicitly the assumption that the phases under consideration admit a mesoscopic TQFT limit.
\begin{assumption}[Mesoscopic TQFT regime]
\label{assumption:mesoscopic_TQFT_description}
If the model is in a gapped, invertible phase, we assume that there is a mesoscopic regime of length scales
\begin{equation*}
    a ,\; \zeta \;\ll\; \ell \;\ll\; L,
\end{equation*}
where $\zeta$ is the correlation length of the microscopic fields, $\ell$ is a coarse–graining scale, and $L$ is the system size. Integrating out the massive matter fields and coarse–graining the lattice to a triangulation $M^{d+1}_{\Delta}$ with simplices of linear size $\sim \ell$ yields an effective description in which the only remaining physical data are flat $G$–valued $1$–cocycles
\begin{equation*}
    {\mathbf g}_{\Delta} \in Z^{1}\bigl(M^{d+1}_{\Delta}, G\bigr),
\end{equation*}
which were described in Sec.~\ref{sec:basics} (see Fig.~\ref{fig:graph}).\footnote{This assumption is expected to apply broadly to conventional gapped, short-range-entangled lattice phases. Exceptional systems with UV/IR mixing may require a different mesoscopic framework; see Sec.~\ref{sec:discussion_extension}(iii).}
\end{assumption}

The partition function reduces to a topological functional of the pair $(M^{d+1}_{\Delta},  {\mathbf g}_{\Delta})$. In the group-cohomology model, this functional is a Dijkgraaf-Witten action determined by a cocycle representative of the class
\begin{equation*}
    [x_G]\in H^{d+1}(G,\U(1)_{w_1}).
\end{equation*}
To see the action concretely we take the standard cocycle description where a representative of $[x_{G}]$ is given by a phase/$\U(1)$-valued function of $(d{+}1)$ group variables, $x_G:G^{d+1}\to \U(1)$, satisfying the cocycle condition, see App.~\ref{app:grpcoh_def}. Equivalently, we may rewrite this in the more familiar additive notation used in physics, by viewing the cocycle as valued in $\mathbb{R}/2\pi\mathbb{Z}$ instead (see Remark \ref{rmk:exp_convention}):
\begin{equation*}
    \widetilde{x}_G:G^{d+1}\to \mathbb{R}/2\pi\mathbb{Z},
\end{equation*}
related to $x_G$ by $x_G=\exp(i\widetilde{x}_G)$. Then the flat background ${\mathbf g}_{\Delta}$ assigns group elements to the oriented edges of each 
$(d+1)$-dimensional simplex, see Fig.~\ref{fig:graph}. Once a branching structure is fixed, each simplex $\sigma$ has a canonical ordering of its vertices, and hence determines an ordered $(d{+}1)$-tuple of group labels on which the cocycle $\widetilde{x}_G$ is evaluated. The corresponding mesoscopic lattice TQFT partition function may then be written schematically as
\begin{equation}
    \label{eq:DW}
    \mathcal{Z}^{\mathrm{DW}}_{x_G}\bigl[M^{d+1}_{\Delta}, {\mathbf g}_{\Delta}\bigr]
    =
    \exp\!\left\{
    i\sum_{\sigma\in \Delta^{d+1}(M^{d+1}_{\Delta})}
    \epsilon(\sigma)\widetilde{x}_G\bigl({\mathbf g}_{\Delta}(\sigma)\bigr)
    \right\},
\end{equation}
where, in the oriented case, $\epsilon(\sigma)=\pm1$ records whether the orientation induced by the branching\footnote{In the triangulation, the edge orientations are part of a chosen branching structure on the triangulation; namely a choice of orientation for every edge such that no $2$-simplex contains an oriented loop. Equivalently, on each $n$-simplex the branching canonically orders the vertices by the number of incoming edges, so that they may be labeled $v_0,\dots,v_n$. This is the standard combinatorial input used in simplicial cocycle formulas for Dijkgraaf-Witten theory and related lattice constructions.} agrees or disagrees with the ambient orientation of $M^{d+1}_{\Delta}$.\footnote{More generally, when orientation-reversing symmetries are present, the simplex amplitudes are valued in the twisted coefficient system $\U(1)_{w_1}$ rather than ordinary $\U(1)$. The twist keeps track of the sign picked up when local orientation patches are glued together, so the full product of simplex weights remains globally well defined on non-orientable backgrounds; see App.~\ref{app:twisted_coefficients}.}

\bigskip 

In practice, we will mostly use the equivalent classifying-space description of the same response because the algebraic group-cohomology and the cohomology of the classifying space $BG$ compute the same cohomology groups, see Eq.~\ref{eq:tor_equiv_first} and App.~\ref{app:grpcoh_relation} for details.
Viewing the flat $G$–bundle defined by ${\mathbf g}_{\Delta}$ as a classifying map
\begin{equation*}
   f_G : M^{d+1}_{\Delta} \longrightarrow BG,
\end{equation*}
and regarding $x_G$ as a cocycle representing $[x_{G}]\in H^{d+1}(BG,\U(1)_{w_{1}})$, this partition function can be written as the evaluation pairing
\begin{equation*}    \mathcal{Z}_{x_G}\bigl[M^{d+1}_{\Delta}, f_{G}\bigr]
    \;=\;
    \big\langle f_G^{*}[x_G],\,[M^{d+1}_{\Delta}]\big\rangle,
\end{equation*}
where $\langle -,-\rangle$ is the canonical pairing between cohomology classes $H^{d+1}(M^{d+1}_{\Delta},U(1))$ and homology classes $H_{d+1}(M^{d+1}_{\Delta},\mathbb Z)$ (see Eq.~\ref{eq:evaluation_pairing}). This is a genuine $(d{+}1)$–dimensional topological field theory of Dijkgraaf–Witten type: the partition function depends on only the cobordism class of $(M^{d+1}_{\Delta},  {\mathbf g}_{\Delta}),$ and not on the microscopic details of the triangulation. In our lattice setting, however, it arises as the effective description of the phase in the mesoscopic window $a,\zeta \ll \ell \ll L$, so we will refer to it as a \emph{mesoscopic lattice TQFT}. This regime is also called the “spatially dependent TQFT” (sTQFT) limit of Ref.~\cite{ThorngrenElse}. Note that from this point on we will refer to the symmetry background by reference to only a classifying map $f_{G}$ and not defect configurations.

\subsubsection{Continuum systems and $\mathrm{O}(d)$–symmetric TQFT classes.} 

In contrast to the above, one may start from a  UV continuum QFT defined directly on a smooth manifold, for which the gapped infrared limit is a \emph{bona fide} TQFT with no reference to any coarse–graining scale.
Consider a smooth $(d{+}1)$–manifold $M^{d+1}$, equipped with an $O$–structure (see Def.~\ref{def:tangential_structure_general}) on the tangent bundle, and coupled to a topological $\mathrm{O}(d)$ background is specified by a classifying map
\begin{equation*}
    f_{\mathrm{O}(d)}: M^{d+1} \longrightarrow B\mathrm{O}(d).
\end{equation*}
Here, we note an important distinction between backgrounds in lattice and continuum theories. In the continuum, symmetry backgrounds are often given a geometric refinement, which is a connection $A_{\mathrm{O}(d)}$ on the corresponding principal $\mathrm{O}(d)$-bundle $P_{\mathrm{O}(d)}\to M^{d+1}$ classified by $f_{\mathrm{O}(d)}$. This connection plays an important role in providing a smooth differential form representation of cohomology classes, as we will see later in the examples. 

Let $\psi$ denote the bosonic matter fields, and let $A_{\mathrm{O}(d)}$ be the background connection associated to the classifying map $f_{\mathrm{O}(d)}: M^{d+1}\to B\mathrm{O}(d)$. A schematic imaginary–time partition function has the form
\begin{equation}
    Z_{\mathrm{Cont}}^{\mathrm{UV}}\bigl[M^{d+1}, f_{\mathrm{O}(d)}, A_{\mathrm{O}(d)}\bigr]
    =
    \int \mathcal D\psi\;
    \exp \left\{
      i \int_{M^{d+1}} \mathcal L\bigl[\psi,\,A_{\mathrm{O}(d)}\bigr]
    \right\},
    \label{eq:PF_continuum_UV}
\end{equation}
where $\mathcal L$ is a local Lagrangian density coupling $\psi$ covariantly to the topological $\mathrm{O}(d)$–background through the connection $A_{\mathrm{O}(d)}$. Here $f_{\mathrm{O}(d)}$ records the underlying topological bundle, while $A_{\mathrm{O}(d)}$ supplies the additional local geometric data familiar from gauge theory.
\medskip

If the theory is gapped and invertible, then, at scales large compared to the correlation length, the gapped degrees of freedom can be integrated out. The resulting effective theory depends on only the topological $\mathrm{O}(d)$–background encoded by a classifying map $f_{\mathrm{O}(d)} : M^{d+1} \to B\mathrm{O}(d)$, and the universal topological response is characterized by a class
\begin{equation*}
    [x_{\mathrm{O}(d)}] \in \Theta^{d+1}(B\mathrm{O}(d)).
\end{equation*} We refer to $\Theta^{d+1}(B\mathrm{O}(d))$ as the continuum classification.

In the group cohomology model in Eq.~\ref{eq:grpcohtormodel}, the long–distance partition function takes the form
\begin{equation*}
    Z_{\mathrm{UV}}^{\mathrm{cont}}\bigl[M^{d+1}, f_{\mathrm{O}(d)}, A_{\mathrm{O}(d)}\bigr]
    \;\xrightarrow[\text{IR}]{\text{gapped}}\; 
    \big\langle f^{*}[x_{\mathrm{O}(d)}],\;[M^{d+1}] \big\rangle,
\end{equation*}
where $\langle -,-\rangle$ denotes the evaluation pairing with the fundamental class (see Eq.~\ref{eq:evaluation_pairing}).  For a genuine strict TQFT, the infrared response depends on only the underlying topological $\mathrm{O}(d)$-background encoded by $f_{\mathrm{O}(d)}$; any choice of connection $A_{\mathrm{O}(d)}$ is additional geometric data and is not needed to define the topological response itself. When the class $x_{\mathrm{O}(d)}$ admits a differential refinement, however, such a connection may be used to write a local smooth representative of the same infrared response in terms of a smooth functional: 
\begin{equation}
    Z_{\mathrm{UV}}^{\mathrm{cont}}\bigl[M^{d+1}, f_{\mathrm{O}(d)}, A_{\mathrm{O}(d)}\bigr]
    =\exp\left\{i\int_{M^{d+1}} \omega_{x_{\mathrm{O}(d)}}\left(A_{\mathrm{O}(d)}\right)\right\}.
    \label{eq:PF_continuum_IR}
\end{equation}

\bigskip 

A popular example of a differential refinement comes from Chern-Simons theory where it is known that the $n$-th Chern class associated to a $U(1)$-symmetric system $[c_{n}]\in H^{2n}(B\U(1), \Z)$ can be represented locally in terms of invariant polynomials of a connection $A$ via the formula:
\begin{equation*}
    \left\langle f_{\U(1)}^{*}c_{n},\; [M^{2n}]\right \rangle \; = \; \int_{M^{2n}}\dfrac{1}{n!} \left(\dfrac{dA}{2\pi}\right)^{n}
\end{equation*}
where $f_{\U(1)}: M^{2n}\to B\U(1)$ is a classifying map corresponding to a $\U(1)$-bundle $\zeta_{\U(1)}\to M^{2n}$ which we call the $\U(1)$-background, and $A\in \Omega^{1}(M)\otimes i\mathbb{R}$ is a local representation of a connection on $\zeta_{U(1)}\to M^{2n}$. We note that we only consider theories which do not depend on geometric refinements like smooth connections; if such a refinement exists, the connections serve merely as extra data which help in reformulation in terms of smooth forms. For further details, see App.~\ref{app:diff_refinement}.

\subsection{Topological response: probing $G$-symmetric TQFTs}
\label{sec:top_resp}
We will see in this subsection that it is possible to characterize a $G$-TQFT phase entirely in terms of its response to \textit{all} $G$-symmetry probes. This idea is formalized in Lemma~\ref{lemma:probe}, and will provide a useful tool for analysis in the following sections. We will work with the group cohomology classification model,
\begin{equation*}
    \Theta^{d+1}(BG) := \tor_{2}\bigl(H^{d+1}(BG,\U(1)_{w_{1}})\bigr),
\end{equation*}
as in Eq.~\ref{eq:grpcohtormodel}. Our goal is to explain how a class
$[x_{G}] \in \Theta^{d+1}(BG)$ produces a universal topological response for any $(d{+}1)$-dimensional system with symmetry $G$. The key tool is the \emph{evaluation pairing} between cohomology and homology classes, which plays the role of ``integrating a topological density over spacetime" in the cohomological language.

\bigskip

Let $M^{d+1}$ be a closed, generally unoriented, spacetime and let $f_G : M^{d+1} \to BG$ be a $G$-symmetry background on $M^{d+1}$. Then, a class 
\begin{equation*}
[x_{G}] \in H^{d+1}(BG,\U(1)_{w_{1}}),
\end{equation*}
is a ``universal topological density" for $G$-backgrounds, in the following sense: it is defined once on the classifying space $BG$, and its realization on any particular spacetime background is obtained by pullback along the corresponding classifying map. Thus, for the specific background $f_G:M^{d+1}\to BG$, pulling back $[x_G]$ gives
\begin{align*}
    f^{*}_{G}: H^{d+1}(BG,\U(1)_{w_{1}}) &\longrightarrow H^{d+1}(M^{d+1},\U(1)_{f^{*}_{G}w_{1}}),\\
     [x_{G}]\qquad  &\longmapsto \qquad f_{G}^{*}[x_{G}].
\end{align*}
This is simply the naturality of cohomology with local coefficients: the classifying map $f_G:M^{d+1}\to BG$ pulls back both the cohomology class and the coefficient local system. The pulled-back class $f_G^{*}[x_G]$ is the corresponding topological density on the specific spacetime.
 
\medskip 

The spacetime $M^{d+1}$ has a fundamental class with coefficients in its orientation local system,
\begin{equation*}
    [M^{d+1}] \in H_{d+1}(M^{d+1},\mathbb Z_{w_{1}(TM)}),
\end{equation*} where $TM$ is the tangent bundle of $M.$ For admissible orientation-reversing backgrounds, the coefficient twist is required to agree with the orientation twist on the space $M^{d+1}$, $f_G^{*}w_{1}=w_{1}(TM)$. This is discussed in formal detail in App.~\ref{app:bordism_spatial_sym}. Physically, an orientation-reversing symmetry defect, such as a reflection domain wall, reverses the local choice of orientation used to write the response density. The condition $f_G^{*}w_{1}=w_{1}(TM)$ says that the sign picked up from the background symmetry twist is the same sign picked up by the local orientation of spacetime. This is what allows the locally defined response density to glue to a globally well-defined topological term.

\medskip 

 Under this identification, cohomology and homology pair by the canonical evaluation map
\begin{equation}
    \langle\;,\;\rangle: H^{d+1}(M^{d+1},\U(1)_{f_G^{*}w_{1}}) \times 
    H_{d+1}(M^{d+1},\mathbb Z_{w_{1}(TM)}) \longrightarrow \U(1).
    \label{eq:evaluation_pairing}
\end{equation}
When $M^{d+1}$ is oriented and the twist is trivial, this reduces to the ordinary evaluation pairing with $[M^{d+1}]\in H_{d+1}(M^{d+1},\mathbb Z)$. Here and below, we often suppress pullbacks in the notation for twisted coefficients: $w_1$ is understood from the base space appearing in the cohomology group. See Remark~\ref{rem:abuse_pulledback_twists}. 

\medskip

\noindent Physically, the pairing plays the role of 
\begin{equation*}
  \exp\left( i \int_{M^{d+1}} \omega_{x_{G}} \right),
\end{equation*}
where $\omega_{x_{G}}$ is a representative topological $(d{+}1)$-form, but written in a coordinate-free, cohomological language.

\medskip

 Given these ingredients, a TQFT class $[x_{G}]\in\Theta^{d+1}(BG)$ has \emph{topological response} on a spatial manifold with symmetry background $(M^{d+1},f_G)$ defined by
\begin{equation}
    \mathcal Z_{x_{G}}(M^{d+1}, f_{G})
    \;:=\;
    \big\langle f_{G}^{*}[x_{G}],\; [M^{d+1}] \big\rangle
    \;\in\; U(1).
    \label{eq:top-response}
\end{equation}
This partition function phase gives the response of the invertible $G$-symmetric TQFT associated to $[x_{G}]$ in the given background $(M^{d+1}, f_{G})$.

\medskip

\begin{remark}
\label{rmk:exp_convention}
 Eq.~\ref{eq:top-response} is often written in the exponentiated form $\exp\bigl(i\,\langle f_{G}^{*}[x_{G}],\; [M^{d+1}] \rangle\bigr)$. This uses the identification $U(1)\cong \mathbb R/2\pi \mathbb Z$ via $t\mapsto e^{i t}$, so that one may view $f_G^*[x_{G}]\in H^{d+1}(M^{d+1},\U(1)_{w_{1}})$ additively as a class in $H^{d+1}(M^{d+1},\mathbb R/2\pi \mathbb Z_{w_{1}})$. We will use this notation in the example section~\ref{sec:examples}.
\end{remark}

\bigskip

\noindent\textit{Probing with subgroup backgrounds.}
A $G$-symmetric phase can be probed using background fields for any subgroup $P\le G$. Physically, one allows symmetry twists and defects restricted to $P$, which is still a legitimate probe of the original $G$-invariant system. Given a $P$-background $f_P:M^{d+1}\to BP,$ and an inclusion $i:P\hookrightarrow G$, the induced map $Bi:BP\to BG$ upgrades $f_P$ to a $G$-background $f_G:=Bi\circ f_P$. For a universal class $[x_{G}]\in H^{d+1}(BG,\U(1)_{w_{1}})$, the
naturality of the pullback gives $(Bi\circ f_P)^*[x_{G}]=f_P^*(Bi)^*[x_{G}]$. 

\medskip 

\noindent We define the \emph{restriction} of $[x_{G}]$ to $P$ by
\begin{equation*}
\res^G_P([x_{G}])\;:=\;(Bi)^*[x_{G}]\in H^{d+1}(BP,\U(1)_{w_{1}}),
\end{equation*}
and the response to the $P$-probe is simply the evaluation of this restricted class on the background $f_P$:
\begin{equation}
\label{eq:subgroup_probe}
\mathcal Z_{x_{G}}(M^{d+1},f_P)
\;:=\;
\big\langle f_P^*\bigl(\res^G_P[x_{G}]\bigr),\,[M^{d+1}]\big\rangle.
\end{equation}
Equivalently, probing a $G$-phase by a $P$-background is the same as first restricting the universal class from $G$ to $P$ and then evaluating it on $(M^{d+1},f_P)$.

\bigskip

\noindent \textit{Detectability lemma.}
We now formalize the idea that, within a fixed classification model, a nontrivial phase must be detectable by \emph{some} symmetry probe. In the cohomology model $\Theta^{d+1}(BG)=H^{d+1}(BG,\U(1)_{w_{1}})$, a phase is specified by a class 
$[x_{G}]\in \Theta^{d+1}(BG)$, and its response to a background
$f_G : M^{d+1}\to BG$ is given by
\begin{equation*}
  \mathcal Z_{x_{G}}(M^{d+1},f_G)
  \;=\;
  \big\langle f_G^{*}[x_{G}],\,[M^{d+1}]\big\rangle
  \;\in\; U(1),
\end{equation*}
as in Eq.~\ref{eq:top-response}. Intuitively, if this response is trivial, i.e., $\mathcal Z_{x_{G}}(M^{d+1},f_G)=1$, for \emph{every} possible spacetime and background, then the corresponding phase should be regarded as trivial in the classification.

\bigskip 

\begin{lemma}[Probe detectability for $G$-TQFTs]
\label{lemma:probe}
Let  $[x_{G}]\in \Theta^{d+1}(BG)$ be a TQFT class. Suppose that for every closed oriented $(d{+}1)$-manifold $M^{d+1}$ and every symmetry background
$f_{G}:M^{d+1}\to BG$ the response
\begin{equation*}
  \mathcal Z_{x_{G}}(M^{d+1},f_G)
  \;=\;
  \big\langle f_G^{*}[x_{G}],\,[M^{d+1}]\big\rangle
  \;=\; 1 \in U(1)
\end{equation*}
is trivial. Then $[x_{G}] = [0]$ in $\Theta^{d+1}(BG)$.
\end{lemma}

\bigskip

A formal proof for the group cohomology classification model can be found in App.~\ref{app:probe_detectibility}. As an immediate consequence, responses determine the TQFT class: if $[x_1],[x_2]\in \Theta^{d+1}(BG)$ satisfy
\begin{equation*}
  \mathcal Z_{x_1}(M^{d+1},f_G)=\mathcal Z_{x_2}(M^{d+1},f_G),
\end{equation*}
for every closed oriented $(d{+}1)$-manifold $M^{d+1},$ and every background $f_G:M^{d+1}\to BG,$ then $x_1=x_2$ in $\Theta^{d+1}(BG)$. This follows because invertibility implies that stacking corresponds to addition in $\Theta^{d+1}(BG)$, and responses multiply: $\mathcal Z_{x_1-x_2}=\mathcal Z_{x_1}\,\mathcal Z_{x_2}^{-1}$. By assumption this equals $1$ for all $(M^{d+1},f_G)$, so the lemma gives $x_1-x_2=0$, hence $[x_1]=[x_2]$. 

\medskip

Finally, this lemma should always be read \emph{relative to the fixed classification model}. It does not rule out the possibility that two genuinely distinct physical phases become indistinguishable after passing to a coarser model (for example, different cobordism classes mapping to the same group-cohomology class). Once a model has been chosen, however, the lemma states that within that model any nontrivial class must be detected by some symmetry probe.

\bigskip 

This concludes the preliminary discussion of TQFTs, symmetry backgrounds, and classification models. We now turn to the continuum-limit problem itself. In the next section we begin from the perspective of a single lattice symmetry group $G_{\Lambda}$ and formulate what it means for a lattice phase to admit a continuum limit. This leads naturally to the orbit categories of lattice symmetries, which will be used in Sec.~\ref{sec:inverselimit} to construct the universal cSPT object, and later in Sec.~\ref{sec:continuum_limit} to formulate continuum limit theories.

\section{Symmetry and the Continuum Limit}
\label{sec:motivation}

\smallskip

We have introduced the language used to describe $G$-symmetric TQFTs and we are now ready to move from an abstract symmetry group $G$ to crystalline settings. For the crystal we take a microscopic model living on a lattice $\Lambda\subset \mathbb{R}^{d}$ with point group $G_{\Lambda}\subset \mathrm{O}(d)$. We again emphasize that the model does not need to obey translation symmetry, and hence the allowed point groups can lie beyond the conventional crystalline set. Not every lattice TQFT class is compatible with passage to the continuum. For a lattice spatial symmetry group $G_{\Lambda}\subset \mathrm{O}(d)$, we denote  \begin{equation*} \Theta_{\mathrm c}^{d+1}(BG_{\Lambda}) \subseteq \Theta^{d+1}(BG_{\Lambda}), \end{equation*} i.e., by a subscripted $\Theta_c$, the subgroup of $G_{\Lambda}$-symmetric lattice TQFT classes that admit a continuum realization. For $[x_{G_{\Lambda}}]\in\Theta_{\mathrm c}^{d+1}(BG_{\Lambda})$, we provisionally write $\kappa_{G_{\Lambda}}([x_{G_{\Lambda}}])$ for its continuum image (anticipating the notation discussed in detail below). 

\noindent Two questions guide this section:
\begin{enumerate}
\item When does a continuum limit exist for a given lattice phase? 
\item How should we understand the combined behavior of all such limits across lattices?
\end{enumerate}
The first question is addressed in Sec.~\ref{sec:lattice_compatible_continuum_limit}, where we identify isotropy and conjugation coherence as necessary physical constraints on the continuum-admitting domain $\Theta_{\mathrm c}^{d+1}(BG_{\Lambda})$. Section~\ref{sec:RG_flow} then relates the resulting continuum-limit maps to microscopic RG/EFT procedures and motivates the requirement that different admissible prescriptions agree on their common domains. The second question is addressed in Sec.~\ref{sec:orbit_categories}, where the lattice symmetry groups and the relations among them are organized into a common categorical framework. The precise construction of the continuum-limit maps, together with their domains, codomains, and compatibility properties, will be developed in Sec.~\ref{sec:continuum_limit}.

\subsection{Lattice Phases Compatible with the Continuum Limit}
\label{sec:lattice_compatible_continuum_limit}

In this subsection we identify two necessary conditions for a lattice TQFT phase to admit a continuum realization: invariance under changes of lattice embedding, and compatibility with symmetry enlargement toward $\mathrm{O}(d)$. In passing from a lattice model to a continuum description, one should not treat every rigid embedding $\iota:\Lambda\hookrightarrow\mathbb R^d$ of the same microscopic lattice system as defining a new intrinsic phase. For example, a square or rectangular lattice carrying a fixed microscopic phase may be placed in the continuum plane in many different orientations. Two such embeddings generally determine different embedded point groups $G_{\Lambda}\subset \mathrm{O}(d), \;\; G_{\Lambda}'=\lambda G_{\Lambda}\lambda^{-1}\subset \mathrm{O}(d)$, where $\lambda\in \mathrm{O}(d)$ is the rigid rotation or reflection relating the two placements.\footnote{Here ``embedded'' simply means that the abstract lattice symmetry group has been realized as an actual subgroup of the continuum orthogonal group $\mathrm{O}(d)$. Thus its elements are concrete spatial operations, such as rotations or reflections of $\mathbb R^d$.} These embeddings are two spatial presentations of the same lattice phase. Their symmetry operations, and  symmetry defect labels are identified by the rule $G_{\Lambda}\ni g\longmapsto \lambda g\lambda^{-1}\in G_{\Lambda}'$. Thus, changing the placement of the lattice changes the names of the symmetry operations and defects, but not the intrinsic phase being described.

\bigskip 

Let $[x_{G_{\Lambda}}]\in\Theta^{d+1}(BG_{\Lambda})$ and $[x_{G_{\Lambda}'}]\in\Theta^{d+1}(BG_{\Lambda}')$ denote the lattice TQFT classes associated with the two embeddings. If they represent the same intrinsic lattice phase, their responses must agree after the corresponding symmetry backgrounds have been matched. More precisely, let $f:M\rightarrow BG_{\Lambda}$ be a symmetry background for the first embedding. Conjugation by $\lambda$ defines an isomorphism $c_{\lambda}:G_{\Lambda}\rightarrow G_{\Lambda}', \;\; g\mapsto \lambda g\lambda^{-1}$, and hence determines the corresponding background for the second embedding, $f' = Bc_{\lambda}\circ f: M\to BG_{\Lambda}'$. The assertion that the two classes represent the same intrinsic phase means that 
\begin{equation*} 
\left\langle f^{*}[x_{G_{\Lambda}}],[M] \right\rangle = \left\langle (f')^{*}[x_{G_{\Lambda}'}],[M] \right\rangle = \left\langle f^{*}(Bc_{\lambda})^{*}[x_{G_{\Lambda}'}],[M] \right\rangle, \end{equation*} 
for every admissible $G_{\Lambda}$-background probe $(M,f)$. By the probe-detectability Lemma~\ref{lemma:probe}, this implies the constraint\footnote{Note that $(Bc_{\Lambda})^{*}=\Theta^{d+1}(Bc_{\lambda})$ is a map $\Theta^{d+1}(Bc_{\lambda}): \Theta^{d+1}(BG_{\Lambda}')\to \Theta^{d+1}(BG_{\Lambda})$, and thus $(Bc_{\lambda})^{*}[x_{G_{\Lambda}'}]\in \Theta^{d+1}(BG_{\Lambda})$.} 
\begin{equation*} 
[x_{G_{\Lambda}}] = (Bc_{\lambda})^{*}[x_{G_{\Lambda}'}], 
\end{equation*} 
which we call conjugation coherence. Thus the TQFT classes assigned to the two placements are not independent: they are presentations of the same intrinsic phase identified under conjugation. Note that, even when $G_{\Lambda}'=G_{\Lambda}=G$, this relation should still be understood as relating two embedding-dependent presentations of the phase, $[x_{G}^{(1)}] = (Bc_{\lambda})^{*}\bigl([x_{G}^{(2)}]\bigr)$, and not as identifying the two classes $[x_{G}^{(1)}]$ and $[x_{G}^{(2)}]$.

\bigskip

Because the continuum limit forgets the microscopic lattice texture, its response cannot depend on the particular rigid embedding used to present the lattice TQFT. In particular, conjugation-related embeddings of the same lattice phase must determine the same infrared response.  This statement can be made more precise at the level of symmetry backgrounds. Let $i_{G_{\Lambda}}:G_{\Lambda}\hookrightarrow \mathrm{O}(d)$ and $i_{G_{\Lambda}'}:G_{\Lambda}'\hookrightarrow \mathrm{O}(d)$ denote the subgroup inclusions, where $G_{\Lambda}'=\lambda G_{\Lambda}\lambda^{-1}$. Let $\widehat f:=Bi_{G_{\Lambda}}\circ f, \;\; \widehat f':=Bi_{G_{\Lambda}'}\circ f',$ denote the continuum backgrounds induced by the two conjugation-related lattice backgrounds which satisfy
\begin{equation*} 
\widehat f' = Bc_{\lambda,\mathrm{O}(d)}\circ \widehat f \quad \text{where} \quad 
\begin{tikzcd}[]
    BG_{\Lambda} \arrow[r, "Bi_{G_{\Lambda}}"] \arrow[d, "Bc_{\lambda}"] & B\mathrm{O}(d)\arrow[d, "Bc_{\lambda, \mathrm{O}(d)}"]\\
    BG_{\Lambda}' \arrow[r, "Bi_{G_{\Lambda}'}"] & B\mathrm{O}(d)
\end{tikzcd}
\end{equation*} 
commutes, and $c_{\lambda,\mathrm{O}(d)}$ denotes conjugation by $\lambda$ in $\mathrm{O}(d)$. Since $c_{\lambda,\mathrm{O}(d)}$ is an inner automorphism of $\mathrm{O}(d)$, the induced map $Bc_{\lambda,\mathrm{O}(d)}:B\mathrm{O}(d)\to B\mathrm{O}(d)$ is homotopic to the identity:
\begin{equation*}
    Bc_{\lambda, \mathrm{O}(d)} \simeq \mathrm{id}\vert_{B\mathrm{O}(d)}\implies \widehat f' \simeq \widehat f.
\end{equation*}
The two lattice backgrounds therefore induce equivalent continuum backgrounds, and a physically admissible continuum-limit prescription must assign them the same infrared response:
\begin{align} 
\left\langle \widehat{f'}^{*}\kappa_{G_{\Lambda}'}([x_{G_{\Lambda}'}]),[M] \right\rangle = \left\langle \widehat{f\;}^{*}\kappa_{G_{\Lambda}}([x_{G_{\Lambda}}]),[M] \right\rangle.
\label{eq:conjugation_stability_continuum_response} 
\end{align} 

\bigskip 

A particularly important case occurs when the conjugation-related embeddings determine the same concrete subgroup, $G_{\Lambda}'=\lambda G_{\Lambda}\lambda^{-1}=G$. The elements $\lambda\in \mathrm{O}(d)$ with this property form the normalizer 
\begin{equation*} 
N_{\mathrm{O}(d)}(G) := \left\{ \lambda\in \mathrm{O}(d) \;\middle|\; \lambda G\lambda^{-1}=G \right\}. 
\end{equation*} 
Conjugation by $\lambda\in N_{\mathrm{O}(d)}(G)$ induces an automorphism $c_{\lambda}: G \to G,\; g\mapsto \lambda g\lambda^{-1}$ and hence an action on the lattice classification $\Theta^{d+1}(BG)$ by $[x_{G}]\mapsto (Bc_{\lambda})^{*}[x_{G}]$. Since elements that centralize $G$ induce the identity automorphism, the effective group of such normalizer-induced automorphisms is the Weyl group \begin{equation*} W_{\mathrm{O}(d)}(G) := N_{\mathrm{O}(d)}(G)/C_{\mathrm{O}(d)}(G), \end{equation*} where $C_{\mathrm{O}(d)}(G):=\{\lambda\in \mathrm{O}(d)\mid \lambda g\lambda^{-1}=g \text{ for all }g\in G\}$.\footnote{Because every $\lambda\in C_{\mathrm{O}(d)}(G)$ induces $c_\lambda=\mathrm{id}_G$, it acts trivially on $\Theta^{d+1}(BG)$.} For $w\in W_{\mathrm{O}(d)}(G)$ represented by $\lambda\in N_{\mathrm{O}(d)}(G)$, we write \begin{equation*} w\cdot[x] := (Bc_{\lambda})^{*}([x]). \end{equation*}
Equation~\eqref{eq:conjugation_stability_continuum_response} then gives 
\begin{equation} 
\left\langle \widehat f^{*}\kappa_G(w\cdot[x]),[M] \right\rangle = \left\langle \widehat f^{*}\kappa_G([x]),[M] \right\rangle, 
\label{eq:weyl_stability_on_probes} 
\end{equation} 
for every continuum background $\widehat f$ induced from a lattice probe.

\medskip 

At this stage, Eq.~\eqref{eq:weyl_stability_on_probes} establishes equality only of the responses accessible through lattice-induced continuum backgrounds. These are the continuum backgrounds obtained from lattice probes through $M\longrightarrow BG\hookrightarrow B\mathrm{O}(d)$. Such probes need not detect every class in $\Theta^{d+1}(B\mathrm{O}(d))$, and therefore probe detectability cannot yet be used to deduce equality in the full continuum classification. We denote by $K_{\mathcal O(G)}\subseteq\Theta^{d+1}(B\mathrm{O}(d))$ the subgroup of continuum phases invisible to all such $G$-probes and define the \emph{$G$-visible continuum quotient}
\begin{equation}
\Theta^{d+1}(B\mathrm{O}(d))_{G} := \frac{\Theta^{d+1}(B\mathrm{O}(d))}{K_{\mathcal O(G)}}.
\label{eq:G_visible_continuum_quotient}
\end{equation}
Its formal construction and explicit description in terms of restriction kernels will be given in Sec.~\ref{sec:refine_domain}. Passing to this quotient promotes the probe-level equality to an equality of the classes themselves:
\begin{equation*}
\kappa_G(w\cdot[x])= \kappa_G([x]) \quad\text{in}\quad \Theta^{d+1}(B\mathrm{O}(d))_{G}.
\end{equation*}

We now translate this embedding independence into a condition on the continuum limit-admitting domain of lattice classifications. The input to the continuum-limit construction is meant to be the intrinsic lattice phase itself, and therefore should not depend on the particular embedding used to present it. When $\lambda\in N_{\mathrm{O}(d)}(G)$, the classes $[x]$ and $w\cdot[x]$ arise from two normalizer-related embeddings of the same lattice phase with the same concrete symmetry group $G\subset \mathrm{O}(d)$. Thus any difference between them reflects the auxiliary choice of embedding, rather than an intrinsic distinction between lattice phases. A class which admits a continuum realization as a single intrinsic lattice phase should therefore be independent of this choice of embedding. For $[x]\in\Theta^{d+1}(BG)$, this condition is exactly
\begin{equation*}
w\cdot[x]=[x]
\qquad
\text{for every }w\in W_{\mathrm{O}(d)}(G).
\end{equation*}
We are therefore led to the necessary condition
\begin{equation*}
\Theta^{d+1}_c(BG)
\subseteq
\Theta^{d+1}(BG)^{W_{\mathrm{O}(d)}(G)}.
\end{equation*}
We refer to this necessary condition as \textit{isotropy}, or \textit{Weyl invariance}.\footnote{After generalized continuum limit theories are defined in Def.~\ref{def:gen_con_lim}, we verify that the isotropy requirement is compatible with the physical axioms entering that definition and with the resulting maximality statement for the universal continuum-limit domain; see Remark~\ref{rmk:univ_isotropy_consistency} and App.~\ref{app:proof_isotropy}.}

\bigskip

Isotropy, however, is only a necessary (but not sufficient) condition for a lattice phase to be compatible with the continuum. A lattice TQFT admitting a continuum limit must also be compatible with symmetry enlargement to the full continuum symmetry $\mathrm{O}(d)$. Concretely, if a phase $[x_{G_{\Lambda}}]\in \Theta^{d+1}(BG_{\Lambda})$ has a continuum limit, then it should fit into a coherent symmetry enlargement pattern along any chain of symmetry groups
\begin{equation*}
    G_{\Lambda}\leq P \leq \mathrm{O}(d),
\end{equation*}
meaning that $[x_{G_{\Lambda}}]$ extends to a phase $[x_{P}]\in \Theta^{d+1}(BP)$ for each intermediate $P$, and that all such extensions map to the same continuum limit class under further symmetry enlargement to $\mathrm{O}(d)$. The same compatibility requirement applies when restricting to subgroups $Q\leq G_{\Lambda}$. We formalize this compatibility requirement as a physical condition on continuum-limit theories in Sec.~\ref{sec:strong_continuum_limit}.

\bigskip 

Having identified these necessary conditions, we may write the continuum-limit assignment introduced above as
\begin{equation*}
  \kappa_{G_\Lambda}: \Theta^{d+1}_{\mathrm c}(BG_{\Lambda})
  \longrightarrow \Theta^{d+1}(B\mathrm{O}(d))_{G_{\Lambda}},
\end{equation*}
which we call a $G_{\Lambda}$-continuum limit map. This map takes a lattice TQFT phase defined by a class $[x_{G_{\Lambda}}]\in\Theta^{d+1}_{\mathrm c}(BG_{\Lambda})$ and returns its continuum TQFT realization in the $G_{\Lambda}$-visible continuum quotient $\Theta^{d+1}(B\mathrm{O}(d))_{G_{\Lambda}}$ defined in Eq.~\eqref{eq:G_visible_continuum_quotient}. The precise determination of the domain $\Theta^{d+1}_{\mathrm c}(BG_{\Lambda})$, the formal construction of the codomain, and the compatibility conditions that the maps $\{\kappa_{G_{\Lambda}}\}$ must satisfy as $G_{\Lambda}$ varies will be developed in Sec.~\ref{sec:continuum_limit}.

\bigskip 

To be explicit, $\kappa_{G_\Lambda}$ provides a concrete map from lattice phases to continuum phases. This can be made clear by considering, for example, the cohomology classification model in which a lattice TQFT class $[x_{G_{\Lambda}}]\in H^{d+1}(BG_{\Lambda}, \U(1))$ is represented by a $\mathbb{R}/2\pi\Z$-valued $G_{\Lambda}$-cocycle, $x_{G_{\Lambda}}\in Z^{d+1}(G_{\Lambda},\mathbb{R}/2\pi\Z)$ (see Sec.~\ref{sec:scales}). In the mesoscopic regime, the corresponding lattice TQFT has partition function
\begin{equation*}    \mathcal{Z}^{\mathrm{DW}}_{x_{G_{\Lambda}}}\bigl[M^{d+1}_{\Delta}, {\mathbf g}_{\Delta}\bigr]
    =
    \exp\!\left\{
    i\sum_{\sigma\in \Delta^{d+1}(M^{d+1}_{\Delta})}
    \epsilon(\sigma)\widetilde{x}_{G_{\Lambda}}\bigl({\mathbf g}_{\Delta}(\sigma)\bigr)
    \right\},
\end{equation*}
where $M^{d+1}_{\Delta}$ is a triangulated manifold with characteristic length scale $\ell$ in the mesoscopic regime, $\epsilon(\sigma)$ is defined as in Eq. \ref{eq:DW}, and ${\bf g}_{\Delta}\in Z^{1}(M^{d+1}_\Delta,G_{\Lambda})$ is a coarse-grained symmetry background (equivalently described by a classifying map $f_{G_{\Lambda}}: M^{d+1}\to BG_{\Lambda}$).

\bigskip 

Applying the continuum limit map $\kappa_{G_\Lambda}$ sends $[x_{G_{\Lambda}}]$ to the $G_{\Lambda}$-visible continuum class
\begin{equation*}
\kappa_{G_\Lambda}\left([x_{G_{\Lambda}}]\right)
\in H^{d+1}(B\mathrm{O}(d),\U(1))_{G_{\Lambda}},
\end{equation*} 
with the quotient notation defined in Eq.~\eqref{eq:G_visible_continuum_quotient}. Choose a representative $[x_{\mathrm{O}(d)}]\in H^{d+1}(B\mathrm{O}(d),\U(1))$ of this quotient class. When this continuum class admits a differential refinement, the continuum partition function can be written as an integral of a smooth density
\begin{equation*}
\mathcal Z_{\mathrm{Cont}}^{\mathrm{IR}}[M^{d+1}, A_{\mathrm{O}(d)}]
=\exp \left\{ i\int_{M^{d+1}} 
\omega_{\left([x_{\mathrm{O}(d)}]\right)}
\left(A_{\mathrm{O}(d)}\right)\right\},
\end{equation*}
where $A_{\mathrm{O}(d)}$ is a connection on a principal $\mathrm{O}(d)$-bundle defined by a topological background $f_{\mathrm{O}(d)}: M^{d+1}\to B\mathrm{O}(d)$, and $\omega_{\left([x_{\mathrm{O}(d)}]\right)}(A_{\mathrm{O}(d)})\in \Omega^{d+1}(M_{d+1})$ is a differential form built from this connection. \emph{In general,} a local density on $M^{d+1}$ need not exist (e.g., for purely $2$-torsion classes like Stiefel-Whitney classes). In those cases the continuum response is specified by a \emph{global} invariant—such as the Arf-invariant~\cite{Karch2019} for Spin-TQFTs(or its generalization for $\mathrm{Pin}^{-}$-TQFTs called the Arf-Brown-Kervaire invariant~\cite{Debray2018}), the Rokhlin-invariant~\cite{Putrov2017}, or the $\eta$-invariant~\cite{Witten2016, Witten2016_2}. We will exhibit both kinds of presentations in Sec.~\ref{sec:examples}.

\subsection{Microscopic motivation from RG/EFT Flows}
\label{sec:RG_flow}

The preceding subsection characterized continuum-limit compatibility directly at the level of low-energy lattice and continuum TQFT classes. We now explain how such continuum-limit maps may arise from microscopic RG/Effective field theory (EFT)  procedures and motivate the requirement that different physically admissible RG prescriptions agree whenever they are defined on the same lattice phase. 

One can envision a continuum-limit procedure acting on UV theories with matter, taking a UV lattice theory $\mathcal{Z}^{\mathrm{UV}}_{\mathrm{Lat}}$ as in Eq.~\eqref{eq:PF_lattice_UV} to a UV continuum theory $\mathcal{Z}^{\mathrm{UV}}_{\mathrm{Cont}}$ as in Eq.~\eqref{eq:PF_continuum_UV}. For example, in the regime of slowly varying spin configurations, the lattice $XY$ model admits a continuum spin-wave approximation. Writing
$s_i=(\cos\theta_i,\sin\theta_i)\in S^1$,
\begin{equation*}
S_{\mathrm{Lat}}^{XY}=-J\sum_{\langle i,j\rangle}\cos(\theta_i-\theta_j)
\;\Rightarrow\;
S_{\mathrm{Cont}}^{XY}\approx \frac{J}{2}\int_{\mathbb{R}^d} d^dx\,(\nabla\theta)^2,
\end{equation*}
where $(|\theta_i-\theta_j|\ll 1)$, and the RHS is a UV continuum description. In a gapped phase the microscopic matter carries a finite correlation length set by the gap: on the lattice we write $\zeta_{\Lambda}\sim m_{\Lambda}^{-1}$, while in the continuum description with matter we write $\zeta_c\sim m_c^{-1}$. Thus the UV lattice theory is characterized by the microscopic lattice scale $a$ together with $\zeta_{\Lambda}$, whereas the corresponding UV continuum description has no lattice spacing but still retains the finite mass scale $\zeta_c$.

Passing to longer length scales then produces effective descriptions across a range of scales. On the lattice side, coarse-graining produces a mesoscopic description at a superlattice scale $\ell$, with $a,\zeta_{\Lambda}\ll \ell \ll L$, and the corresponding effective response is described by a lattice TQFT on a triangulation of scale $\ell$. On the continuum side, one instead integrates out
the massive matter fields of correlation scale $\zeta_c\sim m_c^{-1}$ to reach the
deep infrared, where the remaining response is captured by an invertible continuum TQFT. The point of the present construction is that the induced continuum limit on
these low-energy descriptions is described by continuum-limit maps between TQFTs.
\begin{equation*}
    \begin{tikzcd}
\mathcal{Z}^{\mathrm{UV}}_{\mathrm{Lat}}\bigl[M^{d+1}_{\Lambda},{\mathbf g}_{\Lambda};a,\zeta_{\Lambda}\bigr]
\arrow[d, squigarrow, "\mathrm{RG/EFT}"']
\arrow[r, "\kappa_{\mathrm{UV}}"] &
\mathcal{Z}^{\mathrm{UV}}_{\mathrm{Cont}}\bigl[M^{d+1}, A_{\mathrm{O}(d)};\zeta_{c}\bigr]
\arrow[d, squigarrow, "\mathrm{EFT}"]\\
\mathcal{Z}^{\mathrm{DW}}_{x_{G_{\Lambda}}}\bigl[M^{d+1}_{\Delta},{\mathbf g}_{\Delta};\ell\bigr]
\arrow[r, "\kappa_{G_{\Lambda}}"] &
\mathcal{Z}^{\mathrm{IR}}_{x_{\mathrm{O}(d)}}\bigl[M^{d+1}, A_{\mathrm{O}(d)}\bigr].
\end{tikzcd}
\end{equation*}
Here the left vertical arrow represents lattice coarse-graining to a mesoscopic scale $\ell$, while the right vertical arrow represents integrating out continuum matter fields to reach the deep infrared. The top horizontal arrow is a heuristic continuum realization of the microscopic theory with matter, whereas the bottom horizontal arrow is the continuum-limit map on the induced topological effective descriptions that we formalize in this article. 

\bigskip 

This separation of regimes is important because different microscopic RG/EFT procedures may induce different candidate continuum-limit maps on low-energy TQFT data. For example, one may employ Kadanoff block decimation, Wilsonian coarse-graining, or another method. A more flexible microscopic prescription applies to a larger class of UV lattice systems and therefore induces a continuum-limit map on a larger class of low-energy lattice TQFT phases. One is thus led to candidate maps
\begin{gather*}
\kappa_G^{\mathrm{Kad}}:\Theta^{d+1}_{\mathrm{Kad}}(BG)\to \Theta^{d+1}(B\mathrm{O}(d))_G,\\
\kappa_G^{\mathrm{Wil}}:\Theta^{d+1}_{\mathrm{Wil}}(BG)\to \Theta^{d+1}(B\mathrm{O}(d))_G,
\end{gather*}
where the different lattice TQFT domains, $\Theta^{d+1}_{\mathrm{Kad}}(BG),\; \Theta^{d+1}_{\mathrm{Wil}}(BG)\subseteq \Theta^{d+1}(BG)$, reflect the different microscopic classes of systems for which the corresponding RG/EFT constructions are available. Here the common codomain is the $G$-visible continuum quotient defined in Eq.~\eqref{eq:G_visible_continuum_quotient}.
These examples motivate a physical consistency requirement on continuum-limit prescriptions. Although different RG/EFT procedures may have different domains of applicability, their continuum images should not depend on the operational details of the procedure whenever both prescriptions apply to the same lattice phase. We therefore impose the following assumption.

\bigskip 

\begin{assumption}[Prescription independence of the continuum image]
\label{assumption:prescription_independence}
Let
\begin{equation*}
\bigl(\Theta^{d+1}_{c_1},\kappa^{c_1}\bigr),
\qquad
\bigl(\Theta^{d+1}_{c_2},\kappa^{c_2}\bigr)
\end{equation*}
be two physically admissible continuum-limit prescriptions. Then, for every
$G\in\ofin(\mathrm{O}(d))$ and every lattice phase
\begin{equation*}
[x_G]\in
\Theta^{d+1}_{c_1}(BG)
\cap
\Theta^{d+1}_{c_2}(BG),
\end{equation*}
the two prescriptions assign the same continuum image:
\begin{equation}
\kappa^{c_1}_G([x_G])
=
\kappa^{c_2}_G([x_G])
\in
\Theta^{d+1}(B\mathrm{O}(d))_G.
\label{eq:prescription_independence}
\end{equation}
\end{assumption}
\noindent For example, taking $c_1=\mathrm{Kad}$ and $c_2=\mathrm{Wil}$ gives $\kappa_G^{\mathrm{Kad}}([x_G]) = \kappa_G^{\mathrm{Wil}}([x_G])$ whenever $[x_G]\in \Theta^{d+1}_{\mathrm{Kad}} (BG)\cap\Theta^{d+1}_{\mathrm{Wil}}(BG)$.

\bigskip

The universal continuum limit we construct in Sec.~\ref{sec:inverse_limit_construction}
should therefore be understood in two complementary ways. First, it identifies the maximal domain on which a continuum limit can be defined. For each $G$, it determines the largest subgroup $\Theta^{d+1}_{\mathrm{univ}}(BG)\subseteq \Theta^{d+1}(BG)$ of lattice phases admitting a continuum limit compatible with the full restriction structure. The domain of any other admissible continuum-limit prescription is therefore obtained by restricting this universal domain to a smaller subgroup $\Theta^{d+1}_{c}(BG)\subseteq \Theta^{d+1}_{\mathrm{univ}}(BG)$.  By Assumption~\ref{assumption:prescription_independence}, the corresponding continuum-limit map is the restriction of the universal continuum-limit map to this smaller domain. For example,
\begin{equation*}
\kappa_G^{\mathrm{Kad}} = \kappa_{G}^{\mathrm{univ}}
\big\vert_{\Theta^{d+1}_{\mathrm{Kad}}(BG)} :
\Theta^{d+1}_{\mathrm{Kad}}(BG) \longrightarrow \Theta^{d+1}(B\mathrm{O}(d))_G,
\end{equation*}
where $\Theta^{d+1}_{\mathrm{Kad}}(BG)\subseteq \Theta^{d+1}_{\mathrm{univ}}(BG)$ and similarly for $\kappa_{G}^{\mathrm{Wil}}=\kappa_{G}^{\mathrm{univ}}\vert_{\Theta^{d+1}_{\mathrm{Wil}}(BG)}$. 

\medskip

Second, whenever two admissible continuum-limit prescriptions are simultaneously defined on the same lattice phase, their continuum images agree by Assumption~\ref{assumption:prescription_independence}, and this common continuum limit image is canonically realized by the universal continuum limit. Thus, the universal construction removes dependence on the operational details of a particular prescription and isolates the continuum TQFT determined solely by the lattice phase and its symmetry structure. In this sense, distinct operational realizations or coarse-graining prescriptions may differ in detail, but whenever they describe the same lattice phase, they induce the same continuum topological response.

\bigskip 

The discussion above concerns a fixed lattice symmetry group $G_{\Lambda}$. However, $G_{\Lambda}$-continuum-limit maps should not be studied in isolation, since lattice phases with different microscopic symmetries are related by natural symmetry operations inside $\mathrm{O}(d)$. We therefore begin by organizing these groups and relations into an orbit category.

\subsection{Orbit categories of $\mathrm{O}(d)$}\label{sec:orbit_categories}

Crystalline SPT phases in flat space $\mathbb{R}^{d}$ are protected by finite point groups, each a finite subgroup of $\mathrm{O}(d)$, which we take as our model Lie group. The collection of such subgroups carries additional structure: some point groups are related by rigid re-embeddings of the same lattice, implemented by \textit{conjugation} in $\mathrm{O}(d)$, and many are related by \emph{subgroup inclusion} (one symmetry sits inside another). Both relations are made precise below and will lead naturally to the definition of an orbit category.\footnote{The orbit categories as presented here were motivated by Conner and Floyd's approach of ``families" to equivariant bordisms in Ref.~\cite{Conner1979}, see Stong's Ref.~\cite{Stong1970} for brief introduction.} By Convention~\ref{rmk:convention} (see also note~\cite{note_functor}), we keep the classifying space explicit and write $\Theta^{d+1}(BG)$ when referring to the classification corresponding to symmetry $G$ throughout.

\bigskip

\begin{figure}[H]
    \centering
    \hspace*{1cm}
    \begin{tikzcd}[column sep=2em, row sep=3em]
	&& G && \\
	P \\
	&& BG && {\Theta^{d+1}(BG)} \\
	BP &&& {\Theta^{d+1}(BP)}
	\arrow[from=1-3, to=3-3, "B(-)"]
	\arrow[""{name=0, anchor=center, inner sep=0}, from=2-1, to=1-3, "i_{\lambda }"]
	\arrow[from=2-1, to=4-1, "B(-)"]
	\arrow[from=3-3, to=3-5, "\Theta^{d+1}(-)"]
	\arrow[""{name=1, anchor=center, inner sep=0}, from=3-5, to=4-4, "\Theta^{d+1}(Bi_{\lambda })"]
	\arrow[""{name=2, anchor=center, inner sep=0}, from=4-1, to=3-3, "Bi_{\lambda }", swap]
	\arrow[from=4-1, to=4-4, "\Theta^{d+1}(-)"]
	\arrow[Rightarrow, from=0, to=2, shorten >=10pt, shorten <=10pt, "B(-)"]
	\arrow[Rightarrow, from=2, to=1, shorten >=15pt, shorten <=15pt, "\Theta^{d+1}(-)"]
\end{tikzcd}
    \begin{tikzpicture}[overlay, remember picture]
    \node at (-7, 1.6) {\textbf{Symmetry}};
    \node at (-7, 1.2) {\textbf{groups}};
    \node at (-7.7,-1.6) {\textbf{Classifying}};
    \node at (-7.7,-2) {\textbf{spaces}};
    \node at (-1, 0.4) {\textbf{Classification}};
    \node at (-1, 0) {\textbf{groups}};
    \end{tikzpicture}
    \caption{The classifying-space functor $B:\mathcal O\to\mathbf{hoCW}$ sends a morphism $i_\lambda :P\to G$ (inclusion modulo conjugation) to $B i_\lambda :BP\to BG$. The classification model $\Theta:\mathbf{hoCW}^{\mathrm{op}}\to\mathbf{Ab}$ is contravariant, so its action on $B i_\lambda $ is the restriction
$\Theta^{d+1}(B i_\lambda )=(Bi_{\lambda})^{*}=:\res^{G}_{P}[\lambda ]:\Theta^{d+1}(BG)\to \Theta^{d+1}(BP)$. $i_\lambda $ generally factors as $P \xrightarrow{\,c_\lambda \,} \lambda P\lambda ^{-1} \xhookrightarrow{\,i\,} G$, then
$\res^{G}_{P}[\lambda ]=(Bc_{\lambda})^{*}\circ (Bi)^{*}$.}
    \label{fig:functor_maps}
\end{figure}
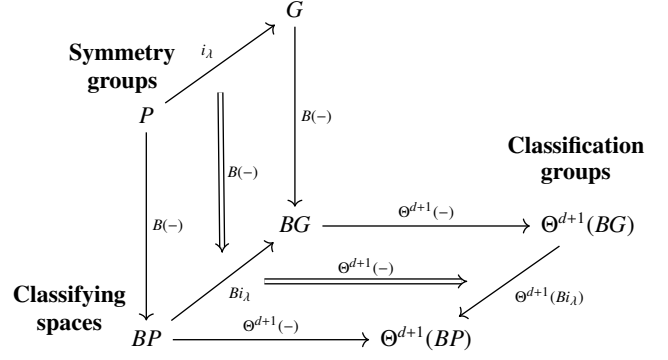

\textit{Conjugation} identifies subgroups that represent the same physical symmetry, up to a passive change of the embedding of the underlying lattice into continuous space. If $P,\,G\subset \mathrm{O}(d)$ and there exists $\lambda\in \mathrm{O}(d)$ with $P=\lambda G\lambda^{-1}$, then the group isomorphism $c_{\lambda }: G \to P,\; g\mapsto \lambda g\lambda ^{-1}$ induces an isomorphism on classifying spaces $B c_{\lambda }: BG \to BP$ and hence, by contravariance, an isomorphism on classifications:
\begin{equation*}
    (Bc_{\lambda})^{*}: \Theta^{d+1}(BP) \to \Theta^{d+1}(BG) ,
\end{equation*}
where $(Bc_{\lambda})^{*}$ is the relation between the classifications induced by the conjugation relation on groups, see Fig.~\ref{fig:functor_maps}. We emphasize that $\Theta^{d+1}(BP)$ and $\Theta^{d+1}(BG)$ remain distinct classification groups, although they are canonically isomorphic via the map $(Bc_{\lambda})^{*}$. 

\bigskip

\textit{Subgroup inclusion} records how one symmetry sits inside another.
If $P\subset G\subset \mathrm{O}(d)$, the inclusion $i:P\hookrightarrow G$ induces $Bi: BP \to BG$ and hence a restriction map
\begin{equation*}
    \res^{G}_{P}=(Bi)^{*}: \Theta^{d+1}(BG)\longrightarrow \Theta^{d+1}(BP),
\end{equation*}
\noindent $\res^{G}_{P} = (Bi)^{*} = \Theta^{d+1}(Bi)$ is the restriction morphism on classification groups induced by subgroup inclusion (see Fig.~\ref{fig:functor_maps}). Physically, restriction corresponds to forgetting part of the symmetry, i.e., probing the same phase using only $P$-backgrounds: a crystalline phase with full point-group symmetry $G$ can always be regarded as a $P$-symmetric phase for any subgroup $P\subset G$. Operationally, this restriction can be implemented by adiabatically adding a $P$-invariant symmetry-breaking perturbation that reduces the exact symmetry from $G$ to $P$ without closing the gap, in which case the quantized topological response to $P$-backgrounds is unchanged and is described by the restricted class, see Sec.~\ref{sec:restriction_morphisms}.

More generally, we consider inclusions modulo conjugation, denoted $i_{\lambda }:P\to G$, and represented by some $\lambda \in \mathrm{O}(d)$ with $\lambda P\lambda ^{-1}\subseteq G$. This map factors canonically as
\begin{equation*}
i_{\lambda }: P \xrightarrow[]{\;c_\lambda \;} \lambda P\lambda ^{-1}\; \xhookrightarrow[]{\,i\,} G.
\end{equation*}
By contravariance, the induced map on classifications is 
\begin{equation}
\res^{G}_{P}[\lambda ]:\Theta^{d+1}(BG)\xrightarrow{(Bi)^{*}}\Theta^{d+1}(B(\lambda P\lambda ^{-1}))
\xrightarrow{(Bc_\lambda )^{*}}\Theta^{d+1}(BP),
\label{eq:res_GP}
\end{equation}
where we say $\res^{G}_{P}[\lambda ]:=(Bc_\lambda )^{*}\circ (Bi)^{*}$ (see Fig.~\ref{fig:functor_maps}) as a composition of group homomorphisms. 

Together, the conjugation and inclusion relations define the structure encoded by an orbit category: its objects are finite subgroups of $\mathrm{O}(d)$, and its morphisms are inclusions modulo conjugation. Working in this category removes redundancies (from conjugate embeddings) and enforces consistency (via restriction along inclusions). We now define the following important orbit categories:

\medskip
\noindent\textbf{(1) Finite–subgroup orbit category $\ofin(\mathrm{O}(d))$.}
Objects are finite subgroups $P\le \mathrm{O}(d)$. For two objects $P,G\in\ofin(\mathrm{O}(d))$, a morphism $P\to G$ is an injective group homomorphism induced by conjugation in $\mathrm{O}(d)$. Explicitly, the morphisms are
\begin{equation*}
\operatorname{Mor}_{\ofin(\mathrm{O}(d))}(P,G)
=
\left\{
c_\lambda:P\to G
\;\middle|\;
\lambda\in \mathrm{O}(d),\;
\lambda P\lambda^{-1}\leq G
\right\},
\end{equation*}
where $c_\lambda(p)=\lambda p\lambda^{-1}$, for $\lambda\in \mathrm{O}(d)$. Defining $N_{\mathrm{O}(d)}(P,G) := \left\{ \lambda\in \mathrm{O}(d) \;\middle|\; \lambda P\lambda^{-1}\leq G \right\}, $ we may equivalently write 
\begin{equation*} 
\operatorname{Mor}_{\ofin(\mathrm{O}(d))}(P,G) \cong N_{\mathrm{O}(d)}(P,G)/C_{\mathrm{O}(d)}(P), 
\end{equation*}
where $N_{\mathrm{O}(d)}(P,G), C_{\mathrm{O}(d)}(P)$ are the normalizer and centralizer respecitvely, because two elements $\lambda,\mu\in N_{\mathrm{O}(d)}(P,G)$ induce the same homomorphism precisely when $\mu^{-1}\lambda\in C_{\mathrm{O}(d)}(P)$.\footnote{The map
$N_{\mathrm{O}(d)}(P,G)\to \operatorname{Mor}_{\ofin(\mathrm{O}(d))}(P,G)$,
$\lambda\mapsto c_\lambda$, is surjective. If
$z\in C_{\mathrm{O}(d)}(P)$, then $c_{\lambda z}=c_\lambda$, so right multiplication
by an element of the centralizer does not change the induced morphism.
Conversely, if $c_\lambda=c_\mu$, then
$\mu^{-1}\lambda\in C_{\mathrm{O}(d)}(P)$.} Thus, the category lists every discrete point-group symmetry and remembers how smaller symmetry groups embed inside larger ones after an appropriate conjugation inside $\mathrm{O}(d)$.

\medskip
\noindent\textbf{(2) 2-primary subcategory $\ofin^{(2)}(\mathrm{O}(d))$.}
This is the full subcategory of $\ofin(\mathrm{O}(d))$ whose objects are the finite $2$-groups (groups with order $2^{n}$ where $n\in \mathbb{N}$).
Morphisms are inherited from $\ofin(\mathrm{O}(d))$.
Focusing on $\ofin^{(2)}(\mathrm{O}(d))$ isolates exactly the sub-category relevant for $\mathbb{Z}_{2}$-valued classifications.

\medskip
\noindent\textbf{(3) Quillen category $\qa(\mathrm{O}(d))$.}
This is a full sub-category of both $\ofin(\mathrm{O}(d)),$ and $\ofin^{(2)}(\mathrm{O}(d))$ with the elementary abelian $2$-subgroups of the form $E\cong(\mathbb{Z}_{2})^{r}$ as the objects. Morphisms are again inherited from the larger categories defined above.

\begin{widetext}

\begin{figure}[H]
  \makebox[\textwidth][c]{%
    \includegraphics[width=0.83\textwidth]{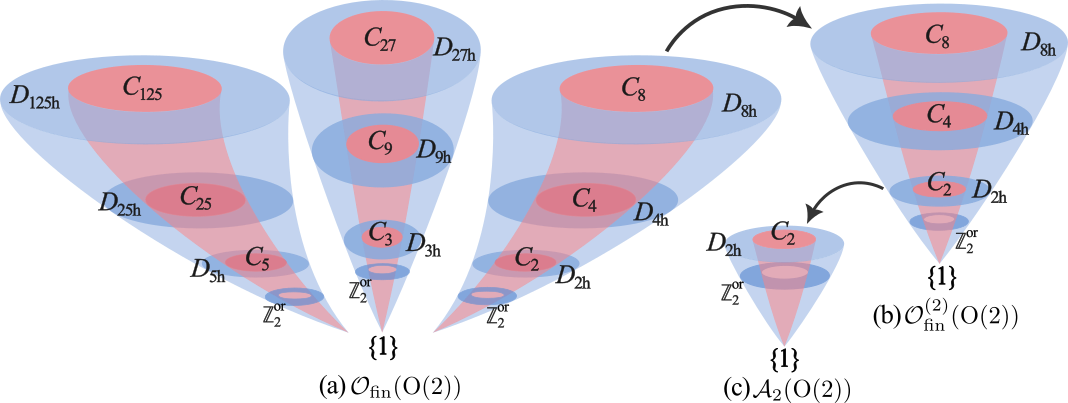}%
  }%
  \caption{Orbit categories for $G=O(2)$. \textbf{(a)} The full finite–subgroup orbit category $\ofin(O(2))$ restricted to primes $p\in\{2,3,5\}$ and orders $p^{s}$ with $s\le 3$. Each cone collects the $C_{p^{s}}$ (red) and $D_{p^{s}h}$ (blue) families for a fixed $p$. One representative per conjugacy class is shown. Upward direction indicates subgroup inclusions; arrows are omitted and morphisms are inclusions up to conjugation. \textbf{(b)} The $2$-primary subcategory $\ofin^{(2)}(O(2))$ with objects as groups of order $2^{n}$. \textbf{(c)} Quillen's elementary abelian subgroup sub-category with groups of the form $E\cong \Z_{2}^{n}$.}
  \label{fig:big}
\end{figure}

\end{widetext}

To make the picture concrete, set $d=2$ so $G = O(2)$ and consult Fig.~\ref{fig:big}.  
The three orbit categories form the nested sequence  
\begin{equation*}
  \qa\bigl(O(2)\bigr)
  \;\subset\;
  \ofin^{(2)}\bigl(O(2)\bigr)
  \;\subset\;
  \ofin\bigl(O(2)\bigr).
\end{equation*}
\noindent Figure \ref{fig:big}(a) shows the full finite–subgroup orbit category $\ofin(O(2))$. The diagram is truncated at exponent $s\le 3$; in other words, for each prime $p\in\{2,3,5\}$ we display only the subgroups of orders $p,\,p^{2},\,p^{3}$. Each cone gathers the cyclic and dihedral families $\{C_{p^{s}},\,D_{p^{s}h}\}$; moving upward in a cone raises $s$ (larger prime power), while different cones correspond to different primes. The apex is the trivial group $\{1\}$. Only one representative from each orbit (under conjugation) is drawn. Figure \ref{fig:big}(b) zooms in on the $2$-primary subcategory $\ofin^{(2)}(O(2))$. In two dimensions the finite $2$-subgroups of $O(2)$ are precisely the cyclic groups $C_{2^{s}}$ and the dihedral groups $D_{2^{s}h}$ (again shown for $s\le 3$). Figure \ref{fig:big}(c) isolates the elementary-abelian core, which is denoted by $\qa(O(2))$. With $d=2$, $\qa(O(2))$ reduces to just three non-trivial orbits: the rotation subgroup $C_{2}$, the reflection or orientation reversing subgroup $\Z_2^{\mathrm{or}},$ and the rank-two dihedral group $D_{2h}\cong \Z_{2}^{x}\times \Z_{2}^{y},$ the latter of which is generated by reflections about two chosen perpendicular axes.

\section{Inverse Limit and the Universal Crystalline SPT Group}\label{sec:inverselimit} We have now developed the machinery needed to define the universal crystalline SPT ($\mathrm{cSPT}^{d+1}_{\mathrm{univ}}$), which is an important ingredient in the mapping procedure outlined in Fig. \ref{fig:funnel}. The construction uses two inputs:

\smallskip
\noindent a) \textit{Classification model.} 
Fix a $(d{+}1)$-dimensional classification model in the sense of Sec.~\ref{sec:classification_models}:
\begin{equation*}
\Theta^{d+1}:\mathbf{hoCW}^{\mathrm{op}}\to \mathbf{Ab}.
\end{equation*}
In particular, for a symmetry group $G$ we write $\Theta^{d+1}(BG)$ for the abelian group classifying invertible $G$-TQFTs in $(d{+}1)$ dimensions (see Convention~\ref{rmk:convention} and note~\cite{note_functor}). In this work our primary examples are the group cohomology and Borel-equivariant cobordism models introduced in Sec.~\ref{sec:classification_models}.

\bigskip

\noindent b) \textit{Orbit category of crystalline subgroups}: 
For a model Lie group $\mathcal{G}$ (in this work $\mathcal{G}=\mathrm{O}(d)$), consider the category whose \emph{objects} are finite subgroups $P,\, Q\le \mathcal{G},$ and whose \emph{morphisms} $i_{g}:P\to Q$ are inclusions up to conjugation, i.e., $i_{\lambda}:P\to \lambda P \lambda^{-1}\hookrightarrow Q$ for some $\lambda \in \mathcal{G}$. 

\bigskip 

The previous section rigorously defined different orbit categories of interest, and now we can put them to work. Think of an orbit category as a directed graph: each \emph{node} is a finite subgroup $P \leq \mathrm{O}(d)$, and each \emph{arrow} represents an inclusion $i_{\lambda }:P \hookrightarrow Q$ up to conjugacy. If we now ``run'' our classification functor $\Theta^{d+1}(-)$ on this graph,  
every node gets assigned an abelian group $\Theta^{d+1}(BP)$ of phases, and every arrow gets assigned the corresponding restriction map  
\begin{equation*}
\res^{Q}_{P}[\lambda ] \colon \Theta^{d+1}(BQ) \longrightarrow \Theta^{d+1}(BP).
\end{equation*} 
See Eq.~\ref{eq:res_GP} and Fig.~\ref{fig:functor_maps} for more details. 

\begin{figure}[H]
    \centering
    \includegraphics[width=0.95\linewidth]{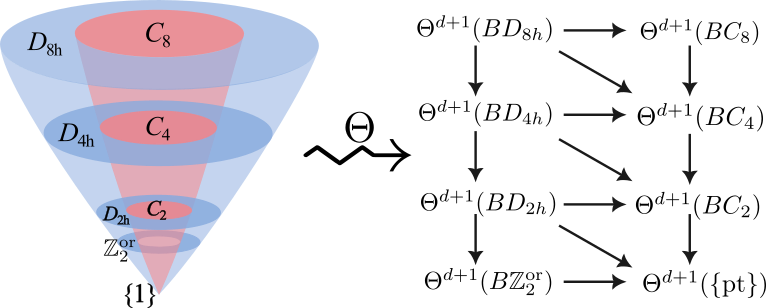}
    \caption{Functorial mapping from the orbit category of 2-subgroups $\ofin^{(2)}(O(2))$(up to degree $n=8$) to a diagram of abelian groups.}
    \label{fig:functor}
\end{figure}

We represent this process in Fig.~\ref{fig:functor} which shows a \emph{functorial mapping} from the orbit category of $2$-subgroups $\ofin^{(2)}(O(2))$ into a network of abelian groups with restriction homomorphisms between them. Note that because the classification model is a \emph{contravariant} functor, the restriction arrows in Fig. \ref{fig:functor} point oppositely to the inclusion maps of the orbit category. We also remark that we will sometimes refer to the orbit categories $\mathcal{O}$ as \textit{indexing categories}, over which we apply the classification functor $\Theta^{d+1}$ to obtain the \textit{diagram}, denoted $\Theta^{d+1}(\mathcal{O})$, which is the RHS in Fig.~\ref{fig:functor}.

\medskip 

Our goal is to distill this entire diagram, i.e., the RHS of Fig.~\ref{fig:functor}, into a \emph{single} classification group that captures all phases detected on \emph{every} subgroup at once, but without over-counting, i.e., if a phase on one subgroup is just the restriction of a phase from another subgroup, those should be identified. Our goal can be achieved using a formal limit defined in category theory called the \emph{inverse limit}. We will make this idea precise by introducing the idea of \emph{compatible families}, which we will now describe.

\subsection{Compatible Families}
\label{sec:compatible_families}

Given an orbit category $\mathcal O(\mathrm{O}(d))$ of subgroups of $\mathrm{O}(d)$, a \emph{compatible family} is a collection of invertible topological phases
\begin{equation*}
   \bigl([x_G] \in \Theta^{d+1}(BG)\bigr)_{G \in\mathrm{Obj}(\mathcal O(\mathrm{O}(d)))},
\end{equation*}
one for each symmetry group $G\in \mathrm{Obj}(\mathcal O(\mathrm{O}(d)))$, such that the collection satisfies compatibility under restriction relations
\begin{equation*}
   [x_P]=\res^{Q}_{P}[\lambda]\bigl([x_Q]\bigr)
   \quad\text{for every arrow }i_{\lambda}: P\longrightarrow Q\text{ in }\;\mathcal O(\mathrm{O}(d)).
\end{equation*} 
In words, a compatible family is an assignment of a phase to each subgroup such that, for every arrow $i_\lambda:P\to Q$, the element at $P$, denoted $[x_{P}]\in \Theta^{d+1}(BP)$, equals the restriction of the element at $Q$, denoted $[x_{Q}]\in \Theta^{d+1}(BQ)$. Assignments that fail any such consistency check are simply not “compatible” and therefore do not form a compatible family. Within the compatible family, denoted as $([x_{G}])_{G\in \mathcal{O}(\mathrm{O}(d))}$, we refer to an individual entry $[x_{G}]\in\Theta^{d+1}(BG)$ as the $G$-component of the family.

\bigskip 

\begin{figure}[H]
    \centering
    \includegraphics[width=0.75\linewidth]{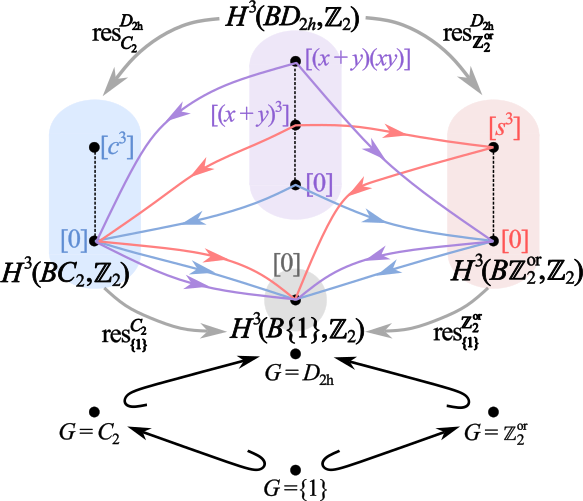}
    \caption{Formation of compatible families. The set of phases connected by a single colored line form one compatible family. We use the pre-TQFT classification model and display only phases in the Weyl-invariant subgroups, i.e., $H^{d+1}(BG, \Z_{2})^{W_{O(2)}(G)}$, for each $G$.}
    \label{fig:quillen_compat}
\end{figure}

Let us demonstrate this concept with an example where $G=O(2)$ and we consider the Quillen category $\qa(O(2))$ which, up to conjugation, has only four symmetry groups. Specifically, the Quillen category is comprised of the dihedral group $D_{2h}=\Z_2^{x}\times \Z_2^{y},$ (generated by reflections across a choice of perpendicular axes denoted $x,\;y$), the cyclic group $C_2$ (generated by $\pi$-rotations in the $xy$-plane), the reflection group about a chosen plane, denoted $\Z_2^{\mathrm{or}}$, and the trivial group denoted $\{1\}$. The structure of the groups is illustrated in the lower graph of Fig. \ref{fig:quillen_compat} where the hooked arrows represent subgroup inclusion. 
Now we can act on the groups with the cohomology classification model for pre-TQFTs with mod-$2$ coefficients, i.e., $H^{d+1}(B-, \Z_{2})$. We use the pre-TQFT classification functor $\Theta^{d+1}(BG)= H^{d+1}(BG, \Z_{2})$ for convenient illustration purposes, but the same discussion can be lifted (see App.~\ref{app:lift} for details) to the full TQFT classification functor that uses integral coefficients $\Theta^{d+1}(BG)=\tor_{2}(H^{d+2}(BG, \Z_{w_{1}}))$ as we do in Sec.~\ref{sec:example_1}.  The classification ring of the dihedral group is generated by two degree one classes associated to two reflection generators, and denoted respectively $x \in H^{1}(BD_{2h}, \Z_2),\; y \in H^{1}(BD_{2h}, \Z_2)$ such that the classification in $d=2$ is given by $H^{3}(BD_{2h}, \Z_{2}) = \mathrm{Span}_{\mathbb{Z}_{2}}\{x^{3}, y^{3}, xy^{2}, x^{2}y\}=\mathbb{Z}_{2}^{4}$.\footnote{For each fixed degree $n$, e.g., $n=d+1$, the mod-$2$ cohomology group of any $BG$, $H^{d+1}(BG, \Z_{2})$, is also a finite vector space over $\Z_{2}$ generated by a basis of cohomology classes.} The cohomology classifications of both the rotation $C_2$ and the reflection $\Z_{2}^{\mathrm{or}}$ groups are each generated by a single generator (denoted $c \in H^{1}(BC_{2}, \Z_2)$ and $s\in H^{1}(B\Z_{2}^{\mathrm{or}}, \Z_2)$ respectively), and are given by $H^{3}(BC_{2}, \Z_2) = \mathrm{Span}_{\mathbb{Z}_{2}}\{c^{3}\} =\Z_2$ and $H^{3}(B\Z_{2}^{\mathrm{or}}, \Z_2) = \mathrm{Span}_{\mathbb{Z}_{2}}\{s^{3}\} =\Z_2$. 

\medskip

To illustrate compatible families we can focus on the top half of Fig. \ref{fig:quillen_compat} where we have restricted the classes for each group to only isotropic (Weyl-invariant) classes. Each compatible family is a full ``loop" of $G$-components which agree under the restriction morphisms.  For example, the red loop represents the $G$-components 
\begin{gather*}
[x_{D_{2h}}]=[(x+y)^3]\in H^{3}(BD_{2h}, \Z_{2}),\;
[x_{C_2}]=[0]\in H^{3}(BC_{2}, \Z_{2}),\\
[x_{\Z_2^{\mathrm{or}}}]=[s^3]\in H^{3}(B\Z_2^{\mathrm{or}}, \Z_{2}), \quad [x_{\{1\}}]=[0]\in H^{3}(B\{1\}, \Z_{2}),
\end{gather*}
and these $G$-components agree pairwise under the restriction maps, e.g., $\res^{D_{2h}}_{\mathbb{Z}_{2}^{\mathrm{or}}}([(x+y)^{3}]) = [s^{3}]$. One conclusion we can draw from this calculation is that the non-trivial phase represented by the class $[(x+y)^3]\in H^{3}(BD_{2h}, \Z_{2})$ becomes a trivial phase if $D_{2h}$ is broken down to the $C_{2}$ rotation symmetry, but remains non-trivial if $D_{2h}$ is broken down to a single reflection $\Z_{2}^{\mathrm{or}}$ symmetry. Furthermore, all non-trivial phases become trivial if the symmetry is fully broken down to the trivial group. In contrast, the class $[c^{3}]\in H^{3}(BC_{2},\Z_{2})$ does not occur as the $C_{2}$-component of any compatible family in the elementary inverse limit over $\qa(O(2))$ shown in Fig.~\ref{fig:quillen_compat}. It therefore cannot be extended coherently over the full Quillen category $\qa(O(2))$ and is obstructed from admitting an $\mathrm O(2)$-structured continuum-limit image.\footnote{This obstruction is specific to the $\mathrm O(2)$-structured continuum limit problem. The corresponding class admits an $\mathrm{SO}(2)$-compatible extension, and therefore admits a continuum limit to an $\mathrm{SO}$-structured continuum limit theory. See Apps.~\ref{app:O_SO_admissible_domain_example}.}

\subsection{Universal cSPT}\label{sec:univCSPT}
With our understanding of compatible families, we are now in a position to  efficiently describe all possible cSPTs in terms of a catagory-theoretic limit. To do so, let us consider the category $\ofin(\mathrm{O}(d))$ which has all discrete point-groups as objects. The set of all compatible families built from this category forms an abelian group under component-wise addition\footnote{If $([x_{G_{\Lambda}}])_{G_{\Lambda}}$ and $([y_{G_{\Lambda}}])_{G_{\Lambda}}$ are two compatible families, we add them subgroup-by-subgroup (component-wise): $([x_{G_{\Lambda}}])_{G_{\Lambda}}+([y_{G_{\Lambda}}])_{G_{\Lambda}}:=([x_{G_{\Lambda}}]+[y_{G_{\Lambda}}])_{G_{\Lambda}}$, where $[x_{G_{\Lambda}}]+[y_{G_{\Lambda}}]$ is the sum in $\Theta^{d+1}(BG_{\Lambda})$.\label{footnote_comp}}. This group is called the \emph{inverse limit} of the classification diagram generated from our chosen orbit category:
\begin{equation*}
   \csptu
   \;:=\;
   \varprojlim_{G_{\Lambda}\in\ofin(\mathrm{O}(d))}\,
      \Theta^{d+1}(BG_{\Lambda}).
\end{equation*}
\noindent This limit can be understood as a collection of all lattice topological phases, with compatibility relations imposed on classes whose symmetries are related by conjugation or inclusion:
\begin{equation}
  \csptu =\dfrac{\left\{([x_{G_{\Lambda_{1}}}], [x_{G_{\Lambda_{2}}}], \cdots) \in \prod\limits_{G_{\Lambda}\in \ofin(\mathrm{O}(d))} \Theta^{d+1}(BG_{\Lambda}) \right\}}{\raisebox{-1.5ex}{$\widetilde{\phantom{XXXX}}$}}.
  \label{eq:cspt_equiv}
\end{equation} 

The category theoretic \textit{universality} of the inverse limit endows it with special properties. First, it comes with canonical projection maps onto all individual lattice classifications: 
\begin{equation*}
\pi_{\Lambda}: \csptu \to \Theta^{d+1}(BG_{\Lambda}),\quad \forall \;\; G_{\Lambda}\in \ofin(\mathrm{O}(d)). 
\end{equation*}
\noindent 
These projections single out a canonical subgroup
\begin{equation*}
    \im(\pi_{\Lambda})\subseteq \Theta^{d+1}(BG_{\Lambda}),
\end{equation*}
consisting of those $G_{\Lambda}$-phases that extend to compatible families over the full orbit category. In Sec.~\ref{sec:inverse_limit_construction} we will show that any phase admitting a continuum-limit must lie inside this subgroup.

\bigskip

\begin{figure}[H]
  \begin{center}
    \includegraphics[width=0.4\textwidth]{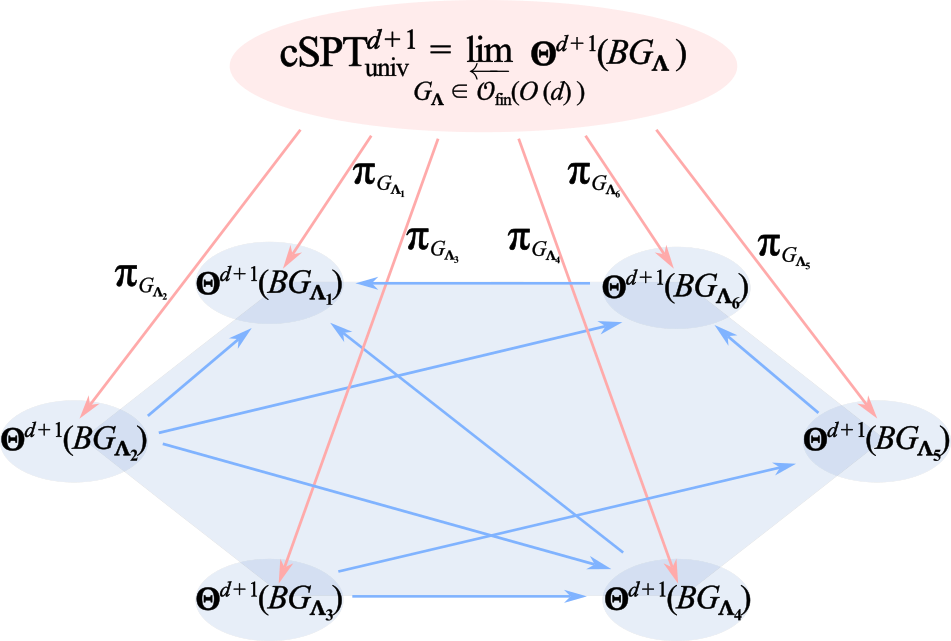}
  \end{center}
  \caption{Schematic illustration of an inverse limit $\csptu$ forming a \textit{cone} over the diagram of lattice classifications. Restriction maps are arrows in blue, and projection maps are arrows in red.}
  \label{fig:cone}
\end{figure}

The second special property is that the inverse limit, along with its projection maps, forms a \textit{cone} over the diagram of lattice classifications, see Fig.~\ref{fig:cone}. This means that the set of projection maps $\{\pi_{\Lambda}: \csptu \to \Theta^{d+1}(BG_{\Lambda})\}$ commute with the restriction relations among lattice classifications. For example, if we have a restriction morphism among lattice classifications $\res^{K}_{H}: \Theta^{d+1}(BK) \to \Theta^{d+1}(BH)$, then projection maps $\pi_{H}, \;\pi_{K}$ satisfy the relation $\pi_{H}= \res^{K}_{H}\circ \pi_{K}$.

The third special property of the inverse limit, also called the \textit{universal property}, states that any cone into the diagram $\Theta^{d+1}$ factors through the inverse limit. 
\begin{equation}
\begin{tikzcd}
  & X
    \arrow[d, "f", "\exists !"']
    \arrow[ddl, "f_{G_{\Lambda_{i}}}"']
    \arrow[ddr, "f_{G_{\Lambda_{j}}}"]
  & \\
  & \csptu
    \arrow[dl, "\pi_{G_{\Lambda_{i}}}"]
    \arrow[dr, "\pi_{G_{\Lambda_{j}}}", swap]
  & \\
  \Theta^{d+1}(BG_{\Lambda_{i}}) \arrow[rr, "f_{ij}"'] & &  \Theta^{d+1}(BG_{\Lambda_{j}})
\end{tikzcd}
\label{eq:universal_property}
\end{equation}
In other words, if there exists an object $X$ with a set of maps $\{f_{G_{\Lambda}}: X \to \Theta^{d+1}(BG_{\Lambda})\}$ for all $G_{\Lambda}\in \ofin(\mathrm{O}(d))$ such that the maps commute with the restriction morphisms, then there exists a unique map $f: X \to \csptu$. An immediate consequence of the universal property is that the inverse limit is unique up to isomorphism.\footnote{If $(L,\{\pi_j\})$ and $(L',\{\pi'_j\})$ are both limits of the same diagram, then the universal property gives unique maps $f:L\to L'$ and $g:L'\to L$ commuting with all projection maps. Applying the universal property once more shows that $g\circ f=\mathrm{id}_L$ and $f\circ g=\mathrm{id}_{L'}$, so $f$ is an isomorphism.} The physical significance of $\csptu$ for the continuum-limit problem is established in Sec.~\ref{sec:inverse_limit_construction}, where Theorem~\ref{thm:finite_to_global_compatibility} shows that every lattice phase admitting a continuum limit can be completed to a compatible family over the full orbit category. Thus, $\csptu$ provides the natural global repository of lattice data relevant to continuum limits, and is the unique such repository.

\par
\vspace{0.5\baselineskip}
\begin{widetext}

\begin{figure}[H]
  \makebox[\textwidth][c]{%
    \includegraphics[width=0.9\textwidth]{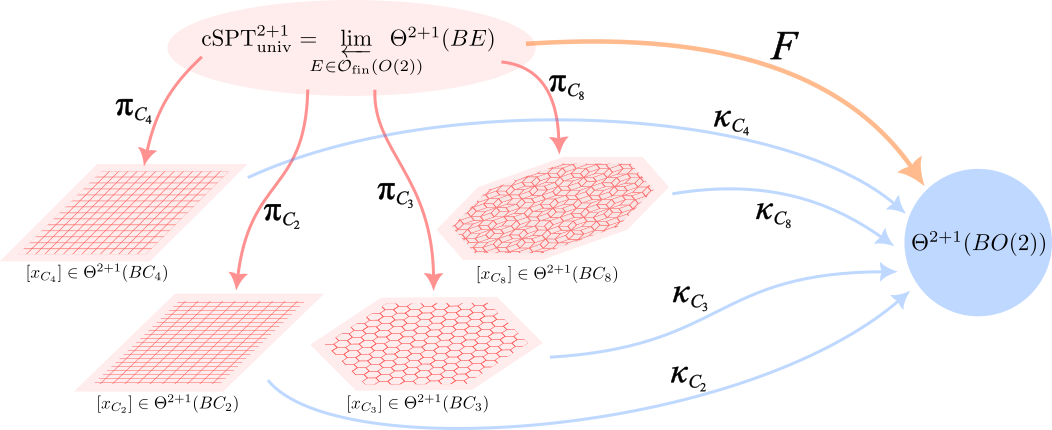}%
  }%
  \caption{\textbf{Universal continuum limit.} For each finite crystalline point group $G_{\Lambda}\le \mathrm{O}(d)$ the lattice classification $\Theta^{d+1}(BG_{\Lambda})$ is depicted (red tiles) and the single-symmetry continuum map $\kappa_{G_{\Lambda}}$ is indicated (blue arrows). The inverse limit over the orbit category $\ofin(\mathrm{O}(d))$ produces the universal object $\csptu$ with canonical projections $\pi_{G_{\Lambda}}:\csptu\to\Theta^{d+1}(BG_{\Lambda})$ (red arrows), see Sec.~\ref{sec:inverselimit}. The universal continuum map $F:\csptu\to\Theta^{d+1}(B\mathrm{O}(d))$ (orange), and its component-wise factorization (see Fig.~\ref{eq:square_factorization}), defines the collection of individual continuum limit maps $\{\kappa_{G_\Lambda}\}$, and equivalently defines the universal theory of continuum limits.}
  \label{fig:continuum_limit}
\end{figure}

\end{widetext}

\subsection{The universal continuum limit map}
\label{sec:univ_cont_limit}
The inverse-limit construction therefore identifies a canonical lattice-side object, $\csptu$ whose elements are families of lattice phases which admit continuum limits. This makes it natural to consider a universal continuum limit map
\begin{equation*}
F:\csptu\to \Theta^{d+1}(B\mathrm{O}(d)),
\end{equation*}
from $\csptu$ to the continuum classification. The detailed construction of this map, together with the proof that it is canonically defined and surjective, will be given in Sec.~\ref{sec:rigor}. For now, we summarize the intended payoff of the construction. The inverse limit assembles the separate lattice classifications into the single universal object $\csptu$, while the map $F$ assigns a continuum class to each compatible family. From the perspective of defining the continuum limit of $G_{\Lambda}$-symmetric lattice phases, the relevant continuum datum is the image of this class in the $G_{\Lambda}$-visible continuum quotient $\Theta^{d+1}(B\mathrm{O}(d))_{G_{\Lambda}}$ defined in Eq.~\eqref{eq:G_visible_continuum_quotient}. Accordingly, we consider the composite
\begin{equation*}
\begin{tikzcd}
\csptu \arrow[r, "F"] \arrow[rr, bend left=30, "\mathrm{pr}_{G_{\Lambda}}\circ F"]& \Theta^{d+1}(B\mathrm{O}(d)) \arrow[r, "\mathrm{pr}_{G_{\Lambda}}"] &\Theta^{d+1}(B\mathrm{O}(d))_{G_{\Lambda}}
\end{tikzcd}.
\end{equation*}
Under the local domain-splitting condition analyzed in App.~\ref{app:continuum_limit_component},\footnote{This condition is verified for $d=1,2$ and assumed in general dimensions. It is needed only for the componentwise factorization into the local maps $\kappa^{\mathrm{univ}}_{G_{\Lambda}}$; the global map $F$ and the main theorem do not depend on it.} this composite depends only on the $G_{\Lambda}$-component $\pi_{G_{\Lambda}}(\mathbf{x})\in\mathrm{Im}(\pi_{G_{\Lambda}})$ of the compatible family $\mathbf{x}\in\csptu$, and therefore factors through the canonical projection $\pi_{G_{\Lambda}}$. This factorization induces the individual continuum-limit map $\kappa^{\mathrm{univ}}_{G_{\Lambda}}$ on $\mathrm{Im}(\pi_{G_{\Lambda}})$, as summarized by the commuting square in Eq.~\eqref{eq:square_factorization}. Thus, the entire family of lattice-by-lattice continuum-limit maps is encoded by the single global map $F$ together with its componentwise factorizations. The overarching structure of the construction is illustrated in Fig.~\ref{fig:continuum_limit}.

More precisely, as we will discuss in Sec.~\ref{sec:continuum_limit}, and formally establish in App.~\ref{app:continuum_limit_component}, for each lattice symmetry group $G_{\Lambda}\in\ofin(\mathrm{O}(d))$, the universal continuum limit map fits into the commuting square:
\begin{equation}
\label{eq:square_factorization}
    \begin{tikzcd}
        \csptu \arrow[r, "F"] \arrow[d, "\pi_{G_{\Lambda}}"] &\Theta^{d+1}(B\mathrm{O}(d)) \arrow[d, "\mathrm{pr}_{G_{\Lambda}}"']\\
        \mathrm{Im}(\pi_{G_{\Lambda}}) \arrow[r, "\kappa^{\mathrm{univ}}_{G_{\Lambda}}"] &\Theta^{d+1}(B\mathrm{O}(d))_{G_{\Lambda}}
    \end{tikzcd}.
\end{equation}
The commutativity of this square uniquely specifies the individual continuum limit map $\kappa^{\mathrm{univ}}_{G_{\Lambda}}$. Here the lower-right entry is the $G_{\Lambda}$-visible continuum quotient recalled above. Its formal construction, together with the definition of the projection map $\mathrm{pr}_{G_{\Lambda}}$, is given in Sec.~\ref{sec:refine_domain}. The commuting square also makes clear that the individual continuum-limit maps are defined on only
\begin{equation*}
\mathrm{Im}(\pi_{G_{\Lambda}}) \subseteq \Theta^{d+1}(BG_{\Lambda}),
\end{equation*}
the subgroup of $G_{\Lambda}$-phases that extend to compatible families over the full orbit category. Thus, the individual maps are not independent continuum-limit prescriptions, but the componentwise manifestations of the single universal map $F$. We refer collectively to the family
\begin{equation*}
\left\{
\kappa_{G_{\Lambda}}^{\mathrm{univ}}:
\mathrm{Im}(\pi_{G_{\Lambda}})
\longrightarrow
\Theta^{d+1}(B\mathrm{O}(d))_{G_{\Lambda}}
\right\}_{G_{\Lambda}\in\ofin(\mathrm{O}(d))}
\end{equation*}
as the \textit{universal continuum limit theory}.

\bigskip 

In the next section we use physical principles to formalize the notion of a generalized continuum limit theory, namely a prescription of continuum-limit maps for all lattice symmetry groups $G_{\Lambda}$ satisfying suitable coherence conditions. We then show that the \emph{universal} continuum limit theory introduced here gives the most general such theory, in the sense that any other continuum limit theory can be understood as a more restricted version of it.

\section{Constructing the continuum limit map}
\label{sec:continuum_limit}

Sections~\ref{sec:motivation} and~\ref{sec:inverselimit} developed the formulation of a continuum limit from a global perspective. By organizing lattice phases over the orbit category and passing to an inverse limit, we obtained a universal lattice classification limit $\csptu$. We then proposed that the map $F: \csptu \to \Theta^{d+1}(B\mathrm{O}(d))$ provides a global continuum limit map from compatible families of lattice classifications to the continuum. In this section we reverse this viewpoint and begin instead with the continuum limit for a single lattice symmetry group $G$. We then ask what physically motivated compatibility conditions such lattice-by-lattice continuum limits must satisfy under symmetry restriction. This leads to the notion of a continuum limit theory: a coherent prescription of continuum-limit maps for all lattice symmetry groups. Our main result is that the universal construction is terminal among continuum limit theories, in the sense that every other admissible prescription can be understood as a more restricted version of it.

\bigskip 

Let $\ofin(\mathrm{O}(d))$ be the category of lattice symmetry groups of systems for which we want to construct the continuum limit. At a first pass, a continuum limit prescription is a choice, for each $G\in \ofin(\mathrm{O}(d))$, of a subgroup
\begin{equation*}
    \Theta^{d+1}_c(BG)\subseteq \Theta^{d+1}(BG),
\end{equation*}
of lattice TQFT classes that admit a continuum limit under the prescription, together with a map
\begin{equation*}
    \kappa_G:\Theta^{d+1}_{c}(BG)\to \Theta^{d+1}(B\mathrm{O}(d)).
\end{equation*}
We will replace this naive codomain in Sec.~\ref{sec:refine_domain} by the $G$-visible continuum quotient introduced in Eq.~\eqref{eq:G_visible_continuum_quotient}, and derive the codomains explicit description in terms of quotienting out phases that are invisible to all $G$-symmetric probes. A central question in what follows is how large the subgroup $\Theta^{d+1}_{c}(BG)$ of phases having admissible continuum limits can be.

\subsection{Refining the co-domain of the continuum limit maps.}
\label{sec:refine_domain}

Fix $G\in\ofin(\mathrm{O}(d))$ and suppose $[x_G]\in \Theta^{d+1}_{c}(BG)$ admits a continuum limit. A natural question is whether the lattice class $[x_G]$ can determine a \emph{unique} continuum class $[x]\in\Theta^{d+1}(B\mathrm{O}(d))$. 
It would stand to reason that this would not be true in general, because a continuum class arises from a larger symmetry group, and therefore a lattice class may be compatible with more than one continuum limit class. Thus,  we must work to understand the possible ambiguities in the continuum image. 
We address this using the probe-response viewpoint of Sec.~\ref{sec:top_resp}. Explicitly, we take the perspective that even though the continuum limit of a lattice TQFT may exhibit enlarged symmetry, the continuum limit phase should be probed \emph{only by background fields compatible with the microscopic lattice symmetry}. Hence, the probes available in this setting are symmetry backgrounds encoded in the maps $f_P:M^{d+1}\to BP$ for subgroups $P\le G$. 

We can now use these physically available probes to specify when two candidate continuum limits should be regarded as equivalent from the perspective of a $G$-symmetric lattice system. The key point is that if two continuum phases give identical responses to every \emph{allowed} probe, then they are operationally indistinguishable. To make this insight precise, let us now show how the continuum target should be quotiented by phases invisible to all such probes.

\bigskip 

Given a candidate continuum limit class $[x]\in \Theta^{d+1}(B\mathrm{O}(d))$, its response to an allowed probe $f_{P}: M^{d+1}\to BP$ is given by the partition function: 
\begin{equation*}
    \mathcal{Z}_{x}(M^{d+1}, f_{P})=\langle f_{P}^{*}\res^{\mathrm{O}(d)}_{P}([x]),\;[M^{d+1}]\;\rangle\;.
\end{equation*}
Now suppose $[x_{1}],[x_{2}]\in \Theta^{d+1}(B\mathrm{O}(d))$ are two candidate continuum limits. If they give identical responses for every probe built from a subgroup $P\le G$, then they are indistinguishable by the available measurements:
\begin{equation*}
    \mathcal{Z}_{x_{1}}(M^{d+1},f_{P})
    =
    \mathcal{Z}_{x_{2}}(M^{d+1},f_{P})
    \quad
    \forall\, (f_{P}:M^{d+1}\to BP,\; P\le G).
\end{equation*}
By the probe-detectability lemma (see Lemma~\ref{lemma:probe}) for $P$-TQFTs, this implies
\begin{equation*}
    \res^{\mathrm{O}(d)}_{P}([x_{1}]-[x_{2}])=0
    \qquad \forall\, P\le G.
\end{equation*}
Equivalently, the ambiguity $[\delta x]:=[x_{1}]-[x_{2}]$ lies in the subgroup $K_{\mathcal{O}(G)}$ of continuum phases that are invisible to all $G$-allowed probes, i.e.,
\begin{equation*}
    [\delta x] \in K_{\mathcal{O}(G)}
    \;:=\;
    \bigcap_{P\le G}\ker\bigl(\res^{\mathrm{O}(d)}_{P}\bigr)\;\subseteq\;\Theta^{d+1}(B\mathrm{O}(d)).
\end{equation*}

Thus, once one continuum representative $[x]\in\Theta^{d+1}(B\mathrm{O}(d))$ is chosen, any shift $[x]\mapsto [x]+[\delta x]$ with $[\delta x]\in K_{\mathcal{O}(G)}$ produces the same responses to all $G$-allowed probes. In other words, the continuum limit for $G$-symmetric lattice TQFTs is well-defined only up to the addition of $G$-invisible phases. Accordingly, the space of continuum data that is detectable from a $G$-symmetric lattice system is the quotient
\begin{equation}
    \Theta^{d+1}(B\mathrm{O}(d))_{G}\;:=\;\dfrac{\Theta^{d+1}(B\mathrm{O}(d))}{K_{\mathcal{O}(G)}},
    \quad K_{\mathcal{O}(G)} = \bigcap_{P\leq G}\ker\bigl(\res^{\mathrm{O}(d)}_{P}\bigr).
    \label{eq:quotient}
\end{equation}

\noindent Hence, it follows that a more meaningful preliminary formulation for a continuum limit prescription is a family of maps
\begin{equation}
    \{\kappa_{G}: \Theta_{c}(BG) \to \Theta^{d+1}(B\mathrm{O}(d))_{G}\;|\; G\in \ofin(\mathrm{O}(d))\},
    \label{eq:continuum_limit}
\end{equation}
where $\Theta^{d+1}(B\mathrm{O}(d))_{G}$ is the quotient of continuum phases by phases invisible to all physically available $P$-symmetry probes with $P\leq G$, as given explicitly in Eq.~\ref{eq:quotient}. We will later refine this prescription by requiring these maps to assemble into a category theoretic natural transformation satisfying the physical conditions {\rm(GL1)} and {\rm(GL2)}.

\bigskip

We see above that the target continuum quotient group, $\Theta^{d+1}(B\mathrm{O}(d))_{G}$, changes as we change the lattice symmetry group, $G\in \ofin(\mathrm{O}(d))$, for which we are defining the continuum limit. It would be convenient to package these quotients into a single functor, denoted $\Phi$, on the orbit category. However, this will take some work to establish as we need to understand the interplay of this quotient with inclusion and conjugation. Let us turn to that now. 

Given $G\in \ofin(\mathrm{O}(d))$, we have already defined the object-wise assignment of the abelian group
\begin{equation*}
    \Phi(G)\;:=\;\Theta^{d+1}(B\mathrm{O}(d))_{G}
    \;=\;
    \dfrac{\Theta^{d+1}(B\mathrm{O}(d))}{K_{\mathcal{O}(G)}}.
\end{equation*}
If $i: G\hookrightarrow P$ is a morphism in $\ofin(\mathrm{O}(d))$ (an inclusion of finite subgroups), then the set of available probes for $P$ includes those for $G$, so
\begin{equation}
    K_{\mathcal{O}(P)}
    \;=\;
    \bigcap_{L\le P}\ker\bigl(\res^{\mathrm{O}(d)}_{L}\bigr)
    \;\subseteq\;
    \bigcap_{L\le G}\ker\bigl(\res^{\mathrm{O}(d)}_{L}\bigr)
    \;=\;
    K_{\mathcal{O}(G)}.
    \label{eq:quotient_res_rel}
\end{equation}
Note that $K_{\mathcal{O}(P)}$ is an intersection over a larger number of groups, and is therefore smaller than $K_{\mathcal{O}(G)}$. This inclusion of kernel subgroups $K_{\mathcal{O}(P)}\subseteq K_{\mathcal{O}(G)}$ induces a canonical surjection
\begin{align}
\label{eq:inclusion}
    \Phi(i)\;:\;
    &\dfrac{\Theta^{d+1}(B\mathrm{O}(d))}{K_{\mathcal{O}(P)}}
    \longrightarrow
    \dfrac{\Theta^{d+1}(B\mathrm{O}(d))}{K_{\mathcal{O}(G)}},\\
    &\;\;[x]+K_{\mathcal{O}(P)} \;\; \longmapsto \;\;[x]+K_{\mathcal{O}(G)},
    \nonumber
\end{align}
where $[x]+K_{\mathcal{O}(P)}$ (resp.\ $[x]+K_{\mathcal{O}(G)}$) denotes a coset representative of the class $[x]\in\Theta^{d+1}(B\mathrm{O}(d))$ in the quotient group $\Theta^{d+1}(B\mathrm{O}(d))_{P}$ (resp.\ $\Theta^{d+1}(B\mathrm{O}(d))_{G}$). 

\bigskip 

For conjugation morphisms in the orbit category, $c_{\lambda}: G \to \lambda G\lambda ^{-1}$ one can show that the kernel subgroups of $G$-invisible and $\lambda G\lambda ^{-1}$-invisible phases in the continuum are the same subgroup of the full continuum group, i.e., $K_{\mathcal{O}(G)}=K_{\mathcal{O}(\lambda G\lambda ^{-1})}\subseteq \Theta^{d+1}(B\mathrm{O}(d))$. Thus $\Phi(G)$ depends on only the conjugacy class of $G$, see~\cite{note_conjugation}. More generally, a morphism in $\ofin(\mathrm{O}(d))$ from $G$ to $Q$ is an inclusion modulo conjugation and can be represented by the factorization  
\begin{equation*}
i_{\lambda}:G\xrightarrow{\;c_{\lambda}\;} \lambda G\lambda ^{-1} \xhookrightarrow[]{\;i\;} Q \;\;\;\text{for some}\;\; \lambda\in \mathrm{O}(d).
\end{equation*} 
Since $\lambda G\lambda^{-1}\le Q$, Eq.~\eqref{eq:quotient_res_rel} gives
\begin{equation*}
K_{\mathcal O(Q)} \subseteq K_{\mathcal O(\lambda G\lambda^{-1})} = K_{\mathcal O(G)}.
\end{equation*}
Therefore, the induced map on quotients, $\Phi(i_\lambda)$, depends only on the inclusion $\lambda G\lambda^{-1}\hookrightarrow Q$ and is given by the same formula as Eq.~\eqref{eq:inclusion}.
One can check immediately that $\Phi(\mathrm{id}_{G})=\mathrm{id}_{\Phi(G)},$ and that for composable inclusions 
\begin{equation*}
G\xrightarrow{i} Q\xrightarrow{j} R\quad  \text{we have} \quad \Phi(j\circ i)=\Phi(i)\circ\Phi(j).
\end{equation*}
Thus
\begin{equation}
    \Phi:\ \ofin(\mathrm{O}(d))^{\mathrm{op}}\longrightarrow \mathbf{Ab},\qquad
    G\longmapsto \Theta^{d+1}(B\mathrm{O}(d))_{G},
    \label{eq:refine_domain}
\end{equation}
is a well–defined functor, which we will call the continuum quotient functor. Note that $\Phi$ is contravariant, just like the classification functors discussed earlier, e.g., an inclusion of symmetry groups $i:G\hookrightarrow P$ induces a map in the opposite direction, $\Phi(i):\Phi(P)\to \Phi(G)$. This reversal of arrows is what the superscript ``$\mathrm{op}$'' indicates.

\subsection{Physical interpretation of the restriction morphisms and the continuum limit} 
\label{sec:restriction_morphisms}
Having defined the continuum quotient functor $\Phi$, we now ask how continuum limits are related for lattice phases whose symmetries are themselves related, for example by an inclusion $i: P\hookrightarrow G.$ At the level of mesoscopic TQFT classes  we will show such phases to be connected by symmetry restriction. For physical intuition it is useful to see this first in a microscopic theory with matter: we start from a $G$-symmetric system, break the symmetry adiabatically to $P$, and then compare the resulting $P$-response and continuum limit with those of the original theory. The point of this discussion is to show that symmetry-related lattice phases give rise to correspondingly related continuum limits. The continuum limit is therefore not just a separate map for each lattice symmetry group, but a coherent prescription across all of them. We will formulate this precisely as a \emph{natural transformation} in Sec.~\ref{sec:natural_transformation}.\footnote{\label{ft:natural_transformation}Informally, just as a functor is a map between categories, a natural transformation is a map between functors. More explicitly, a natural transformation is a \emph{compatible family} of maps between two functors: if $F,G:\mathcal{C}\to\mathcal{D}$ are functors, then a natural transformation $\eta:F\Rightarrow G$ assigns to every object $X\in\mathcal{C}$ a morphism $\eta_X:F(X)\to G(X)$ such that for every morphism $f:X\to Y$ in $\mathcal{C}$ the square $G(f)\circ \eta_X=\eta_Y\circ F(f)$ commutes. In our case, the commutative square in Eq.~\ref{eq:naturality_square} is exactly this compatibility condition.}

\bigskip 

Fix $G\in \ofin(\mathrm{O}(d))$ and let $[x_G]\in \Theta^{d+1}(BG)$ be a deformation class of $G$-symmetric TQFTs, which in the present lattice setting captures the mesoscopic topological response of a gapped $G$-symmetric lattice phase (see Sec.~\ref{sec:scales}). To give this class a microscopic realization, consider a gapped lattice model with imaginary-time partition function
\begin{equation*}
    \mathcal{Z}^{\mathrm{UV}}\bigl[M^{d+1}_\Lambda, {\bf g}\bigr]
    =\sum_{\{\psi\}} \exp\left\{
      i \sum_{X \subset \Lambda}
      \Bigl[\mathcal S_{X}\bigl[\psi\vert_{X}\bigr]
      + \mathcal{S}_{X}^{\mathrm{coup}}\bigl[\psi\vert_{X}, {\bf g}\vert_{X}\bigr] \Bigr]
    \right\},
\end{equation*}
where $\Lambda\subset \mathbb R^{d}$ is the spatial lattice, $X\subset \Lambda$ are local subsets of sites, $\psi\vert_X$ are microscopic bosonic degrees of freedom transforming under $G$, and ${\bf g}\vert_X$ denotes the $G$–background restricted to the oriented edges $\langle ij\rangle\subset X$ (with link variables ${\bf g}^{ij}$). 

\bigskip 

Since the microscopic model is in a gapped, invertible phase, there is a mesoscopic regime of length scales  $a,\zeta \ll \ell \ll L$ (where $a$ is the lattice constant, $\zeta$ the inverse-gap correlation length, $\ell$ is the mesoscopic scale, and $L$ is the system size) in which the massive matter fields can be integrated out, and the lattice can be coarse–grained to a triangulation $M^{d+1}_{\Delta}$. In this regime the effective lattice TQFT response to the background, coarse–grained, classifying map for the $G$–background $f_{G} : M^{d+1}_{\Delta} \to BG$ is given by the evaluation pairing
\begin{equation*}
    \mathcal{Z}_{x_G}\bigl[M^{d+1}_{\Delta},\; f_{G}\bigr]
    \;=\;
    \big\langle f_{G}^{*}[x_G],\,[M^{d+1}_{\Delta}]\big\rangle,
\end{equation*}
see Sec.~\ref{sec:top_resp}, Eq.~\ref{eq:evaluation_pairing}.

\bigskip 

Let us now assume $[x_{G}]$ has a continuum limit, i.e., $[x_{G}]\in \Theta^{d+1}_{c}(BG)$. Recall that for each $G\in\ofin(\mathrm{O}(d))$ we have the continuum quotient group
\begin{equation*}
    \Theta^{d+1}(B\mathrm{O}(d))_{G}
    \;:=\;
    \dfrac{\Theta^{d+1}(B\mathrm{O}(d))}{K_{\mathcal{O}(G)}},
\end{equation*}
and let $\mathrm{pr}_{G}:\Theta^{d+1}(B\mathrm{O}(d))\to\Theta^{d+1}(B\mathrm{O}(d))_{G}$ denote the canonical projection. When we say that the lattice class $[x_G]$ \emph{has a continuum limit}, we mean that there exists a continuum limit class
\begin{equation*}
    \kappa_{G}([x_{G}]) \;\in\; \Theta^{d+1}(B\mathrm{O}(d))_{G},
\end{equation*}
and, equivalently, some representative $[x]\in\Theta^{d+1}(B\mathrm{O}(d))$ such that  
\begin{equation*}
    \kappa_{G}([x_{G}]) \;=\; \mathrm{pr}_{G}([x]) = [x]+K_{\mathcal{O}(G)}.
\end{equation*}
The representative $[x]$ is defined only up to the addition of $G$–invisible continuum phases in $K_{\mathcal{O}(G)}$, but the quotient class $\kappa_{G}([x_{G}])$ is  well-defined, i.e., independent of the choice of representative $[x]$. 

\medskip 

The remainder of this subsection is devoted to understanding how the continuum limit assignment behaves under symmetry restriction from $G$ to a subgroup $P\leq G$. The compatibility of continuum limits under symmetry restriction is the key ingredient needed to promote the continuum limit from a separate assignment for each symmetry group to a coherent prescription across the orbit category.

\medskip 

Let us consider symmetry restriction to a subgroup from a physical perspective.
Take the same microscopic imaginary-time partition function from above, but with a symmetry-breaking perturbation
\begin{equation*}
    \delta S^{\mathrm{SB}}=m\sum_{X\subset \Lambda} S^{\mathrm{SB}}_{X}[\psi\vert_{X}],
\end{equation*}
where each term $S^{\mathrm{SB}}_{X}$ is $P$-invariant but not $G$-invariant, so the exact symmetry is reduced from $G$ down to a subgroup $P\leq G$. Assume that $m$ is turned on adiabatically from $0$ to a small $\epsilon$ without closing the gap. This gives a family of gapped $P$-symmetric microscopic lattice models
\begin{equation*}
    \mathcal{Z}^{\mathrm{UV}}_{m}=\sum_{\{\psi\}} \exp\left\{
      i \sum_{X \subset \Lambda}
      \Bigl[\mathcal S_{X}\bigl[\psi\vert_{X}\bigr]
      + m S^{\mathrm{SB}}_{X}[\psi\vert_{X}] + \mathcal{S}_{X}^{\mathrm{coup}}\bigl[\psi\vert_{X}, {\bf p}\vert_{X}\bigr] \Bigr]
    \right\},
\end{equation*}
coupling to $P$-backgrounds, denoted ${\bf p}$ in the lattice context (see Fig.~\ref{fig:graph}), and as $f_{P}: M^{d+1}\to BP$ as a classifying map in the singular cohomology description. 

\medskip 

Following Chen et. al.~\cite{Chen2013}, such a gapped adiabatic family of bosonic $P$-symmetric microscopic models remains within a single $P$-SPT phase. In other words the low-energy $P$-TQFTs, say $[x^{(m)}]$ for each $m\in [0, \epsilon]$, corresponding to this family of UV theories are representatives of the same class, i.e., $[x^{(m)}]=[x]\in \Theta^{d+1}(BP)\;\;\forall \;\; m$.  
In particular, the two endpoints, $\mathcal{Z}^{\mathrm{UV}}_{0}$ and $\mathcal{Z}^{\mathrm{UV}}_{\epsilon}$, have the same universal response to $P$-backgrounds.

\bigskip 

To be explicit, let $[x_P]\in \Theta^{d+1}(BP)$ denote the mesoscopic topological response class of the theory $\mathcal{Z}_{\epsilon}^{\mathrm{UV}}$ at $m=\epsilon$, whose exact lattice symmetry is $P$. Its response to $P$-backgrounds is 
\begin{equation}
    \mathcal{Z}_{x_P}\bigl[M^{d+1}_{\Delta}, f_{P}\bigr]
    \;=\;
    \big\langle f_{P}^{*}\left([x_P]\right),\; [M^{d+1}_{\Delta}]\big\rangle.
    \label{eq:Psym1}
\end{equation} 
On the other hand, the original $G$-symmetric theory, $\mathcal{Z}^{\mathrm{UV}}_{0}$ at $m=0$, can also be probed by the same $P$-backgrounds. By the subgroup-probe formula~\eqref{eq:subgroup_probe} of Sec.~\ref{sec:top_resp}, its response is
\begin{equation}
    \mathcal{Z}_{x_{G}}\bigl[M^{d+1}_{\Delta}, f_{P}\bigr]
    \;=\;
    \big\langle f_{P}^{*}\bigl(\res^{G}_{P} [x_{G}]\bigr),\; [M^{d+1}_{\Delta}]\big\rangle.
    \label{eq:Psym2}
\end{equation}
By the preceding discussion, the two theories $\mathcal{Z}_{0}^{\mathrm{UV}}$ and $\mathcal{Z}_{\epsilon}^{\mathrm{UV}}$ define the same deformation class of invertible $P$-TQFTs, and hence have the same universal response to every $P$-background. Comparing Eqs.~\eqref{eq:Psym1} and \eqref{eq:Psym2}, the probe-detectability lemma of Sec.~\ref{sec:top_resp} then implies
\begin{equation*}
    [x_{P}]=\res^{G}_{P}([x_{G}]).
\end{equation*}

Recall our initial assumption that $[x_G]\in\Theta^{d+1}_{c}(BG)$ admits a continuum limit represented by
\begin{equation*}
\kappa_G([x_G])=[x]+ K_{\mathcal{O}(G)},
\;\;\;
\text{for some }\;\; [x]\in \Theta^{d+1}(B\mathrm{O}(d)).
\end{equation*}
As shown above, the $P$-symmetric theory $\mathcal{Z}_{\epsilon}^{\mathrm{UV}}$ obtained by adiabatic symmetry breaking has mesoscopic response class $[x_P]=\res^G_P([x_G])$. Thus $[x_G]$ and $[x_P]$ are the mesoscopic response classes of two gapped systems related by symmetry restriction, but they can also be viewed as arising from the same underlying microscopic model $\mathcal{Z}_{0}^{\mathrm{UV}}$ viewed with different sets of allowed probes: with $G$-probes we have $[x_{G}]$ and with restricted $P$-probes we have $\res^{G}_{P}[x_{G}]=[x_{P}]$.

\bigskip 

The relation between their continuum images, then, is exactly the one described in Sec.~\ref{sec:refine_domain}. For the inclusion $i:P\hookrightarrow G$, Eq.~\eqref{eq:quotient_res_rel} gives
\begin{equation*}
K_{\mathcal O(G)}\subseteq K_{\mathcal O(P)},
\end{equation*}
and hence induces the canonical map
\begin{equation*}
\Phi(i):\Theta^{d+1}(B\mathrm{O}(d))_{G}\longrightarrow \Theta^{d+1}(B\mathrm{O}(d))_{P}.
\end{equation*}
Physically, this expresses the fact that every $P$-probe is also a valid $G$-probe, so passing from $G$ to $P$ can only remove distinctions among continuum phases and cannot add distinctions. 

The continuum image of the restricted class $[x_{P}]=\res^{G}_{P}[x_{G}]$ is obtained by applying the canonical quotient map $\Phi(i)$ to the continuum image of $[x_G]$:
\begin{equation}
\kappa_P([x_P])=\Phi(i)\bigl(\kappa_G([x_G])\bigr)
=\Phi(i)\left([x]+ K_{\mathcal{O}(G)}\right)
=[x]+ K_{\mathcal{O}(P)}.
\label{eq:continuum_inclusion_compatibility}
\end{equation}
In this sense, the same continuum representative $[x]$ that realizes the continuum limit of the $G$-symmetric theory also realizes the continuum limit of the restricted $P$-symmetric theory, once both are viewed in their appropriate quotient targets. In particular, the restricted class $[x_{P}]$ also admits a continuum limit, and the restriction maps preserve the continuum-admitting sector:
\begin{equation}
    \res^G_P\bigl(\Theta^{d+1}_{c}(BG)\bigr)\subseteq \Theta^{d+1}_{c}(BP),
    \qquad P\le G.
    \label{eq:restriction_closure}
\end{equation}
This is the desired compatibility statement: lattice phases related by symmetry restriction give rise to correspondingly related continuum limits.

\bigskip 

To extend \eqref{eq:restriction_closure} from subgroup inclusions to arbitrary morphisms in the orbit category, we must also account for conjugation. Let $i_\lambda :P\to G$ be a morphism in $\ofin(\mathrm{O}(d))$, represented by some $\lambda \in \mathrm{O}(d)$ with $\lambda P\lambda ^{-1}\subseteq G$. As recalled in Sec.~\ref{sec:orbit_categories}, it factors canonically as conjugation followed by inclusion
\begin{equation*}
    P \xrightarrow[]{\;c_{\lambda }\;} \lambda P\lambda ^{-1} \xhookrightarrow[]{\,i\,} G.
\end{equation*}
By contravariance, the induced map on classifications is
\begin{equation*}
    \res^{G}_{P}[\lambda ]=
    (Bc_{\lambda })^{*}\circ \res^{G}_{\lambda P\lambda ^{-1}}
    \;:\;
    \Theta^{d+1}(BG)\longrightarrow \Theta^{d+1}(BP),
\end{equation*}
where 
\begin{equation}
\label{eq:conj_isom}
(Bc_{\lambda })^{*}:\Theta^{d+1}(B(\lambda P\lambda ^{-1}))\xrightarrow{\cong}\Theta^{d+1}(BP)
\end{equation}
is an isomorphism. The inclusion part is already controlled by
applying \eqref{eq:restriction_closure} to the subgroup $\lambda P\lambda ^{-1}\le G$:
\begin{equation}
    \label{eq:st1}\res^{G}_{\lambda P\lambda ^{-1}}\bigl(\Theta^{d+1}_{c}(BG)\bigr)
    \;\subseteq\;
    \Theta^{d+1}_{c} \bigl(B(\lambda P\lambda ^{-1})\bigr).
\end{equation}

\noindent The conjugation part follows from the embedding independence established in Sec.~\ref{sec:lattice_compatible_continuum_limit}. 
The groups $P$ and $\lambda P\lambda^{-1}$ describe conjugation-related embeddings of the same abstract lattice symmetry.
Under the conjugation isomorphism~\eqref{eq:conj_isom}, a class
$[x_{\lambda P\lambda^{-1}}]\in \Theta^{d+1}\bigl(B(\lambda P\lambda^{-1})\bigr)$ maps to the conjugate
class $(Bc_\lambda)^{*} \bigl([x_{\lambda P\lambda^{-1}}]\bigr) \in \Theta^{d+1}(BP)$. 
Embedding independence states that one class admits a continuum realization if and only if the corresponding conjugate class does.
Therefore,
\begin{equation} 
(Bc_\lambda)^{*} \Bigl( \Theta^{d+1}_{c}\bigl(B(\lambda P\lambda^{-1})\bigr) \Bigr) = \Theta^{d+1}_{c}(BP). 
\label{eq:st2} 
\end{equation} 
In words, the conjugation isomorphism~\eqref{eq:conj_isom} maps the continuum-admitting subgroup associated with the embedding $\lambda P\lambda^{-1}\subseteq\mathrm{O}(d)$ bijectively onto the continuum-admitting subgroup associated with the embedding $P\subseteq\mathrm{O}(d)$.
At the level of continuum images, the same embedding-independence principle gives, for every $[x_{\lambda P\lambda^{-1}}]\in \Theta_{\mathrm c}^{d+1}\bigl(B(\lambda P\lambda^{-1})\bigr)$, 
\begin{equation} 
\kappa_P\Bigl((Bc_\lambda)^{*}([x_{\lambda P\lambda^{-1}}]) \Bigr) = \Phi(c_\lambda)\Bigl( \kappa_{\lambda P\lambda^{-1}}([x_{\lambda P\lambda^{-1}}]) \Bigr). 
\label{eq:continuum_conjugation_compatibility} 
\end{equation} 
Here $\Phi(c_\lambda): \Phi(\lambda P\lambda^{-1})\to\Phi(P)$ is the canonical identification arising from $K_{\mathcal O(\lambda P\lambda^{-1})}=K_{\mathcal O(P)}$.

\bigskip 

Combining Eqs.~\ref{eq:st1}, \ref{eq:st2} with the factorization of $\res^{G}_{P}[\lambda ]$ shows that $\Theta^{d+1}_{c}$ is closed under restriction along any orbit-category morphism. Indeed, for any $[x_G]\in \Theta^{d+1}_{c}(BG)$, the relevant maps fit into
\begin{equation*}
\begin{tikzcd}
\Theta^{d+1}_{c}(BG)\;\;
\arrow[r, "\res^{G}_{\lambda P\lambda^{-1}}",  "\text{Eq.~\eqref{eq:st1}}"']
\arrow[rr, bend left=25, "\res^{G}_{P}\text{[}\lambda\text{]}"]
&\Theta^{d+1}_{c}\bigl(B(\lambda P\lambda^{-1})\bigr)
\arrow[r, "(Bc_{\lambda})^{*}", "\text{Eq.~\eqref{eq:st2}}"']
&\;\;\Theta^{d+1}_{c}(BP).
\end{tikzcd}
\end{equation*}
Here the bent arrow is the composite $\res^{G}_{P}[\lambda]= (Bc_{\lambda})^{*} \circ \res^{G}_{\lambda P\lambda^{-1}}$. It follows immediately that
\begin{equation}
\label{eq:subfunctor_prop}
    \res^{G}_{P}[\lambda ]\bigl(\Theta^{d+1}_{c}(BG)\bigr)\subseteq \Theta^{d+1}_{c}(BP).
\end{equation}
Hence all morphisms
between classification groups $(Bi_{\lambda })^{*}: \Theta^{d+1}(BG) \to \Theta^{d+1}(BP)$, restrict to morphisms between the continuum limit admitting subgroups, i.e., $\Theta^{d+1}_{c}(Bi_{\lambda }): \Theta^{d+1}_{c}(BG) \to \Theta^{d+1}_{c}(BP)$.
Because these are restrictions of the maps of the functor $\Theta^{d+1}$, they preserve identity morphisms and composition, and therefore define a functor.\footnote{Recall that a functor must preserve
identity morphisms and composition: $F(\mathrm{id}_X)=\mathrm{id}_{F(X)}$ and
$F(\phi\circ\psi)=F(\phi)\circ F(\psi)$ for every composable pair of
morphisms; see \cite[Tag 0013]{stacks-project}.} Equivalently, $\Theta^{d+1}_{c}$ defines a subfunctor 
\begin{equation}
\label{eq:theta_c}
\Theta^{d+1}_{c}\subseteq \Theta^{d+1}:\ofin(\mathrm{O}(d))^{\mathrm{op}}\to\Ab
\end{equation}
via precomposition with $B$ as in Convention~\ref{rmk:convention} (see also Fig.~\ref{fig:functor_maps}).

\subsubsection{Continuum limit as a natural transformation}
\label{sec:natural_transformation}

The preceding discussion shows that the continuum-limit assignment is compatible with symmetry restriction along any morphism $i_\lambda:P\to G$ in the orbit category.
Such a morphism factors as \begin{equation*} P \xrightarrow{\;c_\lambda\;} \lambda P\lambda^{-1} \xhookrightarrow{\;i\;} G,\end{equation*}
i.e., an inclusion up to conjugation. Combining the compatibility under subgroup inclusion established in Eq.~\eqref{eq:continuum_inclusion_compatibility} with the compatibility under conjugation established in Eq.~\eqref{eq:continuum_conjugation_compatibility}, we obtain, for every $[x_G]\in\Theta_{\mathrm c}^{d+1}(BG)$,
\begin{equation*}
  \kappa_P \bigl(\res^{G}_{P}[\lambda ]\left([x_G]\right)\bigr)\;=\;\Phi(i_\lambda )\bigl(\kappa_G\left([x_G]\right)\bigr).
\end{equation*}
By Eq.~\eqref{eq:subfunctor_prop}, the restricted class $\res^G_P[\lambda]([x_G])$ lies in $\Theta_{\mathrm c}^{d+1}(BP)$, so both sides are well defined. Equivalently, the square
\begin{equation}
\begin{tikzcd}
\Theta^{d+1}_{c}(BG) \arrow[r, "\kappa_G"] \arrow[d, "\res^{G}_{P}\text{[}\lambda \text{]}"'] &
\Theta^{d+1}(B\mathrm{O}(d))_{G} \arrow[d, "\Phi(i_\lambda )"]\\
\Theta^{d+1}_{c}(BP) \arrow[r, "\kappa_P"'] &
\Theta^{d+1}(B\mathrm{O}(d))_{P}
\end{tikzcd}
\label{eq:naturality_square}
\end{equation}
commutes. 

In category theoretic terms, the family of continuum-limit maps $\{\kappa_G\}_{G\in\ofin(\mathrm{O}(d))}$ defines a natural transformation $\kappa:\Theta^{d+1}_{\mathrm c}\Rightarrow\Phi$ (see footnote~\ref{ft:natural_transformation}). Here $\Theta^{d+1}_{c}\subseteq \Theta^{d+1}$ is the \emph{domain subfunctor} of the continuum limit: for each lattice symmetry group $G$, the subgroup $\Theta^{d+1}_{c}(BG)\subseteq \Theta^{d+1}(BG)$ consists of those lattice phases that admit a continuum limit, and Eq.~\eqref{eq:subfunctor_prop} shows that these subgroups are preserved by restriction morphisms in the orbit category. The object-wise component map $\kappa_G:\Theta^{d+1}_c(BG)\to \Phi(G)$ is then precisely the continuum-limit map for $G$-symmetric lattice phases: it assigns to each continuum-admitting lattice class its corresponding continuum quotient class visible to $G$-probes. This reformulation is what allows us to move from lattice-by-lattice continuum limits to a single coherent theory across all symmetry groups. In the next subsection, we use this structure as a basic ingredient in defining a continuum limit theory.

\subsection{Strong and generalized continuum limit theories} 
\label{sec:strong_continuum_limit}
The previous subsection showed that continuum limits should be organized as a coherent family of maps across the orbit category. We now ask what additional physical conditions such a family should satisfy. We begin with the simplest case, namely lattice phases that are obtained directly by taking continuum phases having $\mathrm{O}(d)$ symmetry and restricting $\mathrm{O}(d)$ to a finite lattice subgroup $G$. At the microscopic or UV level, one may think of this as a lattice regularization or discretization of a continuum theory, in which the full continuum symmetry is reduced to the symmetry of the lattice.  At the level of the effective low-energy topological description, however, the picture is simpler: the lattice phases are simply symmetry restricted classes obtained by applying the restriction map $\res^{\mathrm{O}(d)}_{G}$ to the continuum phases. Formulating the tautological inverse map from restricted continuum phases to the continuum gives what we call the \emph{strong} continuum limit theory. We will use this case as an anchor for defining a generalized continuum limit theory, which is meant to capture the broader class of lattice phases that admit continuum limits in microscopic settings while still respecting symmetry restriction and symmetry extension.

\bigskip 

In the previous subsections we defined, for each group $G\in\ofin(\mathrm{O}(d))$, the subgroup
\begin{equation*}
    K_{\mathcal{O}(G)}
    \;:=\;
    \bigcap_{P\le G}\ker\bigl(\res^{\mathrm{O}(d)}_{P}\bigr)
    \;\subseteq\;
    \Theta^{d+1}(B\mathrm{O}(d)),
\end{equation*}
consisting of continuum phases that are invisible to all probes built from subgroups $P\le G$.  To define the strong continuum limit, we want to identify the corresponding continuum quotient
\begin{equation*}
\Theta^{d+1}(B\mathrm{O}(d))_{G} = \Theta^{d+1}(B\mathrm{O}(d))/K_{\mathcal O(G)},
\end{equation*}
with the subgroup of lattice classes that actually arise by restricting continuum phases from $\mathrm{O}(d)$ to $G$. The key step is therefore to show that the invisible subgroup $K_{\mathcal O(G)}$ is exactly the kernel of the restriction map
\begin{equation*}
    \res^{\mathrm{O}(d)}_{G}:\ \Theta^{d+1}(B\mathrm{O}(d))\longrightarrow \Theta^{d+1}(BG).
\end{equation*}
Once this is established, the First Isomorphism Theorem identifies the continuum quotient $\Theta^{d+1}(B\mathrm{O}(d))/K_{\mathcal O(G)}$ with the image of $\res^{\mathrm{O}(d)}_{G}$, and the inverse of this identification will be the strong continuum limit map.

\bigskip 

Since $G$ itself appears among the subgroups $P\le G$, we immediately have
\begin{equation}
    K_{\mathcal{O}(G)}
    \;\subseteq\;
    \ker\bigl(\res^{\mathrm{O}(d)}_{G}\bigr).
    \label{eq:K_forward}
\end{equation}
Conversely, if $[x]\in\Theta^{d+1}(B\mathrm{O}(d))$ satisfies $\res^{\mathrm{O}(d)}_{G}([x])=0$, then for any $P\le G$ functoriality of the classification functor $\Theta$
implies
\begin{equation*}
    \res^{\mathrm{O}(d)}_{P}([x])
    \;=\;
    \res^{G}_{P}\bigl(\res^{\mathrm{O}(d)}_{G}([x])\bigr)
    \;=\;
    \res^{G}_{P}(0)
    \;=\;
    0.
\end{equation*}
Hence $[x]\in\ker(\res^{\mathrm{O}(d)}_{P})$ for every $P\le G$, and so
$[x]\in K_{\mathcal{O}(G)}$, giving us:
\begin{equation}
    K_{\mathcal{O}(G)} \; \supseteq \; \mathrm{ker}(\res^{\mathrm{O}(d)}_{G}).
    \label{eq:K_reverse}
\end{equation}
Using Eqs.~\ref{eq:K_forward}, \ref{eq:K_reverse} we conclude that
\begin{equation}
    K_{\mathcal{O}(G)}
    \;=\;
    \ker\bigl(\res^{\mathrm{O}(d)}_{G}\bigr).
    \label{eq:KOH-equals-kernel}
\end{equation}
Physically, a continuum phase is invisible to all probes assembled from
subgroups $P\le G$ if and only if its restriction to the full lattice
symmetry $G$ is trivial in $\Theta^{d+1}(BG)$.

\medskip

\noindent\textit{Strong continuum limit.}
Using \eqref{eq:KOH-equals-kernel}, the quotient
\begin{equation*}
    \Theta^{d+1}(B\mathrm{O}(d))_{G}
    \;=\;
    \dfrac{\Theta^{d+1}(B\mathrm{O}(d))}{K_{\mathcal{O}(G)}}
    \;=\;
    \dfrac{\Theta^{d+1}(B\mathrm{O}(d))}{\ker(\res^{\mathrm{O}(d)}_{G})}
\end{equation*}
is precisely the quotient of the domain of $\res^{\mathrm{O}(d)}_{G}$ by its
kernel. The First Isomorphism Theorem, see note~\cite{note_fit}, therefore gives a canonical isomorphism
\begin{equation}
    \overline{\res}^{\mathrm{O}(d)}_{G}:\ 
    \dfrac{\Theta^{d+1}(B\mathrm{O}(d))}{K_{\mathcal{O}(G)}}
    \xrightarrow[]{\;\;\cong\;\;}
    \mathrm{Im}\bigl(\res^{\mathrm{O}(d)}_{G}\bigr)
    \;\subseteq\;
    \Theta^{d+1}(BG),
    \label{eq:res-bar-iso}
\end{equation}
identifying continuum classes modulo $G$–invisible phases with those lattice
classes that actually arise by restriction from $\mathrm{O}(d)$. Thus, whenever a lattice class lies in the image of $\res^{\mathrm{O}(d)}_{G}$, the isomorphism \eqref{eq:res-bar-iso} can be inverted to recover its continuum origin, uniquely up to the $G$-invisible ambiguity already encoded in the quotient. This is what we call the strong continuum limit.

\medskip

\begin{definition}[Strong continuum limit]
\label{def:strong_con_lim}
For each group $G\in\ofin(\mathrm{O}(d))$, the \emph{strong continuum limit} is the group isomorphism
\begin{equation}
    \overline{\kappa}_{G}:\ 
    \mathrm{Im}\bigl(\res^{\mathrm{O}(d)}_{G}\bigr)
    \xrightarrow[]{\;\;\cong\;\;}
    \Theta^{d+1}(B\mathrm{O}(d))_{G},
    \quad
    \overline{\kappa}_{G}
    :=
    \bigl(\overline{\res}^{\mathrm{O}(d)}_{G}\bigr)^{-1}.
    \label{eq:strong_continuum_limit}
\end{equation}
In other words, any lattice class $[x_{G}]\in\Theta^{d+1}(BG)$ which literally arises
as a restriction $[x_{G}]=\res^{\mathrm{O}(d)}_{G}([x])$ for some
$[x]\in\Theta^{d+1}(B\mathrm{O}(d))$ has a unique continuum limit in $\Theta^{d+1}(B\mathrm{O}(d))_{G}$,
namely
\begin{equation*}
    \overline{\kappa}_{G}([x_{G}])
    \;=\; [x]+ K_{\mathcal{O}(G)}.
\end{equation*}
\end{definition}

The strong continuum limit is automatically compatible with symmetry restriction. If $i:P\hookrightarrow G$ is an inclusion in $\ofin(\mathrm{O}(d))$, then
functoriality of the restriction maps gives
\begin{equation*}
    \res^{\mathrm{O}(d)}_{P}
    \;=\;
    \res^{G}_{P}\circ\res^{\mathrm{O}(d)}_{G}.
\end{equation*}
In particular, if a lattice class lies in the image of $\res^{\mathrm{O}(d)}_{G}$, then its restriction to $P$ lies in the image of $\res^{\mathrm{O}(d)}_{P}$. Equivalently,
\begin{equation}
\res^{G}_{P}\Bigl(\mathrm{Im}\bigl(\res^{\mathrm{O}(d)}_{G}\bigr)\Bigr)
    \;\subseteq\;
    \mathrm{Im}\bigl(\res^{\mathrm{O}(d)}_{P}\bigr).
    \label{eq:image_subfunctor_prop}
\end{equation}
Thus, symmetry restriction preserves the domain on which the strong continuum limit is defined. 

Equivalently, the assignment
\begin{equation}
    \mathrm{Im}\bigl(\res^{\mathrm{O}(d)}_{-}\bigr):G\;\longmapsto\;\mathrm{Im}\bigl(\res^{\mathrm{O}(d)}_{G}\bigr)\subseteq \Theta^{d+1}(BG)
    \label{eq:stong_CL_dom_functor}
\end{equation}
is closed under restriction in the same way as the continuum-limit admitting assignment $\Theta^{d+1}_c$, c.f.,~Eq.~\eqref{eq:subfunctor_prop}, and therefore defines
a domain subfunctor for the strong continuum limit. The corresponding naturality statement then follows exactly as in Sec.~\ref{sec:restriction_morphisms}: for every inclusion $i:P\hookrightarrow G$,
\begin{equation*}
    \overline{\kappa}_{P}\circ
    \res^{G}_{P}\big|_{\mathrm{Im}(\res^{\mathrm{O}(d)}_{G})}
    \;=\;
    \Phi(i)\circ \overline{\kappa}_{G},
\end{equation*}
so the strong continuum limit assembles into a natural transformation
\begin{equation*}
    \overline{\kappa}:\ \mathrm{Im}(\res^{\mathrm{O}(d)}_{-})\Rightarrow \Phi.
\end{equation*}

\bigskip

\noindent \textit{Generalized continuum limit.} The strong continuum limit applies to only the highly constrained subgroup of lattice classes in
$\mathrm{Im}(\res^{\mathrm{O}(d)}_{G})\subseteq\Theta^{d+1}(BG)$. In the microscopic setting, however, it is natural to ask if there exists a larger subgroup
\begin{equation*}
    \mathrm{Im}\bigl(\res^{\mathrm{O}(d)}_{G}\bigr)
    \;\subseteq\;
    \Theta^{d+1}_{c}(BG)
    \;\subseteq\;
    \Theta^{d+1}(BG),
\end{equation*}
consisting of all lattice phases which admit \emph{some} continuum limit in the sense of Eq.~\ref{eq:continuum_limit}. A generalized continuum limit is then obtained by extending the strong continuum limit from the subgroup $\mathrm{Im}(\res^{\mathrm{O}(d)}_{G})$ to a larger continuum-admitting domain, while preserving two physical features of the strong theory. First, the generalized prescription must agree with the strong continuum limit wherever the latter is defined. Second, continuum-admitting phases must continue to fit coherently into finite symmetry-extension networks. These requirements are formalized below as (GL1) and (GL2), respectively. In Sec.~\ref{sec:inverse_limit_construction}, we will show that they imply that every continuum-admitting phase extends to a compatible family over the full orbit category.

\bigskip

\begin{definition}[Generalized continuum limit theory]
\label{def:gen_con_lim}
A \emph{generalized continuum limit theory} consists of
\begin{enumerate}
  \item a subfunctor 
    \begin{equation*}
      \Theta^{d+1}_c\;\subseteq\;\Theta^{d+1}:\ \ofin(\mathrm{O}(d))^{\mathrm{op}}\to\mathbf{Ab},
    \end{equation*}
    called the domain subfunctor (see Sec.~\ref{sec:natural_transformation}) assigning to each group $G$ a subgroup $\Theta^{d+1}_c(BG)\subseteq\Theta^{d+1}(BG)$ of lattice phases that admit a continuum limit, and
  \item a natural transformation
    \begin{equation*}
      \kappa:\ \Theta^{d+1}_c\Rightarrow\Phi
    \end{equation*}
    from the domain subfunctor to the continuum quotient functor $\Phi$ of Eq.~\ref{eq:refine_domain}, called the \emph{continuum limit}, whose component at $G$ is denoted
    \begin{equation*}
      \kappa_G:\Theta^{d+1}_c(BG)\to \Theta^{d+1}(B\mathrm{O}(d))_G.
    \end{equation*}
\end{enumerate}
Such that the following compatibility relations hold for every $G\in\ofin(\mathrm{O}(d))$:
\begin{enumerate}
  \item[(GL1)] (\emph{Strong–limit extension})
  The image of the continuum restriction map, which is the domain of the strong limit, is contained in the continuum–admitting
  subgroup,
  \[
    \mathrm{Im}\bigl(\res^{\mathrm{O}(d)}_{G}\bigr)
      \;\subseteq\;
      \Theta_{c}(BG),
  \]
  and on this subgroup $\kappa_{G}$ agrees with the strong continuum limit,
  \[
      \kappa_{G}\big\vert_{\mathrm{Im}(\res^{\mathrm{O}(d)}_{G})}
      \;=\;
      \overline{\kappa}_{G}.
  \]
  \item[(GL2)] (\emph{Symmetry extension completeness})
  For every $P\in\ofin(\mathrm{O}(d))$ and every finite rooted symmetry neighborhood $\mathcal C$ of $P$, by which we mean a finite, connected full subcategory $\mathcal{C} \subseteq\ofin(\mathrm{O}(d))$ containing $P$, the canonical projection
\begin{equation}
\pi_P^{\mathcal C}:
\varprojlim_{G\in\mathcal C}
\Theta_{\mathrm c}^{d+1}(BG)
\twoheadrightarrow
\Theta_{\mathrm c}^{d+1}(BP)
\label{eq:finite_symmetry_extension_completeness}
\end{equation}
is surjective, where the inverse limit is formed using the restriction maps
induced by all morphisms in $\mathcal C$.\footnote{A full subcategory contains every orbit-category morphism between its chosen objects. It is connected if any two of its objects can be joined by a finite chain of morphisms in the underlying undirected graph. The term ``rooted'' only records that $P$ is the distinguished object whose component is prescribed.}

Equivalently, every $P$-symmetric phase that admits a continuum limit can be completed, over any finite rooted symmetry neighborhood of $P$, to a compatible family of continuum-admitting phases whose $P$-component is the prescribed phase. 
\end{enumerate}
\end{definition}

\noindent \textit{Motivation for (GL1)-strong-limit extension.}
A continuum phase $[x]\in\Theta^{d+1}(B\mathrm{O}(d))$ can be lattice regularized by restricting its symmetry from $\mathrm{O}(d)$ to a finite lattice subgroup $G$, producing
\begin{equation*}
[x_G]=\res^{\mathrm{O}(d)}_G([x]) \in \mathrm{Im}\bigl(\res^{\mathrm{O}(d)}_G\bigr).
\end{equation*}
Such a phase must lie in the domain of any admissible continuum-limit prescription, and taking its continuum limit should recover the original continuum theory up to the $G$-invisible ambiguity:
\begin{equation*}
\kappa_G\bigl(\res^{\mathrm{O}(d)}_G([x])\bigr) = [x]+K_{\mathcal O(G)} = \overline{\kappa}_G\bigl(\res^{\mathrm{O}(d)}_G([x])\bigr).
\end{equation*}
Thus, a generalized continuum limit must contain the strong domain and agree with the strong continuum limit wherever the latter is defined. This is precisely condition {\rm(GL1)}.

\bigskip 

\noindent \textit{Motivation for (GL2)-symmetry-extension completeness.}
Axiom (GL2) is a coherence requirement for treating the continuum limit as a ``symmetry-extension'' procedure. If a symmetry $P$ embeds into a larger lattice symmetry $G$, and ultimately into the ambient model group $\mathrm{O}(d)$,
\begin{equation*}
    P\xhookrightarrow[]{} G \hookrightarrow \mathrm{O}(d),
\end{equation*}
then a $P$-symmetric phase $[x_{P}]\in\Theta^{d+1}_c(BP)$ with continuum avatar $\kappa_P([x_P])=[x]+K_{\mathcal{O}(P)}$  should not ``live in isolation.'' That is, whenever the symmetry extends from $P$ to $G$, one expects $[x_P]$ to arise as the restriction of some $G$-symmetric phase $[x_G]$ admitting the same continuum limit (up to $G$-invisible probes). Equivalently, one should be able to choose $[x_G]$ so that the factorization $P\hookrightarrow G\hookrightarrow \mathrm{O}(d)$ is reflected by
\begin{equation}
\begin{tikzcd}
    \text{[}x_{P}\text{]} \arrow[r, dashed] \arrow[rr, bend left=30, "\kappa_{P}"] & \text{[} x_{G}\text{]}  \arrow[r, "\kappa_{G}"] &\text{[} x\text{]} 
\end{tikzcd},
\label{eq:single_symmetry_extension}
\end{equation}
where the dashed arrow denotes symmetry extension along $P\hookrightarrow G$. 

\medskip 

This extension along a single symmetry tower, that is, along a chain of inclusions $P\hookrightarrow G\hookrightarrow \mathrm{O}(d)$, is not sufficient to capture the full coherence expected of the continuum-limit procedure. There is generally no preferred tower through which a finite lattice symmetry approaches the ambient continuum symmetry. Rather, a given symmetry $P$ may admit several enlargements, and these different symmetry-extension paths may overlap through further subgroup and conjugation relations. 
For example, suppose that $G_1$ and $G_2$ contain two common subgroups $P$ and $R$ such that there is no morphism $R\to P$ in $\ofin(\mathrm{O}(d))$:
\begin{equation*}
\begin{tikzcd}[row sep=0.2em, column sep=2.8em]
    R \arrow[r, hook] \arrow[ddr, hook]  \arrow[dd, hook', "\textcolor{red}{\times}" description] & G_{1} \arrow[dr, hook]\\
    &&\mathrm{O}(d)\\
    P \arrow[r, hook] \arrow[uur, hook] &G_{2} \arrow[ur, hook] 
\end{tikzcd}
\end{equation*}
An arrowwise extension condition would allow one to choose classes $[x_{G_1}]$ and $[x_{G_2}]$ satisfying
\begin{equation*}
\res^{G_1}_P([x_{G_1}]) = [x_P] = \res^{G_2}_P([x_{G_2}]).
\end{equation*}
Although the two lifts agree on the prescribed $P$-component, they need not agree on the other common subgroup $R$. Indeed, because there is no morphism $R\to P$ in the orbit category, contravariance provides no restriction map from the $P$-classification to the $R$-classification that would determine $[x_R]$ from $[x_P]$.
Thus one may have
\begin{equation*}
\res^{G_1}_R([x_{G_1}]) \neq \res^{G_2}_R([x_{G_2}]).
\end{equation*}
The two independently chosen lifts therefore need not define a single compatible assignment on the combined symmetry network containing $P$, $R$, $G_1$, and $G_2$.

\medskip 

The physical issue in this example is the coherent extension of the prescribed $P$-phase toward larger symmetry groups. In contrast, passing from $P$ to its subgroups (perhaps up to conjugation) does not introduce any compatibility problems. If $S$ admits a morphism $\alpha:S\to P$, then its component is already fixed by restriction:
\begin{equation*}
[x_S] = \Theta_{\mathrm c}^{d+1}(B\alpha)([x_P]).
\end{equation*}
The coherence of continuum-limit admitting classes, here $[x_{P}]\in \Theta^{d+1}_{c}(BP)$, under conjugation discussed in Sec.~\ref{sec:lattice_compatible_continuum_limit} and Sec.~\ref{sec:restriction_morphisms} ensure that this assignment is independent of the chosen orbit-category morphism $\alpha,$ and is compatible with all further restrictions. 
Thus, once $[x_P]$ is prescribed, its restriction determines a
compatible family on the full subcategory $\widetilde{\ofin}(P)\subseteq\ofin(\mathrm{O}(d))$ spanned by the subgroups of $P$.\footnote{The ordinary orbit category $\ofin(P)$ has objects $S\le P$ and morphisms induced by conjugation within $P$, i.e. $\operatorname{Mor}_{\ofin(P)}(U,V)
= \left\{ c_\lambda:U\to V \;\middle|\; \lambda\in P,\; \lambda U\lambda^{-1}\leq V \right\}$. Here $\widetilde{\ofin}(P)$ denotes instead the full subcategory of $\ofin(\mathrm{O}(d))$ on the same objects; it therefore also contains morphisms induced by conjugation in the ambient group $\mathrm{O}(d)$. Conjugation independence from Sec.~\ref{sec:lattice_compatible_continuum_limit} ensures that the restrictions of $[x_P]$ are compatible with these additional morphisms.}
The genuinely new requirement is to extend this determined family coherently across the different upward symmetry-extension paths.

\medskip 

To accomplish this, the lifts must therefore be chosen coherently over every finite rooted symmetry neighborhood of $P$. In the terminology of (GL2), such a neighborhood is a finite connected full subcategory $\mathcal C\subseteq\ofin(\mathrm{O}(d))$ containing $P$. For the simplest neighborhood, corresponding to the single symmetry extension in Eq.~\eqref{eq:single_symmetry_extension}, this reduces to the original requirement that $[x_P]$ admit a $G$-symmetric lift. More generally, one should be able to assign a phase to every object of $\mathcal C$ so that, for each morphism $\alpha:R\to G$ in $\mathcal C$, the induced restriction map sends the $G$-component to the $R$-component. Axiom (GL2) formalizes this finite coherence requirement.

\bigskip 

Finally, one further point implicit in the formulation is worth emphasizing: {\rm(GL2)} is a \textit{local} condition imposed separately at each finite stage of the symmetry-extension problem. A finite rooted symmetry neighborhood $\mathcal C$ may contain an arbitrarily large, but finite, number of objects, but {\rm(GL2)} considers one such neighborhood at a time, with only the $P$-component fixed in advance. The phases assigned to the remaining objects of $\mathcal C$ may be chosen jointly so as to form a compatible family on that neighborhood. For a different finite neighborhood $\mathcal C'$, a different compatible extension may be chosen; {\rm(GL2)} does not require the extensions constructed on $\mathcal C$ and $\mathcal C'$ to agree on their overlap. Despite this, in Theorem \ref{thm:finite_to_global_compatibility} below, we prove that the existence of these separate compatible families (with fixed $[x_{P}]$) across separate neighborhoods implies the existence of a global compatible family with fixed $[x_{P}]$.

\medskip

A finite rooted symmetry neighborhood may be visualized as a finite web of overlapping symmetry towers anchored at $P$. Each tower represents one route of symmetry enlargement, while their overlaps record the restriction and conjugation relations that make the different routes coherent. This picture is schematic: not every object need lie above $P$ along a single chain, but connectedness ensures that every object belongs to the same finite symmetry-extension network as $P$. Thus, (GL2) asserts that no obstruction to extending the prescribed $P$-phase appears at any finite stage of the symmetry-extension procedure. It does not assume that a compatible family has already been chosen on the full orbit category. The passage from these separate finite extensions to one global compatible family is a distinct mathematical consequence, established in Sec.~\ref{sec:inverse_limit_construction} using compactness.

\bigskip

The strong continuum limit theory $\bigl(\mathrm{Im}(\res^{\mathrm{O}(d)}_{-}),\overline{\kappa}\bigr)$ is the simplest example of a generalized continuum limit theory. Condition (GL1) holds by construction. To verify (GL2), fix $[x_P]\in\mathrm{Im}(\res^{\mathrm{O}(d)}_P)$ and a finite rooted symmetry neighborhood $\mathcal C$ of $P$. Choose $[x]\in\Theta^{d+1}(B\mathrm{O}(d))$ such that
\begin{equation*}
[x_P]=\res_P^{\mathrm{O}(d)}([x]).
\end{equation*}
For every $G\in\mathcal C$, define
\begin{equation*}
[x_G]:=\res_G^{\mathrm{O}(d)}([x]).
\end{equation*}
Functoriality of restriction implies that these classes form a compatible family over $\mathcal C$ with prescribed $P$-component $[x_P]$. Hence the projection $\pi_P^{\mathcal C}$ is surjective on the strong domain, and (GL2) holds.

\bigskip 

The argument above shows that the strong continuum limit theory is indeed a generalized continuum limit theory.  Moreover, condition {\rm(GL1)} requires, for every $G\in\ofin(\mathrm{O}(d))$,
\begin{equation*}
\mathrm{Im}\bigl(\res^{\mathrm{O}(d)}_{G}\bigr) \subseteq \Theta^{d+1}_{c}(BG).
\end{equation*}
Thus $\mathrm{Im}\bigl(\res^{\mathrm{O}(d)}_{G}\bigr)$ is the minimal continuum-admitting subgroup allowed by the definition at each lattice symmetry group $G$. In the next subsection, we establish the complementary maximality statement: 
for every $G\in\ofin(\mathrm{O}(d))$,
\begin{equation*}
\Theta^{d+1}_{c}(BG) \subseteq \mathrm{Im}(\pi_G),
\end{equation*}
where $\pi_G:\mathrm{cSPT}^{d+1}_{\mathrm{univ}} \to\Theta^{d+1}(BG)$ is the canonical projection defined in Sec.~\ref{sec:inverselimit}. We then use the groups $\mathrm{Im}(\pi_G)$ to construct the universal continuum limit theory.

\subsection{Generalized continuum limit using the universal cSPT construction}
\label{sec:inverse_limit_construction}

We now show that the abstract notion of a generalized continuum limit theory from Sec.~\ref{sec:strong_continuum_limit} forces continuum-admitting lattice phases to organize themselves into compatible families over the entire orbit category. Concretely, let $(\Theta^{d+1}_c,\kappa^{c})$ be any generalized continuum limit theory. Then, for each lattice symmetry group $G\in\ofin(\mathrm{O}(d))$, every class $[x_G]\in \Theta^{d+1}_c(BG)$ must occur as the $G$-component of some compatible family in the universal cSPT group. Equivalently, every such class lies in the image of the canonical projection $
\pi_G:\mathrm{cSPT}^{d+1}_{\mathrm{univ}}\to \Theta^{d+1}(BG)$, i.e., $\mathrm{Im}(\pi_{G})$. This identifies a canonical maximal choice of continuum-admitting domain, which will motivate the universal continuum limit construction developed below.

\medskip

As discussed in Sec.~\ref{sec:inverselimit}, the universal cSPT group 
\begin{equation*}
\mathrm{cSPT}^{d+1}_{\mathrm{univ}}:=\varprojlim_{G\in\ofin(\mathrm{O}(d))}\Theta^{d+1}(BG),
\end{equation*}
is the abelian group of compatible families $\left([x_{G}]\right)_{G\in \ofin(\mathrm{O}(d))}.$ Recall that a compatible family is a family of lattice TQFTs which satisfy compatibility relations $\res^{G}_{P}[g]([x_{G}])=[x_{P}]$ for all symmetry relations $i_{g}: P\to G$. We write $\pi_{G}: \mathrm{cSPT}^{d+1}_{\mathrm{univ}} \to \Theta^{d+1}(BG)$ for the canonical projection maps, and denote the images of the projection maps by $\mathrm{Im}(\pi_{G})\subseteq \Theta^{d+1}(BG)$.\footnote{Recall that $\Theta^{d+1}(BG)$ is the group of ALL lattice TQFT phases, and $\Theta_{c}^{d+1}(BG)$ is the subgroup of continuum-limit admitting phases in some continuum limit. We prove $\Theta^{d+1}_{c}(BG)\subseteq \mathrm{Im}(\pi_{G})\subseteq \Theta^{d+1}(BG)$.}

\bigskip
\noindent \textit{From generalized continuum limits to compatible families.}
Let $(\Theta_{c},\kappa^{c})$ be a generalized continuum limit in the sense of Def.~\ref{def:gen_con_lim}. Fix a symmetry group $P\in\ofin(\mathrm{O}(d))$ and a continuum–admitting lattice phase
\begin{equation*}
    [x_{P}]\in\Theta^{d+1}_{c}(BP)\;\subseteq\;\Theta^{d+1}(BP).
\end{equation*}
We claim that $[x_{P}]$ can be completed to a global compatible family $\left([x_{G}]\right)_{G\in\ofin(\mathrm{O}(d))}$ with $[x_{G}]\in\Theta_{c}(BG)$ for all $G\in \ofin(\mathrm{O}(d))$, and with $[x_{P}]$ as its $P$–component.

\medskip

\begin{theorem}[Finite extension completeness implies global compatibility]
\label{thm:finite_to_global_compatibility} 
Let $(\Theta^{d+1}_{c},\kappa^{c})$ be a generalized continuum limit theory. Then every class $[x_P]\in\Theta^{d+1}_{c}(BP)$ extends to a compatible family
\begin{equation*}
\left([x_G]\right)_{G\in\ofin(\mathrm{O}(d))} \in \varprojlim_{G\in\ofin(\mathrm{O}(d))} \Theta^{d+1}_{c}(BG)\subseteq
\mathrm{cSPT}^{d+1}_{\mathrm{univ}},,
\end{equation*}
whose $P$-component is the prescribed class $[x_P]$.
\end{theorem}

\medskip 

\noindent The physical content of the theorem is that, once a continuum-admitting phase can be extended consistently across every finite-rooted symmetry network $\mathcal{C} \subset \ofin(\mathrm{O}(d))$, there is no additional obstruction that appears when all finite lattice symmetries are considered simultaneously. The proof is a purely mathematical globalization argument and introduces no further physical input, so we defer it to App.~\ref{app:gl2}.

\bigskip 

We now use this result to identify the largest possible continuum-admitting domain. By Thm.~\ref{thm:finite_to_global_compatibility}, every continuum–admitting phase $[x_{G}]\in\Theta^{d+1}_{c}(BG)$ extends to a compatible family over the full orbit category. Recall from Sec.~\ref{sec:inverselimit} that such globally compatible families are precisely the elements of the universal cSPT group. Therefore, there exists
\begin{equation*}
    \left([x_{P}]\right)_{P\in\ofin(\mathrm{O}(d))}\in \mathrm{cSPT}^{d+1}_{\mathrm{univ}},
\end{equation*}
whose $G$-component is the prescribed class $[x_G]$. Since $\pi_G$ is the canonical projection onto the $G$-component, it follows that
\begin{equation}
    [x_G]\;=\;\pi_G\bigl(([x_P])_{P\in\ofin(\mathrm{O}(d))}\bigr)
    \;\in\;
    \mathrm{Im}(\pi_G) \;\subseteq\; \Theta^{d+1}(BG).
    \label{eq:Theta-c-contained-in-im-piH}
\end{equation}
Because this argument applies for every $[x_G]\in \Theta^{d+1}_c(BG)$, and for all $G\in \ofin(\mathrm{O}(d))$, we obtain
\begin{equation}
    \Theta^{d+1}_c(BG) \;\subseteq\;
    \mathrm{Im}(\pi_G) \;\subseteq\; \Theta^{d+1}(BG)
    \label{eq:Theta-c-subfunctor-of-im-pi}
\end{equation}
for all $G$ in $\ofin(\mathrm{O}(d))$. 

\bigskip 

Consider now the assignment
\begin{equation*}
    \mathrm{Im}(\pi_-): \ofin(\mathrm{O}(d))^{\mathrm{op}}
    \longrightarrow \mathbf{Ab}, \quad G\mapsto
    \mathrm{Im}(\pi_G) \subseteq \Theta^{d+1}(BG).
\end{equation*}
To show that this assignment defines a subfunctor of $\Theta^{d+1}$, it is enough to show that every restriction map preserves these subgroups. Thus, let $\alpha:P\to G$ be a morphism in $\ofin(\mathrm{O}(d)),$ and let $[x_G]\in\mathrm{Im}(\pi_G)$. By the definition of $\mathrm{Im}(\pi_G)$, there exists a compatible family
\begin{equation*}
X=\left([x_Q]\right)_{Q\in\ofin(\mathrm{O}(d))}
\in \mathrm{cSPT}^{d+1}_{\mathrm{univ}},
\end{equation*}
whose $G$-component is $\pi_G(X)=[x_G]$. Compatibility of the family along $\alpha:P\to G$ gives
\begin{equation*}
\Theta^{d+1}(B\alpha)([x_G]) = \Theta^{d+1}(B\alpha)\bigl(\pi_G(X)\bigr)
= \pi_P(X) \in \mathrm{Im}(\pi_P).
\end{equation*}
Therefore,
\begin{equation*}
\Theta^{d+1}(B\alpha) \bigl(\mathrm{Im}(\pi_G)\bigr) \subseteq \mathrm{Im}(\pi_P)
\end{equation*}
for every morphism $\alpha:P\to G$. Hence the restriction maps of $\Theta^{d+1}$ restrict to the subgroups $\mathrm{Im}(\pi_G)$, and $G\mapsto\mathrm{Im}(\pi_G)$ defines a subfunctor of $\Theta^{d+1}$.\footnote{For an abelian-group-valued functor $\mathcal F$, a collection of subgroups $\mathcal F'(G)\subseteq\mathcal F(G)$ defines a subfunctor precisely when every morphism map of $\mathcal F$ sends the subgroup at its source into the subgroup at its target. In the case above, this property holds for $\mathrm{Im}(\pi_{-})$ by definition of the compatible family.}

Together with Eq.~\eqref{eq:Theta-c-subfunctor-of-im-pi}, this gives a chain of subfunctor inclusions\footnote{An inclusion of subfunctors $\mathrm{F}\subseteq\mathrm{G}$ means a natural transformation $\mathrm{F}\Rightarrow\mathrm{G}$ whose component at every object is injective.}
\begin{equation}
    \Theta_{c}^{d+1}(B-) \;\subseteq\;
    \mathrm{Im}(\pi_-)
    \;\subseteq\; \Theta^{d+1}(B-).
    \label{eq:Theta-c-subfunctor}
\end{equation}
Thus, for each $G\in\ofin(\mathrm{O}(d))$, the subgroup
\begin{equation}
    \Theta^{d+1}_{\mathrm{univ}}(BG)
    :=
    \mathrm{Im}(\pi_G)
    \subseteq
    \Theta^{d+1}(BG)
    \label{eq:universal-Theta-c1}
\end{equation}
provides a canonical upper bound on every continuum-admitting subgroup that can occur in a generalized continuum limit theory. It consists precisely of those $G$-TQFT phases that extend to globally compatible families over $\ofin(\mathrm{O}(d))$.

\bigskip 

\begin{remark}[Consistency check: isotropy]
\label{rmk:univ_isotropy_consistency}
The maximality statement above is compatible with the isotropy condition identified in Sec.~\ref{sec:lattice_compatible_continuum_limit}. Indeed, combining Eq.~\eqref{eq:Theta-c-subfunctor-of-im-pi} with Theorem~\ref{thm:universal_domain_weyl_invariant} in App.~\ref{app:proof_isotropy}, we obtain, for every $G\in\ofin(\mathrm{O}(d))$,
\begin{equation*}
    \Theta^{d+1}_{c}(BG)
    \subseteq
    \operatorname{Im}(\pi_G)
    \subseteq
    \Theta^{d+1}(BG)^{W_{\mathrm{O}(d)}(G)}.
\end{equation*}
Thus every generalized continuum-admitting domain automatically satisfies the isotropy requirement.
\end{remark}

\bigskip

The preceding argument identifies $\Theta^{d+1}_{\mathrm{univ}}(B-):=\mathrm{Im}(\pi_-)$ as the canonical candidate for the largest continuum-admitting domain. To promote this domain to a generalized continuum limit theory, it remains to construct a natural transformation
\begin{equation*}
\kappa^{\mathrm{univ}}: \Theta^{d+1}_{\mathrm{univ}} \Longrightarrow \Phi.
\end{equation*}
Equivalently, for each $G\in\ofin(\mathrm{O}(d))$, we must construct a component map
\begin{equation*}
\kappa^{\mathrm{univ}}_{G}:
\Theta^{d+1}_{\mathrm{univ}}(BG)
=
\mathrm{Im}(\pi_G)
\longrightarrow
\Theta^{d+1}(B\mathrm{O}(d))_{G}.
\end{equation*}
The guiding idea is that, from the perspective of a $G$-symmetric lattice system, the continuum response should depend only on the lattice phase visible at symmetry $G$, rather than on how that phase is completed to a compatible family over other lattice symmetries. Thus, once the global surjective map
\begin{equation*}
    F:\ \mathrm{cSPT}^{d+1}_{\mathrm{univ}}
    \longrightarrow
    \Theta^{d+1}(B\mathrm{O}(d))
\end{equation*}
has been constructed in Sec.~\ref{sec:rigor}, we ask whether its $G$-visible part depends only on the $G$-component of a compatible family. Under the local domain-splitting condition analyzed in App.~\ref{app:continuum_limit_component},\footnote{As noted in Sec.~\ref{sec:univ_cont_limit}, this condition is verified for $d=1,2$ and assumed in general dimensions, it affects only the local componentwise factorization, not the global map $F$ or the main theorem.} the $G$-visible part of the universal continuum map, namely the composite 
\begin{equation*}
    \mathrm{pr}_G\circ F:\ \mathrm{cSPT}^{d+1}_{\mathrm{univ}}
\longrightarrow \Theta^{d+1}(B\mathrm{O}(d))_{G},
\end{equation*}
should factor through the canonical projection $\pi_G:\ \mathrm{cSPT}^{d+1}_{\mathrm{univ}}\to \mathrm{Im}(\pi_G)$, so that there exists a map
$\kappa_G^{\mathrm{univ}}$ making the following diagram commute:\footnote{Recall, $\pi_{G}: \csptu \to \Theta^{d+1}(BG)$ are the canonical projection maps defined in Sec.~\ref{sec:inverselimit} and $\mathrm{pr}_{G}: \Theta^{d+1}(B\mathrm{O}(d))\to \Theta^{d+1}(B\mathrm{O}(d))_{G}$ are quotient maps defined in Sec.~\ref{sec:refine_domain}.} 
\begin{equation}
\begin{tikzcd}
\csptu \arrow[r, "F"] \arrow[d, "\pi_{G}"'] &
\Theta^{d+1}(B\mathrm{O}(d))  \arrow[d, "\mathrm{pr}_G"]\\
\Theta^{d+1}_{\mathrm{univ}}(BG) \arrow[r, "\kappa_{G}^{\mathrm{univ}}"] &
\Theta^{d+1}(B\mathrm{O}(d))_{G}
\end{tikzcd},\;\; \mathrm{pr}_G\circ F=\kappa^{\mathrm{univ}}_{G}\circ \pi_G.
\label{eq:descent}
\end{equation}
The nontrivial point is the well-definedness of this factorization. App.~\ref{app:continuum_limit_component} gives the formal construction of the component maps $\kappa^{\mathrm{univ}}_{G}$ from the local domain-splitting condition and establishes their compatibility with restriction. Moreover, App.~\ref{app:continuum_limit_component} shows that these maps are compatible with restriction, i.e., the corresponding naturality square, analogous to Eq.~\ref{eq:naturality_square}, commutes for every morphism in $\ofin(\mathrm{O}(d))$. Hence, the family
\begin{equation*}
    \kappa^{\mathrm{univ}}=\{\kappa^{\mathrm{univ}}_{G}\}_{G\in\ofin(\mathrm{O}(d))}
\end{equation*}
defines a natural transformation $\kappa^{\mathrm{univ}}: \Theta^{d+1}_{\mathrm{univ}}\Rightarrow \Phi$, and therefore the universal continuum limit theory
\begin{equation*}
    (\Theta^{d+1}_{\mathrm{univ}},\kappa^{\mathrm{univ}}).
\end{equation*}
We will now argue that every other generalized continuum limit theory is obtained by restricting this universal one to a smaller domain.

\subsection{Comparing generalized continuum-limit prescriptions}
\label{sec:compare_continuum_limits}

As discussed in Sec.~\ref{sec:lattice_compatible_continuum_limit}, one may view different EFT/RG procedures, e.g., Kadanoff block decimation and Wilsonian coarse-graining, as inducing different continuum limit maps on the corresponding low-energy topological theories. Although such procedures may have different domains of applicability, Assumption~\ref{assumption:prescription_independence} requires them to agree on the resulting continuum image whenever they are simultaneously defined on the same lattice phase. The role of the universal continuum limit is precisely to provide a canonical realization of this common image.

\bigskip 

The discussion in Sec.\ref{sec:inverse_limit_construction} identifies, for each $G\in \ofin(\mathrm{O}(d))$, the largest continuum limit admitting domain $\Theta^{d+1}_{\mathrm{univ}}(BG):=\mathrm{Im}(\pi_G)$. This makes it possible to compare different generalized continuum-limit prescriptions in a uniform way. Let $(\Theta^{d+1}_c,\kappa^{c})$ be any generalized continuum-limit theory. By the inclusion
\begin{equation*}
\Theta^{d+1}_c(BG)\subseteq \Theta^{d+1}_{\mathrm{univ}}(BG),
\end{equation*}
we have that $\kappa^c_G$ and $\kappa^{\mathrm{univ}}_G$ are therefore both defined on $\Theta^{d+1}_c(BG)$, Assumption~\ref{assumption:prescription_independence} implies
\begin{equation}
\kappa^c_G = \kappa^{\mathrm{univ}}_G \big\vert_{\Theta^{d+1}_c(BG)}.
\label{eq:universal_restriction_6E}
\end{equation}
Equivalently, for every $G\in \ofin(\mathrm{O}(d))$, one has a factorization
\begin{equation}
\label{eq:factorization_6E}
\begin{tikzcd}
\Theta^{d+1}_c(BG) \arrow[r, hook] \arrow[rr, bend left=18, "\kappa_{G}^{c}"]
& \Theta^{d+1}_{\mathrm{univ}}(BG) \arrow[r, "\kappa_G^{\mathrm{univ}}"']
& \Theta^{d+1}(B\mathrm{O}(d))_G .
\end{tikzcd}
\end{equation}

\noindent Thus every generalized continuum-limit prescription is obtained by restricting the universal continuum limit to a smaller continuum-admitting domain. In App.~\ref{app:maximal} we formulate this categorically by showing that the universal continuum limit is a terminal object in the category of generalized continuum-limit theories. 

\bigskip 

With the universal factorization property of Eq.~\eqref{eq:factorization_6E} in place, the conceptual structure of the continuum-limit theory is complete. The remaining task is to construct the global map $F:\mathrm{cSPT}^{d+1}_{\mathrm{univ}} \longrightarrow \Theta^{d+1}(B\mathrm{O}(d))$  underlying the universal continuum limit and to prove the properties invoked above. In the next section, Sec.~\ref{sec:rigor}, we carry out this construction and establish that $F$ is canonically defined and surjective, and the induced componentwise factorization through the maps $\pi_G$ is verified in App.~\ref{app:continuum_limit_component}.

\section{Stratification theorems}
\label{sec:rigor} 

In this section, we outline the proof of the main theorem~\ref{theorem:main}.
We focus on the pre-TQFT model~\eqref{eq:preTQFT} and the full group-cohomology model~\eqref{eq:grpcohtormodel}, and construct the universal continuum limit map 
\begin{equation*} F_{2}: \mathrm{cSPT}^{d+1}_{\mathrm{univ}} \longrightarrow \Theta^{d+1}(B\mathrm{O}(d)). \end{equation*} 
The goal is to prove that $F_{2}$ is surjective in the classification models considered here. The cases without internal symmetry are treated first in Secs.~\ref{sec:preTQFT_Quillen}-\ref{sec:lifted_quillen}, and involve only $2$-primary torsion continuum classes. The additional bookkeeping required by integral free classes appears only after adding internal symmetry, and is discussed separately in Sec.~\ref{sec:internal_sym}.

\bigskip 

The key insight that makes the construction of the continuum-limit map possible is the elementary-detection principle. Symmetry restriction naturally produces maps from the continuum classification to lattice classifications, whereas a continuum limit must proceed in the opposite direction and reconstruct a continuum class from lattice data. Elementary detection makes this reconstruction possible, i.e.,
in the classification models considered here, an $\mathrm{O}(d)$-symmetric continuum class is completely determined by its restrictions to elementary abelian $2$-subgroups of $\mathrm{O}(d)$. Physically, this means that if we just measure all elementary responses of a continuum class it will identify it uniquely. We prove this statement in the appendices; see Lemma~\ref{lemma:elem_response_detect_pretqft} in App.~\ref{app:elem_detection_pretqft}, and Lemma~\ref{lemma:elem_response_detect_grpcoh} in App.~\ref{app:main_throerem_proof_main_results}. This principle motivates the factorization of the continuum limit through the elementary subcategory (see Eq.~\ref{eq:factorization}): if the continuum class is determined by elementary response data, then the continuum limit assignment should depend uniquely on the elementary part of a universal compatible family.

\bigskip 

Let $\mathcal{A}_{2}(\mathrm{O}(d))$ denote the full subcategory of $\ofin(\mathrm{O}(d))$ spanned by elementary $2$-subgroups; we write $i: \mathcal{A}_{2}(\mathrm{O}(d)) \hookrightarrow \mathcal{O}_{\mathrm{fin}}(\mathrm{O}(d))$ for the sub-category inclusion. We can then consider the inverse limit of the classification functor over this elementary subcategory\footnote{Here we label elementary $2$-subgroups with $E$ instead of the more generic group label $G.$}
\begin{equation*}\mathrm{cSPT}^{d+1}_{\mathrm{elem}}= \varprojlim_{E\in \mathcal{A}_{2}(\mathrm{O}(d))} \Theta^{d+1}(BE)\;.
\end{equation*}
This group records the compatible elementary restriction data of lattice TQFT phases. Since $\mathcal{A}_{2}(\mathrm{O}(d)) \subseteq \mathcal{O}_{\mathrm{fin}}(\mathrm{O}(d))$, restricting a compatible family on $\mathcal{O}_{\mathrm{fin}}(\mathrm{O}(d))$ to the elementary objects defines a canonical projection: 
\begin{equation*}
\Pi: \mathrm{cSPT}^{d+1}_{\mathrm{univ}} \to \mathrm{cSPT}^{d+1}_{\mathrm{elem}}.
\end{equation*}

\bigskip 

To simplify the proof of the main theorem, we prove an intermediate theorem, namely:

\medskip

\begin{theorem}[{\bf Elementary detection theorem}]
\label{theorem:intermediate}
The Quillen restriction map 
\begin{equation}
\res_{\mathcal{A}}:\Theta^{d+1}(B\mathrm{O}(d)) \to \mathrm{cSPT}^{d+1}_{\mathrm{elem}},
\label{eq:quillen_isom}
\end{equation}
is an isomorphism. 
\end{theorem}

\medskip

We prove this result sequentially for the pre-TQFT model and for the full group-cohomology model without internal symmetry in Secs. \ref{sec:preTQFT_Quillen} and \ref{sec:lifted_quillen} respectively. The internal-symmetry extension is treated separately in Sec.~\ref{sec:internal_sym}, where the additional hypotheses needed for the extension are stated, and the possible appearance of new integral/free classes is addressed. 

\bigskip 

Theorem~\ref{theorem:intermediate} immediately implies we can define the elementary continuum limit map as the isomorphism
\begin{equation*}
\widetilde{F}_{2}: \mathrm{cSPT}^{d+1}_{\mathrm{elem}} \xrightarrow[]{\;\;\cong\;\;} \Theta^{d+1}(B\mathrm{O}(d)),\quad \widetilde{F}_{2}:= \res_{\mathcal{A}}^{-1}.
\end{equation*}
The role of the elementary-detection principle is to explain why the universal continuum limit should pass through this elementary reconstruction step. Since a continuum class is uniquely determined by its elementary restriction data, the continuum-limit assignment should depend on only the elementary part of a universal compatible family. Thus, we define the universal continuum-limit map by the factorization
\begin{equation}
\begin{tikzcd}
         \mathrm{cSPT}^{d+1}_{\mathrm{univ}} \arrow[r, "\Pi"] \arrow[rr,bend left=20,"F_{2}"]
        &\mathrm{cSPT}^{d+1}_{\mathrm{elem}} \arrow[r, " \widetilde{F}_{2}"] & \Theta^{d+1}(B\mathrm{O}(d)),
    \end{tikzcd}
    \label{eq:factorization}
\end{equation}
such that the full continuum limit is given by the composition $ F_{2}:= \widetilde{F}_{2}\circ \Pi$. 
The surjectivity of $F_{2}$ then reduces to proving surjectivity of $\Pi$.

\bigskip 

We note that the existence of the Quillen restriction map, Eq.~\ref{eq:quillen_isom}, is a formal consequence of the universal property of the inverse limit (see Eq.~\ref{eq:universal_property}): the restriction maps to elementary abelian 2-subgroups $\res^{\mathrm{O}(d)}_{E}: \Theta^{d+1}(B\mathrm{O}(d)) \to \Theta^{d+1}(BE)$ are compatible with morphisms in $\mathcal{A}_{2}(\mathrm{O}(d))$. Hence they assemble into a cone over the diagram $\Theta^{d+1}: BE\mapsto \Theta^{d+1}(BE),$ and therefore the universal property implies there exists a unique map denoted $\res_{\mathcal{A}}$ from $\Theta^{d+1}(B\mathrm{O}(d))$
into the limit $\mathrm{cSPT}^{d+1}_{\mathrm{elem}}$. Likewise, restricting to all finite subgroups in $\ofin(\mathrm{O}(d))$ yields the well-defined map: 
\begin{equation}
\res_{\mathcal{O}}:\Theta^{d+1}(B\mathrm{O}(d)) \to \mathrm{cSPT}^{d+1}_{\mathrm{univ}}.
\label{eq:resO_injective}
\end{equation}
And the sub-category inclusion $\qa(\mathrm{O}(d))\hookrightarrow \ofin(\mathrm{O}(d))$ implies the factorization: \begin{equation*}\res_{\mathcal{A}}= \Pi \circ \res_{\mathcal{O}}.
\end{equation*}
Since Theorem~\ref{theorem:intermediate} identifies $\res_{\mathcal A}$ as an isomorphism in the cases treated in Secs.~\ref{sec:preTQFT_Quillen}-\ref{sec:lifted_quillen}, it follows that $\Pi$ is surjective and $\res_{\mathcal O}$ is injective. Therefore $F_2=\widetilde F_2\circ \Pi$ is surjective. Thus, in the absence of integral free continuum classes, the proof of the main theorem reduces to the elementary detection theorem.

\subsection{Quillen's isomorphism for pre-TQFTs}
\label{sec:preTQFT_Quillen}
As a starting point, we will prove the elementary detection theorem at the pre-TQFT level. In the pre-TQFT model we work with the mod-$2$ group cohomology classification 
\begin{equation*}
  H^{d+1}(BG,\mathbb Z_{2}),
\end{equation*}
as introduced in Eq.~\ref{eq:preTQFT}. 
Here, we can directly apply Quillen's restriction theorem, stated in App.~\ref{app:quillen_statement}(see Thm.~\ref{theorem:quillen}).
Specializing to $G=\mathrm{O}(d)$ and $p=2$, the theorem identifies the restriction map
\begin{equation}
\res^{\mathrm{O}(d)}_{\mathcal{A}_{2}}=:\res_{\mathcal A}:\ H^{*}(B\mathrm{O}(d),\mathbb Z_{2})
  \longrightarrow
  \varprojlim_{E\in\mathcal A_{2}(\mathrm{O}(d))} H^{*}(BE,\mathbb Z_{2}),
  \label{eq:quillen_mod2}
\end{equation}
as an \emph{$F$-isomorphism}: it is an isomorphism after quotienting by nilpotent elements {(equivalently, its kernel and cokernel are nilpotent).
Intuitively, this says that $\mathbb{Z}_2$ continuum invariants are already detected using just elementary abelian $2$-subgroups in $\mathcal{A}_2(\mathrm{O}(d))$, up to a mild nilpotent ambiguity, which in the present setting is harmless, as we explain below.\footnote{Here nilpotent means that some cup-power of a class vanishes; over $\mathbb Z_2$, that is $x^{2^k}=0$ for some $k> 0$.}

\bigskip 

For the specific cohomology rings appearing in Eq.~\ref{eq:quillen_mod2}, one can strengthen Quillen’s $F$-isomorphism to a genuine isomorphism. Concretely, both the domain $H^{*}(B\mathrm{O}(d),\mathbb Z_{2}),$ and the inverse-limit co-domain of Eq.~\ref{eq:quillen_mod2}, are nil-closed as unstable Steenrod algebras (notions which are defined in App.~\ref{app:refinement}). And by the Gunawardena-Lannes-Zarati criterion (see Prop.~\ref{prop:glz}), which states that an $F$-isomorphism between nil-closed unstable Steenrod algebras is an isomorphism; $\res_{\mathcal A}$ is an isomorphism in the pre-TQFT model. Thus, looking at the ring isomorphism at a fixed degree $(d+1)$ proves the elementary detection theorem~\ref{theorem:intermediate} for the pre-TQFT model. 

\subsection{Lifted Quillen's isomorphism for the group cohomology model}
\label{sec:lifted_quillen}

In this section, we consider the full cohomology classification model of Eq.~\ref{eq:grpcohtormodel}. Using the isomorphism $H^{d+1}(BG,U(1))\cong H^{d+2}(BG,\Z)$, we may equivalently write the $2$-torsion classification as
\begin{equation*}
    \Theta^{d+1}(BG)= \tor_{2}\big(H^{d+2}(BG,\Z_{w_{1}})\big).
\end{equation*}

\medskip

\begin{notation}
\label{notation:torsion}
For any abelian group $A$, we write
\begin{equation}
A[2]:=\ker(\times 2:A\to A)
\label{eq:order2}
\end{equation}
for its order-$2$ subgroup (i.e., the subgroup annihilated by multiplication by $2$), and
\begin{equation}
\tor_{2}(A):=\{a\in A\mid 2^{k}a=0\text{ for some }k\ge 1\}
\label{eq:2primarytorsion}
\end{equation}
for its $2$-primary torsion subgroup. If $f:A\to B$ is a homomorphism, then $f[2]:A[2]\to B[2]$ denotes the induced map on these subgroups obtained by restriction. In general one has $A[2]\subseteq \tor_{2}(A)$, while in the present setting, for $G=\mathrm{O}(d)$ and $G\in \mathcal A_{2}(\mathrm{O}(d))$, we can show the stronger condition that
\begin{equation}
H^{d+2}(BG,\mathbb Z_{w_{1}})[2]=
\tor_{2}\bigl(H^{d+2}(BG,\mathbb Z_{w_{1}})\bigr),
\label{eq:tor_equiv}
\end{equation}
see Theorem.~\ref{thm:Greenblatt_twisted_no_odd_torsion} and App.~\ref{app:elem_integer},~\ref{app:elem_twisted}.
\end{notation}

\bigskip 

Our goal is to prove the elementary detection theorem~\ref{theorem:intermediate} in the full cohomology model, i.e., that the restriction map to the elementary sector
\begin{equation*}
  \tor_{2}(H^{d+2}(B\mathrm{O}(d), \Z_{w_{1}}))\to
  \varprojlim_{E\in\mathcal A_{2}(\mathrm{O}(d))}\tor_{2}(H^{d+2}(BE, \Z_{w_{1}})),
\end{equation*}
is an isomorphism. Using Eq. \ref{eq:quillen_isom} and the identification in Eq.~\ref{eq:tor_equiv}, we denote this map by $\res_{\mathcal A}[2]$.\footnote{Although we view it here as a map on the $2$-primary torsion classification groups, we keep the notation $[2]$ to emphasize that, for the groups appearing in this argument, the $2$-primary torsion subgroup is identified with the order-$2$ subgroup by Eq.~\ref{eq:tor_equiv}.} The central idea is to pass from mod-$2$ classes to integral classes via the Bockstein natural transformation. This natural transformation is associated to the short exact sequence of coefficients $0\to \Z_{w_{1}} \to \Z_{w_{1}} \to \Z_{2}\to 0$, which defines for each mod-2 cohomology classification the integral lift: 
\begin{equation*}
    \beta_{G}: H^{d+1}(BG, \Z_{2}) \twoheadrightarrow H^{d+2}(BG, \Z_{w_{1}})[2],
\end{equation*}
such that for any morphism $P \to G$ in the elementary abelian category $\mathcal{A}_{2}(\mathrm{O}(d))$, we have the commuting square:
\begin{equation*}
\begin{tikzcd}
H^{d+2}(BG, \mathbb{Z}_{w_{1}})[2]
\arrow[r, "\res^{G}_{P}\text{[}2\text{]}"] & H^{d+2}(BP, \mathbb{Z}_{w_{1}})[2]\\
H^{d+1}(BG, \mathbb{Z}_{2}) \arrow[r, "\res_{P}^{G}"] \arrow[u, "\beta_{G}"] &  H^{d+1}(BP, \mathbb{Z}_{2}) \arrow[u, "\beta_{P}"]
\end{tikzcd}.
\end{equation*}
\noindent Hence, taking the Bockstein ``lift'' commutes with restricting the symmetry. We can use the identification in Eq.~\ref{eq:tor_equiv} to infer the same relations on the $2$-primary torsion subgroup of the twisted integral cohomology classification.

\bigskip 

Because of this compatibility, the Bockstein maps may be applied objectwise to the Quillen diagram over $\qa(\mathrm{O}(d))$. This produces a well-defined map on inverse limits:
\begin{equation}
\label{eq:limit_Bockstein}
    \beta_{\lim}: \varprojlim\limits_{E\in\mathcal A_{2}(\mathrm{O}(d))} H^{d+1}(BE,\mathbb Z_{2}) \to \varprojlim\limits_{E\in \qa(\mathrm{O}(d))} \tor_2\left(H^{d+2}(BE, \Z_{w_{1}})\right),
\end{equation}
and the commutative diagram between the mod-$2$ Quillen restriction map of Eq.~\ref{eq:quillen_mod2} and the torsion-level map $\res_{\mathcal A}[2]$:
\begin{equation*}
\begin{tikzcd}
\mathllap{\tor_2\left(H^{d+2}(B\mathrm{O}(d), \Z_{w_{1}})\right)}
\arrow[r, "\res_{\mathcal{A}}\text{[}2\text{]}"] &  \mathrlap{\varprojlim\limits_{E\in \qa(\mathrm{O}(d))} \tor_2\left(H^{d+2}(BE, \Z_{w_{1}})\right)} \\
\mathllap{H^{d+1}(B\mathrm{O}(d),\mathbb Z_{2})} \arrow[u, "\beta_{\mathrm{O}(d)}", shift left=24pt] \arrow[r, "\res_{\mathcal{A}}"]
  &\mathrlap{\varprojlim\limits_{E\in\mathcal A_{2}(\mathrm{O}(d))} H^{d+1}(BE,\mathbb Z_{2})} \arrow[u, "\beta_{\lim}", swap, shift right=54pt]
\end{tikzcd}.
\end{equation*}
Since we already proved in Sec.~\ref{sec:preTQFT_Quillen} that $\res_{\mathcal A}$ is an isomorphism in mod-$2$ cohomology for $G=\mathrm{O}(d)$, the Bockstein comparison diagram above proves surjectivity of $\res_{\mathcal A}[2]$ whenever the inverse-limit Bockstein $\beta_{\lim}$ is surjective. In the degree relevant to the classification, namely degree $d+2$, this surjectivity is proved in Prop.~\ref{prop:limit_bockstein_surjective}(see also the surjectivity part of Lemma~\ref{lemma:lift_quillen}). 
To prove injectivity, one uses the analogous comparison diagram for the mod-$2$ reduction maps (see Claim~\ref{claim:res2_exists_natural}, Claim~\ref{claim:rho_BO_injective}, and the injectivity part of Lemma~\ref{lemma:lift_quillen}). Together these two arguments show that, in the physical degree $(d+2),$ the map $\res_{\mathcal{A}}[2]$ is an isomorphism.\footnote{The restriction to degree $d+2$ is essential. For even $d$, the inverse-limit Bockstein can fail to be surjective in degrees $d+4k$ because of free-shadow classes; see Sec.~\ref{sec:internal_sym} and App.~\ref{app:free_shadow_refinement}. Since $d+2$ is not of this form, the obstruction is absent here. For odd $d$, the inverse-limit Bockstein is surjective in all degrees.} This proves theorem~\ref{theorem:intermediate} for the group cohomology model. The details of the construction can be found in App.~\ref{app:lift}.

\subsection{Adding internal symmetry}
\label{sec:internal_sym}

We now explain when the stratification theorem extends in the presence of an onsite internal symmetry $H$. Throughout we restrict to the \emph{decoupled} setting, meaning that adjoining $H$ does not force a change of tangential structure (Def.~\ref{def:decoupled_internal}). For example, for oriented spacetimes ($\zeta=\SO$), antiunitary symmetries such as time-reversal typically require enlarging $\SO$ to $\mathrm{O}$, while including fermion parity typically requires refining to a spin-type structure; such cases are best handled by choosing the appropriate tangential structure $\zeta$ from the outset.

\bigskip

We incorporate an internal symmetry $H,$ which is generally acted upon by spatial symmetry $\mathrm{O}(d)$ through a homomorphism
$\mu:\mathrm{O}(d)\to\mathrm{Aut}(H)$, so the total symmetry group is 
\begin{equation}
\Gamma_{\mathrm{O}(d)} \;=\; H\rtimes_{\mu} \mathrm{O}(d),\qquad
BH \xrightarrow[]{\jmath_{\mathrm{O}(d)}} B\Gamma_{\mathrm{O}(d)} \xrightarrow[]{p_{\mathrm{O}(d)}} B\mathrm{O}(d).
\label{eq:jmath_O}
\end{equation}
For every finite $G\le \mathrm{O}(d)$, we have the induced extension
\begin{equation}
\Gamma_{G} \;=\; H\rtimes_{\mu\vert_{G}} G,\qquad
BH \xrightarrow[]{\jmath_{G}} B\Gamma_{G} \xrightarrow[]{p_{E}} BG.
\label{eq:jmath_G}
\end{equation}
Here $\jmath_{\mathrm{O}(d)}$ and $\jmath_G$ denote the fiber inclusions, while
$p_{\mathrm{O}(d)}$ and $p_E$ denote the corresponding projections. 

As in the case without internal symmetry, we keep the same classification model
\begin{equation*}
\Theta^{d+1}(BG):=\tor_{2}\bigl(H^{d+2}(BG;\Z_{w_1})\bigr).
\end{equation*}
What changes is not the classification rule, but the symmetry input. The action $\mu:\mathrm{O}(d)\to \mathrm{Aut}(H)$ defines an extension for each spatial symmetry group $G\le \mathrm{O}(d)$ given by the total symmetry group $\Gamma_G:=H\rtimes_{\mu|_G}G$. Thus the $H$-enriched lattice classification is obtained by applying the same classification model to these total symmetry groups:
\begin{equation}
\Theta^{d+1}_{H}(BG):=\Theta^{d+1}(B\Gamma_G)
=\tor_{2}\bigl(H^{d+2}(B\Gamma_G;\Z_{w_1})\bigr),
\label{eq:class_H}
\end{equation}
for all $G\in \ofin(\mathrm{O}(d))$. Equivalently, $\Theta^{d+1}_{H}$ is the composite of $\Theta^{d+1}$ with the assignment $G\mapsto B\Gamma_G$, using the convention of Remark~\ref{rmk:convention}.

\bigskip 

\noindent \textbf{Note}: In the following discussion, we are interested in the crystalline part of the $H$-enriched classification. This includes the purely spatial symmetry protected cSPTs considered before, as well as mixed crystalline-internal symmetry protected SPTs protected jointly by spatial and internal symmetries. In contrast, purely internal SPTs, belonging to the classification $H^{d+2}(BH, \Z)$, play no role in the continuum-limit problem in the decoupled setting: they form a fixed internal sector independent of the spatial subgroup $G\le \mathrm{O}(d)$.

\bigskip

We impose two mild hypotheses\footnote{The extension to internal symmetry is not automatic, since adjoining $H$ can introduce new cohomological classes and torsion structures absent from the purely spatial problem. The hypotheses below identify a broad class of cases in which these additional contributions remain controlled and the stratification argument continues to apply.} (that hold for the standard physical choices $H=\Z_2$, $\U(1)$, $\mathrm{SU}(N)$, $\SO(N)$; see App.~\ref{app:canon_factor_H}): 
\begin{enumerate}[label=(H\arabic*),leftmargin=2.2em]
\item the fibration $BH\xrightarrow[]{\jmath_{\mathrm{O}(d)}} B\Gamma_{\mathrm{O}(d)}\xrightarrow[]{p_{\mathrm{O}(d)}} B\mathrm{O}(d)$ satisfies the mod-$2$ Leray-Hirsch hypothesis, i.e., the fiber classes in $H^{*}(BH,\mathbb Z_{2})$ admit compatible global lifts; equivalently, the restriction map $\jmath_{\mathrm{O}(d)}^{*}:H^{*}(B\Gamma_{\mathrm{O}(d)},\mathbb Z_{2})\to H^{*}(BH,\mathbb Z_{2})$ is surjective giving us the Leray-Hirsch isomorphism:
\begin{equation*}
    \Delta_{\mathrm{O}(d)}: H^{*}(B\mathrm{O}(d),\mathbb Z_{2})\otimes R^{*} \;\xrightarrow{\;\cong\;}\; H^{*}(B\Gamma_{\mathrm{O}(d)},\mathbb Z_{2}),
\end{equation*}
where $R^{*}:=H^{*}(BH,\mathbb Z_{2})$ is the \textit{constant} fiber cohomology piece,
\item in degree $d+2$ cohomology, the $2$-primary torsion~\eqref{eq:2primarytorsion} is exhausted by order-$2$ torsion~\eqref{eq:order2}, i.e., $\tor_2(A)=A[2]$ for $A=H^{d+2}(B\Gamma_{\mathrm{O}(d)}, \Z_{w_1})$ and $A=H^{d+2}(B\Gamma_E,\Z_{w_1})$ with $E\in \mathcal{A}_{2}(\mathrm{O}(d))$. 
\end{enumerate}

\bigskip

\noindent Under (H1), the Leray-Hirsch isomorphism, denoted $\Delta_{\mathrm{O}(d)}$ above, induces Leray-Hirsch isomorphsms for the fibrations $BH \longrightarrow B\Gamma_E \longrightarrow BE$ given by
\begin{equation*}
\Delta_{E}: H^{*}(BE,\mathbb Z_{2})\otimes R^{*}
\;\xrightarrow{\;\cong\;}\; H^{*}(B\Gamma_{E},\mathbb Z_{2}),
\quad R^{*}:=H^{*}(BH,\mathbb Z_{2}),
\end{equation*}
for $E\in\qa(\mathrm{O}(d))$. These isomorphisms show that the dependence on the internal symmetry enters only through the fixed fiber factor $R^{*}$. The compatibility of the chosen Leray-Hirsch generators with restriction\footnote{This compatibility is not an additional assumption imposed separately for each $E$. The Leray-Hirsch isomorphism for $BH\to B\Gamma_{\mathrm{O}(d)}\to B\mathrm{O}(d)$ induces the corresponding isomorphisms for the pullback fibrations $BH\to B\Gamma_E\to BE$. The naturality of this induced system over $\mathcal A_2(\mathrm{O}(d))$ is proved in App.~\ref{app:leray_hirsch}.} implies that the $H$-enriched elementary restriction map is identified with 
\begin{equation*}
\res_{\mathcal A}\otimes \mathrm{id}_{R^{*}}: H^{*}(B\mathrm{O}(d);\mathbb Z_2)\otimes R^{*} \rightarrow \varprojlim_{E\in\qa(\mathrm{O}(d))} \bigl(H^{*}(BE;\mathbb Z_2)\otimes R^{*}\bigr).
\end{equation*}
Since $\res_{\mathcal A}$ is an isomorphism in the pre-TQFT model by Sec.~\ref{sec:preTQFT_Quillen}, it follows that the $H$-enriched pre-TQFT restriction map is also an isomorphism. Thus the internal symmetry sector factors off as the fixed coefficient factor $R^{*}$, and the elementary restriction problem reduces to the purely spatial Quillen isomorphism. The precise construction and proofs are given in App.~\ref{app:canon_factor_H}.

\medskip

We now lift this statement to the full group-cohomology model. Let 
\begin{gather*} 
\mathrm{cSPT}^{d+1}_{\mathrm{elem},H} := \varprojlim_{E\in\mathcal A_2(\mathrm{O}(d))} \Theta^{d+1}_{H}(BE), \\
\mathrm{cSPT}^{d+1}_{\mathrm{univ},H} := \varprojlim_{G\in \ofin(\mathrm{O}(d))} \Theta^{d+1}_{H}(BG), 
\end{gather*} 
be the $H$-enriched versions of the elementary and universal cSPT groups, respectively. Our goal is to construct the $H$-enriched stratification map
\begin{equation*}
F_{2,H}: \mathrm{cSPT}^{d+1}_{\mathrm{univ},H} \longrightarrow \Theta^{d+1}_{H}(B\mathrm{O}(d)).
\end{equation*}
As in the cases without internal symmetry, the construction proceeds through the elementary subcategory. The new feature is that the elementary restriction map must first be corrected by quotienting out the free-shadow classes defined below.

In analogy with Secs.~\ref{sec:preTQFT_Quillen}-\ref{sec:lifted_quillen}, one might try to prove the elementary detection theorem~\ref{theorem:intermediate} directly for the $H$-enriched classification model in Eq.~\ref{eq:class_H}, with the analogous elementary and finite subgroup restriction maps given by
\begin{gather}
\label{eq:strict_elem_res_H}
\res_{H,\mathcal A}: \tor_2\left(H^{d+2}(B\Gamma_{\mathrm{O}(d)},\mathbb Z_{w_1})\right) \longrightarrow \mathrm{cSPT}^{d+1}_{\mathrm{elem},H},\\
\res_{H,\mathcal O}: \tor_2\left(H^{d+2}(B\Gamma_{\mathrm{O}(d)},\mathbb Z_{w_1})\right) \longrightarrow \mathrm{cSPT}^{d+1}_{\mathrm{univ},H}.
\label{eq:strict_fin_res_H}
\end{gather}
However, the addition of internal symmetry introduces a new complication: the map~\eqref{eq:strict_elem_res_H} need not be an isomorphism, because free continuum classes can have nontrivial torsion images after restriction to finite crystalline symmetry groups. We refer to these torsion images as \emph{free-shadow classes}: they are ``shadows'' of free continuum classes seen inside the crystalline symmetry torsion classifications. 
We first identify the obstruction to the isomorphism in Eq.~\eqref{eq:strict_elem_res_H} in Sec.~\ref{sec:obs_to_isom}. We then quotient out the elementary free-shadow subgroup and construct the surjective stratification map in Sec.~\ref{sec:constr}. Finally, in Sec.~\ref{sec:phys_int_of_quot}, we explain why this quotient is also required by the physical condition {\rm(GL1)}. 

\subsubsection{Obstruction to isomorphism}
\label{sec:obs_to_isom}
We first observe that the continuum cohomology group with internal symmetry generally contains a non-trivial subgroup of free classes:
\begin{equation*}
    \mathrm{Free}\; H^{d+2}(B\Gamma_{\mathrm{O}(d)}, \Z_{w_{1}})\;\subseteq \; H^{d+2}(B\Gamma_{\mathrm{O}(d)};\mathbb Z_{w_1}).
\end{equation*}
Since we focus on the continuum-limit of crystalline phases, the relevant part of this free subgroup is the subgroup of free classes which are protected by spatial symmetry, and which we denote by\footnote{More precisely, we define
\begin{equation*}
\mathrm{Free}_{\mathrm{cr}}\,
H^{d+2}(B\Gamma_{\mathrm{O}(d)},\mathbb Z_{w_1})
:=
\ker\!\left(
\jmath_{\mathrm{O}(d)}^{*}\big|_{\mathrm{Free}}
\right),
\end{equation*}
where
$\jmath_{\mathrm{O}(d)}^{*}:H^{d+2}(B\Gamma_{\mathrm{O}(d)},\mathbb Z_{w_1})
\to H^{d+2}(BH,\mathbb Z)$
forgets the spatial symmetry. Thus these classes vanish when only the
internal symmetry is retained.} 
\begin{equation*}
\mathrm{Free}_{\mathrm{cr}}\, H^{d+2}(B\Gamma_{\mathrm{O}(d)},\Z_{w_1}) \subseteq
\mathrm{Free}\, H^{d+2}(B\Gamma_{\mathrm{O}(d)},\Z_{w_1}).
\end{equation*}
This is the first place where adding internal symmetry changes the argument in an essential way. Without internal symmetry, the free twisted integral classes supported on $B\mathrm{O}(d)$ are generated by products of the twisted Euler class $e_d\in H^d(B\mathrm{O}(d),\Z_{w_1})$ with untwisted Pontryagin classes $p_i\in H^{4i}(B\mathrm{O}(d),\Z)$. Thus, free classes having no internal symmetries occur in degrees $d+4i$, and in particular do not appear in the physical degree $d+2$ relevant to the torsion TQFT classification. With internal symmetry present, however, the fiber $BH$ may contribute degree-two integral classes. For example, when $H$ contains a $\U(1)$ factor, the first Chern class $c_1\in H^2(BH,\Z)$ gives rise to mixed crystalline-internal classes of the form
\begin{equation*}
e_d\cup c_1 \in H^{d+2}(B\Gamma_{\mathrm{O}(d)},\Z_{w_1}).
\end{equation*}
These classes are free in the continuum classification, and, because of the presence of the Euler class, require spatial symmetry for their protection.

\bigskip 

We now consider the restriction of classes in $\mathrm{Free}_{\mathrm{cr}}\, H^{d+2}(B\Gamma_{\mathrm{O}(d)},\Z_{w_1})$ to finite crystalline subgroup classifications. For such a class, ordinary restriction gives
\begin{equation*}
\res^{\Gamma_{\mathrm{O}(d)}}_{\Gamma_G}\big|_{\mathrm{Free}_{\mathrm{cr}}}:
\mathrm{Free}_{\mathrm{cr}}\,
H^{d+2}(B\Gamma_{\mathrm{O}(d)},\Z_{w_1}) \longrightarrow H^{d+2}(B\Gamma_G,\Z_{w_1}).
\end{equation*}
If $G$ is finite, then every class in the image of this restriction map is torsion by the restriction-corestriction argument in App.~\ref{app:res_cores}(see Eq.~\eqref{eq:internal_crystalline_torsion}). Its canonical $2$-primary component therefore defines an element of the crystalline classification group
\begin{equation*}
\Theta^{d+1}_{H}(BG) =
\tor_2\left(H^{d+2}(B\Gamma_G,\Z_{w_1})\right).
\end{equation*}
For elementary subgroups $E\in\mathcal A_2(\mathrm{O}(d))$, Hypothesis~{\rm(H2)} further implies that this relevant $2$-primary torsion is order-$2$.

\medskip 

Using the universal property of the inverse limit, these crystalline restrictions assemble into the full-domain elementary and universal restriction maps
\begin{gather}
\label{eq:elem_res_H}
\res^{\mathrm{full}}_{\mathcal A,H}: \Theta^{d+1}_{H}(B\mathrm{O}(d)) \oplus \mathrm{Free}_{\mathrm{cr}}\,
H^{d+2}(B\Gamma_{\mathrm{O}(d)},\Z_{w_1}) \longrightarrow \mathrm{cSPT}^{d+1}_{\mathrm{elem},H},
\\
\res^{\mathrm{full}}_{\mathcal O,H}: \Theta^{d+1}_{H}(B\mathrm{O}(d)) \oplus
\mathrm{Free}_{\mathrm{cr}}\, H^{d+2}(B\Gamma_{\mathrm{O}(d)},\Z_{w_1}) \longrightarrow \mathrm{cSPT}^{d+1}_{\mathrm{univ},H}.
\label{eq:finite_res_H}
\end{gather}
Here ``full-domain'' means that the domain includes both the strict torsion continuum classes and the spatially protected free continuum classes whose restrictions to finite crystalline sectors become torsion. For each finite $G\leq\mathrm{O}(d)$, let
\begin{equation*}
\pi_{(2)}^{G}: \tor H^{d+2}(B\Gamma_G,\mathbb Z_{w_1})
\longrightarrow \tor_2 H^{d+2}(B\Gamma_G,\mathbb Z_{w_1}),
\end{equation*}
denote the canonical projection onto the $2$-primary torsion subgroup. Then, for $[a]\in\Theta_H^{d+1}(B\mathrm{O}(d)), \; [\widetilde x]\in \mathrm{Free}_{\mathrm{cr}}\, H^{d+2}(B\Gamma_{\mathrm{O}(d)},\mathbb Z_{w_1}), $ the $G$-component of the universal full-domain restriction is
\begin{equation}
\pi_G\!\left( \res^{\mathrm{full}}_{H,\mathcal O} ([a],[\widetilde x]) \right)
= \res^{\Gamma_{\mathrm{O}(d)}}_{\Gamma_G}([a]) +
\pi_{(2)}^{G}\!\left( \res^{\Gamma_{\mathrm{O}(d)}}_{\Gamma_G}([\widetilde x]) \right).
\label{eq:full_domain_restriction_component}
\end{equation}
Thus the torsion summand is restricted ordinarily, whereas the crystalline free summand is first restricted and then projected onto its $2$-primary torsion component.}\footnote{\label{fn:full_domain_restriction} For finite $G$, restriction-corestriction implies that $\res^{\Gamma_{\mathrm{O}(d)}}_{\Gamma_G}([\widetilde x])$ is torsion, so $\pi_{(2)}^{G}$ is defined. Naturality of the primary decomposition makes these projections compatible with further restriction. The elementary map is obtained by restricting to $E\in\mathcal A_2(\mathrm{O}(d))$; see App.~\ref{app:op_def_proj}.}

\bigskip 

Let us now denote the images of the crystalline free classes under the full-domain restriction maps by
\begin{gather} 
\label{eq:free_shadow_A}
S_{\mathrm{free},H}^{\mathcal A, d+1} := \operatorname{Im}\left( \res_{H,\mathcal A}^{\mathrm{full}}\big|_{\mathrm{Free}_{\mathrm{cr}} H^{d+2}(B\Gamma_{\mathrm{O}(d)};\mathbb Z_{w_1})} \right) \subseteq \mathrm{cSPT}^{d+1}_{\mathrm{elem},H},\\
S_{\mathrm{free},H}^{\mathcal O, d+1} := \operatorname{Im}\left( \res_{H,\mathcal O}^{\mathrm{full}}\big|_{\mathrm{Free}_{\mathrm{cr}} H^{d+2}(B\Gamma_{\mathrm{O}(d)};\mathbb Z_{w_1})} \right) \subseteq \mathrm{cSPT}^{d+1}_{\mathrm{univ},H}.
\label{eq:free_shadow_O}
\end{gather}
We call these subgroups the \textit{free shadow classes}. Note, in particular, that the elementary free shadow classes $S_{\mathrm{free},H}^{\mathcal A}$ are in fact order-2 torsion because of Hypothesis (H2). Now consider the $H$-enriched version of the Bockstein map~\eqref{eq:limit_Bockstein} on the elementary inverse limit:
\begin{equation*}
    \beta^{H}_{\lim}: \varprojlim\limits_{E\in\mathcal A_{2}(\mathrm{O}(d))} H^{d+1}(B\Gamma_{E},\mathbb Z_{2}) \longrightarrow \mathrm{cSPT}^{d+1}_{\mathrm{elem}, H}.
\end{equation*}
One can show that the free shadow classes constructed in Eq.~\ref{eq:free_shadow_A} are not mapped on to by the Bockstein image in the elementary inverse limit. In fact, one can prove that the $H$-enriched version of the elementary cSPT group admits a splitting: 
\begin{equation}
\label{eq:cspt_splitting}
\mathrm{cSPT}_{\mathrm{elem},H}^{d+1} = \operatorname{Im}\left(\beta_{\lim}^{H}\right)
\oplus S_{\mathrm{free},H}^{\mathcal A, d+1},
\end{equation}
where $\beta_{\lim}^{H}$ only maps on to the first summand
\begin{equation} \operatorname{Im}(\beta_{\lim}^{H}) = \operatorname{Im}\left( \res_{H,\mathcal A}\right)=\operatorname{Im}\left( \res_{H,\mathcal A}^{\mathrm{full}}\bigr\vert_{\Theta^{d+1}_{H}(B\mathrm{O}(d))}\right).
\label{eq:bock_image_summand}
\end{equation} 
This is proven in Claim~\ref{claim:Sfree_H_splitting}. Therefore, Eq.~\eqref{eq:strict_elem_res_H} cannot be an isomorphism in general: its image is the Bockstein-detected summand~\eqref{eq:bock_image_summand}, while the elementary cSPT group also contains the free-shadow summand~\eqref{eq:free_shadow_A}. Thus the elementary free-shadow subgroup $S_{\mathrm{free},H}^{\mathcal A,d+1}$ is precisely the additional summand that obstructs $\res_{H,\mathcal A}$ from being an isomorphism. We therefore quotient out this summand before carrying out the
elementary reconstruction.

\subsubsection{Constructing the $H$-enriched stratification map}
\label{sec:constr}
We now construct the $H$-enriched stratification map $F_{2,H}$ mentioned above. 
By the splitting in Eq.~\eqref{eq:cspt_splitting}, quotienting by the elementary free-shadow subgroup removes the free-shadow summand:
\begin{equation*}
\frac{\mathrm{cSPT}^{d+1}_{\mathrm{elem},H}}
{S_{\mathrm{free},H}^{\mathcal A,d+1}}
\cong
\operatorname{Im}(\beta_{\lim}^{H}).
\end{equation*}
Let
\begin{equation*}
q_{\mathrm{free}}^{\mathcal A,H}: \mathrm{cSPT}^{d+1}_{\mathrm{elem},H}
\longrightarrow \frac{\mathrm{cSPT}^{d+1}_{\mathrm{elem},H}}{S_{\mathrm{free},H}^{\mathcal A,d+1}}
\end{equation*}
denote the quotient map. Moreover, by Eq.~\eqref{eq:bock_image_summand}, the remaining Bockstein-detected summand is exactly the image of the torsion continuum classification under elementary restriction. Accordingly, the appropriate replacement for the obstructed isomorphism in Eq.~\eqref{eq:strict_elem_res_H} is obtained by quotienting out the elementary free-shadow summand. The quotient-corrected elementary-detection theorem states that the resulting map
\begin{equation}
\label{eq:mod_detection_thm}
q_{\mathrm{free}}^{\mathcal A,H}
\circ
\res_{H,\mathcal A}
\big|_{\Theta^{d+1}_{H}(B\mathrm{O}(d))}
:
\Theta^{d+1}_{H}(B\mathrm{O}(d))
\longrightarrow
\frac{\mathrm{cSPT}^{d+1}_{\mathrm{elem},H}}
{S_{\mathrm{free},H}^{\mathcal A,d+1}}
\end{equation}
is an isomorphism. Thus the elementary subcategory still detects the torsion continuum classification, but only after the finite-sector shadows of free continuum classes have been removed. This is proved rigorously in App.~\ref{app:proof_internal}.

\bigskip 

We therefore define the elementary $H$-enriched stratification map by
\begin{gather}
\nonumber
\widetilde F_{2,H} : \mathrm{cSPT}^{d+1}_{\mathrm{elem},H}\to \Theta^{d+1}_{H}(B\mathrm{O}(d)),\\
\widetilde F_{2,H} :=
\left( q_{\mathrm{free}}^{\mathcal A,H} \circ
\res_{H,\mathcal A} \big|_{\Theta^{d+1}_{H}(B\mathrm{O}(d))} \right)^{-1}
\circ q_{\mathrm{free}}^{\mathcal A,H}.
\label{eq:widetilde_F_2H}
\end{gather}
Thus, $\widetilde F_{2,H}$ quotients out the elementary free-shadow subgroup and recovers the continuum torsion class on the Bockstein-detected summand. Finally, define the universal $H$-enriched stratification map by 
\begin{equation*}
F_{2,H} := \widetilde F_{2,H}\circ \Pi_H\quad \text{where}\quad \Pi_{H}: \mathrm{cSPT}^{d+1}_{\mathrm{univ},H}\to \mathrm{cSPT}^{d+1}_{\mathrm{elem},H}.
\end{equation*}
Equivalently,
\begin{equation}
F_{2,H} = \left( q_{\mathrm{free}}^{\mathcal A,H} \circ \res_{H,\mathcal A} \big|_{\Theta^{d+1}_{H}(B\mathrm{O}(d))} \right)^{-1} \circ q_{\mathrm{free}}^{\mathcal A,H}
\circ \Pi_H.
\label{eq:F_2H}
\end{equation}
Finally, the factorization
\begin{equation*}
\res_{H,\mathcal A} = \Pi_H\circ\res_{H,\mathcal O}
\end{equation*}
implies
\begin{align*}
F_{2,H}\circ\res_{H,\mathcal O} &=
\left( q_{\mathrm{free}}^{\mathcal A,H} \circ \res_{H,\mathcal A}
\big|_{\Theta^{d+1}_{H}(B\mathrm{O}(d))} \right)^{-1} \circ
q_{\mathrm{free}}^{\mathcal A,H}
\circ \Pi_H \circ \res_{H,\mathcal O}
\\
&= \left( q_{\mathrm{free}}^{\mathcal A,H} \circ \res_{H,\mathcal A}
\big|_{\Theta^{d+1}_{H}(B\mathrm{O}(d))} \right)^{-1} \circ q_{\mathrm{free}}^{\mathcal A,H} \circ \res_{H,\mathcal A}
\\
&= \mathrm{id}_{\Theta^{d+1}_{H}(B\mathrm{O}(d))}.
\end{align*}
Thus $\res_{H,\mathcal O}$ is a right inverse of $F_{2,H}$. In
particular, for every $[a]\in\Theta^{d+1}_{H}(B\mathrm{O}(d))$, $[a] = F_{2,H}\bigl(\res_{H,\mathcal O}([a])\bigr)$, so $F_{2,H}$ is surjective. Moreover, this identity verifies condition {\rm(GL1)} on the strict torsion continuum classes.

\subsubsection{Physical interpretation of the quotient} 
\label{sec:phys_int_of_quot}

Having to quotient out the free-shadow classes is not simply a mathematical detail, it has a physical interpretation. We now explain why compatibility with the physical condition {\rm(GL1)} from Sec. \ref{sec:strong_continuum_limit} requires the free-shadow classes to be discarded in the construction of the elementary and universal stratification maps $\widetilde F_{2,H}$ and $F_{2,H}$ defined in Eqs.~\eqref{eq:widetilde_F_2H} and~\eqref{eq:F_2H}, respectively.

Suppose, as a heuristic enlargement of the torsion TQFT classification, that there exists a full continuum-limit prescription which remembers both torsion and free response classes:
\begin{equation*}
    \mathcal{F}^{H}: \varprojlim_{G\in \ofin(\mathrm{O}(d))} H^{d+2}(B\Gamma_G, \Z_{w_{1}}) \to H^{d+2}(B\Gamma_{\mathrm{O}(d)}, \Z_{w_{1}}).
\end{equation*}
This full map should satisfy the analog of condition {\rm(GL1)} (see Def.~\ref{def:gen_con_lim}), i.e., it should satisfy that whenever a crystalline compatible family is obtained by restricting a continuum response, the continuum-limit map should recover that response. Let
\begin{align*}
\res_{H,\mathcal O}^{\mathrm{coh}}: \Theta^{d+1}_{H}(B\mathrm{O}(d)) \oplus
\mathrm{Free}_{\mathrm{cr}}\, &H^{d+2}(B\Gamma_{\mathrm{O}(d)},\Z_{w_1})\\
&\longrightarrow \varprojlim_{G\in\ofin(\mathrm{O}(d))} H^{d+2}(B\Gamma_G \Z_{w_1}),
\end{align*}
denote ordinary restriction in full integral cohomology, retaining both the free and torsion components of each restricted class. Recall that $\res^{\mathrm{full}}_{H,\mathcal O}$, defined in Eq.~\eqref{eq:finite_res_H}, has the same torsion-plus-crystalline-free domain but takes values in $\mathrm{cSPT}^{d+1}_{\mathrm{univ},H}$ by retaining the $2$-primary torsion component of each finite-subgroup restriction; see footnote~\ref{fn:full_domain_restriction} for the componentwise definition in terms of $\pi_{(2)}^{G}$.
In particular, if
\begin{equation*}
[x]\in \mathrm{Free}_{\mathrm{cr}}\, H^{d+2}(B\Gamma_{\mathrm{O}(d)},\Z_{w_1}),
\end{equation*}
then the full continuum-limit prescription should satisfy
\begin{equation*}
\mathcal F^H\left( \res_{H,\mathcal O}^{\mathrm{coh}}([x]) \right) = [x],
\end{equation*}
where the subscript $\mathcal{O}$ indicates that this statement must be formulated over the full category of finite-subgroups of $\mathrm{O}(d)$: $\ofin(\mathrm{O}(d))$, rather than only over the elementary subcategory. This is because elementary restriction does not, in general, retain enough information to reconstruct a free continuum class. The full compatible family of finite-subgroup restrictions is therefore the appropriate input for the heuristic reconstruction above.\footnote{The elementary-detection principle fails for free continuum classes: their elementary restrictions need not determine the original free class. In contrast, the compatible family over all finite subgroups can retain additional $2$-adic information and may therefore permit such a reconstruction. We use this only as heuristic motivation and do not develop the full construction here.}

\bigskip

Now we pass from the full cohomology classification to the torsion TQFT classification. Consider the following subgroup of the full continuum cohomology group:
\begin{equation*}
\Theta^{d+1}_{H}(B\mathrm{O}(d)) \oplus \mathrm{Free}_{\mathrm{cr}}\,
H^{d+2}(B\Gamma_{\mathrm{O}(d)},\Z_{w_1})
\subseteq H^{d+2}(B\Gamma_{\mathrm{O}(d)},\Z_{w_1}).
\end{equation*}
It consists of the torsion continuum classes together with the crystalline free classes whose finite-subgroup restrictions produce the free-shadow classes. On this subgroup, define the projection
\begin{equation*}
\tau_{\mathrm{O}(d)}^{H}:
\Theta^{d+1}_{H}(B\mathrm{O}(d))
\oplus
\mathrm{Free}_{\mathrm{cr}}\,
H^{d+2}(B\Gamma_{\mathrm{O}(d)},\Z_{w_1})
\longrightarrow
\Theta^{d+1}_{H}(B\mathrm{O}(d))
\end{equation*}
by $ \tau_{\mathrm{O}(d)}^{H}(a,x)=a$. Thus $\tau_{\mathrm{O}(d)}^{H}$ is the identity on the torsion summand and vanishes on the crystalline free summand.

\bigskip

Compatibility between the full continuum-limit prescription
$\mathcal F^H$ and the continuum-limit map on torsion TQFTs,
$F_{2,H}$, means that
\begin{equation*}
F_{2,H}\circ\res_{H,\mathcal O}^{\mathrm{full}}
=\tau_{\mathrm{O}(d)}^{H} \circ\mathcal F^H \circ\res_{H,\mathcal O}^{\mathrm{coh}}
\end{equation*}
on the torsion-plus-crystalline-free subgroup above.
Now let
\begin{equation*}
[x]\in
\mathrm{Free}_{\mathrm{cr}}\,
H^{d+2}(B\Gamma_{\mathrm{O}(d)},\Z_{w_1}).
\end{equation*}
The full analog of condition {\rm(GL1)} gives
\begin{equation*}
\mathcal F^H\left( \res_{H,\mathcal O}^{\mathrm{coh}}([x]) \right) =[x].
\end{equation*}
Consequently,
\begin{align*}
F_{2,H}\left( \res_{H,\mathcal O}^{\mathrm{full}}([x]) \right)
&= \tau_{\mathrm{O}(d)}^{H}\left(
\mathcal F^H\left( \res_{H,\mathcal O}^{\mathrm{coh}}([x]) \right) \right)\\
&= \tau_{\mathrm{O}(d)}^{H}([x])=0,
\end{align*}
where the final equality follows directly from the definition of
$\tau_{\mathrm{O}(d)}^{H}$. Since $S_{\mathrm{free},H}^{\mathcal O,d+1}$ is the image of the crystalline free subgroup under
$\res_{H,\mathcal O}^{\mathrm{full}}$, this shows that
\begin{equation*}
\left.F_{2,H}\right|_{S_{\mathrm{free},H}^{\mathcal O,d+1}} =0.
\end{equation*}
Hence $S_{\mathrm{free},H}^{\mathcal O,d+1} \subseteq \ker(F_{2,H})$, and the universal property of the quotient gives a unique factorization\footnote{More generally, if $S\subseteq\ker(F)$ for a homomorphism $F:A\to B$, then $F\circ\iota=0$, where $\iota:S\hookrightarrow A$ is the inclusion. Since $A/S=\operatorname{coker}(\iota)$, the universal property of the cokernel gives a unique homomorphism $\overline F:A/S\to B$ such that $F=\overline F\circ q$, where $q:A\to A/S$ is the quotient map; see
\cite[Tag 0106]{stacks-project}.}
\begin{equation}
\label{eq:quotient_O}
\begin{tikzcd}
\mathrm{cSPT}^{d+1}_{\mathrm{univ},H}
\arrow[r,"q_{\mathrm{free}}^{\mathcal O,H}"]
\arrow[rr,bend left=30,"F_{2,H}"]
& \dfrac{\mathrm{cSPT}^{d+1}_{\mathrm{univ},H}}
{S_{\mathrm{free},H}^{\mathcal O,d+1}}
\arrow[r,"\overline F_{2,H}"]
& \Theta^{d+1}_{H}(B\mathrm{O}(d)).
\end{tikzcd}
\end{equation}
By the elementary-detection principle, the continuum-limit assignment is
determined by the elementary restriction data. Since
\begin{equation*}
\Pi_H\left(
S_{\mathrm{free},H}^{\mathcal O,d+1}
\right)
=
S_{\mathrm{free},H}^{\mathcal A,d+1},
\end{equation*}
the preceding argument analogously implies that
\begin{equation*}
\left.\widetilde F_{2,H}\right|_
{S_{\mathrm{free},H}^{\mathcal A,d+1}}
=
0.
\end{equation*}
Thus the elementary free-shadow subgroup lies in the kernel of the
elementary reconstruction map, which provides the physical motivation
for the quotient construction in Sec.~\ref{sec:constr}.

\subsection{Proof of the corollaries} \label{sec:proof_corollaries} 

We briefly explain how the corollaries stated after Theorem~\ref{theorem:main} follow from the stratification theorem. We present the argument without internal symmetry; the $H$-enriched statement follows by applying the same reasoning to the map $F_{2,H}$ in Eq.~\eqref{eq:maintheorem_H}.

\medskip 

\noindent\emph{Lattice realizability.} 
Let $[x]\in \Theta^{d+1}(B\mathrm{O}(d))$ be a torsion continuum TQFT class. Restricting $[x]$ to all finite subgroups of $\mathrm{O}(d)$ gives a compatible family $\res_{\mathcal O}([x])\in \mathrm{cSPT}^{d+1}_{\mathrm{univ}}$. By construction of the continuum-limit map, 
\begin{equation*}
F_2\circ \res_{\mathcal O} = \operatorname{id} \quad \text{on}\;\;\; \Theta^{d+1}(B\mathrm{O}(d)). 
\end{equation*}
Hence $F_2\bigl(\res_{\mathcal O}([x])\bigr)=[x]$. Thus every strict continuum TQFT class has a lattice representative, namely its compatible family of finite-subgroup restrictions. 

A more concrete representative can be obtained from the elementary detection theorem. Let $E_{\max}\cong(\mathbb Z_2)^d$ be a maximal elementary abelian subgroup of $\mathrm{O}(d)$, generated by coordinate reflections. Since \[ \res_{\mathcal A}: \Theta^{d+1}(B\mathrm{O}(d)) \xrightarrow{\;\cong\;} \mathrm{cSPT}^{d+1}_{\mathrm{elem}}, \] and since the elementary inverse limit is identified with the Weyl-invariant subgroup \[ \mathrm{cSPT}^{d+1}_{\mathrm{elem}} \cong \Theta^{d+1}(BE_{\max})^{W_{\mathrm{O}(d)}(E_{\max})}, \] the class $X$ is already represented by its restriction to $E_{\max}$, subject to the Weyl-invariance condition. Therefore one may choose a lattice representative with maximal elementary reflection symmetry which captures all elementary response data of $X$ at once.

\medskip 

\noindent\emph{Continuum-limit admissibility and collapse of lattice data.} The converse of lattice realizability is generally false. Fix a finite lattice symmetry group $G\leq \mathrm{O}(d)$. A $G$-lattice TQFT class admits a continuum image from the universal construction only if it occurs as the $G$-component of a universal compatible family. Thus the admissible subgroup is \[ \operatorname{Im}(\pi_G) \subseteq \Theta^{d+1}(BG), \] where $\pi_G$ denotes projection of a universal compatible family to its $G$-component. Classes outside this subgroup do not extend coherently over the finite-subgroup system, and hence have no continuum TQFT image in the universal construction. The continuum limit is also not faithful in general. If two universal compatible families differ by an element of $\ker(F_2)$, then they have the same continuum image. Therefore the continuum limit forgets precisely the lattice information lying in $\ker(F_2)$. In this sense, the continuum limit has two effects: it excludes lattice classes which are not continuum-admissible, and it collapses distinct continuum-admissible lattice families whose difference is invisible to $F_2$. The same statements hold in the $H$-enriched setting after replacing $F_2$, $\pi_G$, and $\Theta^{d+1}(BG)$ by their $H$-enriched analogues. Finally, all statements here concern the strict, $2$-torsion TQFT sector. Free continuum classes are not part of the target of Theorem~\ref{theorem:main}; when they appear through internal symmetries, their elementary free shadows are quotiented out before reconstructing the strict continuum class.

In summary, Secs.~\ref{sec:preTQFT_Quillen}, \ref{sec:lifted_quillen}, and \ref{sec:internal_sym} outlined the proof that the universal continuum limit maps (Eqs.~\ref{eq:maintheorem},~\ref{eq:maintheorem_H} in main theorem~\ref{theorem:main}) are canonically defined and surjective, first in the pre-TQFT model, then in the full group-cohomology model, and finally with decoupled onsite internal symmetry. 
The construction being canonical is important because it means that the map does not depend on arbitrary choices, such as a choice of specific methods to obtain the continuum limit, and is defined entirely by the overarching symmetry structure which follows from the definition of the TQFTs. Surjectivity, together with the framework of generalized continuum limits developed in Sec.~\ref{sec:continuum_limit}, yields the main theorem and its corollaries. The analogous statements for the cobordism classification model are proved in App.~\ref{app:cob_proof} and App.~\ref{app:cob_int}. We also discuss how the proofs generalize under a change of the tangential structure of the classification models. In the next section, we apply the construction to several example systems.

\section{Examples}
\label{sec:examples}

In this section we present three examples that illustrate how our framework applies to invertible lattice phases and their continuum limits in physically relevant settings. In Sec.~\ref{sec:example_1} we analyze an explicit $(2{+}1)$D example with ambient continuum symmetry group $\mathcal G=O(2)$. This focuses on crystalline phases whose lattice point-group symmetries are finite subgroups of $O(2).$ Explicitly, we show how their lattice data assemble into compatible families, and explain how the resulting continuum responses may be interpreted in terms of anomaly inflow. 
In Sec.~\ref{sec:example_2}, we investigate whether nontrivial lattice TQFT data can be lost under the continuum-limit map. We begin with an order-$4$ $C_4$-protected higher-order SPT defined for a single lattice symmetry group and ask whether it extends to a universal compatible family. We show that either no such extension exists, or, if an extension exists, two copies of the system lies in a nontrivial compatible family with trivial continuum image. We then give a more direct, but structurally different, example: we show that restrictions of free continuum classes form compatible families of torsion lattice phases that are annihilated by the strict torsion-valued continuum-limit map. 
Finally, in Sec.~\ref{sec:afm_vbs}, we use the universal restriction framework to formulate a general criterion for determining whether a continuum anomaly is detected by a given lattice point group. We then apply this criterion to the mixed spin-spatial anomaly of the $(2{+}1)$D N\'eel-VBS transition and show that the corresponding bulk class is detected precisely when the rotational subgroup has even order. This recovers the distinction between square and honeycomb lattices discovered in previous work and extends the result to general two-dimensional point groups.

\subsection{$(2{+}1)$D crystalline responses and boundary inflow}
\label{sec:example_1}

The aim of this subsection is to illustrate the continuum-limit construction for concrete $(2{+}1)$D crystalline systems having discrete rotation/reflection symmetry and the corresponding continuum symmetry group $O(2)$. By Theorem~\ref{theorem:intermediate}, the continuum classification is already saturated by crystal point groups that are elementary abelian $2$-subgroups of $O(2)$. Hence, we can work in the Quillen category $\qa(O(2))$, where, conveniently, everything can be computed and illustrated explicitly. 
\begin{figure}[H]
    \centering
    \includegraphics[width=0.75\linewidth]{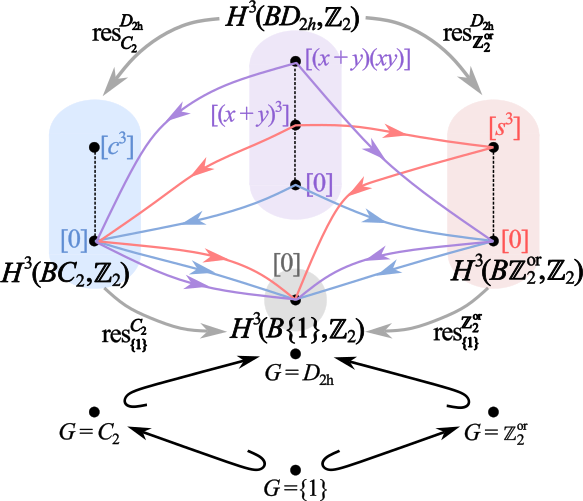}
    \caption{ 
    Formation of compatible families in the inverse limit over $\qa(O(2))$. Phases connected by lines of the same color form a single compatible family. We use the pre-TQFT classification model and display only phases in the (isotropic) Weyl-invariant subgroups, i.e., $H^{d+1}(BG, \Z_{2})^{W_{O(2)}(G)}$, for each $G,$ as they are the only ones that have a continuum limit.}
    \label{fig:compatible-families}
\end{figure}

\subsubsection{Pre-TQFT model: mod-2 classes and compatible families}
\label{sec:example_1_preTQFT}

As discussed in Sec.~\ref{sec:inverselimit}, up to conjugation, the Quillen orbit category $\qa(O(2))$ has four elementary abelian objects: $D_{2h}=\mathbb Z_2^{x}\times \mathbb Z_2^{y}$,
$C_2$, $\mathbb Z_2^{\mathrm{or}}$, and the trivial group $\{1\}$. Recall that the group $\mathbb Z_2^{\mathrm{or}}$ contains a single orientation-reversing reflection, while $D_{2h}$ is generated by two such reflections. We can compute the classification in the pre-TQFT model defined in Eq.~\ref{eq:preTQFT}. Let $x,\; y$ be the two degree-$1$ generators in $H^1(BD_{2h},\mathbb Z_2)$, and let $c,\; s$ be the degree-$1$ generators in $H^{1}(BC_{2}, \Z_{2}),$ and $H^{1}(B\Z_{2}^{\mathrm{or}}, \Z_{2})$ respectively. Hence, 
\begin{gather*}
H^{3}(BD_{2h},\mathbb Z_2)= \mathrm{Span}_{\mathbb{Z}_{2}}\left\{x^{3}, y^{3}, xy^{2}, x^{2}y\right\}=\mathbb{Z}_{2}^{4},
\\
H^{3}(BC_2,\mathbb Z_2)=\mathrm{Span}_{\mathbb{Z}_{2}}\left\{c^{3}\right\}=\mathbb{Z}_{2},\;
\\ H^{3} \big(B\mathbb Z_2^{\mathrm{or}},\mathbb Z_2\big)=\mathrm{Span}_{\mathbb{Z}_{2}}\left\{s^{3}\right\}=\mathbb{Z}_{2}.
\end{gather*}
Not all classes in these groups are isotropic, so we restrict to isotropic classes $H^{3}(BG, \mathbb{Z}_{2})^{W_{O(2)}(G)}$ (Weyl-invariant under conjugation in $O(2)$).\footnote{Recall that only isotropic phases can have continuum limits, i.e., $\Theta_{c}(BE)\subseteq \Theta^{d+1}(BE)^{W_{\mathrm{O}(d)}(E)}$. See Sec.~\ref{sec:motivation} and App.~\ref{app:proof_isotropy}.} 

In Fig.~\ref{fig:compatible-families} we show restriction relations among these classes and the compatible families for the Quillen category. Recall that compatible families are choices of phases for each object in an orbit category that agree under restriction maps, see Sec.~\ref{sec:inverselimit}.  In fact, each different colored loop in Fig.~\ref{fig:compatible-families} represents a compatible family. We can see that some lattice phases are not connected by a colored loop and hence do not form a component of a compatible family (an example being the non-trivial class $[c^3]$ for the $C_2$ group).

As discussed in Sec. \ref{sec:univCSPT}, the inverse limit over $\qa(O(2))$ is the abelian group formed from the set of compatible families
\begin{equation*}
([x_{D_{2h}}],\,[x_{C_2}],\,[x_{\mathbb Z_2^{\mathrm{or}}}],\,[x_{\{1\}}])\in \varprojlim_{E\in \qa(O(2))} H^{2+1}(BE, \Z_{2}),
\end{equation*}
with group operation given by component-wise addition (c.f., Footnote~\ref{footnote_comp}). Since, up to conjugation, $D_{2h}$ is the unique maximal object of $\qa(O(2))$, every compatible family is completely determined by its $D_{2h}$-component. That is, once a class on $D_{2h}$ is chosen, the classes on $C_2$, $\mathbb Z_2^{\mathrm{or}}$, and $\{1\}$ that enter the compatible family are determined by its restrictions along the corresponding morphisms, e.g., $([x_{D_{2h}}],\,\res^{D_{2h}}_{C_{2}}[x_{D_{2h}}],\,\res^{D_{2h}}_{\Z_{2}^{\mathrm{or}}}[x_{D_{2h}}],\,\res^{D_{2h}}_{\{1\}}[x_{D_{2h}}])$. Thus, to determine the inverse-limit group, it is enough to identify those isotropic classes on $D_{2h}$ that generate distinct compatible families. 

In Fig. \ref{fig:compatible-families} we see three compatible families, one of which is trivial. Thus, there are two nontrivial generators of the inverse limit group, which we label by their $D_{2h}$-components:
\begin{equation}
\begin{gathered}
    [(x{+}y)^3] \in H^{3}\big(BD_{2h}, \mathbb Z_{2}\big)^{W_{O(2)}(D_{2h})},
    \\
    [(x{+}y)(xy)] \in H^{3}\big(BD_{2h}, \mathbb Z_{2}\big)^{W_{O(2)}(D_{2h})}.
    \label{eq:ex1_mod2_gen}
\end{gathered}
\end{equation}
Their associated compatible families are obtained by restricting these classes along the morphisms of $\qa(O(2))$, for example along $\res^{D_{2h}}_{C_2}$ and $\res^{D_{2h}}_{\mathbb Z_2^{\mathrm{or}}}$, as shown by the colored loops in Fig.~\ref{fig:compatible-families}.

\bigskip 

To identify the continuum representatives of these two families at the pre-TQFT (i.e., mod-$2$) stage, we compare with the continuum classification (see Ch.7, Theorem 7.1 in Ref.~\cite{Milnor}):
\begin{equation*}
    H^{3}(BO(2), \mathbb{Z}_{2}) = \mathrm{Span}_{\mathbb{Z}_{2}}\{w_{1}^{3}, \; w_{1}w_{2}\}=\mathbb{Z}_{2}^{2},
\end{equation*}
where $w_{1},\; w_{2}$ are the first and second Stiefel-Whitney classes respectively. Under the 
mod-$2$ Quillen $F$-isomorphism correspondence (see Thm.~\ref{theorem:quillen}, and Sec.~\ref{sec:preTQFT_Quillen}), the two generators on each side are connected via 
\begin{equation}
\begin{gathered}
    \res^{O(2)}_{D_{2h}}  :\; w_2\,w_1 \longmapsto [(x{+}y)(xy)],\\
    \res^{O(2)}_{D_{2h}}  :\; w_1^{3} \longmapsto [(x{+}y)^3].
    \label{eq:ex1_quillen}
    \end{gathered}
\end{equation} Thus, the two compatible families \eqref{eq:ex1_mod2_gen} correspond to the continuum generators $w_1^3$ and $w_1w_2$ via \eqref{eq:ex1_quillen}.

\subsubsection{Group cohomology: Bockstein lift and surviving families}
\label{sec:example_1_grpcoh}

To obtain results for the full group cohomology model defined in Eq.~\ref{eq:grpcohtormodel} we lift the mod-$2$ classification to the 2-torsion subgroup of the integral cohomology with twisted coefficients via the Bockstein map\footnote{We prove that this Bockstein map is surjective onto $H^{d+2}(BG, \Z_{w_{1}})[2]$ in App.~\ref{app:lift}, see Claim~\ref{claim:objectwise_bockstein_surjective}. Recall from notation~\ref{notation:torsion} that $H^{d+2}(BG,\Z_{w_{1}})[2]=\tor_{2}\left(H^{d+2}(BG,\Z_{w_{1}})\right)$ for the cases considered here, so applying the map to the mod-$2$ classification produces the full cohomology classification.} 
\begin{equation}
\beta_{w_{1}}: H^{d+1}(BG, \mathbb{Z}_{2}) \twoheadrightarrow \tor_{2}(H^{d+2}(BG, \mathbb{Z}_{w_{1}})).
\label{eq:ex1_Bockstein}
\end{equation}
Recall that the twisted coefficients $\mathbb Z_{w_1}$ encode the sign action of orientation-reversing symmetries on local response amplitudes; see Def.~\ref{def:grpcoh_classmodel} and Apps.~\ref{app:grpcoh_def},~\ref{app:twisted_coefficients} for details.
In the present example, one finds
\begin{equation}
\beta_{w_{1}}\left(\left[w_{1}^{3}\right]\right)=0,
\qquad
\beta_{w_{1}}(\left[w_{1}w_{2}\right])\neq 0.
\label{eq:vanishing_w13}
\end{equation}
See Note ~\cite{note_kill} for details. Correspondingly, for the two compatible families identified above, the $D_{2h}$-components satisfy\footnote{The lattice-side statements follow from Eq.~\ref{eq:ex1_quillen} together with the naturality of the Bockstein map under restriction.}
\begin{equation*}
\beta_{w_{1}}\left(\left[(x+y)^{3}\right]\right)=0,
\qquad
\beta_{w_{1}}(\left[(x+y)xy\right])\neq 0.
\end{equation*}
Thus, after passing from the pre-TQFT model to the full group-cohomology model, the family generated by $[(x+y)^3]$ becomes trivial, while the family generated by $[(x+y)xy]$ survives. Equivalently, there remains a unique nontrivial elementary cSPT family, together with its corresponding nontrivial continuum response class, i.e., $\beta_{w_{1}}([w_{1}w_{2}])$.

\bigskip 

Finally, we represent these classes in terms of the group cohomology model with $\U(1)_{w_{1}}$-coefficients, as defined in Def.~\ref{def:grpcoh_classmodel}, using the isomorphism (see Eq.~\ref{eq:U1_to_Z_finite} in App.~\ref{app:U1_coeff_to_Z}): 
\begin{equation}
    e: \tor_{2}\left(H^{d+1}(BG, \U(1)_{w_{1}}) \right)\rightarrow \tor_{2}\left(H^{d+2}(BG, \Z_{w_{1}})\right).
    \label{eq:e}
\end{equation}
Here, we use the notation $\overline{x}:= e^{-1}(\beta_{w_{1}}(x))$, that is, we represent the $\U(1)$-valued group cohomology classes corresponding to the lift of the mod-$2$ classes $x$ as $\overline{x}$. 
In these terms, the unique surviving compatible family is represented by
\begin{equation*}
\left[\overline{xy(x+y)}\right] \in H^{3}(BD_{2h}, \U(1)_{w_{1}}),
\end{equation*} 
with the compatible family defined over the category $\mathcal{A}_{2}(O(2))$ associated to the class given explicitly by the collection:
\begin{equation*}
    \left(\left[\overline{xy(x+y)}\right], \;[\overline{0}_{C_{2}}],\;[\overline{0}_{\Z_{2}^{\mathrm{or}}}],\;[\overline{0}_{\{1\}}]\right)\in \varprojlim_{E\in \qa(O(2))} H^{3}(BE, \U(1)_{w_{1}}),
\end{equation*}
where we see that the $G$-components for $G=C_{2},\Z_{2}^{\mathrm{or}}, \{1\}$ are trivial and represented by $[\overline{0}_{G}]$. The continuum limit of this compatible family is given by
\begin{equation*}
    \left[\overline{w_1w_2}\right]\in H^{3}(BO(2), \U(1)_{w_{1}}).
\end{equation*}

Thus, among the elementary symmetry groups in $\qa(O(2))$, only $D_{2h}$ supports a nontrivial class in the full group-cohomology model. Accordingly, for lattice structures whose point-group symmetry is $D_{2h}$—for example $pmm$-type lattices, when only the point-group part is taken as the protecting symmetry—there is a single nontrivial deformation class of lattice TQFTs admitting a continuum limit, represented by $[\overline{(x+y)xy}]$. In contrast, for lattice structures whose point-group symmetry is $C_{2}$ or $\Z_{2}^{\mathrm{or}}$—for example $p2$- and $pm$-type lattices—the corresponding classes are trivial in the full group-cohomology model. Hence only the $D_{2h}$-symmetric lattice phase survives, and it flows to the continuum response class $[\overline{w_{1}w_{2}}]$.\footnote{This conclusion is specific to the purely spatial classification considered in this subsection. In the presence of internal symmetry, purely internal classes and additional mixed crystalline-internal classes may arise from spatial cohomology classes in different degrees, potentially admitting a continuum limit.}

\bigskip 

An explicit microscopic realization of the corresponding crystalline phase can be obtained from Horinouchi's commuting-projector model for the $(2{+}1)$D bosonic $\U(1)\rtimes\mathbb Z_2^T$ SPT by applying the crystalline equivalence principle \cite{Horinouchi2020,ThorngrenElse}. Under this correspondence, the internal $\U(1)$ symmetry is reinterpreted as spatial $\SO(2)$ rotation, while the antiunitary $\mathbb Z_2^T$ symmetry is reinterpreted as an orientation-reversing mirror symmetry. This produces an $\mathrm O(2)$-protected crystalline SPT whose topological response is the class considered here.

\bigskip 

\noindent In the next two subsections, we analyze the topological response of the continuum limit and show how this response theory can be used to construct the anomalous boundary theory of the corresponding continuum SPT phase.

\subsubsection{Continuum response theory}

Given a closed $(2{+}1)$-manifold $M$ equipped with an $O(2)$-background $f_{O(2)}:M\to BO(2)$, the invertible TQFT response for the non-trivial class is defined by the general formula~\eqref{eq:top-response}:
\begin{equation}
  \mathcal Z_{\;\overline{w_1w_2}}\left[M,f\right]
  \;:=\;
  \left\langle f_{O(2)}^{*}([\overline{w_1w_2}]),\, [M] \right\rangle
  \;\in\; U(1).
\end{equation}
Equivalently, using the exponentiated convention explained in remark~\ref{rmk:exp_convention} (i.e.,\ identifying $U(1)\cong \mathbb R/2\pi \mathbb Z$), we may write
\begin{equation*}
\mathcal Z_{\;\overline{w_1w_2}}\left[M,f_{O(2)}\right]
= \exp\left(i\,\left\langle f_{O(2)}^{*}(\left[\overline{w_1w_2}\right]),\,[M]\right\rangle\right).
\end{equation*} This expression is defined entirely by the topological response class $[\overline{w_1w_2}]$ and the homotopy class of the background map $f_{O(2)}:M\to BO(2)$. In general, such topological responses need not admit geometric representation as the integral of a smooth differential-form density. In particular, torsion classes such as $w_1^3$ and $w_1w_2$ do not admit such a refinement. Nevertheless, these topological responses are more general than differential-form actions, and in some situations they admit smooth geometric representatives after reduction to lower-dimensional subspaces. In the present example, this occurs in the boundary description below, where the surviving response reduces to an oriented boundary theory involving only $w_2$, which in turn admits a Chern-Weil representative in terms of an $\SO(2)$ connection and its curvature; see, e.g., Sec.~\ref{sec:example_1_grav} and Eq.~\ref{eq:bdry_geom}.

\medskip

To make contact with the cochain-level anomaly inflow computation below, it is convenient to temporarily revert to the underlying $\Z_2$-valued Stiefel-Whitney classes. When exponentiating a $\Z_2$-valued pairing we use the standard embedding $\Z_2\hookrightarrow U(1)$ sending $0\mapsto 1$ and $1\mapsto -1$, i.e.,
\begin{equation*}
x\in \Z_2
\quad\longmapsto\quad
\exp(i\pi \,x)\in U(1).
\end{equation*}
Accordingly, for a closed $(2{+}1)$-manifold $M$ with  principal $O(2)$-bundle $\pi_f: P\to M$ defined by the classifying map $f:M \to BO(2)$, we may write
\begin{equation}
\mathcal Z_{\overline{w_1w_2}}\left[M, f\right]=
\exp\Big(i\pi \,\left\langle w_1(P)\smile w_2(P),\,[M]\right\rangle\Big)
\in \{\pm1\}\subset U(1),
\label{eq:Z2_to_U1_embedding}
\end{equation}
where $w_{1}(P)=f^{*}(w_{1}),\; w_{2}(P)=f^{*}(w_{2}) \in H^{*}(M, \Z_{2})$\footnote{We use notation $w_{i}(P)\in H^{*}(M,\Z_{2})$ to distinguish from $w_{i}(M)=w_{i}(TM)$ which refers to Stiefel-Whitney classes on the tangent space of the manifold.} are pull-backs of the universal Stiefel-Whitney classes $w_{1},w_{2}$ on $BO(2)$ (see Sec.~\ref{sec:top_resp} for more details).

\subsubsection{Boundary response and anomaly inflow}
\label{sec:example_1_grav}

Let $M$ now be a compact $(2{+}1)$D manifold with (possibly empty) boundary $\partial M$.  Choose cocycle representatives\footnote{We use standard terminology from algebraic topology for cohomology $H^*(M,A)=Z^{*}(M,A)/B^{*}(M,A)$ (cochain group $C^{*}(M, A)$, $\delta: C^{*}(M,A) \to C^{*+1}(M,A)$ for coboundary operator, cocycle group $Z^{*}(M, A)=\{c\in C^{*}(M, A)\;|\; \delta c=0\}$, and coboundary group $B^{*}(M, A)=\{b\in C^{*}(M,A)\;|\; \exists \;c\in C^{*-1}(M,A)\;\; \text{s.t.}\; b=\delta c\}$) throughout this subsection. For a mathematical reference see Hatcher~\cite{Hatcher_AT}. For applications of cochain formulations to topological response actions and invertible TQFTs, see e.g., Ref.~\cite{Tiwari2018} App.C.} $r\in Z^{1}(M,\Z_2)$ and $u\in Z^{2}(M,\Z_2)$ for $w_1(P)$ and $w_2(P)$,
respectively.  Using these cocycle representatives in Eq. \ref{eq:Z2_to_U1_embedding} we find the bulk topological action 
\begin{equation}
  \mathcal Z_{\mathrm{bulk}}\left[M;\;r,u\right]
  \;:=\;
  \exp \Big(i\pi \int_{M} r\smile u\Big).
  \label{eq:bulk_ru}
\end{equation}
When $\partial M\neq\varnothing$ this expression is well-defined but not gauge
invariant by itself: under a $\Z_2$ gauge transformation
$r\mapsto r+\delta\lambda$ with $\lambda\in C^{0}(M,\Z_2)$, and because $\delta u=0$ one finds
\begin{equation}
  \frac{\mathcal Z_{\mathrm{bulk}}\left[M;\;r+\delta\lambda,u\right]}
       {\mathcal Z_{\mathrm{bulk}}[M;\;r,u]}
  \;=\;
  \exp \Big(i\pi \int_{\partial M}\lambda\smile u\Big).
  \label{eq:bulk_variation}
\end{equation}
This is the inflow phase that must be canceled by a $(1{+}1)$D boundary theory.

\medskip

We now impose a \emph{boundary condition} appropriate to the $O(2)$ setting: we
assume that the bundle restricted over the boundary has vanishing first Stiefel-Whitney class,
\begin{equation}
  w_1(P|_{\partial M})=0\in H^{1}(\partial M,\Z_2),
  \label{eq:boundary_w1zero}
\end{equation}
so that $P|_{\partial M}$ admits an $\SO(2)$-reduction.  Equivalently,
$r|_{\partial M}$ is a coboundary, hence there exists a (locally constant)
$0$-cochain $\sigma\in C^{0}(\partial M,\Z_2)$ such that
\begin{equation}
  r|_{\partial M}=\delta\sigma.
  \label{eq:r_boundary_trivialization}
\end{equation}
The choice of $\sigma$ is part of the boundary data; on each connected component
of $\partial M$ it is constant, reflecting the two possible orientations of the boundary. 

With this choice of boundary orientation, we define a boundary partition function
\begin{equation}
  \mathcal Z_{\mathrm{bdry}}\left[\partial M;\;\sigma,u\right]
  \;:=\;
  \exp \Big(i\pi \int_{\partial M}\sigma\smile u\Big).
  \label{eq:boundary_counterterm}
\end{equation}
Under the induced transformation $\sigma\mapsto \sigma+\lambda|_{\partial M}$,
its variation is
\begin{equation}
  \frac{\mathcal Z_{\mathrm{bdry}}\left[\partial M;\;\sigma+\lambda,u\right]}
       {\mathcal Z_{\mathrm{bdry}}\left[\partial M;\;\sigma,u\right]}
  \;=\;
  \exp \Big(i\pi \int_{\partial M}\lambda\smile u\Big).
  \label{eq:bdry_variation}
\end{equation}
Comparing \eqref{eq:bulk_variation} and \eqref{eq:bdry_variation}, we see that the
combined partition function
\begin{equation}
  \mathcal Z_{\mathrm{tot}}[M,\partial M;\;r,u;\sigma]
  \;:=\;
  \mathcal Z_{\mathrm{bulk}}[M;r,u]\cdot
  \mathcal Z_{\mathrm{bdry}}[\partial M;\sigma,u]^{-1}
  \label{eq:total_partition}
\end{equation}
is gauge invariant.  This is the anomaly inflow mechanism for the response class $\overline{w_1w_2}$: the boundary theory depends on the choice of boundary orientation $\sigma$, and its anomalous gauge variation is precisely canceled by the bulk inflow phase.\footnote{This type of construction is analogous to similar constructions done in Chern-Simons theory. See for example, the well known anomaly inflow construction for Chern-Simons theory of fractional quantum hall(FQH) liquids in Ch.7.4.5, Ref.~\cite{wen2004quantum}.} 

\bigskip

Finally, as promised, we write the boundary theory in terms of smooth differential forms.  Assume $\partial M$ has a single connected component, so that the boundary orientation $\sigma\in C^{0}(\partial M,\Z_{2})$ is constant (hence $\sigma=0$ or $1$).  The boundary condition $w_{1}(P|_{\partial M})=0$ allows us to choose an $\SO(2)$-reduction $\widetilde L\to \partial M$ of the restricted $O(2)$-bundle
$P|_{\partial M}$.  For such a reduction one has 
\begin{equation*}
  w_{2}(P|_{\partial M}) \;=\; c_{1}(\widetilde L)   \pmod 2,
\end{equation*}
where $c_{1}(\widetilde L)\in H^{2}(\partial M, \Z)$ is the first Chern class of the bundle $\widetilde L \to \partial M$. Now, to define the smooth geometric response we provide $\omega$ as a $\SO(2)\cong U(1)$ connection on $\widetilde L$ with curvature $F_{\omega}.$ Then the Chern-Weil homomorphism (see Ch.11 in Ref.~\cite{Nakahara2018}) gives us the differential refinement
\begin{equation}
\label{eq:ChernWeil}
  \big\langle c_{1}(\widetilde L),[\partial M]\big\rangle
  \;=\;
  \int_{\partial M}\frac{F_{\omega}}{2\pi}
  \;\in\;\Z.
\end{equation}
Thus, the boundary partition function may be written as 
\begin{equation}
\label{eq:bdry_geom}
  \mathcal Z_{\mathrm{bdry}}(\partial M;\sigma,\omega)
  \;=\;
  \exp \left\{\, i\pi\,\sigma  \int_{\partial M}\frac{F_{\omega}}{2\pi}\right\}.
\end{equation} 

Although Eq.~\ref{eq:bdry_geom} evaluates to $1$ when $\sigma=0$, the anomaly is captured by the \emph{relative} transformation between the two boundary orientations: for $\lambda\in H^{0}(\partial M,\Z_{2})$,
\begin{equation}
  \frac{\mathcal Z_{\mathrm{bdry}}(\partial M;\sigma+\lambda,\omega)}
       {\mathcal Z_{\mathrm{bdry}}(\partial M;\sigma,\omega)}
  \;=\;
  \exp \left\{\, i\pi\,\lambda  \int_{\partial M}\frac{F_{\omega}}{2\pi}\right\}.
  \label{eq:bdry_geom_variation}
\end{equation}
This nontrivial dependence on the choice of boundary orientation is the
boundary anomaly, and it is precisely canceled by the bulk inflow phase
\eqref{eq:bulk_variation}.

\bigskip 

In summary: in the pre-TQFT model, the inverse limit over $\qa(O(2))$ yields two nontrivial compatible families. These two families are generated on the maximal subgroup $D_{2h}$ by $[(x{+}y)^3]$ and $[(x{+}y)(xy)]$ respectively in Eq.~\ref{eq:ex1_mod2_gen}, and correspond (under the Quillen correspondence \eqref{eq:ex1_quillen}) to the continuum classes $w_1^3$ and $w_1w_2$. After lifting via the Bockstein map \eqref{eq:ex1_Bockstein}, the $w_1^3$ family becomes trivial~\eqref{eq:vanishing_w13}, so only the $w_1w_2$ family survives in the full group-cohomology model, and is represented by the continuum response class $[\overline{w_1w_2}]\in H^3(BO(2),U(1)_{w_1})$. Finally, we exhibit the response of class $[\overline{w_1w_2}]$ by a boundary construction: the bulk response on $M$ reduces to a $(1{+}1)$D theory on the oriented boundary $\partial M$, which is (gravitationally) anomalous and is canceled by anomaly inflow from the $(2{+}1)$D bulk (Sec.~\ref{sec:example_1_grav}).

\subsection{Higher torsion phases}
\label{sec:example_2}

In this section, we examine higher-order lattice SPT phases from two complementary perspectives and explain their relation to lattice TQFT data that is erased/forgotten by the continuum-limit map. We first study an explicit second-order SPT protected by $C_4$ rotation and internal symmetry $H=\mathbb Z_4\times\mathbb Z_4$. We show in Sec.~\ref{sec:example_2_highertorsion}, this phase either fails to extend to a compatible universal family, and hence admits no continuum limit, or its extension produces, upon doubling, a nontrivial lattice phase with trivial continuum image. We then turn in Sec.~\ref{sec:free_shadow_example} to a more systematic mechanism: restrictions of free continuum TQFT classes to finite spatial symmetries form compatible families of torsion lattice phases, but are annihilated by the strict torsion-valued continuum-limit map. 

Together, these examples show how higher-order torsion phenomena on the lattice are naturally associated with information that is either obstructed from reaching the continuum or becomes invisible after taking the continuum limit. Heuristically this might have been expected as many higher order phases are tightly linked to the discrete lattice structure, e.g., gapless boundary states can appear on sharp corners, but they spread out and annihilate each other if the corners are softened and smooth as would be the case for a boundary that approximates a continuous rotation symmetry. Beyond our discussions below, we will leave a detailed understanding about higher order phases to future work.

\subsubsection{Second order SPT}
\label{sec:example_2_highertorsion}

We now examine the continuum-limit admissibility of a higher-order lattice phase with genuine higher $2$-primary torsion. Consider a bosonic system in $(2+1)$ dimensions with internal symmetry
$H:=\mathbb Z_{4}^{(1)}\times\mathbb Z_{4}^{(2)}$ and fourfold crystal rotational symmetry $G_c=C_{4}\subset \SO(2)$. Since the spatial symmetry is orientation preserving and acts trivially on $H$, the total symmetry group is the direct product $\Gamma_{C_{4}}
=H\times C_{4}$, and the orientation local system restricts to the constant coefficient system on $B\Gamma_{C_{4}}$. The corresponding group-cohomology classification is therefore  
\begin{equation}
\Theta^{2+1}(B\Gamma_{C_{4}})
=
\mathrm{Tor}_{2}
\left(
H^{4}(B\Gamma_{C_{4}},\mathbb Z)
\right).
\end{equation}

Because the total symmetry is the direct product $H\times G_c$, one can apply the insight of Rasmussen and Lu~\cite{Rasmussen2020} to use the K\"unneth formula to separate the full group-cohomology classification into purely crystalline, purely internal, and mixed crystalline-internal sectors. Within the mixed sector, the $n$th-order bosonic SPT phases in $d$ spatial dimensions are classified by
\begin{equation*}
H^{n-1}\left( G_c, H^{d-n+3}(BH,\mathbb Z_{w_1}) \right).
\end{equation*}
Specializing to second-order phases in $d=2$, we obtain\footnote{Rasmussen and Lu write the mixed part of the K\"unneth decomposition as
\begin{equation*}
\bigoplus_{k=0}^{d} H^{k}\left( G_c^{*}, H^{d-k+1}(H,U(1)) \right).
\end{equation*}
The $n$th-order HOSPT sector is the term obtained by setting $k=n-1$, namely $H^{n-1}\left(G_c^{*};H^{d-n+2}(H,U(1))\right)$. Here $G_c^{*}$ is obtained by representing orientation-reversing elements of $G_c$ as antiunitary elements. In our notation, the same sign action is encoded by the coefficient system $\mathbb Z_{w_1}$, and for finite $H$ the equivalence with the expression in the main text follows from $H^{q}(H,U(1))\cong H^{q+1}(BH,\mathbb Z)$. See Ref.~\cite{Rasmussen2020}.}
\begin{equation}
\mathcal C_{\mathrm{HOSPT}}(G_c,H) := H^{1}\left( G_c, H^{3}(BH,\mathbb Z_{w_1}) \right),
\label{eq:HOSPT_general_classification}
\end{equation}
where, following Ref.~\cite{Rasmussen2020}, $\mathcal C_{\mathrm{HOSPT}}(G_c,H)$ denotes the second-order SPT sector of the full group-cohomology classification. In the present case, $G_c=C_4\subset\SO(2)$ is orientation preserving, so the coefficient twist is trivial.

\bigskip 

For $H=\mathbb Z_4^{(1)}\times\mathbb Z_4^{(2)}$, the standard classification of projective representations of finite abelian groups gives 
\begin{equation} 
H^{2}(H,U(1)) \cong H^{3}(BH,\mathbb Z) \cong \mathbb Z_4. \label{eq:Z4Z4_projective_classification} \end{equation} 
And more generally, $H^{2}(\mathbb Z_a\times\mathbb Z_b,U(1)) \cong\mathbb Z_{\gcd(a,b)}$. Since the $C_4$ action on the internal symmetry is trivial, it also acts trivially on the coefficient group in Eq.~\eqref{eq:Z4Z4_projective_classification}. Consequently,\footnote{Here $\operatorname{Hom}(A,B)$ denotes the abelian group of group homomorphisms from $A$ to $B$. The identification
$H^{1}(C_4,\mathbb Z_4)\cong\operatorname{Hom}(C_4,\mathbb Z_4)$ holds because the action of $C_4$ on the coefficient group $\mathbb Z_4$ is trivial.} 
\begin{equation} 
\mathcal C_{\mathrm{HOSPT}} \left( C_4,\mathbb Z_4\times\mathbb Z_4 \right) \cong H^{1}(C_4,\mathbb Z_4) \cong \operatorname{Hom}(C_4,\mathbb Z_4) \cong \mathbb Z_4. 
\label{eq:C4_Z4Z4_HOSPT_classification} 
\end{equation} 
This agrees with the Rasmussen-Lu formula $H^{1}(C_n,\mathbb Z_{\gcd(a,b)}) \cong\mathbb Z_{\gcd(n,a,b)}$ upon setting $n=a=b=4$; see Ref.~\cite{Rasmussen2020}.

Now, let
\begin{equation}
[x_{C_{4}}] \in \mathcal C_{\mathrm{HOSPT}}
\left( C_{4}, \mathbb Z_{4}\times\mathbb Z_{4}
\right) \subseteq \Theta^{2+1}(B\Gamma_{C_{4}}),
\label{eq:x_C4}
\end{equation}
denote a generator of this $\mathbb Z_4$ higher-order SPT sector.\footnote{The generated group of HOSPTs $\Z_4\langle [x_{C_{4}}] \rangle$ is one mixed summand in the K\"unneth decomposition of the full group-cohomology classification $\Theta^{2+1}(B\Gamma_{C_4})$, which also contains purely spatial, purely internal, and other mixed contributions.} Equation~\eqref{eq:C4_Z4Z4_HOSPT_classification} implies that $[x_{C_{4}}]$ has exact order $4$:
\begin{equation}
4[x_{C_{4}}]=0, \qquad 2[x_{C_{4}}]\neq0.
\label{eq:C4_HOSPT_order_four}
\end{equation}
Thus the lattice classification contains a well-defined higher-order phase whose nontriviality persists after stacking two copies and disappears only after stacking four copies.

\bigskip

\begin{figure}[H]
    \centering
    \includegraphics[width=0.8\linewidth]{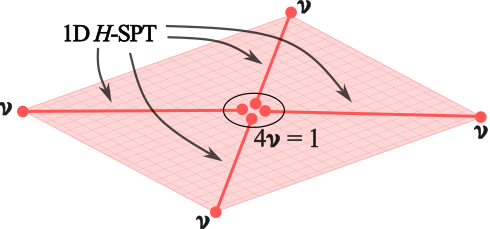}
    \caption{Dimensional-reduction picture of the $C_4$-protected HOSPT:
    The four radial $r_{\pi/2}$-domain walls are related by $C_4$ rotation and are each decorated by the one-dimensional $H$-SPT class
    $\nu\in H^{2}(H,U(1))\cong\mathbb Z_4$. Their outer endpoints furnish projective corner representations, while the four inner endpoint classes combine to $4\nu=0$ and may be symmetrically gapped at the rotation center.}
    \label{fig:C4_HOSPT_dimensional_reduction}
\end{figure}

The physical meaning of the class $[x_{C_{4}}]$ is most transparent through the dimensional-reduction construction of Rasmussen and Lu~\cite{Rasmussen2020}. Let $r_{\pi/2}$ denote the generator of $C_4$. Since the $C_4$ action on $H^{2}(H,U(1))$ is trivial, we have
\begin{equation*}
H^{1}\left(C_4;H^{2}(H,U(1))\right)
\cong
\operatorname{Hom}\left(C_4,H^{2}(H,U(1))\right),
\end{equation*}
so the class $[x_{C_{4}}]$ is determined by the one-dimensional $H$-SPT phase assigned to an $r_{\pi/2}$-domain wall. For the generator,
\begin{equation*}
[x_{C_{4}}](r_{\pi/2})=\nu=1 \in H^{2}(H,U(1)) \cong \mathbb Z_4.
\end{equation*}
In the decorated-domain-wall picture, the four radial segments in Fig.~\ref{fig:C4_HOSPT_dimensional_reduction} are the four $C_4$-related $r_{\pi/2}$-domain walls, each decorated by the same one-dimensional $H$-SPT phase $\nu$. Each endpoint of a decorated segment carries a projective representation of $H=\mathbb Z_4^{(1)}\times\mathbb Z_4^{(2)}$. Let $h_1$ and $h_2$ denote the generators of the two internal $\mathbb Z_4$ factors. The endpoint representation associated to $\nu\in\mathbb Z_4$ is characterized by
\begin{equation}
\widehat U_{h_1}\widehat U_{h_2}
=
\exp\left(\frac{2\pi i\nu}{4}\right)
\widehat U_{h_2}\widehat U_{h_1}.
\label{eq:Z4Z4_corner_projective_rep}
\end{equation}
For $\nu=1$, the internal symmetry operators therefore commute projectively with a phase $i$.

The order-$4$ stacking law is reflected directly in the corner representation. Stacking two copies gives $\nu=2$, for which
\begin{equation*}
\widehat U_{h_1}\widehat U_{h_2}
=
-\widehat U_{h_2}\widehat U_{h_1},
\end{equation*}
so the stacked commutator phase (and thus the higher order SPT) remains nontrivial. After stacking four copies, the commutator phase becomes unity and the corner representation becomes linear, allowing the phase to be symmetrically trivialized. The corner modes are protected as a $C_4$-symmetric configuration, although their precise locations may change under crystalline-symmetric deformations of the boundary termination.

\bigskip

We now ask whether the lattice phase $[x_{C_{4}}]$ admits a continuum limit in the sense of the universal construction. Let
\begin{equation}
\pi_{C_4}: \mathrm{cSPT}^{2+1}_{\mathrm{univ},H}
\longrightarrow \Theta^{2+1}(B(H\times C_4))
\label{eq:C4_component_projection}
\end{equation}
denote the canonical projection onto the $C_4$ component of a compatible family. By definition, $[x_{C_{4}}]$ is continuum-limit admissible precisely when $[x_{C_{4}}]\in\operatorname{Im}(\pi_{C_4})$.
This leads to two mutually exclusive possibilities.

\bigskip 

\noindent (i) There may be no compatible family 
\begin{equation*}
X=\left([x_{G}]\right)_{G\in \ofin(\mathrm{O}(d))}\in \mathrm{cSPT}^{2+1}_{\mathrm{univ},H},
\end{equation*}
satisfying $\pi_{C_4}(X)=[x_{C_{4}}]$. Note that here $X$ is a shorthand denoting the compatible family as a whole, while
\begin{equation*}
[x_G]:=\pi_G(X) \in \Theta_H^{2+1}(BG) = \Theta^{2+1}\bigl(B(H\times G)\bigr)
\end{equation*}
denotes its $G$-component. In this case, $[x_{C_{4}}]$ is intrinsically crystalline: although it defines a valid $C_4$-symmetric lattice phase, it does not extend coherently across the finite-subgroup category, and therefore admits no continuum image under the universal construction.

\bigskip 

\noindent (ii) Instead suppose that such an extension $X\in \mathrm{cSPT}_{\mathrm{univ}}^{2+1}$ exists, with 
$[x_{C_4}]=\pi_{C_4}(X)$. Consider the doubled compatible family
\begin{equation}
\gamma:=2X\in \mathrm{cSPT}^{2+1}_{\mathrm{univ}}.
\label{eq:doubled_HOSPT_family}
\end{equation}
Since the component projections are homomorphisms,
\begin{equation}
\pi_{C_4}(\gamma) = 2\pi_{C_4}(X) = 2[x_{C_{4}}]
\neq0.
\label{eq:doubled_HOSPT_nonzero}
\end{equation}
Thus $\gamma$ is a nontrivial element of the universal cSPT classification.

\medskip 

From~\eqref{eq:x_C4}, $[x_{C_4}]$ belongs to the mixed spatial-internal HOSPT sector and becomes trivial when the spatial symmetry is forgotten. Since $\Gamma_{C_4}=H\times C_4$, the corresponding map is the pullback along the fiber inclusion $\jmath_{C_4}:BH\to B\Gamma_{C_4}$ defined in Sec.~\ref{sec:internal_sym} (see Eq.~\ref{eq:jmath_G}), so that
\begin{equation*}
\jmath_{C_4}^*= \res^{H\times C_4}_{H},
\qquad \jmath_{C_4}^*\bigl([x_{C_4}]\bigr)=\res^{H\times C_4}_{H}\bigl([x_{C_4}]\bigr)=0.
\end{equation*}
Let $[x_{\{1\}}]=\pi_{\{1\}}(X)\in\Theta^{2+1}(BH)$, denote the component of the compatible family at the trivial spatial point group (but keeping the $H$ symmetry). Compatibility with the inclusion $H\cong H\times\{1\}\hookrightarrow H\times C_4$ gives
\begin{equation*}
[x_{\{1\}}] = \res^{H\times C_4}_{H}\bigl([x_{C_4}]\bigr) = 0.
\end{equation*} For each $E\in\mathcal A_2(\mathrm O(2))$, let $[x_E]=\pi_E(X) \in \Theta^{2+1}\bigl(B(H\times E)\bigr)$ be part of the same compatible family as $[x_{C_4}].$ 
Compatibility with the fiber inclusion $\jmath_{E}:H\cong H\times\{1\}\hookrightarrow H\times E$ implies
\begin{equation}
\jmath_E^*([x_E])
=\res^{H\times E}_{H}([x_E])=[x_{\{1\}}]=0.
\label{eq:HOSPT_elementary_no_internal}
\end{equation}
Thus, every elementary component $[x_E]$ has no purely internal contribution. Now, define the subgroup of classes which become trivial when the spatial symmetry is forgotten by 
\begin{equation*}
\Theta^{2+1}_{\mathrm{sp}}\bigl(B(H\times E)\bigr) :=
\ker\left( \res^{H\times E}_{H}: \Theta^{2+1}\bigl(B(H\times E)\bigr)
\to \Theta^{2+1}(BH) \right).
\end{equation*} Thus, $[x_E]\in\Theta^{2+1}_{\mathrm{sp}}\bigl(B(H\times E)\bigr)$. 

\medskip 

\noindent Next, one can prove, as shown in Lemma~\ref{lem:Z4Z4_spatial_kernel_order2} of App.~\ref{app:higher_order_example}, that for elementary abelian $E$, the subgroup of classes 
$\Theta^{2+1}_{\mathrm{sp}}\bigl(B(H\times E)\bigr)$ is annihilated by $\times 2$, i.e., stacking two copies of any class in this subgroup gives the trivial class.\footnote{Restriction-corestriction shows that classes in $\ker(\res^{H\times E}_{H})$ are torsion arising from the finite spatial factor $E$. For elementary abelian $2$-groups, the relevant positive-degree spatial cohomology has exponent $2$, so this kernel is annihilated by multiplication by $2$; see Lemma~\ref{lem:Z4Z4_spatial_kernel_order2}.}
Consequently,
\begin{equation}
2[x_E]=0 \qquad \forall\;\;E\;\in\mathcal A_2(O(2)).
\label{eq:doubled_HOSPT_elementary_vanishing}
\end{equation}
It follows that the doubled family restricts trivially to the
elementary subcategory:
\begin{equation}
\Pi_H(\gamma) = \bigl(2[x_E]\bigr)_{E\in\mathcal A_2(O(2))} = 0.
\label{eq:HOSPT_Pi_vanishing}
\end{equation}
Recall $\Pi_{H}:\mathrm{cSPT}_{\mathrm{univ},H}^{d+1}\to \mathrm{cSPT}_{\mathrm{elem},H}^{d+1}$.
Since its $C_4$ component is nonzero by Eq.~\eqref{eq:doubled_HOSPT_nonzero}, the family $\gamma$ itself is nontrivial. Hence
\begin{equation}
0\neq\gamma=2X\in\ker(\Pi_H).
\label{eq:HOSPT_nontrivial_kernel_Pi}
\end{equation}
The specialized higher-Bockstein construction in App.~\ref{app:higher_order_example} defines the continuum-limit map by the factorization
\begin{equation*}
F_{2,H} =
\widetilde F_{2,H}\circ\Pi_H
\end{equation*}
and proves $\ker(\Pi_H)\subseteq\ker(F_{2,H})$.\footnote{In App.~\ref{app:higher_order_example}, these specialized degree-$4$ maps on the full $2$-primary torsion sector are denoted $\Pi_H^{4,(2)}$ and $\widetilde F_{2,H}^{4,(2)}$.} We therefore conclude that
\begin{equation}
0\neq\gamma=2X
\in
\ker(\Pi_H)
\subseteq
\ker(F_{2,H}).
\label{eq:HOSPT_nontrivial_kernel_class}
\end{equation}

\noindent We thus obtain the continuum-extension dichotomy
\begin{equation}
\begin{aligned}
&\text{Either}\;\;\;[x_{C_{4}}]\notin\operatorname{Im}(\pi_{C_4}) \implies [x_{C_{4}}]\, \text{does not have CL},\\
&\text{or}\;\;\exists\,X\ \text{with}\ 
\pi_{C_4}(X)=[x_{C_{4}}]
\ \text{and}\ 
0\neq2X\in\ker(F_{2,H}).
\end{aligned}
\label{eq:HOSPT_continuum_extension_dichotomy}
\end{equation}
In the first case, the order-$4$ HOSPT has no continuum-limit description. In the second, it admits a compatible extension, but doubling that extension produces nontrivial lattice information which is erased by the continuum limit. 
That is, in the second case, the doubled compatible family is nontrivial as lattice TQFT data but lies in the kernel of the continuum-limit map, and therefore has trivial continuum image.

\bigskip 

Several effective continuum descriptions of higher-order phases have been proposed, including boundary Dirac theories in which corner or hinge modes arise from symmetry-related mass domain walls, and multipolar or topological-crystal response theories that retain spatially varying masses, facet decompositions, or other crystalline geometric data~\cite{Khalaf2018,You2021,Huang2022}.\footnote{These works are formulated largely in free-fermion or massive-Dirac settings, with Ref.~\cite{You2021} also treating an interacting bosonic phase. They therefore relate only indirectly to the bosonic invertible TQFT framework considered here.} Such descriptions are useful once these additional choices have been made. However, the auxiliary mass textures, facet decompositions, and domain-wall configurations are not uniquely determined by the bulk phase, its symmetry, or its background geometry. This dependence is harmless only if different admissible choices yield equivalent continuum theories (up to deformations). Without such choice-independence, the construction does not canonically assign a continuum theory to the phase. As with a choice of coordinates or reference frame, these data may aid a particular description but should not affect the theory itself. Our framework therefore requires a continuum image determined functorially by the compatible symmetry data. The obstruction we identify in Eq.~\ref{eq:HOSPT_continuum_extension_dichotomy} potentially rules out this canonical TQFT image, rather than model-dependent effective descriptions built from additional non-intrinsic choices.

\subsubsection{Free continuum restrictions and higher-order lattice phases}
\label{sec:free_shadow_example}

A second source of lattice phases with trivial continuum image is provided by restrictions of free continuum classes. Consider, for example, the $(2+1)$D system with internal symmetry $H=U(1)$. For each $G\leq \mathrm O(2)$, write $\Gamma_G:=\U(1)\rtimes_{\rho|_G}G$. Consider now the free continuum class
\begin{equation}
[\widetilde x] := c_{1}(\xi_{\U(1)})\smile e_{2}(\xi_{\mathrm{O}(2)}) \in
H^{4}\left( B\Gamma_{\mathrm O(2)}, \mathbb Z_{w_{1}} \right),
\label{eq:free_shadow_continuum_class}
\end{equation}
where $c_{1}(\xi_{\U(1)})$ is the internal first Chern class, and $e_{2}(\xi_{\mathrm{O}(2)})$ is the twisted Euler class of the spatial $\mathrm{O}(2)$ bundle. 
Recall from Sec.~\ref{sec:internal_sym} (see Eq.~\ref{eq:jmath_O}) that
\begin{equation*}
\jmath_{\mathrm O(2)}:
B\U(1)\longrightarrow B\Gamma_{\mathrm O(2)}
\end{equation*}
is the internal-symmetry fiber inclusion, so that $\jmath_{\mathrm O(2)}^*$ is the spatial-symmetry-forgetting map. The subgroup $\mathrm{Free}_{\mathrm{cr}}\, H^{4}(B\Gamma_{\mathrm O(2)},\mathbb Z_{w_1})$ is defined as the kernel of this map on the free subgroup.
Since $\jmath_{\mathrm O(2)}^{*}(e_2)=0$, we have $\jmath_{\mathrm O(2)}^{*}([\widetilde x])=0$, and therefore
\begin{equation*}
[\widetilde x] \in \mathrm{Free}_{\mathrm{cr}}\,
H^{4}\left(B\Gamma_{\mathrm O(2)},\mathbb Z_{w_1}\right).
\end{equation*}

Applying the full-domain restriction map introduced in Sec.~\ref{sec:internal_sym} (see Eq.~\ref{eq:finite_res_H}) to the crystalline free class $[\widetilde x]$, define
\begin{equation}
X_{\widetilde x}
:=\res^{\mathrm{full}}_{H,\mathcal O}(0,[\widetilde x])
= \left([x_G]\right)_{G\in\ofin(\mathrm O(2))}
\in \mathrm{cSPT}^{2+1}_{\mathrm{univ},H}.
\label{eq:free_shadow_components}
\end{equation}
Here the first entry denotes the vanishing torsion continuum component. Thus $X_{\widetilde x}$ is a compatible family of torsion lattice TQFT classes. By the componentwise definition of $\res^{\mathrm{full}}_{H,\mathcal O}$ given in Eq.~\ref{eq:full_domain_restriction_component}, each $[x_G]$ is the $2$-primary torsion component of the ordinary cohomological restriction of $[\widetilde x]$ to $B\Gamma_G$, i.e., $[x_{G}]=\pi_{G}^{(2)}\left(\res^{B\Gamma_{\mathrm{O}(d)}}_{B\Gamma_{G}}[\widetilde{x}]\right)$.

\smallskip 

\noindent The nonzero components of this family have the usual higher-order interpretation on rotational lattices. In particular, $\res^{O(2)}_{C_4}(e_2)$ generates $H^{2}(BC_4,\mathbb Z)\cong\mathbb Z_4$, so the $C_4$ component is a nontrivial order-$4$ mixed rotation-$U(1)$ lattice TQFT phase, represented physically by a second-order SPT with fractional corner charge. This is almost exactly the same mechanism as discussed in Sec.~\ref{sec:example_2_highertorsion}, with the charges now fractional $\U(1)$ charges representing the projective representation of the $\U(1)$-symmetry. 

\bigskip 

By construction, $X_{\widetilde x}$ is the universal free-shadow family associated to the free continuum class $[\widetilde x]$:
\begin{equation*}
X_{\widetilde x} \in
S_{\mathrm{free},H}^{\mathcal O, 2+1} \subseteq \mathrm{cSPT}^{2+1}_{\mathrm{univ},H}.
\end{equation*}
Its restriction to the elementary subcategory therefore satisfies $\Pi_H(X_{\widetilde x}) \in S_{\mathrm{free},H}^{\mathcal A, 2+1}$. Recall from Sec.~\ref{sec:internal_sym} that the elementary
$H$-enriched stratification map
\begin{equation*}
\widetilde F_{2,H}: \mathrm{cSPT}^{2+1}_{\mathrm{elem},H} \longrightarrow \Theta_H^{2+1}(BO(2)),
\end{equation*}
is defined by quotienting out the elementary free-shadow subgroup. Consequently,
\begin{equation*}
S_{\mathrm{free},H}^{\mathcal A, 2+1} = \ker(\widetilde F_{2,H}).
\end{equation*}
Using the factorization $F_{2,H} = \widetilde F_{2,H}\circ\Pi_H$, we obtain
\begin{equation}
F_{2,H}(X_{\widetilde x})=0.
\label{eq:free_shadow_trivial_continuum_limit}
\end{equation}
Equation~\eqref{eq:free_shadow_trivial_continuum_limit} expresses the loss of information that occurs when the free continuum response is viewed only through the strict torsion TQFT model. The compatible family $X_{\widetilde x}$ retains the torsion shadows of $[\widetilde x]$ supported by the finite crystalline symmetries, but it does not retain the free continuum data from which those shadows originate. In particular, the response $c_1(\xi_{\U(1)})\smile e_2(\xi_{\mathrm O(2)})$ belongs to the free continuum sector rather than to the torsion continuum classification $\Theta_H^{2+1}(B\mathrm O(2))$. The torsion-valued continuum-limit map therefore sends its free-shadow family to the trivial torsion continuum class.\footnote{At the level of physical response theories, free integral classes generally require differential refinements containing connection and curvature data. Such geometric information is not retained by the torsion finite-subgroup restrictions; see App.~\ref{app:diff_refinement}.}

More generally, every crystalline free continuum class having nonzero $2$-primary restrictions to finite spatial subgroups produces a compatible free-shadow family in $\ker(F_{2,H})$. This reflects a limitation of the strict torsion TQFT model: it retains the torsion lattice shadows of free continuum responses, but not the full continuum information required to reconstruct those responses.

\subsection{Quantum criticality: AFM-VBS transition}
\label{sec:afm_vbs}

\begin{figure}[H]
    \centering
    \includegraphics[width=0.9\linewidth]{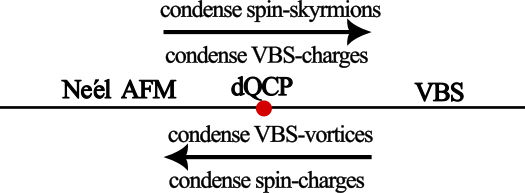}
    \caption{Schematic illustration of a phase diagram between N\'eel and VBS order at the square-lattice dQCP. The topological defects of one order parameter carry the quantum numbers of the competing order, so disordering one phase naturally drives the system toward the other order rather than toward a trivial symmetric phase.}
    \label{fig:AFMVBS}
\end{figure}

Although our main focus and results concern gapped invertible phases, our work also has implications for gapless critical theories whose low-energy descriptions carry anomaly data. Indeed, we can use the bulk-boundary correspondence between bulk TQFTs and boundary anomalies to study the boundary anomalies using their corresponding bulk TQFT classes.\footnote{Note that here we call a theory with the t'Hooft anomaly a ``boundary" theory, but it need not be the literal physical boundary of a TQFT phase. The bulk-boundary correspondence is a conceptual tool used in analysis: the boundary theory can be well defined on its own as long as we do not gauge the global symmetry with the t'Hooft anomaly.} A particularly important example is the $(2{+}1)$D N\'eel-VBS deconfined critical point (dQCP) in quantum antiferromagnets, where, guided by theory predictions\cite{ScienceSenthil,Senthil_dqcp_prb, Levin2004}, numerics on square-lattice spin models found an apparently direct and continuous\footnote{This should be understood as a statement valid on the accessible length scales in these works. The deconfinement transition may in fact be the wrong low-energy description, and confinement may occur at the longest distance scales.} transition between two distinct symmetry-breaking phases—the N\'eel antiferromagnet and the valence-bond solid—in a setting where the conventional Landau picture would instead suggest a first-order transition or an intermediate phase~\cite{Sandvik2007,Melko2008,Lou2009,Sandvik2010,Nahum2015}. This unusual critical behavior  can be understood through an effective continuum theory in which the two orders are \emph{intertwined}: the topological defects of one order parameter carry the quantum numbers of the other, see Fig.~\ref{fig:AFMVBS}. In particular, skyrmions of the N\'eel order carry VBS quantum numbers, while VBS vortices carry spin-$\tfrac12$, so condensing the defects of one phase drives the system naturally into the competing ordered phase rather than into a trivial disordered state~\cite{ScienceSenthil,Senthil_dqcp_prb,Levin2004,ChongWang_dqcp}.\footnote{A standard microscopic Hamiltonian realization is the square-lattice spin-$\tfrac12$ $J$-$Q$ model, in which nonfrustrating four-spin interactions tune the Heisenberg antiferromagnet into a valence-bond-solid phase~\cite{Sandvik2007}.}
The corresponding critical theory at such a transition is therefore not a conventional Landau theory of the two order parameters, but the noncompact CP$^{1}$ gauge theory, whose fractionalized spinons and emergent gauge field provide a local continuum description of this defect intertwinement~\cite{Levin2004,ScienceSenthil,Senthil_dqcp_prb}. In this continuum description, mixed spin-lattice anomalies explain the square-lattice transition, which suggests that the square lattice allows a form of the anomaly. This gives rise to the natural question: can all 2D lattices realize a non-trivial form of the continuum anomaly and such a transition? And if not, what is the general principle to understand this? 

The first question was addressed for the honeycomb lattice in Ref.~\cite{spatial_symm_3}, where it was shown that the corresponding mixed anomaly is absent. Consequently, unlike on the square lattice, the continuum anomaly does not enforce dQCP behavior on the honeycomb lattice. Instead, an intermediate trivial symmetric phase is allowed in principle, and is consistent with the existence of trivial symmetric gapped honeycomb states constructed in Refs.~\cite{Kim2016, Jian2016}. Our goal in this subsection is to place that line of reasoning on a sharper and more general footing: we answer the first question for all 2D lattices in Sec.~\ref{sec:afm_vbs_lattice} using the universal continuum limit construction as our guiding principle.

To answer the questions raised above—namely, on which lattice point groups the bulk class associated to the continuum anomaly can be detected non-trivially—we use some results derived in Sec.~\ref{sec:rigor}. 
In the explicit computations of Sec.~\ref{sec:rigor} (see Eq.~\ref{eq:resO_injective}), we proved that the map
\begin{equation*}
    \res_{\mathcal{O}}:\Theta^{d+1}(B(H\rtimes_{\rho} \mathrm{O}(d)))\hookrightarrow 
    \mathrm{cSPT}^{d+1}_{\mathrm{univ},H},
\end{equation*}
is injective. This map encodes the cone/collection of all restriction maps from the continuum classification to individual lattice classifications via composition with projection maps, i.e., $\res^{\mathrm{O}(d)}_{G}= \pi_{G}\circ \res_{\mathcal{O}}$:
\begin{equation*}
    \begin{tikzcd}
        \Theta^{d+1}(B(H\rtimes_{\rho} \mathrm{O}(d))) \arrow[dr, "\res^{\mathrm{O}(d)}_{G}"]\arrow[r, "\res_{\mathcal{O}}"] &\;\;\mathrm{cSPT}^{d+1}_{\mathrm{univ},H} \arrow[d, "\pi_{G}"]\\
        & \Theta^{d+1}(B(H\rtimes_{\rho} G))
    \end{tikzcd}.
\end{equation*}

Intuitively, the injectivity of $\res_{\mathcal{O}}$ expresses a global statement: a given continuum class $[x]$ is faithfully encoded in a full compatible family of lattice restrictions $(\res^{\mathrm{O}(d)}_{G})_{G\in \ofin(\mathrm{O}(d))}$. 
For the present application, however, we need a more local and operational statement, namely a criterion for a fixed lattice symmetry $G_{\Lambda}\in \ofin(\mathrm{O}(d))$. 
A continuum class $[x]$ is detected non-trivially on lattices with point group $G_{\Lambda}$ precisely when its restriction along the inclusion $G_{\Lambda}\hookrightarrow \mathrm{O}(d)$ is nonzero:
\begin{equation*}
    \res^{\mathrm{O}(d)}_{G_{\Lambda}}([x])\neq 0.
\end{equation*}
This gives a direct criterion for deciding which lattice symmetries detect the bulk class associated to a given continuum anomaly.

\bigskip 

While our construction may seem a bit formal, it offers a nice benefit. Namely, our construction is natural with respect to symmetry restriction: meaning that one can directly move from one concrete setting (e.g., honeycomb lattice with $G_{\Lambda}=C_{3}$) to another (e.g., triangular lattice with $G_{\Lambda}'=C_{6}$) related by symmetry inclusion $i: C_{3}\hookrightarrow C_{6}$ by applying the associated restriction morphism 
\begin{equation*}
\res^{C_{6}}_{C_{3}}: \Theta^{2+1}(BC_{6}) \to \Theta^{2+1}(BC_{3}), 
\end{equation*}
on bulk TQFT classification data. This offers an advantage in contrast to individual case by case inspections of the effective field theories corresponding to specific lattice systems.

\bigskip 

In the remainder of this subsection we will work in the group cohomology model~\eqref{eq:grpcohtormodel}. Specifically, we will apply our framework to lattice spin systems in $(2+1)$D, which have internal $H=\SO(3)_{\mathrm{s}}$-spin rotation symmetry\footnote{This internal symmetry can be shown to satisfy the hypotheses~\ref{hyp:1}-~\ref{hyp:2} required for the extension of the classification theorem to work.}, and lattice point group symmetry $G_{\Lambda}=C_{n}\rtimes \Z_{2}^{\mathrm{or}}$ where $C_{n}$ is the rotational symmetry of the lattice, and $\Z_{2}^{\mathrm{or}}$ is an orientation reversing symmetry which we identify as reflection across the $xz$-plane with action $y\to-y$. The full symmetry of the lattice system is $G_{\mathrm{lat}}=\SO(3)_{\mathrm{s}}\times G_{\Lambda}$. 

In Sec.~\ref{sec:afm_vbs_cont} we take the continuum dQCP description as input, discuss its effective anomalous response, and identify the $(3+1)$D bulk classification term whose boundary carries this anomaly. Then, in Sec.~\ref{sec:afm_vbs_lattice}, we determine exactly which lattice systems in $(2+1)$D detect this bulk class non-trivially and can therefore allow for the possibility of a dQCP transition.

\subsubsection{The continuum theory: mixed anomaly and classification}
\label{sec:afm_vbs_cont}
In this subsection we take the proposed continuum description of the anomalous square-lattice dQCP as input, and use it to identify the corresponding bulk SPT classification term. For convenience we employ a Lagrangian formulation that makes the local field content and projective symmetry action explicit. The corresponding bulk cohomology class records the intrinsic, deformation-invariant TQFT data. Our goal is to identify this class so that the anomaly can be analyzed using the continuum-limit and lattice-detection framework developed above.

The continuum limit of a $2+1$D lattice with lattice symmetry $G_{\Lambda}$ is obtained by the symmetry embedding 
\begin{equation*}
    i_{\Lambda}: G_{\Lambda} \hookrightarrow O(2)_{\mathrm{rot}}.
\end{equation*}
Here $\mathrm{O}(2)_{\mathrm{rot}}=\SO(2)_{\mathrm{rot}}\rtimes \Z_{2}^{\mathrm{or}}$ where $\SO(2)_{\mathrm{rot}}$ is the continuous rotation symmetry in 2D, and $\Z_{2}^{\mathrm{or}}$ is an orientation reversing symmetry, like mirror symmetry in the plane. In addition to the lattice symmetry, we also have the spin rotation symmetry $H=\SO(3)_{\mathrm{s}}$, thus giving the full symmetry in the continuum to be $G_{\mathrm{Cont}}=\SO(3)_{\mathrm{s}}\times \mathrm{O}(2)_{\mathrm{rot}}$. With these symmetries in mind, the proposed continuum description of the square-lattice dQCP is the dual-vortex model~\cite{SenthilBTI, Senthil_dqcp_prb, ScienceSenthil}
\begin{align}
    \mathcal{L}_{\mathrm{Cont}}^{\mathrm{UV}}[\psi_{\pm}, a, A_{s}, &A_{r}]= \sum_{s=\pm}\bigg\{\biggr\rvert \left(\partial_{\mu}-ia_{\mu}-is \frac{A_{s\mu}}{2}\right) \psi_{s} \biggr\rvert^{2} \nonumber \\
    &\qquad \qquad \quad  + V(|\psi_{s}|^{2})\bigg\}
  +\dfrac{1}{2\pi} A_{r} \wedge da,
  \label{eq:dual_vortex}
\end{align}
where $\psi_{s}$ transforms projectively under $\SO(3)_{\mathrm{s}}$, with the composite $\Psi= \psi^{\dagger}_{+}\psi_{-}$ carrying unit $\SO(3)_{\mathrm{s}}$-charge. The field $a$ is an emergent $\mathrm{U}(1)_{g}$ gauge field whose vortex defects $\mathcal{M}_{a}\sim da$ carry unit $\SO(2)_{\mathrm{rot}}$-charge.  The orientation-reversing symmetry $\mathbb Z_{2}^{\mathrm{or}}$ acts by exchanging the two vortex species, $\psi_{s}\rightarrow (-1)^{s}\psi_{-s}$, so that the full continuum symmetry is realized as $\mathrm{SO}(3)_{\mathrm{s}}\times \mathrm{O}(2)_{\mathrm{rot}}$. And finally, $A_{r}, A_{s}$ are $\SO(2)_{\mathrm{rot}}$-background and the diagonal part of $\SO(3)_{\mathrm{s}}$-background gauge fields respectively.\footnote{It is known that this theory has an equivalent dual description, in which we flip the labels $s \leftrightarrow r$, see Ref.~\cite{ChongWang_dqcp}.} 

It can be shown that integrating out the matter\footnote{One way to see this is by constructing a $\Z_{2}^{\mathrm{or}}$-symmetry($\psi_{+}\leftrightarrow\psi_{-}$) breaking boundary termination: gap out $\psi_{+}$, and condense $\psi_{-}$, and use the Higgs mechanism to integrate out $a$ which acquires a Higgs mass giving $\langle\psi_{+}\rangle=0,\; \langle\psi_{-}\rangle\neq 0,\; \langle a+A_{s}/2\rangle=0$.} in this UV continuum theory gives us the effective IR response:
\begin{equation}
    \mathcal{L}_{\mathrm{Cont}}^{\mathrm{IR}}[A_{s}, A_{r}] =\dfrac{1}{2}\cdot \dfrac{1}{2\pi} A_{s}\wedge dA_{r}
    \label{eq:afm_vbs_anomaly}
\end{equation}
where the extra factor of $1/2$, which is the remnant of charge fractionalization in the UV theory, signals a mixed anomaly between the spin and spatial symmetries.\footnote{Note that one can obtain the response without the factor of $1/2$ from a $2+1$D theory with a linear representation of the symmetry; so the factor of $1/2$ is a signature of projective implementation of the symmetry, as we see below.} 
The mixed anomaly~\eqref{eq:afm_vbs_anomaly} is encoded in the global definition of the topological coupling term
\begin{equation}
    \frac{i}{2\pi}\int_{M^{2+1}} A_r\wedge da,
    \label{eq:afm_vbs_bdry_coup}
\end{equation}
in the dual vortex theory~\eqref{eq:dual_vortex}.  To identify the bulk SPT term corresponding to the mixed anomaly~\eqref{eq:afm_vbs_anomaly}, we follow Appendix~B of Ref.~\cite{spatial_symm_3}. The coupling~\eqref{eq:afm_vbs_bdry_coup} is only locally meaningful, and on topologically nontrivial bundles it is not globally well defined on the $(2+1)$D manifold $M^{2+1}$. One therefore defines the theory~\eqref{eq:afm_vbs_bdry_coup} by choosing an extension to a four-manifold $Y^{3+1}$ with boundary $\partial Y^{3+1}=M^{2+1}$\footnote{This is similar to the definition of Chern-Simons terms in $(2+1)$D for gauge groups $G$ which are not connected and simply connected, see Sec.1~\cite{Dijkgraaf1990}. In this case a similar extension is necessary to define the CS term appropriately.}:
\begin{equation}
    \frac{i}{2\pi}\int_{M^{2+1}} A_r\wedge da
    \;:=\;
    2\pi i\int_{Y^{3+1}} \frac{dA_r}{2\pi}\wedge \frac{da}{2\pi}.
    \label{eq:bulk_ext}
\end{equation}
The anomaly is then the statement that this definition can depend on the choice of extension $Y^{3+1}$.\footnote{\label{fn:gauge}If $A_r\mapsto A_r+\alpha$ with $[\alpha/2\pi]\in H^1(M,\mathbb Z)$, then on $M\times[0,1]$ with $\widehat A_r=A_r+t\alpha$ and $\widehat a=a$,
\begin{equation*}
2\pi i\int_{M\times[0,1]} \frac{d\widehat A_r}{2\pi}\wedge \frac{d\widehat a}{2\pi}
= \frac{i}{2\pi}\int_M \alpha\wedge da,
\end{equation*}
so the large gauge variation is realized as a special case of extension dependence.}

\bigskip 

The corresponding bulk $(3+1)$D SPT term is the cohomology class whose inflow cancels this extension dependence. The fluxes $dA_r/2\pi$ and $da/2\pi$ represent integral degree-$2$ classes: the first Chern classes, or more generally the twisted Euler classes, of the corresponding $\mathrm{O}(2)$ bundles. Their product therefore gives the degree-$4$ class suggested by the four-dimensional extension term. Furthermore, the factor of $1/2$ in the anomalous response encodes the projective half-charge carried by the vortex field. Doubling the response removes this projective phase and yields a linearly realized symmetry. Thus, the anomaly, as manifested in this projective phase, depends on only the parity of the integral degree-$4$ class, so we can consider its mod-$2$ reduction. 

Since the mod-$2$ reduction of the Euler class of a rank-$2$ real bundle is its second Stiefel-Whitney class, the extension term suggests the mod-$2$ bulk class
\begin{equation}
    w_2(\xi_{\mathrm{r}})\smile w_2(\widetilde{\xi}_g).
    \label{eq:candidate_bulk}
\end{equation}
Here $\xi_{\mathrm{r}}$ is the physical
$\mathrm{O}(2)_{\mathrm{rot}}$ background bundle, and
$\widetilde{\xi}_g$ is the auxiliary
$\mathrm{O}(2)_g=\mathrm{U}(1)_g\rtimes\mathbb Z_2^{\mathrm{or}}$
bundle associated with the emergent gauge sector. Equation~\eqref{eq:candidate_bulk} provides a useful intermediate description inherited from the local gauge-theory formulation of the boundary anomaly. However, the intrinsic bulk class, whose evaluation on physical symmetry backgrounds gives the observable response of the theory, must be expressed entirely in the cohomology of the physical $\mathrm{SO}(3)_{\mathrm{s}}\times\mathrm{O}(2)_{\mathrm{rot}}$ symmetry. We therefore use the projective symmetry transformation of the vortex field $\psi$ to express $w_2(\widetilde{\xi}_g)$ in terms of the characteristic classes of the physical spin and spatial bundles.

To achieve this we identify that the vortex field $\psi$ transforms under the double cover
\begin{equation*}
    \widehat G :=\frac{\mathrm{SU}(2)_{\mathrm{s}}\times \mathrm{Pin}^{-}(2)_{g}}{\mathbb Z_{2}} \longrightarrow
    \mathrm{SO}(3)_{\mathrm{s}}\times \mathrm{O}(2)_{g},
\end{equation*}
rather than directly under the $\mathrm{SO}(3)_{\mathrm{s}}\times \mathrm{O}(2)_{\mathrm{g}}$.\footnote{\label{note:doublecover}This is because $\psi$ carries a half $\mathrm{SO}(3)_{\mathrm{s}}$ charge, so it is an honest representation only of the double cover $\mathrm{SU}(2)_{\mathrm{s}}$, while the orientation-reversing generator $C\in \mathrm{O}(2)_{g}=\mathrm{U(1)}_{g}\rtimes \Z_{2}^{\mathrm{or}}$ acts projectively with $C^{2}:\psi\mapsto -\psi$. The quotient identifies this $-1$ action with the $2\pi$ rotation in $\mathrm{SU}(2)_{\mathrm{s}}$, which also acts by $-1$.}
Accordingly, $\psi$ is a section of a vector bundle associated to a principal $\widehat G$-bundle. Equivalently, the $\mathrm{SO}(3)_{\mathrm{s}}$ background bundle $\xi_{\mathrm{s}}$ and the auxiliary $\mathrm{O}(2)_{g}$ gauge bundle $\widetilde{\xi}_{g}$ must admit a combined lift through the cover $\widehat G$. The existence of this lift implies
\begin{equation}
    w_{2}(\widetilde{\xi}_{g}) + w_{1}(\widetilde{\xi}_{g})^{2}
    = w_{2}(\xi_{\mathrm{s}}),
    \label{eq:rel_1}
\end{equation}
because the two central $\mathbb Z_{2}$ actions are identified in $\widehat G$ (see Footnote \ref{note:doublecover}), the obstruction to the $\mathrm{Pin}^{-}(2)_{g}$ lift must agree with the obstruction to the $\mathrm{SU}(2)_{\mathrm{s}}$ lift.\footnote{The obstruction to lifting an $\mathrm{SO}(n)$ bundle $\xi$ to a $\mathrm{Spin}(n)$ bundle is $w_{2}(\xi)$, whereas the obstruction to lifting an $\mathrm{O}(n)$ bundle $\xi$ to a $\mathrm{Pin}^{-}(n)$ bundle is $w_{2}(\xi)+w_{1}(\xi)^{2}$; see Ref.~\cite{Kirby1991}.} Finally, $\mathrm{O}(2)_{\mathrm{rot}}$ and $\mathrm{O}(2)_{\mathrm{g}}$ share the same orientation reversing symmetry. Hence, if $\xi_{\mathrm{r}}$ denotes
the physical $\mathrm{O}(2)_{\mathrm{rot}}$ background bundle, then
\begin{equation}
    w_{1}(\widetilde{\xi}_{g}) = w_{1}(\xi_{\mathrm{r}}).
    \label{eq:rel_2}
\end{equation}
Substituting relations~\eqref{eq:rel_1},~\eqref{eq:rel_2} into the candidate bulk class~\eqref{eq:candidate_bulk} gives
\begin{align}
    [x_{\mathrm{bulk}}]
    &= w_{2}(\xi_{\mathrm{r}})\smile\bigl(w_{2}(\xi_{\mathrm{s}})+w_{1}(\xi_{\mathrm{r}})^{2}\bigr),
    \label{eq:afm_vbs_class}
\end{align}
which is the corresponding $(3+1)$D bulk SPT classification term. Hence, the full anomaly free bulk-boundary composite theory can be written as:
\begin{gather}
    S\!_{\mathrm{bulk}+\mathrm{bdry}} = S\!_{\mathrm{bdry}} + S\!_{\mathrm{bulk}},\label{eq:full_th}\\
    S\!_{\mathrm{bdry}} = \int_{M^{2+1}} \left|\left(\partial_{\mu}-ia_{\mu}-is\dfrac{A_s}{2}\right)\psi_{s}\right|^{2}+2\pi i\int_{Y^{3+1}} \dfrac{dA_{r}}{2\pi}\wedge \dfrac{da}{2\pi},\nonumber\\
    S\!_{\mathrm{bulk}} =\pi i \int_{X^{3+1}\sqcup_{M^{2+1}} (-Y^{3+1})} w_{2}(\xi_{\mathrm{r}})\smile\bigl(w_{2}(\xi_{\mathrm{s}})+w_{1}(\xi_{\mathrm{r}})^{2}\bigr).\nonumber
\end{gather}
Here, the bulk term exactly cancels the anomaly associated to the boundary coupling term, e.g., one can be test the cancellation by monitoring the change under large gauge transformations of the background field $A_{r}$ (see footnote~\ref{fn:gauge}). The crucial point of this construction is that the full theory~\eqref{eq:full_th} is dependent \textit{only} on the value of the background on the physical bulk and boundary $(X^{3+1},\; \partial X^{3+1}=M^{2+1})$.

\begin{figure}[H]
    \centering
    \includegraphics[width=0.55\linewidth]{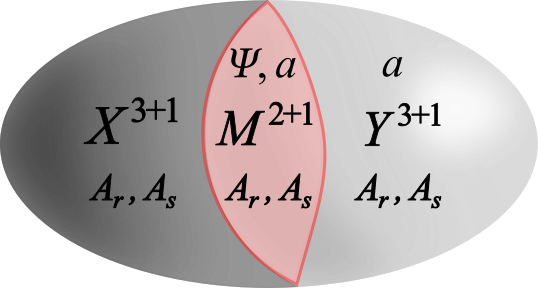}
    \caption{Schematic of anomaly inflow with an auxiliary extension. The physical bulk $X^{3+1}$ has boundary $M^{2+1}$, which supports the anomalous boundary theory. To define the boundary topological coupling globally, one chooses an auxiliary four-manifold $Y^{3+1}$ with the same boundary $\partial Y=M$. Gluing $X$ and $-Y$ along $M$ produces the closed four-manifold $X^{3+1}\cup_{M^{2+1}}(-Y^{3+1})$, on which the bulk SPT term is evaluated.}
\end{figure}

It can be shown that the bulk class in Eq. \ref{eq:afm_vbs_class} can be lifted from mod-$2$ coefficients to $\mathrm{U}(1)$-coefficients giving us the classification term in the full group cohomology classification model (see Eq.~\ref{eq:grpcohtormodel})
\begin{equation}
    \left[\;\overline{w_{2}(\xi_{\mathrm{r}})\cup (w_{2}(\xi_{s})+w_{1}^{2}(\xi_{r}))}\;\right] \in H^{4}(B\SO(3)_{\mathrm{s}}\times B\mathrm{O}(3)_{\mathrm{rot}}, \U(1)_{w_{1}}),
    \label{eq:afm_vbs_class2}
\end{equation}
where we use the notation defined in the previous subsection (see discussion in previous section around Eq.~\eqref{eq:e} and App.~\ref{app:U1_coeff_to_Z}). This completes the translation from the local Lagrangian description of the anomaly to its intrinsic bulk TQFT classification class, on which the continuum-limit and lattice-detection maps developed above can now be applied.

\subsubsection{Lattice realization of $(2+1)$D mixed anomaly}
\label{sec:afm_vbs_lattice}
Our task now is to understand in which lattice systems the anomaly (Eq.~\ref{eq:afm_vbs_anomaly}) is realized faithfully, following the discussion in the introduction to this subsection. The bulk SPT associated to this anomaly is represented by the mixed class in Eq.~\ref{eq:afm_vbs_class} (and its lifted $\U(1)$-valued refinement in Eq.~\ref{eq:afm_vbs_class2}) for the continuum symmetry $\SO(3)_{\mathrm{s}}\times \mathrm{O}(3)_{\mathrm{rot}}$, where planar spatial symmetries $g$ are embedded into $\mathrm{O}(3)$ by $\rho(g)=\mathrm{diag}(g,1)$. For a lattice point group $G_{\Lambda}$, the question of faithful lattice realization reduces to whether this continuum class is detected after restriction along $G_{\Lambda}\hookrightarrow \mathrm{O}(3)$:
\begin{equation}
    \res^{O(3)}_{G_{\Lambda}}(\,[x_{\mathrm{bulk}}]\,)\neq 0.
    \label{eq:detection_criterion}
\end{equation}
In what follows, we compute this restriction and extract the general criterion on lattice symmetries required to support the anomaly.

\bigskip 

To evaluate the detection condition it is convenient to factor the inclusion of the lattice point group through planar symmetries:
\begin{equation*}
    G_{\Lambda}\xrightarrow{i_{\Lambda}} \mathrm{O}(2)_{\mathrm{rot}}
    \xrightarrow{\ \rho\ } \mathrm{O}(3)_{\mathrm{rot}},
    \qquad \rho(g)=\mathrm{diag}(g, 1).
\end{equation*}
Since we care about the background rotation piece of the bulk response, let us fix a non-vanishing internal background with $w_{2}(\xi_{\mathrm{s}})\neq 0.$ Then the bulk response is detected on a lattice with point group $G_{\Lambda}$ if and only if the spatial degree-$2$ class $w_{2}(\xi_{r})$ is nontrivial after restriction:
\begin{equation}\res_{G_\Lambda}^{O(2)}\bigl(w_{2}(\xi_{r})\bigr)\;\neq\;0
    \quad\text{in}\;\;\;H^{2}(BG_{\Lambda},\Z_{2}),
    \label{eq:criterion_w2_w1sq}
\end{equation}
\textit{and} we can lift this class to a non-trivial $\mathrm{U}(1)$-valued class in the full group cohomology model. We now evaluate~\eqref{eq:criterion_w2_w1sq} for the two families of planar point groups relevant here:

\medskip

\noindent\textit{(i) Cyclic groups $G_{\Lambda}=C_{n}$.}
Since $C_{n}\subset \SO(2)$ is orientation preserving we have $\res^{\mathrm{O(2)}}_{C_{n}}(w_{1}(\xi_{r}))=0$. 
Thus the restriction is controlled by the degree-$2$ class $w_2(\xi_r)$. Under the standard rotational embedding $C_n\hookrightarrow\mathrm{SO}(2)$, this class restricts to a generator of the corresponding cohomology group:\footnote{Let $i_n:C_n\hookrightarrow\mathrm{SO}(2)\cong\mathrm{U}(1)$ be the standard rotation representation, and let $L_n\to BC_n$ be the associated complex line bundle. The class $c_1(L_n)$ generates $H^2(BC_n,\mathbb Z)\cong\mathbb Z_n$, and the underlying real two-plane bundle is the restriction of $\xi_r$. Hence $\res_{C_n}^{\mathrm{O}(2)}(w_2(\xi_r)) =\rho_2(c_1(L_n))$. This class vanishes when $n$ is odd and is the unique nonzero element of $H^2(BC_n,\mathbb Z_2)\cong\mathbb Z_2$ when $n$ is even.}
\begin{equation}
    \res^{\mathrm{O(2)}}_{C_n}(w_{2}(\xi_{r}))\;\; \text{ generates}\;\; H^{2}(BC_{n},\Z_{2})\cong \Z_{\gcd(2,n)}.
    \label{eq:cyclic_w2_gcd}
\end{equation}
Because $\res^{\mathrm{O(2)}}_{C_n}(w_{2}(\xi_{r}))\neq 0$ for even $n$, we have
\begin{align}
    &\res^{\mathrm{O(2)}}_{G_{\Lambda}}\left(w_{2}(\xi_{\mathrm{r}})\smile\bigl(w_{2}(\xi_{\mathrm{s}})+w_{1}(\xi_{\mathrm{r}})^{2}\bigr)\right )\nonumber \\
     =\;\;&\res^{\mathrm{O(2)}}_{G_{\Lambda}}\left(w_{2}(\xi_{\mathrm{r}})\smile w_{2}(\xi_{\mathrm{s}})\right )\neq 0
    \quad\text{iff}\;\; n \; \text{even}.
    \label{eq:cyclic_detection1}
\end{align}
Finally, we ask if this class can be lifted to $\mathrm{U}(1)$-valued group cohomology. The answer is yes and essentially follows from the existence of the lift in the continuum version of the anomaly. Thus:
\begin{equation}
    \left[\overline{\res^{\mathrm{O(2)}}_{C_{n}}\left(w_{2}(\xi_{\mathrm{r}})\smile\bigl(w_{2}(\xi_{\mathrm{s}})+w_{1}(\xi_{\mathrm{r}})^{2}\bigr)\right )}\right]\neq 0 \;\;\text{iff} \;\; n\;\; \text{even}.
    \label{eq:cyclic_detection2}
\end{equation}
Therefore, 2D lattice systems with symmetries including only rotation symmetry $C_{n}$ can realize the bulk SPT class in Eq.~\ref{eq:afm_vbs_class2} if and only if the order of the rotation symmetry $n$ is even.

\bigskip

\noindent\textit{(ii) Dihedral groups $G_{\Lambda}=D_{nh}=C_{n}\rtimes \Z_{2}^{\mathrm{or}}$.}
The presence of the reflection subgroup $\Z_{2}^{\mathrm{or}}\subset D_{nh}$ implies both restrictions 
\begin{equation*}
\res^{\mathrm{O(2)}}_{G_{\Lambda}}(w_{1}(\xi_{r})),\qquad 
\res^{\mathrm{O(2)}}_{G_{\Lambda}}(w_{1}^{2}(\xi_{r})).
\end{equation*}
are non-zero.\footnote{To verify this statement, restrict further to the subgroup generated by a single reflection. If $u$ denotes the nonzero class in $H^{1}(B\mathbb Z_{2},\mathbb Z_{2})$, then $w_{1}(\xi_r)$ restricts to $u$, while its square restricts to $u^{2}\neq 0$ in $H^{*}(B\mathbb Z_{2},\mathbb Z_{2})\cong\mathbb Z_{2}[u]$.} Meanwhile, further restriction of $\res^{\mathrm{O}(2)}_{D_{nh}}(w_{2}(\xi_{r}))$ to the rotation subgroup $C_n\subset D_{nh}$ reproduces the cyclic restriction computed in Eq.~\eqref{eq:cyclic_w2_gcd}. It is therefore nonzero when $n$ is even, and one can show that it vanishes when $n$ is odd. Hence
\begin{equation*}
    \res^{\mathrm{O}(2)}_{D_{nh}}\bigl(w_{2}(\xi_{r})\bigr)
    \neq 0
    \quad\text{if and only if}\quad
    n\ \text{is even}.
\end{equation*}Therefore, at the level of mod-$2$ classes:
\begin{equation}
    \res^{\mathrm{O(2)}}_{G_{\Lambda}}\left(w_{2}(\xi_{\mathrm{r}})\smile\bigl(w_{2}(\xi_{\mathrm{s}})+w_{1}(\xi_{\mathrm{r}})^{2}\bigr)\right )\neq 0 
    \quad\text{iff}\;\; n \; \text{even}.
    \label{eq:dihedral_detection1}
\end{equation}
Similar to before, the $\mathrm{U}(1)$-valued lift exists and \begin{equation}
    \left[\overline{\res^{\mathrm{O(2)}}_{D_{nh}}\left(w_{2}(\xi_{\mathrm{r}})\smile\bigl(w_{2}(\xi_{\mathrm{s}})+w_{1}(\xi_{\mathrm{r}})^{2}\bigr)\right )}\right]\neq 0 \;\;\text{iff} \;\; n\;\; \text{even}.
    \label{eq:dihedral_detection2}
\end{equation}
Therefore, 2D lattice systems with symmetry given by dihedral group $D_{nh}$ can realize the bulk SPT class in Eq.~\ref{eq:afm_vbs_class2} if and only if the order of the rotation symmetry $n$ is even.

\medskip
Combining~\eqref{eq:cyclic_detection2} and~\eqref{eq:dihedral_detection2}, we obtain a simple lattice criterion for the mixed anomaly: it can be detected only on
lattice point groups containing a rotation symmetry of \emph{even order} $n.$ In particular, we see that the square lattice can realize the continuum anomaly because its rotation symmetry is of even order $n=4$, whereas the honeycomb lattice cannot realize the symmetry because its rotation symmetry is of odd order $n=3$. Hence, we reproduce the results from Ref.~\cite{spatial_symm_3}. The result further generalizes these predictions for all 2D lattices with point group protecting symmetries.

\subsubsection{Application of lattice detection result}

An interesting application of the result can be seen by considering the previously discussed symmetry relation between the honeycomb and the triangular lattice. The triangular lattice has a $C_{6}$ symmetry, and the honeycomb lattice has a $C_{3}$ symmetry. Since the order of the triangular lattice rotation symmetry is even, it can support a realization of the bulk anomaly computed in Eq.~\ref{eq:afm_vbs_class}. The honeycomb lattice, however, cannot, as discussed previously. We also note that the two lattice symmetries are related by inclusion $i: C_{3}\hookrightarrow C_{6}$, which induces the following restriction morphism on classification data:
\begin{equation*}
    \res^{C_{6}}_{C_{3}}: \Theta^{2+1}(BC_{6}) \to \Theta^{d+1}(BC_{3}).
\end{equation*}
To make the implication of this restriction morphism explicit, we can understand the morphism through possible microscopic realizations:
\begin{equation*}
    \mathcal{Z}^{\mathrm{UV}}_{m}=\sum_{\{\psi\}} \exp\left\{
      i \sum_{X \subset \Lambda}
      \Bigl[\mathcal S_{X}\bigl[\psi\vert_{X}\bigr]
      + m S^{\mathrm{SB}}_{X}[\psi\vert_{X}] + \mathcal{S}_{X}^{\mathrm{coup}}\bigl[\psi\vert_{X}, {\bf p}\vert_{X}\bigr] \Bigr]
    \right\},
\end{equation*}
where we understand the $m$-coupling to be a symmetry breaking coupling breaking the symmetry down from $C_{6}\to C_{3}$. Physically, this could be induced through a phase transition in which atoms on a triangular lattice develop a bipartite charge condensate, or through mechanical strains/deformations which enforce this symmetry reduction. When this symmetry breaking perturbation is turned on, the anomaly will vanish, and a dQCP will not appear.

\section{Discussion}
\label{sec:discussion}
In this section we will review and synthesize the results of our article and then motivate some possible extensions, applications, and open questions. 

\subsection{Summary of Main Results and Extensions}
\label{sec:discussion_main_results}

In our article we answered the following questions (in the context of crystalline invertible TQFTs with finite-order lattice symmetries and decoupled internal symmetries): 
\begin{enumerate}
    \item \underline{Lattice realizability:} Can all continuum TQFTs be regularized to a lattice TQFT while preserving their topological properties?\\
    The answer here is yes: every continuum TQFT phase is faithfully represented by a compatible family of lattice TQFTs. This follows from the injectivity of the restriction map (c.f.,~\eqref{eq:resO_injective}):
    \begin{equation*}
        \res_{\mathcal{O}}: \Theta^{d+1}(B\mathrm{O}(d)) \to \csptu.
    \end{equation*}
    For a continuum phase $[x]\in\Theta^{d+1}(B\mathrm{O}(d))$, this map assigns the compatible family
    \begin{equation*}
    \res_{\mathcal O}([x]) = \left\{ \res^{\mathrm{O}(d)}_G([x])
    \right\}_{G\in\ofin(\mathrm{O}(d))},
    \end{equation*}
    of its restrictions to all finite lattice point groups. Injectivity means that this \textit{global} family retains all the information needed to \textit{uniquely} determine $[x]$. However, it does not by itself imply that any single member of the family does so.\\
    This global result can be sharpened to a \textit{local}, single-lattice statement: every $(d+1)$-dimensional $\mathrm{O}(d)$-symmetric continuum TQFT admits a lattice realization in a lattice with a maximal elementary abelian group symmetry $E_{\mathrm{max}}\cong \Z_{2}^{d}$ generated by reflections of the coordinate axes in a chosen frame.
    This \textit{local} statement is proven in Sec.~\ref{sec:proof_corollaries} using the elementary detection theorem~\ref{theorem:intermediate} which gives the isomorphism: 
    \begin{equation}
        \res_{\mathcal{A}}: \Theta^{d+1}(B\mathrm{O}(d)) \to \varprojlim_{E\in \mathcal{A}_{2}(\mathrm{O}(d))} \Theta^{d+1}(BE),
        \label{eq:disc1}
    \end{equation}
    together with the identification
    \begin{equation}
        \varprojlim_{E\in \mathcal{A}_{2}(\mathrm{O}(d))} \Theta^{d+1}(BE)\cong  \Theta^{d+1}(BE_{\mathrm{max}})^{W_{\mathrm{O}(d)}(E_{\mathrm{max}})}.
        \label{eq:disc2}
    \end{equation} 
    These two results are established in App.~\ref{app:lift}, (Lemma~\ref{lemma:lift_quillen}), and App.~\ref{app:symm_ring}(Claim~2), respectively. \\
    \noindent Together, Eqs.~\eqref{eq:disc1} and~\eqref{eq:disc2} show that the restriction of a continuum phase $[x]\in\Theta^{d+1}(B\mathrm{O}(d))$ to $E_{\mathrm{max}}$ retains enough information to reconstruct the original continuum phase uniquely. Thus, every continuum TQFT phase admits a single $E_{\mathrm{max}}$-symmetric lattice representative that captures all of its topological data at once.
    \item \underline{Continuum limit:} Do all lattice TQFTs admit a continuum limit? And when a continuum limit exists, does it preserve all lattice topological data? The answer to both questions is negative: some lattice phases admit no continuum limit, while others admit a continuum limit that forgets nontrivial lattice topological data.\\
    Explicitly, the continuum-limit map is defined on only a subset of lattice TQFT phases, $\mathrm{Im}(\pi_G)\subseteq\Theta^{d+1}(BG)$, and phases outside this subset do not have a well-defined continuum limit. For example, lattice phases whose responses depend on the choice of embedding of the lattice into the continuum cannot have a well-defined continuum limit. This is discussed in Sec.~\ref{sec:lattice_compatible_continuum_limit}. Furthermore, even when a lattice phase admits a continuum limit, the continuum-limit map may erase topological information that can be supported only by the lattice structures. Examples of this information loss are discussed in Sec.~\ref{sec:example_2}.
\end{enumerate}

\noindent The two results above establish the lattice realizability of continuum torsion TQFTs and characterize the possible obstructions and information loss encountered in the continuum-limit map. Before turning to broader applications, we record three direct extensions of the present framework. These extensions modify specific assumptions used in the analysis, but do not require a fundamentally different formulation of the lattice-continuum problem.

\subsubsection{Extensions and cases beyond the present framework}
\label{sec:discussion_extension}

Our results are formulated at the level of bosonic, \emph{torsion}-classified, invertible TQFTs protected by finite point-group symmetry and decoupled internal symmetries. Three assumptions can be relaxed in relatively direct ways.

\smallskip

\noindent \textit{(i) Integer-quantized (“free”) responses.} While we showed that all \emph{torsion} continuum TQFT data admit lattice realizations, we have excluded integer-quantized response terms that are \emph{almost} topological (see ft.~\ref{ft:almost}) in the sense that their partition functions require extra geometric choices (e.g., a framing) and can exhibit framing anomalies. Since we showed all continuum, torsion-classified TQFTs satisfying our assumptions have a faithful lattice regularization, it is natural to conjecture that any obstruction to lattice realizability of continuum \emph{(almost) TQFT} data, if it occurs, is confined to this integer-quantized sector.

\smallskip

\noindent \textit{(ii) Translation symmetry and nontrivially coupled spatial/internal symmetries.} We have ignored translation symmetry and focused on finite point-group data. Incorporating translations, and more general spatially coupled internal symmetries, is known to produce qualitatively new constraints (e.g., Lieb-Schultz-Mattis-type anomalies), and may obstruct the existence of a continuum limit in ways not visible in our setting. Extending the present framework to include these ingredients is an important direction for future work.

\smallskip

\noindent \textit{(iii) Beyond the crystalline liquid mesocsopic limit.} A structural assumption throughout this paper was the existence of a mesoscopic regime whose response is captured by standard topological lattice actions (e.g., Dijkgraaf-Witten-type theories). There are lattice phases for which no such smooth mesoscopic description is available; prominent examples include fractonic and related UV/IR-mixed phases~\cite{ShirleySlagleChen2019,DuaSarkarWilliamsonCheng2020,Slagle2021FQFT,GorantlaLamSeibergShao2021}. Understanding whether, and in what sense, such systems admit continuum limits requires additional ideas beyond those developed here.

\subsection{Efficient characterization by elementary responses}
\label{sec:efficient_response_characterization}

The preceding subsection summarized the main results and their immediate extensions. An important tool used in the framework leading to the main results is the probe-detectability Lemma~\ref{lemma:probe}, which states that an invertible $G$-TQFT is uniquely determined by its partition-function responses on all admissible spacetime and background-field configurations. The elementary detection theorem~\ref{theorem:intermediate}, which also plays an important role in the proofs of the main results, shows that this full collection of probes is highly redundant. Under mild assumptions, it is enough to consider backgrounds valued in elementary abelian $2$-subgroups of $G$.\footnote{The precise assumptions are that the relevant $2$-primary TQFT classification $\Theta^{d+1}(BG)$ is detected by mod-$2$ cohomology, this holds for most physically relevant compact Lie groups(see App.~\ref{app:no_odd_torsion} for example) and that the Gunawardena-Lannes-Zarati criterion, stated in Prop.~\ref{prop:glz}, holds, so that Quillen restriction is an isomorphism rather than only an $F$-isomorphism.} Moreover, because $\Theta^{d+1}(BG)$ is a finite group for the models considered here,, a finite collection of such elementary probes determines the TQFT class uniquely. Here $G$ may be either a spatial or an internal symmetry group, provided its classification satisfies these elementary-detection assumptions.

\bigskip 

The probe-detectability Lemma~\ref{lemma:probe} gives an operational characterization of an invertible $G$-TQFT in terms of its responses to background fields. A probe is represented by a closed spacetime manifold $M^{d+1}$ equipped with a $G$-symmetry background field,\footnote{ More precisely, probes related by a bordism over $BG$ give the same response, so the independent probes are bordism classes $\left[M^{d+1},f\right]\in\Omega_{d+1}^{\mathrm O}(BG)$. Here, $\Omega_{d+1}^{\mathrm O}(BG)$ denotes the unoriented bordism group of closed $(d+1)$-manifolds equipped with maps to $BG$; its group operation is induced by disjoint union. For a general tangential structure $\xi$, this group is replaced by $\Omega_{d+1}^{\xi}(BG)$.}
\begin{equation}
\left(M^{d+1},f:M^{d+1}\longrightarrow BG\right).
\label{eq:probe_bordism_group}
\end{equation}
The deformation class of the theory is uniquely determined by the collection of partition-function responses
\begin{equation*}
\mathcal{Z}_x\left(M^{d+1},f\right), \qquad
\forall\; \left(M^{d+1},f\right).
\end{equation*}
This response-based notion of identification is analogous to quantum state tomography, in which an unknown quantum state is reconstructed from an informationally complete collection of measurement outcomes \cite{VogelRisken1989,JamesEtAl2001,LvovskyRaymer2009}.

\bigskip 

For a compact Lie group $G$, there are generally infinitely many manifold-background pairs representing possible probes, and the full collection of responses is highly redundant. Let $\mathcal A_2(G)$ denote the Quillen category of elementary abelian $2$-subgroups $E\leq G,$ where $E\cong (\mathbb Z_2)^r$. Whenever the mod-$2$ cohomology of $BG$ satisfies the Gunawardena-Lannes-Zarati criterion (see Prop.~\ref{prop:glz} in App.~\ref{app:gl2}), Quillen's restriction map is promoted from an $F$-isomorphism to a genuine isomorphism. In the corresponding TQFT classification, this gives an exact elementary detection isomorphism
\begin{equation}
\res_{\mathcal{A}_{2}}^{G}: \Theta^{d+1}(BG) \xrightarrow{\;\cong\;}
\varprojlim_{E\in\mathcal A_2(G)} \Theta^{d+1}(BE).
\label{eq:general_elementary_detection}
\end{equation}
Thus, a $G$-TQFT is uniquely determined by its restrictions to elementary abelian $2$-subgroups, and hence by its responses to elementary symmetry backgrounds.\footnote{If the Gunawardena-Lannes-Zarati criterion is not satisfied, Quillen's map remains an $F$-isomorphism. Elementary restriction then still controls the classification up to the corresponding nilpotent ambiguities, although it need not determine the complete TQFT class exactly.}

\bigskip 

Because for a fixed spacetime dimension $(d+1)$ the classification group $\Theta^{d+1}(BG)$ is finite in the models considered here, exact elementary detection can be reduced further to a finite family of probes. Indeed, for every pair of distinct TQFT classes, elementary detection guarantees the existence of an elementary probe on which their responses differ. Choosing one such probe for each pair produces a finite collection whose joint response map is injective.\footnote{The finiteness used here follows from the finiteness hypotheses appearing in Theorem~\ref{theorem:quillen}. For a compact Lie group $G$, $H^*(BG,\Z_2)$ is a finitely generated graded $\Z_2$-algebra, and hence each fixed-degree group $H^{d+1}(BG,\Z_2)$ is finite-dimensional over $\Z_2$  and therefore finite. 
If the classification contains $m$ classes, the pairwise argument gives the crude upper bound $\binom{m}{2}$ probes, although a substantially smaller separating family may exist.} Consequently, one may choose elementary subgroups $E_i\leq G$, inclusions $\iota_i:E_i\hookrightarrow G$, and probe classes
\begin{equation*}
p_i= \left( M_i^{d+1}, f_i:M_i^{d+1}\longrightarrow BE_i \right)
\quad i=1,\ldots,N,
\end{equation*}
such that
\begin{equation}
[x]\longmapsto \big( \mathcal{Z}_x\left(M_1,B\iota_1\circ f_1\right),\;\cdots,\; \mathcal{Z}_x\left(M_N,B\iota_N\circ f_N\right) \big)
\label{eq:finite_elementary_response_map}
\end{equation}
is injective. The probes may therefore be chosen once for the classification problem, independently of the unknown phase being tested. The full TQFT class can then be identified from a finite, informationally complete collection of elementary responses rather than from measurements over arbitrary $G$-backgrounds. 

\bigskip 

This reduction may be useful in settings where only a limited number of response measurements can be implemented. The elementary detection theorem provides a mathematical basis for resource-efficient topological-response characterization: once the detecting probes have been identified, only finitely many experiments are required to distinguish the phases in the classification.  This goal is closely related to experimental and quantum-information approaches that extract topological invariants or other physically relevant data from a restricted collection of measurements~\cite{AidelsburgerEtAl2014,ElbenEtAl2020,SatzingerEtAl2021}. These methods likewise seek to extract informationally complete or task-relevant data from substantially fewer measurements than required by full tomography. More broadly, compressed quantum-state tomography, direct fidelity estimation, and classical-shadow methods similarly reduce the measurement resources needed to identify selected properties of a quantum system \cite{GrossEtAl2010,FlammiaLiu2011,RiofrioEtAl2017, HuangKuengPreskill2020}.

 Finally, we remark that characterization using only the elementary abelian 2-subgroups is analogous in spirit to compressed sensing where a complex signal can be reconstructed from far fewer measurements than would ordinarily be expected, provided that the signal has a sufficiently simple underlying structure---for example, only a small number of nonzero components---and that the measurements are chosen appropriately~\cite{CandesRombergTao2006,Donoho2006,CandesWakin2008}. The underlying mechanisms are nevertheless different. Compressed sensing relies on the sparsity of the signal being measured in some domain (like the Fourier domain) and typically gives statistical or approximate reconstruction. In contrast, elementary detection follows from exact algebraic and category-theoretic structure and gives exact reconstruction at the TQFT level. Translating the elementary probes into practical experimental protocols, and determining their robustness under noise and imperfect symmetry implementation, remain important open questions.

\subsection{Simulation and lattice regularization of topological phases and anomalies}
Since our results are general, we can use them to make progress answering fundamental questions associated to the relation between lattice and continuum phases satisfying our assumptions. In addition to the concrete examples discussed in Sec.~\ref{sec:examples}, we will point out some important conceptual applications to broader questions.

For example, any attempt to simulate or discretize a continuum phase must choose a UV regulator. The UV regulator is often a lattice, and different regulators may preserve different subsets of the quantized continuum response data. Our results give a sharp way to pose, and often answer, the question: which continuum properties survive under lattice regularization? 

To this end, given a continuum response or anomaly data, and a proposed lattice symmetry environment, the restriction maps determine exactly what remains well-defined, quantized, and detectable by the corresponding background probes after passing to the lattice environment. In this sense, the framework provides a screening principle for lattice regularizations: it identifies those lattice symmetry settings in which a targeted continuum invariant survives as an unambiguous topological response. It also flags when the invariant becomes invisible or requires additional structure to be present in the regulated description. 

A well-known arena where such ``compatibility” questions become unavoidable is the nonperturbative regularization of anomalous continuum theories, e.g., gauge theories with matter, where one must decide which aspects of the desired symmetry or anomaly data can remain manifest in a given regulated setting. For example, the long-running program of lattice formulations of chiral fermions illustrates how regulator choice constrains which symmetry and anomaly features can be implemented faithfully and transparently~\cite{Ginsparg1982, Kaplan1992}. Likewise, modern anomaly-based formulations of Lieb–Schultz–Mattis–type constraints highlight how microscopic lattice symmetries (notably translations, often intertwined with internal symmetries) restrict the possible IR topological responses~\cite{Seiberg_Cheng, Zou2026}. While these settings involve ingredients beyond our present assumptions (e.g., free/integer-quantized response sectors or translation symmetries), they underscore the broader point: our methods provide a general framework for analyzing ``what survives discretization” as a functorial restriction problem. Moreover, our results suggest a systematic route to extending the analysis once those additional ingredients are incorporated.

\subsection{Conjectural extension to microscopic lattice theories}
\label{sec:microscopic_continuum_conjecture}

Our results throughout have been formulated at the level of invertible TQFTs, which  describe the universal, low-energy topological response obtained after the massive microscopic degrees of freedom of a gapped phase have been integrated out (see Sec.~\ref{sec:scales}). The use of invertible field theories as long-distance descriptions of gapped, short-range entangled phases is standard in modern classifications of topological phases \cite{Freed2014,Freed2021,FreedHopkins}. 
Accordingly, the universal low-energy response of a gapped short range entangled crystalline phase with finite spatial symmetry $G\subset\mathrm{O}(d)$ is described by an invertible $G$-TQFT, whose deformation class is an element of the classification group $\Theta^{d+1}(BG)$.\footnote{In the group-cohomology model, a cocycle
$\alpha\in Z^{d+1}(BG,\mathbb R/\mathbb Z_{w_1})$ defines an
invertible $G$-TQFT(functor) $\mathcal Z_\alpha$, while cohomologous cocycles define equivalent TQFTs. Thus the deformation class is labeled by $[\alpha]\in H^{d+1}(BG,\mathbb R/\mathbb Z_{w_1})$; see App.~\ref{app:grpcoh_functor}.} This description follows from the classification framework introduced in Sec.~\ref{sec:basics} and is consistent with the crystalline equivalence principle and related treatments of invertible phases with spatial symmetry \cite{ThorngrenElse,FreedHopkins}. Here, we ask whether the main conclusions established at the TQFT level - lattice realizability, the possible nonexistence of a continuum limit, and the loss of lattice topological information under that limit - extend to the underlying microscopic lattice theories. We conjecture that they do for the class of phases considered here.

\bigskip

Let $\mathfrak P_{G}^{\mathrm{UV}}$ denote the class of bosonic, gapped, short range entangled $G$-symmetric lattice phases that admit a mesoscopic lattice TQFT description discussed in Sec.~\ref{sec:scales}.\footnote{Recall that we formulate the lattice topological response at an intermediate scale $\ell$ satisfying $a,\zeta\ll \ell\ll L$, where $a$ is the lattice spacing, $\zeta$ is the correlation length, and $L$ is the relevant macroscopic scale. At this scale, the massive microscopic degrees of freedom have been integrated out, while the crystalline origin of the spatial symmetry data remains manifest.} 
We write $\mathcal P_G^{\mathrm{UV}}\in\mathfrak P_G^{\mathrm{UV}}$ for an individual microscopic phase, which we take to be an equivalence class of microscopic $G$-symmetric Hamiltonians, rather than a particular Hamiltonian model.\footnote{For example, if two $G$-symmetric Hamiltonians $H_0$ and $H_1$ are connected by a $G$-symmetric gapped path, then they represent the same microscopic phase $\mathcal P_G^{\mathrm{UV}}$.}
Integrating out the massive matter fields, and then coarse-graining within the mesoscopic window, assigns to each microscopic phase the deformation class of its invertible mesoscopic TQFT. We denote this assignment by the mesoscopic classification map
\begin{equation}
\operatorname{cl}^{\mathrm{mes}}_G: \mathfrak P_{G}^{\mathrm{UV}} \longrightarrow \Theta^{d+1}(BG), \qquad
\operatorname{cl}^{\mathrm{mes}}_G
(\mathcal P_G^{\mathrm{UV}}) =[x_G],
\label{eq:microscopic_to_mesoscopic_classification}
\end{equation}
where $\Theta^{d+1}(BG)$ is the group of deformation classes of invertible $G$-TQFTs, with a group operation induced by stacking. Quasi-adiabatic continuation and local-unitary renormalization imply that different Hamiltonian representatives of the same microscopic phase have the same long-distance topological data \cite{HastingsWen2005,ChenGuWen2010}. Thus, $\operatorname{cl}^{\mathrm{mes}}_G$ is defined on the phase $\mathcal P_G^{\mathrm{UV}}$, rather than on a chosen Hamiltonian representative. Phases outside the class $\mathfrak P_G^{\mathrm{UV}}$, including fractonic and other UV/IR-mixed phases without a mesoscopic invertible-TQFT description mentioned in Sec.~\ref{sec:discussion_extension} (iii), are not considered here. 

\medskip 

Now we introduce the microscopic continuum-limit prescriptions to
which the conjecture below will apply. Let $\mathfrak P_{\mathrm{O}(d)}^{\mathrm{cont}}$ denote the class of $\mathrm{O}(d)$-symmetric bosonic, gapped, short range entangled continuum phases. Passing to the low-energy topological sector assigns to each such phase the deformation class of an invertible $\mathrm{O}(d)$-TQFT. We denote this assignment by the continuum classification map
\begin{equation}
\operatorname{cl}^{\mathrm{cont}}_{\mathrm{O}(d)}:
\mathfrak P_{\mathrm{O}(d)}^{\mathrm{cont}}
\longrightarrow
\Theta^{d+1}(B\mathrm{O}(d)).
\label{eq:continuum_phase_classification}
\end{equation}
For a fixed finite lattice symmetry
group $G\leq\mathrm O(d)$, a microscopic continuum-limit prescription
is a map
\begin{equation}
\mathcal C_G:
\mathfrak P_{\mathcal C,G}^{\mathrm{UV}}
\longrightarrow
\mathfrak P_{\mathrm O(d)}^{\mathrm{cont}}.
\label{eq:microscopic_CL}
\end{equation}
where we have denoted its domain of
applicability by
\begin{equation*}
\mathfrak P_{\mathcal C,G}^{\mathrm{UV}}
:=
\operatorname{Dom}(\mathcal C_G)
\subseteq
\mathfrak P_G^{\mathrm{UV}},
\end{equation*} since a particular continuum limit prescription need not be applicable to every
$G$-symmetric microscopic phase.

The microscopic argument of
Sec.~\ref{sec:restriction_morphisms} shows that continuum limits must
also be compatible with the restriction and conjugation relations
between different lattice symmetry sectors. Thus, the maps
$\mathcal C_G$ cannot be chosen independently for each $G$. We define
a microscopic continuum-limit prescription to be a coherent family
\begin{equation}
\mathcal C
=
\left\{
\mathcal C_G:
\mathfrak P_{\mathcal C,G}^{\mathrm{UV}}
\longrightarrow
\mathfrak P_{\mathrm O(d)}^{\mathrm{cont}}
\right\}_{G\in\ofin(\mathrm O(d))},
\label{eq:microscopic_continuum_family}
\end{equation}
where coherence means that the prescriptions respect the symmetry
restriction and conjugation relations established in
Sec.~\ref{sec:restriction_morphisms}.\footnote{More precisely, for every inclusion $P\hookrightarrow G$, let $\operatorname{res}^{G,\mathrm{UV}}_{P}: \mathfrak P_G^{\mathrm{UV}}\to\mathfrak P_P^{\mathrm{UV}}$ denote the symmetry-restriction map on microscopic phases. We require
\begin{equation*}
\operatorname{res}^{G,\mathrm{UV}}_{P} \left(
\mathfrak P_{\mathcal C,G}^{\mathrm{UV}} \right) \subseteq \mathfrak P_{\mathcal C,P}^{\mathrm{UV}}, \qquad \mathcal C_P\circ \operatorname{res}^{G,\mathrm{UV}}_{P} = \mathcal C_G,
\end{equation*}
on $\mathfrak P_{\mathcal C,G}^{\mathrm{UV}}$. The mesoscopic classification maps are likewise required to satisfy
\begin{equation*}
\operatorname{cl}^{\mathrm{mes}}_P \circ \operatorname{res}^{G,\mathrm{UV}}_{P}
= \operatorname{res}^{G}_{P} \circ \operatorname{cl}^{\mathrm{mes}}_G.
\end{equation*}
Consequently, the induced mesoscopic domains are preserved under symmetry restriction.} Applying the mesoscopic classification map to the domain of $\mathcal C_G$ gives
\begin{equation}
\Theta_{\mathcal C}^{d+1}(BG) := \operatorname{cl}^{\mathrm{mes}}_G
\left( \mathfrak P_{\mathcal C,G}^{\mathrm{UV}} \right)
\subseteq \Theta^{d+1}(BG).
\label{eq:induced_mesoscopic_domain}
\end{equation}
The coherence imposed on the microscopic domains makes the induced domains $\Theta_{\mathcal C}^{d+1}(BG)$ compatible under symmetry restriction.

\bigskip

The construction above associates to each microscopic continuum-limit prescription~\eqref{eq:microscopic_CL} a domain of mesoscopic TQFT classes~\eqref{eq:induced_mesoscopic_domain}. To relate this induced domain to the universal continuum-limit construction developed in Secs.~\ref{sec:motivation},~\ref{sec:inverselimit}, and \ref{sec:continuum_limit}, we assume that taking the continuum limit commutes, at the level of topological data, with integrating out massive degrees of freedom and flowing to the low-energy EFT.

\begin{assumption}[RG/EFT compatibility]
\label{assumption:RG_EFT_compatibility}
For every finite lattice symmetry group
$G\in\ofin(\mathrm{O}(d)),$ and every microscopic continuum-limit
prescription $\mathcal C$, there exists a well-defined map
\begin{equation}
\kappa_G^{\mathcal C}:
\Theta_{\mathcal C}^{d+1}(BG)
\longrightarrow
\Theta^{d+1}(B\mathrm{O}(d))/K_{\mathcal O(G)},
\label{eq:induced_mesoscopic_continuum_map}
\end{equation}
such that the diagram
\begin{equation}
\label{eq:microscopic_mesoscopic_descent}
\begin{tikzcd}
\mathfrak P_{\mathcal C,G}^{\mathrm{UV}}
\arrow[r,"\mathcal C_G"]
\arrow[d,"\operatorname{cl}^{\mathrm{mes}}_G"']
&
\mathfrak P_{\mathrm{O}(d)}^{\mathrm{cont}}
\arrow[d,"\operatorname{pr}_G\circ
\operatorname{cl}^{\mathrm{cont}}_{\mathrm{O}(d)}"]
\\
\Theta_{\mathcal C}^{d+1}(BG)
\arrow[r,"\kappa_G^{\mathcal C}"']
&
\Theta^{d+1}(B\mathrm{O}(d))/K_{\mathcal O(G)}
\end{tikzcd}
\end{equation}
commutes.
\end{assumption}

\medskip 
In words, Eq.~\eqref{eq:microscopic_mesoscopic_descent} states that taking the microscopic continuum limit, and then extracting its low-energy topological sector, agrees with integrating out the massive microscopic degrees of freedom first, and then passing to the mesoscopic TQFT and applying the TQFT continuum limit $\kappa_G^{\mathcal C}$.

Equations~\eqref{eq:induced_mesoscopic_domain} and \eqref{eq:induced_mesoscopic_continuum_map} together define the mesoscopic continuum-limit prescription
\begin{equation*}
\left( \Theta_{\mathcal C}^{d+1}, \kappa^{\mathcal C} \right),
\end{equation*}
induced by the microscopic prescription $\mathcal C$. The physical principles used to define continuum-limit prescriptions at the TQFT level apply equally to the mesoscopic prescription induced by $\mathcal C$. In particular, it must extend the strong continuum limit and be complete under symmetry extension, as required by conditions {\rm(GL1)}-{\rm(GL2)} of Def.~\ref{def:gen_con_lim}. Together with the coherence imposed above, these conditions make $\left(\Theta_{\mathcal C}^{d+1},\kappa^{\mathcal C}\right)$ a generalized continuum-limit prescription. 

\bigskip

\par
\vspace{0.5\baselineskip}
\begin{widetext}

\begin{figure}[H]
\makebox[\textwidth][c]{%
    \includegraphics[width=0.8\textwidth]{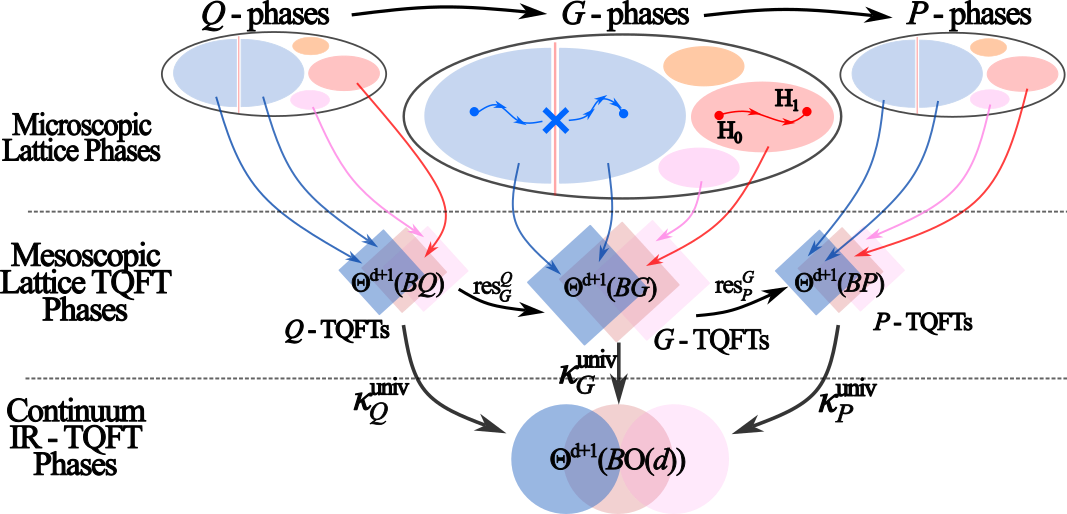}%
  }%
  \caption{Schematic relation between microscopic lattice phases, mesoscopic lattice TQFT classes, and continuum infrared TQFT classes. At the microscopic level, each colored region represents a deformation class of gapped Hamiltonians: Hamiltonians belonging to the same phase are related by a symmetry-preserving gapped adiabatic deformation, as illustrated by the path from $H_0$ to $H_1$ in the red region. The arrows from the microscopic to the mesoscopic level represent the microscopic-to-mesoscopic continuum-limit maps, e.g. $\mathrm{cl}^{\mathrm{mes}}_{G}$. These maps may be many-to-one, as illustrated by the blue regions, so that distinct microscopic descriptions can determine the same mesoscopic TQFT class; dual microscopic theories with the same t'Hooft anomaly provide a familiar example of this possibility. The figure depicts nested lattice symmetries $Q\supset G\supset P$. Accordingly, the microscopic phases in the top row are related by successive partial symmetry breaking, while at the mesoscopic level the corresponding TQFT classes are related by the restriction maps $\res^Q_G$ and $\res^G_P$. Finally, the local universal continuum-limit maps $\kappa_Q^{\mathrm{univ}}$, $\kappa_G^{\mathrm{univ}}$, and $\kappa_P^{\mathrm{univ}}$ send continuum-admissible mesoscopic classes to $\mathrm O(d)$-structured infrared TQFT classes in $\Theta^{d+1}(B\mathrm O(d))$. Colors track corresponding data through the three levels; the diagram is schematic and suppresses additional compatibility relations.}
  \label{fig:section9}
\end{figure}

\end{widetext}

For every microscopic continuum-limit prescription $\mathcal C$, Assumption~\ref{assumption:RG_EFT_compatibility} produces a generalized continuum-limit prescription on the corresponding mesoscopic TQFT classes. The global-compatibility theorem, Theorem~\ref{thm:finite_to_global_compatibility}, therefore implies
\begin{equation}
\mathfrak P_{\mathcal C,G}^{\mathrm{UV}}
\subseteq
\left(\operatorname{cl}^{\mathrm{mes}}_G\right)^{-1}
\left(\operatorname{Im}(\pi_G)\right).
\label{eq:necessary_UV_domain_containment}
\end{equation}
Since $\mathcal C$ was arbitrary, this gives the prescription-independent obstruction
\begin{equation}
\operatorname{cl}^{\mathrm{mes}}_G
\left(\mathcal P_G^{\mathrm{UV}}\right)
\notin\operatorname{Im}(\pi_G)
\quad\Longrightarrow\quad
\mathcal P_G^{\mathrm{UV}}
\ \text{has no continuum limit}.
\label{eq:microscopic_continuum_obstruction}
\end{equation}
The conjecture below asserts that this necessary condition is also sufficient.

\bigskip 

\begin{conjecture}[Microscopic inheritance of the universal continuum limit]
\label{conj:microscopic_obstruction_inheritance}
For every finite lattice symmetry group $G\in\ofin(\mathrm O(d))$, the following statements hold for the bosonic, gapped, short-range-entangled lattice phases satisfying Assumption~\ref{assumption:mesoscopic_TQFT_description}.
\begin{enumerate}
    \item[{\rm (i)}] \emph{Existence of the continuum limit.}
    A microscopic phase admits an $\mathrm{O}(d)$-symmetric continuum limit if and only if its mesoscopic classification lies in the universal continuum-admitting domain, which we define by:
    \begin{equation}
    \mathfrak P_{\mathrm{univ},G}^{\mathrm{UV}}
    :=
    \left(\operatorname{cl}^{\mathrm{mes}}_G \right)^{-1}
    \left(\operatorname{Im}(\pi_G)\right).
    \label{eq:microscopic_universal_domain}
    \end{equation}

    \item[{\rm (ii)}] \emph{Universality of the continuum image.}
    Let $\mathcal P_G^{\mathrm{UV}}$ be a microscopic phase with mesoscopic classification
    \begin{equation*}
    \operatorname{cl}^{\mathrm{mes}}_G
    \left( \mathcal P_G^{\mathrm{UV}} \right)
    = [x_G]\in\operatorname{Im}(\pi_G).
    \end{equation*}
    For every microscopic continuum-limit prescription $\mathcal C$ whose domain contains $\mathcal P_G^{\mathrm{UV}}$, the low-energy topological sector of the resulting continuum phase satisfies
    \begin{equation}
    \operatorname{pr}_G
    \left[\operatorname{cl}^{\mathrm{cont}}_{\mathrm{O}(d)}
    \left( \mathcal C_G(\mathcal P_G^{\mathrm{UV}})
    \right) \right] = \kappa_G^{\mathrm{univ}}([x_G]).
    \label{eq:microscopic_universal_image}
    \end{equation}
    Thus, microscopic continuum-limit prescriptions may differ in nonuniversal details, but they determine the same universal continuum class visible from the original lattice symmetry sector.
\end{enumerate}
\end{conjecture}
\bigskip

\noindent Three consequences follow. First, Eq.~\eqref{eq:microscopic_continuum_obstruction} implies that any microscopic realization of a mesoscopic class outside $\operatorname{Im}(\pi_G)$ gives a bosonic, gapped, short-range-entangled lattice phase with no $\mathrm{O}(d)$-symmetric continuum limit. Hence, if such mesoscopic classes admit microscopic realizations, there exist microscopic lattice theories that do not admit continuum limits. Second, if a nonzero class $0\neq[k_G]\in\ker(\kappa_G^{\mathrm{univ}})$
is realized by a microscopic phase, then that phase admits a continuum limit whose response to $G$-symmetry probes is trivial. Thus, topological information present in the microscopic lattice phase may be lost in the continuum.\footnote{Note here that $G$-symmetry probes are the only probes available in the physical setting and therefore trivial response to $G$-symmetry probes implies triviality of the phase in a physically relevant context, i.e. $[0]\in \Theta^{d+1}(B\mathrm{O}(d))/K_{\mathcal{O}(G)}$. If additionally $K_{\mathcal O(G)}=0$, then the resulting continuum TQFT class is itself unambiguously trivial. See Sec.~\ref{sec:refine_domain} for more details.} Finally, provided that the mesoscopic lattice representatives constructed here admit microscopic realizations, every continuum invertible TQFT class has at least one microscopic crystalline lattice representative. Thus, assuming Conjecture~\ref{conj:microscopic_obstruction_inheritance} and microscopic realizability of the relevant mesoscopic classes, all of the TQFT-level conclusions established in Sec.~\ref{sec:discussion_main_results} extend to microscopic lattice theories.

\bigskip 

A more formal treatment of Conjecture~\ref{conj:microscopic_obstruction_inheritance} may require refining the mesoscopic and continuum TQFT classifications used above from groups of deformation classes to spaces or higher groupoids of the underlying extended field theories. The cobordism hypothesis provides a framework for organizing this extended field-theoretic data~\cite{Baez1995,Lurie,FreedCobordism,SchommerPries2017}, while related classifying-space constructions incorporate geometric background structures and families of functorial field theories~\cite{MTW2010,GradyPavlov2022}. These frameworks could therefore refine the mesoscopic and continuum sides of the RG/EFT-compatibility diagram, but they do not by themselves construct the microscopic-to-mesoscopic RG assignment. 
A complete treatment would still require a mathematical construction of the microscopic-to-mesoscopic classification map and a proof of the RG/EFT compatibility assumed above. We leave these problems to future work.

\begin{acknowledgments}
RC thanks Barry Bradlyn, Michael Stone, and Fahad Mahmood for insightful comments and discussions that helped clarify the broader perspective and presentation of this work. RC also thanks Matthew O'Brien, Akash Vijay, Daniel Belkin, Shubhang Goswami, and Souvik Dutta for helpful conversations and feedback. TLH and RC thank ARO MURI
W911NF2020166 for support. 
\end{acknowledgments}

\addtocontents{toc}{\protect\AppendixTOCSectionsOnly}
\appendix

\section{Axiomatic TQFT and symmetry backgrounds}
\label{app:tqft}

\subsection{Bordism category and the Atiyah–Segal axioms}

\begin{definition}[(Unoriented) Bordism category]
\label{def:bordismcategory}
We denote $\bordd$ as the symmetric monoidal category whose:\\
\noindent 1. objects are closed $d$-dimensional manifolds $Y^{d}$.\\
\noindent 2. morphisms $Y^{d}_{\mathrm{in}}\to Y^{d}_{\mathrm{out}}$ are diffeomorphism classes of triples $(X^{d+1},\;\varphi_{\mathrm{in}},\;\varphi_{\mathrm{out}})$ where $X^{d+1}$ is a compact $(d{+}1)$-manifold and
\begin{equation*}
\varphi_{\mathrm{in}}:Y^{d}_{\mathrm{in}}\xrightarrow{\cong}\partial_{\mathrm{in}}X^{d+1},\qquad
\varphi_{\mathrm{out}}:Y^{d}_{\mathrm{out}}\xrightarrow{\cong}\partial_{\mathrm{out}}X^{d+1},
\end{equation*}
identify the boundary as a disjoint union
$\partial X^{d+1}=\partial_{\mathrm{in}}X^{d+1}\sqcup \partial_{\mathrm{out}}X^{d+1}$,
together with collar embeddings
\begin{equation*}
\iota_{\mathrm{in}}:[0,1)\times \partial_{\mathrm{in}}X^{d+1} \hookrightarrow X^{d+1},\quad
\iota_{\mathrm{out}}:(-1,0]\times \partial_{\mathrm{out}}X^{d+1} \hookrightarrow X^{d+1},
\end{equation*}
whose restrictions to $\{0\}\times \partial_{\mathrm{in/out}}X^{d+1}$ are the inclusions into $\partial X^{d+1}$. Two such triples are equivalent if there is a diffeomorphism $F:X^{d+1}\to (X')^{d+1}$ preserving the boundary identifications and commuting with collar embeddings, i.e. $F\circ \iota_{\mathrm{in/out}}=\iota'_{\mathrm{in/out}}\circ (\mathrm{id}\times F_{\partial})$.\\
\noindent 3. symmetric monoidal structure is defined using the disjoint union:
    \begin{equation*}
        \sqcup: \bordd\times \bordd\to \bordd
    \end{equation*}
\noindent with monoidal unit the empty manifold, denoted $\emptyset$.
\end{definition}

\medskip 

\begin{remark}
Def.~\ref{def:bordismcategory} is the bordism category of smooth manifolds with no additional tangential structure specified. Equivalently, it is the $\xi=O$ case of Def.~\ref{def:bordismcategory_zeta}.
\end{remark}

\bigskip

Before moving on, we note that the bordism category defined in Def.~\ref{def:bordismcategory} implicitly allows all smooth $n$-manifolds. For such a manifold $M$, the tangent bundle $TM$ has an associated principal $\mathrm{O}(n)$-bundle of orthonormal frames. Equivalently, $TM$ is classified (up to isomorphism) by a classifying map
\begin{equation*}
    c_{TM}: M \to B\mathrm{O}(n)
\end{equation*}
There are, however, special classes of manifolds which admit a reduction in structure group. For example, the class of oriented manifolds have tangent bundles with associate frame bundles which are principal $\SO(n)$-bundles. Similarly, geometric conditions on manifolds (orientation, spin, pin, etc.) can be understood as additional structure on the tangent bundle. The correct framework for systematically encoding such choices is that of \emph{tangential structures}, which we now define.

\bigskip 

\begin{definition}[Tangential structure]
\label{def:tangential_structure_general}
An \emph{$n$-dimensional tangential structure} is a pointed space $\xi_n$
together with a pointed fibration
\begin{equation}
    \alpha_n:\xi_n \longrightarrow B\mathrm{O}(n).
\end{equation}
A \emph{$\xi_n$-structure} on an $n$-manifold $M$ is a choice of lift of the tangent space classifying map along $\alpha_n$, i.e.\ a map $\widetilde c_{TM}:M\to \xi_n$ and a homotopy $\alpha_n\circ \widetilde c_{TM}\simeq c_{TM}$.
\end{definition}

\bigskip 

\noindent For example, orientation corresponds to the lift across the fibration $B\SO(n)\rightarrow BO(n)$: a lift $M\to B\SO(n)$ is exactly a reduction of the frame bundle to an $\SO(n)$-bundle. Similarly, a spin structure is a lift of $c_{TM}$ through $B\mathrm{Spin}(n)\to B\SO(n)\rightarrow BO(n)$. In many applications it is convenient to stabilize in dimension, so we do not have to keep track of the dimension. Let $BO:=\mathrm{colim}_{n} BO(n)$, then we can define a stable tangential structure.

\bigskip 

\begin{definition}[Stable tangential structure]
\label{def:stable_tangential_structure}
A \emph{stable tangential structure} is a pointed space $\xi$ together with a
pointed fibration
\begin{equation}
    \alpha:\xi \longrightarrow BO.
\end{equation}
For each $n$, this determines an $n$-dimensional tangential structure
$\xi_n\to BO(n)$ by pullback along $BO(n)\to BO$. A \emph{$\xi$-structure} on an $n$-manifold means a $\xi_n$-structure.
\end{definition}

\bigskip 

\noindent When we say $\mathrm{O}$-structure, or $\SO$-structure, we mean the stable tangential structure which applies to all dimensions. The notion of the bordism category naturally generalizes to that of bordism category of $\xi$-manifolds.

\bigskip 

\begin{definition}[$\xi$-bordism category]
\label{def:bordismcategory_zeta}
Fix a stable tangential structure $\xi$ and a dimension $d$.
We define $\bordd^{\xi}$ to be the symmetric monoidal category obtained from Def.~\ref{def:bordismcategory} by requiring that
\begin{enumerate}[label=\roman*,leftmargin=1.7em]
\item objects are closed $d$-manifolds $Y^{d}$ equipped with $\xi$-structure,
\item morphisms $Y^{d}_{\mathrm{in}}\to Y^{d}_{\mathrm{out}}$ are diffeomorphism classes of of triples $(X^{d+1},\;\varphi_{\mathrm{in}},\;\varphi_{\mathrm{out}})$ as in Def.~\ref{def:bordismcategory}, together with a $\xi$-structure on $X^{d+1}$ restricting to the given boundary $\xi$-structures.
\end{enumerate}
Here all diffeomorphisms are required to preserve the $\xi$-structures. The monoidal structure is disjoint union.
\end{definition}

\bigskip 

\noindent Tangential structures come with a natural notion of ``finer'' versus ``coarser''
structure: a finer structure carries more data, and there is a canonical
forgetful map to a coarser one.

\bigskip 

\begin{definition}[Refinement / forgetting tangential data]
\label{def:refinement_tangential}
Let $\xi$ and $\zeta$ be stable tangential structures with maps
$\alpha:\xi\to BO$ and $\beta:\zeta\to BO$. A \emph{map of tangential structures} $\xi\to \zeta$ is a pointed map
$f:\xi\to \zeta$ such that $\beta\circ f\simeq \alpha$.
In this case, we say that $\xi$ is \emph{finer} than $\zeta$ and $f$ \emph{forgets}
part of the tangential data.
\end{definition}

\bigskip 

\noindent Importantly, such an $f$ induces a symmetric monoidal \emph{forgetful functor} on bordism categories
\begin{equation}
\label{eq:forgetful_bord_functor}
    U_f:\bordd^{\xi}\longrightarrow
    \bordd^{\zeta}.
\end{equation}
\noindent This notion of forgetting structure will be important in our discussion of relating results of systems defined on unoriented manifolds, and systems defined on strictly oriented manifolds in App.~\ref{app:so_str}.

\bigskip 

\noindent Let us denote the symmetric monoidal category of complex vector spaces with monoidal operation the tensor product $\otimes$, and with monoidal unit the complex line $\mathbb{C}$ as $\vect$. 
For objects in the oriented Bordism category $M^{d}\in \bordd^{\mathrm{O}}$ one denotes the oriented reversed copy as $-M^{d}$.

\bigskip

\begin{definition}[Atiyah-Segal (oriented)TQFT, ~\cite{Atiyah1988}]
\label{def:AtiyahSegal}
A TQFT is defined as a symmetric monidal functor from the oriented bordism category to the category of vector spaces:
\begin{equation*}
    \mathcal{Z}:  \bordd^{\SO} \to \vect
\end{equation*}
\end{definition}

\noindent We now record the four Atiyah–Segal axioms that such a functor must satisfy.  Let $M^{d},\;N^{d}$ be closed oriented $d$-manifolds and $X_1^{d+1},\; X_2^{d+1}$ be oriented bordisms from $M^{d}\to N^{d}$ and $N^{d}\to P^{d}$ respectively, with $\partial X_1^{d+1}= (-M^{d})\sqcup N^{d}$ and $\partial X_2^{d+1}=(-N^{d})\sqcup P^{d}$.

\begin{enumerate}[label=(A\arabic*)]
  \item Functoriality.\\
        (a) diffeomorphism invariance: If $f: X^{d}\to Y^{d}$ is a diffeomorphism on objects, then it induces an isomorphism $\mathcal{Z}(f): \mathcal{Z}(X^{d})\to \mathcal{Z}(Y^{d})$.\\
        (b) composition law: For bordisms $X_{1}^{d+1},X_{2}^{d+1}$ composed by gluing along their common boundary $N^{d}$, denoted $X_2^{d+1} \circ X_1^{d+1}$, we have $\mathcal Z(X_2^{d+1} \circ X_1^{d+1})=\mathcal Z(X_2^{d+1})\circ \mathcal Z(X_1^{d+1})$, where $\circ$ in the $\vect$ category means composition of linear maps.
  \item Duality (finite dimensionality).
        Every oriented $M^{d}$ has dual $(-M^d)$, the orientation reversed copy in $\bordd^{\SO}$, then functoriality implies $\mathcal Z(M^{d})$ is dualizable and hence finite-dimensional, with non-degenerate pairing
        $\mathcal Z(M^{d})\otimes \mathcal Z(-M^{d})\to\mathbb{C}$.
  \item Symmetric–monoidal structure.\\
        (a) On objects: $\mathcal Z(M^{d}\sqcup N^{d})=\mathcal Z(M^{d})\otimes \mathcal Z(N^{d})$.\\
        (b) On morphisms: for bordisms $X_1^{d+1}\sqcup X_2^{d+1}$, $\mathcal Z(X_1^{d+1}\sqcup X_2^{d+1})=\mathcal Z(X_1^{d+1})\otimes \mathcal Z(X_2^{d+1})$.\\
        (c) Gluing(sewing): if $X^{d+1}=X_2^{d+1}\circ X_1^{d+1}$ is obtained by gluing along
$N^{d}$, then
\[
\mathcal Z(X^{d+1}) \;=\;
\mathrm{Tr}_{\mathcal Z(N^{d})}
\bigl(\mathcal Z(X_2^{d+1})\circ \mathcal Z(X_1^{d+1})\bigr)
:\mathcal Z(M^{d})\to \mathcal Z(P^{d}).
\]
  \item Unit. $\mathcal Z(\varnothing)=\mathbb{C}$.
\end{enumerate}

Atiyah originally formulated the axioms for TQFTs on the oriented bordism category $\bordd^\SO$~\cite{Atiyah1988}. The same axioms apply verbatim to bordism categories equipped with additional or reduced tangential structure. In particular, Turaev and Turner~\cite{Turaev2005} describe the unoriented case by suppressing orientation data in Atiyah’s axioms. More generally, for any stable tangential structure $\xi$ we define a TQFT to be a symmetric monoidal functor
\begin{equation*}
    \mathcal{Z}: \bordd^{\xi} \to \vect,
\end{equation*}
satisfying the analogues of the Atiyah-Segal axioms discussed in Def.~\ref{def:AtiyahSegal}. Finally, we also note that if the target category is that of $1$-dimensional complex vector spaces, or complex lines $\Line$, then we call the TQFT an invertible TQFT.

\bigskip

\begin{remark}[Picard groupoids]
\label{rmk:picard}
Let $(C, \otimes, 1)$ be some symmetric monoidal category with monoidal operation $\otimes$ and unit $1$. We say an element $x\in (C, \otimes, 1)$ is invertible if there exists $x^{-1}\in C$ such that $x\otimes x^{-1}\cong 1$. A \textit{Picard groupoid} is a symmetric monoidal category in which every object is $\otimes$-invertible and all morphisms are invertible under composition. At this point, we note that the sub-category of complex lines $\Line \subset \vect$ forms a Picard groupoid. Furthermore, one can "complete" a symmetric monoidal category $(C, \otimes,1)$ by formally adjoining inverses of objects under $\otimes$, and morphisms under composition to obtain a \textit{Picard completion} denoted $\overline{C}$. For every symmetric monoidal functor $C$ there exists functor $\eta: C \to \overline{C}$, with the Picard completion characterized by the universal property that for every Picard groupoid $\mathcal{P}$, composition with $\eta$ induces an equivalence of functors $\mathrm{Fun}^{\otimes}(\overline{C}, \mathcal{P})\simeq \mathrm{Fun}^{\otimes}(C, \mathcal{P})$ which means
\begin{equation*}
    \forall \; F: C \to \mathcal{P},\;\; \exists\;\; \overline{F} : \overline{C}\to \mathcal{P}\;\;\text{st}\;\; F=\overline{F}\circ \eta.
\end{equation*}
This property can be viewed as the terminal property of the Picard completion: every functor from $C$ into a Picard groupoid must factor through the Picard completion $\overline{C}$. We use this discussion to define invertible TQFTs.
\end{remark}

\bigskip

\begin{definition}[Freed-Hopkins~\cite{Freed2021}, factorization of invertible TQFTs] An invertible TQFT $\mathcal{Z}: \bordd^{\xi} \to \vect$ factors uniquely through a symmetric monoidal functor:
\begin{equation*}
\hat{\mathcal{Z}}: \overline{\bordd^{\xi}}\to \Line
\end{equation*}
\noindent where $\overline{\bordd^{\xi}}$ is the Picard completion of the bordism category. Conversely, every such $\hat{\mathcal Z}$ corresponds to an invertible TQFT:
\begin{equation*}
    \begin{tikzcd}
        \bordd^{\xi} \arrow[d] \arrow[rr, "\mathcal{Z}"] &&\vect\\
        \overline{\bordd^{\xi}} \arrow[rr, "\hat{\mathcal{Z}}"] &&\Line \arrow[u, hookrightarrow]
    \end{tikzcd}
\end{equation*}
\end{definition}

\bigskip 

\noindent Because $\overline{\bordd^{\xi}}$ and $\Line$ are Picard groupoids, the above factorization identifies invertible TQFTs with a map between Picard groupoids.

\subsection{Introducing symmetry data in the bordism category}
\label{sec:bordism_with_symmetry}

Fix a stable tangential structure $\xi=O$. In this section, we want to incorporate symmetry data in our bordism categories. It will be important for us to treat internal symmetries and spatial symmetries separately, because as mentioned in the Preliminaries section~\ref{sec:basics}, spatial symmetries act on space whereas internal symmetries do not. We will go over the easier case of internal symmetries first, and then generalize to add spatial symmetries.

\subsubsection{Internal symmetries}
\label{app:bordism_internal_sym}

Let $G$ be a compact Lie group of internal symmetries. The standard way to encode a background $G$-field in the Atiyah-Segal bordism category is to decorate manifolds by a map (defined up to homotopy) to $BG$ (equivalently, by a principal $G$-bundle).

\bigskip 

\begin{definition}[Bordism category with (unitary) internal $G$-backgrounds]
\label{def:bordism_BG_background}
Define $\bordd^{\xi}(BG)$ to be the symmetric monoidal bordism category whose
\begin{enumerate}[label=\roman*,leftmargin=1.7em]
\item objects are pairs $(Y^{d}, Q)$ with closed $d$-manifolds $Y^{d}$ equipped with a $\xi$-structure together with a principal $G$-bundle $Q\to Y^{d}$;
\item morphisms $(Y^{d}_{\mathrm{in}},Q_{\mathrm{in}})\to (Y^{d}_{\mathrm{out}},Q_{\mathrm{out}})$ are diffeomorphism classes of compact $(d{+}1)$-dimensional $\xi$-bordisms $(X^{d+1}, \varphi_{\mathrm{in}}, \varphi_{\mathrm{out}})$, as defined in Def.~\ref{def:bordismcategory_zeta}, together with $G$-bundle $P\to X^{d+1}$ such that the restrictions on the boundary give bundle isomorphisms 
\begin{equation*}
\varphi_{\mathrm{in}}^{*}(P\vert_{\partial_{\mathrm{in}}X^{d+1}}) \xrightarrow[]{\cong} Q_{\mathrm{in}},\quad \varphi_{\mathrm{out}}^{*}(P\vert_{\partial_{\mathrm{out}}X^{d+1}}) \xrightarrow[]{\cong} Q_{\mathrm{out}};\end{equation*}
\end{enumerate}
and where diffeomorphisms are required to preserve the $\xi$-structure, and carry bundles over via bundle isomorphisms covering the diffeomorphisms. The monoidal structure is disjoint union.
\end{definition}

\bigskip 

\begin{remark}
\label{remark:pgb}
Equivalently, a principal $G$-bundle $Q\to Y^{d}$ is the same data as a classifying map(up to homotopy) on objects $a:Y^{d}\to BG$, and similarly for bordisms. So, in principal, we can replace pairs $(Y^{d}, Q)$ in the definition above by pairs $(Y^{d}, [f])$ where $[f]$ refers to homotopy classes of maps $f: M \to BG$ such that $f^{*}EG\cong Q$. We will freely switch between these equivalent pictures, depending on convenience, while being very careful to not define particular choices of classifying map but using homotopy classes of classifying maps instead. 
\end{remark}

\bigskip

\noindent For internal symmetries it is sometimes convenient to fold the $G$-background into the tangential structure itself. Concretely, suppose we are given an extension
\begin{equation}
\label{eq:FH_extension}
1 \longrightarrow G \longrightarrow H_{n}
\xrightarrow{\;\rho_{n}\;} \mathrm{O}(n) \longrightarrow 1.
\end{equation}
Then $BH_n$ comes equipped with a map $B\rho_n:BH_n\to B\mathrm{O}(n)$, hence determines an $n$-dimensional tangential structure in the sense of Definition~\ref{def:tangential_structure_general}. We denote the associated bordism category of $BH_n$-structured manifolds by $\bordd^{H_n}$. When $G$ is purely internal (i.e.\ it does not act on spacetime), the $BH_n$-structure
description is equivalent to working with $G$-backgrounds together with the underlying $\mathrm{O}(n)$-structure; accordingly one may replace $\bordd^{H_n}$ by $\bordd^{\mathrm{O}}(BG)$ whenever convenient.

\medskip

\noindent
In the rest of this appendix we use the background-field presentation
$\Bord^{\xi}(BG)$, since it makes the physical interpretation of a fixed $G$-background manifest.

\medskip

\subsubsection{Spatial symmetries}
\label{app:bordism_spatial_sym}

If $G$ acts on space by diffeomorphisms (e.g.\ reflections or rotations), then $G$ is not purely internal and one cannot in general encode a $G$-background simply by a classifying map to $BG$. To take into account the action of symmetry on space, we first define separately the pristine defect free space $\Sigma$, and keep track of the (left)action of $G$ on $\Sigma$ through 
\begin{equation*}
\sigma: G \to \mathrm{Diff}(\Sigma).
\end{equation*}
For our paper, we choose $\Sigma=\mathbb{R}^{d}$ to approximate a condensed matter experiment setting. Then, the corresponding generalization of the classifying space $BG$ is called the Borel construction on $\Sigma$:
\begin{equation}
\label{eq:borel_construction}
    \Sigma_{G}:= \dfrac{\Sigma\times EG}{\sim_{G}}=\{(x, e)\in \Sigma\times EG\;|\; (x,e)\sim (\sigma(g^{-1})\cdot x,\; e\lhd g)\}
\end{equation}
where $\lhd: EG \times G\to EG$ is the canonical free right action of the group $G$ on the total space $EG$. We will denote elements of $\Sigma_{G}$ by equivalence classes $[x,e]\in \Sigma_{G}$. We now propose a definition of a "spatial gauge field":

\bigskip 

\begin{definition}[Spatial gauge field]
\label{def:sgf}
    Given a model space $\Sigma$ on which spatial symmetry $G$ acts through the action $\sigma: G\to \mathrm{Diff}(\Sigma)$, we define a spatial gauge field on a spacetime $M$ as being defined, up to homotopy, by a classifying map
    \begin{equation}
        F: M \to \Sigma_{G}.
    \end{equation}
\end{definition}

\medskip

\noindent We want to compare this definition to the one given in Ref.~\cite{ThorngrenElse}. Before we do this, we look at some immediate implications of the definition above. First, we note that there exists a Cartan mixing diagram given by:
\begin{equation}
    \begin{tikzcd}
        \Sigma \arrow[d, "\mathrm{mod}\;G", swap] &\Sigma\times EG \arrow[d, "\sim_{G}"] \arrow[l, "\mathrm{pr}_{1}", swap] \arrow[r, "\mathrm{pr}_{2}"]& EG \arrow[d, "\pi_{G}"]\\
        \Sigma/G &\Sigma_{G}  \arrow[l, "\underline{\mathrm{pr}}_{1}"] \arrow[r, "\underline{\mathrm{pr}}_{2}", swap]& BG
    \end{tikzcd}
\end{equation}
We do not elaborate on the properties of this diagram, but only note that it commutes and provides the maps defined in the diagram in a canonical fashion.\footnote{The ordinary quotient $\Sigma/G$ may be singular (e.g. an orbifold) when the action has fixed points. In contrast, the homotopy quotient $\Sigma_{G}$ is always well-defined and is the natural target for classifying spatial gauge fields.} For more details see Ch.4 in Ref.~\cite{Tu2020}. Thus, using the diagram and definition~\ref{def:sgf} we can construct the map:
\begin{equation*}
    a: M \to BG, \quad a:=\underline{\mathrm{pr}}_{2} \circ F\;.
\end{equation*}
which can be used to pull-back the universal bundle $EG$, to a principal $G$-bundle on $M$, given by $\pi: P\to M$ where $P=a^{*}(EG)$. 

\bigskip 

\noindent We can also view this bundle $P$ as the pull-back of the principal $G$-bundle $\Sigma\times EG \to \Sigma_{G}$ across the map $F$ such that $P=F^*(\Sigma\times EG)$ is given explicitly as a pullback bundle:
\begin{align}
\label{eq:pgb}
    P:&=M \times_{\Sigma_{G}}(\Sigma\times EG)\\
    &= \{(m,(x,e))\in M\times(\Sigma\times EG)\;|\; F(m)=[x,e]\}.
    \nonumber
\end{align}
where $[x,e]\in \Sigma_{G}$ are equivalence classes of points as defined in Eq.~\ref{eq:borel_construction}. The projection map $\pi: P \to M$ is then given by $\pi(m, (x,e))=m$. The diagonal action of $G$ on $\Sigma\times EG$ induces a free right action on $P$:
\begin{equation*}
    (m,(x,e))\cdot g = (m,(\sigma(g^{-1})\;x, e\lhd g)).
\end{equation*}
With these conventions, we can define the $G$-equivariant map
\begin{equation*}
    \hat{f}: P \to \Sigma, \qquad \hat{f}(m,(x,e)):=x.
\end{equation*}
To check equivariance, note that the diagonal right action  induces the action:
\begin{equation*}
    \hat{f}(p\cdot g)= \sigma(g^{-1})\cdot \hat{f}(p).
\end{equation*}
Therefore, using the Cartan mixing diagram we were able to define the maps $(\pi,\hat{f})$ using the classifying map $F$ in Def.~\ref{def:sgf}. The pair $(\pi, \hat{f})$, however, are the defining maps of spatial gauge fields in Ref.~\cite{ThorngrenElse}, see their Def.2. One can also prove the inverse: given the pair $(\pi, \hat{f})$, the map $F: M \to \Sigma_{G}$ commuting with the mixing diagram is uniquely defined. Therefore, Def.~\ref{def:sgf} is equivalent to the Def.2 in Ref.~\cite{ThorngrenElse}.

\bigskip 

To complete the definition we need one additional piece of data: a homomorphism
\begin{equation*}
\mu: G\to \{\pm 1\}\cong \mathbb Z_2,
\end{equation*}
which we call the twist character of the symmetry. In the crystalline setting, since we are primarily interested in point group symmetries where $G\subset \mathrm{O}(d)$, there exists a canonical group homomorphism:
\begin{equation*}
\mu(g)=
\begin{cases}
+1\quad \text{if}\;\; \sigma(g)\;\; \text{preserves orientation,}\\
-1\quad \text{if}\;\; \sigma(g)\;\; \text{reverses orientation.}
\end{cases}
\end{equation*}
Recall $\sigma: G \to \mathrm{Diff}(\Sigma)$ is the map encoding the action of $G$ on the space $\Sigma$. More generally, we allow $\mu$ as independent input (not necessarily determined by $\sigma$) when $G$ contains an internal symmetry, so that the special case $\sigma=1$ but $\mu\neq 1$ describes internal symmetries with a sign twist (e.g. time-reversal-type symmetries). We then require that the bundle associated to $P\to M$, as defined in Eq.~\ref{eq:pgb}, via $\mu$ given by:
\begin{equation*}
    P_{\mu}:= P\times_{\mu}\Z_{2} \longrightarrow M
\end{equation*}
agree with the orientation double cover over $M$, i.e. there exists a bundle isomorphism:
\begin{equation*}
    P_{\mu}\cong \mathrm{Or}(M).
\end{equation*}
Recall, that the orientation double cover is classified by the first Stiefel-Whitney class of the tangent bundle $TM \to M$, denoted $w_{1}(TM)$, and the requirement above is equivalent to enforcing $w_{1}(P_{\mu})=w_{1}(TM)$. This constraint appears in Def. 2 of Ref.~\cite{ThorngrenElse} as well and, as we will see in App.~\ref{app:grpcoh_functor}, plays a crucial role in relating the geometric and gauge aspects of the bordism category. The full definition is then given by:

\bigskip 

\begin{definition}[Spatial gauge field]
\label{def:sgf_full}
Fix a $G$-space $(\Sigma,\sigma)$ and its Borel construction $\Sigma_G$. Fix also a twist character $\mu:G\to \mathbb Z_2$. A spatial gauge field on $M$ is a homotopy class of maps 
\begin{equation*}
F:M\longrightarrow  \Sigma_G
\end{equation*}
such that, if $P\to M$ is the induced principal $G$-bundle (equivalently, the pullback $P=F^{*}(\Sigma\times EG)\to M$), then the associated $\mathbb Z_2$-bundle
\begin{equation*}
    P_{\mu}:=P\times_{\mu}\mathbb Z_2 \;\longrightarrow\; M
\end{equation*}
is isomorphic to the orientation double cover $\mathrm{Or}(M)\to M$. Equivalently, $w_1(P_{\mu})=w_1(TM)\in H^1(M;\mathbb Z_2)$.
\end{definition}

\bigskip

\begin{remark}[Internal symmetry backgrounds as a special case]
\label{remark:int_simplification}
Assume $\sigma$ is trivial, so that $\Sigma_G\simeq \Sigma\times BG$. If in addition $\Sigma$ is contractible (e.g.\ $\Sigma=\mathbb{R}^d$), then $\Sigma_G\simeq BG$ and a homotopy class of maps $F:M\to \Sigma_G$ is equivalent to an ordinary $G$-bundle $P\to M$. In this case Def.~\ref{def:sgf_full} reduces to the condition $w_1(P_\mu)=w_1(TM)$.
Thus $\mu=1$ recovers the usual unitary internal background, while $\mu\neq 1$ describes an internal symmetry with a sign twist. A time-reversal-type symmetry is of this latter form: it acts trivially on the space $\Sigma$, but its nontrivial twist character records the reversal of spacetime orientation.
\end{remark}

\bigskip 

We now use the definition of spatial gauge fields in Def.~\ref{def:sgf_full} to rigorously define bordism categories with spatial $G$-backgrounds:

\bigskip 

\begin{definition}[Bordism category with generalized $G$-backgrounds]
\label{def:bordism_spatial_BG_background}
Given a $G$-space $(\Sigma, \sigma)$ and its Borel construction $\Sigma_{G}$, and where $G$ has twist character defined by $\mu: G \to\Z_{2}$, we define $\bordd^{\xi}(\Sigma_{G},\mu )$ to be the symmetric monoidal bordism category whose
\begin{enumerate}[label=\roman*,leftmargin=1.7em]
\item objects are pairs $(Y^{d}, [f])$ with closed $d$-manifolds $Y^{d}$ equipped with a $\xi$-structure together with a homotopy class of spatial gauge field $[f]\in [Y^{d}, \Sigma_{G}]$ represented by classifying maps $f:Y^{d}\to \Sigma_{G}$ such that the the associated $\Z_{2}$-bundle $P_{\mu}\cong a^{*}(EG)\times_{\mu} \Z_{2}$, where $a:=\underline{\mathrm{pr}}_{2}\circ f$, is bundle isomorphic to the orientation double cover of $Y^{d}$;
\item morphisms $(Y^{d}_{\mathrm{in}},[f_{\mathrm{in}}])\to (Y^{d}_{\mathrm{out}},[f_{\mathrm{out}}])$ are diffeomorphism classes of compact $(d{+}1)$-dimensional $\xi$-bordisms $(X^{d+1}, \varphi_{\mathrm{in}}, \varphi_{\mathrm{out}})$ together with a \textit{relative} homotopy class of spatial gauge field $[F]\in [X^{d+1}, \Sigma_{G}]_{\partial}$ represented by classifying maps $F:X^{d+1}\to \Sigma_{G}$ such that the restrictions on the boundary satisfy $[F]\vert_{Y_{\mathrm{in}}} = [f_{\mathrm{in}}]$ and $[F]\vert_{Y_{\mathrm{out}}} = [f_{\mathrm{out}}]$ such that the the associated $\Z_{2}$-bundle $Q_{\mu}\cong A^{*}(EG)\times_{\mu} \Z_{2}$, where $A:=\underline{\mathrm{pr}}_{2}\circ F$, is bundle isomorphic to the orientation double cover of $X^{d+1}$.
\end{enumerate}
Given morphisms represented by $F_1:X_1\to\Sigma_G$ from $(Y_{\mathrm{in}},[f_{\mathrm{in}}])$ to $(Y,[f])$ and $F_2:X_2\to\Sigma_G$ from $(Y,[f])$ to $(Y_{\mathrm{out}},[f_{\mathrm{out}}])$, choose representatives which restrict to the same boundary representative on $Y$ (and are fixed on collars). Then the glued bordism $X_2\circ X_1$ carries a glued map $F_2\cup F_1:X_2\circ X_1\to\Sigma_G$, well-defined up to relative homotopy, and defines the composite.
\end{definition}

\bigskip

\noindent Here, in contrast to Def.~\ref{def:bordism_BG_background}, we used the representation of bundles in terms of homotopy classes of the classifying maps because this gives a more compact definition. One could equivalently use pairs $(\pi, \hat{f})$ to define spatial symmetry backgrounds and define the bordism category using this description instead, but it will be equivalent to the one given above. 

\subsection{Defining $G$-TQFTs}
\label{app:GTQFT}

We can now define the corresponding definitions of TQFTs with $G$-backgrounds we saw earlier in this section. 

\bigskip 

\begin{definition}[Unitary internal $G$-TQFT]
\label{def:internal_GTQFT}
Fix $\xi=\mathrm{O}$ and a group $G$ which is purely unitary (no sign twist). An internal $G$-TQFT is a symmetric monoidal functor
\begin{equation*}
\mathcal Z:\bordd^{\xi}(BG)\longrightarrow \vect .
\end{equation*}
It is invertible if it factors through $\Line\subset\vect$.
\end{definition}

\bigskip

\begin{definition}[Generalized $G$-TQFT]
\label{def:spatial_GTQFT}
Fix a $G$-space $(\Sigma,\sigma)$ and let $\Sigma_G$ be its Borel construction. Fix also the twist character $\mu: G\to \Z_{2}$. A generalized $G$-TQFT on $\Sigma$ (with $\xi=\mathrm{O}$) is a symmetric monoidal functor
\begin{equation*}
    \mathcal Z:\bordd^{\xi}(\Sigma_G, \mu)\longrightarrow \vect,
\end{equation*}
where $\bordd^{\xi}(\Sigma_G, \mu)$ is the bordism category with generalized $G$-backgrounds from Def.~\ref{def:bordism_spatial_BG_background}. It is invertible if it factors through $\Line\subset\vect$.
\end{definition}

\bigskip 

\begin{remark}
This general notion of a $G$-TQFT includes the following cases: (a) the unitary internal $G$-TQFT in which both $\sigma,\mu$ are trivial and $\Sigma$ is contractible, (b) twisted internal $G$-TQFT in which $\sigma$ is trivial and $\Sigma$ is contractible, but $\mu$ is non-trivial. Time-reversal symmetry is a good example of this case.\footnote{The assumption that $\Sigma$ is contractible is what removes the extra $\Sigma$-valued background data. If $\sigma=1$ but $\Sigma$ is not contractible, then $\Sigma_G\simeq\Sigma\times BG$, so a generalized background $F:M\to\Sigma_G$ is equivalently a pair $(\phi,f_G)$ with $\phi:M\to\Sigma$ and $f_G:M\to BG$. Thus the construction is richer than the ordinary internal symmetry background, although it still canonically projects to one.} (c) And finally, the case of spatial symmetry when both $\sigma,\mu$ are non-trivial. In the case of spatial $G$-backgrounds, we do not write $\mu$ explicitly in the argument of the bordism category since it is defined implicitly by the spatial action $\sigma$. 
\end{remark}

\bigskip 

We also define the category of $G$-TQFTs which refers to the family of all $G$-TQFTs which satisfy the Atiyah-Segal axioms. For us, in particular, it can also mean the sub-category(of the full category which is exhaustive) of $G$-TQFTs given to us by a particular construction, for example, the Freed-Quinn construction in App.~\ref{app:grpcoh_functor}.

\medskip 

\begin{definition}[Category of generalized $G$-TQFTs]
Define
\begin{equation*}
   \tqftd^{G}(\Sigma,\mu)
   \;:=\;
   \underline{\mathrm{Fun}}^{\otimes} 
   \bigl(
      \bordd^{\xi}(\Sigma_G,\mu),\;
      \vect
   \bigr),
\end{equation*}
the $(\infty,1)$-category whose objects are symmetric–monoidal functors
$\mathcal Z:\bordd^{\xi}(\Sigma_G,\mu)\to\vect$, and whose morphisms are symmetric–monoidal
natural transformations. The monoidal product is pointwise.
\end{definition}

\subsection{Continuum TQFTs: why $\mathrm{O}(d)$ symmetry?}
\label{app:why_Od_continuum}

Here, we explain why the continuum target used in the main text is $\Theta^{d+1}(B\mathrm{O}(d))$, or equivalently why, after translations are discarded, the intrinsic continuum spatial symmetry is taken to be $\mathrm{O}(d)$. The central physical point is that continuum fields are defined relative to the local geometry of space. Their components and index structures transform under changes of local frame, and are therefore naturally organized as sections of bundles associated to the frame bundle of the underlying manifold. Thus the possible transformation laws of continuum fields are controlled by the structure group of the tangent bundle.

\medskip 

\noindent In this paper, continuum space is modeled by flat Euclidean space $\mathbb R^{d}$ with its standard Euclidean metric. We do not retain translations as protecting symmetry data, so the relevant intrinsic spatial symmetry is the remaining orthogonal frame symmetry. Moreover, we assume that the continuum background carries no additional structure such as a preferred lattice, axis, foliation, or anisotropic tensor. With only the Euclidean metric specified, the relevant frame bundle is the bundle of orthonormal frames, whose structure group is $\mathrm{O}(d)$. Equivalently, tangent and cotangent indices of continuum fields transform under changes of orthonormal frame. This is the sense in which $\mathrm{O}(d)$ is the intrinsic spatial symmetry of the bare continuum background.

\bigskip 

A continuum theory may of course have only a subgroup $K\subset \mathrm{O}(d)$ as an unbroken spatial symmetry. Such a reduction, however, requires extra background data. A standard example is an anisotropic gradient term for a scalar field,
\begin{equation}
    \delta\mathcal{L}_{\mathrm{anis}}[\phi^i]\sim \lambda\, n^{i}n^{j}\,\partial_{i}\phi\,\partial_{j}\phi ,
\end{equation}
where $n^{i}$ is a fixed unit vector selecting a preferred spatial direction. If $n^{i}$ is treated as a background field that transforms under $\mathrm{O}(d)$, the term is $\mathrm{O}(d)$-covariant. Once a particular value of $n^{i}$ is fixed, the visible spatial symmetry is reduced to the stabilizer subgroup of $n^{i}$ inside $\mathrm{O}(d)$. Thus the reduced symmetry is not part of the bare Euclidean continuum background alone, but arises after adjoining symmetry-reducing background data.

\medskip 

\noindent We therefore call continuum theories with explicitly reduced spatial symmetry \emph{derived} in the following restricted sense. They are obtained from the intrinsic $\mathrm{O}(d)$-structured continuum problem by adjoining additional symmetry-reducing background data, or equivalently by restricting along an inclusion $K\hookrightarrow \mathrm{O}(d)$ and using the naturality of the classification functor. Thus $\Theta^{d+1}(B\mathrm{O}(d))$ is the continuum classification with no additional symmetry-reducing geometric data. Theories with only subgroup symmetry encode further background choices and are treated as restrictions of this intrinsic continuum target.

\section{The Cobordism theoretic classification}
\label{app:cob}

In this appendix, we collect and derive in a compact form, for completeness, the unoriented bordism classification developed in the work of Freed and Hopkins. Starting from the geometric definition of bordism, we review the construction of the unoriented Thom spectrum $\mathrm{MO}$, its relation to unoriented bordism through the Pontryagin-Thom theorem, and the stable-homotopy-theoretic classification of invertible TQFTs. This allows us to compare the bordism-character classification proposed by Kapustin with the spectrum-level classification of Freed and Hopkins, and to show that both agree with the bordism classification model used in Eq.~\eqref{eq:cob_model}. The Freed-Hopkins framework is formulated for stable fully extended invertible TQFTs, so this agreement also shows that the bordism classification employed here applies in this more refined extended setting. Together with the subsequent applications in Apps.~\ref{app:cob_proof} and~\ref{app:so_str}, this establishes that the general stratification and continuum-limit methods developed in this manuscript apply to extended invertible TQFTs as well.

\medskip 

\noindent The appendix is organized as follows. In Sec.~\ref{app:cob_prelim}, we introduce the geometric unoriented bordism groups, their ring structure, and their Pontryagin duals. In Sec.~\ref{app:cob_spectrum_description}, we construct the Thom spectrum $\mathrm{MO}$ and explain how it represents unoriented bordism as a generalized homology theory. Section~\ref{app:thom_unoriented_bordism} reviews Thom's computation of the unoriented bordism ring and the stable splitting of $\mathrm{MO}$. In Sec.~\ref{app:cob1}, we relate invertible TQFT functors to maps of spectra, review the Freed-Hopkins classification of stable fully extended theories, and compare it with the bordism-character classification. Finally, Sec.~\ref{app:cob2} develops the Borel-equivariant version relevant to spatial symmetries and derives the classification $\Omega_{\mathrm O}^{d+1}(BG)$ used throughout the manuscript.

\subsection{The geometric origin of the unoriented bordism ring $\Omega^{\mathrm O}_{*}$}
\label{app:cob_prelim}

A bordism between two $n$-dimensional manifolds, $M^{n},\;N^{n}$, is an equivalence relation defined by diffeomorphism classes of $(n+1)$-dimensional manifolds, $[W^{n+1}]$, which interpolates smoothly between $M^{n}$ and $N^{n}$ such that $\partial W^{n+1} = M^{n}\sqcup N^{n}$. This can be represented pictorially as follows:

\begin{figure}[H]
    \centering
    \includegraphics[width=0.7\linewidth]{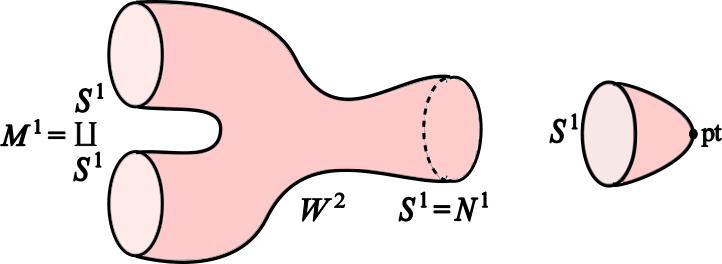}
    \caption{(Left)Bordism from $M^{1}=S^{1}\amalg S^{1}$ to $N^{1}=S^{1}$. The equivalence is provided by $W^{2}$, often called a pair of pants; (right) Bordism from $S^{1}$ to a pt given by a disk.}
    \label{fig:enter-label}
\end{figure}

\noindent It is easy to show that bordisms satisfy the requirements to qualify them as equivalence relations. The equivalence class of $n$-dimensional manifolds under bordisms is denoted $\Omega_{n}^{O}$. 
For example, the "pair of pants" equivalence above can be used to show that $\Omega^{\mathrm{O}}_{1}=0$: every closed $1$-manifold is a disjoint union of circles, and each circle bounds a disk.

\begin{definition}[Unoriented bordism ring $\Omega^{O}_{*}$]
We define an additive structure on bordism classes using the disjoint union or connected sum, i.e., $[M]+[N]:=[M\sqcup N]=[M\#N]$, and a multiplicative structure using the Cartesian product of manifolds, i.e., $[M]\times [N]:=[M\times N]$. This makes
\begin{equation}
    \Omega^{O}_{*}:=\bigoplus_{n\geq 0} \Omega_{n}^{O}    
\end{equation}
\noindent a graded abelian ring with grading defined by the dimension. Furthermore, one can show that $2[M]= [M\sqcup M]= [\partial(M\times I)]=0$ making $\Omega_{*}^{O}$ a graded commutative algebra over $\mathbb{Z}_{2}$.
\end{definition}

\bigskip

\begin{definition}[Pontryagin dual of the bordism group]
For each degree $n$, define
\begin{equation}
    \widehat{\Omega}_{\mathrm O}^{\,n}
    :=
    \operatorname{Hom}
    \bigl(\Omega_n^{\mathrm O},\mathrm U(1)\bigr).
\end{equation}
Since $\Omega_n^{\mathrm O}$ is a finite-dimensional $\mathbb Z_2$-vector
space, there is an isomorphism
\begin{equation*}
    \widehat{\Omega}_{\mathrm O}^{\,n}
    \cong
    \operatorname{Hom}_{\mathbb Z_2}
    \bigl(\Omega_n^{\mathrm O},\mathbb Z_2\bigr).
\end{equation*}
\end{definition}

\noindent For orientation, the first few unoriented bordism groups are listed in
Table~\ref{tab:unoriented-cobordism}; see, for example,
Refs.~\cite{Thom1954, Stong_unorCob}. The Stiefel-Whitney monomials
shown in the third column give characteristic numbers that distinguish
the indicated bordism classes.

\begin{table}[H]
\footnotesize
\setlength{\tabcolsep}{4pt}
\centering
\caption{Low-dimensional bordism groups
$\Omega_n^{\mathrm O}$ for $n\leq 4$.}
\label{tab:unoriented-cobordism}
\begin{tabular}{@{}c c l l@{}}
\toprule
$n$ &
$\Omega_n^{\mathrm O}$ &
Detecting Stiefel-Whitney numbers &
Manifold representatives \\[2pt]
\midrule
0 & $\mathbb Z_2$ &
$1$ &
point \\

1 & $0$ &
--- &
--- \\

2 & $\mathbb Z_2$ &
$w_2$ &
$\mathbb{RP}^2$ \\

3 & $0$ &
--- &
--- \\

4 & $\mathbb Z_2^2$ &
$w_4,\;w_2^2$ &
$\mathbb{RP}^4,\;
\mathbb{RP}^2\times\mathbb{RP}^2$ \\
\bottomrule
\end{tabular}
\end{table}

\subsection{Thom spectrum and the unoriented bordism ring}
\label{app:cob_spectrum_description}

\begin{notation}
\label{notation:mod2_coefficients_cobordism}
Throughout the remainder of App.~\ref{app:cob} all ordinary homology and cohomology groups are taken with coefficients in $\mathbb Z_{2}$ unless stated otherwise. We suppress the coefficient group and write $H^{*}(X)$ in place of $H^{*}(X,\mathbb Z_{2})$.\\
\noindent For spectra $E$ and $G$, we write \begin{equation*} [E,G] := \pi_0F(E,G) \end{equation*} for the abelian group of homotopy classes of maps of spectra. Thus $F(E,G)$ denotes the function spectrum, whereas $[E,G]$ denotes its set of connected components. The notation $\operatorname{Hom}_{\mathrm{sp}}(E,G)$, when used, denotes the same group. For pointed spaces $X$ and $Y$, the notation $[X,Y]_*$ denotes based homotopy classes of maps.
\end{notation}

\bigskip 

We now describe how the unoriented bordism groups are organized into a generalized homology theory. The relevant spectrum is the unoriented Thom spectrum, denoted $\mathrm{MO}$. We use $\mathrm{MO}(k)$ for the finite-stage Thom spaces used to construct the spectrum, and reserve notation $\underline{\mathrm{MO}}_{q}$ for the infinite-loop spaces representing the associated generalized cohomology theory.

\medskip 

\noindent Let
\begin{equation*}
    B\mathrm{O}(k)
    :=
    \operatorname{Gr}_{k}(\mathbb{R}^{\infty})
\end{equation*}
be the Grassmannian of $k$-dimensional linear subspaces of $\mathbb{R}^{\infty}$ (see Ch 5, Ref.~\cite{Milnor} for reference). It carries the universal rank-$k$ real vector bundle
\begin{equation*}
    \gamma_{k}
    :=
    \left\{
        (V,v)\in B\mathrm{O}(k)\times\mathbb{R}^{\infty}
        \,\middle|\,
        v\in V
    \right\}
    \longrightarrow
    B\mathrm{O}(k).
\end{equation*}
The $k$th Thom space is defined by
\begin{equation}
    \mathrm{MO}(k)
    :=
    \operatorname{Th}(\gamma_{k})
    =
    D(\gamma_{k})/S(\gamma_{k}),
    \label{eq:MO_k_Thom_space}
\end{equation}
where $D(\gamma_k)$ and $S(\gamma_k)$ denote the unit disk and unit sphere bundles of $\gamma_k$, respectively. To construct the spectrum structure, consider the inclusion
\begin{equation*}
    \iota_{k}:
    B\mathrm{O}(k)\longrightarrow B\mathrm{O}(k+1),
    \qquad
    V\longmapsto\mathbb{R}\oplus V,
\end{equation*}
together with a bundle map
\begin{equation*}
\begin{tikzcd}
\epsilon^{1}\oplus\gamma_{k}
    \arrow[r,"\widetilde{\iota}_{k}"]
    \arrow[d]
&
\gamma_{k+1}
    \arrow[d]
\\
B\mathrm{O}(k)
    \arrow[r,"\iota_{k}"]
&
B\mathrm{O}(k+1),
\end{tikzcd}
\end{equation*}
where $\epsilon^{1}$ is the trivial real line bundle. Applying the Thom-space
construction gives
\begin{equation}
    \sigma_{k}:
    \Sigma\mathrm{MO}(k)
    \cong
    \operatorname{Th}(\epsilon^{1}\oplus\gamma_{k})
    \longrightarrow
    \operatorname{Th}(\gamma_{k+1})
    =
    \mathrm{MO}(k+1).
    \label{eq:MO_spectrum_structure_map}
\end{equation}
The sequence
\begin{equation*}
    \mathrm{MO}
    :=
    \left\{
        \mathrm{MO}(k),\sigma_{k}
    \right\}_{k\geq0}
\end{equation*}
is the unoriented Thom spectrum.

\bigskip 

The spectrum $\mathrm{MO}$ determines a generalized homology theory
\begin{equation}
\begin{aligned}
    \mathrm{MO}_{n}(X)
    &:=
    \pi_{n}\left(
        \mathrm{MO}\wedge\Sigma^{\infty}X_{+}
    \right)
\\
    &\cong
    \underset{k\to\infty}{\operatorname{colim}}\,
    \pi_{n+k}\left(
        \mathrm{MO}(k)\wedge X_{+}
    \right),
\end{aligned}
\label{eq:MO_generalized_homology}
\end{equation}
where $X_{+}$ denotes $X$ with a disjoint basepoint. The
Pontryagin-Thom theorem identifies this generalized homology theory with
geometric unoriented bordism:
\begin{equation}
    \Omega_{n}^{\mathrm{O}}(X)
    \xrightarrow{\;\cong\;}
    \mathrm{MO}_{n}(X).
    \label{eq:PT_general_isomorphism}
\end{equation}

\noindent We briefly review the geometric construction of this isomorphism. Let
$(M^{n},f)$ represent a class in $\Omega_{n}^{\mathrm{O}}(X)$, where
$M^{n}$ is a closed smooth manifold and $f:M\to X$ is a continuous map.
Choose, for sufficiently large $k$, an embedding
\begin{equation*}
    e:M^{n}\hookrightarrow\mathbb{R}^{n+k}.
\end{equation*}
The normal bundle $\nu_{M}\to M$ is a rank-$k$ real vector bundle and is
therefore classified by a map
\begin{equation*}
    c_{\nu}:M\longrightarrow B\mathrm{O}(k).
\end{equation*}
By the tubular neighborhood theorem, a tubular neighborhood $N(M)$ of
$M$ in $\mathbb{R}^{n+k}$ is identified with the disk bundle
$D(\nu_{M})$.

The one-point compactification
$\mathbb{R}^{n+k}\cup\{\infty\}\cong S^{n+k}$ can now be collapsed outside
the tubular neighborhood. This gives the Pontryagin-Thom collapse map
\begin{equation}
    c_{M}:
    S^{n+k}
    \longrightarrow
    \operatorname{Th}(\nu_{M}).
    \label{eq:PT_collapse_map}
\end{equation}
The classifying map $c_{\nu}$ induces a bundle map
$\nu_{M}\to\gamma_{k}$ and hence a map of Thom spaces
\begin{equation*}
    \operatorname{Th}(\nu_{M})
    \longrightarrow
    \operatorname{Th}(\gamma_{k})
    =
    \mathrm{MO}(k).
\end{equation*}
Together with the map $f:M\to X$, this produces a based map
\begin{equation}
    S^{n+k}
    \xrightarrow{\;c_{M}\;}
    \operatorname{Th}(\nu_{M})
    \longrightarrow
    \mathrm{MO}(k)\wedge X_{+}.
    \label{eq:PT_map_with_background}
\end{equation}
Its homotopy class defines an element
\begin{equation*}
    \theta_{k}(M,f)
    \in
    \pi_{n+k}\left(
        \mathrm{MO}(k)\wedge X_{+}
    \right).
\end{equation*}
Different choices of embedding, tubular neighborhood, and classifying map
give the same class after stabilization. Passing to the colimit in
Eq.~\eqref{eq:MO_generalized_homology} therefore gives a well-defined map
\begin{equation}
    \theta_{X}:
    \Omega_{n}^{\mathrm{O}}(X)
    \longrightarrow
    \mathrm{MO}_{n}(X).
    \label{eq:PT_map}
\end{equation}
The construction is invariant under bordism. The Pontryagin-Thom theorem states that $\theta_{X}$ is an isomorphism.

\noindent For $X=\{\mathrm{pt}\}$, this reduces to
\begin{equation}
    \Omega_{n}^{\mathrm{O}}
    \xrightarrow{\;\cong\;}
    \mathrm{MO}_{n}(\{\mathrm{pt}\})
    =
    \pi_{n}(\mathrm{MO})
    \cong
    \underset{k\to\infty}{\operatorname{colim}}\,
    \pi_{n+k}\bigl(\mathrm{MO}(k)\bigr).
    \label{eq:PT_point_isomorphism}
\end{equation}
Thus, the Thom spectrum $\mathrm{MO}$ directly represents unoriented
bordism as a generalized homology theory.

\bigskip 

For completeness, the same spectrum also determines a generalized
cohomology theory
\begin{equation}
    \mathrm{MO}^{q}(X)
    :=
    \left[
        \Sigma^{\infty}X_{+},
        \Sigma^{q}\mathrm{MO}
    \right].
    \label{eq:MO_generalized_cohomology}
\end{equation}
The degree-$q$ representing space is denoted
\begin{equation}
\label{eq:MO_underlined_space}
    \underline{\mathrm{MO}}_{q}
    :=
    \Omega^{\infty}\Sigma^{q}\mathrm{MO},\quad \text{s.t.}\quad
    \mathrm{MO}^{q}(X)
    \cong
    \left[
        X_{+},
        \underline{\mathrm{MO}}_{q}
    \right]_{*}.
\end{equation}
We will make little use of this cobordism cohomology theory in what follows. Following terminology used in parts of the physics literature, we separately define the Pontryagin-dual bordism group
\begin{equation}
    \widehat{\Omega}_{\mathrm{O}}^{\,n}(X)
    :=
    \operatorname{Hom}\left(
        \Omega_{n}^{\mathrm{O}}(X),
        \mathrm{U}(1)
    \right).
    \label{eq:dual_bordism_group}
\end{equation}
This is the object that appears in the torsion classification proposed by Kapustin.

\subsection{Thom's computation of the unoriented bordism ring, and stable splitting of $\mathrm{MO}$}
\label{app:thom_unoriented_bordism}

The coefficient ring of the unoriented bordism theory was computed by Thom.

\begin{theorem}[Thom]
\label{thm:unoriented_bordism_ring}
The unoriented bordism ring is a polynomial algebra with one generator in every positive degree not of the form $2^{s}-1$:
\begin{equation}
    \Omega_{*}^{\mathrm O} \cong
    \mathbb Z_{2} [x_i\mid i\geq2,\; i\neq2^{s}-1],
    \qquad |x_i|=i.
    \label{eq:unoriented_bordism_polynomial}
\end{equation}
\end{theorem}

\noindent The generators $x_i$ are bordism classes represented by closed manifolds; they should not be identified with the Stiefel-Whitney classes $w_i\in H^i(B\mathrm O)$.

\bigskip 

\noindent For a closed unoriented $n$-manifold $M$ and a degree-$n$ polynomial $P\in H^n(B\mathrm O)$, the associated Stiefel-Whitney number is
\begin{equation}
    \left\langle
        P\bigl(w_1(TM),\ldots,w_n(TM)\bigr), [M]_2
    \right\rangle \in\mathbb F_2.
    \label{eq:Stiefel_Whitney_number}
\end{equation}
These numbers depend only on the bordism class of $M$, and Thom's theorem implies that they detect unoriented bordism:
\begin{equation}
    [M]=[N]\in\Omega_n^{\mathrm O} \;\Longleftrightarrow\;
    \left\langle P(w(TM)),[M]_2\right\rangle
    = \left\langle P(w(TN)),[N]_2\right\rangle
\end{equation}
for every $P\in H^n(B\mathrm O)$. Equivalently, evaluation on Stiefel-Whitney numbers gives a surjection
\begin{equation}
    H^n(B\mathrm O) \twoheadrightarrow
    \operatorname{Hom}_{\mathbb Z_2}
    \bigl(\Omega_n^{\mathrm O},\mathbb Z_2\bigr)
    \cong
    \widehat{\Omega}_{\mathrm O}^{\,n}.
    \label{eq:SW_polynomials_surject_characters}
\end{equation}
Thus every character of $\Omega_n^{\mathrm O}$ is represented by a degree-$n$ polynomial in Stiefel-Whitney classes, although the representing polynomial need not be unique.

\bigskip 

Choose polynomial generators
\begin{equation*}
    \Omega_*^{\mathrm O} \cong \mathbb Z_2[x_i\mid i\geq2,\;i\neq2^s-1],
\end{equation*}
and let $\mathcal B$ be the associated homogeneous monomial basis.

\begin{theorem}[Stable splitting of $\mathrm{MO}$]
\label{thm:MO_stable_splitting}
There is an equivalence of spectra
\begin{equation}
    \mathrm{MO} \simeq \bigvee_{\alpha\in\mathcal B}
    \Sigma^{|\alpha|}H\mathbb Z_2.
    \label{eq:MO_stable_splitting}
\end{equation}
The equivalence is additive and is not asserted to preserve the multiplicative structure of $\mathrm{MO}$.
\end{theorem}

\begin{proof}
This is the classical generalized Eilenberg-Mac Lane splitting of the unoriented Thom spectrum. Thom's calculation identifies $H^*(\mathrm{MO})$ as a free module over the mod-$2$ Steenrod algebra, with one generator for each homogeneous basis element of $\Omega_*^{\mathrm O}$. Choosing representatives of these generators produces a map from $\mathrm{MO}$ to the wedge in Eq.~\eqref{eq:MO_stable_splitting}; the resulting map is an isomorphism on mod-$2$ cohomology and hence a stable equivalence. See Ref.~\cite[Ch.~25, Sec.~6]{JPMay}.
\end{proof}

\begin{corollary}
\label{cor:MO_additive_decomposition}
After choosing the splitting in
Eq.~\eqref{eq:MO_stable_splitting}, there is a natural additive
isomorphism
\begin{equation}
    \mathrm{MO}_n(X)
    \cong
    \bigoplus_{p+q=n}
    H_p(X;\mathbb Z_2)\otimes_{\mathbb Z_2}\Omega_q^{\mathrm O}.
\end{equation}
\end{corollary}

\begin{remark}[Choices in the stable splitting]
\label{rem:MO_splitting_noncanonical}
The stable splitting involves two choices. First, Thom's polynomial presentation
\begin{equation*}
    \Omega_*^{\mathrm O}
    \cong
    \mathbb F_2[x_i\mid i\geq2,\;i\neq2^s-1]
\end{equation*}
does not canonically specify the polynomial generators $x_i$: a generator may be modified by decomposable classes of the same degree. Once a system of polynomial generators is fixed, however, its associated homogeneous monomial basis $\mathcal B$ is determined. Second, after fixing $\mathcal B$, one must choose a stable equivalence
\begin{equation*}
    \omega: \mathrm{MO}
    \xrightarrow{\;\simeq\;}
    \bigvee_{\alpha\in\mathcal B}
    \Sigma^{|\alpha|}H\mathbb F_2.
\end{equation*}
The Steenrod operations and their action on $H^*(\mathrm{MO})$ are canonical. The additional choice lies only in selecting free $\mathcal A$-module generators, or equivalently a particular equivalence $\omega$ realizing the splitting.

\noindent We fix these choices throughout. The decomposition in Corollary~\ref{cor:MO_additive_decomposition} therefore depends on the chosen splitting, but is natural in $X$. These choices enter only the proof of Proposition~\ref{prop:cobordism_restriction_isomorphism}. The map $\res_{\mathrm{cob}}^{\mathcal A}$ itself is defined canonically by restriction along subgroup inclusions. Hence the statement that it is an isomorphism, and its uniquely determined inverse, are independent of the chosen splitting.
\end{remark}

\subsection{Classification of invertible TQFTs}
\label{app:cob1}

In this section we briefly review the stable homotopy-theoretic classification of invertible continuum TQFTs developed by Freed and Hopkins in Refs.~\cite{Freed2014,FreedHopkins}, and show that, in the stable unoriented case, the resulting classification group agrees with the cobordism classification used in our paper (see Eq.~\ref{eq:cob_model}). We specialize throughout to theories with unoriented tangential structure. To connect with the notation of the main text, the spacetime dimension $n$ should be replaced by $d+1$. See App.~\ref{app:tqft} for the general definition of an invertible TQFT.

\subsubsection{From invertible TQFT functors to spectrum maps}
\label{sec:cob_functors_to_spectra}

We next explain how the spectrum-level formulation of Freed and Hopkins relates to the definition of invertible TQFTs given in App.~\ref{app:tqft}. In appendix~\ref{app:tqft}, an unextended unoriented TQFT is defined as a symmetric monoidal functor
\begin{equation*}
    \mathcal Z:
    \bordd^{\mathrm O}
    \longrightarrow
    \vect.
\end{equation*}
An invertible theory factors through the Picard completion of the
bordism category and the Picard groupoid of complex lines:
\begin{equation}
    \widehat{\mathcal Z}:
    \overline{\bordd^{\mathrm O}}
    \longrightarrow
    \Line.
    \label{eq:unextended_invertible_TQFT}
\end{equation}
\noindent The Picard groupoids and Picard completions appearing here were defined in Remark~\ref{rmk:picard}. We use the following standard relation between Picard groupoids and spectra.

\bigskip

\begin{proposition}[Picard groupoids and spectra]
\label{prop:Picard_groupoids_spectra}
A Picard $\infty$-groupoid $\mathcal P$ determines a connective
spectrum $\operatorname{sp}(\mathcal P)$ whose infinite-loop space
satisfies
\begin{equation*}
    \Omega^\infty\operatorname{sp}(\mathcal P)
    \simeq
    \lVert\mathcal P\rVert.
\end{equation*}
Moreover, a symmetric monoidal functor
$\Phi:\mathcal P\to\mathcal Q$ induces a map of spectra
\begin{equation*}
    \operatorname{sp}(\Phi):
    \operatorname{sp}(\mathcal P)
    \longrightarrow
    \operatorname{sp}(\mathcal Q).
\end{equation*}
For an ordinary Picard groupoid, the associated spectrum has vanishing
homotopy groups above degree $1$.
\end{proposition}

\begin{proof}
See Ref.~\cite[Sec.~1.1]{SchommerPries2017}.
\end{proof}

\bigskip 

Applying Proposition~\ref{prop:Picard_groupoids_spectra} to
Eq.~\eqref{eq:unextended_invertible_TQFT}, the invertible TQFT functor
induces a map between the associated spectra,
\begin{equation}
    \operatorname{sp}\bigl(\widehat{\mathcal Z}\bigr):
    \operatorname{sp}
    \bigl(\overline{\bordd^{\mathrm O}}\bigr)
    \longrightarrow
    \operatorname{sp}(\Line).
    \label{eq:unextended_TQFT_spectrum_map}
\end{equation}
Passing to their infinite-loop spaces recovers the classifying-space
map
\begin{equation*}
    \bigl\lVert\widehat{\mathcal Z}\bigr\rVert:
    \bigl\lVert\overline{\bordd^{\mathrm O}}\bigr\rVert
    \longrightarrow
    \lVert\Line\rVert.
\end{equation*}

\bigskip

The functor in Eq.~\eqref{eq:unextended_invertible_TQFT} is defined on the unextended Atiyah-Segal bordism category $\bordd^{\mathrm O}$: its objects are closed $d$-manifolds and its morphisms are $(d+1)$-dimensional bordisms. Freed and Hopkins instead formulate their classification using the fully extended unoriented bordism $(\infty,d+1)$-category, which we denote
\begin{equation*}
    {\bf Bord}_{d+1}^{\mathrm O}.
\end{equation*}
A fully extended theory includes bordisms of all codimensions and is therefore not the same object as the unextended functor in Eq.~\eqref{eq:unextended_invertible_TQFT}. The passage from invertible functors to spectra extends to this fully
extended setting. Restricting the bordism and target categories to
their invertible objects and morphisms produces Picard
$\infty$-groupoids. By Proposition~\ref{prop:Picard_groupoids_spectra}, these Picard
$\infty$-groupoids determine connective spectra. Consequently, a fully
extended invertible TQFT is encoded by a map between the corresponding
spectra; see Refs.~\cite[Sec.~1.1]{SchommerPries2017} and~\cite[Sec.~5.2]{FreedHopkins}.

\bigskip

\begin{proposition}[The spectrum of the bordism category]
\label{prop:bordism_category_MTO}
The classifying space of the fully extended unoriented bordism
$(\infty,n)$-category is naturally equivalent, as an infinite-loop
space, to
\begin{equation}
    \bigl\lVert{\bf Bord}_{n}^{\mathrm O}\bigr\rVert
    \simeq
    \Omega^\infty\bigl(\Sigma^nMTO_n\bigr)
    =
    \Omega^{\infty-n}MTO_n.
    \label{eq:bordism_category_MTO}
\end{equation}
Consequently, the connective spectrum associated to
${\bf Bord}_{n}^{\mathrm O}$ is $\Sigma^nMTO_n$.
\end{proposition}

\begin{proof}
This is the unoriented case of the identification of extended bordism categories with Madsen-Tillmann spectra; see Ref.~\cite{SchommerPries2017}. For the unextended bordism category, see Ref.~\cite{MTW2010}.
\end{proof}

\noindent Proposition~\ref{prop:bordism_category_MTO} explains the source in the
Freed-Hopkins classification. After passing from the invertible TQFT
functor to the associated map of spectra, the source bordism category
is represented by $\Sigma^nMTO_n$. Geometrically, the Thom spectrum
arises from the Pontryagin-Thom construction applied to the stable
normal bundles of unoriented $n$-manifolds.

\subsubsection{Madsen-Tillmann spectra and stablilization}
\label{sec:cob_MTO_stability}

We now describe the Madsen-Tillmann spectra appearing in Proposition~\ref{prop:bordism_category_MTO} and explain how they stabilize to the unoriented Thom spectrum $\mathrm{MO}$. Let
\begin{equation*}
    V_n\longrightarrow B\mathrm O(n)
\end{equation*}
be the tautological rank-$n$ bundle, which is the bundle denoted $\gamma_n$ in Sec.~\ref{app:cob_spectrum_description}. The Madsen-Tillmann spectrum is the Thom spectrum of the virtual bundle $-V_n$, 
\begin{equation} MTO_n := \operatorname{Th}(-V_n), \qquad \Sigma^nMTO_n = \operatorname{Th}(\mathbb R^n-V_n). \label{eq:MTO_definition} \end{equation} 
The appearance of the negative tautological bundle reflects the fact that the Pontryagin-Thom construction is naturally expressed in terms of the normal-bundle rather than the tangent-bundle.\footnote{If an $n$-manifold $M$ is embedded in $\mathbb R^{n+q}$, then its tangent and normal bundles satisfy $TM\oplus\nu_M\cong\epsilon^{n+q}$, and hence $[\nu_M]=[\epsilon^{n+q}]-[TM]$ as a virtual bundle. Thus the virtual bundle $-V_n$ encodes the stable complement of the universal tangent bundle. More concretely, over $\operatorname{Gr}_n(\mathbb R^{n+q})$, the orthogonal complement of the tautological bundle represents $\epsilon^{n+q}-V_n$, and its Thom space gives the $q$th space of $\Sigma^nMTO_n$.} The inclusions $\mathrm O(n)\hookrightarrow\mathrm O(n+1)$ induce a canonical sequence of spectra 
\begin{equation} 
\Sigma^nMTO_n \longrightarrow \Sigma^{n+1}MTO_{n+1} \longrightarrow \Sigma^{n+2}MTO_{n+2} \longrightarrow\cdots . 
\label{eq:MTO_filtration} 
\end{equation}   
The colimit of this sequence in the category of spectra is the unoriented Thom spectrum: \begin{equation} \mathrm{MO} \simeq \operatorname*{colim}_{m\to\infty} \Sigma^{n+m}MTO_{n+m}. \label{eq:MTO_colimit_MO} \end{equation} See Ref.~\cite[Sec.~7.1]{FreedHopkins}. In particular, there is a canonical map
\begin{equation}
    \iota_n: \Sigma^nMTO_n
    \longrightarrow \mathrm{MO},
    \label{eq:MTO_to_MO}
\end{equation}
which is not generally an equivalence. This map compares the dimension-$n$ bordism spectrum with its stable counterpart and will be used below to define stable invertible theories.

\subsubsection{Universal targets and the bordism classification}
\label{sec:cob_classification_computation}

The passage from the target category, here $\Line \subset \vect$, of an invertible TQFT to a target spectrum is more subtle than the corresponding identification of the source spectrum. The appropriate target depends on whether theories are classified up to isomorphism or only up to continuous deformation\footnote{Two TQFTs are isomorphic if they are related by an invertible symmetric monoidal natural transformation. They are deformation equivalent if they lie in the same path component of the space of theories, that is, if they can be connected by a continuous one-parameter family of TQFTs. Deformation equivalence is therefore a coarser relation than isomorphism.}, and therefore on whether the $\mathbb{C}^{\times}$-valued partition function amplitudes of the theory are treated discretely or with their natural continuous topology. Freed and Hopkins formulate these choices using universal target spectra; see Ref.~\cite[Sec.~5.3]{FreedHopkins} for the detailed construction. To classify topological theories up to isomorphism, they regard $\mathbb C^\times$ as a discrete abelian group and use the Brown-Comenetz dual $I\mathbb C^\times$. To classify deformation classes, theories connected by a continuous family are identified, and the corresponding construction uses the shifted Anderson dual spectrum, denoted $\Sigma I\mathbb{Z}$. We first review the Brown-Comenetz construction.

\bigskip 

\begin{definition}[Brown-Comenetz duality {\cite{BrownComenetz1976}}]
\label{def:Brown_Comenetz_duality}
Let $A$ be an injective abelian group. Brown-Comenetz duality is the
contravariant functor
\begin{equation}
\begin{aligned}
    I_A:
    \mathbf{Sp}^{\mathrm{op}}
    &\longrightarrow
    \mathbf{Sp},\\
    E
    &\longmapsto
    I_A(E),
\end{aligned}
\label{eq:Brown_Comenetz_functor}
\end{equation}
characterized by the natural isomorphisms
\begin{equation}
    \pi_{-n}\bigl(I_A(E)\bigr)
    \cong
    \operatorname{Hom}_{\mathrm{Ab}}
    \bigl(\pi_n(E),A\bigr)
    \label{eq:Brown_Comenetz_defining_property}
\end{equation}
for every spectrum $E$ and every integer $n$.
\end{definition}

\noindent The Brown-Comenetz dual of the sphere spectrum,
\begin{equation*}
    I_A:=I_A(\mathbb S),
\end{equation*}
is called the Brown-Comenetz dualizing spectrum. It represents the
degree-zero character functor in the sense that
\begin{equation}
    [E, I_{A}]=\operatorname{Hom}_{\mathrm{sp}}(E,I_A)
    \cong
    \operatorname{Hom}_{\mathrm{Ab}}
    \bigl(\pi_0(E),A\bigr).
    \label{eq:Brown_Comenetz_representing_property}
\end{equation}
The Brown-Comenetz dual of an arbitrary spectrum is recovered from
this single dualizing spectrum by the function spectrum
\begin{equation}
    I_A(E)
    \simeq
    F(E,I_A).
    \label{eq:Brown_Comenetz_function_spectrum}
\end{equation}
Equivalently, suspension gives
\begin{equation}
    \operatorname{Hom}_{\mathrm{sp}}
    \bigl(E,\Sigma^nI_A\bigr)
    \cong
    \operatorname{Hom}_{\mathrm{Ab}}
    \bigl(\pi_n(E),A\bigr).
    \label{eq:Brown_Comenetz_shifted_property}
\end{equation}
This property is also called the defining property of the Brown-Comenetz dual.

\bigskip 

\noindent Taking $A=\mathbb C^\times$, regarded as a discrete abelian group,
gives the dualizing spectrum $I\mathbb C^\times$. Freed and Hopkins
refer to this as the Pontryagin dual of the sphere spectrum. By
Eq.~\eqref{eq:Brown_Comenetz_shifted_property}, maps
\begin{equation*}
    E\longrightarrow\Sigma^nI\mathbb C^\times
\end{equation*}
are precisely $\mathbb C^\times$-valued characters of $\pi_n(E)$.
When $E$ is a bordism spectrum, these characters assign multiplicative
partition functions to bordism classes. This is the universal target adopted by Freed and Hopkins for
isomorphism classes of invertible topological theories.

\noindent The Freed-Hopkins Ansatz therefore represents an invertible
$n$-dimensional fully extended unoriented theory by a spectrum map
\begin{equation}
    F_n:
    \Sigma^nMTO_n
    \longrightarrow
    \Sigma^nI\mathbb C^\times.
    \label{eq:FH_extended_invertible_theory}
\end{equation}
The space of such theories is the zero space of the corresponding
mapping spectrum, while their isomorphism classes form the abelian
group
\begin{equation}
    \operatorname{Hom}_{\mathrm{sp}}
    \bigl(
        \Sigma^nMTO_n,
        \Sigma^nI\mathbb C^\times
    \bigr).
    \label{eq:FH_extended_isomorphism_classes}
\end{equation}

\bigskip 

The Ansatz in Eq.~\eqref{eq:FH_extended_invertible_theory} describes an invertible theory at a fixed spacetime dimension $n$. Freed and Hopkins impose an additional stability condition requiring such a
theory to be compatible with the canonical enlargement of the tangential data. Under the inclusion $\mathrm O(n)\hookrightarrow\mathrm O(n+1)$, the rank-$n$ tangential bundle is enlarged by a trivial real line,
\begin{equation*}
    V_n\longmapsto V_n\oplus\mathbb R.
\end{equation*}
Since this operation adds no nontrivial information, a stable topological theory should be insensitive to it and depend only on the stable tangent bundle, or equivalently stable normal bundle, structure. This requires the theory to be compatible with the stabilization maps in the sequence
\begin{equation*}
    \Sigma^nMTO_n
    \longrightarrow
    \Sigma^{n+1}MTO_{n+1}
    \longrightarrow\cdots .
\end{equation*}
Following Ref.~\cite[Def.~7.16]{FreedHopkins}, this compatibility is expressed by requiring the fixed-dimensional theory to be the restriction of a theory defined on the colimit spectrum $\mathrm{MO}$. Explicitly, the map in Eq.~\eqref{eq:FH_extended_invertible_theory} must extend along the canonical map in Eq.~\eqref{eq:MTO_to_MO}; that is, there must exist
\begin{equation}
    \widetilde F:
    \mathrm{MO}
    \longrightarrow
    \Sigma^nI\mathbb C^\times,
    \qquad
    F_n=\widetilde F\circ\iota_n.
    \label{eq:stable_theory_extension}
\end{equation}
Thus the isomorphism classes of stable fully extended unoriented
theories form the group
\begin{equation}
    [MO, \Sigma^nI\mathbb C^\times]=\operatorname{Hom}_{\mathrm{sp}}
    \bigl(
        \mathrm{MO},
        \Sigma^nI\mathbb C^\times
    \bigr).
    \label{eq:FH_stable_unoriented_classification}
\end{equation}
For deformation classes, Freed and Hopkins instead use the shifted
Anderson dual, which we discuss below.

\bigskip 

We now show that the Freed-Hopkins classification group for stable fully extended unoriented theories agrees with the bordism-character group used by Kapustin and in this manuscript. Applying Eq.~\ref{eq:Brown_Comenetz_shifted_property} to $E=\mathrm{MO}$ gives
\begin{equation}
\begin{aligned}
    \left[
        \mathrm{MO},
        \Sigma^nI\mathbb C^\times
    \right]
    &\cong
    \operatorname{Hom}
    \bigl(\pi_n(\mathrm{MO}),\mathbb C^\times\bigr)\\
    &\cong
    \operatorname{Hom}
    \bigl(\Omega_n^{\mathrm O},\mathbb C^\times\bigr),
\end{aligned}
\label{eq:FH_to_bordism_characters}
\end{equation}
where the second isomorphism follows from the Pontryagin-Thom
isomorphism in Eq.~\eqref{eq:PT_point_isomorphism}.

\bigskip 

Since $\Omega_n^{\mathrm O}$ has exponent $2$, it is naturally an $\Z_2$-vector space, and every character has image in the subgroup $\{\pm1\}\subset\mathbb C^\times$. Identifying $\{\pm1\}\cong\mathbb Z_2$, we obtain
\begin{equation}
\begin{aligned}
    \left[
        \mathrm{MO},
        \Sigma^nI\mathbb C^\times
    \right]
    &\cong
    \operatorname{Hom}
    \bigl(\Omega_n^{\mathrm O},\mathrm U(1)\bigr)\\
    &=\operatorname{Hom}_{\Z_{2}}
    \bigl(\Omega_n^{\mathrm O},\Z_{2}\bigr)\cong
    \widehat{\Omega}_{\mathrm O}^{\,n}.
\end{aligned}
\label{eq:FH_Kapustin_bordism_classification}
\end{equation}
Thus the Freed-Hopkins classification group for stable fully extended unoriented theories agrees with the bordism-character group used by Kapustin~\cite{Kapustin2014}. It is also the $G=\{1\}$ specialization of the classification model in Eq.~\eqref{eq:cob_model}. The extension to nontrivial internal symmetry $G$, which replaces $\mathrm{MO}$ by the corresponding $G$-equivariant bordism spectrum, is discussed in Sec.~\ref{app:cob2}.

\bigskip 

\begin{remark}[Relation between the universal targets.]
The passage from isomorphism classes to deformation classes is governed by the exponential sequence of abelian groups
\begin{equation}
    0 \longrightarrow
    \mathbb Z(1)
    \longrightarrow
    \mathbb C
    \xrightarrow{\exp}
    \mathbb C^\times
    \longrightarrow
    1,
    \label{eq:exponential_sequence_coefficients}
\end{equation}
where $\mathbb Z(1)=2\pi i\mathbb Z$. This sequence induces a fiber
sequence of spectra
\begin{equation}
    H\mathbb C
    \xrightarrow{\exp}
    I\mathbb C^\times
    \longrightarrow
    \Sigma I\mathbb Z(1),
    \label{eq:BC_Anderson_fiber_sequence}
\end{equation}
where $H\mathbb C$ is the Eilenberg-MacLane spectrum of the additive
group $\mathbb C$. After suspending and mapping out of a spectrum $E$, the fiber sequence
induces
\begin{equation}
    [E,\Sigma^nH\mathbb C\bigr]
    \longrightarrow
    [E,\Sigma^nI\mathbb C^\times]
    \longrightarrow [E,\Sigma^{n+1}I\mathbb Z(1)].
\label{eq:BC_to_Anderson_map}
\end{equation}
Two isomorphism classes have the same image under the second map
precisely when their difference is induced by a
$\mathbb C$-valued class through the exponential map. This implements
the passage from the discrete topology on $\mathbb C^\times$ to its
usual topology: such differences can be connected through a continuous
family and are therefore identified under deformation. Consequently,
the shifted Anderson dual records deformation classes rather than
individual isomorphism classes; see Ref.~\cite[Sec.~5.3]{FreedHopkins}.
\end{remark}

\bigskip 

\noindent \textit{Deformation classes of TQFTs.} Applying the preceding construction to stable unoriented theories, Freed and Hopkins identify their deformation classes with the torsion subgroup
\begin{equation}
    \left[
        \mathrm{MO},
        \Sigma^{n+1}I\mathbb Z(1)
    \right]_{\mathrm{tor}}.
    \label{eq:FH_deformation_classification}
\end{equation}
The Anderson-dual universal coefficient sequence 
\begin{align}
\nonumber
0\longrightarrow
\operatorname{Ext}^{1}
\bigl(\pi_n(\mathrm{MO}),\mathbb Z(1)\bigr)
\longrightarrow
&\left[
    \mathrm{MO},
    \Sigma^{n+1}I\mathbb Z(1)
\right]
\\
\longrightarrow&\operatorname{Hom}
\bigl(\pi_{n+1}(\mathrm{MO}),\mathbb Z(1)\bigr)
\longrightarrow 0.
\label{eq:Anderson_dual_UCT_MO}
\end{align}
Since $\pi_{n+1}(\mathrm{MO})\cong\Omega_{n+1}^{\mathrm O}$ is finite,
the rightmost term in Eq.~\eqref{eq:Anderson_dual_UCT_MO} vanishes.
Consequently, the Anderson-dual group is itself torsion in the
unoriented case, and
\begin{equation}
\begin{aligned}
    \left[
        \mathrm{MO},
        \Sigma^{n+1}I\mathbb Z(1)
    \right]_{\mathrm{tor}}
    &\cong
    \operatorname{Ext}^{1}
    \bigl(\pi_n(\mathrm{MO}),\mathbb Z(1)\bigr)\\
    &\cong
    \operatorname{Hom}
    \bigl(\pi_n(\mathrm{MO}),\mathbb Q/\mathbb Z(1)\bigr).
\end{aligned}
\label{eq:Anderson_dual_torsion_computation}
\end{equation}
Using
$\pi_n(\mathrm{MO})\cong\Omega_n^{\mathrm O}$ and the fact that
$\Omega_n^{\mathrm O}$ is a finite group of exponent $2$, we obtain
\begin{equation}
\begin{aligned}
    \left[
        \mathrm{MO},
        \Sigma^{n+1}I\mathbb Z(1)
    \right]_{\mathrm{tor}}
    &\cong
    \operatorname{Hom}
    \bigl(\Omega_n^{\mathrm O},\mathrm U(1)\bigr)\\
    &=\operatorname{Hom}_{\Z_{2}}
    \bigl(\Omega_n^{\mathrm O},\Z_{2}\bigr)\cong
    \widehat{\Omega}_{\mathrm O}^{\,n}.
\end{aligned}
\label{eq:FH_deformation_Kapustin_comparison}
\end{equation}
Therefore, in the unoriented case, the Freed-Hopkins classifications of isomorphism classes and deformation classes both reduce to the classification $\widehat{\Omega}_{\mathrm O}^{\,n} = \operatorname{Hom}_{\Z_{2}} \bigl(\Omega_n^{\mathrm O},\Z_{2}\bigr)$. This is the $G=\{1\}$ specialization of the classification model in Eq.~\eqref{eq:cob_model}. The extension to nontrivial internal symmetry is discussed in Sec.~\ref{app:cob2}.

\subsection{Equivariant cobordism theory based classification}
\label{app:cob2}

In the previous section, we reviewed the stable unoriented classification of invertible TQFTs. We now incorporate a spatial action of a compact Lie group $G$ on the spatial manifold. To formulate an equivariant classification, the spatial manifold must appear explicitly as an input carrying a $G$-action. Since the case of primary interest is Euclidean space $\mathbb R^d$, which is noncompact, Freed and Hopkins use generalized Borel-Moore homology to define the classification on locally compact spatial manifolds.

\bigskip 

\noindent Atiyah duality then rewrites this homological classification as
$E$-cohomology twisted by the tangent bundle, a form that admits a
Borel-equivariant refinement. For a linear action of $G$ on
$V\cong\mathbb R^d$, the equivariant classification initially contains
a twist by the associated representation bundle over $BG$.\footnote{Here the twist means that the source is the Thom spectrum $\operatorname{Th}(BG,\varepsilon^d-V_\lambda)$, rather than the ordinary suspension spectrum $BG_+$. The virtual bundle $\varepsilon^d-V_\lambda$ records the action of $G$ on the spatial directions. An $\mathrm{MO}$-orientation supplies a Thom isomorphism which, after smashing with $\mathrm{MO}$, identifies this twisted source with $BG_+\wedge\mathrm{MO}$; see Sec.~\ref{app:cob_B5b}.} We show
that the universal $\mathrm{MO}$-orientation removes this twist,
reducing the source spectrum to $BG_+\wedge\mathrm{MO}$. Finally, we
identify the resulting $G$-equivariant classification group with the
bordism-character group $\Omega_{\mathrm O}^{d+1}(BG)$ used in
Eq.~\eqref{eq:cob_model}.

\subsubsection{Road map to equivariantizing classification}

We first package the nonequivariant classification into a generalized
cohomology theory. Let
\begin{equation}
    E := F\bigl(\mathrm{MO}, \Sigma^2I\mathbb Z(1)\bigr)
    \label{eq:unoriented_phase_spectrum}
\end{equation}
be the spectrum of stable unoriented invertible phases. Its homology
and cohomology coefficient groups at a point satisfy
\begin{equation}
\begin{aligned}
    E_{-d}(\mathrm{pt})
    &=\pi_{-d}
    \bigl(
        E\wedge\Sigma^\infty\mathrm{pt}_+
    \bigr)\\
    &= \bigl[
        \mathbb S,
        \Sigma^dE
    \bigr]= E^d(\mathrm{pt})\\
    &\cong \bigl[
        \mathrm{MO},
        \Sigma^{d+2}I\mathbb Z(1)
    \bigr].
\end{aligned}
\label{eq:unoriented_phase_coefficients}
\end{equation}
For spacetime dimension $n=d+1$, the final group in
Eq.~\eqref{eq:unoriented_phase_coefficients} is precisely the
Freed-Hopkins deformation classification computed in
Eq.~\eqref{eq:FH_deformation_Kapustin_comparison}.
The equality $E_{-d}(\mathrm{pt})=E^d(\mathrm{pt})$ holds because both sides are
equal to the homotopy group $\pi_{-d}(E)$.
Freed and Hopkins extend the classification from flat Euclidean space
to arbitrary locally compact spatial manifolds using generalized
Borel-Moore homology. If a locally compact space $Y$ is presented as
the complement
\begin{equation*}
    Y=\overline Y\setminus Y_0
\end{equation*}
of a subcomplex $Y_0$ in a compact space $\overline Y$, then
\begin{equation*}
    E_{0,\mathrm{BM}}(Y)
    \cong
    E_0(\overline Y,Y_0).
\end{equation*}
Here, relative $E$-homology(RHS above) provides a concrete model for Borel-Moore $E$-homology(LHS above) once a compactification $Y=\overline Y\setminus Y_0$ has been chosen. Although the classification to be equivariantized is the Borel-Moore homology theory $E_{0,\mathrm{BM}}(Y)$, the subsequent manipulations are carried out using its relative-homology presentation $E_0(\overline Y,Y_0)$. This allows us to apply Spanier-Whitehead and Atiyah duality and thereby rewrite the classification in a form suitable for the Borel-equivariant construction. For Euclidean space, we take
$\overline Y=S^d$ and $Y_0=\{\infty\}$. Hence
\begin{equation}
\begin{aligned}
    E_{0,\mathrm{BM}}(\mathbb R^d)
    &\cong
    E_0(S^d,\{\infty\})\\
    &=
    \pi_0(E\wedge S^d)\\
    &=
    \pi_{-d}(E)
    =
    E_{-d}(\mathrm{pt}).
\end{aligned}
\label{eq:BM_Euclidean_classification}
\end{equation}
Thus the Borel-Moore formulation reproduces the original
classification of phases on $\mathbb R^d$ while extending the
classification to more general spatial manifolds; see
Ref.~\cite[Eq.~(2.2)]{Freed2021}.

\bigskip 

To prepare for the equivariant construction, we next rewrite relative $E$-homology, $E_{0}(Y, \partial Y)$, as $E$-cohomology twisted by the tangent bundle of $Y$.

\begin{remark}[Spanier-Whitehead and Atiyah duality]
\label{rem:Spanier_Whitehead_Atiyah_duality}
For a dualizable spectrum $A$, its Spanier-Whitehead dual is the
function spectrum
\begin{equation*}
    DA
    :=
    F(A,\mathbb S).
\end{equation*}
There is then a natural adjunction
\begin{equation}
    [\mathbb S,
        E\wedge A]
    \cong
    [DA, E].
    \label{eq:Spanier_Whitehead_duality_adjunction}
\end{equation}
For a compact smooth manifold $Y$ with boundary, Atiyah duality
identifies the Spanier-Whitehead dual of its relative suspension
spectrum with the Thom spectrum of the negative tangent bundle:
\begin{equation}
    D\bigl(\Sigma^\infty(Y/\partial Y)\bigr)
    \simeq
    \operatorname{Th}(Y,-TY).
    \label{eq:Atiyah_duality_relative}
\end{equation}
See Ref.~\cite{AtiyahThomComplexes} and
Ref.~\cite[Remark~2.6]{Freed2021}.
\end{remark}

\bigskip 

\noindent For a compact smooth $d$-manifold $Y$ with boundary, its relative
$E$-homology is
\begin{equation*}
    E_0(Y,\partial Y)
    =
    [\mathbb S, E\wedge\Sigma^\infty(Y/\partial Y)].
\end{equation*}
Applying Spanier-Whitehead duality and then Atiyah duality gives
\begin{equation}
\begin{aligned}
    E_0(Y,\partial Y)
    &\cong [\operatorname{Th}(Y,-TY),
        E]\\
    &\cong [\Sigma^{\mathbb R^d-TY}(Y),
        \Sigma^dE].
\end{aligned}
\label{eq:relative_homology_twisted_cohomology}
\end{equation}
Here $\Sigma^{\mathbb R^d-TY}(Y)$ is the Thom spectrum of the
virtual rank-zero bundle $\varepsilon^d-TY\to Y$.

\noindent Thus Atiyah duality rewrites relative $E$-homology as
$E$-cohomology twisted by the tangent bundle of the spatial manifold.
This is the form that admits the Borel-equivariant refinement used
below. We now state the corresponding equivariant classification from
Ref.~\cite{Freed2021}.

\bigskip

\begin{ansatz}[Freed-Hopkins ansatz~{\cite[Ansatz~3.3]{Freed2021}}]
\label{ansatz:equivariant_BM_classification}
Let $Y$ be a locally compact spatial manifold equipped with an action
of a compact Lie group $G$. Suppose that $Y$ admits a
$G$-equivariant compactification
\begin{equation*}
    Y=\overline Y\setminus Y_0,
\end{equation*}
where $\overline Y$ is compact and $Y_0\subset\overline Y$ is a
$G$-invariant subspace. In the unoriented bosonic case, the
classification of invertible phases on $Y$ is the Borel-Moore
$G$-equivariant homology group
\begin{equation}
\begin{aligned}
    E_{0,\mathrm{BM}}^{hG}(Y)
    &\cong
    E_0^{hG}(\overline Y,Y_0)\\
    &:=
    \left[
        \mathbb S,
        E\wedge
        \Sigma^\infty(\overline Y/Y_0)
    \right]^{hG},
\end{aligned}
\label{eq:equivariant_BM_classification}
\end{equation}
where $E$ is the spectrum defined in
Eq.~\eqref{eq:unoriented_phase_spectrum}.\footnote{If $Y$ is compact,
then Borel-Moore equivariant homology reduces to ordinary
equivariant homology:
\begin{equation*}
    E_{0,\mathrm{BM}}^{hG}(Y)
    \cong
    E_0^{hG}(Y)
    =
    \left[
        \mathbb S,
        E\wedge\Sigma^\infty Y_+
    \right]^{hG}.
\end{equation*}
Thus the formula involving $Y_+$ is the compact special case of the
general Borel-Moore classification. See
Ref.~\cite[Eq.~(3.2) and Ansatz~3.3]{Freed2021}; the corresponding
non-equivariant statement appears immediately after
Eq.~(2.2) therein.}
\end{ansatz}

\bigskip

\noindent We now apply Ansatz~\ref{ansatz:equivariant_BM_classification} to flat Euclidean
space with a linear spatial action of $G$. Let
$V\cong\mathbb R^d$ be the real representation defined by $\lambda:G\longrightarrow\mathrm O(V)$.
Its one-point compactification is the representation sphere $S^V$,
whose distinguished point $\infty$ is fixed by $G$. Thus
\begin{equation*}
    V
    =
    S^V\setminus\{\infty\}
\end{equation*}
is a $G$-equivariant compactification of the form appearing in
Ansatz~\ref{ansatz:equivariant_BM_classification}. Therefore,
\begin{equation}
\begin{aligned}
    E_{0,\mathrm{BM}}^{hG}(V)
    &\cong
    E_0^{hG}(S^V,\{\infty\})\\
    &=
    \left[
        \mathbb S,
        E\wedge S^V
    \right]^{hG}\cong
    \left[
        S^{-V},
        E
    \right]^{hG}.
\end{aligned}
\label{eq:equivariant_Euclidean_BM_classification}
\end{equation}
Here $S^{-V}=D(S^V)$ is the Spanier-Whitehead dual, and hence the
smash-product inverse, of $S^V$ in the stable homotopy category of
$G$-spectra. We use the following property of Borel $G$-equivariant spectra:

\bigskip 

\begin{proposition}[Freed-Hopkins~{\cite[Sec.~4.1]{Freed2021}}]
\label{prop:B4}
Let $M$ be a $G$-spectrum and let $N$ be a spectrum regarded as a
$G$-spectrum with trivial $G$-action. Then
\begin{equation}
    [M,N]^{hG}
    \cong
    [EG_+\wedge_GM,N].
    \label{eq:Borel_maps_orbit_spectrum}
\end{equation}
\end{proposition}

\bigskip

\noindent Applying Proposition~\ref{prop:B4} to
Eq.~\eqref{eq:equivariant_Euclidean_BM_classification} gives
\begin{equation}
\begin{aligned}
    E_{0,\mathrm{BM}}^{hG}(V)
    &\cong
    \left[
        EG_+\wedge_GS^{-V},
        E
    \right]\\
    &\cong
    \left[
        \operatorname{Th}(BG,-V_\lambda)\wedge\mathrm{MO},
        \Sigma^2I\mathbb Z(1)
    \right]\\
    &\cong
    \left[
        \operatorname{Th}
        \bigl(BG,\varepsilon^d-V_\lambda\bigr)
        \wedge\mathrm{MO},
        \Sigma^{d+2}I\mathbb Z(1)
    \right],
\end{aligned}
\label{eq:equivariant_Euclidean_Thom_classification}
\end{equation}
where
\begin{equation*}
    V_\lambda := EG\times_GV
    \longrightarrow BG
\end{equation*}
is the vector bundle associated with the representation $\lambda:G\to\mathrm O(V)$. The second line in~\eqref{eq:equivariant_Euclidean_Thom_classification} uses
\begin{equation*}
    EG_+\wedge_GS^{-V} \simeq \operatorname{Th}(BG,-V_\lambda)
\end{equation*}
together with the function-spectrum adjunction and the definition of $E$. The final line in~\eqref{eq:equivariant_Euclidean_Thom_classification} suspends both the source and target by $d$, so that the spatial action is recorded by the virtual rank-zero bundle $(\varepsilon^d-V_\lambda)\to BG$. This is the bundle twist removed in Sec.~\ref{app:cob_B5b}.

\subsubsection{Simplifications to the equivariant classification}
\label{app:cob_B5b}

In this subsection, we simplify the equivariant classification by
removing the twist associated with the spatial representation bundle $V_\lambda$.
We first recall how a ring spectrum represents a multiplicative
generalized cohomology theory and how an orientation in such a theory
gives a Thom isomorphism. We then show that every real vector bundle
admits a natural $\mathrm{MO}$-orientation. Finally, we apply the
spectrum-level Thom isomorphism to the virtual rank-zero bundle
$\varepsilon^d-V_\lambda\to BG$, reducing the domain spectrum to
$BG_+\wedge\mathrm{MO}$.

\bigskip

\noindent\textit{Ring spectra and generalized Thom orientations.}
A spectrum $R$ represents the reduced generalized cohomology theory
\begin{equation}
    \widetilde R^{\,k}(X)
    := [\Sigma^\infty X,
        \Sigma^kR]
    \label{eq:ring_spectrum_represented_cohomology}
\end{equation}
on pointed spaces $X$. If $R$ is a ring spectrum, its multiplication
and unit maps
\begin{equation*}
    \mu:
    R\wedge R\longrightarrow R,
    \qquad
    \eta:
    \mathbb S\longrightarrow R
\end{equation*}
induce products and units in the represented cohomology theory. Thus
$R^*(X)$ is a generalized cohomology ring rather than only a graded
abelian group.

Let $\xi\to B$ be a real vector bundle of rank $r$. An
$R$-orientation of $\xi$ is a class
\begin{equation}
    u_{\xi}
    \in
    \widetilde R^{\,r}
    \bigl(\operatorname{Th}(\xi)\bigr)
    \label{eq:R_Thom_class}
\end{equation}
whose restriction to the Thom space of every fiber,
\begin{equation*}
    S^r
    \longrightarrow
    \operatorname{Th}(\xi),
\end{equation*}
is the suspension of the unit of $R$. The class $u_{\xi}$ is called an
$R$-Thom class.

\bigskip 

\begin{proposition}[Generalized Thom isomorphism]
\label{prop:generalized_Thom_isomorphism}
If a rank-$r$ real vector bundle $\xi\to B$ admits an $R$-orientation,
multiplication by its Thom class gives an isomorphism
\begin{equation}
    R^k(B)
    \xrightarrow{\;\cong\;}
    \widetilde R^{\,k+r}
    \bigl(\operatorname{Th}(\xi)\bigr).
    \label{eq:generalized_Thom_isomorphism}
\end{equation}
Equivalently, there is an equivalence of $R$-module spectra
\begin{equation}
    \Sigma^\infty\operatorname{Th}(\xi)\wedge R
    \simeq
    \Sigma^r\Sigma^\infty B_+\wedge R.
    \label{eq:spectrum_level_Thom_isomorphism}
\end{equation}
\end{proposition}

\begin{proof}
The Thom diagonal
\begin{equation*}
    \operatorname{Th}(\xi)
    \longrightarrow
    B_+\wedge\operatorname{Th}(\xi)
\end{equation*}
followed by the Thom class $u_\xi$ defines the corresponding map of
$R$-module spectra. The defining fiberwise condition on $u_\xi$
implies that this map is an equivalence. See
Ref.~\cite[Prop.~2.18]{ABGHR}.
\end{proof}

\bigskip

\begin{proposition}[Universal $\mathrm{MO}$-orientation
{\cite[Chs.~4-5]{Stong_unorCob}}]
\label{prop:universal_MO_orientation}
Every real vector bundle admits a natural $\mathrm{MO}$-orientation.
\end{proposition}

\begin{proof}
Let
\begin{equation*}
    \gamma_r
    \longrightarrow
    B\mathrm O(r)
\end{equation*}
be the universal real vector bundle of rank $r$. By construction, the
$r$th space of the Thom spectrum $\mathrm{MO}$ is
\begin{equation*}
    \mathrm{MO}(r)
    =
    \operatorname{Th}(\gamma_r).
\end{equation*}
The canonical map from this Thom space into the spectrum $\mathrm{MO}$
defines a class
\begin{equation*}
    u_r
    \in
    \widetilde{\mathrm{MO}}^{\,r}
    \bigl(\operatorname{Th}(\gamma_r)\bigr).
\end{equation*}
Its restriction to the Thom space $S^r$ of each fiber of $\gamma_r$ is
the suspension of the unit
$\mathbb S\to\mathrm{MO}$. Hence $u_r$ is an
$\mathrm{MO}$-Thom class for the universal bundle.

Now let $\xi\to B$ be any real vector bundle of rank $r$, classified by
a map
\begin{equation*}
    f_{\xi}:
    B\longrightarrow B\mathrm O(r).
\end{equation*}
The induced map of Thom spaces pulls $u_r$ back to a class
\begin{equation*}
    u_{\xi}
    :=
    f_{\xi}^{*}u_r
    \in
    \widetilde{\mathrm{MO}}^{\,r}
    \bigl(\operatorname{Th}(\xi)\bigr).
\end{equation*}
Since the restriction of $u_{\xi}$ to each fiber is the unit of
$\mathrm{MO}$, this class is an $\mathrm{MO}$-orientation of $\xi$.
\end{proof}

\bigskip 

\begin{claim}
\label{claim:MO_representation_bundle_untwisting}
Let
\begin{equation*}
    V_{\lambda}
    :=
    EG\times_G\mathbb R^d
    \longrightarrow
    BG
\end{equation*}
be the representation bundle induced by
$\lambda:G\to\mathrm O(d)$. Then
\begin{equation}
    \operatorname{Th}
    \bigl(BG,\varepsilon^d-V_{\lambda}\bigr)
    \wedge\mathrm{MO}
    \simeq
    BG_+\wedge\mathrm{MO}.
    \label{eq:MO_representation_bundle_untwisting}
\end{equation}
\end{claim}

\begin{proof}
By Proposition~\ref{prop:universal_MO_orientation}, the real vector
bundle $V_\lambda\to BG$ admits an $\mathrm{MO}$-orientation. The
generalized Thom isomorphism is stable under the addition of trivial
bundles and therefore extends to virtual real vector bundles. For a
virtual bundle $\zeta\to B$ of virtual rank $q$, it gives
\begin{equation*}
    \operatorname{Th}(B,\zeta)\wedge\mathrm{MO}
    \simeq
    \Sigma^qB_+\wedge\mathrm{MO}.
\end{equation*}
The virtual bundle
\begin{equation*}
    \left(\varepsilon^d-V_\lambda\right)
    \longrightarrow
    BG
\end{equation*}
has virtual rank $d-d=0$. Applying the virtual Thom isomorphism
therefore gives
\begin{equation*}
    \operatorname{Th}
    \bigl(BG,\varepsilon^d-V_\lambda\bigr)
    \wedge\mathrm{MO}
    \simeq
    BG_+\wedge\mathrm{MO},
\end{equation*}
as claimed.
\end{proof}

\noindent Claim~\ref{claim:MO_representation_bundle_untwisting} therefore removes
the twist by the spatial representation bundle. The resulting
classification agrees with the internal-symmetry classification, as
expected from the crystalline-equivalence principle; see
Ref.~\cite[Example~3.5]{FreedHopkins}. Consequently,
\begin{equation*}
    \text{cSPT}^{d+1}_{G} = [BG_{+}\wedge MO, \Sigma^{d+2}I\mathbb{Z}].
\end{equation*}

\subsubsection{Recovering the bordism-character classification}

We now express the spectrum-level classification obtained above in
terms of the Borel-equivariant unoriented bordism groups used in
Eq.~\eqref{eq:cob_model}. The Anderson-dual universal coefficient
sequence gives
\begin{equation}
\begin{aligned}
0&\longrightarrow
\operatorname{Ext}^{1}
\bigl(
    \pi_{d+1}(BG_+\wedge\mathrm{MO}),
    \mathbb Z(1)
\bigr)\\
&\longrightarrow
\operatorname{Hom}_{\mathrm{sp}}
\bigl(
    BG_+\wedge\mathrm{MO},
    \Sigma^{d+2}I\mathbb Z(1)
\bigr)
\\
&\longrightarrow
\operatorname{Hom}
\bigl(
    \pi_{d+2}(BG_+\wedge\mathrm{MO}),
    \mathbb Z(1)
\bigr)
\longrightarrow 0.
\end{aligned}
\label{eq:equivariant_Anderson_UCT}
\end{equation}
By the stable splitting of $\mathrm{MO}$ established in Sec.~\ref{app:cob_B5b}, the groups $\pi_*(BG_+\wedge\mathrm{MO})$ are $\mathbb Z_2$-vector spaces.
Therefore, the rightmost term in
Eq.~\eqref{eq:equivariant_Anderson_UCT} vanishes, and
\begin{equation}
\begin{aligned}
\operatorname{Hom}_{\mathrm{sp}}
\bigl(
    BG_+\wedge\mathrm{MO},
    \Sigma^{d+2}I\mathbb Z(1)
\bigr)
\cong
\operatorname{Ext}^{1}
\bigl(
    \pi_{d+1}(BG_+\wedge\mathrm{MO}),
    \mathbb Z(1)
\bigr).
\end{aligned}
\label{eq:equivariant_Anderson_Ext}
\end{equation}
Using the Pontryagin-Thom identification
\begin{equation*}
    \pi_{d+1}(BG_+\wedge\mathrm{MO})
    \cong
    \Omega_{d+1}^{\mathrm O}(BG)
\end{equation*}
and the coefficient sequence
\begin{equation*}
    0
    \longrightarrow
    \mathbb Z(1)
    \longrightarrow
    \mathbb Q(1)
    \longrightarrow
    \mathbb Q/\mathbb Z(1)
    \longrightarrow 0,
\end{equation*}
we obtain the canonical identification
\begin{equation*}
    \operatorname{Ext}^{1}
    \bigl(
        \Omega_{d+1}^{\mathrm O}(BG),
        \mathbb Z(1)
    \bigr)
    \cong
    \operatorname{Hom}
    \bigl(
        \Omega_{d+1}^{\mathrm O}(BG),
        \mathbb Q/\mathbb Z(1)
    \bigr).
\end{equation*}
Since $\Omega_{d+1}^{\mathrm O}(BG)$ has exponent $2$, every $\mathrm U(1)$-valued character has image in
$\{\pm1\}\subset\mathbb Q/\mathbb Z(1)$. Hence
\begin{equation}
\begin{aligned}
\operatorname{Hom}_{\mathrm{sp}}
\bigl(
    BG_+\wedge\mathrm{MO},
    \Sigma^{d+2}I\mathbb Z(1)
\bigr)
&\cong
\operatorname{Hom}
\bigl(
    \Omega_{d+1}^{\mathrm O}(BG),
    \mathbb Q/\mathbb Z(1)
\bigr)
\\
&\cong
\operatorname{Hom}
\bigl(
    \Omega_{d+1}^{\mathrm O}(BG),
    \mathrm U(1)
\bigr)
\\
&=
\Omega_{\mathrm O}^{d+1}(BG).
\end{aligned}
\label{eq:equivariant_bordism_character_classification}
\end{equation}
Thus the Borel-equivariant Freed-Hopkins construction reproduces the
bordism-character classification used in Eq.~\eqref{eq:cob_model},
which in the unoriented case is entirely $2$-torsion.

\section{Group cohomology classification I}
\label{app:grpcoh}

\subsection{Group cohomology: algebraic and topological definitions}
\label{app:grpcoh_def}

In the main text we frequently use the group cohomology classification model
\begin{equation}
    \tcspt = H^{d+1}(G,U(1))
    \;\simeq\; H^{d+2}(BG,\mathbb{Z}),
\end{equation}
where $BG$ is the classifying space of $G$, and the second description emphasizes that all symmetry backgrounds are encoded by classifying maps $f:M^{d+1}\to BG$. In this appendix we briefly recall the algebraic definition of group cohomology with coefficients $A$, denoted $H^\ast(G,A)$, and its equivalence with the topological definition via $BG$. We will then generalize the definitions with twisted coefficients in App.~\ref{app:twisted_coefficients}.

\subsubsection{Algebraic definition}

Let $G$ be a (discrete) group and let $R$ be a commutative ring (in this paper the main cases are $R=\mathbb{Z}$, $R=\mathbb{Z}_2$, or $R=\mathbb{R}/\mathbb{Z}$). Write $RG$ for the group ring of $G$ over $R$. An $R$-module equipped with an $R$-linear action of $G$ is called an $RG$-module, and we will use the terminology $G$-module (over $R$) to mean a left $RG$-module. Let $A$ be an $G$-module, then the module is called trivial if $g\cdot a=a$ for all $g\in G$ and $a\in A$.

\bigskip 

Maps between $G$-modules which respect the $G$-linear action, or equivalently maps that are $G$-equivariant, are called $G$-module maps. Given two $G$-modules $A,B$ we can define the $G$-module $\mathrm{Hom}_{R}(A,B)$ via the action
\begin{equation*}
    (g\cdot f)(a) = g\cdot f(g^{-1}\cdot a),\quad \forall \; f\in \mathrm{Hom}_{R}(A,B),\;\; a\in A, \;g\in G.
\end{equation*}

\noindent We can also define a $G$-chain complex as a chain complex of $G$-modules
\begin{equation*}
    \cdots \;X_{n} \xrightarrow[]{\partial_{n}} X_{n-1} \xrightarrow[]{\partial_{n-1}} \cdots \xrightarrow[]{\partial_{2}} X_{1} \xrightarrow[]{\partial_{1}} X_{0} \to 0
\end{equation*}
in which the differentials are $G$-module maps. Recall that the homology of a chain complex is simply defined at each node as the quotient module:
\begin{equation}
    H_{n}(X) = \dfrac{\mathrm{ker}(\partial_{n})}{\mathrm{Im}(\partial_{n+1})}
    \label{eq:homology}
\end{equation}
If the $G$-complex $X$ is supplied with an augmentation map $\epsilon: X_{0}\to A$, then we define the $G$-complex 
\begin{equation*}
    \cdots \;X_{n} \xrightarrow[]{\partial} X_{n-1} \xrightarrow[]{\partial} \cdots \xrightarrow[]{\partial} X_{1} \xrightarrow[]{\partial} X_{0} \to A
\end{equation*}
called the $G$-resolution of $A$ provided the complex is exact. If each $X_{i}$ is projective (or free) as an $RG$-module, then this augmentation is called a projective (or free) resolution. 

\bigskip 

\begin{definition}[Algebraic group cohomology]
\label{def:grpcoh_alg}
The group cohomology of $G$ with coefficients in a $G$-module $A$ is
\begin{equation}
    H^n(G,A)\;:=\;\mathrm{Ext}^n_{RG}(R,A),
\end{equation}
where $R$ is regarded as a trivial $RG$-module. Concretely, one computes this as follows: choose a projective (or free) resolution of $R$ by $RG$-modules,
\begin{equation}
    \cdots \longrightarrow P_2 \longrightarrow P_1 \longrightarrow P_0 \longrightarrow R \longrightarrow 0,
\end{equation}
apply $\mathrm{Hom}_{RG}(-,A)$ to obtain a cochain complex
\begin{equation}
\label{eq:hom_cochain}
    0 \longrightarrow \mathrm{Hom}_{RG}(P_0,A) \longrightarrow \mathrm{Hom}_{RG}(P_1,A)  \longrightarrow \cdots,
\end{equation}
and define $H^n(G,A)$ to be the cohomology of that complex. 
\end{definition}

\bigskip

\begin{remark}
    The $n$th cohomology group of the ``derived Hom'' cochain complex in Eq.~\eqref{eq:hom_cochain} is, by definition, $\mathrm{Ext}^{n}_{RG}(R,A)$.
\end{remark}

\bigskip 

\noindent Although the construction above appears to depend on the choice of projective resolution, the resulting cohomology groups are canonically independent of this choice. In particular, $H^\ast(G,A)$ is well-defined and functorial: group homomorphisms $\varphi:H\to G$ induce natural maps $H^\ast(G,A)\to H^\ast(H,A)$ by pullback (restriction). See \cite{evens1991cohomology}. These are the same restriction maps that appear in the main text when we restrict a symmetry background (or defect data) from $G$ to a subgroup $H$.

\bigskip 

Finally, to make this more explicit, we use the standard representation of cocycles in which an $n$-cochain with coefficients in a $G$-module $A$ is simply a function
\begin{equation*}
    x:G^{n}\to A.
\end{equation*}
If we use additive notation for the coefficient module $A$, the coboundary of such a cochain is
\begin{equation}
\label{eq:bar_differential}
\begin{aligned}
(\delta x)(g_{1},\dots,g_{n+1})
&=
g_{1}\cdot x(g_{2},\dots,g_{n+1}) \\
&\quad + \sum_{i=1}^{n}(-1)^{i}\,
x(g_{1},\dots,g_{i}g_{i+1},\dots,g_{n+1}) \\
&\quad +(-1)^{n+1}x(g_{1},\dots,g_{n}).
\end{aligned}
\end{equation}
An $n$-cocycle is a function $x:G^{n}\to A$ satisfying $\delta x=0$, and two cocycles differing by a coboundary define the same cohomology class in $H^{n}(G,A)$. In the untwisted case $A=\U(1)$ with trivial $G$-action, this is the familiar cocycle description used in Dijkgraaf-Witten theory. More generally, for twisted coefficients the same formula applies, but now the first term involves the nontrivial $G$-action on $A$. This is the concrete model underlying the schematic Dijkgraaf-Witten partition function in Eq.~\eqref{eq:DW}.

\subsubsection{Topological definition and local coefficients}

We briefly recall the topological definition of cohomology in a form that makes contact with group cohomology. Throughout we assume that the reader is comfortable with simplicial structures and the idea that a (co)homology theory is computed as the cohomology of a differential graded cochain complex.

\bigskip 

\noindent \textit{Ordinary cohomology}:
Let $X$ be a space (for us, typically a CW complex or simplicial complex) and let $A$ be an abelian group. The ordinary cohomology groups $H^n(X;A)$ may be computed from the cochain complex $C^\ast(X;A):=\mathrm{Hom}(C_\ast(X),A)$, where $C_\ast(X)$ is the singular (or simplicial) chain complex:
\begin{equation*}
    \cdots \;C_{n}(X) \xrightarrow[]{\partial_{n}} C_{n-1}(X) \xrightarrow[]{\partial_{n-1}} \cdots \xrightarrow[]{\partial_{2}} C_{1}(X) \xrightarrow[]{\partial_{1}} C_{0}(X) \to 0
\end{equation*}
Here, each element $C_{n}(X)$ is the $\mathbb{Z}$-linear group of $n$-dimensional chains in $X$, and the maps $\partial_{n}$ are boundary maps. Concretely, an $n$-cochain assigns an element of $A$ to each $n$-chain $\sigma_{n}\in C_{n}(X)$, and the coboundary map $\delta:C^n(X;A)\to C^{n+1}(X;A)$ is the dual to the boundary map on chains.

\bigskip

\noindent \textit{Cohomology with local coefficients}:
In many applications the coefficient group is not globally constant over the base space. Instead, one attaches a copy of an abelian group $A$ locally over points in local patches of $X$, and parallel transport around a nontrivial loop may act on this copy by an automorphism of $A$.
This is encoded by a \emph{local system} $\underline{A}$ on $X$. Concretely, if $X$ is path-connected and $\widetilde{X}\to X$ is the universal cover, then specifying a local system with fiber $A$ is equivalent to specifying a representation
\begin{equation}
    \rho:\pi_1(X,x_0)\to \mathrm{Aut}(A),
    \label{eq:local_sys_rho}
\end{equation}
i.e.\ an action of $\pi_1(X)$ on $A$.

\medskip 

One then defines cochains with local coefficients using $\pi_1(X)$-equivariant cochains on $\widetilde{X}$. For example, if $C_\ast(\widetilde{X})$ denotes the simplicial (or singular) chain
complex of the universal cover, then the cochain complex with local coefficients is
\begin{equation}
    C^n(X;\underline{A})
    \;:=\;
    \mathrm{Hom}_{\mathbb{Z}[\pi_1(X)]}\big(C_n(\widetilde{X}),A\big),
\end{equation}
where $\mathbb{Z}[\pi_1(X)]$ acts on $C_n(\widetilde{X})$ by deck transformations, and it acts on $A$ through the representation $\rho$. The coboundary map is dual to the boundary on chains,
and the resulting cohomology groups are by definition
\begin{equation}
    H^n(X;\underline{A}) \;:=\; H^n\big(C^\ast(X;\underline{A})\big).
\end{equation}
When the $\pi_1(X)$-action on $A$ is trivial, this reduces to ordinary cohomology $H^n(X;A)$. With this in hand, we can realize group cohomology topologically as follows:

\bigskip 

\begin{definition}[Topological group cohomology]
    Let $G$ be a discrete group and let $BG$ be its classifying space.
A $G$-module $M$ canonically determines a local system $\underline{M}$ on $BG$ (via the identification $\pi_1(BG)\cong G$ and the given action of $G$ on $M$). Then we define the topological group cohomology of $G$ with coefficients $M$ as
\begin{equation}
    H^n(BG;\underline{M}).
\end{equation}
\end{definition}

\subsubsection{Relating the algebraic and topological constructions}
\label{app:grpcoh_relation}

The algebraic and topological definitions of group cohomology are related by a canonical choice of projective resolution coming from a universal cover. 
Consider $X=BG$, so that $X$ is a model for $K(G,1)$ where its universal cover $\widetilde X=EG$ is known to be contractible. We keep the notation $X$ for the moment only to display the standard construction. Let $X$ be a connected CW complex with fundamental group $\pi_1(X)\cong G$, and let $\widetilde{X}\to X$ denote its universal cover.
The group $G$ acts on $\widetilde{X}$ by deck transformations, hence the cellular (or simplicial) chain groups $C_n(\widetilde{X})$ carry natural structures of left $\mathbb{Z}G$-modules. Moreover, each $C_n(\widetilde{X})$ is a free $\mathbb{Z}G$-module: choosing a set of $n$-cells in $X$ and lifting them to $\widetilde{X}$ identifies $C_n(\widetilde{X})$ with a direct sum of copies of $\mathbb{Z}G$, one for each $n$-cell of $X$. The boundary map is $G$-equivariant, so $C_\ast(\widetilde{X})$ is a chain complex of $\mathbb{Z}G$-modules
\begin{equation}
    \cdots \longrightarrow C_2(\widetilde{X}) \longrightarrow C_1(\widetilde{X})
    \longrightarrow C_0(\widetilde{X}) \longrightarrow 0 .
\end{equation} 
There is also an augmentation map $\epsilon:C_0(\widetilde{X})\to \mathbb{Z}$ sending each lifted $0$-cell to $1$, where $\mathbb{Z}$ is regarded as the trivial $\mathbb{Z}G$-module. This yields an augmented complex
\begin{equation}
    \cdots \longrightarrow C_2(\widetilde{X}) \longrightarrow C_1(\widetilde{X})
    \longrightarrow C_0(\widetilde{X}) \xrightarrow{\ \epsilon\ } \mathbb{Z} \longrightarrow 0.
\end{equation}

\bigskip 

Specializing now to $X=BG\simeq K(G,1)$, the universal cover is $\widetilde X=EG$, which is contractible. Hence the above augmented complex is exact. Hence $C_\ast(\widetilde{X})$ is a free $\mathbb{Z}G$-resolution of $\mathbb{Z}$. Applying $\mathrm{Hom}_{\mathbb{Z}G}(-,M)$ produces a cochain complex
\begin{equation}
    0 \longrightarrow \mathrm{Hom}_{\mathbb{Z}G}(C_0(\widetilde{X}),M)
    \longrightarrow \mathrm{Hom}_{\mathbb{Z}G}(C_1(\widetilde{X}),M)
    \longrightarrow \cdots,
\end{equation}
and its cohomology computes $\mathrm{Ext}^\ast_{\mathbb{Z}G}(\mathbb{Z},M)$ by definition. In other words, for $X=K(G,1)$ one obtains the canonical identification
\begin{equation}
    H^n(G,M)=
    \mathrm{Ext}^n_{\mathbb{Z}G}(\mathbb{Z},M)\cong H^n\!\big(\mathrm{Hom}_{\mathbb{Z}G}(C_\ast(\widetilde{X}),M)\big).
\end{equation}
This is precisely the same complex that appears in the topological definition of cohomology with local coefficients, since a $G$-module $M$ defines a local system $\underline{M}$ on $X$ and one may take
\begin{equation}
    C^n(X;\underline{M})
    \;:=\;
    \mathrm{Hom}_{\mathbb{Z}G}(C_n(\widetilde{X}),M).
\end{equation}
Thus, after taking $X=BG$, the algebraic Ext-definition and the topological cohomology of $BG$ with the corresponding local system compute the same group:
\begin{equation*}
    H^n(G,M)\cong H^n(BG;\underline M).
\end{equation*}
They differ only in the choice of resolution used to compute the same derived functor.

\subsection{Twisted coefficients and local systems}
\label{app:twisted_coefficients}

In the main text we frequently use twisted integral coefficients, denoted $\Z_{w_{1}}$. Conceptually, this twist is nothing more than the sign local system determined by orientation reversal. In this subsection we explain (i) how $\Z_{w_{1}}$ is defined on $B\mathrm{O}(d)$, (ii) how it pulls back to any manifold equipped with an $\mathrm{O}(d)$-bundle (in particular the tangent bundle), and (iii) how it restricts to $BG$ for any symmetry group $G$ equipped with a homomorphism $i:G\hookrightarrow \mathrm{O}(d)$.

\bigskip

\noindent \textit{The universal twist on $B\mathrm{O}(d)$.}
Recall that $\mathrm{O}(d)$ has two connected components, hence
\begin{equation*}
    \pi_{1}(B\mathrm{O}(d))\cong \pi_{0}(\mathrm{O}(d))\cong \Z_{2}.
\end{equation*}
There is therefore a canonical homomorphism
\begin{equation}
\label{eq:w1_pi1}
    w_{1}:\pi_{1}(B\mathrm{O}(d))\longrightarrow \{\pm 1\}\cong \Z_{2},
\end{equation}
which records whether a loop in $B\mathrm{O}(d)$ corresponds to an orientation-preserving or orientation-reversing element of $\mathrm{O}(d)$. Composing with the sign action on $\Z$, given by $\{\pm 1\}\hookrightarrow \mathrm{Aut}(\Z),$ defines a representation \begin{equation}
\label{eq:sign_coeff}
w_{1}:\pi_{1}(B\mathrm{O}(d))\to \{\pm 1\} \hookrightarrow \mathrm{Aut}(\Z).
\end{equation} 
By definition, the associated local system of abelian groups on $B\mathrm{O}(d)$ is the twisted integral local system $\Z_{w_{1}}$. 

\bigskip

\noindent \textit{Pullback to manifolds with $\mathrm{O}(d)$-structure.}
Let $P\to M$ be a principal $\mathrm{O}(d)$-bundle with classifying map
\begin{equation}
\label{eq:classifying_map_O}
    f_{P}:M\longrightarrow B\mathrm{O}(d).
\end{equation}
Since cohomology with local coefficients is functorial, pulling back along $f_{P}$ automatically
pulls back the coefficient system. In particular, the universal twisted local system
$\Z_{w_{1}}$ on $B\mathrm{O}(d)$ induces a twisted local system on $M$ defined by
\begin{equation}
\label{eq:pullback_local_system}
    \Z_{w_{1}(P)} \;:=\; f_{P}^{*}\big(\Z_{w_{1}}\big),
\end{equation}
equivalently the twist on $M$ is the pullback class
\begin{equation*}
    w_{1}(P):=f_{P}^{*}(w_{1})\in H^{1}(M,\Z_{2}).
\end{equation*}
When $P=F(TM)$ is the frame bundle of the tangent bundle, this recovers the usual orientation local system $\Z_{w_{1}(TM)}$. In particular, if $M$ is orientable then $w_{1}(TM)=0$ and the system $\Z_{w_{1}(TM)}$ is canonically identified with the constant coefficients $\Z$.

\medskip

\noindent With this convention, a class $\alpha\in H^{n}\big(B\mathrm{O}(d),\Z_{w_{1}}\big)$ pulls back to a class on $M$ valued in the induced local system $\Z_{w_{1}(P)}$:
\begin{equation}
\label{eq:pullback_class}
    f_{P}^{*}(\alpha)\in H^{n}\big(M,\Z_{w_{1}(P)}\big),
\end{equation}
as used throughout the main text (see Sec.~\ref{sec:top_resp}).

\bigskip

\noindent \textit{Restriction maps and coefficient bookkeeping.}
Let $i_{G}:G\hookrightarrow \mathrm{O}(d)$ be a homomorphism and let $Bi_{G}:BG\to B\mathrm{O}(d)$ be the induced map on classifying spaces. Since our coefficients on $B\mathrm{O}(d)$ are twisted by the sign local system $\Z_{w_{1}}$, pulling back a class along $Bi_{G}$ necessarily pulls back the local system as well. Equivalently, the induced twist on $BG$ is
\begin{equation*}
    (Bi_{G})^*(\Z_{w_1})=\Z_{(Bi_{G})^*w_1}, \qquad 
    (Bi_{G})^*w_1=\det\circ i_{G}:G\to\{\pm1\}\cong\Z_2.
\end{equation*}
With this convention, the restriction morphism is simply the pullback map with these coefficients:
\begin{equation}
\label{eq:restriction_twisted}
    \res^{\mathrm{O}(d)}_{G}:=(Bi_{G})^{*}:
    H^{n}\big(B\mathrm{O}(d),\Z_{w_{1}}\big)
    \longrightarrow
    H^{n}\big(BG,\Z_{(Bi_{G})^{*}w_{1}}\big).
\end{equation}

\bigskip

\begin{remark}[Abuse of notation for pulled-back twists]
\label{rem:abuse_pulledback_twists}
Throughout the paper, we use the same symbol $w_1$ for the universal class on $B\mathrm{O}(d)$ and for its various pullbacks. More precisely, if $f:M\to B\mathrm{O}(d)$ is a classifying map, then the local system denoted $\Z_{w_1}$ on $M$ means the pulled-back local system $f^*(\Z_{w_1})$, equivalently the local system associated to the class $f^*w_1\in H^1(M,\Z_2)$.

\bigskip 

\noindent Similarly, if $i_{G}:G\hookrightarrow \mathrm{O}(d)$ is a homomorphism and $Bi_{G}:BG\to B\mathrm{O}(d)$ is the induced map, then in expressions such as $ H^n(BG,\Z_{w_1})$, the coefficient system $\Z_{w_1}$ should be understood as the pulled-back local system
\begin{equation}
(Bi_{G})^*(\Z_{w_1})=\Z_{(Bi_{G})^*w_1}.
\end{equation}
Equivalently, the induced twist on $BG$ is the sign representation $\det\circ i_{G}:G\to\{\pm1\}\cong \Z_2$. The intended meaning of $w_1$ should always be read from the base space appearing in the cohomology group.
\end{remark}

\subsection{Group cohomology classification model}
\label{sec:grpcohmodel_def}

In the main text, we define the group cohomology classification model(see Def.~\ref{def:grpcoh_classmodel}) for $(d+1)$-dimensional systems with protecting symmetry $G$ as
\begin{equation}
    \tcspt := H^{d+1}(G, \U(1)_{w_{1}}) \cong H^{d+2}(BG, \Z_{w_{1}}).
    \label{eq:coh_iden}
\end{equation}
For finite groups, $H^{*}(G,-)$ denotes ordinary group cohomology, as in Def.~\ref{def:grpcoh_alg}, and the identification with cohomology of $BG$ follows from App.~\ref{app:grpcoh_relation}. For compact Lie groups, $H^{*}(G,-)$ denotes Borel group cohomology, discussed below in App.~\ref{app:Borel_compact_Lie}. In this subsection we explain the coefficient twist and the degree shift from $\U(1)$ to integral coefficients.

\subsubsection{Coefficient twists (orientation-reversing symmetries)}
When the symmetry group contains elements that act with a sign on the topological response, one works with twisted coefficients. Concretely, one specifies a homomorphism
\begin{equation}
    w_1:G\to \{\pm 1\}\cong \mathbb{Z}_2,
\end{equation}
and regards $\mathbb{Z}$ as a $G$-module via the sign action
$g\cdot n := (-1)^{w_1(g)}n$. We denote the resulting local coefficient system on $BG$ by
$\mathbb{Z}_{w_1}$. Equivalently, $\mathbb{Z}_{w_1}$ is the local system associated to the
representation $\pi_1(BG)\cong G \to \mathrm{Aut}(\mathbb{Z})\cong \{\pm 1\}$.

\bigskip 

\noindent Now, the same dictionary as above applies:
\begin{equation}
    H^{n}(G,\mathbb{Z}_{w_1}) \;\cong\; H^{n}(BG;\mathbb{Z}_{w_1}).
\end{equation}
This is the natural cohomological language for response terms on unoriented manifolds, and it is the form we use when discussing orientation-reversing crystalline symmetries such as reflections.

\subsubsection{From $U(1)$ coefficients to integral cohomology.}
\label{app:U1_coeff_to_Z}

For finite groups, the classification model in the main text we often write the cohomology group $H^{d+1}(G,U(1)_{w_1})$. Following the standard argument reviewed in Appendix~J of Ref.~\cite{Chen2013}, one starts from the (twisted) exponential short exact sequence of $G$-modules
\begin{equation}
    0 \longrightarrow \mathbb{Z}_{w_1} \longrightarrow \mathbb{R}_{w_1}
    \longrightarrow U(1)_{w_1} \longrightarrow 0,
\end{equation}
which induces a long exact sequence in (Borel) group cohomology, and in particular a connecting homomorphism
\begin{equation}
\label{eq:delta_exp}
    \delta:\;H^{d+1}(G,U(1)_{w_1}) \longrightarrow H^{d+2}(G,\mathbb{Z}_{w_1}).
\end{equation}

\medskip

\noindent If $G$ is a finite group, then $H^{n>0}(G,\mathbb{R}_{w_1})=0$, and therefore the connecting homomorphism $\delta$ is an isomorphism in positive degree. Concretely, one obtains
\begin{equation}
\label{eq:U1_to_Z_finite}
    H^{d+1}(G,U(1)_{w_1}) \cong H^{d+2}(G,\mathbb{Z}_{w_1}),
    \;\; (G\text{ finite},\; d+1>0),
\end{equation}
in complete analogy with Eqs.~(J37)-(J44) of Ref.~\cite{Chen2013}. Combining this with the topological realization of group cohomology from App.~\ref{app:grpcoh_relation} gives
\begin{equation}
    H^{d+1}(G,\U(1)_{w_1}) \cong H^{d+2}(BG,\mathbb{Z}_{w_1}), 
    \qquad (G\text{ finite}).
    \label{eq:G_finite}
\end{equation}

\subsubsection{Compact Lie groups and Borel cohomology.}
\label{app:Borel_compact_Lie}

If $G$ is a compact Lie group, the appropriate replacement for discrete group cohomology is the \emph{Borel} group cohomology, denoted $H^\ast_B(G,-)$, which can be computed from the classifying space $BG$. In particular, Ref.~\cite{Chen2013} recalls the identification
\begin{equation}
    H^n_B(G,\mathbb{Z}_{w_1}) \;\cong\; H^n(BG,\mathbb{Z}_{w_1}),
\end{equation}
valid also when $G$ acts nontrivially on the coefficient modules. Together with the same exponential-sequence argument as above, this gives the canonical degree shift
\begin{equation}
    H^{d+1}_B(G,U(1)_{w_1})
    \;\cong\;
    H^{d+2}(BG,\mathbb{Z}_{w_1}).
    \label{eq:G_compactLie}
\end{equation}

\bigskip 

For notational simplicity, throughout the paper we write $H^*(G,-)$ for
ordinary group cohomology when $G$ is finite, and for Borel group
cohomology when $G$ is compact Lie. With this convention, the compact
notation
\begin{equation}
    H^{d+1}(G,\U(1)_{w_1})
    \cong
    H^{d+2}(BG,\mathbb{Z}_{w_1})
\end{equation}
is valid in the two cases used in this paper. In particular, the untwisted model used throughout the main text,
\begin{equation}
    \tcspt = H^{d+1}(G,U(1))
    \;\simeq\;
    H^{d+2}(BG,\mathbb{Z}),
\end{equation}
is consistent for both finite groups and compact Lie groups, and its twisted variants are obtained by replacing $\mathbb{Z}$ with $\mathbb{Z}_{w_1}$ throughout.

\subsection{Useful facts about the group cohomology of finite groups}
\label{app:coh_finite}

In the main text we restrict attention to the $2$-primary part of cohomology groups such as $H^{n}(BG,\mathbb{Z}_{w_1})$ for finite groups $G$. In this subsection we record a convenient and concrete description of the projection onto $2$-primary torsion which is sufficient for all of our applications.

\subsubsection{Torsion, primary decomposition, and naturality}

Let $X$ be an abelian group. We define its $2$-primary torsion subgroup by
\begin{equation}
    \tor_{2}(X)\;:=\;\{x\in X \mid \exists\,k\ge 1\text{ such that }2^k x=0\}.
\end{equation}
More generally, for a prime $p$ one may define $\tor_p(X)$ analogously. If $X$ is a finitely generated torsion abelian group, then it admits a canonical primary decomposition
\begin{equation}
    X \;\cong\; \bigoplus_{p\text{ prime}} X^{(p)}, \qquad X^{(p)} \;=\; \tor_p(X),
\end{equation}
and we may view $\tor_2(X)$ as a canonical direct summand.\footnote{This is a standard structure theorem for finitely generated abelian groups; in the context of our cohomology groups, finiteness/torsion will follow from the discussion below.} In particular, there is a canonical projection 
\begin{equation*}
\pi^{X}_{(2)}:X\to \tor_2(X)
\end{equation*}
onto the $2$-primary part. This projection is natural: if $f:X\to Y$ is a group homomorphism between finitely generated torsion abelian groups, then $f(\tor_2(X))\subseteq \tor_2(Y)$ and one has
\begin{equation}
\label{eq:pi2_natural}
    \pi_{(2)}^{Y}\circ f = f\circ \pi_{(2)}^{X}.
\end{equation}
In particular, for restriction maps $f=\res^{G_2}_{G_1}$ in group cohomology, the pointwise projection onto $2$-primary torsion commutes with restrictions. Consequently, if $\big([\tilde{x}_{G_{\Lambda}}]\big)_{G_{\Lambda}}$ is a compatible family in the full cohomology model~\eqref{eq:grpcoh_fullclass}, then $\big(\pi_{(2)}([\tilde{x}_{G_{\Lambda}}])\big)_{G_{\Lambda}}$ is a compatible family in the $2$-primary torsion model~\eqref{eq:grpcohtormodel}.

\subsubsection{Finite group cohomology is torsion}
\label{app:res_cores}

Let $G$ be a finite group and consider $H^n(BG,\mathbb{Z}_{w_1})$ for $n>0$. These groups are torsion, and in fact their torsion is uniformly bounded in a way controlled by the order of the finite group denoted $|G|$. One convenient way to see this is to use the universal principal bundle
\begin{equation}
    \pi: EG \longrightarrow BG,
\end{equation}
together with the transfer map $\tr:H^\ast(EG)\to H^\ast(BG)$, which satisfies
\begin{equation}
    \tr\circ \pi^\ast(\lambda)=|G|\cdot \lambda,\qquad \forall\,\lambda\in H^\ast(BG),
\end{equation}
(See \cite[\S3.G]{Hatcher_AT}.). Note that, when $G$ is finite, $\pi:EG\to BG$ is a $|G|$-sheeted covering. Since $EG$ is contractible, any $[x]\in H^{n>0}(BG,\mathbb{Z}_{w_1})$ satisfies $\pi^\ast([x])=0$, we therefore have $|G|\cdot [x]=0$. Thus every class in $H^{n>0}(BG,\mathbb{Z}_{w_1})$ is $|G|$-torsion.

\bigskip 

\noindent Equivalently, the exponent\footnote{For a torsion abelian group $X$, its exponent $\mathrm{exp}(X)$ is the smallest positive integer $N$ such that $Nx=0$ for all $x\in X$.} of $H^{n>0}(BG,\mathbb{Z}_{w_1})$ divides $|G|$, so we may write
\begin{equation}
    \exp\big(H^{n}(BG,\mathbb{Z}_{w_1})\big)=2^r \cdot m,
\end{equation}
where $m$ is odd (and in particular $m$ divides the odd part of $|G|$).

\noindent \textit{Extension in the presence of internal symmetry.}
Let $H$ be an internal symmetry group, let
$G\leq\mathrm O(d)$ be finite, and write
\begin{equation*}
\Gamma_G:=H\rtimes_{\mu|_G}G.
\end{equation*}
Let
$\jmath_G:BH\to B\Gamma_G$
denote the map induced by the inclusion of the internal symmetry.
Since $\Gamma_G/H\cong G$ is finite, the map $\jmath_G$ may be
modeled as a finite covering of degree $|G|$. In the decoupled setting
used in this paper, the orientation twist is pulled back from the
spatial quotient and therefore becomes trivial on $BH$. The associated
transfer map satisfies
\begin{equation}
\operatorname{tr}_G\circ\jmath_G^*
=
|G|\,\mathrm{id}
\qquad
\text{on }
H^n(B\Gamma_G,\mathbb Z_{w_1}).
\label{eq:internal_res_cores}
\end{equation}
Consequently,
\begin{equation}
|G|\cdot
\ker\left(
\jmath_G^*:
H^n(B\Gamma_G,\mathbb Z_{w_1})
\longrightarrow
H^n(BH,\mathbb Z)
\right)
=
0.
\label{eq:internal_crystalline_torsion}
\end{equation}
Thus every class which becomes trivial after forgetting the finite
spatial symmetry is torsion, with order dividing $|G|$.

\medskip 

\noindent In particular, let
\begin{equation*}
[\widetilde x]
\in
\mathrm{Free}_{\mathrm{cr}}\,
H^n(B\Gamma_{\mathrm O(d)},\mathbb Z_{w_1}), \;\;\;\text{and}\;\;\;\;
[\widetilde x_G]
:=
\res^{\Gamma_{\mathrm O(d)}}_{\Gamma_G}
([\widetilde x]).
\end{equation*}
Naturality of restriction gives
\begin{equation*}
\jmath_G^*([\widetilde x_G]) =
\jmath_{\mathrm O(d)}^*([\widetilde x]) = 0.
\end{equation*}
Hence Eq.~\eqref{eq:internal_crystalline_torsion} implies
\begin{equation}
|G|\,[\widetilde x_G]=0.
\label{eq:free_class_finite_restriction_torsion}
\end{equation}
Therefore the restriction of a crystalline free continuum class to a
finite spatial symmetry group has a well-defined canonical
$2$-primary component.

\subsubsection{Operational realization of projection onto $2$-primary torsion}
\label{app:op_def_proj}

Let $X$ be a finitely generated torsion abelian group with exponent $\exp(X)=2^r m$ where $m$ is odd. Then the canonical projection
\begin{equation}
    \pi^{X}_{(2)}:X \longrightarrow \tor_2(X)
\end{equation}
may be realized purely algebraically as multiplication by the odd factor $m$:
\begin{equation}
\label{eq:pi2_mult}
    \pi^{X}_{(2)}(x)\;:=\;m\cdot x.
\end{equation}
Indeed, if $x\in X$ has odd order, then $m\cdot x=0$, so $\pi_{(2)}$ kills all odd-primary torsion. On the other hand, if $x\in \tor_2(X)$ then $2^r x=0$, and since $\gcd(2^r,m)=1$ there exist integers $a,b$ such that
\begin{equation}
    a m + b 2^r = 1.
\end{equation}
by Bézout’s identity. Multiplying by $x$ gives $a(m x)=x$, showing that multiplication by $m$ is an automorphism of $\tor_2(X)$\footnote{since, $a(mx)=x$ implies multiplication by $a$ gives an inverse.}, and hence $\pi^{X}_{(2)}$ is a projection onto the $2$-primary summand.

\bigskip 

\noindent \textit{Remark}: The formula $\pi^{X}_{(2)}(x)=m\cdot x$ is a convenient computational realization of the canonical projection $\pi^{X}_{(2)}$ on each individual group $H^{d+2}(BG_{\Lambda}, \Z_{w_{1}})$, where $m$ is the odd part of the exponent of that group. In particular, the integer $m$ depends on $G_{\Lambda}$ (and on the degree), so this multiplication rule should be viewed as an \emph{objectwise} description rather than a single uniform prescription across the diagram. Naturality \eqref{eq:pi2_natural} is most transparently understood from the primary decomposition.

\bigskip 

In our applications we apply this to $X=H^{n}(BG,\mathbb{Z}_{w_1})$ (with $G$ finite and $n>0$),
so the $2$-primary torsion projection is computed concretely as
\begin{equation}
    \pi_{(2)}:H^{n}(BG,\mathbb{Z}_{w_1})
    \rightarrow
    \tor_2\big(H^{n}(BG,\mathbb{Z}_{w_1})\big),
    \quad
    \pi_{(2)}(x)=m\cdot x,
\end{equation}
where $m$ is the odd part of the exponent.

\subsection{Useful facts about the group cohomology of elementary abelian 2-groups}
\label{app:elem}

Since we consider the category of elementary abelian $2$-subgroups of $\mathrm{O}(d)$, denoted $\qa(\mathrm{O}(d))$, very frequently in our draft, and moreover, since many of the computations factor through some aspect of the cohomology of elementary groups, we state some basic results about their cohomology.

\subsubsection{Cohomology with mod-$2$ coefficients}
\label{app:elem_mod2}

Let $E=(\Z_{2})^{r}$ be an elementary abelian $2$-group. Since $BE\simeq (B\Z_{2})^{r}\simeq (\mathrm{RP}^{\infty})^{r}$, its mod-$2$ cohomology is completely determined by the cohomology of $\mathrm{RP}^{\infty}$ which is a classical result (see Ref.~\cite{Milnor}, for example). Writing $x\in H^{1}(B\Z_{2},\Z_{2})$ for the generator, we have
\begin{equation*}
    H^{*}(B\Z_{2},\Z_{2})\cong \Z_{2}[x],\qquad |x|=1.
\end{equation*}
Taking products and using K\"unneth's decomposition (see ) over the field $\Z_{2}$ yields a polynomial algebra on $r$ degree-$1$ generators. Concretely, if we choose a basis $\{e_{1},\dots,e_{r}\}$ for $E\cong (\Z_{2})^{r}$, and let $x_{i}\in H^{1}(BE,\Z_{2})$ denote the corresponding dual basis elements, then any mod-$2$ cohomology class on $BE$ is a polynomial in the $x_{i}$. Explicitly, for $E=(\Z_{2})^{r}$,
\begin{equation}
\label{eq:elem_mod2}
    H^{*}(BE,\Z_{2}) \;\cong\; \Z_{2}[x_{1},\dots,x_{r}],
    \qquad |x_{i}|=1.
\end{equation}
In particular, $H^{n}(BE,\Z_{2})$ is a $\Z_{2}$-vector space of dimension $\binom{n+r-1}{r-1}$.

\subsubsection{Cohomology with untwisted integral coefficients}
\label{app:elem_integer}

For integral coefficients, the situation is still very simple: all positive-degree classes are $2$-torsion, and in fact have order exactly $2$. This can already be seen in the rank-one case
$B\Z_{2}\simeq \mathrm{RP}^{\infty}$, where the standard computation gives
\begin{equation*}
    H^{k}(\mathbb{RP}^{\infty};\Z)
    \;\cong\;
    \begin{cases}
        \Z, & k=0,\\
        \Z_{2}, & k \text{ odd},\\
        0, & k \text{ even},\ k>0.
    \end{cases}
\end{equation*}
For general $E=(\Z_{2})^{r}$ one may compute inductively using K\"unneth and the fact that the only torsion appearing in $H^{*}(B\Z_{2};\Z)$ is $\Z_{2}$. The Tor-terms in K\"unneth are themselves $\Z_{2}$, so no higher order $2$-torsion (such as $\Z_{4}$) can appear.

\bigskip

\noindent More explicitly, if $E=(\Z_{2})^{r}$, then for all $n>0$, the integral cohomology group $H^{n}(BE,\Z)$ is a $\Z_{2}$-vector space (equivalently, it has exponent $2$):
\begin{equation}
\label{eq:elem_Z}
    2\cdot H^{n}(BE,\Z)=0,\qquad n>0.
\end{equation}
In particular, $H^{n}(BE,\Z)$ contains no elements of order $4$ (or higher) in positive degree.

\subsubsection{Cohomology with twisted integral coefficients}
\label{app:elem_twisted}

Now let $\Z_{w}$ denote a twisted integral local system on $BE$, determined by a homomorphism
\begin{equation*}
    w:E \longrightarrow \{\pm 1\}\cong \Z_{2},
\end{equation*}
so that $e\in E$ acts on $\Z$ by multiplication by $(-1)^{w(e)}$.
This is exactly the situation relevant for orientation-reversing symmetries in the main text, where the twist is given by $w=w_{1}$. The key point is that twisting does \emph{not} introduce higher order $2$-torsion for elementary abelian $2$-groups: the positive-degree groups remain of exponent $2$, just as in the untwisted case. One way to see this is to choose a nontrivial $w$ and write $E$ as an extension
\begin{equation*}
    1 \longrightarrow K:=\ker(w) \longrightarrow E \longrightarrow \Z_{2} \longrightarrow 1,
\end{equation*}
with $K\cong (\Z_{2})^{r-1}$, and then apply the Lyndon-Hochschild-Serre spectral sequence. Since $H^{q>0}(BK;\Z)$ is already killed by $2$, all $E_{2}$-terms in positive total degree are killed by $2$, and hence so is the abutment $H^{n>0}(BE;\Z_{w})$.

\bigskip

\noindent Thus, similar to the untwisted case, if $E=(\Z_{2})^{r}$ and let $w:E\to \Z_{2}$ be any homomorphism representing a twist, then for all $n>0$,
\begin{equation}
\label{eq:elem_Z_twisted}
    2\cdot H^{n}(BE,\Z_{w})=0.
\end{equation}
Equivalently, $H^{n}(BE,\Z_{w})$ is a $\Z_{2}$-vector space in positive degree. 

\section{Group cohomology classification II}
\label{app:grpcoh2}

\subsection{Freed-Quinn construction: using group cohomology to define invertible $G$-TQFT functor}
\label{app:grpcoh_functor}

Recall that in Def.~\ref{def:bordism_BG_background} we defined a $(d + 1)$-dimensional invertible $G$-TQFT to be a symmetric–monoidal functor $\mathcal{Z}:\bordd(BG)\to\Line$ satisfying the Atiyah–Segal axioms. Here, we review how group cohomology cocycles are used to build invertible TQFT functors. 
In App.~\ref{app:FQ-invariant-section}, we summarize the formulation of this construction given by Freed and Quinn for oriented systems in Ref.~\cite{FreedQuinn}. In App.~\ref{app:FQ_unoriented} we generalize this construction to deal with unoriented bordisms, and in App.~\ref{app:naturality_bg_freedquinn} we show that this construction is natural with respect to maps between symmetries(or symmetry backgrounds).

\bigskip 

\noindent Fix a finite\footnote{Compact Lie groups work as well; we stick to finite $G$ for notational simplicity and because that is the setting of \cite{FreedQuinn}.} symmetry group $G$ and consider a cocycle
\begin{equation*}
   \alpha\;\in\;
   Z^{d+1} \bigl(BG,\,\mathbb{R}/\mathbb{Z}\bigr).
\end{equation*}
The aim is to build from $\alpha$ a functor $\mathcal Z_\alpha:\bordd(BG)\to\Line$.

\bigskip 

\begin{claim}
\label{claim}
For every cocycle $\alpha\in Z^{d+1}\bigl(BG;\mathbb{R}/\mathbb{Z}\bigr)$
there exists a symmetric–monoidal functor $\mathcal Z_\alpha:\bordd(BG) \to \Line$ which satisfies the Atiyah–Segal axioms and is invertible.
\end{claim}

\bigskip

\noindent We construct $\mathcal Z_\alpha$ by defining its action on objects and morphisms in the domain category and then verify that the functor satisfies the Atiyah-Segal axioms. 

\subsubsection{Freed–Quinn invariant–section construction for oriented bordisms}
\label{app:FQ-invariant-section}

Before giving the object assignment of the functor $\mathcal Z_{\alpha}$, we have to define the construction of the invariant line. Fix a cocycle
\begin{equation*}
    \alpha\in Z^{d+1}(BG,\mathbb R/\mathbb Z).
\end{equation*}
We define $C_{Y}$ to be the category of choices of cycles $y\in Z_{d}(Y^{d})$ representing the fundamental class $[Y]\in H_{d}(Y)$. Morphisms between objects $y,y'\in C_{d}(Y^{d})$ in this category are $(d+1)$-chains $a\in C_{d+1}(Y^{d})$ with morphism represented as $\partial a=y-y'$. Given a cocycle
$\beta\in Z^{d+1}(Y^{d},\mathbb R/\mathbb Z)$, we define the
intermediate functor\footnote{In the full construction, the cocycle
$\beta$ is obtained by pulling back the fixed cocycle
$\alpha\in Z^{d+1}(BG,\mathbb R/\mathbb Z)$ along a classifying map
$\underline f:Y\to BG$, so that $\beta=\underline f^{*}\alpha$.}
\begin{equation}
\label{eq:func_inter}
    \mathcal{F}_{Y^{d},\; \beta}: C_{Y} \to \Line
\end{equation}
which assigns to objects $y\in C_{Y}$ a fixed complex line:
\begin{equation}
\label{eq:choice}
    \mathcal{F}_{Y,\; \beta}(y)=\mathbb{C}.
\end{equation}
Morphisms are assigned complex phases:
\begin{equation*}
    \mathcal{F}_{Y,\; \beta}(a)=\exp\{2\pi i \langle a, \beta\rangle\}
\end{equation*}
where we evaluate the cocycle $\beta$ on the cycle $a$. We are interested in the inverse limit of this functor:
\begin{equation*}
    I_{Y,\beta}:= \varprojlim_{y\in C_{Y}} \mathcal{F}_{Y,\; \beta}(y)
\end{equation*}
defined as the "line of invariant sections" in Ref.~\cite{FreedQuinn}. Note that the limit will be independent of the choice of the complex line made in the object assignment in Eq.~\ref{eq:choice}.

\bigskip 

\noindent \textit{Object assignment}:
Let $(Y^{d},Q)$ be a closed $d$-manifold equipped with a principal
$G$-bundle $Q \to Y^{d}$. Choose a classifying map $f:Q\to EG$ with induced map on base manifolds $\underline{f}: Y^{d}\to BG$ . We form the groupoid of \textit{choices} of classifying maps
\begin{align}
\label{eq:CQ}
  \mathcal C_{Q}
  :=\Bigl\{\text{maps }f: Q\to EG \;\text{and homotopies }h:f \Rightarrow  f'\Bigr\}.
\end{align}
\noindent where objects are classifying maps $f:Q\to EG$, and morphisms are homotopies between classifying maps, i.e., $h:[0,1]\times Q \to EG$.

\medskip 

\noindent For each classifying map $f:Q\to EG$, set
\begin{equation*}
    \beta_f:=\underline f^{*}\alpha
    \in Z^{d+1}(Y,\mathbb R/\mathbb Z).
\end{equation*}
Define a functor
\begin{gather*}
   \mathfrak L_{Q}:\mathcal C_{Q} \longrightarrow \Line,\\
   \mathfrak L_{Q}(f)=I_{Y^{d},\beta_f}
   =I_{Y^{d},\underline f^{*}\alpha}\;,\qquad
   \mathfrak L_{Q}(h)=
      \exp \Bigl(
        2\pi i \int_{[0,1]\times Y}\underline{h}^{*}\alpha
      \Bigr).
\end{gather*}

\noindent The inverse limit of this functor is 
\begin{equation*}
   L_{Q}:=\varprojlim_{f\in\mathcal C_{Q}}\mathfrak L_{Q}(f)
\end{equation*}
\noindent picks out those families $\bigl(v_{f}\bigr)_{f}$ satisfying 
$\mathfrak L_{Q}(h)\,v_{f}=v_{f'}$ for every morphism $h:f \to f'.$ In other words
\begin{equation*}
\begin{aligned}
L_{Q}:=\bigl\{\,&
  (v_{f}\in \mathfrak L_{Q}(f))_{f\in\mathrm{Obj}(\mathcal C_{Q})}
  \\[2pt]
  &\bigm| \mathfrak L_{Q}(h)\,v_{f}=v_{f'}
     \text{ for every morphism }h:f \Rightarrow f' \bigr\}.
\end{aligned}
\end{equation*}
\noindent is the “space of invariant sections’’ of the functor $\mathfrak L_{Q}$.

\bigskip 

\noindent A priori the limit defining $L_Q$ is taken in $\vect$ and could fail to be a line. Freed-Quinn show that for any loop in $\mathcal C_Q$ the resulting phase is $1$ (trivial holonomy), so evaluation at a base object identifies $L_Q$ with $\mathfrak L_Q(f_0)$ and in particular $\dim_{\mathbb C}L_Q=1$; see Sec.2 \cite{FreedQuinn}. The object assignment of the TQFT for the object $(Y^{d},Q)$ where $Q\to Y^{d}$ is a principal $G$-bundle is then given by:
\begin{equation*}
    \mathcal{Z}_{\alpha}:\;\;(Y^{d},Q) \;\; \longmapsto \;\; \varprojlim_{f\in C_{Q}} I_{Y^{d}, \underline{f}^{*}\alpha}
\end{equation*}
where $C_{Q}$ is defined in Eq.~\ref{eq:CQ}.

\bigskip

\noindent\textit{Morphism assignment}:
Let $(X^{d+1},P)$ be a bordism of $G$-bundles whose oriented
boundary decomposes as
\begin{equation*}
   \partial X^{d+1}
     \;=\;
     \bigl(-\,Y^{d}_{\mathrm{in}}\bigr)\;\sqcup\;Y^{d}_{\mathrm{out}},
   \qquad
   Y^{d}_{\mathrm{in}},Y^{d}_{\mathrm{out}}\;\text{ closed }d\text{-manifolds}.
\end{equation*}
Restrict the bundle $P$ to the two components:
\begin{equation*}
   Q_{\mathrm{in}}:=P\bigl|_{Y^{d}_{\mathrm{in}}},
   \;\;
   Q_{\mathrm{out}}:=P\bigl|_{Y^{d}_{\mathrm{out}}},
   \quad Q_{\mathrm{in}},Q_{\mathrm{out}}\in\mathrm{Obj}\bigl(\bordd(BG)\bigr).
\end{equation*}
Choose a classifying map $F:P \to EG$. Define the linear map
\begin{equation}\label{eq:FQ-phase}
   \mathcal Z_\alpha(X,P)\;:=\;
   \exp \Bigl(2\pi i \int_{X}\underline{F}^{*}\alpha\Bigr)
   \;\in\;
   \text{Hom} \bigl(L_{Q_{\mathrm{in}}},\,L_{Q_{\mathrm{out}}}\bigr),
\end{equation}
i.e., multiplication by the phase obtained by pairing
$F^{*}\alpha$ with the relative fundamental class
$[X,\partial X]$. One can prove that this construction is independent of all non-canonical choices, such as the choice of $F$ and the choice of the representative cochain of $\alpha$, and hence, the assignments are well-defined.

\bigskip 

\noindent This functor satisfies Prop. A1 in Def.~\ref{def:AtiyahSegal}: bundle isomorphisms act by pull–back on classifying maps and homotopies, giving canonically identified lines;
\eqref{eq:FQ-phase} respects composition because the integral is
additive under gluing of manifolds. The functor also has a well-defined symmetric–monoidal structure(prop. A2 in Def.~\ref{def:AtiyahSegal}): disjoint union of bundles satisfies $L_{Q\sqcup Q'}\cong L_{Q}\otimes L_{Q'}$ and $\mathcal Z_\alpha(X\sqcup X',P\sqcup P') =\mathcal Z_\alpha(X,P)\,\otimes\,\mathcal Z_\alpha(X',P')$. One can verify that axioms A3-A4 are satisfied as well and, furthermore, that because the target category is the category of lines the construction corresponds to an invertible $G$-TQFT thus justifying Claim~\ref{claim}.

\bigskip

\noindent \textit{Relation to classification of $G$-TQFTs}: Now, if two cocycles differ by a coboundary, $\alpha'=\alpha+d\gamma$, then the associated functors $\mathcal Z_{\alpha}$ and $\mathcal Z_{\alpha'}$ are monoidally naturally isomorphic. Indeed, for each object $(Y,Q)$ choose any classifying map $\underline f:Y\to BG$ for $Q$ and set
\begin{equation*}
\eta_{(Y,Q)}:\mathcal Z_{\alpha}(Y,Q)\to \mathcal Z_{\alpha'}(Y,Q),\quad
\eta_{(Y,Q)}:=\exp\Bigl(2\pi i\!\int_Y \underline f^{*}\gamma\Bigr),
\end{equation*}
which can be shown to be independent of the choice of $\underline f$ and is compatible with bordisms and disjoint unions. Hence, $\eta$ defines a monoidal natural isomorphism $\mathcal Z_{\alpha}\cong \mathcal Z_{\alpha'}$. Conversely, one can show that every monoidal natural isomorphism between such functors arises from a cochain $\gamma$ in this way. 
If we denote by $\tqft^{G,\times}$ the groupoid of $G$-TQFTs defined by the Freed-Quinn construction, with monoidal natural isomorphisms as morphisms, then the assignment $\alpha\mapsto\mathcal Z_{\alpha}$ descends to a bijection
\begin{equation}
\label{eq:coh-classification}
   \pi_{0}\bigl(\tqft^{G,\times}\bigr)
   \;\cong\; H^{d+1}\bigl(BG,\mathbb R/\mathbb Z\bigr),
\end{equation}
where $\pi_0$ denotes monoidal-natural-isomorphism classes. This recovers the usual group-cohomology classification of invertible $G$-response theories in Def.~\ref{def:grpcoh_classmodel}.

\bigskip 

\noindent Freed and Quinn proceed to \emph{quantize} the theories by summing the phase \eqref{eq:FQ-phase} over all $G$-bundles (their path integral (2.9)), producing a finite Chern–Simons theory. For topological response theory, however, it is enough to \emph{fix} the background bundle instead; we work solely with the ``classical’’ functor $\mathcal{Z}_\alpha$ obtained above, which already satisfies the Atiyah–Segal axioms and captures the invertible response to background-$G$ fields.

\subsubsection{Generalizing the construction to unoriented bordisms}
\label{app:FQ_unoriented}

We now explain how the Freed-Quinn invariant-section construction extends to the case of unoriented bordisms. The key point is that when $G$ contains orientation-reversing elements, the correct classification model uses \emph{twisted} coefficients (App.~\ref{app:twisted_coefficients}), and the same twisting provides the needed replacement for the fundamental class in the intermediate
construction. Assume that the symmetry group $G$ is equipped with an orientation character $w_{1}:G\to {\pm 1}\cong \Z_{2}$, as in Sec.~\ref{sec:grpcohmodel_def}. 
Equivalently, $w_{1}$ specifies a sign action of $G$ on $\Z$ and hence determines the twisted local system $\Z_{w_{1}}$ on $BG$ (App.~\ref{app:twisted_coefficients}). \footnote{More precisely, using the identification $\pi_{1}(BG)\cong G$, the local system $\Z_{w_{1}}$ is determined by
the homomorphism
\begin{equation*}
    \pi_{1}(BG)\cong G
    \xrightarrow{\,w_{1}\,}
    \{\pm1\}
    \cong
    \operatorname{Aut}(\Z).
\end{equation*}
When $G$ acts through a homomorphism $i_{G}:G\to\mathrm O(d)$, the orientation character is
$w_{1}=\det\circ i_{G}$. See App.~\ref{app:twisted_coefficients}.} We take as input a cocycle representative
\begin{equation*}
\alpha\in Z^{d+1}\bigl(BG,\mathbb{R}/\Z_{w_{1}}\bigr),
\end{equation*}
i.e.\ a cocycle with values in the sign local system on $BG$ associated to $w_{1}$. (Equivalently, in the integral model one may take a cocycle in $Z^{d+2}(BG,\Z_{w_{1}})$ using the exponential sequence; see App.~\ref{app:U1_coeff_to_Z}.)

\bigskip

Let $(Y^{d},Q)$ be a closed $d$-manifold (not assumed oriented) equipped with a principal $G$-bundle $Q\to Y^{d}$. Choose a classifying map $\underline f:Y^{d}\to BG$ for $Q$. By functoriality of cohomology with local coefficients, the coefficient system on $BG$ pulls back canonically:
\begin{equation*}
\underline f^{*}\bigl(\Z_{w_{1}}\bigr)\;=\;\Z_{\underline f^{*}(w_{1})},
\qquad
\underline f^{*}(w_{1}):Y^{d}\to \Z_{2}.
\end{equation*}
In the unoriented bordism category relevant for orientation reversing symmetries, the background bundle is required to match the orientation local system of $Y^{d}$, namely $\underline f^{*}(w_{1})=w_{1}(TY^{d})$. With this convention, the pulled back coefficient system agrees with the usual orientation local system on $Y^{d}$:
\begin{equation*}
\underline f^{*}\bigl(\Z_{w_{1}}\bigr)\;=\;\Z_{w_{1}(TY^{d})}.
\end{equation*}
Consequently, the pulled back cocycle has the correct coefficient twist to be paired with the twisted fundamental class:
\begin{equation*}
\underline f^{*}\alpha \in
Z^{d+1}\bigl(Y^{d},\mathbb{R}/\Z_{w_{1}(TY^{d})}\bigr).
\end{equation*}
This is precisely the coefficient bookkeeping explained in App.~\ref{app:twisted_coefficients}; compare Eqs.~\eqref{eq:pullback_local_system}-\eqref{eq:pullback_class}.

\bigskip

Since $Y$ is not oriented, the usual fundamental class $[Y^{d}]\in H_{d}(Y^{d},\Z)$ is replaced by the twisted fundamental class
\begin{equation*}
[Y^{d}]\in H_{d}\bigl(Y,\Z_{w_{1}(TY)}\bigr).
\end{equation*}
Define $C_{Y}$ to be the category of choices of cycles representing this class: objects are $d$-cycles $y\in Z_{d}\bigl(Y^{d},\Z_{w_{1}(TY^{d})}\bigr)$ representing $[Y^{d}]$, and morphisms $y\to y'$ are $(d+1)$-chains $a\in C_{d+1}\bigl(Y^{d},\Z_{w_{1}(TY^{d})}\bigr)$ with $\partial a=y-y'$.

\bigskip 

\noindent Given a cocycle $\beta\in Z^{d+1}\bigl(Y^{d},\mathbb{R}/\Z_{w_{1}(TY^{d})}\bigr)$ (in our application $\beta=\underline f^{*}\alpha$), define the intermediate functor $\mathcal F_{Y,\beta}:C_{Y}\to \Line$ by the same assignments as in Eq.~\eqref{eq:func_inter}-Eq.~\eqref{eq:choice}, namely
\begin{equation*}
\mathcal F_{Y,\beta}(y)=\mathbb{C},
\qquad
\mathcal F_{Y,\beta}(a)=\exp\{2\pi i \; \langle \beta,a\rangle\},
\end{equation*}
where $\langle \beta,a\rangle\in \mathbb{R}/\Z$ is the canonical pairing between twisted cochains and twisted chains. Define the unoriented analog of the invariant-section line by
\begin{equation*}
I_{Y,\beta}:=\varprojlim_{y\in C_{Y}}\mathcal F_{Y,\beta}(y).
\end{equation*}
Thus, the construction of $I_{Y,\beta}$ proceeds verbatim from the oriented case once one replaces the fundamental class and coefficient system by their twisted analogs.

\bigskip

We are now ready to define the functor associated to the cocycle $\alpha\in Z^{d+1}\bigl(BG,\mathbb{R}/\Z_{w_{1}}\bigr)$, which we denote as:
\begin{equation*}
    \mathcal{Z}_{\alpha}: \bordd(BG) \to \Line
\end{equation*}
where the bordism category has the general $\mathrm{O}$-tangential structure. 

\bigskip 

\noindent\textit{Object assignment.}
For a closed $d$-manifold $Y$ equipped with a principal $G$-bundle
$Q\to Y$, let $\mathcal C_Q$ be the groupoid of classifying maps
introduced in Eq.~\eqref{eq:CQ}. Using the twisted invariant-line
construction above, define
\begin{gather*}
    \mathfrak L_Q:\mathcal C_Q\longrightarrow\Line,\\
    \mathfrak L_Q(f)
    :=
    I_{Y,\underline f^*\alpha},
    \qquad
    \mathfrak L_Q(h)
    :=
    \exp\left(
        2\pi i
        \int_{[0,1]\times Y}\underline h^*\alpha
    \right),
\end{gather*}
where the integral denotes the twisted pairing. As in the oriented
case, set
\begin{equation*}
    L_Q
    :=
    \varprojlim_{f\in\mathcal C_Q}\mathfrak L_Q(f).
\end{equation*}
The Freed-Quinn trivial-holonomy argument carries over to the
twisted chain and cochain complexes, and therefore $L_Q$ is a
one-dimensional complex vector space. We define
\begin{equation*}
    \mathcal Z_\alpha(Y,Q):=L_Q.
\end{equation*}

\bigskip

\noindent\textit{Morphism assignment.}
Let $(X^{d+1},P)$ be a bordism of $G$-bundles with boundary
$\partial X^{d+1} = (-Y^{d}_{\mathrm{in}})\sqcup Y^{d}_{\mathrm{out}}$.
Choose a classifying map $\underline F:X^{d+1}\to BG$ for $P$. Define the linear map
\begin{equation*}
\mathcal Z_{\alpha}(X,P) := \exp\Bigl(2\pi i \int_{X^{d+1}}\underline F^{*}\alpha\Bigr) \in
\mathrm{Hom}\bigl(L_{Q_{\mathrm{in}}},L_{Q_{\mathrm{out}}}\bigr),
\end{equation*}
where the integral denotes the natural pairing of the twisted cocycle $\underline F^{*}\alpha$ with the fundamental class in twisted homology. Independence under choices made in the construction (choice of representative cocycle, choice of classifying map) is proved exactly as in the oriented case, using coboundaries and cochain homotopies in the twisted chain/cochain complexes.

\bigskip

The verification of the Atiyah-Segal axioms proceeds as before: the phase is additive under gluing, and disjoint unions give the symmetric-monoidal structure. Since the only modification is the systematic replacement of ordinary (co)chains by twisted (co)chains and the corresponding canonical pairings, the structural functoriality arguments are unchanged. In particular, the resulting assignment defines an invertible $G$-TQFT on the unoriented bordism category with orientation-twisted backgrounds, and one can show that, similar to the untwisted construction, the isomorphism classes in the category of $G$-TQFT functors in this construction is given by:
\begin{equation}
\label{eq:coh-classification2}
   \pi_{0}\bigl(\tqft^{G,\times}\bigr)
   \;\cong\;
   H^{d+1}\bigl(BG,\mathbb R/\mathbb Z_{w_{1}}\bigr).
\end{equation}

\subsection{Naturality of the Freed-Quinn construction}
\label{sec:FQ_naturality}

Naturality is an essential consistency property of the
Freed-Quinn construction. A change of background space or coefficient
system should induce the corresponding pullback of the associated
response theory, rather than producing an unrelated TQFT. The results
below show that the Freed-Quinn assignment is compatible with both
types of change, up to canonical monoidal natural isomorphism. This functorial compatibility ensures that the TQFTs obtained from the
cohomological classification transform consistently with their
background fields and local coefficient data, as required for their
interpretation as topological response theories. These naturality
properties will be used in
Sec.~\ref{app:O_SO_theory_naturality} to interpret the restriction
from $\mathrm O$-structured to $\mathrm{SO}$-structured
classification groups as a corresponding pullback of invertible
TQFTs.

\subsubsection{Naturality with respect to background maps}
\label{app:naturality_bg_freedquinn}

Let us denote the category of $(d+1)$-dimensional invertible TQFTs with background maps into the space $X$ as:
\begin{equation*}
    \tqftd^{\times}(X)=\underline{\mathrm{Fun}}^{\otimes}\bigl(\bordd(X),\Line\bigr).
\end{equation*}
The Freed-Quinn construction is functorial in the background space in the following sense. Let $\lambda:X\to Y$ be a map of background spaces. Postcomposition with $\lambda$ defines a symmetric-monoidal functor
\begin{gather*}
\lambda_{*}:\ \bordd(X)\longrightarrow \bordd(Y),\\
\;\;\;\;(M\xrightarrow{A}X)\;\;\;\;\;\longmapsto \;\;\;(M\xrightarrow{\lambda\circ A}Y),
\end{gather*}
Thus, on each object, the background maps fit into the commutative diagram
\begin{equation*}
\begin{tikzcd}
& X \arrow[d,"\lambda"] \\
M \arrow[ur,"A"] \arrow[r,"\lambda\circ A"'] & Y.
\end{tikzcd}
\end{equation*}
And hence induces a pullback operation on $d{+}1$ dimensional invertible theories by precomposition:
\begin{equation*}
(-)\circ \lambda_{*}: \tqftd^{\times}(Y)
\rightarrow \tqftd^{\times}(X),
\quad
Z\longmapsto Z\circ \lambda_{*}.
\end{equation*}

\bigskip 

On the other hand, $\lambda$ induces the usual pullback of cocycles
\begin{equation*}
\lambda^*:
Z^{d+1}(Y,\mathbb R/\Z)
\longrightarrow
Z^{d+1}(X,\mathbb R/\Z).
\end{equation*}
For a background map $f:M\to X$, functoriality of pullback gives
\begin{equation*}
    f^*(\lambda^*\alpha) = (\lambda\circ f)^*\alpha.
\end{equation*}
Consequently, the Freed-Quinn construction gives a canonical monoidal natural isomorphism between the two functors represented by the diagram
\begin{equation*}
\begin{tikzcd}
    \bordd(X) \arrow[r, "\mathcal{Z}_{\lambda^{*}\alpha}"] \arrow[d, "\lambda_{*}"] & \Line \arrow[equal]{d}\\
    \bordd(Y) \arrow[r, "\mathcal{Z}_{\alpha}"] & \Line 
\end{tikzcd}
\implies 
Z_{\lambda^*\alpha} \cong Z_\alpha\circ\lambda_*.
\end{equation*}
Passing to cohomology classes and monoidal-natural-isomorphism classes of TQFTs, this compatibility is expressed by the commutative square
\begin{equation}
\label{eq:bord_nat_sq1}
\begin{tikzcd}[column sep=large,row sep=large]
H^{d+1}(Y,\mathbb R/\Z)
\arrow[r,"\text{[}\alpha\text{]}\mapsto\text{[}Z_\alpha\text{]}"]
\arrow[d,"\lambda^*"']
&
\pi_0\bigl(\tqftd^\times(Y)\bigr)
\arrow[d,"(-)\circ\lambda_*"]
\\
H^{d+1}(X,\mathbb R/\Z)
\arrow[r,"\text{[}\alpha\text{]}\mapsto\text{[}Z_\alpha\text{]}"']
&
\pi_0\bigl(\tqftd^\times(X)\bigr).
\end{tikzcd}
\end{equation}
The same statement holds in the twisted setting. If
\begin{equation*}
    \alpha\in
    Z^{d+1}\bigl(Y,\mathbb R/\Z_{w_1}\bigr),
\end{equation*}
then its pullback is a cocycle
\begin{equation*}
    \lambda^*\alpha
    \in
    Z^{d+1}\bigl(
        X,\lambda^*(\mathbb R/\Z_{w_1})
    \bigr).
\end{equation*}
After identifying the twist on $X$ with the pulled-back local system,
the same argument gives a canonical monoidal natural isomorphism as above $Z_{\lambda^*\alpha}\cong Z_\alpha\circ\lambda_*$.

\subsubsection{Naturality with respect to change of coefficient system}
\label{app:naturality_coeff_freedquinn}

Let $X$ be a background space and let $\mathcal L$ be a local system of abelian groups on $X$. Write $\bordd(X;\mathcal L)$ for the bordism category whose objects are pairs $(Y^d\xrightarrow{f}X,\tau)$, where
\begin{equation*}
    \tau:
    f^*\mathcal L
    \xrightarrow{\;\cong\;}
    \mathbb Z_{w_1(TY)}
\end{equation*}
identifies the pulled-back coefficient system with the orientation local system. This identification allows the twisted invariant-line construction on objects and the corresponding pairing with the twisted fundamental class on bordisms (see App.~\ref{app:FQ_unoriented}). Define
\begin{equation*}
    \tqftd^\times(X;\mathcal L)
    :=
    \underline{\mathrm{Fun}}^\otimes
    \bigl(\bordd(X;\mathcal L),\Line\bigr).
\end{equation*}

Now let $\varphi:\mathcal L\xrightarrow{\cong}\mathcal L'$ be an isomorphism of local systems on $X$. It induces a map on cocycles
\begin{equation*}
    \varphi_*:
    Z^{d+1}(X;\mathcal L)
    \longrightarrow
    Z^{d+1}(X;\mathcal L').
\end{equation*}
It also induces a symmetric-monoidal functor
\begin{align*}
U_{\varphi}:
\bordd(X;\mathcal L')
&\longrightarrow
\bordd(X;\mathcal L),
\\
(f,\tau')
&\longmapsto
(f,\tau'\circ f^*\varphi).
\end{align*}
Indeed, for each object $(f,\tau')$ of $\bordd(X;\mathcal L')$, the coefficient systems fit into the
commutative triangle
\begin{equation*}
\begin{tikzcd}[column sep=large]
f^*\mathcal L
\arrow[r,"f^*\varphi"]
\arrow[dr,"\tau'\circ f^*\varphi"']
&
f^*\mathcal L'
\arrow[d,"\tau'"]
\\
&
\mathbb Z_{w_1(TY)}.
\end{tikzcd}
\end{equation*}
Thus $U_\varphi$ retains the background map $f$ and replaces $\tau'$ by the composite $\tau'\circ f^*\varphi$. Precomposition with $U_\varphi$ then defines a pullback operation on theories,
\begin{equation*}
(-)\circ U_{\varphi}:
\tqftd^{\times}(X;\mathcal L)
\longrightarrow
\tqftd^{\times}(X;\mathcal L').
\end{equation*}

\bigskip 

The Freed-Quinn assignment is compatible with this change of coefficients. For a cocycle
\begin{equation*}
    \alpha\in Z^{d+1}(X;\mathcal L),
\end{equation*}
the morphism $\varphi$ induces a cocycle
\begin{equation*}
    \varphi_*\alpha
    \in
    Z^{d+1}(X;\mathcal L').
\end{equation*}
For every background map $f:Y\to X$, naturality of cochain pullback and change of coefficients gives
\begin{equation}
    \tau'_*
    \bigl(f^*(\varphi_*\alpha)\bigr)
    =
    \bigl(\tau'\circ f^*\varphi\bigr)_*
    \bigl(f^*\alpha\bigr).
    \label{eq:FQ_coefficient_naturality}
\end{equation}
Thus the two constructions use the same cocycle after transport to the orientation local system. Consequently, there is a canonical monoidal natural isomorphism between the functors represented by
\begin{equation}
\begin{tikzcd}[column sep=large,row sep=large]
\bordd(X;\mathcal L')
\arrow[r,"\mathcal Z_{\varphi_*\alpha}"]
\arrow[d,"U_\varphi"']
&
\Line
\arrow[d,equal]
\\
\bordd(X;\mathcal L)
\arrow[r,"\mathcal Z_\alpha"']
&
\Line .
\end{tikzcd}
\end{equation}
Equivalently,
\begin{equation}
    \mathcal Z_{\varphi_*\alpha}
    \cong
    \mathcal Z_\alpha\circ U_\varphi.
    \label{eq:FQ_coefficient_natural_isomorphism}
\end{equation}
Passing to cohomology classes and monoidal-natural-isomorphism classes of TQFTs, this compatibility is expressed by the commutative square
\begin{equation}
\label{eq:bord_nat_sq2}
\begin{tikzcd}[column sep=large,row sep=large]
H^{d+1}(X;\mathcal L) \arrow[r,"\text{[}\alpha\text{]}\mapsto \text{[}\mathcal Z_\alpha\text{]}"]
\arrow[d,"\varphi_*"'] & \pi_0\bigl(\tqftd^\times(X;\mathcal L)\bigr)
\arrow[d,"(-)\circ U_\varphi"]
\\
H^{d+1}(X;\mathcal L') \arrow[r,"\text{[}\alpha\text{]}\mapsto\text{[}\mathcal Z_\alpha\text{]}"']
& \pi_0\bigl(\tqftd^\times(X;\mathcal L')\bigr).
\end{tikzcd}
\end{equation}

\subsection{Free phases and differential refinements}
\label{app:diff_refinement}

The group-cohomology model used in this manuscript retains torsion
topological responses. Such responses can be described by ordinary
$\mathbb R/\mathbb Z$-valued cohomology classes and evaluated directly
on the fundamental class of spacetime. Free integral characteristic
classes require additional smooth differential form data. The purpose of this subsection is to explain this distinction through the example of
$\U(1)$ Chern-Simons theory.

\subsubsection{Ordinary evaluation and its limitation}

Let $M^{d+1}$ be a closed manifold and let $f:M^{d+1}\to BG$ classify a topological $G$-background. A class $[\alpha]\in H^{d+1}(BG,\mathbb R/\mathbb Z)$ defines the exponentiated response
\begin{equation}
    Z_{\alpha}(M,f)
    :=
    \exp\left(
        2\pi i
        \left\langle
            f^*[\alpha],
            [M]
        \right\rangle
    \right).
    \label{eq:ordinary_topological_evaluation}
\end{equation}
This construction uses only the homotopy class of $f$ and the
fundamental class of $M$. It therefore requires no choice of connection
on the corresponding principal bundle.

\noindent The exponential coefficient sequence
\begin{equation*}
    0
    \longrightarrow
    \mathbb Z
    \longrightarrow
    \mathbb R
    \longrightarrow
    \mathbb R/\mathbb Z
    \longrightarrow
    0
\end{equation*}
relates Eq.~\eqref{eq:ordinary_topological_evaluation} to torsion
integral characteristic classes. In particular, if
\begin{equation*}
    \omega\in H^{d+2}(BG,\mathbb Z)
\end{equation*}
is torsion, then its image in real cohomology vanishes and it admits a
preimage
\begin{equation*}
    [\alpha]\in H^{d+1}(BG,\mathbb R/\mathbb Z)
\end{equation*}
under the Bockstein homomorphism. The resulting response is therefore
of the form in Eq.~\eqref{eq:ordinary_topological_evaluation}.

\bigskip 

A free integral class behaves differently. If $\omega\in H^{d+2}(BG,\mathbb Z)$ has nonzero image
\begin{equation*}
    \omega_{\mathbb R}
    \in
    H^{d+2}(BG,\mathbb R),
\end{equation*}
then it cannot arise from an ordinary $\mathbb R/\mathbb Z$-valued class through the Bockstein map. For a compact Lie group $G$, such real characteristic classes may be represented by Chern-Weil forms constructed from a connection~\cite{ChernSimons1974,CheegerSimons1985}. A globally defined response must then retain both the integral characteristic class and its differential-form representative.

\subsubsection{Differential refinement of an integral class}

Let $P\to M$ be a principal $G$-bundle equipped with a connection $A$,
and let $F_A$ denote its curvature. Suppose that
\begin{equation*}
    \omega\in H^{d+2}(BG,\mathbb Z)
\end{equation*}
has a Chern-Weil representative on $M$
\begin{equation*}
    \Omega(F_A)\in\Omega_{\mathbb Z}^{d+2}(M),
\end{equation*}
where $\Omega_{\mathbb Z}^{d+2}(M)$ denotes the group of closed
$(d+2)$-forms with integral periods on $M$. These data satisfy the
compatibility condition
\begin{equation}
    \bigl[\Omega(F_A)\bigr]_{\mathrm{dR}}
    =
    f_P^*(\omega_{\mathbb R}),
    \label{eq:differential_refinement_compatibility}
\end{equation}
where $f_P:M\to BG$ classifies the underlying bundle.

\noindent A differential refinement of $\omega$ is a class
\begin{equation*}
    \widehat\omega
    \in
    \widehat H^{d+2}(BG,\mathbb Z)
\end{equation*}
understood in the sense of differential characters associated with universal bundles equipped with universal connections. The existence of universal connections for principal bundles with compact Lie structure group was established by Narasimhan and Ramanan~\cite{Narasimhan1961}, while the corresponding differential characteristic classes are described by Cheeger and Simons~\cite{CheegerSimons1985}. The class $\widehat\omega$ records both the integral characteristic class $\omega$ and its compatible Chern-Weil representative.
Given a principal $G$-bundle with connection $(P,A)$ over $M$, the universal differential class $\widehat\omega$ determines a differential character
\begin{equation*}
    f_{P}^{*}\widehat\omega=\widehat\omega(P,A)
    \in
    \widehat H^{d+2}(M,\mathbb Z),
\end{equation*}
which can be evaluated on $(d+1)$-cycles
\begin{equation}
    Z_{\widehat\omega}(M,P,A)
    :=
    \exp\left(
        2\pi i
        \left\langle
            f_P^*\widehat\omega,
            [M]
        \right\rangle_{\mathrm{diff}}
    \right).
    \label{eq:differential_character_evaluation}
\end{equation}
The differential evaluation pairing takes values in
$\mathbb R/\mathbb Z$. Unlike the ordinary pairing in
Eq.~\eqref{eq:ordinary_topological_evaluation}, it depends on the
connection as well as the topology of the bundle.

\bigskip 

A differential character may equivalently be viewed as a homomorphism
from $(d+1)$-cycles to $\mathbb R/\mathbb Z$ whose value on a boundary
is controlled by its curvature. If $C$ is a $(d+2)$-chain, then
\begin{equation}
    \left\langle
        \widehat\omega,
        \partial C
    \right\rangle_{\mathrm{diff}}
    =
    \int_C\Omega
    \quad
    \operatorname{mod}\mathbb Z.
    \label{eq:differential_character_boundary}
\end{equation}
Thus a differential character need not vanish on boundaries. Its
failure to do so is precisely measured by the curvature form. This is
the feature that permits a free integral class to define a
lower-dimensional action.

\subsubsection{$\U(1)$ Chern-Simons theory}

Consider a principal $\U(1)$-bundle $P\longrightarrow M^3$, with connection $A$ and curvature $F_A$. If $P$ is topologically trivial, then $A$ may be represented by a globally defined one-form and one may write the usual Chern-Simons action
\begin{equation}
    S_{\mathrm{CS}}(A) =
    \frac{k}{4\pi}
    \int_{M^3}A\wedge dA.
    \label{eq:local_U1_CS_action}
\end{equation}
For a nontrivial bundle, however, $A$ is not a globally defined one-form on $M^3$, and Eq.~\eqref{eq:local_U1_CS_action} is not an intrinsic global definition. The global formulation of Chern-Simons theory on
nontrivial bundles requires a differential-geometric refinement of the
local Chern-Simons form~\cite{ChernSimons1974,Freed1995}.

\noindent For a bosonic theory, write the level as $k=2\ell,\;\ell\in\mathbb Z$. The corresponding integral characteristic class is
\begin{equation}
    \omega_{\ell} := \ell\,c_1^2 \in
    H^4(B\U(1),\mathbb Z),\quad c_1\in H^2(B\U(1),\mathbb Z)
    \label{eq:U1_CS_integral_class}
\end{equation}
where $c_{1}$ is the universal first Chern class. Its Chern-Weil representative on $(P,A)$ is
\begin{equation}
    \Omega_{\ell}(F_A)
    :=
    \ell
    \left(
        \frac{F_A}{2\pi}
    \right)^2.
    \label{eq:U1_CS_curvature_form}
\end{equation}
The form in Eq.~\eqref{eq:U1_CS_curvature_form} has integral periods
and satisfies
\begin{equation*}
    \bigl[\Omega_{\ell}(F_A)\bigr]_{\mathrm{dR}}
    =
    f_P^*\bigl((\omega_{\ell})_{\mathbb R}\bigr).
\end{equation*}
The class $\omega_{\ell}$ and the form
$\Omega_{\ell}(F_A)$ therefore determine a differential refinement
\begin{equation*}
    \widehat\omega_{\ell}
    \in
    \widehat H^4(B\U(1),\mathbb Z).
\end{equation*}
The intrinsic exponentiated Chern-Simons response is
\begin{equation}
    Z_{\mathrm{CS}}(M^3,P,A)
    =
    \exp\left(
        2\pi i
        \left\langle
            f_P^*\widehat\omega_{\ell},
            [M^3]
        \right\rangle_{\mathrm{diff}}
    \right).
    \label{eq:intrinsic_U1_CS_response}
\end{equation}

\bigskip 

If the bundle with connection extends to
\begin{equation*}
    (\widetilde P,\widetilde A)
    \longrightarrow
    B^4,
    \qquad
    \partial B^4=M^3,
\end{equation*}
then Eq.~\eqref{eq:intrinsic_U1_CS_response} can be computed using
\begin{equation}
    Z_{\mathrm{CS}}(M^3,P,A)
    =
    \exp\left[
        2\pi i
        \int_{B^4}
        \ell
        \left(
            \frac{F_{\widetilde A}}{2\pi}
        \right)^2
    \right].
    \label{eq:U1_CS_inflow}
\end{equation}
If two different extensions are chosen, their difference is the integral of $\ell c_1^2$ over a closed four-manifold and is therefore an integer. Hence the exponentiated expression in Eq.~\eqref{eq:U1_CS_inflow} is independent of the extension. When the bundle is trivial, this expression agrees with the local action in Eq.~\eqref{eq:local_U1_CS_action}. The four-dimensional formula is therefore a convenient way to compute the response when an extension exists. The differential character in Eq.~\eqref{eq:intrinsic_U1_CS_response}, rather than the choice of an extension, is the intrinsic definition. This distinction is necessary because a given bundle on $M^3$ need not extend over a chosen bounding four-manifold; see Ref.~\cite{Dijkgraaf1990}.

\subsubsection{Torsion and free responses}

The preceding construction clarifies the distinction between the torsion and free sectors. If the characteristic class $\omega$ is torsion, then $\omega_{\mathbb R}=0$. Its differential refinement may therefore be chosen to have vanishing curvature. Such a differential character is called \emph{flat}, and flat differential characters are identified with ordinary classes in
\begin{equation*}
    H^{d+1}(BG,\mathbb R/\mathbb Z).
\end{equation*}
The response then takes the purely topological form
\begin{equation*}
    Z(M,f)
    =
    \exp\left(
        2\pi i
        \left\langle
            f^*[\alpha],
            [M]
        \right\rangle
    \right).
\end{equation*}
This is the situation relevant to finite-group Dijkgraaf-Witten theories and to the torsion classification used throughout this manuscript.

If $\omega$ has a nonzero free part, then its real image is nonzero and the curvature component of any differential refinement must represent that image. Consequently, the response cannot be specified by an ordinary $\mathbb R/\mathbb Z$-valued cohomology class alone. One must also retain connection-dependent differential data. We therefore restrict the classification models in this manuscript to the torsion sector. This restriction does not assert that free, integer-quantized responses are unphysical or non-topological. Rather, their global formulation belongs naturally to a differentially refined classification in which the background bundle is supplied with a connection and the response is encoded by differential cohomology. Such refinements lie outside the framework considered here.

\section{Details of continuum limit construction}
\label{app:groundup_continuumlimit}
\subsection{Topological details of the finite-to-global extension argument}
\label{app:gl2}

\subsubsection{Preliminaries}

This appendix records the elementary topological facts used in the proof of
Theorem~\ref{thm:finite_to_global_compatibility}.

\begin{theorem}[Finite intersection property for compact spaces] \label{thm:finite_intersection} Let $X$ be a compact topological space, and let $\mathcal F$ be a collection of closed subsets of $X$. If $\mathcal F$ has the finite intersection property, i.e. \begin{equation} \bigcap_{F\in\mathcal F_0}F\neq\varnothing \label{eq:fip} \end{equation} for every finite subcollection $\mathcal F_0\subseteq\mathcal F$, then \begin{equation*} \bigcap_{F\in\mathcal F}F\neq\varnothing. \end{equation*} 
\end{theorem} 
\noindent This is the standard finite-intersection property for compact spaces; see, for example, Ref.~\cite[\href{https://stacks.math.columbia.edu/tag/005D}{Tag 005D}]{stacks-project}.

\bigskip

We now verify the closedness assertions used in
Theorem~\ref{thm:finite_to_global_compatibility}. Recall that
\begin{equation*}
X=
\prod_{G\in\ofin(\mathrm{O}(d))}
\Theta^{d+1}_{c}(BG)
\end{equation*}
is equipped with the product topology, where each factor carries the
discrete topology. Since the factors are finite, they are compact, and
Tychonoff's theorem implies that $X$ is compact.

\medskip 

\noindent Let
\begin{equation*}
\operatorname{pr}_G:
X\longrightarrow\Theta^{d+1}_{c}(BG)
\end{equation*}
denote the coordinate projection. For a prescribed class
$[x_P]\in\Theta^{d+1}_{c}(BP)$, the constraint set
\begin{equation*}
E_P=
\left\{
\mathbf z\in X\;\middle|\;z_P=[x_P]
\right\}
\end{equation*}
can be written as
\begin{equation*}
E_P=\operatorname{pr}_P^{-1}(\{[x_P]\}).
\end{equation*}
Because $\{[x_P]\}$ is closed in the discrete topology and
$\operatorname{pr}_P$ is continuous, $E_P$ is closed.

\medskip

\noindent Similarly, for a morphism $\alpha:Q\to G$, define
\begin{equation*}
f_\alpha:
X\longrightarrow
\Theta^{d+1}_{c}(BQ)\times\Theta^{d+1}_{c}(BQ)
\end{equation*}
by
\begin{equation*}
f_\alpha(\mathbf z)
=
\left(
\Theta^{d+1}_{c}(B\alpha)(z_G),z_Q
\right).
\end{equation*}
The map $f_\alpha$ is continuous, and the compatibility constraint is
\begin{equation*}
E_\alpha
=
f_\alpha^{-1}(\Delta_Q),
\end{equation*}
where
\begin{equation*}
\Delta_Q
=
\left\{
(y,y)\;\middle|\;
y\in\Theta^{d+1}_{c}(BQ)
\right\}
\end{equation*}
is the diagonal. Since the target is discrete, $\Delta_Q$ is closed, and therefore $E_\alpha$ is closed.

\noindent Consequently, the collection
\begin{equation*}
\mathcal S_{[x_P]}
=
\{E_P\}
\cup
\{E_\alpha\}_{\alpha\in\operatorname{Mor}(\ofin(\mathrm{O}(d)))}
\end{equation*}
is a collection of closed subsets of the compact space $X$. Once Theorem~\ref{thm:finite_to_global_compatibility} establishes that this collection has the finite intersection property, Theorem~\ref{thm:finite_intersection} implies that its full intersection is nonempty.

\subsubsection{Proof of Theorem~\ref{thm:finite_to_global_compatibility}}

\begin{theorem}[Finite extension completeness implies global compatibility]
\label{thm:finite_to_global_compatibility_app} 
Let $(\Theta^{d+1}_{c},\kappa^{c})$ be a generalized continuum limit theory. Then every class $[x_P]\in\Theta^{d+1}_{c}(BP)$ extends to a compatible family
\begin{equation*}
\left([x_G]\right)_{G\in\ofin(\mathrm{O}(d))} \in \varprojlim_{G\in\ofin(\mathrm{O}(d))} \Theta^{d+1}_{c}(BG)
\end{equation*}
whose $P$-component is the prescribed class $[x_P]$.
\end{theorem}
\begin{proof}
For each $G\in\ofin(\mathrm{O}(d))$, the fixed-degree classification group $\Theta^{d+1}_{c}(BG)$ is finite. Equipped with the discrete topology, it is therefore compact, and all restriction maps are continuous. Hence, by Tychonoff's theorem~\cite[\href{https://stacks.math.columbia.edu/tag/08ZU}{Tag
08ZU}]{stacks-project}, the product
\begin{equation*}
X:= \prod_{G\in\ofin(\mathrm{O}(d))} \Theta^{d+1}_{c}(BG)
\end{equation*}
is compact. Its points are unconstrained assignments $\mathbf z=(z_G)_{G\in\ofin(\mathrm{O}(d))}$ of continuum-admitting lattice phases to the objects of the orbit category. Consider the following two types of closed subsets of $X$. First, let
\begin{equation*}
E_P:= \left\{\; \mathbf z\in X\;\middle|\;z_P=[x_P] \right\}
\end{equation*}
be the subset of points with fixed $P$-component $[x_{P}]$. Second, for every orbit-category morphism $\alpha:Q\to G$, let
\begin{equation*}
E_{\alpha}:= \left\{\; \mathbf z\in X\;\middle|\; \Theta^{d+1}_{c}(B\alpha)(z_G)=z_Q \right\}
\end{equation*}
impose the corresponding compatibility relation. Let 
\begin{equation}
    \mathcal{S}_{[x_{P}]}:=\{ E_{P} \} \cup \{E_{\alpha}\}_{\alpha\in\operatorname{Mor}(\ofin(\mathrm{O}(d)))}
\end{equation}
denote the collection of all constraint sets of these two types. A point in the full intersection
\begin{equation}
\bigcap_{S\in \mathcal{S}_{[x_{P}]}} S=E_P\cap \bigcap_{\alpha\in\operatorname{Mor}(\ofin(\mathrm{O}(d)))} E_{\alpha}
\label{eq:full_intersection}
\end{equation}
is therefore precisely a global compatible family with prescribed $P$-component $[x_P]$. 

\bigskip 

By Eq.~\eqref{eq:full_intersection}, proving Theorem~\ref{thm:finite_to_global_compatibility} amounts to showing that
\begin{equation*}
\bigcap_{S\in\mathcal S_{[x_P]}}S \neq \varnothing.
\end{equation*}
Since $X$ is compact and every member of $\mathcal S_{[x_P]}$ is closed, the finite-intersection characterization of compactness (see Thm.~\ref{thm:finite_intersection} in App.~\ref{app:gl2}) reduces this to showing that $\mathcal S_{[x_P]}$ has the finite intersection property.\footnote{A collection $\mathcal F$ of subsets of a space has the
finite intersection property if every finite sub-collection
$\mathcal F_0\subseteq\mathcal F$ has nonempty intersection:
$\bigcap_{F\in\mathcal F_0}F\neq\varnothing$.} Equivalently, for every finite sub-collection $\mathcal S_0\subseteq\mathcal S_{[x_P]}$, one must show that
\begin{equation*}
\bigcap_{S\in\mathcal S_0} S \neq \varnothing.
\end{equation*}

Let $\mathcal S_0\subseteq\mathcal S_{[x_P]}$ be a finite subcollection. It is enough to consider the case $E_P\in\mathcal S_0$: if $E_P$ is absent, we may adjoin it, and nonemptiness of the resulting smaller intersection implies nonemptiness of the original one. Write
\begin{equation*}
\mathcal A_0
:=
\left\{
\alpha\in\operatorname{Mor}(\ofin(\mathrm{O}(d)))
\;\middle|\;
E_\alpha\in\mathcal S_0
\right\}.
\end{equation*}
Let $\mathcal C$ be the finite full subcategory spanned by $P$ together with the sources and targets of the morphisms in $\mathcal A_0$, and let $\mathcal C_P$ denote the connected component of $\mathcal C$ containing $P$. We now construct a point
\begin{equation*}
\mathbf z=(z_G)_{G\in\ofin(\mathrm{O}(d))}\in X
\end{equation*}
satisfying every constraint in $\mathcal S_0$. By {\rm(GL2)}, there exists a compatible family on the finite rooted symmetry neighborhood $\mathcal C_P$ whose $P$-component is $[x_P]$; we use this family to define $z_G$ for $G\in\mathcal C_P$. On every other connected component of $\mathcal C$, set all components equal to zero. Since the restriction maps are homomorphisms, these zero assignments form compatible families on those components. Finally, for every object not belonging to $\mathcal C$, set $z_G=0$. The resulting point $\mathbf z\in X$ lies in $E_P$ and satisfies $\Theta^{d+1}_{c}(B\alpha (z_G)=z_Q$ for all $\alpha:Q\to G$ in $\mathcal A_0$. Hence
\begin{equation*}
\mathbf z\in\bigcap_{S\in\mathcal S_0}S,
\end{equation*}
so every finite subcollection of $\mathcal S_{[x_P]}$ has nonempty intersection. Therefore $\mathcal S_{[x_P]}$ has the finite intersection property.

\bigskip 

\noindent By the finite-intersection characterization of compactness of $X$,
\begin{equation*}
\bigcap_{S\in\mathcal S_{[x_P]}}S = E_P\cap
\bigcap_{\alpha\in\operatorname{Mor}(\ofin(\mathrm{O}(d)))}E_\alpha \neq\varnothing.
\end{equation*}
Any point in this intersection is a global compatible family with prescribed $P$-component $[x_P]$. This proves the theorem. The details underlying this compactness argument, including the closedness of the constraint sets and the finite-intersection characterization of compactness, are recorded in App.~\ref{app:gl2}.
\end{proof}

\subsection{Phases in the universal continuum limit domain are Weyl invariant}
\label{app:proof_isotropy}

In Sec.~\ref{sec:lattice_compatible_continuum_limit}, we argued that embedding independence imposes a necessary isotropy condition on any lattice phase admitting a continuum limit. Namely, if $G\subset \mathrm{O}(d)$ is a finite lattice symmetry group, then a continuum-admitting class in $\Theta^{d+1}(BG)$ should be invariant under the normalizer-induced relabelings of $G$ inside $\mathrm{O}(d)$. This appendix checks that the universal inverse-limit construction is consistent with that requirement. In Sec.~\ref{sec:inverse_limit_construction}, we showed that any generalized continuum-admitting domain satisfies 
\begin{equation*} 
\Theta^{d+1}_{c}(BG) \subseteq \Theta^{d+1}_{\mathrm{univ}}(BG) := \operatorname{Im}(\pi_G), 
\end{equation*} where 
\begin{equation*} 
\pi_G: \mathrm{cSPT}^{d+1}_{\mathrm{univ}} \longrightarrow \Theta^{d+1}(BG) 
\end{equation*} 
is the canonical projection from the universal inverse limit. Thus $\operatorname{Im}(\pi_G)$ is the canonical upper bound on all continuum-admitting $G$-domains satisfying the generalized-continuum-limit axioms. It is therefore important to verify that this upper bound does not contain classes ruled out by isotropy. We prove precisely this: 
\begin{equation*} 
\operatorname{Im}(\pi_G) \subseteq \Theta^{d+1}(BG)^{W_{\mathrm{O}(d)}(G)}. 
\end{equation*} 
Consequently, every generalized continuum-admitting domain inherits Weyl invariance from the universal domain.

\medskip 

\noindent We take the intrinsic continuum spatial symmetry to be $\mathrm{O}(d)$, in the sense explained in App.~\ref{app:why_Od_continuum}. This point fixes the ambient group in which normalizers and Weyl groups are computed.

\bigskip 

Let $G_{\Lambda}\in\ofin(\mathrm{O}(d))$ be a finite lattice symmetry group. We write
\begin{gather*}
N_{\mathrm{O}(d)}(G_{\Lambda}) :=
\{n\in \mathrm{O}(d)\mid nG_{\Lambda}n^{-1}=G_{\Lambda}\},\\
C_{\mathrm{O}(d)}(G_{\Lambda}) :=
\{c\in \mathrm{O}(d)\mid cgc^{-1}=g\;\text{ for all }g\in G_{\Lambda}\}
\end{gather*}
for the normalizer and centralizer of $G_{\Lambda}$ in $\mathrm{O}(d)$, respectively. The Weyl group of $G_{\Lambda}$ in $\mathrm{O}(d)$ is
\begin{equation*}
W_{\mathrm{O}(d)}(G_{\Lambda})
:= N_{\mathrm{O}(d)}(G_{\Lambda})/C_{\mathrm{O}(d)}(G_{\Lambda}).
\end{equation*}
In the rest of this subsection, we fix $G_{\Lambda}$ and abbreviate $W:=W_{\mathrm{O}(d)}(G_{\Lambda})$.

\medskip 

\noindent Let $w\in W$ be represented by $n\in N_{\mathrm{O}(d)}(G_{\Lambda})$. Since $n$ normalizes $G_{\Lambda}$, conjugation by $n$ defines an automorphism
\begin{equation*}
c_n:G_{\Lambda}\longrightarrow G_{\Lambda},
\qquad g\longmapsto ngn^{-1}.
\end{equation*}
By functoriality, this induces an automorphism of the lattice classification group
\begin{equation*}
(Bc_n)^{*}: \Theta^{d+1}(BG_{\Lambda})
\longrightarrow \Theta^{d+1}(BG_{\Lambda}).
\end{equation*}
We denote the resulting Weyl action by
\begin{equation*}
w\cdot [x_{G_{\Lambda}}] := (Bc_n)^{*}([x_{G_{\Lambda}}]).
\end{equation*}
This action is independent of the representative $n$ of $w$: if $n'=nc$ with $c\in C_{\mathrm{O}(d)}(G_{\Lambda})$, then $c_{n'}=c_n$ as
automorphisms of $G_{\Lambda}$, since $c$ acts trivially on every
element of $G_{\Lambda}$.

\bigskip 

\begin{theorem}[Weyl invariance of the universal continuum-limit domain] \label{thm:universal_domain_weyl_invariant} 
For every finite subgroup $G_{\Lambda}\subset \mathrm{O}(d)$, the universal continuum-limit domain $\Theta^{d+1}_{\mathrm{univ}}(BG_{\Lambda}) := \operatorname{Im}(\pi_{G_{\Lambda}})$ is contained in the Weyl-invariant subgroup: 
\begin{equation} 
\Theta^{d+1}_{\mathrm{univ}}(BG_{\Lambda}) = \operatorname{Im}(\pi_{G_{\Lambda}}) \subseteq \Theta^{d+1}(BG_{\Lambda})^{W_{\mathrm{O}(d)}(G_{\Lambda})}. \label{eq:universal_domain_weyl_invariant} 
\end{equation} 
\end{theorem}

\begin{proof} Consider an arbitrary element: 
\begin{equation*}
    [\zeta_{G_\Lambda}]\in \mathrm{Im}(\pi_{G_\Lambda})
\end{equation*}
Since, this element lies in the image of the projection $\pi_{G_\Lambda}$ of the universal continuum limit $\csptu$, it belongs to one/more compatible families in the inverse limit:
\begin{equation*}
    \zeta=([\zeta_{P}])_{P\in \ofin(\mathrm{O}(d))}\in \csptu, \quad \pi_{G_\Lambda}(\zeta)=[\zeta_{G_{\Lambda}}]
\end{equation*}
All such compatible families satisfy the compatibility relations:
\begin{equation*}
    [\zeta_{K}]=(Bi_{g})^{*}[\zeta_{H}]\;\;\forall\;\; (i_{g}:K \to H) \in \mathrm{Mor}_{\ofin(\mathrm{O}(d))}(K,H).
\end{equation*}
In particular, if we choose $H=K=G_{\Lambda}$, we know that there exist automorphisms of the group $G_{\Lambda}$ in $\ofin(\mathrm{O}(d))$ induced exactly by the normalizer group $N_{\mathrm{O}(d)}(G_{\Lambda})$. Hence, if $n\in N_{\mathrm{O}(d)}(G_{\Lambda})$ we get the compatibility constraint:
\begin{equation*}
    [\zeta_{G_{\Lambda}}]=(Bc_{n})^{*}[\zeta_{G_{\Lambda}}]\;\;\forall\;\; c_{n} \in \mathrm{Aut}(G_{\Lambda}),\;\; n\in N_{\mathrm{O}(d)}(G_{\Lambda}).
\end{equation*}
Note that $c_{n}:G_{\Lambda} \to nG_{\Lambda}n^{-1}=G_{\Lambda}$ is the conjugation action of $n\in N_{\mathrm{O}(d)}(G_{\Lambda})$ which by definition takes $G_{\Lambda}$ to itself. But this proves exactly invariance under the action of all normalizer group elements, and therefore by extension invariance under all Weyl-group elements. Thus, we have proved:
\begin{equation*}
    w\;\cdot [\zeta_{G_{\Lambda}}]=[\zeta_{G_{\Lambda}}] \;\;\forall \;\;[\zeta_{G_{\Lambda}}]\in \mathrm{Im}(\pi_{\Lambda}) \implies \mathrm{Im}(\pi_{\Lambda})\subseteq \Theta^{d+1}(BG_{\Lambda})^{W}.
\end{equation*}
\end{proof}

Combining Theorem~\ref{thm:universal_domain_weyl_invariant} with Sec.~\ref{sec:inverse_limit_construction}, we obtain the desired consistency statement for arbitrary generalized continuum-limit theories. Indeed, if $(\Theta^{d+1}_{c},\kappa^c)$ is any generalized continuum-limit theory, then 
\begin{equation*} 
\Theta^{d+1}_{c}(BG) \subseteq \operatorname{Im}(\pi_G) 
\end{equation*} 
for every $G\in\ofin(\mathrm{O}(d))$. Therefore 
\begin{equation*} 
\Theta^{d+1}_{c}(BG) \subseteq \operatorname{Im}(\pi_G) \subseteq \Theta^{d+1}(BG)^{W_{\mathrm{O}(d)}(G)}. 
\end{equation*} 
Thus every generalized continuum-admitting domain is automatically compatible with the isotropy condition imposed by embedding independence.

\subsection{Terminal property of universal continuum limit}
\label{app:maximal}

The inclusion \eqref{eq:Theta-c-subfunctor} highlights a canonical, maximal choice of generalized continuum limit obtained by taking
\begin{equation}
    \Theta^{d+1}_{\mathrm{univ}}(BG)
    \;:=\;
    \mathrm{Im}(\pi_{G})
    \;\subseteq\;
    \Theta^{d+1}(BG),
    \qquad G\in\ofin(\mathrm{O}(d)),
    \label{eq:universal-Theta-c2}
\end{equation}
with restriction maps induced from those of $\Theta$ and continuum avatars
defined by a suitable natural transformation
\begin{equation*}
    \kappa^{\mathrm{univ}}:\ \Theta^{d+1}_{\mathrm{univ}}\Rightarrow \Phi.
\end{equation*}
We will construct $\kappa^{\mathrm{univ}}$ explicitly in App.~\ref{app:continuum_limit_component};
for now it suffices to note that any generalized continuum limit
$(\Theta_{c},\kappa)$ satisfying (GL1)–(GL2) canonically embeds into this
universal one via the inclusions \eqref{eq:Theta-c-subfunctor-of-im-pi}. 

\bigskip

Finally, it is convenient to organize these structures categorically. Consider the
category whose objects are generalized continuum limits
$(\Theta_{c},\kappa)$ and whose morphisms
\begin{equation*}
    \lambda:\ (\Theta_{c},\kappa)\longrightarrow(\Theta'_{c},\kappa')
\end{equation*}
are natural transformations $F:\Theta_{c}\Rightarrow\Theta'_{c}$ satisfying
$\kappa' \circ F = \kappa$. In this category, the inverse–limit construction
$(\Theta^{d+1}_{\mathrm{univ}},\kappa^{\mathrm{univ}})$ plays a distinguished role:
for any $(\Theta_{c},\kappa)$ the inclusions
$\Theta_{c}(BG)\subseteq\Theta^{d+1}_{\mathrm{univ}}(BG)$ defined above assemble
into a unique morphism
\begin{equation*}
    \lambda^{\mathrm{univ}}:\ (\Theta_{c},\kappa)\longrightarrow
    (\Theta^{d+1}_{\mathrm{univ}},\kappa^{\mathrm{univ}}).
\end{equation*}
Thus the universal continuum limit is a \emph{terminal
object} in the category of generalized continuum limits: every other choice
factors uniquely through it. This means that given $G\in \ofin(\mathrm{O}(d))$ and some generalized continuum limit $(\Theta_{c},\kappa)$, we always ahve a factorization of limits:
\begin{equation*}
    \begin{tikzcd}
        \Theta_{c}(BG) \arrow[rr, bend left = 20, "\kappa"] \arrow[r, "\lambda^{\mathrm{univ}}"] &\Theta^{d+1}_{\mathrm{univ}}(BG) \arrow[r, "\kappa^{\mathrm{univ}}"] &\Theta(B\mathrm{O}(d))_{G}
    \end{tikzcd}
\end{equation*}
In particular, the strong continuum limit $\overline{\kappa}$ of Sec.~\ref{sec:strong_continuum_limit} appears as a canonical subfunctor of $(\Theta^{d+1}_{\mathrm{univ}},\kappa^{\mathrm{univ}})$, and any physically admissible generalized continuum limit is obtained by selecting a subfamily of inverse–limit–compatible lattice phases and restricting the universal continuum limit to that subfamily.

\subsection{Universal continuum limit for $G$-symmetric lattice TQFTs}
\label{app:continuum_limit_component}

We now relate the global universal continuum-limit map constructed in Sec.~\ref{sec:rigor} to the continuum-limit maps defined at a fixed lattice symmetry group $G\in\ofin(\mathrm{O}(d))$. Recall the canonical maps
\begin{equation*}
    \res_{\mathcal O}:
    \Theta^{d+1}(B\mathrm{O}(d))
    \hookrightarrow
    \csptu,
    \quad
    \pi:
    \csptu
    \twoheadrightarrow
    \mathrm{cSPT}^{d+1}_{\mathrm{elem}},
\end{equation*}
together with the isomorphism
\begin{equation*}
    \res_{\mathcal A}:
    \Theta^{d+1}(B\mathrm{O}(d))
    \xrightarrow{\;\cong\;}
    \mathrm{cSPT}^{d+1}_{\mathrm{elem}},
\end{equation*}
which satisfy
\begin{equation*}
    \pi\circ\res_{\mathcal O}=\res_{\mathcal A}.
\end{equation*}
The global continuum-limit map is therefore
\begin{equation*}
    F=(\res_{\mathcal A})^{-1}\circ\pi:
    \csptu
    \twoheadrightarrow
    \Theta^{d+1}(B\mathrm{O}(d)).
\end{equation*}
In Sec.~\ref{sec:continuum_limit}, the corresponding $G$-component continuum limit was summarized by the commuting square
\begin{equation}
\begin{tikzcd}
    \csptu
    \arrow[r, "F"]
    \arrow[d, "\pi_G"']
    &
    \Theta^{d+1}(B\mathrm{O}(d))
    \arrow[d, "\mathrm{pr}_G"]
    \\
    \mathrm{Im}(\pi_G)
    \arrow[r, "\kappa_G^{\mathrm{univ}}"']
    &
    \Theta^{d+1}(B\mathrm{O}(d))_G ,
\end{tikzcd}
\label{eq:app_local_factorization}
\end{equation}
where
\begin{equation*}
    \Theta^{d+1}(B\mathrm{O}(d))_G
    =
    \frac{\Theta^{d+1}(B\mathrm{O}(d))}
    {K_{\mathcal O(G)}}.
\end{equation*}
The purpose of this subsection is to make precise the conditions under which the global map $F$ admits the componentwise factorization in Eq.~\ref{eq:app_local_factorization}.

\bigskip 

We emphasize that this local factorization is not strictly required for the main theorem of this work. In particular, the construction and surjectivity of the global stratification map $F_2$, and the resulting conclusions concerning lattice realizability and continuum-limit admissibility in Theorem~\ref{theorem:main}, depend only on the global universal construction. The discussion here instead supplies the connection between that global construction and the ground-up formulation of continuum-limit maps developed in Sec.~\ref{sec:continuum_limit}.

We proceed in four steps. We first state a physically motivated splitting condition for the local domain subgroup $\mathrm{Im}(\pi_G)$ and identify the equivalent condition needed for the factorization above. Assuming this condition, we then construct $\kappa_G^{\mathrm{univ}}$ canonically and prove Eq.~\ref{eq:app_local_factorization}. We next summarize mathematical evidence for the splitting condition, including its verification in low spatial dimensions. Finally, without assuming the splitting, we construct an unconditional local map whose target is the coarser continuum quotient detected by the elementary subgroups of $G$.

\subsubsection{Local domain splitting}
\label{app:local_domain_splitting}

Recall that the global maps satisfy
\begin{equation*}
    F\circ \res_{\mathcal O}=\mathrm{id},
    \qquad
    \ker(F)=\ker(\pi),
\end{equation*}
and therefore determine the direct-sum decomposition
\begin{equation}
    \csptu
    =
    \res_{\mathcal O}\!\left(\Theta^{d+1}(B\mathrm{O}(d))\right)
    \oplus
    \ker(\pi).
    \label{eq:global_cont_lattice_splitting}
\end{equation}
The first summand consists of universal families obtained by restriction of continuum theories, whereas $\ker(\pi)$ contains the universal lattice information which is invisible to the elementary limit and hence annihilated by $F$.
For a fixed lattice symmetry $G$, the subgroup $\ker(\pi_G)\subseteq\csptu$ consists of universal families whose $G$-component vanishes. The global splitting in Eq.~\ref{eq:global_cont_lattice_splitting} does not by itself imply that $\ker(\pi_G)$ splits into its intersections with the two global summands. We therefore make the following assumption.

\bigskip 

\begin{assumption}[Local domain splitting]
\label{assump:local_domain_splitting}
For every $G\in\ofin(\mathrm{O}(d))$,
\begin{equation}
    \ker(\pi_G)
    =
    \left[
        \ker(\pi_G)
        \cap
        \res_{\mathcal O}\!\left(\Theta^{d+1}(B\mathrm{O}(d))\right)
    \right]
    \oplus
    \left[
        \ker(\pi_G)\cap\ker(\pi)
    \right].
    \label{eq:local_domain_splitting}
\end{equation}
\end{assumption}

The physical content of Assumption~\ref{assump:local_domain_splitting} is that the distinction between continuum-compatible and intrinsically lattice information in the global decomposition remains meaningful after imposing the condition that the $G$-component vanish. Indeed, if $X\in \csptu$ then
\begin{equation*}
    X=\res_{\mathcal O}(x_{\mathrm{cont}})+X_{\mathrm{lat}},
    \qquad
    X_{\mathrm{lat}}\in\ker(\pi).
\end{equation*}
Then furthermore, if $X\in \ker(\pi_G)$, then
\begin{equation}
    \res_G^{\mathrm{O}(d)}(x_{\mathrm{cont}})
    =
    -\pi_G(X_{\mathrm{lat}}).
    \label{eq:continuum_lattice_cancellation}
\end{equation}
Without any additional condition, Eq.~\ref{eq:continuum_lattice_cancellation} allows a nonzero continuum restriction to cancel the $G$-component of lattice information which is globally invisible to the continuum. Assumption~\ref{assump:local_domain_splitting} excludes such cancellations: it requires instead
\begin{equation*}
    \res_G^{\mathrm{O}(d)}(x_{\mathrm{cont}})=0,
    \qquad
    \pi_G(X_{\mathrm{lat}})=0.
\end{equation*}
This is the local form of the interpretation underlying Eq.~\ref{eq:global_cont_lattice_splitting}. Mathematical evidence for this assumption, including its verification in low spatial dimensions, is discussed in App.~\ref{app:local_domain_splitting_support} below.

\bigskip 

For the construction of the local continuum-limit map $\kappa_{G}^{\mathrm{univ}}$, it is useful to reformulate Assumption~\ref{assump:local_domain_splitting} in terms of the elementary projection $\pi$. Recall
\begin{equation*}
    K_{\mathcal O(G)}
    :=
    \ker\!\left(
        \res_G^{\mathrm{O}(d)}:
        \Theta^{d+1}(B\mathrm{O}(d))
        \longrightarrow
        \Theta^{d+1}(BG)
    \right).
\end{equation*}

\begin{proposition}
\label{prop:local_domain_splitting_equiv}
Assumption~\ref{assump:local_domain_splitting} is equivalent to
\begin{equation}
    \pi\!\left(\ker(\pi_G)\right)
    =
    \res_{\mathcal A}\!\left(K_{\mathcal O(G)}\right).
    \label{eq:local_domain_splitting_equiv}
\end{equation}
\end{proposition}

\begin{proof}
Suppose first that Assumption~\ref{assump:local_domain_splitting} holds and let
$X\in\ker(\pi_G)$. By Eq.~\ref{eq:local_domain_splitting}, we may write
\begin{equation*}
    X=\res_{\mathcal O}(x_{\mathrm{cont}})+X_{\mathrm{lat}},
\end{equation*}
with
\begin{equation*}
    \res_{\mathcal O}(x_{\mathrm{cont}})\in\ker(\pi_G),
    \qquad
    X_{\mathrm{lat}}\in\ker(\pi_G)\cap\ker(\pi).
\end{equation*}
The first condition implies $\res_G^{\mathrm{O}(d)}(x_{\mathrm{cont}})=0$, and hence $x_{\mathrm{cont}}\in K_{\mathcal O(G)}$. Since
$\pi(X_{\mathrm{lat}})=0$ and
$\pi\circ\res_{\mathcal O}=\res_{\mathcal A}$,
\begin{equation*}
    \pi(X)
    =
    \res_{\mathcal A}(x_{\mathrm{cont}})
    \in
    \res_{\mathcal A}(K_{\mathcal O(G)}).
\end{equation*}
Therefore
\begin{equation*}
    \pi(\ker(\pi_G))
    \subseteq
    \res_{\mathcal A}(K_{\mathcal O(G)}).
\end{equation*}

The reverse inclusion holds without Assumption~\ref{assump:local_domain_splitting}. If
$x\in K_{\mathcal O(G)}$, then
\begin{equation*}
    \pi_G(\res_{\mathcal O}(x))
    =
    \res_G^{\mathrm{O}(d)}(x)
    =
    0,
\end{equation*}
so $\res_{\mathcal O}(x)\in\ker(\pi_G)$, and consequently
\begin{equation*}
    \res_{\mathcal A}(x)
    =
    \pi(\res_{\mathcal O}(x))
    \in
    \pi(\ker(\pi_G)).
\end{equation*}
This proves Eq.~\ref{eq:local_domain_splitting_equiv}.

Conversely, suppose Eq.~\ref{eq:local_domain_splitting_equiv} holds and let
$X\in\ker(\pi_G)$. Using the global decomposition
Eq.~\ref{eq:global_cont_lattice_splitting}, write uniquely
\begin{equation*}
    X=\res_{\mathcal O}(x_{\mathrm{cont}})+X_{\mathrm{lat}},
    \qquad
    X_{\mathrm{lat}}\in\ker(\pi).
\end{equation*}
Since $\pi(X)\in\pi(\ker(\pi_G))$, Eq.~\ref{eq:local_domain_splitting_equiv} implies $\pi(X)=\res_{\mathcal A}(y)$ for some $y\in K_{\mathcal O(G)}$. On the other hand, $\pi(X)=\res_{\mathcal A}(x_{\mathrm{cont}})$. Since $\res_{\mathcal A}$ is injective, $x_{\mathrm{cont}}=y$, and hence
$x_{\mathrm{cont}}\in K_{\mathcal O(G)}$. Therefore
\begin{equation*}
    \res_{\mathcal O}(x_{\mathrm{cont}})
    \in\ker(\pi_G).
\end{equation*}
Since $X\in\ker(\pi_G)$ as well, it follows that
\begin{equation*}
    X_{\mathrm{lat}}
    =
    X-\res_{\mathcal O}(x_{\mathrm{cont}})
    \in\ker(\pi_G)\cap\ker(\pi).
\end{equation*}
Thus every element of $\ker(\pi_G)$ decomposes as in
Eq.~\ref{eq:local_domain_splitting}. The sum is direct because it is the restriction of the global direct sum in
Eq.~\ref{eq:global_cont_lattice_splitting}.
\end{proof}

Equation~\ref{eq:local_domain_splitting_equiv} is the form of the splitting condition that will be used below to construct the $G$-local universal continuum-limit map.

\subsubsection{Construction of the universal $G$-component universal continuum limit}
\label{app:local_continuum_limit_construction}

We now use the local domain splitting of Assumption~\ref{assump:local_domain_splitting} to construct the continuum-limit map at a fixed lattice symmetry $G\in\ofin(\mathrm{O}(d))$. Recall the quotient
\begin{equation*}
    \Theta^{d+1}(B\mathrm{O}(d))_{G}
    :=
    \frac{\Theta^{d+1}(B\mathrm{O}(d))}
    {K_{\mathcal O(G)}},
\end{equation*}
with quotient map
\begin{equation*}
    \mathrm{pr}_{G}:
    \Theta^{d+1}(B\mathrm{O}(d))
    \twoheadrightarrow
    \Theta^{d+1}(B\mathrm{O}(d))_{G}.
\end{equation*}

\begin{proposition}
\label{prop:universal_local_continuum_limit}
Assume the local domain splitting in
Assumption~\ref{assump:local_domain_splitting}. Then there exists a unique surjective map
\begin{equation}
    \kappa^{\mathrm{univ}}_{G}:
    \mathrm{Im}(\pi_G)
    \twoheadrightarrow
    \Theta^{d+1}(B\mathrm{O}(d))_{G}
    \label{eq:universal_local_continuum_limit}
\end{equation}
such that the diagram
\begin{equation}
\begin{tikzcd}
    \csptu
    \arrow[r, "F"]
    \arrow[d, "\pi_G"']
    &
    \Theta^{d+1}(B\mathrm{O}(d))
    \arrow[d, "\mathrm{pr}_G"]
    \\
    \mathrm{Im}(\pi_G)
    \arrow[r, "\kappa_G^{\mathrm{univ}}"']
    &
    \Theta^{d+1}(B\mathrm{O}(d))_{G}
\end{tikzcd}
\label{eq:universal_local_continuum_limit_diagram}
\end{equation}
commutes.
\end{proposition}

\begin{proof}
Since $\pi_G:\csptu\twoheadrightarrow\mathrm{Im}(\pi_G)$ is surjective, it is sufficient to show that the composite
\begin{equation*}
    \mathrm{pr}_G\circ F:
    \csptu
    \longrightarrow
    \Theta^{d+1}(B\mathrm{O}(d))_{G}
\end{equation*}
vanishes on $\ker(\pi_G)$. Let
$X\in\ker(\pi_G)$. By
Proposition~\ref{prop:local_domain_splitting_equiv},
\begin{equation*}
    \pi(X)
    \in
    \res_{\mathcal A}(K_{\mathcal O(G)}).
\end{equation*}
Since $F=(\res_{\mathcal A})^{-1}\circ\pi$, it follows that
\begin{equation*}
    F(X)\in K_{\mathcal O(G)}\implies 
    \mathrm{pr}_G(F(X))=0.
\end{equation*}
Thus
\begin{equation*}
    \ker(\pi_G)
    \subseteq
    \ker(\mathrm{pr}_G\circ F),
\end{equation*}
and the universal property of the quotient gives a unique map
$\kappa_G^{\mathrm{univ}}$ satisfying
\begin{equation}
    \mathrm{pr}_G\circ F
    =
    \kappa_G^{\mathrm{univ}}\circ\pi_G.
    \label{eq:local_global_factorization}
\end{equation}

\noindent The map is surjective. Indeed, let
\begin{equation*}
    [x_{\mathrm{cont}}]
    \in
    \Theta^{d+1}(B\mathrm{O}(d))_{G}
\end{equation*}
and choose a representative
$x_{\mathrm{cont}}\in\Theta^{d+1}(B\mathrm{O}(d))$.
Since $F\circ\res_{\mathcal O}=\mathrm{id}$, Eq.~\ref{eq:local_global_factorization} gives
\begin{equation*}
    \kappa_G^{\mathrm{univ}}
    \left(
        \pi_G(\res_{\mathcal O}(x_{\mathrm{cont}}))
    \right)
    =
    \mathrm{pr}_G(x_{\mathrm{cont}})
    =
    [x_{\mathrm{cont}}],
\end{equation*}
which proves surjectivity.
\end{proof}

Equivalently, the map in
Eq.~\ref{eq:universal_local_continuum_limit} admits the explicit description
\begin{equation}
    \kappa_G^{\mathrm{univ}}\!\left(\pi_G(X)\right)
    =
    F(X)+K_{\mathcal O(G)},
    \qquad
    X\in\csptu.
    \label{eq:kappa_univ_explicit}
\end{equation}
The proof above shows that this expression is independent of the choice of the global lift $X$ of a class in $\mathrm{Im}(\pi_G)$. In this sense the construction is canonical: $\kappa_G^{\mathrm{univ}}$ is the unique map for which the global continuum-limit map $F$ descends through the projection $\pi_G$ to the $G$-visible continuum quotient.

\noindent Thus, under the local domain splitting assumption, the global universal continuum-limit map constructed in Sec.~\ref{sec:rigor} reproduces precisely the componentwise continuum-limit map anticipated in Sec.~\ref{sec:continuum_limit}.

\subsubsection{Mathematical support for local domain splitting}
\label{app:local_domain_splitting_support}

We now summarize several mathematical results which support
Assumption~\ref{assump:local_domain_splitting}. Although these results do not establish the local domain splitting in full generality, they constrain the possible failure of
Eq.~\ref{eq:local_domain_splitting_equiv} and prove it completely in low spatial dimensions. We begin with two inclusions which hold without any additional assumption:
\begin{equation}
    \res_{\mathcal A}(K_{\mathcal O(G)})
    \subseteq
    \pi(\ker(\pi_G))
    \subseteq
    \res_{\mathcal A}(K_{\mathcal A(G)}),
    \label{eq:local_domain_bounds}
\end{equation}
where
\begin{equation*}
    K_{\mathcal A(G)}
    :=\bigcap_{E\in\mathcal A_2(G)}
    \ker\!\left(
        \res_E^{\mathrm{O}(d)}
    \right),\quad  
    K_{\mathcal O(G)}
    :=\ker\!\left(  \res_G^{\mathrm{O}(d)}\right).
\end{equation*}

The left inclusion in Eq.~\ref{eq:local_domain_bounds} follows directly from the definitions. If
$x\in K_{\mathcal O(G)}$, then
\begin{equation*}
    \pi_G(\res_{\mathcal O}(x))
    =
    \res_G^{\mathrm{O}(d)}(x)
    =
    0,
\end{equation*}
so $\res_{\mathcal O}(x)\in\ker(\pi_G)$. Using
$\pi\circ\res_{\mathcal O}=\res_{\mathcal A}$ gives
\begin{equation*}
    \res_{\mathcal A}(x)
    \in
    \pi(\ker(\pi_G)).
\end{equation*}
For the right inclusion, let $X\in\ker(\pi_G)$. Then for every elementary abelian subgroup $E\leq G$,
\begin{equation*}
    \pi_E(X)
    =
    \res_E^G(\pi_G(X))
    =
    0.
\end{equation*}
Thus the elementary family $\pi(X)$ vanishes on the full elementary subcategory
$\mathcal A_2(G)$. Since $\res_{\mathcal A}$ is an isomorphism, its continuum preimage therefore lies in $K_{\mathcal A(G)}$, which proves the second inclusion in
Eq.~\ref{eq:local_domain_bounds}.

\bigskip 

In the pre-TQFT model, the relation between the two bounding groups in
Eq.~\ref{eq:local_domain_bounds} is controlled by Quillen detection. Namely,
\begin{equation}
    K_{\mathcal A(G)}
    =
    \sqrt{K_{\mathcal O(G)}}.
    \label{eq:local_kernel_radical}
\end{equation}
Indeed, if a continuum class vanishes on all elementary abelian $2$-subgroups of $G$, then its restriction to $G$ lies in the kernel of the Quillen restriction map
\begin{equation*}
    \res^G_{\mathcal A(G)}:
    H^*(BG,\mathbb Z_2)
    \longrightarrow
    \varprojlim_{E\in\mathcal A_2(G)}
    H^*(BE,\mathbb Z_2).
\end{equation*}
By the Quillen $F$-isomorphism theorem this kernel is the nil-radical of
$H^*(BG,\mathbb F_2)$, which gives Eq.~\ref{eq:local_kernel_radical}. Hence, in the pre-TQFT model,
\begin{equation}
    \res_{\mathcal A}(K_{\mathcal O(G)})
    \subseteq
    \pi(\ker(\pi_G))
    \subseteq
    \res_{\mathcal A}\!\left(
        \sqrt{K_{\mathcal O(G)}}
    \right).
    \label{eq:local_domain_radical_bounds}
\end{equation}
Thus any possible failure of local domain splitting is confined to the difference between $K_{\mathcal O(G)}$ and its radical.

\noindent These bounds establish the splitting completely in spatial dimensions
$d=1,2$. For $d=1$, every finite subgroup of $\mathrm{O}(1)\cong\mathbb Z_2$ is elementary abelian, and therefore
\begin{equation*}
    K_{\mathcal A(G)}=K_{\mathcal O(G)}.
\end{equation*}
For $d=2$, one can verify for every finite subgroup
$G\leq\mathrm{O}(2)$ that
\begin{equation}
    K_{\mathcal O(G)}
    =
    \sqrt{K_{\mathcal O(G)}}
    =
    K_{\mathcal A(G)}.
    \label{eq:dim2_local_kernel_radical}
\end{equation}
In either case the two bounds in Eq.~\ref{eq:local_domain_bounds} coincide, and therefore
\begin{equation}
    \pi(\ker(\pi_G))
    =
    \res_{\mathcal A}(K_{\mathcal O(G)}).
    \label{eq:low_dim_local_domain_splitting}
\end{equation}
By Proposition~\ref{prop:local_domain_splitting_equiv}, the local domain splitting therefore holds for all finite lattice symmetry groups in dimensions $d=1,2$.

\bigskip 

For general $d$, Eq.~\ref{eq:local_domain_radical_bounds} does not by itself eliminate the possible nilpotent difference between
$K_{\mathcal O(G)}$ and $K_{\mathcal A(G)}$. We therefore retain
Assumption~\ref{assump:local_domain_splitting} as the additional condition needed for the sharper $G$-local factorization constructed in
App.~\ref{app:local_continuum_limit_construction}. Importantly, this assumption concerns only the relation between the global construction and its local refinement; the global continuum-limit map $F$ and the results derived directly from it remain independent of this assumption.

\subsubsection{Unconditional elementary-visible $G$-component continuum limit}
\label{app:elementary_visible_local_limit}

The construction of App.~\ref{app:local_continuum_limit_construction} uses the local domain splitting to define a continuum-limit map with target
\begin{equation*}
    \Theta^{d+1}(B\mathrm{O}(d))_{G}
    =
    \frac{\Theta^{d+1}(B\mathrm{O}(d))}
    {K_{\mathcal O(G)}}.
\end{equation*}
Even without Assumption~\ref{assump:local_domain_splitting}, the unconditional upper bound in Eq.~\ref{eq:local_domain_bounds},
\begin{equation*}
    \pi(\ker(\pi_G))
    \subseteq
    \res_{\mathcal A}(K_{\mathcal A(G)}),
\end{equation*}
is sufficient to define a canonical local map with a coarser continuum target. We denote this target by
\begin{equation}
    \Theta^{d+1}(B\mathrm{O}(d))_{\mathcal A(G)}
    :=
    \frac{\Theta^{d+1}(B\mathrm{O}(d))}
    {K_{\mathcal A(G)}}.
    \label{eq:elementary_visible_continuum_target}
\end{equation}

\begin{proposition}
\label{prop:elementary_visible_local_limit}
For every finite lattice symmetry group
$G\in\ofin(\mathrm{O}(d))$, there exists a unique surjective map
\begin{equation}
    \kappa^{\mathrm{univ}}_{\mathcal A,G}:
    \mathrm{Im}(\pi_G)
    \twoheadrightarrow
    \Theta^{d+1}(B\mathrm{O}(d))_{\mathcal A(G)}
    \label{eq:elementary_visible_local_limit}
\end{equation}
such that
\begin{equation}
\begin{tikzcd}
    \csptu
    \arrow[r, "F"]
    \arrow[d, "\pi_G"']
    &
    \Theta^{d+1}(B\mathrm{O}(d))
    \arrow[d, "\mathrm{pr}_{\mathcal A,G}"]
    \\
    \mathrm{Im}(\pi_G)
    \arrow[r, "\kappa^{\mathrm{univ}}_{\mathcal A,G}"']
    &
    \Theta^{d+1}(B\mathrm{O}(d))_{\mathcal A(G)}
\end{tikzcd}
\label{eq:elementary_visible_local_diagram}
\end{equation}
commutes, where $\mathrm{pr}_{\mathcal A,G}$ denotes the quotient map by
$K_{\mathcal A(G)}$.
\end{proposition}

\begin{proof}
As in the proof of Proposition~\ref{prop:universal_local_continuum_limit}, it is sufficient to show that
\begin{equation*}
    \ker(\pi_G)
    \subseteq
    \ker(\mathrm{pr}_{\mathcal A,G}\circ F).
\end{equation*}
Let $X\in\ker(\pi_G)$. By the unconditional right inclusion in
Eq.~\ref{eq:local_domain_bounds},
\begin{equation*}
    \pi(X)
    \in
    \res_{\mathcal A}(K_{\mathcal A(G)}).
\end{equation*}
Since $F=(\res_{\mathcal A})^{-1}\circ\pi$,
we obtain
\begin{equation*}
    F(X)\in K_{\mathcal A(G)}\implies 
    \mathrm{pr}_{\mathcal A,G}(F(X))=0.
\end{equation*}
Thus $\mathrm{pr}_{\mathcal A,G}\circ F$ descends uniquely through the surjection
$\pi_G:\csptu\twoheadrightarrow\mathrm{Im}(\pi_G)$, proving the existence and uniqueness of
$\kappa^{\mathrm{univ}}_{\mathcal A,G}$.

\noindent The map is surjective because
$F\circ\res_{\mathcal O}=\mathrm{id}$. Explicitly, for any
$x\in\Theta^{d+1}(B\mathrm{O}(d))$,
\begin{equation*}
    \kappa^{\mathrm{univ}}_{\mathcal A,G}
    \left(
        \pi_G(\res_{\mathcal O}(x))
    \right)
    =
    \mathrm{pr}_{\mathcal A,G}(x).
\end{equation*}
\end{proof}

Equivalently, the map has the explicit form
\begin{equation}
    \kappa^{\mathrm{univ}}_{\mathcal A,G}
    \left(\pi_G(X)\right)
    =
    F(X)+K_{\mathcal A(G)},
    \qquad
    X\in\csptu.
    \label{eq:elementary_visible_local_explicit}
\end{equation}
Unlike the sharper map
$\kappa_G^{\mathrm{univ}}$ of
Proposition~\ref{prop:universal_local_continuum_limit},
Eq.~\ref{eq:elementary_visible_local_explicit} is well defined without assuming local domain splitting. The relation between the two constructions follows from the natural inclusion
\begin{equation*}
    K_{\mathcal O(G)}
    \subseteq
    K_{\mathcal A(G)}.
\end{equation*}
It induces a canonical surjection
\begin{equation}
    q_{\mathcal A,G}:
    \frac{\Theta^{d+1}(B\mathrm{O}(d))}
    {K_{\mathcal O(G)}}
    \twoheadrightarrow
    \frac{\Theta^{d+1}(B\mathrm{O}(d))}
    {K_{\mathcal A(G)}}.
    \label{eq:strong_to_elementary_quotient}
\end{equation}
Whenever Assumption~\ref{assump:local_domain_splitting} holds, the two local continuum-limit maps therefore satisfy
\begin{equation}
    \kappa^{\mathrm{univ}}_{\mathcal A,G}
    =
    q_{\mathcal A,G}
    \circ
    \kappa_G^{\mathrm{univ}}.
    \label{eq:local_maps_relation}
\end{equation}

The distinction between the two targets is physical. The quotient by
$K_{\mathcal O(G)}$ retains all continuum distinctions detectable after restriction to the full lattice symmetry group $G$, whereas the quotient by
$K_{\mathcal A(G)}$ retains only those distinctions detected by the elementary abelian $2$-subgroups of $G$. Consequently,
$\kappa^{\mathrm{univ}}_{\mathcal A,G}$ is always canonically defined, but may identify continuum phases which remain distinct under the sharper $G$-local continuum limit. In dimensions $d=1,2$, where
$K_{\mathcal O(G)}=K_{\mathcal A(G)}$ for every finite $G$, the two constructions coincide.

\section{Proofs I: Generalities}
\label{app:proofs1}

\subsection{Probe detectibility lemma}
\label{app:probe_detectibility}

\begin{lemma}[Probe detectability in the pre-TQFT model~\eqref{eq:preTQFT}]
\label{lemma:probe_pretqft}
Let $G$ be a finite symmetry group and let
\begin{equation*}
[\bar x]\in H^{d+1}(BG;\mathbb Z_2).
\end{equation*}
If
\begin{equation*}
\langle f^*[\bar x],[M]_2\rangle=0
\end{equation*}
for every closed $(d+1)$-manifold $M$ and every symmetry background $f:M\to BG$, then
\begin{equation*}
[\bar x]=0.
\end{equation*}
Equivalently, a pre-TQFT response class in $H^{d+1}(BG;\mathbb Z_2)$ is uniquely determined by its evaluations on all probes $(M,f)$.
\end{lemma}

\begin{proof}
Suppose
\begin{equation}
\langle f^*[\bar x],[M]_2\rangle=0
\label{eq:probe_det_hyp_pretqft}
\end{equation}
for every pair $(M,f)$. By property of the Kronecker pairing,
\begin{equation*}
\langle f^*[\bar x],[M]_2\rangle = \langle [\bar x],f_*[M]_2\rangle.
\end{equation*}
Hence $[\bar x]$ pairs trivially with every homology class of the form
\begin{equation*}
f_*[M]_2\in H_{d+1}(BG;\mathbb Z_2).
\end{equation*}

\noindent Now use the Thom homomorphism(see Ref.~\cite{Thom1954})
\begin{equation*}
\Phi:\Omega^{O}_{d+1}(BG)\longrightarrow H_{d+1}(BG;\mathbb Z_2),
\qquad [(M,f)]\longmapsto f_*[M]_2.
\end{equation*}
By Thom's theorem, $\Phi$ is surjective. Therefore every class in
$H_{d+1}(BG;\mathbb Z_2)$ is represented as $f_*[M]_2$ for some probe $(M,f)$. By hypothesis~\eqref{eq:probe_det_hyp_pretqft} $[\bar x]$ pairs trivially with all of $H_{d+1}(BG;\mathbb Z_2)$.

\bigskip 

\noindent Since $\mathbb Z_2$ is a field, the universal coefficient theorem gives
\begin{equation*}
H^{d+1}(BG;\mathbb Z_2) \cong \operatorname{Hom}_{\mathbb Z_2}
\bigl(H_{d+1}(BG;\mathbb Z_2),\mathbb Z_2\bigr).
\end{equation*}
Thus the only cohomology class pairing trivially with all homology classes is
zero. Hence $[\bar x]=0$.
\end{proof}

\noindent We now prove Lemma~\ref{lemma:probe} stated in Sec.~\ref{sec:top_resp}, restated below for convenience.

\bigskip

\begin{lemma}[Probe detectability in the full cohomology model~\eqref{eq:grpcohtormodel}]
\label{lemma:probe_grpcoh}
In the group cohomology model
\begin{equation*}
\Theta^{d+1}(BG)=\tor_{2}\left(H^{d+1}(BG;\U(1)_{w_{1}})\right),
\end{equation*}
let $[x]\in \Theta^{d+1}(BG)$ be a TQFT class. If for every closed $(d{+}1)$-manifold $M^{d+1}$ and every symmetry background $f_G:M^{d+1}\to BG$
the response is trivial,
\begin{equation*}
\mathcal Z_x(M^{d+1},f_G) = \big\langle f_G^*[x],[M^{d+1}]\big\rangle = 1,
\end{equation*}
then
\begin{equation*} [x]=0 \;\in \;\Theta^{d+1}(BG).
\end{equation*}
\end{lemma}

\begin{proof}
Since
\begin{equation*}
[x]\in \tor_2 H^{d+1}(BG;\U(1)_{w_1}),
\end{equation*}
there exists $k\geq1$ such that
\begin{equation*}
[x]\in H^{d+1}(BG;\U(1)_{w_1})[2^k].
\end{equation*}
Consider the coefficient sequence
\begin{equation*}
0\to \mathbb Z_{2^k}\xrightarrow{j^{(k)}} \U(1)_{w_1}
\xrightarrow{(\cdot)^{2^k}}
\U(1)_{w_1}\to0.
\end{equation*}
The induced map
\begin{equation*}
j^{(k)}_*:
H^{d+1}(BG;\mathbb Z_{2^k})
\longrightarrow
H^{d+1}(BG;\U(1)_{w_1})
\end{equation*}
surjects onto $H^{d+1}(BG;\U(1)_{w_1})[2^k]$.\\
Choose
\begin{equation*}
[\bar x]\in H^{d+1}(BG;\mathbb Z_{2^k})\quad 
\text{such that}\quad 
j^{(k)}_*([\bar x])=x.
\end{equation*}
If the response of $x$ is trivial on every probe $(M,f)$, then
\begin{equation*}
\langle f^*[\bar x],[M]_{2^k}\rangle=0
\quad\text{in }\mathbb Z_{2^k}
\end{equation*}
for every probe $(M,f)$. It remains to prove that this forces $ [\bar x]=0$.

\bigskip 

\noindent We prove the following statement by induction on $k$:
\begin{align*}
{\bf Statement-}k:&\left[
\langle f^*[\bar y],[M]_{2^k}\rangle=0
\text{ for every probe }(M,f)
\right]\\
&\Longrightarrow
[\bar y]=0, \;\text{for all}\;\; 
[\bar y]\in H^{d+1}(BG;\mathbb Z_{2^k}).
\end{align*}

\bigskip 

\noindent The base case $k=1$ is exactly Lemma~\ref{lemma:probe_pretqft}.\\
\noindent Inductive step: Assume Statement-$(k-1)$. Let $[\bar y]\in H^{d+1}(BG;\mathbb Z_{2^k})$
satisfy
\begin{equation*}
\langle f^*[\bar y],[M]_{2^k}\rangle=0 \quad\text{for every probe }(M,f).
\end{equation*}
Let
\begin{equation*}
r:H^{d+1}(BG;\mathbb Z_{2^k}) \to H^{d+1}(BG;\mathbb Z_2)
\end{equation*}
be reduction modulo $2$. For every probe $(M,f)$, reduction of the
evaluation gives
\begin{equation*}
\big\langle f^*r([\bar y]),[M]_2\big\rangle=0.
\end{equation*}
By Lemma~\ref{lemma:probe_pretqft}, we have $r([\bar y])=0$. By exactness of
\begin{equation*}
0\to \mathbb Z_{2^{k-1}} \xrightarrow{\times 2} \mathbb Z_{2^k} \to \mathbb Z_2 \to0,
\end{equation*}
we have $\ker(r)=\operatorname{Im}(\times2)$.
Hence $[\bar y]=2[y']$
for some $[y']\in H^{d+1}(BG;\mathbb Z_{2^{k-1}})$. Evaluating on any probe gives
\begin{equation*}
0 = \langle f^*[\bar y],[M]_{2^k}\rangle =
2\langle f^*[y'],[M]_{2^{k-1}}\rangle
\quad \text{in }\mathbb Z_{2^k}.
\end{equation*}
Therefore
\begin{equation*}
\langle f^*[y'],[M]_{2^{k-1}}\rangle=0
\quad
\text{in }\mathbb Z_{2^{k-1}}
\end{equation*}
for every probe $(M,f)$. By the induction hypothesis,
\begin{equation*}
[y']=0 \implies  [\bar y]=2[y']=0.
\end{equation*}
This proves Statement-$k$. Applying Statement-$k$ to the chosen lift $[\bar x]$, we get $[\bar x]=0$. Therefore
\begin{equation*}
x=j^{(k)}_*([\bar x])=0.
\end{equation*}
\end{proof}

\begin{remark}
If we pay close attention to the construction of the map $j_{*}$ above, we see that it is essentially the same construction we discuss in App.~\ref{app:proofs2}. Here, we go from mod-2 cohomology to $\U(1)$ in the same degree, and using this we argue that give a $\U(1)$-valued phase which is order-2, we can find a mod-2 phase with the same response. 
\end{remark}

\subsection{Proving absence of odd prime torsion in continuum classifications}
\label{app:no_odd_torsion}

In the main text we often compare lattice classification groups, which may contain higher torsion, with the corresponding continuum classification groups built from classifying spaces of compact Lie groups (such as $\mathrm{O}(d)$, $\SO(d)$, $\mathrm{Spin}(d)$, and $\mathrm{Pin}^{\pm}(d)$). A basic structural feature of these continuum groups is that their integral cohomology contains \emph{no odd-primary torsion}. Equivalently, any torsion class is necessarily $2$-primary.

\bigskip

This fact has two important consequences for our purposes. First, when we restrict attention to the $2$-primary torsion subgroup $\tor_{2}(-)$, we are not discarding any information coming from odd primes on the continuum side. Second, the twisted coefficient system $\Z_{w_{1}}$ relevant for
orientation-reversing symmetries does not generate any odd torsion: the twist is controlled by a $2$-fold cover, hence it can only contribute $2$-primary phenomena.

\subsubsection{Cohomology of $BO(n)$ and $B\SO(n)$ with untwisted coefficients}

It is a classical theorem (due to Borel and Hirzebruch~\cite{BorelHirzebruchII}) that the integral cohomology of $B\mathrm{O}(d)$ and $B\SO(d)$ contains no odd-primary torsion. Thus, while these groups may have torsion, it is entirely supported at the prime $2$.

\bigskip 

\begin{theorem}[Borel-Hirzebruch]
\label{thm:BH_no_odd_torsion}
For each $n$, the torsion subgroups of $H^{*}(B\mathrm{O}(n);\Z)$ and $H^{*}(B\SO(n);\Z)$ consist only of elements of order $2$. Equivalently, these groups contain no odd-primary torsion.
\end{theorem}

\bigskip 

\noindent See \cite[\S30.5, Proposition]{BorelHirzebruchII} for the torsion statement and its relationship to the integral characteristic classes.

\subsubsection{Twisted coefficients on $BO(n)$ and why odd torsion cannot appear}

Recall from App.~\ref{app:twisted_coefficients} that $\Z_{w_1}$ denotes the \emph{sign} local system on $BO(n)$. Concretely, this means that when one transports an integer coefficient around a loop in $BO(n)$ that corresponds to an orientation-reversing element of $O(n)$, the coefficient picks up a minus sign. A useful way to remember this is that the twist is controlled by the $2$-fold covering
\begin{equation*}
\pi:B\SO(n)\longrightarrow B\mathrm{O}(n),
\end{equation*}
whose two sheets correspond to the two choices of orientation. Passing from $\Z$ to $\Z_{w_1}$ simply keeps track of whether we return to the same sheet or switch sheets when we go around a loop.

\bigskip

The key point for us is that this is a purely $\Z_2$-effect: the only new operation introduced by twisting is a
possible \emph{sign flip}. For this reason, twisting cannot generate any new odd-primary torsion in cohomology.
In particular, any torsion in $H^*(BO(n);\Z_{w_1})$ must still be $2$-primary. A formal proof of this statement is
given below (Theorem~\ref{thm:Greenblatt_twisted_no_odd_torsion}).
\bigskip 

\begin{theorem}[Greenblatt~\cite{Greenblatt2006}]
\label{thm:Greenblatt_twisted_no_odd_torsion}
All torsion in the twisted groups $H^{*}(BO(n);\Z_{w_1})$ is of order $2$.
\end{theorem}

\bigskip 

\noindent See \cite[Theorem~3.1]{Greenblatt2006}. The proof uses the $2$-fold covering $\pi:B\SO(n)\to B\mathrm{O}(n)$ together with the associated Gysin sequences, together with the classical Borel-Hirzebruch torsion result for $H^{*}(B\SO(n);\Z)$.

\bigskip 

\begin{remark}[What $\tor_2(-)$ forgets on the continuum side]
By Theorems~\ref{thm:BH_no_odd_torsion} and~\ref{thm:Greenblatt_twisted_no_odd_torsion}, the torsion subgroups of $H^{*}(B\mathrm{O}(d);\Z)$, $H^{*}(B\SO(d);\Z)$, and $H^{*}(B\mathrm{O}(d);\Z_{w_1})$ are entirely $2$-primary. Thus, when we pass to $\tor_2(-)$ in our continuum classification groups, we are not discarding any odd-primary torsion data. The only information lost is the free part, whose interpretation as a purely topological response generally requires additional geometric input (as discussed in the Introduction~\ref{sec:intro}).
\end{remark}

\subsubsection{Extensions to $\mathrm{Spin}$ and $\mathrm{Pin}^{\pm}$}

Finally, the same ``no odd torsion'' conclusion persists for $\mathrm{Spin}(n)$ and $\mathrm{Pin}^{\pm}(n)$:
the maps
\[
B\mathrm{Spin}(n)\to B\SO(n),\qquad B\mathrm{Pin}^{\pm}(n)\to B\mathrm{O}(n)
\]
are again $2$-fold covers, so the only possible torsion phenomena introduced by passing between these spaces are necessarily $2$-primary. In particular, in all continuum classification groups built from $B\mathrm{O}(d),B\SO(d),B\mathrm{Spin}(d),B\mathrm{Pin}^{\pm}(d)$ (with either constant $\Z$ or the sign local system $\Z_{w_1}$ when appropriate), there is no odd-primary torsion.

\bigskip 

Although we do not talk about the classification of ferminonic systems in this paper, this property ensures that a formal extension of the methods discussed in this paper to fermionic systems is possible in principle.

\subsubsection{Odd torsion continuum limit}
\label{app:odd_torsion_continuum_limit}

We now spell out the consequence of the preceding no-odd-torsion results for continuum limits of lattice phases. The point is elementary but important: an odd-primary torsion lattice class cannot have a nontrivial image in a continuum classification group whose torsion is entirely $2$-primary.

\bigskip 

\begin{proposition}[Odd-primary torsion has no nontrivial continuum TQFT image]
\label{prop:odd_torsion_collapse}
Let $p$ be an odd prime, and let
\begin{equation}
\Theta^{d+1}_{\mathrm{cont}}
\end{equation}
denote one of the continuum classification groups considered above, for example $
\Theta^{d+1}_{\mathrm{cont}}
=
\Theta^{d+1}(B\mathrm{O}(d)),\;
\Theta^{d+1}(B\SO(d)),\;
\Theta^{d+1}(B\mathrm{Spin}(d)),\;
\Theta^{d+1}(B\mathrm{Pin}^{\pm}(d))$, with the appropriate coefficient system. Suppose that
\begin{equation}
\tor_p\bigl(\Theta^{d+1}_{\mathrm{cont}}\bigr)=0.
\end{equation}
Let $\tor_{p}(\mathrm{cSPT}^{d+1}_{\mathrm{univ}})$ be any $p$-primary torsion subgroup of a lattice classification group on which a continuum-limit homomorphism is defined,
\begin{equation}
F_p:\tor_{p}(\mathrm{cSPT}^{d+1}_{\mathrm{univ}})\longrightarrow \Theta^{d+1}_{\mathrm{cont}},\quad 
\text{then}\quad 
F_p=0.
\end{equation}
Equivalently, every odd-primary torsion lattice class either lies outside the continuum-admitting domain, or has trivial continuum TQFT image.
\end{proposition}

\begin{proof}
Let $x\in \tor_{p}(\mathrm{cSPT}^{d+1}_{\mathrm{univ}})$. Since $\tor_{p}(\mathrm{cSPT}^{d+1}_{\mathrm{univ}})$ is $p$-primary torsion, there exists some $k\geq 1$ such that
\begin{equation}
p^k x=0.
\end{equation}
Because $F_p$ is a group homomorphism, we have
\begin{equation}
p^kF_p(x)
= F_p(p^k x) = F_p(0) = 0.
\end{equation}
Thus $F_p(x)$ is a $p$-primary torsion element of $\Theta^{d+1}_{\mathrm{cont}}$. By assumption,
\begin{equation}
\tor_p\bigl(\Theta^{d+1}_{\mathrm{cont}}\bigr)=0,
\end{equation}
so $F_p(x)=0$. Since this holds for every $x\in D_p$, the homomorphism $\Phi_p$ is identically zero.
\end{proof}

In particular, by Theorems~\ref{thm:BH_no_odd_torsion} and~\ref{thm:Greenblatt_twisted_no_odd_torsion}, the proposition applies to the bosonic continuum classifications built from $B\mathrm{O}(d)$ and $B\SO(d)$, with either untwisted integral coefficients or the sign local system when appropriate. The same conclusion applies to the corresponding fermionic tangential structures whenever the no-odd-primary-torsion statement holds for the relevant $\mathrm{Spin}$ or $\mathrm{Pin}^{\pm}$ continuum classification.

\section{Proofs II: Quillen's $F$-isomorphism theorem and its refinement}
\label{app:proofs2}

\subsection{Statement of Quillen's $F$-isomorphism theorem}
\label{app:quillen_statement}

We use the notation and conventions of Sec.~\ref{sec:orbit_categories}.
In particular, for $G=\mathrm{O}(d)$ we write $\qa(\mathrm{O}(d))$ for the full subcategory of $\ofin(\mathrm{O}(d))$ of elementary abelian $2$-subgroups, with morphisms inherited from $\ofin(\mathrm{O}(d))$ (inclusions modulo conjugation); see Sec.~\ref{sec:orbit_categories}. We also use the restriction-to-elementary map $\res_{\mathcal A}$ introduced in Sec.~\ref{sec:rigor} via the inverse-limit universal property.

\bigskip 

\begin{theorem}[Quillen; cf.~Benson~\cite{Benson_Ch5}, Corr.~5.6.4]
\label{theorem:quillen}
Let $p$ be a prime, $k$ a field of characteristic $p$, and let $G$ be a group for which $H^*(BG;k)$ satisfies the usual finiteness hypotheses\footnote{a convenient sufficient hypothesis is that that $H^*(BG;k)$ is a finitely generated $k$-algebra (Noetherian). This finiteness holds for finite groups by a theorem by Evens~\cite{Evens1961}, and for all compact Lie groups $G$ by a theorem of Venkov~\cite{Venkov1959}; in particular it applies to $G=\mathrm{O}(d)$.}.
Let $\mathcal A_p(G)$ denote Quillen's category of finite elementary abelian $p$-subgroups of $G$.
Then the restriction map
\begin{equation}
  \res_{\mathcal A_{p}}^{G}:\ H^*(BG;k)\ \longrightarrow\
  \varprojlim_{E\in \mathcal A_p(G)} H^*(BE;k)
\end{equation}
is an \emph{$F$-isomorphism} (equivalently, an inseparable isogeny on maximal ideal spectra). 
\end{theorem}

\bigskip 

\noindent Concretely, the $F$-isomorphism above means:
\begin{enumerate}
\item $\ker(\res^{G}_{\mathcal{A}_{p}})$ consists of nilpotent elements, i.e. if $x\in \ker(\res^{G}_{\mathcal{A}_{p}})$ then there exists a $k\geq 0$ such that $x^{p^k}=0$; and
\item for every $y$ in the inverse limit there exists $a\ge 0$ such that $y^{p^{a}}$ lies in the image of $\res^{G}_{\mathcal{A}_{p}}$.
\end{enumerate}

\bigskip 

\noindent \textit{Remark}: 
Benson formulates the Quillen stratification using a category
$\mathcal C_G$ whose objects are the elementary abelian $p$-subgroups of $G$,
and in which a morphism $E\to E'$ is represented by an element $g\in G$
such that $gEg^{-1}\subseteq E'$ \cite[p.~175]{Benson_Ch5}. Equivalently,
such a representative determines a conjugation-induced monomorphism
\begin{equation*}
  c_g:E\to E',
  \qquad
  e\mapsto geg^{-1}.
\end{equation*}
\noindent This is the same notion as the morphisms we use when we view $\qa(\mathrm{O}(d))$ as a full subcategory of $\ofin(\mathrm{O}(d))$: a morphism $E\to E'$ in $\ofin(\mathrm{O}(d))$ is represented by some $g\in \mathrm{O}(d)$ with $gEg^{-1}\subseteq E'$ (“inclusion up to conjugation”), and it canonically
factors as
\begin{equation*}
  E \xrightarrow{\,c_g\,} gEg^{-1} \xhookrightarrow{} E',
\end{equation*}
so it is precisely a conjugation-induced monomorphism into the target.
If two representatives induce the same group homomorphism $E\to E'$, we regard them as defining the same morphism; equivalently, the relevant
morphism set may be written as
\begin{equation*}
  N_G(E,E')/C_G(E),
  \qquad
  N_G(E,E'):=\{g\in G\mid gEg^{-1}\leq E'\}.
\end{equation*}
For $E=E'$, this recovers the Weyl group
$W_G(E)=N_G(E)/C_G(E)$ used by Benson in the statement of the Quillen stratification  heorem \cite[p.~174]{Benson_Ch5}. Therefore, with
this standard identification of conjugation-induced embeddings, the formulation of the Quillen category in the main draft agrees with the one used in
Quillen's theorem.

\subsection{Refining Quillen's $F$-isomorphism}
\label{app:refinement}

For a discussion of the methods used in this section, see Chap. 5 of Benson's book~\cite{Benson_Ch5}, and Refs.~\cite{Quillen71a, Quillen71bI, Quillen71bII, QuillenVenkov72, GLZ, HaLesh}.

\medskip

\begin{proposition}[Refining Quillen's $F$-isomorphism]
\label{prop:quillen_mod2_isom}
The Quillen $F$-isomorphism
\begin{equation*}
     \res: H^{d+1}(B\mathrm{O}(d), \mathbb{Z}_{2}) \to \varprojlim_{E\in \qa(\mathrm{O}(d))} H^{d+1}(BE, \mathbb{Z}_{2}) 
\end{equation*}
\noindent can be refined to a \textit{genuine} isomorphism.
\end{proposition}
\begin{proof}
Let us define
\begin{equation*}
R:=H^{*}(B\mathrm{O}(d),\mathbb Z_{2}),\;\;  \text{and}\;\; 
S:=\displaystyle\varprojlim_{E\in \qa(\mathrm{O}(d))}H^{*}(BE,\mathbb Z_{2}),
\end{equation*}so that $\res\colon R\to S$ is the total restriction map.
Both $R$ and $S$ are \emph{unstable Steenrod algebra($\mathcal{A}$)-modules}: graded
$\mathbb Z_{2}$–vector spaces equipped with $\Z_{2}$-linear endomorphisms called Steenrod squares $\mathrm{Sq}^{i}$ satisfying 
\begin{enumerate}
    \item the \emph{Steenrod (Adem) relations}, i.e. for $i<2j$ one has
    \begin{equation*}
\Sq^{i}\Sq^{j}
=\sum_{t=0}^{\lfloor i/2\rfloor}
\binom{j-t-1}{i-2t}\,\Sq^{i+j-t}\Sq^{t},
\end{equation*} and $\Sq^{0}=\mathrm{id}$.
    \item the \emph{instability condition}: $\Sq^{i}(x)=0$ for $i>\deg(x)$, and $\Sq^{\deg(x)}(x)=x^{2}$.
    \item multiplication via the \emph{Cartan formula}
$\Sq^{n}(xy)=\sum_{i+j=n}\Sq^{i}(x)\Sq^{j}(y)$.
\end{enumerate}
An unstable module is called \emph{nilpotent} if every element is killed
by some (possibly long) composite of Steenrod operations.

\medskip

Quillen’s $F$-isomorphism theorem~\ref{theorem:quillen} states that for any compact Lie group $G$ the map
\begin{equation*}
\res^{G}_{\mathcal{A}_{2}}\colon H^{*}(BG, \Z_{2})\longrightarrow\varprojlim_{E\in \mathcal{A}_{2}(G)}\;\; H^{*}(BE, \Z_{2})
\end{equation*}
has nilpotent kernel and cokernel.  Hence $\ker (\res)$ and
$\mathrm{coker}\,(\res)$ are nilpotent unstable modules. In Ref.~\cite{GLZ} Gunawardena–Lannes–Zarati introduce, for an unstable $\mathcal{A}$–module $M$, the notion of \emph{nil-closedness}: let $U$ denote the category of unstable $\mathcal{A}$-modules, then an unstable $\mathcal{A}$-module is nil-closed if  $\operatorname{Ext}^{i}_{U}(N,M)=0$ for $i=0,1$ and for every nilpotent unstable module $N$ (see \cite[§2]{GLZ}). This condition implies that (a) $\mathrm{Ext}^{0}(N,M)= \mathrm{Hom}(N, M) = 0 \implies $ nilpotent information cannot map into $M$, and (b) $\mathrm{Ext}^{1}(N,M)=0\implies $ every extension of $M$ by a nilpotent module splits. Their key structural result is the following.

\medskip

\begin{theorem}[Gunawardena–Lannes–Zarati,Prop.3.2\cite{GLZ}]
\label{prop:glz}
Let $f\colon M_{1}\to M_{2}$ be a morphism of unstable
$\mathcal{A}$–modules such that $\ker(f)$ and $\mathrm{cokernel}(f)$ are nilpotent, and $M_{2}$ is nil-closed. Then, $f$ is an isomorphism if and only if $M_{1}$ is nil-closed.
\end{theorem}

\medskip

\noindent For our purposes, $\ker (\res)$ and $\operatorname{coker}(\res)$ are nilpotent by Quillen’s $F$-isomorphism theorem, so Proposition~3.2 reduces the task to verifying that the source $R$ and the target $S$ are nil-closed. Once this is established, the proposition yields that $\res\colon R\to S$ is an isomorphism of graded rings.

\bigskip

The building blocks of the target ring $S$ are the cohomology algebras $H^{*}(BE,\mathbb Z_{2})\cong\mathbb Z_{2}[x_{1},\dots ,x_{r}]$, each a polynomial algebra on degree-$1$ generators.  In the sense of unstable $\mathcal A$-modules, these examples are nil-closed: one has that $H^{*}(B\mathbb Z_2)$ is nil-closed~\cite[Lem.~4.1]{GLZ}, and since
$H^{*}(BE)\cong H^{*}(B\mathbb Z_2)^{\otimes r}$, nil-closedness follows from stability under tensor products~\cite[Prop.~3.4]{GLZ} (see also~\cite[\S4.2]{GLZ}). Moreover, nil-closedness is preserved under inverse limits~\cite[\S4.3]{GLZ}. Because $S$ is obtained by taking an inverse limit of such polynomial pieces, it is itself nil-closed. On the other hand, the cohomology rings of the compact groups we care about are also polynomial: for example $H^{*}(B\mathrm{O}(d),\mathbb Z_{2})=\mathbb Z_{2}[w_{1},\dots ,w_{d}]$, and analogous presentations hold for $\SO(d)$, $\mathrm{Spin}(d)$ and $\mathrm{Pin}^{\pm}(d)$. In particular, for compact Lie groups $G$, nil-closedness of $H^{*}(BG)$ is equivalent to the Quillen map being an isomorphism~\cite[Prop.~4.5]{GLZ}; we will use this criterion for the source ring $R=H^{*}(B\mathrm{O}(d),\mathbb Z_{2})$. Since $\res$ possesses nilpotent kernel and cokernel while both source and target are nil-closed, Proposition 3.2 of GLZ implies that
$\res \colon R\to S$ is an isomorphism of graded rings.  Passing to the
$(d + 1)$-st graded component yields the desired bijection
\begin{equation}
\res\colon H^{d+1}(B\mathrm{O}(d),\mathbb Z_{2})\xrightarrow{\;\cong\;}
\varprojlim_{E\in \qa(G)}H^{d+1}(BE,\mathbb Z_{2}),
\label{eq:qisom_mod2}
\end{equation}
so the original $F$-isomorphism refines to a genuine isomorphism.
\end{proof}

\subsection{Explicit construction for the inverse limit over $\qa(\mathrm{O}(d))$}
\label{app:symm_ring}

\begin{notation} We use the notation $\mathcal{A}=\qa(\mathrm{O}(d))$, and $H^{*}(X):=H^{*}(X, \Z_{2})$ for mod-$2$ cohomology. 
\end{notation}
\bigskip 

To demonstrate that the Eq.~\ref{eq:qisom_mod2} is indeed true, we will explicitly construct the inverse limit and show that $F$ is a bijection. The idea is to construct a finite co-initial category with a terminal object. Consider the transporter category over some maximal group $E_{\mathrm{max}}=\mathbb{Z}_{2}^{d}$:$$
\mathcal{T}(E_{\mathrm{max}}),\quad \mathrm{Obj}(\mathcal{T}(E_{\mathrm{max}}))=\{ E \;|\; E\leq E_{\mathrm{max}} \}
$$with morphisms $E \to E'$ generated by:$$
\mathrm{Hom}_{\mathcal{T}(E_{\mathrm{max}})}(E, F) = \{n\in W_{\mathrm{O}(d)}(E_{\mathrm{max}})\;|\; nEn^{-1}\leq F\}.
$$The transporter category is to be understood as a way to make rigid the orbit category of a chosen subgroup $E_{\mathrm{max}}$. 

\bigskip 
\noindent \textbf{Note}: that the transporter category is defined up to isomorphism with respect to conjugation, i.e., given any other $E_{\mathrm{max}}' = E_{\mathrm{max}}^{g}$, we have an isomorphism of categories: $$\Phi: \mathcal{T}(E_{\mathrm{max}}) \to \mathcal{T}(E_{\mathrm{max}}').$$
We can construct the forgetful functor from this transporter category $\mathcal{T}(E_{\mathrm{max}})$ to the Quillen category $\mathcal{A}$:$$
u: \mathcal{T}(E_{\mathrm{max}}) \to \mathcal{A};\quad (E\leq E_{\mathrm{max}}) \mapsto E\in \mathrm{O}(d).
$$where we forget the choice $E_{\mathrm{max}}$.

\bigskip 

\noindent Given two small categories $\mathbf{I},\;\mathbf{A}$, and a functor $$
i:\mathbf{I} \to \mathbf{A}
$$then we can ask if the category $\mathbf{I}$ is co-initial in $\mathbf{A}$. If this condition indeed holds, then we can say that given any contra-variant functor $$
F: \mathbf{A}^{\mathrm{op}} \to \mathbf{Ab}
$$the inverse limit of this functor over $\mathbf{A}$ is the same as the inverse limit of the functor $$
F\vert_{\mathbf{I}}=F\circ i:\; \mathbf{I}^{\mathrm{op}} \to \mathbf{Ab}
$$over $\mathbf{I}$. In other words, we have: $$\varprojlim_{E_{S}\in \mathbf{I}} F(E_{S}) = \varprojlim_{E\in \mathbf{A}} F(E).$$

\bigskip 

\noindent \textbf{Claim 1}: the transporter category $\mathcal{T}(E_{\mathrm{max}})$ is co-initial in the Quillen category $\mathcal{A}_{2}(\mathrm{O}(d))$. 

\noindent \textbf{Proof}: Recall that we define $u: \mathcal{T}(E_{\mathrm{max}}) \to \mathcal{A}$ to be the forgetful functor where we forget the choice of $E_{\mathrm{max}}$ in $\mathcal{A}$. To prove co-initiality, we must show that for every object $E=\mathbb{Z}_{2}^{r}\in \mathcal{A}$, the comma-category $(u\downarrow E)$ is non-empty and has a contractible nerve:

\begin{enumerate}
    \item \textbf{non-empty:} choose $g\in \mathrm{O}(d)$ with $gEg^{-1}\leq E_{\mathrm{max}}$, which is possible because every elementary abelian subgroup is conjugate into $E_{\mathrm{max}}$. Now, consider the object $$
(E^{g}, c_{g^{-1}})\in (u\downarrow E);\quad c_{g^{-1}}: u(E^{g})=E^{g} \to E
$$which clearly exists, and therefore makes $(u\downarrow E)$ non-empty. Intuitively, we want to take $E\in \mathcal{A}$, and produce an object in $\mathcal{T}$, and the obvious way to do this is through conjugation $E^{g}\leq E_{\mathrm{max}}$, and thus $E^{g}\in \mathcal{T}$.
    \item \textbf{terminal object:} We prove $(E^{g}, c_{g^{-1}})$ is a terminal object in $(u\downarrow E)$. Clearly, it should be terminal because it is the largest possible object in an \textit{ascending} category. Let $(F, c_{h})\in(u\downarrow E)$ be an object in the comma-category. Then, we need to show that there exists a morphism $f_{n}: F \to E^{g}$ which induces morphism in the comma-category which is represented by some $n\in W_{\mathrm{O}(d)}(E_{\mathrm{max}})$ such that $nFn^{-1}\leq E^{g}$. 
   By definition, $c_{h}: F \to E$ is a morphism in $\mathcal{A}$, and thus it should be represented by some $h\in \mathrm{O}(d)$ such that $hFh^{-1} \leq E$. 
   
   Since, we know $c_{g^{-1}}(E^{g})=E$, this implies: $$
c_{h}(F)\leq c_{g^{-1}}(E^{g})\implies c_{g}\circ c_{h}(F)\leq E^{g}\implies c_{gh}(F)\leq E^{g}
$$which gives us the require $n=gh$ which induces the morphism $f_{n}:F \to E^{g}$. Since, this holds irrespective of the object $F\in (u\downarrow E)$, this implies $(E^{g}, c_{g^{-1}})$ is a terminal object in the comma category.
\end{enumerate}

\noindent A category with a terminal object has a contractible nerve, and hence $\mathcal{T}(E_{\mathrm{max}})$ is co-initial in the Quillen category $\mathcal{A}_{2}(\mathrm{O}(d))$.
$\square$.

\bigskip  

\noindent This implies $u: \mathcal{T}(E_{\mathrm{max}}) \to \mathcal{A}_{2}(\mathrm{O}(d))$ is co-initial. And consequently, we have that given any contra-variant functor $F: \mathcal{A}^{op} \to \mathbf{Ab}$, the equality: $$\varprojlim_{F\in \mathcal{T}(E_{\mathrm{max}})} F(u(F)) \; \cong \; \varprojlim_{E\in \mathcal{A}} F(E).$$

\bigskip 

\noindent \textbf{Claim 2}: Let 
\begin{equation*}
H\mathbb{Z}_{2}: \mathbf{hoCW}^{\mathrm{op}} \to \mathbf{Ab}; \quad BG \mapsto H^{*}(BG) 
\end{equation*}
be the mod-$2$ cohomology functor. Then, 
\begin{equation*}
\varprojlim\limits_{E\in \mathcal{T}(E_{\mathrm{max}})} (H\mathbb{Z}_{2}\circ u) (E) = H^{*}(BE_{\mathrm{max}})^{W_{\mathrm{O}(d)}(E_{\mathrm{max}})}
\end{equation*}
\begin{proof}[\textnormal{\textbf{Proof}}] First, we observe that every object $E\in \mathcal{T}(E_{\mathrm{max}})$ sits under $E_{\mathrm{max}}$, i.e., $E\leq E_{\mathrm{max}}$. Hence, a compatible family 
\begin{equation*}
\{x_{E}\in H^{*}(BE)\}_{E\leq E_{\mathrm{max}}}
\end{equation*}
is completely determined by its value on the maximal object $x_{E_{\mathrm{max}}}\in H^{*}(BE_{\mathrm{max}})$: 
\begin{equation*}
x_{E} = \res^{E_{\mathrm{max}}}_{E}(x_{E_{\mathrm{max}}})\quad \forall\;\; E\leq E_{\mathrm{max}}.
\end{equation*}
\noindent Second, compatibility with conjugation by $n\in N_{\mathrm{O}(d)}(E_{\mathrm{max}})$ imposes: $$(c_{n})^{*}(x_{E_{\mathrm{max}}}) = x_{E_{\mathrm{max}}} \implies x_{E_{\mathrm{max}}} \in H^{*}(BE_{\mathrm{max}})^{W_{\mathrm{O}(d)}(E_{\mathrm{max}})}.$$
Thus, a compatible family is uniquely determined by a class $x_{E_{\mathrm{max}}}\in H^{*}(BE_{\mathrm{max}})^{W_{\mathrm{O}(d)}(E_{\mathrm{max}})}$. The converse holds as well, but we do not give a proof.
\end{proof}

\bigskip

\subsubsection{Demonstration that $\res$ is an isomorphism}

\noindent The refined Quillen isomorphism is: 
\begin{equation*}
\res: H^{*}(B\mathrm{O}(d)) \to \varprojlim_{E\in \mathcal{A}_{2}(\mathrm{O}(d))} H^{*}(BE).
\end{equation*}
We have just shown very explicitly above that the RHS is equivalent to the Weyl-invariant subring of the cohomology of a maximal group $E=\mathbb{Z}_{2}^{d}$, and therefore: 
\begin{equation*}
\varprojlim_{E\in \mathcal{A}_{2}(\mathrm{O}(d))} H^{*}(BE) = (H^{*}(BE))^{W} = \mathbb{Z}_{2}[x_{1},\dots, x_{d}]^{W}
\end{equation*}
For the standard maximal subgroup $E_{\mathrm{max}}\cong(\mathbb Z_2)^d$ generated by the coordinate reflections, the Weyl group is $W_{\mathrm O(d)}(E_{\mathrm{max}})\cong S_d$, acting on $H^*(BE_{\mathrm{max}}) =\mathbb Z_2[x_1,\ldots,x_d]$ by permuting the generators $x_i$. Therefore the invariant subring is that of symmetric polynomials: $$
\varprojlim_{E\in \mathcal{A}_{2}(\mathrm{O}(d))} H^{*}(BE)  = \mathbb{Z}_{2}[e_{1},\dots, e_{d}]
$$where $e_{1}= \sum_{i}x_{i}$, $e_{2}=\sum_{i<j}x_{i}x_{j}$ and so on until $e_{d}=x_{1}\dots x_{d}$. And therefore, the inverse limit is simply a polynomial ring with $d$-generators each of ranks from $1$ to $d$.  
The LHS side is of course the cohomology generated by Stiefel-Whitney classes and is given by: $$
H^{*}(B\mathrm{O}(d)) = \mathbb{Z}_{2}[w_{1},\dots, w_{d}].
$$and therefore we have a canonical identification by observing that we have: $$\res: w_{i} \mapsto e_{i}.$$Hence, re-affirming our confidence in the Quillen map being an isomorphism.

\subsection{Elementary detection in the pre-TQFT model}
\label{app:elem_detection_pretqft}

The refined Quillen isomorphism has the following immediate consequence: in the pre-TQFT model~\eqref{eq:preTQFT}, an $\mathrm{O}(d)$-symmetric continuum class is uniquely determined by
its elementary restrictions.

\begin{lemma}[Elementary detection in the pre-TQFT model]
\label{lemma:elem_detection_pretqft}
Let
\begin{equation*}
[\bar x],[\bar y]\in H^{d+1}(B\mathrm{O}(d);\mathbb Z_2).
\end{equation*}
Suppose that for every elementary abelian $2$-subgroup $E\leq \mathrm{O}(d)
$ one has
\begin{equation*}
\res^{\mathrm{O}(d)}_{E}([\bar x])
= \res^{\mathrm{O}(d)}_{E}([\bar y])
\quad
\text{in}\;\; H^{d+1}(BE;\mathbb Z_2).
\end{equation*}
Then $[\bar x]=[\bar y]$. Equivalently, an $\mathrm{O}(d)$-symmetric pre-TQFT continuum class is completely determined by its restrictions to elementary abelian $2$-subgroups of $\mathrm{O}(d)$.
\end{lemma}

\begin{proof}
Consider the difference
\begin{equation*}
[\bar z]:=[\bar x]-[\bar y].
\end{equation*}
By assumption,
\begin{equation*}
\res^{\mathrm{O}(d)}_{E}([\bar z])=0
\quad \forall \;\; E\in\qa(\mathrm{O}(d)).
\end{equation*}
Therefore the compatible family of elementary restrictions of $[\bar z]$ is zero:
\begin{equation*}
\res_{\mathcal A}([\bar z])=0\; \in 
\varprojlim_{E\in\qa(\mathrm{O}(d))}
H^{d+1}(BE;\mathbb Z_2).
\end{equation*}
By Prop.~\ref{prop:quillen_mod2_isom}, the map
\[
\res_{\mathcal A}:
H^{d+1}(B\mathrm{O}(d);\mathbb Z_2)
\longrightarrow
\varprojlim_{E\in\qa(\mathrm{O}(d))}
H^{d+1}(BE;\mathbb Z_2)
\]
is an isomorphism. In particular, it is injective. Hence
\begin{equation*}
[\bar z]=0 \implies  [\bar x]=[\bar y].
\end{equation*}
\end{proof}

\begin{lemma}[Elementary responses determine pre-TQFT continuum classes]
\label{lemma:elem_response_detect_pretqft}
Let
\begin{equation*}
[\bar x],[\bar y]\in H^{d+1}(B\mathrm{O}(d);\mathbb Z_2).
\end{equation*}
Suppose that for every elementary abelian $2$-subgroup $E\leq \mathrm{O}(d)$, for every closed $(d+1)$-manifold $M$, and for every background $f:M\to BE$, one has equality of responses
\begin{equation*}
\big\langle f^*\res^{\mathrm{O}(d)}_{E}([\bar x]),[M]_2\big\rangle =
\big\langle f^*\res^{\mathrm{O}(d)}_{E}([\bar y]),[M]_2\big\rangle .
\end{equation*}
Then $[\bar x]=[\bar y]$. Equivalently, the elementary responses of an $\mathrm{O}(d)$-symmetric pre-TQFT continuum class determine the class uniquely.
\end{lemma}

\begin{proof}
Let
\begin{equation*}
[\bar z]:=[\bar x]-[\bar y].
\end{equation*}
By assumption, for every elementary abelian $2$-subgroup $E\leq \mathrm{O}(d)$,
every closed $(d+1)$-manifold $M$, and every background $f:M\to BE$, one has
\begin{equation*}
\big\langle f^*\res^{\mathrm{O}(d)}_{E}([\bar z]),[M]_2\big\rangle=0.
\end{equation*}
Applying Lemma~\ref{lemma:probe_pretqft} to the finite group $E$, we obtain
\begin{equation*}
\res^{\mathrm{O}(d)}_{E}([\bar z])=0
\qquad
\text{in }H^{d+1}(BE;\mathbb Z_2).
\end{equation*}
Since this holds for every elementary abelian $2$-subgroup $E\leq \mathrm{O}(d)$,
Lemma~\ref{lemma:elem_detection_pretqft} implies
\begin{equation*}
[\bar z]=0 \implies 
[\bar x]=[\bar y].
\end{equation*}
\end{proof}

\section{Proofs III: Proof of main theorem in the group cohomology model}
\label{app:lift}

\begin{notation}
By Remark~\ref{rem:abuse_pulledback_twists}, we write $\Z_{w_1}$ for the pulled-back local system on $BE$ induced by $Bi_E:BE\to B\mathrm{O}(d)$. All restriction and Bockstein maps appearing in the overview below are defined precisely in Sec.~\ref{app:main_theorem_prep}; in particular, $\res^{\mathrm{full}}$ denotes restriction before passing to torsion(see Eq.~\ref{eq:res_full}), $\res[2]$ denotes its restriction to order-$2$ torsion classes(see Eq.~\ref{eq:res_tor}), and $\beta_{\lim}$ denotes the inverse-limit Bockstein induced from the objectwise twisted Bockstein maps(see Eq.~\ref{eq:beta_lim}). \\
\noindent We also follow {\bf Notation}~\ref{notation:torsion} to denote various types of torsion subgroups.
\end{notation}

\bigskip 

In this section we prove the main theorem in the group cohomology model~\eqref{eq:grpcohtormodel} defined by the classification functor:
\begin{equation*}
    \Theta^{d+1}: G \mapsto \tor_2 \left(H^{d+2}(BG, \Z_{w_{1}})\right).
\end{equation*}
The proof proceeds by passing through the elementary subcategory $\mathcal{A}:=\mathcal A_2(\mathrm{O}(d))$ of elementary abelian $2$-subgroups. In mod-$2$ cohomology, Quillen’s theorem identifies the continuum cohomology of $B\mathrm{O}(d)$ with the inverse limit over this elementary category (see Prop.~\ref{prop:quillen_mod2_isom}). We study the corresponding lift to twisted integral cohomology using the Bockstein associated to the short exact sequence 
\begin{equation*}
0\to \mathbb Z_{w_1}\xrightarrow{\times 2}\mathbb Z_{w_1}\to\mathbb Z_2\to 0.
\end{equation*}
giving us in particular the commuting square:
\begin{equation*}
    \begin{tikzcd}
        \tor_{2}\left(H^{d+2}(B\mathrm{O}(d), \Z_{w_{1}})\right) \arrow[r, "\res\text{[}2\text{]}"] & \varprojlim\limits_{E\in \mathcal{A}} \;  \tor_{2} \left(H^{d+2}(BE,\Z_{w_1})\right)\\
        H^{d+1}(B\mathrm{O}(d), \Z_{2}) \arrow[r, "\res", "\cong"'] \arrow[u, "\beta_{w_{1}}"] &\varprojlim\limits_{E\in \mathcal{A}} H^{d+1}(BE, \Z_{2}) \arrow[u, "\beta_{\lim}"]
    \end{tikzcd}
\end{equation*}
There is one subtlety with this approach. The lifted Quillen restriction map can be shown to be an isomorphism through this Bokstein lift if and only if the Bockstein maps are surjective. The inverse-limit Bockstein $\beta_{\lim}$ is not surjective in all degrees when $d$ is even\footnote{the twisted Euler class is \textit{free}, i.e. integer quantized, only when $d$ is even. So, the problems associated to the subtlety do not occur when $d$ is odd.}: the twisted Euler class, denoted $E_{d}\in H^{d+2}(B\mathrm{O}(d), \Z_{w_{1}})$, and its product with untwisted Pontryagin classes, denoted $p_{i}\in H^{d+2}(B\mathrm{O}(d), \Z)$, are free continuum classes whose restrictions to elementary subgroups become order-$2$ torsion classes. These classes belong in the elementary inverse limit, i.e. $\res^{\mathrm{full}}(E_{d}\cup f(p_i))\in \varprojlim_{E\in \mathcal{A}}  \tor_{2} \left(H^{d+4n}(BE,\Z_{w_1})\right)$ for some $n>1$, but are not Bockstein-detected. We call this subgroup of images of free continuum classes under $\res^{\mathrm{full}}$ the free-shadow subgroup $S_{\mathrm{free}}^{n}\subseteq \varprojlim_{E\in \mathcal{A}}  \tor_{2} \left(H^{n}(BE,\Z_{w_1})\right)$.

\bigskip 

The result we prove is the following restricted form of the main theorem.

\begin{theorem}[Main theorem in the group cohomology model]
\label{appG:main_theorem}
For every spatial dimension $d\geq 1$, the stratification map
\begin{equation}
F_2:\csptu \twoheadrightarrow \tor_{2}\left(H^{d+2}(B\mathrm{O}(d),\Z_{w_1})\right),
\end{equation}
is surjective. Here 
\begin{equation*}
\csptu=\varprojlim\limits_{G\in \ofin(\mathrm{O}(d))} \; \tor_{2} \left(H^{d+2}(BG,\Z_{w_1})\right).
\end{equation*}
\end{theorem}

\bigskip 

\noindent The proof has two parts. First, we show that in the degree relevant to the main theorem, namely cohomological degree $d+2$, the mod-$2$ Quillen isomorphism admits a faithful lift to order-$2$ twisted integral classes, and is an isomorphism(Lemma~\ref{lemma:lift_quillen}). This allows to prove the main theorem for specific degree $(d+2)$. Second, in general degree, we prove a canonical splitting of the elementary inverse limit $\varprojlim_{E\in \mathcal{A}} \;  \tor_{2} \left(H^{n}(BE,\Z_{w_1})\right)$ into a Bockstein-detected summand $\mathrm{Im}(\beta_{\lim}^{n-1})$ and the free-shadow summand $S_{\mathrm{free}}^{n}$. This splitting defines a canonical retraction of the elementary restriction map,
\begin{equation*}
\widetilde F_2: \varprojlim_{E\in \mathcal{A}} \;  \tor_{2} \left(H^{n}(BE,\Z_{w_1})\right)\twoheadrightarrow \tor_{2}\left(H^{n}(B\mathrm{O}(d),\Z_{w_1})\right)
\end{equation*}
from the elementary sector to the continuum order-$2$ torsion group in arbitrary degree $n>1$. Composing $\widetilde F_2$ with the projection from $\csptu$ gives the stratification map $F_2$ in general degree, and the retraction property implies that $F_2$ is surjective.

\subsection{Preparatory results}
\label{app:main_theorem_prep}

\begin{claim}[Objectwise Bockstein surjectivity]
\label{claim:objectwise_bockstein_surjective}
Let $X$ be a space equipped with a local system $\mathbb Z_{w_1}$. Then the twisted Bockstein
\begin{equation*}
\beta^{n}_{w_1}:H^n(X,\mathbb Z_2)\to H^{n+1}(X,\mathbb Z_{w_1})
\end{equation*}
surjects onto $H^{n+1}(X,\mathbb Z_{w_1})[2]$.
\end{claim}

\begin{proof}[\textnormal{\textbf{Proof of Claim~\ref{claim:objectwise_bockstein_surjective}}}]
We can define the SES of coefficients 
\begin{equation*}
    0\to \Z_{w_{1}} \xrightarrow[]{\times 2} \Z_{w_{1}} \xrightarrow[]{\mathrm{mod}\; 2} \Z_{2} \to 0,
\end{equation*}
using which we obtain the LES in cohomology
\begin{equation}
    \begin{tikzcd}
    \cdots H^{d+1}(X,\Z_{w_{1}}) \rar[]{(\times 2)^{d+1}} & H^{d+1}(X,\Z_{w_{1}}) \rar[]{\rho^{d+1}}
             \ar[draw=none]{d}[name=X, anchor=center]{}
    & H^{d+1}(X,\Z_{2}) \ar[rounded corners,
            to path={ -- ([xshift=2ex]\tikztostart.east)
                      |- (X.center) \tikztonodes
                      -| ([xshift=-2ex]\tikztotarget.west)
                      -- (\tikztotarget)}]{dll}[at end]{\beta^{d+1}_{w_{1}}} \\      
  H^{d+2}(X,\Z_{w_{1}}) \rar[]{(\times 2)^{d+2}} & H^{d+2}(X,\Z_{w_{1}}) \rar[]{\rho^{d+2}} & H^{d+2}(X,\Z_{2})\cdots 
\end{tikzcd}
\label{eq:LES}
\end{equation}
Exactness then implies: 
\begin{equation*}
    \mathrm{Im}\left(\beta^{d+1}_{w_{1}}\right) = \mathrm{ker}\left((\times 2)^{d+2}    \right)
\end{equation*}
And therefore, every Bockstein image is killed by the $\times 2$ map, and the 2-torsion subgroup is the full image of the map $\mathrm{Im}(\beta^{d+1}_{w_{1}}) = H^{d+2}(X, \mathbb{Z}_{w_{1}})[2]$. In other words, the map with target restricted to the order-$2$ torsion subgroup
\begin{equation*}
    \beta_{w_{1}}^{d+1}: H^{d+1}(X, \mathbb{Z}_{2}) \twoheadrightarrow H^{d+2}(X, \mathbb{Z}_{w_{1}})[2].    
\end{equation*}
is a surjection.
\end{proof}

\bigskip 

\begin{claim}[Injectivity of reduction on order-$2$ torsion]
\label{claim:rho_injective_order2}
Suppose $H^n(X,\Z_{w_1})$ has no exact $4$-torsion, i.e.
\begin{equation}
H^n(X,\Z_{w_1})[4]=H^n(X,\Z_{w_1})[2].
\end{equation}
Then the restricted mod-$2$ reduction map
\begin{equation}
\rho_{X}^{n}\vert_{[2]}:
H^n(X,\Z_{w_1})[2]\longrightarrow H^n(X,\Z_2)
\end{equation}
is injective.
\end{claim}

\begin{proof}[\textnormal{\textbf{Proof of Claim~\ref{claim:rho_injective_order2}}}]
We use the same short exact sequence of coefficients
\begin{equation}
0\to \Z_{w_1}\xrightarrow{\times 2}\Z_{w_1}\xrightarrow{\rho}\Z_2\to 0.
\end{equation}
The associated long exact sequence in cohomology contains the segment
\begin{equation}
H^n(X,\Z_{w_1})
\xrightarrow{(\times 2)^{n}}
H^n(X,\Z_{w_1})
\xrightarrow{\rho_{X}^{n}}
H^n(X,\Z_2).
\end{equation}
Exactness implies $\ker(\rho_{X}^{n})=\operatorname{Im}\left((\times 2)^{n}\right)$.

\bigskip 

\noindent Now let $x\in H^n(X,\Z_{w_1})[2]$ and suppose $\rho_{X}^{n}(x)=0$. Since $x\in \ker(\rho_{X}^{n})$, exactness implies that there exists some $u\in H^n(X,\Z_{w_1})$ such that $x=2u$. But $x$ is order-$2$, so $2x=0$, and hence $4u=0$. Thus $u\in H^n(X,\Z_{w_1})[4]$. By the no exact $4$-torsion assumption, $u\in H^n(X,\Z_{w_1})[2]$, so $2u=0$. Therefore $x=2u=0$. Hence $\ker(\rho_{X}^{n}\vert_{[2]})=0$, and the restricted reduction map $\rho_{X}^{n}\vert_{[2]}$ is injective.
\end{proof}

\medskip 

\begin{claim}[Injectivity of reduction on $B\mathrm{O}(d)$]
\label{claim:rho_BO_injective}
The restricted mod-$2$ reduction map
\begin{equation}
\rho_{B\mathrm{O}(d)}\vert_{[2]}:
H^{d+2}(B\mathrm{O}(d),\Z_{w_1})[2]
\longrightarrow
H^{d+2}(B\mathrm{O}(d),\Z_2)
\end{equation}
is injective.
\end{claim}

\begin{proof}[\textnormal{\textbf{Proof of Claim~\ref{claim:rho_BO_injective}}}]
By Theorem~\ref{thm:Greenblatt_twisted_no_odd_torsion}, all torsion in
$H^*(B\mathrm{O}(d),\Z_{w_1})$ is of order $2$. In particular,
\begin{equation}
H^{d+2}(B\mathrm{O}(d),\Z_{w_1})[4]
=
H^{d+2}(B\mathrm{O}(d),\Z_{w_1})[2].
\end{equation}
Thus Claim~\ref{claim:rho_injective_order2} applies with $X=B\mathrm{O}(d)$ and $n=d+2$, proving that $\rho_{B\mathrm{O}(d)}\vert_{[2]}$ is injective.
\end{proof}

\medskip 

\noindent \textbf{Convention.}
In the remainder of this subsection, we write $\rho:=\rho_{B\mathrm{O}(d)}$ when the source is $H^*(B\mathrm{O}(d),\Z_{w_1})$. For every $E\in\mathcal A_2(\mathrm{O}(d))$ and every positive degree $n>0$, the twisted integral cohomology group
\begin{equation}
H^n(BE,\mathbb Z_{(Bi_E)^*w_1})
\end{equation}
is entirely order-$2$, as shown in App.~\ref{app:elem_mod2},~\ref{app:elem_twisted}. Hence
\begin{equation}
H^n(BE,\mathbb Z_{w_1}) = H^n(BE,\mathbb Z_{w_1})[2].
\end{equation}
In what follows, we suppress the symbol $[2]$ on the elementary-group side. We keep the notation $[2]$ on the $B\mathrm{O}(d)$ side, where twisted integral cohomology may contain non-torsion classes.

\noindent Consequently on elementary subgroups in positive degree, restricting the reduction map to the order-$2$ subgroup does not change the domain:
\begin{equation}
\rho^{n}_E\vert_{[2]} := \rho^{n}_E\big\vert_{H^{n}(BE,\Z_{w_1})[2]}=\rho^{n}_{E}.
\end{equation}

\bigskip 

\noindent For each $E\in\mathcal A_2(\mathrm{O}(d))$, let
\begin{equation}
\beta_E^n: H^n(BE,\mathbb Z_2) \longrightarrow H^{n+1}(BE,\mathbb Z_{w_1})
\label{eq:Bockstein_E}
\end{equation}
denote the twisted Bockstein with codomain restricted to its image $H^{n+1}(BE,\mathbb Z_{w_1})[2]$, and let
\begin{equation}
\rho_E^n: H^n(BE,\mathbb Z_{w_1}) \longrightarrow H^n(BE,\mathbb Z_2)
\label{eq:reduction_E}
\end{equation}
denote the mod-$2$ reduction map. These maps are natural in $E$. Therefore, as $E$ varies over $\mathcal A_2(\mathrm{O}(d))$, they define natural transformations of the corresponding cohomology diagrams. By the universal property of the inverse limit, they induce degreewise maps on inverse limits:
\begin{gather}
    \rho^{n}_{\lim}: \varprojlim_{E\in\mathcal A_2(\mathrm{O}(d))} H^n(BE,\Z_{w_1})
\longrightarrow \varprojlim_{E\in\mathcal A_2(\mathrm{O}(d))} H^n(BE,\Z_2),
    \label{eq:rho_lim}\\
    \beta^{n}_{\lim}: \varprojlim_{E\in\mathcal A_2(\mathrm{O}(d))} H^n(BE,\Z_{2})
\longrightarrow \varprojlim_{E\in\mathcal A_2(\mathrm{O}(d))} H^{n+1}(BE,\Z_{w_{1}}).
    \label{eq:beta_lim}
\end{gather}

\medskip 

\begin{claim}[Injectivity of $\rho^{n}_{\lim}$]
\label{claim:rho_lim_injective}
For every elementary abelian subgroup $E\in\mathcal A_2(\mathrm{O}(d))$ and every $n>0$, the mod-$2$ reduction map
\begin{equation}
\rho^{n}_E:
H^n(BE,\Z_{w_1}) \longrightarrow H^n(BE,\Z_2)
\end{equation}
is injective. Consequently, the induced inverse-limit map $\rho^{n}_{\lim}$ in Eq.~\eqref{eq:rho_lim} is injective.
\end{claim}

\begin{proof}[\textnormal{\textbf{Proof of Claim~\ref{claim:rho_lim_injective}}}]
For each $E\in\mathcal A_2(\mathrm{O}(d))$, the group $H^n(BE,\Z_{w_1})$ has no exact $4$-torsion by App.~\ref{app:elem_mod2},~\ref{app:elem_twisted}. Therefore Claim~\ref{claim:rho_injective_order2} applies objectwise. Since $H^n(BE,\Z_{w_1})=H^n(BE,\Z_{w_1})[2]$ in the present range, it follows that $\rho^{n}_E$ itself is injective.

\bigskip 

\noindent It remains to pass to the inverse limit. Let
$x\in \varprojlim_{E\in\mathcal A_2(\mathrm{O}(d))}H^n(BE,\Z_{w_1})$
and suppose $\rho^{n}_{\lim}(x)=0$. Applying the projection to each $E$ gives $\rho^{n}_E(\pi_E x)=0$. Since each $\rho^{n}_E$ is injective, we get $\pi_E x=0$ for every $E$. An element of an inverse limit is determined by all of its projections, so $x=0$. Hence $\rho^{n}_{\lim}$ is injective.
\end{proof}

\bigskip 

\noindent Rewriting, the map in terms of the simplified Weyl-invariant subgroup of the maximal group notation, we have: $$\rho^{n}_{\lim}: H^{n}(BE_{\mathrm{max}}, \mathbb{Z}_{w_{1}})^{W} \to H^{n}(BE_{\mathrm{max}})^{W},$$ 
and this is injective. 

\bigskip 

\begin{notation}
Before moving on, we set up the degreewise notation used in the proof. Define
\begin{equation}
\Lambda^n := \varprojlim_{E\in\mathcal A_2(\mathrm{O}(d))} H^n(BE,\mathbb Z_2), 
\end{equation}
and
\begin{equation}
\widetilde{\Lambda}^n := \varprojlim_{E\in\mathcal A_2(\mathrm{O}(d))}
H^n(BE,\mathbb Z_{w_1}).
\end{equation}
The objectwise Bockstein maps~\eqref{eq:Bockstein_E} and mod-$2$ reduction maps~\eqref{eq:reduction_E} induce degreewise maps
\begin{equation}
\beta_{\lim}^n: \Lambda^n \longrightarrow \widetilde{\Lambda}^{n+1},\qquad 
\rho_{\lim}^n: \widetilde{\Lambda}^n \longrightarrow \Lambda^n.
\end{equation}
We define
\begin{equation}
D_{\lim}^n := \rho_{\lim}^{n+1}\circ \beta_{\lim}^n:
\Lambda^n \longrightarrow \Lambda^{n+1}.
\end{equation}
\end{notation}
Objectwise exactness of the LES in Eq.~\ref{eq:LES} implies that, for every
$E\in\mathcal A_2(\mathrm{O}(d))$,
\begin{equation}
\beta_E^{n+1}\circ \rho_E^{n+1}=0.
\end{equation}
Since $\beta_{\lim}^{n+1}$ and $\rho_{\lim}^{n+1}$ are induced by the objectwise maps, the same relation holds after passing to the inverse limit:
\begin{equation}
\beta_{\lim}^{n+1}\circ \rho_{\lim}^{n+1}=0.
\label{eq:LES_limit_exactness}
\end{equation}
Indeed, after projecting to each $E$, this composite is $\beta_E^{n+1}\circ \rho_E^{n+1}$,
which vanishes objectwise. Since elements of an inverse limit are determined by their projections, the inverse-limit composite~\eqref{eq:LES_limit_exactness} vanishes.
Therefore
\begin{equation}
D_{\lim}^{n+1}\circ D_{\lim}^{n} = \rho_{\lim}^{n+2} \circ
\bigl(\beta_{\lim}^{n+1}\circ \rho_{\lim}^{n+1}\bigr) \circ \beta_{\lim}^{n} = 0,
\end{equation}
i.e. $D_{\lim}^{2}=0$ and the maps $D_{\lim}^{n}$ assemble into a cochain complex.
\begin{equation}
\label{eq:Lambda_complex}
(\Lambda,D_{\lim}), \qquad \Lambda:=\bigoplus_{n\geq 0}\Lambda^n, \qquad
D_{\lim}:=\bigoplus_{n\geq 0}D_{\lim}^n.
\end{equation}

\bigskip

We now make the operator action of $D_{\lim}$ explicit. For any space $X$ equipped with a local system $\Z_{w_1}$, the composite
\begin{equation}
D_X:=\rho_X\circ \beta_{w_1}
\end{equation}
is given on mod-$2$ cohomology by
\begin{equation}
D_X(\alpha)=w_1(X)\cup \alpha+\Sq^1(\alpha).
\label{eq:def_DX_w1_Sq1}
\end{equation}
Moreover, this construction is natural in $X$. Under Quillen's restriction isomorphism, we identify
\begin{equation}
\Lambda = \varprojlim_{E\in\mathcal A_2(\mathrm{O}(d))}H^*(BE,\Z_2) \cong \mathbb F_2[e_1,\dots,e_d],
\end{equation}
where $e_i$ is the image of the Stiefel-Whitney class $w_i\in H^i(B\mathrm{O}(d),\Z_2)$ under restriction to the Weyl-invariant subring of a maximal elementary abelian subgroup. On $B\mathrm{O}(d)$, using the Wu formula
\begin{equation}
\Sq^1(w_i)=w_1w_i+(i-1)w_{i+1},
\end{equation}
we get
\begin{equation}
D_{B\mathrm{O}(d)}(w_i) = w_1w_i+\Sq^1(w_i) = (i-1)w_{i+1}.
\end{equation}
By naturality, the same formula holds after passing to the inverse limit:
\begin{equation}
D_{\lim}(e_i) = (i-1)e_{i+1},
\label{eq:Dlim_on_ei}
\end{equation}
where we set $e_{d+1}=0$. Equivalently,
\begin{equation}
D_{\lim}(e_i) = \begin{cases}
e_{i+1}, & i\ \mathrm{even},\\
0, & i\ \mathrm{odd}.
\end{cases}
\end{equation}
In addition, we note $D_{\lim}(1)=e_1$ and a important corner case: when we have even $d=2m$ the top class $e_d=e_{2m}$ satisfies $D_{\lim}(e_d)=e_{d+1}=0$ even though $d$ is even.

\bigskip 

\noindent Let
\begin{equation}
m_o:=\left\lfloor\frac{d-1}{2}\right\rfloor,
\qquad
m_e:=\left\lfloor\frac{d}{2}\right\rfloor.
\end{equation}
We introduce generators $\{z_t\}_{t=0}^{m_o}$ and $\{y_t\}_{t=1}^{m_e}$ by
\begin{equation}
z_t:=e_{2t+1}\quad (t=0,\dots,m_o), \qquad y_t:=e_{2t}\quad (t=1,\dots,m_e).
\end{equation}
Using Eq.~\eqref{eq:Dlim_on_ei}, the action on the new generators is
\begin{gather}
D_{\lim}(z_t)=0 \quad (t=0,\dots,m_o),\quad D_{\lim}(y_t)=z_t \quad (t=1,\dots,m_o).
\end{gather}
If $d$ is even, there is an additional top even generator $y_{m_e}=e_d$, and since we set $e_{d+1}=0$, it satisfies $D_{\lim}(y_{m_e})=0$.
Equivalently,
\begin{equation}
\Lambda \cong \mathbb F_2[z_0,y_1,z_1,\dots,y_{m_e},(z_{m_e})],
\end{equation}
where $z_{m_e}$ is present only when $d$ is odd. In particular,
\begin{equation}
|z_t|=2t+1, \qquad |y_t|=2t.
\end{equation}

\noindent Together with $D_{\lim}(1)=z_0$ and the Cartan formula for $\Sq^1$, this gives the operator formula
\begin{equation}
D_{\lim}=\delta+T, \qquad \delta:=\sum_{t=1}^{m_o}z_t\partial_{y_t},
\qquad T:=z_0(1+\mathcal E),
\end{equation}
where
\begin{equation}
\mathcal E
:=
z_0\partial_{z_0}
+
\sum_{t=1}^{m_e}y_t\partial_{y_t}
+
\sum_{t=1}^{m_o}z_t\partial_{z_t}.
\end{equation}

\bigskip 

\begin{claim}[Degree-$d+2$ acyclicity]
\label{claim:degree_dplus2_acyclic}
The degree-$d+2$ cohomology of $(\Lambda,D_{\lim})$ vanishes:
\begin{equation}
H^{d+2}(\Lambda,D_{\lim})=0.
\end{equation}
More precisely, if $d=2m+1$, then
\begin{equation}
H^{*}(\Lambda,D_{\lim})=0,
\end{equation}
whereas if $d=2m$, then
\begin{equation}
H^{*}(\Lambda,D_{\lim})
\cong
y_m\,\mathbb F_2[y_1^2,\dots,y_m^2].
\end{equation}
In particular, in the even-dimensional case the nonzero cohomology begins in degree $d$ and then occurs only in degrees $d+4k$, so it does not appear in degree $d+2$.
\end{claim}

\begin{proof}[\textnormal{\textbf{Proof of Claim~\ref{claim:degree_dplus2_acyclic}}}]
We use the decomposition
\begin{equation}
D_{\lim}=\delta+T,
\qquad
\delta=\sum_{t=1}^{m_o}z_t\partial_{y_t},
\qquad
T=z_0(1+\mathcal E),
\end{equation}
where $m_o=\lfloor (d-1)/2\rfloor$ and $m_e=\lfloor d/2\rfloor$. Recall that the essential difference between odd and even $d$ is that, when $d$ is even, the top even generator $y_{m_e}=e_d$ has no partner $z_{m_e}=e_{d+1}$.

\medskip

Define a decreasing filtration by powers of $z_0$:
\begin{equation}
F^p\Lambda:=z_0^p\Lambda,
\qquad
\Lambda=F^0\Lambda\supseteq F^1\Lambda\supseteq F^2\Lambda\supseteq \cdots .
\label{eq:z0_filtration}
\end{equation}
Since $\delta$ does not change the $z_0$-degree and $T$ increases it by one, we have
\begin{equation}
\delta(F^p\Lambda)\subseteq F^p\Lambda,
\qquad
T(F^p\Lambda)\subseteq F^{p+1}\Lambda\subseteq F^p\Lambda.
\end{equation}
Thus $D_{\lim}(F^p\Lambda)\subseteq F^p\Lambda$, so $(\Lambda,D_{\lim})$ is a filtered complex. In each fixed total degree this filtration is finite, and hence the associated spectral sequence converges to $H^*(\Lambda,D_{\lim})$.

\medskip

Let $A:=\Lambda/(z_0)$. Then $\Lambda\cong \bigoplus_{p\geq 0}z_0^p A$, and
\begin{equation}
\mathrm{gr}^p\Lambda:=\frac{F^p\Lambda}{F^{p+1}\Lambda}
\cong z_0^p A.
\label{eq:associated_graded_identification}
\end{equation}
\noindent The filtration above produces a spectral sequence whose first pages are the iterated cohomology groups
\begin{equation}
E_1\cong H_\delta(\Lambda),
\qquad
E_2\cong H_T\bigl(H_\delta(\Lambda)\bigr).
\label{eq:E1_E2_iterated}
\end{equation}
We now compute these pages, separating the odd and even cases.

\medskip

\noindent \textit{Odd-dimensional case.}
Suppose $d=2m+1$. Then $m_o=m_e=m$, and
$\Lambda\cong \mathbb F_2[z_0,y_1,z_1,\dots,y_m,z_m]$. In this case every $y_t$ is paired with a corresponding $z_t$. Since $\delta$ does not involve $z_0$, we first compute $H_\delta(A)$ and then adjoin $z_0$. For a single pair $(y,z)$, every polynomial can be written uniquely as
\begin{equation}
f(y,z)=f_0(y^2,z)+y f_1(y^2,z).
\end{equation}
Over $\mathbb F_2$, we have $\partial_y(y^{2k})=0$ and $\partial_y(y^{2k+1})=y^{2k}$, so
\begin{equation}
z\partial_y f=z f_1(y^2,z).
\end{equation}
Therefore $\ker(z\partial_y)=\mathbb F_2[y^2,z]$ and $\operatorname{Im}(z\partial_y)=z\cdot \mathbb F_2[y^2,z]$, which gives
\begin{equation}
H\bigl(\mathbb F_2[y,z],z\partial_y\bigr)
\cong
\mathbb F_2[y^2].
\end{equation}
Tensoring over the $m$ independent pairs $(y_t,z_t)$ gives
\begin{equation}
H_\delta(A)\cong \mathbb F_2[y_1^2,\dots,y_m^2],
\qquad
H_\delta(\Lambda)\cong \mathbb F_2[z_0,y_1^2,\dots,y_m^2].
\label{eq:Hdelta_result_odd}
\end{equation}

Now compute the induced $T$-cohomology. A class in $H_\delta(\Lambda)$ is represented by $z_0^p f(y_1^2,\dots,y_m^2)$. Since each $y_t$ appears only with even exponent in $f$, the Euler operator satisfies $\mathcal E(f)\equiv 0\pmod 2$. Hence
\begin{equation}
T\bigl(z_0^p f\bigr)
=
z_0(1+\mathcal E)\bigl(z_0^p f\bigr)
=
(1+p)z_0^{p+1}f.
\end{equation}
Thus the induced differential sends even powers of $z_0$ isomorphically onto the odd powers and vanishes on odd powers. Equivalently, if
$E_1^{\mathrm{even}}=\mathbb F_2[z_0^2,y_1^2,\dots,y_m^2]$ and
$E_1^{\mathrm{odd}}=z_0\mathbb F_2[z_0^2,y_1^2,\dots,y_m^2]$, then the induced differential is multiplication by $z_0$ from $E_1^{\mathrm{even}}$ to $E_1^{\mathrm{odd}}$ and vanishes on $E_1^{\mathrm{odd}}$. Since $E_1^{\mathrm{even}}$ has no zero divisors, this gives
\begin{equation}
E_2=H_T\bigl(H_\delta(\Lambda)\bigr)=0.
\end{equation}
Therefore the spectral sequence collapses to zero, and
\begin{equation}
H^*(\Lambda,D_{\lim})=0
\end{equation}
when $d$ is odd. In particular, $H^{d+2}(\Lambda,D_{\lim})=0$.

\medskip

\noindent \textit{Even-dimensional case.}
Now suppose $d=2m$. Then $m_e=m$ and $m_o=m-1$, so
\begin{equation}
\Lambda\cong \mathbb F_2[z_0,y_1,z_1,\dots,z_{m-1},y_m].
\end{equation}
The odd-dimensional computation fails in exactly one place: the top even generator $y_m=e_d$ has no partner $z_m=e_{d+1}$. Thus $\delta$ pairs $y_1,\dots,y_{m-1}$ with $z_1,\dots,z_{m-1}$, but leaves $y_m$ unpaired.

\noindent Repeating the same $\delta$-cohomology computation gives
\begin{equation}
H_\delta(\Lambda)
\cong
\mathbb F_2[z_0,y_1^2,\dots,y_{m-1}^2,y_m].
\label{eq:Hdelta_result_even}
\end{equation}
The induced $T$-differential is still $z_0(1+\mathcal E)$ on this ring. The same parity computation as above now leaves precisely the odd powers of the unpaired generator $y_m$. Thus
\begin{equation}
H^*(\Lambda,D_{\lim})
\cong
y_m\,\mathbb F_2[y_1^2,\dots,y_{m-1}^2,y_m^2].
\label{eq:even_cohomology_result}
\end{equation}
This is the precise failure of full acyclicity in even dimension.

\bigskip 

\noindent However, this failure does not affect the degree needed here. Since $|y_m|=d$ and the remaining generators in \eqref{eq:even_cohomology_result} have degrees divisible by $4$, the nonzero cohomology occurs only in degrees $d,d+4,d+8,\dots$. In particular, there is no cohomology in degree $d+2$. Therefore
\begin{equation}
H^{d+2}(\Lambda,D_{\lim})=0
\end{equation}
also in the even-dimensional case. Combining the odd and even cases proves the claim.
\end{proof}

\medskip

\begin{claim}
\label{claim:imD_equal_imrho}
If
\begin{equation}
\operatorname{Im}(D_{\lim}^{d+1})
= \operatorname{Im}(\rho_{\lim}^{d+2}),
\end{equation}
then $\beta_{\lim}^{d+1}$ is surjective.
\end{claim}

\begin{proof}[\textnormal{\textbf{Proof of Claim~\ref{claim:imD_equal_imrho}}}]
Let
\begin{equation}
b\in  \varprojlim_{E\in\mathcal A_2(\mathrm{O}(d))} H^{d+2}(BE,\Z_{w_1})
\end{equation}
be arbitrary. Then $ \rho_{\lim}^{d+2}(b) \in \operatorname{Im}(\rho_{\lim}^{d+2})$.
By the assumption $\operatorname{Im}(D_{\lim}^{d+1}) = \operatorname{Im}(\rho_{\lim}^{d+2})$, there exists
\begin{equation}
a\in  \varprojlim_{E\in\mathcal A_2(\mathrm{O}(d))} H^{d+1}(BE,\Z_2)
\end{equation}
such that
\begin{equation}
D_{\lim}^{d+1}(a) = \rho_{\lim}^{d+2}(b).
\end{equation}
Since $D_{\lim}^{d+1} = \rho_{\lim}^{d+2}\circ \beta_{\lim}^{d+1}$,
we obtain
\begin{equation}
\rho_{\lim}^{d+2}\bigl(\beta_{\lim}^{d+1}(a)\bigr) = \rho_{\lim}^{d+2}(b).
\end{equation}
By injectivity of $\rho_{\lim}^{d+2}$, it follows that $\beta_{\lim}^{d+1}(a)=b$.
Since $b$ was arbitrary, $\beta_{\lim}^{d+1}$ is surjective.
\end{proof}

\bigskip

\begin{claim}
\label{claim:acyclic_implies_imD_equal_imrho}
If $ H^{d+2}(\Lambda,D_{\lim})=0$, i.e.
\begin{equation}
\ker\left(D_{\lim}^{d+2}:\Lambda^{d+2}\to\Lambda^{d+3}\right)
= \operatorname{Im}\left(D_{\lim}^{d+1}:\Lambda^{d+1}\to\Lambda^{d+2}\right),
\end{equation}
then
\begin{equation}
\operatorname{Im}(D_{\lim}^{d+1})
= \operatorname{Im}(\rho_{\lim}^{d+2}).
\end{equation}
\end{claim}
\begin{proof}[\textnormal{\textbf{Proof of Claim~\ref{claim:acyclic_implies_imD_equal_imrho}}}]
Since $D_{\lim}^{d+1}=\rho_{\lim}^{d+2}\circ \beta_{\lim}^{d+1}$, we immediately have
\begin{equation}
\operatorname{Im}(D_{\lim}^{d+1})
\subseteq
\operatorname{Im}(\rho_{\lim}^{d+2}).
\end{equation}
It remains to prove the reverse inclusion. Let $x\in \operatorname{Im}(\rho_{\lim}^{d+2})$. 
Since $x\in\operatorname{Im}(\rho_{\lim}^{d+2})$, write $x=\rho_{\lim}^{d+2}(b)$; then $\beta_{\lim}^{d+2}(x)=\beta_{\lim}^{d+2}\rho_{\lim}^{d+2}(b)=0$ (see~\eqref{eq:LES_limit_exactness}).
Hence
\begin{equation}
D_{\lim}^{d+2}(x)
=
\rho_{\lim}^{d+3}\vert_{[2]}\bigl(\beta_{\lim}^{d+2}(x)\bigr)
=
0.
\end{equation}
Thus $x\in \ker(D_{\lim}^{d+2})$. By the assumption $H^{d+2}(\Lambda,D_{\lim})=0$, we have
\begin{equation}
\ker(D_{\lim}^{d+2})
=
\operatorname{Im}(D_{\lim}^{d+1}).
\end{equation}
Therefore $x\in \operatorname{Im}(D_{\lim}^{d+1})$. Since $x$ was arbitrary, this proves
$\operatorname{Im}(\rho_{\lim}^{d+2})\subseteq \operatorname{Im}(D_{\lim}^{d+1})$. Combining this with the first inclusion gives
\begin{equation}
\operatorname{Im}(D_{\lim}^{d+1})
=
\operatorname{Im}(\rho_{\lim}^{d+2}).
\end{equation}
\end{proof}

\bigskip 

\begin{notation}
For each $E\in \mathcal A_2(\mathrm{O}(d))$, let $i_E:E\hookrightarrow \mathrm{O}(d)$ denote the inclusion, and let $Bi_E:BE\longrightarrow B\mathrm{O}(d)$ be the induced map on classifying spaces. We write
\begin{equation}
(Bi_E)^*:H^*(B\mathrm{O}(d),\Z_2)\longrightarrow H^*(BE,\Z_2)
\end{equation}
for the induced restriction map in mod-$2$ cohomology. Similarly, the twist $w_1\in H^1(B\mathrm{O}(d),\Z_2)$ pulls back to $(Bi_E)^*w_1\in H^1(BE,\Z_2)$, and restriction gives
\begin{equation}
(Bi_E)^*:H^*(B\mathrm{O}(d),\Z_{w_1}) \longrightarrow H^*(BE,\Z_{(Bi_E)^*w_1})
\end{equation}
with local coefficients.
\end{notation}

These restriction maps are compatible with morphisms in $\mathcal A_2(\mathrm{O}(d))$. Indeed, if
$u:E\to E'$ is a morphism in $\mathcal A_2(\mathrm{O}(d))$, then
$Bi_E\simeq Bi_{E'}\circ Bu$, and hence
\begin{equation}
(Bu)^*\circ (Bi_{E'})^*=(Bi_E)^*.
\end{equation}
Therefore, by the universal property of the inverse limit, the compatible family
$\{(Bi_E)^*\}_E$ defines degreewise maps
\begin{equation}
\res^n:
H^n(B\mathrm{O}(d),\Z_2)
\longrightarrow
\varprojlim_{E\in\mathcal A_2(\mathrm{O}(d))}
H^n(BE,\Z_2),
\end{equation}
and
\begin{equation}
\res_{\mathcal A}^{\mathrm{full},n}:
H^n(B\mathrm{O}(d),\Z_{w_1})
\longrightarrow
\varprojlim_{E\in\mathcal A_2(\mathrm{O}(d))}
H^n(BE,\Z_{(Bi_E)^*w_1}).
\label{eq:res_full}
\end{equation}
The first map is the mod-$2$ restriction map. The second map is the full twisted-integral restriction map; it is defined before restricting to torsion.

Restricting the domain of $\res_{\mathcal A}^{\mathrm{full},n}$ to the order-$2$ torsion subgroup gives
\begin{equation}
\res_{\mathcal A}^{n}[2]:
H^n(B\mathrm{O}(d),\Z_{w_1})[2]
\longrightarrow
\varprojlim_{E\in\mathcal A_2(\mathrm{O}(d))}
H^n(BE,\Z_{(Bi_E)^*w_1}).
\label{eq:res_tor}
\end{equation}
In the main degree $n=d+2$, we write
\begin{equation}
\res:=\res^{d+2}\;,\quad \res[2]:=\res_{\mathcal A}^{d+2}[2].
\end{equation}

\bigskip 

\begin{claim}[Naturality of $\res\text{[}2\text{]}$]
\label{claim:res2_exists_natural}
The map $\res[2]=\res_{\mathcal A}^{d+2}[2]$ is compatible with the Bockstein maps: \begin{equation} \res[2]\circ \beta_{w_1}^{d+1} = \beta_{\lim}^{d+1}\circ \res^{d+1}, \end{equation} and with mod-$2$ reduction: \begin{equation} \res^{d+2}\circ \rho^{d+2}\vert_{[2]} = \rho_{\lim}^{d+2}\circ \res[2]. \end{equation}
\end{claim}

\begin{proof}[\textnormal{\textbf{Proof of Claim~\ref{claim:res2_exists_natural}}}]
By construction, \begin{equation} \res[2] = \res_{\mathcal A}^{\mathrm{full},d+2} \big\vert_{H^{d+2}(B\mathrm{O}(d),\Z_{w_1})[2]}. \end{equation} 
First, we prove compatibility with the Bockstein. Since elements in the inverse limit are determined by their projections, it suffices to check the projections after applying $\pi_E$ for every $E$. Using the defining property of $\res[2]$ and naturality of the Bockstein, we obtain
\begin{equation}
\pi_E\circ \res[2]\circ \beta^{d+1}_{w_1}
=
(Bi_E)^*\circ \beta^{d+1}_{w_1}
=
\beta^{d+1}_{(Bi_E)^*w_1}\circ (Bi_E)^*.
\end{equation}
On the other hand, by the definition of $\beta_{\lim}^{d+1}$ and the usual mod-$2$ restriction map $\res$, we also have
\begin{equation}
\pi_E\circ \beta_{\lim}^{d+1}\circ \res
=
\beta^{d+1}_{(Bi_E)^*w_1}\circ (Bi_E)^*.
\end{equation}
Therefore the two maps $\res[2]\circ \beta^{d+1}_{w_1}$ and $\beta_{\lim}^{d+1}\circ \res$ have the same projection to every $E$, and hence are equal.

\medskip

The compatibility with mod-$2$ reduction is analogous. Again, it suffices to check after projecting to each $E$. The projection of $\res\circ \rho^{d+2}\vert_{[2]}$ to the $E$-component is $(Bi_E)^*\circ \rho^{d+2}\vert_{[2]}$, while the projection of $\rho_{\lim}^{d+2}\circ \res[2]$ is $\rho^{d+2}_E\vert_{[2]}\circ (Bi_E)^*$. These two maps agree by naturality of mod-$2$ reduction. Therefore
\begin{equation}
\res\circ \rho^{d+2}\vert_{[2]}
=
\rho_{\lim}^{d+2}\circ \res[2].
\end{equation}
This proves the existence and the stated naturality properties of $\res[2]$.
\end{proof} 

\subsection{Main results}
\label{app:main_throerem_proof_main_results}

\begin{proposition}[Surjectivity of the inverse-limit Bockstein]
\label{prop:limit_bockstein_surjective}
The inverse-limit Bockstein
\[
\beta_{\lim}^{d+1}:
\varprojlim_{E\in\mathcal A_2(\mathrm{O}(d))} H^{d+1}(BE,\mathbb Z_2) \longrightarrow 
\varprojlim_{E\in\mathcal A_2(\mathrm{O}(d))} H^{d+2}(BE,\mathbb Z_{w_1})
\]
is surjective.
\end{proposition}

\begin{proof}[\textnormal{\textbf{Proof of Proposition~\ref{prop:limit_bockstein_surjective}}}]
By Claim~\ref{claim:degree_dplus2_acyclic}, we have
\begin{equation}
H^{d+2}(\Lambda,D_{\lim})=0.
\end{equation}
Then Claim~\ref{claim:acyclic_implies_imD_equal_imrho} implies
\begin{equation}
\operatorname{Im}(D_{\lim}^{d+1})
=
\operatorname{Im}(\rho_{\lim}^{d+2}).
\end{equation}
Finally, Claim~\ref{claim:imD_equal_imrho} implies that $\beta_{\lim}^{d+1}$ is surjective.
\end{proof}

\bigskip 

\begin{lemma}[Lifted Quillen isomorphism]
\label{lemma:lift_quillen}
The Quillen restriction isomorphism
\begin{equation*}
    \mathrm{res}:H^{d+2}(B\mathrm{O}(d), \mathbb{Z}_{2})\to \varprojlim\limits_{E\in \mathcal{A}_{2}(\mathrm{O}(d))}H^{d+2}(BE, \mathbb{Z}_{2})
\end{equation*}
lifts to an isomorphism on order-$2$ twisted integral classes: 
\begin{equation*}
    \mathrm{res}[2]:H^{d+2}(B\mathrm{O}(d), \mathbb{Z}_{w_{1}})[2]\to \varprojlim\limits_{E\in \mathcal{A}_{2}(\mathrm{O}(d))}H^{d+2}(BE, \mathbb{Z}_{w_{1}})[2].
\end{equation*}
\end{lemma}

\begin{proof}[\textnormal{\textbf{Proof of Lemma~\ref{lemma:lift_quillen}}}]
We prove that $\res[2]$ is both surjective and injective.

\bigskip 

\noindent First, we prove surjectivity. Let
\begin{equation*}
t\in \varprojlim_{E\in\mathcal A_2(\mathrm{O}(d))}
H^{d+2}(BE,\mathbb Z_{(Bi_E)^*w_1})
\end{equation*}
be arbitrary. By Proposition~\ref{prop:limit_bockstein_surjective}, there exists
\begin{equation*}
y\in \varprojlim_{E\in\mathcal A_2(\mathrm{O}(d))}H^{d+1}(BE,\mathbb Z_2)
\end{equation*}
such that $\beta_{\lim}^{d+1}(y)=t$. Since the mod-$2$ Quillen restriction map is an isomorphism in degree $d+1$, there exists $x\in H^{d+1}(B\mathrm{O}(d),\mathbb Z_2)$ such that $\res(x)=y$. Set
$s:=\beta^{d+1}_{w_1}(x)\in H^{d+2}(B\mathrm{O}(d),\mathbb Z_{w_1})[2]$. Then, using the Bockstein compatibility from Claim~\ref{claim:res2_exists_natural}, we get
\begin{equation}
\res[2](s) = \res[2](\beta^{d+1}_{w_1}(x)) = \beta_{\lim}^{d+1}(\res(x)) = \beta_{\lim}^{d+1}(y) = t.
\end{equation}
Thus $\res[2]$ is surjective.

\bigskip 

\noindent We now prove injectivity. Let $s\in H^{d+2}(B\mathrm{O}(d),\mathbb Z_{w_1})[2]$ and suppose $\res[2](s)=0$. Using the reduction compatibility from Claim~\ref{claim:res2_exists_natural}, we have
\begin{equation}
\res\bigl(\rho\vert_{[2]}(s)\bigr) = \rho_{\lim}^{d+2}(\res[2](s)) = 0.
\end{equation}
Since the mod-$2$ Quillen restriction map is injective in degree $d+2$, it follows that $\rho^{d+2}\vert_{[2]}(s)=0$. By Claim~\ref{claim:rho_BO_injective}, the map $\rho^{d+2}\vert_{[2]}$ is injective, so $s=0$. Therefore $\res[2]$ is injective.

\noindent Since $\res[2]$ is both surjective and injective, it is an isomorphism. This proves the lifted Quillen isomorphism.
\end{proof}

\begin{lemma}[Elementary responses determine continuum TQFT classes]
\label{lemma:elem_response_detect_grpcoh}
Let
\begin{equation*}
x,y\in \tor_{2}H^{d+2}(B\mathrm{O}(d), \Z_{w_{1}})
\end{equation*}
be continuum classes in the full group-cohomology model~\eqref{eq:grpcohtormodel}. Suppose that for every elementary abelian $2$-subgroup $E\leq \mathrm{O}(d)$,
for every closed $(d+1)$-manifold $M$, and for every background $f:M\to BE$, one has equality of elementary responses
\begin{equation*}
\big\langle f^*\res^{\mathrm{O}(d)}_{E}(x),[M]\big\rangle
=
\big\langle f^*\res^{\mathrm{O}(d)}_{E}(y),[M]\big\rangle .
\end{equation*}
Then $x=y$. Equivalently, the elementary responses of an $\mathrm{O}(d)$-symmetric continuum TQFT class determine the class uniquely.
\end{lemma}

\begin{proof}
Let
\begin{equation*}
z:=x-y.
\end{equation*}
By assumption, for every elementary abelian $2$-subgroup $E\leq \mathrm{O}(d)$,
every closed $(d+1)$-manifold $M$, and every background $f:M\to BE$, one has
\begin{equation*}
\big\langle f^*\res^{\mathrm{O}(d)}_{E}(z),[M]\big\rangle=1.
\end{equation*}
Equivalently, the TQFT class
\begin{equation*}
\res^{\mathrm{O}(d)}_{E}(z)\in  \tor_{2}H^{d+2}(BE, \Z_{w_{1}})
\end{equation*}
has trivial response on every probe $(M,f:M\to BE)$. By the probe
detectability lemma in the full cohomology model, Lemma~\ref{lemma:probe_grpcoh},
this implies
\begin{equation*}
\res^{\mathrm{O}(d)}_{E}(z)=0 \; \in  \tor_{2}\left(H^{d+2}(BE, \Z_{w_{1}})\right).
\end{equation*}
Since this holds for every elementary abelian $2$-subgroup $E\leq \mathrm{O}(d)$,
the compatible family of elementary restrictions of $z$ is zero:
\begin{equation*}
\res[2](z)=0\;\in \mathrm{cSPT}^{d+1}_{\mathrm{elem}}.
\end{equation*}
By Lemma.~\ref{lemma:lift_quillen}, the Quillen restriction map
\begin{equation*}
\res[2]:
\Theta^{d+1}(B\mathrm{O}(d))
\longrightarrow
\mathrm{cSPT}^{d+1}_{\mathrm{elem}}
\end{equation*}
is injective. Hence
\begin{equation*}
z=0 \implies  x=y.
\end{equation*}
\end{proof}

\subsubsection{Canonical factorization and surjectivity of $\Pi$}
\label{app:canon_factor}

In the previous subsection we proved the lifted Quillen isomorphism over the elementary abelian category $\mathcal A_2(\mathrm{O}(d))$. We now compare this elementary restriction map $\res[2]$(see Lemma~\ref{lemma:lift_quillen}) with the corresponding restriction map over the full finite-subgroup orbit category $\mathcal O_{\mathrm{fin}}(\mathrm{O}(d))$. The distinction between these two indexing categories is important. For elementary abelian subgroups $E\in\mathcal A_2(\mathrm{O}(d))$, the twisted integral cohomology groups $H^n(BE,\Z_{w_1})$ are entirely order-$2$ in positive degrees. Similarly, by Theorem~\ref{thm:Greenblatt_twisted_no_odd_torsion}, all torsion in $H^n(B\mathrm{O}(d),\Z_{w_1})$ is order-$2$. Thus, on the continuum side and on the elementary side, the $2$-primary torsion subgroup coincides with the order-$2$ subgroup:
\begin{equation}
\operatorname{Tor}_2 H^n(B\mathrm{O}(d),\Z_{w_1})
=
H^n(B\mathrm{O}(d),\Z_{w_1})[2],
\end{equation}
and
\begin{equation}
\operatorname{Tor}_2 H^n(BE,\Z_{w_1})
=
H^n(BE,\Z_{w_1})
\quad
(E\in\mathcal A_2(\mathrm{O}(d)),\ n>0).
\end{equation}
Here, as throughout this appendix, $\Z_{w_1}$ on $BE$ denotes the pulled-back local system. By contrast, for a general finite subgroup $G\subset \mathrm{O}(d)$, the group $H^n(BG,\Z_{w_1})$ may contain higher $2$-primary torsion. Therefore, when working over the full finite-subgroup orbit category, we keep the notation $\operatorname{Tor}_2$ rather than replacing it by order-$2$ torsion.

\bigskip 

We now use the torsion group-cohomology classification model from Definition~\ref{def:grpcoh_classmodel}. In the integral-cohomology presentation, this assigns to a finite subgroup $G\subset \mathrm{O}(d)$ the group
\begin{equation}
\Theta^{d+1}(BG)
\cong
\operatorname{Tor}_2 H^{d+2}(BG,\Z_{w_1}).
\end{equation}
Thus the assignment $G\mapsto \Theta^{d+1}(BG)$ is the classification functor whose inverse limits we consider below.

\bigskip 

\noindent Let
\begin{equation}
\mathcal A:=\mathcal A_2(\mathrm{O}(d)), \qquad \mathcal O:=\mathcal O_{\mathrm{fin}}(\mathrm{O}(d)),
\end{equation}
and let $j:\mathcal A\hookrightarrow \mathcal O$ denote the inclusion. With this notation, the lifted Quillen isomorphism from Lemma~\ref{lemma:lift_quillen} may be rewritten as
\begin{equation}
\res_{\mathcal A}[2]: \Theta^{d+1}(B\mathrm{O}(d)) \longrightarrow \varprojlim_{E\in\mathcal A} \Theta^{d+1}(BE).
\end{equation}
This is the same map previously denoted $\res[2]$; the subscript $\mathcal A$ is added only to emphasize that the inverse limit is taken over the elementary abelian category.

\bigskip

We next define the analogous restriction map over the full finite-subgroup category. This is the same universal-property construction used in Claim~\ref{claim:res2_exists_natural}. For each $G\in\mathcal O$, let $i_G:G\hookrightarrow \mathrm{O}(d)$ denote the inclusion and let $Bi_G:BG\to B\mathrm{O}(d)$ be the induced map on classifying spaces. Restriction along $Bi_G$ gives a map
\begin{equation}
\res^{\mathrm{O}(d)}_{G}:=(Bi_G)^*: \Theta^{d+1}(B\mathrm{O}(d)) \longrightarrow \Theta^{d+1}(BG).
\end{equation}
This map is well-defined because restriction preserves $2$-primary torsion: if $x$ is killed by some power of $2$, then so is $(Bi_G)^*x$.

\noindent These restriction maps are compatible with morphisms in $\mathcal O$. Indeed, if $u:G\to G'$ is a morphism in $\mathcal O$, then the corresponding classifying maps satisfy $Bi_G\simeq Bi_{G'}\circ Bu$, and hence
\begin{equation}
(Bu)^*\circ (Bi_{G'})^*=(Bi_G)^*, \quad \text{or}\quad (Bu)^*\circ \res^{\mathrm{O}(d)}_{G'}=\res^{\mathrm{O}(d)}_{G}.
\end{equation}
Therefore the family $\{\res^{\mathrm{O}(d)}_{G}\}_{G\in\mathcal O}$ defines a cone from $\Theta^{d+1}(B\mathrm{O}(d))$ to the diagram $G\mapsto \Theta^{d+1}(BG)$ over $\mathcal O$. By the universal property of the inverse limit, this cone determines a unique map
\begin{equation}
\res_{\mathcal O}[2]:
\Theta^{d+1}(B\mathrm{O}(d)) 
\longrightarrow 
\varprojlim_{G\in\mathcal O}\Theta^{d+1}(BG),
\end{equation}
characterized by the projection identities
\begin{equation}
p_G^{\mathcal O}\circ \res_{\mathcal O}[2]=\res^{\mathrm{O}(d)}_{G}
\qquad
\text{for every }G\in\mathcal O.
\end{equation}

\bigskip 

The inclusion $j:\mathcal A\hookrightarrow \mathcal O$ lets us restrict the classification diagram over $\mathcal O$ to the elementary subcategory. Since $\Theta^{d+1}$ is contravariant, this restricted diagram is
\begin{equation}
\Theta^{d+1}\vert_{\mathcal A} := \Theta^{d+1}\circ j^{\mathrm{op}}: \mathcal A^{\mathrm{op}}\longrightarrow \mathbf{Ab}.
\end{equation}
Concretely, for $E\in\mathcal A$ one has
\begin{equation}
(\Theta^{d+1}\vert_{\mathcal A})(E)=\Theta^{d+1}(BE),
\end{equation}
and the morphisms are precisely the restriction maps inherited from the full finite-subgroup orbit category. Therefore every compatible family over $\mathcal O$ restricts canonically to a compatible family over $\mathcal A$. This defines the projection
\begin{equation}
\Pi: \varprojlim_{G\in\mathcal O}\Theta^{d+1}(BG)
\longrightarrow \varprojlim_{E\in\mathcal A}\Theta^{d+1}(BE).
\end{equation}
Equivalently, if $p_G^{\mathcal O}$ and $p_E^{\mathcal A}$ denote the inverse-limit projections, then $\Pi$ is characterized by
\begin{equation}
p_E^{\mathcal A}\circ \Pi
=
p_E^{\mathcal O}
\end{equation}
for every $E\in\mathcal A$, where on the right-hand side $E$ is regarded as an object of $\mathcal O$.

\begin{claim}[Canonical factorization]
\label{claim:canonical_factorization_res}
Let $j:\mathcal A\hookrightarrow \mathcal O$ be the inclusion of the elementary abelian category into the full finite-subgroup orbit category, and let
\begin{equation}
\Pi:
\varprojlim_{G\in\mathcal O}\Theta^{d+1}(BG)
\longrightarrow
\varprojlim_{E\in\mathcal A}\Theta^{d+1}(BE)
\end{equation}
be the canonical projection obtained by restricting a compatible family over $\mathcal O$ to the subcategory $\mathcal A$. Then
\begin{equation}
\res_{\mathcal A}[2]
=
\Pi\circ \res_{\mathcal O}[2].
\end{equation}
\end{claim}

\begin{proof}[\textnormal{\textbf{Proof of Claim~\ref{claim:canonical_factorization_res}}}]
We compare the two maps $\res_{\mathcal A}[2]$ and $\Pi\circ \res_{\mathcal O}[2]$ after projecting to each object $E\in\mathcal A$. Let $p_G^{\mathcal O}$ denote the projection
\begin{equation*}
\varprojlim_{G\in\mathcal O}\Theta^{d+1}(BG)\to \Theta^{d+1}(BG),
\end{equation*}
and let $p_E^{\mathcal A}$ denote the projection 
\begin{equation*}
\varprojlim_{E\in\mathcal A}\Theta^{d+1}(BE)\to \Theta^{d+1}(BE).
\end{equation*}
By the defining property of $\Pi$, we have
\begin{equation}
p_E^{\mathcal A}\circ \Pi = p_E^{\mathcal O}, \quad \forall \; E\in \mathcal{A}.
\end{equation}

\noindent Using this identity and the defining property of $\res_{\mathcal O}[2]$, we get
\begin{equation}
p_E^{\mathcal A}\circ \Pi\circ \res_{\mathcal O}[2]
=
p_E^{\mathcal O}\circ \res_{\mathcal O}[2]
=
(Bi_E)^*.
\end{equation}
On the other hand, by the defining property of $\res_{\mathcal A}[2]$,
\begin{equation}
p_E^{\mathcal A}\circ \res_{\mathcal A}[2]
=
(Bi_E)^*.
\end{equation}
Thus $p_E^{\mathcal A}\circ \Pi\circ \res_{\mathcal O}[2]
=
p_E^{\mathcal A}\circ \res_{\mathcal A}[2]$ for every $E\in\mathcal A$. Since elements in an inverse limit are determined by their projections, the two maps agree:
\begin{equation}
\res_{\mathcal A}[2]
=
\Pi\circ \res_{\mathcal O}[2].
\end{equation}
\end{proof}

\begin{proposition}[Surjectivity of $\Pi$ in degree $(d+1)$]
\label{prop:Pi_surjective}
The projection map
\begin{equation}
\Pi:
\varprojlim_{G\in\mathcal O}\Theta^{d+1}(BG)
\longrightarrow
\varprojlim_{E\in\mathcal A}\Theta^{d+1}(BE)
\end{equation}
is surjective. Moreover, $\res_{\mathcal O}[2]$ is injective.
\end{proposition}

\begin{proof}[\textnormal{\textbf{Proof}}]
By Claim~\ref{claim:canonical_factorization_res}, we have the canonical factorization
\begin{equation}
\res_{\mathcal A}[2] = \Pi\circ \res_{\mathcal O}[2].
\end{equation}
By Lemma~\ref{lemma:lift_quillen}, the map $\res_{\mathcal A}[2]$ is an isomorphism. We first prove that $\res_{\mathcal O}[2]$ is injective. Suppose $x\in \Theta^{d+1}(B\mathrm{O}(d))$ satisfies $\res_{\mathcal O}[2](x)=0$. Then
\begin{equation}
\res_{\mathcal A}[2](x) = \Pi\bigl(\res_{\mathcal O}[2](x)\bigr) = 0.
\end{equation}
Since $\res_{\mathcal A}[2]$ is injective, it follows that $x=0$. Hence $\res_{\mathcal O}[2]$ is injective.

\noindent We now prove that $\Pi$ is surjective. Let
$t\in \varprojlim_{E\in\mathcal A}\Theta^{d+1}(BE)$
be arbitrary. Since $\res_{\mathcal A}[2]$ is surjective, there exists $x\in \Theta^{d+1}(B\mathrm{O}(d))$ such that $\res_{\mathcal A}[2](x)=t$. Using the factorization again, we get
\begin{equation}
t = \res_{\mathcal A}[2](x) = \Pi\bigl(\res_{\mathcal O}[2](x)\bigr).
\end{equation}
Thus $t$ lies in the image of $\Pi$. Since $t$ was arbitrary, $\Pi$ is surjective.
\end{proof}

\subsubsection{Proof of the main theorem}
\label{app:main_theorem_proof}

\begin{proof}[\textnormal{\textbf{Proof of Theorem~\ref{appG:main_theorem}}}]
We first prove the statement without internal symmetry. Let
\begin{equation}
\mathcal O:=\mathcal O_{\mathrm{fin}}(\mathrm{O}(d)), \qquad \mathcal A:=\mathcal A_2(\mathrm{O}(d)),
\end{equation}
and write
\begin{equation}
\mathrm{cSPT}^{d+1}_{\mathrm{univ}} := \varprojlim_{G\in\mathcal O}\Theta^{d+1}(BG),
\quad
\mathrm{cSPT}^{d+1}_{\mathrm{elem}} := \varprojlim_{E\in\mathcal A}\Theta^{d+1}(BE),
\end{equation}
for the full finite-subgroup and elementary inverse limits in the $2$-primary group-cohomology model. By Lemma~\ref{lemma:lift_quillen}, the elementary restriction map is an isomorphism:
\begin{equation}
\res_{\mathcal A}[2]:
\Theta^{d+1}(B\mathrm{O}(d)) \xrightarrow{\;\cong\;} \mathrm{cSPT}^{d+1}_{\mathrm{elem}}.
\end{equation}
We therefore define
\begin{equation}
\widetilde F_2 := \res_{\mathcal A}[2]^{-1}:
\mathrm{cSPT}^{d+1}_{\mathrm{elem}} \longrightarrow \Theta^{d+1}(B\mathrm{O}(d)).
\end{equation}
By Proposition~\ref{prop:Pi_surjective}, the projection
\begin{equation}
\Pi: \mathrm{cSPT}^{d+1}_{\mathrm{univ}}
\longrightarrow \mathrm{cSPT}^{d+1}_{\mathrm{elem}}
\end{equation}
is surjective. Therefore the composite
\begin{equation}
F_2 := \widetilde F_2\circ \Pi:
\mathrm{cSPT}^{d+1}_{\mathrm{univ}} \longrightarrow \Theta^{d+1}(B\mathrm{O}(d))
\end{equation}
is surjective, since it is the composition of a surjective map with an isomorphism. This proves that the full stratification map $F_{2}$ is surjective.
\end{proof}

\noindent The corollaries stated after Theorem~\ref{theorem:main} follow formally from the construction above and are proved in Sec.~\ref{sec:proof_corollaries}; we do not repeat their proofs here.

\subsection{General-degree refinement and free shadow classes}
\label{app:free_shadow_refinement} 

The proof above used a special feature of cohomological degree $d+2$, namely 
\begin{equation} 
H^{d+2}(\Lambda,D_{\lim})=0. 
\end{equation} 
In arbitrary cohomological degree this vanishing need not hold. The purpose of this subsection is to record the resulting general-degree correction. This refinement is not needed for the no-internal-symmetry theorem proved above, but it becomes important when additional internal symmetry classes are included. For $n>0$, define 
\begin{equation} 
\mathrm{cSPT}_{\mathrm{elem}}^n := \varprojlim_{E\in\mathcal A_2(\mathrm{O}(d))} H^n(BE,\mathbb Z_{w_1}). \end{equation} 
Recall that on elementary subgroups, the twisted integral cohomology is entirely order-$2$.

\medskip 

\noindent Recall from Eq.~\eqref{eq:res_full} the full twisted-integral restriction map $\res_{\mathcal A}^{\mathrm{full},n}$. We now use its restriction to free continuum classes to define the corresponding elementary free-shadow subgroup. If $d$ is odd, set 
\begin{equation} 
S_{\mathrm{free}}^n:=0. 
\end{equation}
This is because for odd $d$, the primary free generator of twisted integral cohomology, the twisted Euler class, is order-2. If $d=2m$ is even, let 
\begin{equation*} 
E_d\in H^d(B\mathrm{O}(d),\mathbb Z_{w_1}) 
\end{equation*}
denote the universal twisted Euler class. The free twisted integral classes relevant to the obstruction are generated by the Euler class and Pontryagin polynomials\footnote{this is because a product of a twisted class with an untwisted class is twisted. Whereas the product of a twisted class with another twisted class is untwisted.}: 
\begin{equation} 
E_d\cdot \mathbb Z[p_1,\dots,p_m] \subseteq H^*(B\mathrm{O}(d),\mathbb Z_{w_1}). 
\end{equation} Define 
\begin{equation} 
S_{\mathrm{free}}^n := \operatorname{Im} \left( \res_{\mathcal A}^{\mathrm{full},n} \big|_{\left(E_d\cdot\mathbb Z[p_1,\dots,p_m]\right)^n} \right) \subseteq \mathrm{cSPT}_{\mathrm{elem}}^n . 
\label{eq:Sfree_definition} 
\end{equation} 
We call $S_{\mathrm{free}}^n$ the free-shadow subgroup. Its elements are torsion classes in the elementary inverse limit which arise as restrictions of free continuum classes. The mod-$2$ reduction satisfies 
\begin{equation} 
\rho_{\lim}^n(S_{\mathrm{free}}^n) = \left( e_d\mathbb F_2[e_2^2,\dots,e_d^2] \right)^n \label{eq:Sfree_reduction} 
\end{equation} 
when $d$ is even, and $\rho_{\lim}^n(S_{\mathrm{free}}^n)=0$ when $d$ is odd.

\bigskip 

\begin{claim}[Exact sequence measuring the failure of Bockstein surjectivity] \label{claim:bockstein_failure_sequence} There is a short exact sequence 
\begin{equation} 
\label{eq:ses_Bockstein_failure}
0 \longrightarrow \operatorname{Im}(\beta_{\lim}^{n-1}) \longrightarrow \mathrm{cSPT}_{\mathrm{elem}}^n \xrightarrow{\overline{\rho}_{\lim}^{\,n}} \dfrac{\operatorname{Im}(\rho_{\lim}^n)} {\operatorname{Im}(D_{\lim}^{n-1})} \longrightarrow 0 . 
\end{equation} 
Here 
\begin{equation} 
\overline{\rho}_{\lim}^{\,n}(x) := [\rho_{\lim}^n(x)],\quad \text{where}\quad [\;\;]: \operatorname{Im}(\rho_{\lim}^n) \xrightarrow{} \dfrac{\operatorname{Im}(\rho_{\lim}^n)} {\operatorname{Im}(D_{\lim}^{n-1})}
\end{equation} 
is the quotient map.
\end{claim}

\begin{proof}[\textnormal{\textbf{Proof}}]
The map $\overline{\rho}_{\lim}^{\,n}$ is surjective by definition of the target. Suppose $\overline{\rho}_{\lim}^{\,n}(x)=0 $, then 
\begin{equation*}
\rho_{\lim}^n(x)\in\operatorname{Im}(D_{\lim}^{n-1}).
\end{equation*}
So there exists $a\in\Lambda^{n-1}$ such that 
\begin{equation} 
\rho_{\lim}^n(x) = D_{\lim}^{n-1}(a) = \rho_{\lim}^n\beta_{\lim}^{n-1}(a). 
\end{equation} 
Since $\rho_{\lim}^n$ is injective, $x=\beta_{\lim}^{n-1}(a).$ 
Hence 
\begin{equation*}
\ker(\overline{\rho}_{\lim}^{\,n}) \subseteq \operatorname{Im}(\beta_{\lim}^{n-1}). 
\end{equation*}
Conversely, if $x=\beta_{\lim}^{n-1}(a)$, then 
\begin{equation} 
\rho_{\lim}^n(x) = D_{\lim}^{n-1}(a) \in \operatorname{Im}(D_{\lim}^{n-1}), 
\end{equation} 
so $\overline{\rho}_{\lim}^{\,n}(x)=0 $. Therefore 
\begin{equation} 
\ker(\overline{\rho}_{\lim}^{\,n}) = \operatorname{Im}(\beta_{\lim}^{n-1}), 
\end{equation} 
which proves exactness. 
\end{proof}

\bigskip 

\begin{claim}[Free shadows identify the quotient obstruction]
\label{claim:Sfree_quotient_iso}
The restriction of $\overline{\rho}_{\lim}^{\,n}$ to $S_{\mathrm{free}}^n$ is an isomorphism:
\begin{equation}
\overline{\rho}_{\lim}^{\,n}\big\vert_{S_{\mathrm{free}}^n}:
S_{\mathrm{free}}^n
\xrightarrow{\cong}
\frac{\operatorname{Im}(\rho_{\lim}^{n})}
{\operatorname{Im}(D_{\lim}^{n-1})}.
\end{equation}
\end{claim}

\begin{proof}[\textnormal{\textbf{Proof of Claim~\ref{claim:Sfree_quotient_iso}}}]
If $d$ is odd, then $H^{*}(\Lambda,D_{\lim})=0$. Therefore using Claim~\ref{claim:acyclic_implies_imD_equal_imrho}
\begin{equation*}
\operatorname{Im}(\rho_{\lim}^{n}) =
\operatorname{Im}(D_{\lim}^{n-1}),\quad \text{if}\;\; d\;\;\text{odd}
\end{equation*}
so the quotient is zero. Since $S_{\mathrm{free}}^n=0$ by definition in odd spatial dimension, the claim follows. Now suppose $d=2m$ is even. By the computation of $H^{*}(\Lambda,D_{\lim})$, we have the normal form
\begin{equation}
\ker(D_{\lim}^{n})
=
\operatorname{Im}(D_{\lim}^{n-1})
\oplus
\left(e_d\mathbb F_2[e_2^2,\dots,e_d^2]\right)^n.
\label{eq:normal_form_Sfree}
\end{equation}
By the definition of $S_{\mathrm{free}}^n$ and naturality of mod-$2$ reduction,
\begin{equation}
\rho_{\lim}^{n}(S_{\mathrm{free}}^n)
=
\left(e_d\mathbb F_2[e_2^2,\dots,e_d^2]\right)^n.
\label{eq:Sfree_reduction_again}
\end{equation}

\bigskip 

We first prove surjectivity. Let
\begin{equation}
[z]\in
\frac{\operatorname{Im}(\rho_{\lim}^{n})}
{\operatorname{Im}(D_{\lim}^{n-1})}
\end{equation}
be arbitrary, with $z\in \operatorname{Im}(\rho_{\lim}^{n})$. Since $z$ lies in the image of $\rho_{\lim}^{n}$, we have
\begin{equation}
D_{\lim}^{n}(z)=0.
\end{equation}
Thus $z\in\ker(D_{\lim}^{n})$. By the normal form in Eq.~\eqref{eq:normal_form_Sfree}
\begin{equation}
\exists\;a\in \Lambda^{n-1},\; f\in
\left(e_d\mathbb F_2[e_2^2,\dots,e_d^2]\right)^n\;\;\text{st}\;\;z=D_{\lim}^{n-1}(a)+f.
\end{equation}
By Eq.~\eqref{eq:Sfree_reduction_again}, there exists $s\in S_{\mathrm{free}}^n$ such that $\rho_{\lim}^{n}(s)=f$.
Therefore
\begin{equation}
\overline{\rho}_{\lim}^{\,n}(s)
=
[\rho_{\lim}^{n}(s)]
=
[f]
=
[z],
\end{equation}
because $[D_{\lim}^{n-1}(a)]=0$ in the quotient. Hence
$\overline{\rho}_{\lim}^{\,n}\big\vert_{S_{\mathrm{free}}^n}$ is surjective.

\bigskip

\noindent We now prove injectivity. Suppose $s\in S_{\mathrm{free}}^n$ satisfies $ \overline{\rho}_{\lim}^{\,n}(s)=0$. By definition of the quotient, this means
\begin{equation}
\rho_{\lim}^{n}(s)\in \operatorname{Im}(D_{\lim}^{n-1}).
\end{equation}
On the other hand, since $s\in S_{\mathrm{free}}^n$, Eq.~\eqref{eq:Sfree_reduction_again} gives
\begin{equation}
\rho_{\lim}^{n}(s)
\in
\left(e_d\mathbb F_2[e_2^2,\dots,e_d^2]\right)^n.
\end{equation}
The direct-sum decomposition in Eq.~\eqref{eq:normal_form_Sfree} implies
\begin{equation}
\operatorname{Im}(D_{\lim}^{n-1})
\cap
\left(e_d\mathbb F_2[e_2^2,\dots,e_d^2]\right)^n
=
0.
\end{equation}
Therefore $\rho_{\lim}^{n}(s)=0$. Since $\rho_{\lim}^{n}$ is injective by Claim~\ref{claim:rho_lim_injective}, it follows that $s=0$. Hence $\overline{\rho}_{\lim}^{\,n}\big\vert_{S_{\mathrm{free}}^n}$ is injective.

\noindent Thus $\overline{\rho}_{\lim}^{\,n}\big\vert_{S_{\mathrm{free}}^n}$ is both surjective and injective, so it is an isomorphism.
\end{proof}

\bigskip 

\begin{claim}[Free-shadow splitting] 
\label{claim:Sfree_splitting} There is a direct-sum decomposition 
\begin{equation} \mathrm{cSPT}_{\mathrm{elem}}^n = \operatorname{Im}(\beta_{\lim}^{n-1}) \oplus S_{\mathrm{free}}^n . 
\end{equation} 
\end{claim} 
\begin{proof}[\textnormal{\textbf{Proof of Claim~\ref{claim:Sfree_splitting}}}]
By Claim~\ref{claim:bockstein_failure_sequence}, we have a short exact sequence
\begin{equation}
0
\longrightarrow
\operatorname{Im}(\beta_{\lim}^{n-1})
\longrightarrow
\mathrm{cSPT}_{\mathrm{elem}}^n
\xrightarrow{\overline{\rho}_{\lim}^{\,n}}
\frac{\operatorname{Im}(\rho_{\lim}^{n})}
{\operatorname{Im}(D_{\lim}^{n-1})}
\longrightarrow
0.
\end{equation}
By Claim~\ref{claim:Sfree_quotient_iso}, the restriction of $\overline{\rho}_{\lim}^{\,n}$ to $S_{\mathrm{free}}^n$ is an isomorphism onto the quotient term. Therefore $S_{\mathrm{free}}^n$ gives the complement to $\operatorname{Im}(\beta_{\lim}^{n-1})$, and hence
\begin{equation}
\mathrm{cSPT}_{\mathrm{elem}}^n
=
\operatorname{Im}(\beta_{\lim}^{n-1})
\oplus
S_{\mathrm{free}}^n.
\end{equation}
\end{proof}

\bigskip 

\begin{claim}[Identification of the Bockstein-detected summand] 
\label{claim:bockstein_summand} 
The map $\res_{\mathcal A}^{n}[2]$ induces an isomorphism 
\begin{equation} 
\res_{\mathcal A}^{n}[2]: H^n(B\mathrm{O}(d),\mathbb Z_{w_1})[2] \xrightarrow{\cong} \operatorname{Im}(\beta_{\lim}^{n-1}) \subseteq \mathrm{cSPT}_{\mathrm{elem}}^n . 
\end{equation}
\end{claim} 
\begin{proof}[\textnormal{\textbf{Proof}}]
We first identify the image of $\res_{\mathcal A}^{n}[2]$. Naturality of the Bockstein gives the commutative diagram 
\begin{equation} 
\begin{tikzcd} 
H^{n-1}(B\mathrm{O}(d),\mathbb Z_2) \arrow[r,"\beta_{w_1}^{n-1}"] \arrow[d,"\res"',"\cong"] & H^n(B\mathrm{O}(d),\mathbb Z_{w_1})[2] \arrow[d,"\res_{\mathcal A}^{n}\text{[}2\text{]}"] \\ 
\Lambda^{n-1} \arrow[r,"\beta_{\lim}^{n-1}"] & \mathrm{cSPT}_{\mathrm{elem}}^n . 
\end{tikzcd} 
\end{equation} 
The top Bockstein is surjective onto the order-$2$ torsion subgroup, and the left vertical map is the mod-$2$ Quillen isomorphism. Therefore the image of $\res_{\mathcal A}^{n}[2]$ is exactly $\operatorname{Im}(\beta_{\lim}^{n-1})$. 

\bigskip 

It remains to prove that $\res_{\mathcal A}^{n}[2]$ is injective. For this we use the naturality square for mod-$2$ reduction:
\begin{equation}
\begin{tikzcd}
H^n(B\mathrm{O}(d),\mathbb Z_{w_1})[2]
    \arrow[r,"\res_{\mathcal A}^{n}\text{[}2\text{]}"]
    \arrow[d,"\rho^n\vert_{[2]}"']
&
\mathrm{cSPT}_{\mathrm{elem}}^n
    \arrow[d,"\rho_{\lim}^{n}"]
\\
H^n(B\mathrm{O}(d),\mathbb Z_2)
    \arrow[r,"\res^{n}"',"\cong"]
&
\Lambda^n .
\end{tikzcd}
\label{eq:reduction_summand_square}
\end{equation}
Suppose $a\in H^n(B\mathrm{O}(d),\mathbb Z_{w_1})[2]$ satisfies $\res_{\mathcal A}^{n}[2](a)=0$. Applying the reduction square~\eqref{eq:reduction_summand_square}, we get
\begin{equation}
\res^{n}\bigl(\rho^n\vert_{[2]}(a)\bigr)
=\rho_{\lim}^{n}\bigl(\res_{\mathcal A}^{n}[2](a)\bigr) = 0.
\end{equation}
Since $\res^{n}$ is injective by Quillen's mod-$2$ theorem, it follows that
\begin{equation}
\rho^n\vert_{[2]}(a)=0.
\end{equation}
By injectivity of mod-$2$ reduction on order-$2$ torsion for $B\mathrm{O}(d)$, we conclude $a=0$. Thus $\res_{\mathcal A}^{n}[2]$ is injective. 

\noindent Therefore $\res_{\mathcal A}^{n}[2]$ is an isomorphism onto its image, and since its image is
$\operatorname{Im}(\beta_{\lim}^{n-1})$, the claim follows.
\end{proof}

\begin{claim}[Quotient by free shadows] 
\label{cor:free_shadow_quotient} There is a canonical isomorphism 
\begin{equation} \frac{\mathrm{cSPT}_{\mathrm{elem}}^n} {S_{\mathrm{free}}^n} \cong H^n(B\mathrm{O}(d),\mathbb Z_{w_1})[2]. \end{equation} 
\end{claim} 
\begin{proof}[\textnormal{\textbf{Proof}}] 
Using Claim~\ref{claim:Sfree_splitting}, \[ \frac{\mathrm{cSPT}_{\mathrm{elem}}^n} {S_{\mathrm{free}}^n} \cong \operatorname{Im}(\beta_{\lim}^{n-1}). \] The result follows from Claim~\ref{claim:bockstein_summand}. 
\end{proof} 

\begin{proposition}[General-degree elementary retraction] 
\label{prop:general_degree_retraction} There exists a canonical surjective homomorphism \begin{equation} 
\widetilde F_2^n: \mathrm{cSPT}_{\mathrm{elem}}^n \twoheadrightarrow H^n(B\mathrm{O}(d),\mathbb Z_{w_1})[2] 
\end{equation} 
with kernel $S_{\mathrm{free}}^n$, satisfying 
\begin{equation} 
\widetilde F_2^n\circ\res_{\mathcal A}^{n}[2] = \mathrm{id}_{H^n(B\mathrm{O}(d),\mathbb Z_{w_1})[2]} . 
\end{equation} 
\end{proposition} 
\begin{proof}[\textnormal{\textbf{Proof}}] 
By Claim~\ref{claim:Sfree_splitting}, every element $x\in\mathrm{cSPT}_{\mathrm{elem}}^n$ has a unique decomposition 
\begin{equation} x=x_{\mathrm{Bock}}+x_{\mathrm{free}}, 
\end{equation} 
where \[ x_{\mathrm{Bock}}\in\operatorname{Im}(\beta_{\lim}^{n-1}), \qquad x_{\mathrm{free}}\in S_{\mathrm{free}}^n . \] Define 
\begin{equation} \widetilde F_2^n(x) := \left(\res_{\mathcal A}^{n}[2]\right)^{-1} (x_{\mathrm{Bock}}). 
\end{equation} 
This is well-defined by Claim~\ref{claim:bockstein_summand}. Its kernel is exactly $S_{\mathrm{free}}^n$, and it is a left inverse to $\res_{\mathcal A}^{n}[2]$, proving the proposition. 
\end{proof}

\bigskip 

\begin{proposition}[General-degree stratification map]
\label{prop:general_degree_stratification}
For every cohomological degree $n>0$, there is a canonical surjective homomorphism
\begin{equation}
F_2^n:
\mathrm{cSPT}_{\mathrm{univ}}^n
\longrightarrow
H^n(B\mathrm{O}(d),\mathbb Z_{w_1})[2].
\end{equation}
Moreover,
\begin{equation}
F_2^n\circ \res_{\mathcal O}^{n}[2]
=
\mathrm{id}_{H^n(B\mathrm{O}(d),\mathbb Z_{w_1})[2]}.
\end{equation}
In particular, $\res_{\mathcal O}^{n}[2]$ is injective.
\end{proposition}

\begin{proof}[\textnormal{\textbf{Proof of Proposition~\ref{prop:general_degree_stratification}}}]
Let
\begin{equation}
\mathcal O:=\mathcal O_{\mathrm{fin}}(\mathrm{O}(d)),
\qquad
\mathcal A:=\mathcal A_2(\mathrm{O}(d)).
\end{equation}
Define
\begin{equation}
\mathrm{cSPT}_{\mathrm{univ}}^n
:=
\varprojlim_{G\in\mathcal O}
\operatorname{Tor}_2 H^n(BG,\mathbb Z_{w_1}),
\end{equation}
and
\begin{equation}
\mathrm{cSPT}_{\mathrm{elem}}^n
:=
\varprojlim_{E\in\mathcal A}
H^n(BE,\mathbb Z_{w_1}).
\end{equation}
Let
\begin{equation}
\Pi^n:
\mathrm{cSPT}_{\mathrm{univ}}^n
\longrightarrow
\mathrm{cSPT}_{\mathrm{elem}}^n
\end{equation}
be the canonical projection obtained by restricting a compatible family over
$\mathcal O$ to the elementary subcategory $\mathcal A$. By Proposition~\ref{prop:general_degree_retraction}, we have a canonical surjection
\begin{equation}
\widetilde F_2^n:
\mathrm{cSPT}_{\mathrm{elem}}^n
\twoheadrightarrow
H^n(B\mathrm{O}(d),\mathbb Z_{w_1})[2],
\end{equation}
satisfying
\begin{equation}
\widetilde F_2^n\circ \res_{\mathcal A}^{n}[2]
=
\mathrm{id}_{H^n(B\mathrm{O}(d),\mathbb Z_{w_1})[2]}.
\end{equation}
We define
\begin{equation}
F_2^n
:=
\widetilde F_2^n\circ \Pi^n.
\end{equation}
It remains to prove that $F_2^n$ is surjective. By the canonical factorization of restriction maps, we have
\begin{equation}
\res_{\mathcal A}^{n}[2]
=
\Pi^n\circ \res_{\mathcal O}^{n}[2].
\end{equation}
Therefore
\begin{align}
F_2^n\circ \res_{\mathcal O}^{n}[2]
&=
\widetilde F_2^n\circ \Pi^n\circ \res_{\mathcal O}^{n}[2] \\
&=
\widetilde F_2^n\circ \res_{\mathcal A}^{n}[2] \\
&=
\mathrm{id}_{H^n(B\mathrm{O}(d),\mathbb Z_{w_1})[2]}.
\end{align}
Thus $F_2^n$ has a right section, namely $\res_{\mathcal O}^{n}[2]$, and hence $F_2^n$ is surjective.

\bigskip

\noindent The same identity also implies that $\res_{\mathcal O}^{n}[2]$ is injective. Indeed, if
\begin{equation}
\res_{\mathcal O}^{n}[2](x)=0,
\end{equation}
then applying $F_2^n$ gives
\begin{equation}
x
=
F_2^n\bigl(\res_{\mathcal O}^{n}[2](x)\bigr)
=
0.
\end{equation}
This proves the proposition.
\end{proof}

\bigskip 

\begin{corollary}[Degree-$d+2$ specialization]
\label{cor:degree_dplus2_retraction_isomorphism}
In cohomological degree $n=d+2$, the free-shadow subgroup vanishes $S_{\mathrm{free}}^{d+2}=0$. Consequently, the elementary retraction of Proposition~\ref{prop:general_degree_retraction} is an isomorphism:
\begin{equation}
\widetilde F_2^{d+2}:
\mathrm{cSPT}_{\mathrm{elem}}^{d+2}
\xrightarrow{\cong}
H^{d+2}(B\mathrm{O}(d),\mathbb Z_{w_1})[2].
\end{equation}
Moreover,
\begin{equation}
\widetilde F_2^{d+2}
=
\left(\res_{\mathcal A}^{d+2}[2]\right)^{-1}.
\end{equation}
\end{corollary}

\bigskip 

\noindent \noindent This follows immediately from Proposition~\ref{prop:general_degree_retraction} together with the fact that $S_{\mathrm{free}}^{d+2}=0$. Indeed, if $d$ is odd then $S_{\mathrm{free}}^*=0$, while if $d$ is even the free-shadow sector is supported only in degrees $d+4k$. Hence it does not appear in degree $d+2$. 

\bigskip 

\begin{corollary}[Main theorem in the group cohomology model]
\label{cor:main_theorem_group_cohomology}
For every spatial dimension $d\geq 1$, the stratification map
\begin{equation}
F_2:
\mathrm{cSPT}_{\mathrm{univ}}^{d+1}
\longrightarrow
\tor_2\left(H^{d+2}(B\mathrm{O}(d),\mathbb Z_{w_1})\right)
\end{equation}
is surjective.
\end{corollary}

\bigskip 
\noindent This follows simply by substituting $n=d+2$ in Proposition~\ref{prop:general_degree_stratification}.

\section{Main theorem: adding internal symmetries}
\label{app:canon_factor_H}

This appendix proves the statements made in Sec.~\ref{sec:internal_sym} for classes of internal symmetries which are physically relevant. We incorporate internal symmetries $H$ which generally couples to spatial symmetry $\mathrm{O}(d)$ by a homomorphism $\mu_{\mathrm{O}(d)}:\mathrm{O}(d)\to\mathrm{Aut}(H)$, so the total group is the semi-direct product $\Gamma_{\mathrm{O}(d)} \;=\; H\rtimes_{\mu_{\mathrm{O}(d)}} \mathrm{O}(d)$ with the corresponding fibration on classifying spaces
\begin{equation*}
BH \longrightarrow B\Gamma_{\mathrm{O}(d)} \xrightarrow{\;p_{\mathrm{O}(d)}\;} B\mathrm{O}(d),
\end{equation*}
and, for every finite $E\le \mathrm{O}(d)$, we have the induced fibration
\begin{equation*}BH \longrightarrow B\Gamma_{E} \xrightarrow{\;p_{E}\;} BE,\qquad \Gamma_E:=H\rtimes_{\mu_{E}}E.
\end{equation*}
Note that in this appendix, we assume internal symmetries are decoupled from the spatial structure, as defined below.

\bigskip 

\begin{definition}[Decoupled internal symmetry]\label{def:decoupled_internal}
Fix a tangential structure group $\zeta$ for spacetime (see Def.~\ref{def:tangential_structure_general}). An onsite internal symmetry $H$ is decoupled from the spatial structure if the full symmetry group is of the form $\Gamma_{\mathrm{O}(d)} = H\rtimes_{\mu_{\mathrm{O}(d)}} \mathrm{O}(d)$ with $H$ interacting trivially with the $\zeta$-structure (equivalently, adjoining $H$ does not force a change of tangential structure).
\end{definition}

\medskip

\noindent In the case of $\zeta=\SO$, for example, this excludes anti-unitary onsite symmetries; equivalently, all onsite internal symmetries are assumed unitary. This is a key consideration for identifying the correct domain bordism categories used to define TQFT functors. For example, if we consider bosonic systems on oriented manifolds, that is with $\SO$-tangential structure\footnote{Importantly, this excludes un-oriented manifolds.}, and then consider adding time-reversal symmetry, it will necessitate a coarser $\text{O}$-tangential structure because insertion of time-reversal domain walls generally ruins any notion of spacetime orientability. On the other hand, if we consider adding fermion parity, denoted $\Z_{2}^{F}$, as an internal symmetry, then a proper analysis requires a refinement of the tangential structure to $\mathrm{Spin}$-structure. We do not consider these cases because they would, in principle, be better handled by starting out with the correct tangential structure for the systems from the outset. 

\bigskip 

\noindent Now, recall the orbit categories
\begin{equation*}
\mathcal{O}:=\mathcal{O}_{\mathrm{fin}}(\mathrm{O}(d)),
\qquad
\mathcal{A}:=\mathcal{A}_{2}(\mathrm{O}(d)),
\end{equation*}
where we note that $\mathcal{A}$ is the usual Quillen category defined as the category of finite elementary 2-subgroups of $\mathrm{O}(d)$, and $\mathcal{O}$ is the category of all finite subgroups of $\mathrm{O}(d)$. We work with the group-cohomology classification model from Sec.~\ref{sec:classification_models}, viewed as a contravariant functor
\begin{equation*}
\Theta^{d+1}:\mathbf{hoCW}^{\mathrm{op}}\to\mathbf{Ab}, \quad \Theta^{d+1}(BG):=\tor_{2}\big(H^{d+2}(BG;\Z_{w_1})\big),
\end{equation*} 
on classifying spaces of groups $BG$ equipped with the orientation local system $w_{1}$ defined with respect to the twist character of the symmetry(see App.~\ref{app:bordism_spatial_sym}). We define the \emph{$H$-enriched} lattice classification functor on spatial symmetry groups to be the precomposition of $\Theta^{d+1}$ with the assignment $E\mapsto B\Gamma_E$:
\begin{gather}
\label{eq:internal_functor}
\Theta^{d+1}_{H}\ :=\ \Theta^{d+1}\circ (B\Gamma_{(-)})^{\mathrm{op}}
\;:\;\ofin(\mathrm{O}(d))^{\mathrm{op}}\to\mathbf{Ab},\\
E\longmapsto \Theta^{d+1}_{H}(BE):=\Theta^{d+1}(B\Gamma_E).\nonumber 
\end{gather}
In the group-cohomology model, this is
\begin{equation*}
\Theta^{d+1}_{H}(BE) \;=\; \tor_{2}\big(H^{d+2}(B\Gamma_E;\Z_{w_1})\big).
\end{equation*}
Note that the coefficient should really be $p_{E}^{*}\Z_{w_{1}}$ as discussed below, but by abuse of notation we will often simply write $\Z_{w_{1}}$ where it should be understood from context that the local system is pulled back on the total space $B\Gamma_{E}$ from that defined on $BE$.

\bigskip 

\noindent We now define the analogs of the restriction morphisms: 
\begin{gather}
\label{eq:internal_resA}
\res_{H,\mathcal{A}}[2]: \Theta^{d+1}_{H}(B\mathrm{O}(d)) \to \varprojlim_{E\in \mathcal{A}} \Theta^{d+1}_{H}(BE)\;,\\
\res_{H,\mathcal{O}}[2]: \Theta^{d+1}_{H}(B\mathrm{O}(d)) \to \varprojlim_{E\in \mathcal{O}} \Theta^{d+1}_{H}(BE)\;.
\label{eq:internal_resO}
\end{gather}
We also note that the projection map of spaces: 
\begin{equation*}
p_{E}: B\Gamma_{E} \to BE
\end{equation*}
can be used to induce maps between local coefficient systems. Consider on $BE$ the SES of local coefficient systems 
\begin{equation*}
    0\to \mathbb{Z}_{w_{1}} \xrightarrow[]{\times 2} \mathbb{Z}_{w_{1}} \xrightarrow[]{\rho} \mathbb{Z}_{2} \to 0   
\end{equation*}
Its pullback along $p_{E}$ can be used to induce a SES of local coefficient systems for the cohomology of $B\Gamma_{E}$ via $p_{E}^{*}$ and denoted as: 
\begin{equation}
\label{eq:coeff_pullback}
0\to p_{E}^{*}(\mathbb{Z}_{w_{1}}) \xrightarrow[]{\times 2} p_{E}^{*}(\mathbb{Z}_{w_{1}}) \xrightarrow[]{\rho} p_{E}^{*}(\mathbb{Z}_{2})\to 0.
\end{equation}

\noindent The short exact sequences induce long exact sequences in cohomology given by\footnote{{\bf Notation}: we use $\beta_{E}^{H}=\beta_{\Gamma_{E}}$ and $\rho_{E}^{H}=\rho_{\Gamma_{E}}$.}: 
\begin{equation*}
    \dots \xrightarrow[]{\times 2} H^{d+1}(B\Gamma_{E}, p_{E}^{*}\Z_{w_{1}}) \xrightarrow[]{\rho_{E}^{H}} H^{d+1}(B\Gamma_{E}, p_{E}^{*}\Z_{2}) \xrightarrow[]{\beta^{H}_{E}} \dots
\end{equation*}
on the total space $B\Gamma_{E}$, and 
\begin{equation*}
    \dots \xrightarrow[]{\times 2} H^{d+1}(BE, \Z_{w_{1}}) \xrightarrow[]{\rho_{E}} H^{d+1}(BE, \Z_{2}) \xrightarrow[]{\beta_{E}} \dots
\end{equation*}
on the base space $BE$. We now claim that there exists a Leray-Hirsch isomorphism in the mod-$2$ cohomology of the fiber space $p_{E}: B\Gamma_{E}\to BE$. 

\subsection{Leray-Hirsch isomorphism}
\label{app:leray_hirsch}

The first hypothesis we impose is that the fibration 
\begin{equation*}
    BH \to B\Gamma_{\mathrm{O}(d)}\to B\mathrm{O}(d)    
\end{equation*}
satisfies pre-conditions required for the application of the Leray-Hirsch isomorphism, which we state below.

\bigskip 

\begin{theorem}[Leray-Hirsch; cf. Hatcher~\cite{Hatcher_AT}, Thm. 4D.1]
Let $\pi:E\to B$ be a fibration and $R$ a commutative ring. Assume $B$ is path-connected and, for each $q$, $H^{q}(F;R)$ is a free $R$-module of finite rank for the fiber $F=\pi^{-1}(b)$. If the restriction map $f^*$ induced by the inclusion $f:F \hookrightarrow E$, and given by
\begin{equation*}
f^*:H^{*}(E;R)\longrightarrow H^{*}(F;R)
\end{equation*}
is surjective (equivalently: there exist classes $u_1,\dots,u_m\in H^{*}(E;R)$ whose restrictions form an $R$-basis of $H^{*}(F;R)$), then the $H^{*}(B;R)$-linear map
\begin{align*}
\Delta : H^{*}(B;R&)\otimes_{R} H^{*}(F;R)\longrightarrow H^{*}(E;R),\\
&b\otimes r\ \qquad \qquad \longmapsto\ \pi^{*}b\smile \widetilde r
\end{align*}
is an isomorphism, where $\widetilde r$ is any lift of $r$ whose fiberwise restriction equals $r$. 
\end{theorem}

\bigskip 

\noindent All except the last of the pre-conditions specified above are satisfied for the fibration $BH \to B\Gamma_{\mathrm{O}(d)}\to B\mathrm{O}(d)$ for compact Lie group (or finite group) $H$ and commutative ring $R=\Z_{2}$. We assume the final condition also holds, and state it as a hypothesis:

\bigskip 

\begin{hypothesis}\label{hyp:1}
Given the fibration $BH\to B\Gamma_{\mathrm{O}(d)}\to B\mathrm{O}(d)$, with the fiber inclusion $f_{\mathrm{O}(d)}: BH \hookrightarrow B\Gamma_{\mathrm{O}(d)}$, we assume the induced restriction map 
\begin{equation*}
    f_{\mathrm{O}(d)}^{*}: H^{*}(B\Gamma_{\mathrm{O}(d)}, \Z_{2}) \to H^{*}(BH, \Z_{2}) 
\end{equation*}
is surjective. Equivalently, there exists a (non-canonical) lift:
\begin{equation*}
    s_{\mathrm{O}(d)}: H^{*}(BH, \Z_{2}) \to H^{*}(B\Gamma_{\mathrm{O}(d)}, \Z_{2})
\end{equation*}
which takes a chosen homogeneous $\Z_{2}$-basis denoted $\{r_{i}\}$ of $H^{*}(BH, \Z_{2})$ to global lifts, denoted $\{\widetilde{r}_i:=s_{\mathrm{O}(d)}(r_i)\}$, in $H^*(B\Gamma_{\mathrm{O}(d)},\Z_{2})$ such that $f_{\mathrm{O}(d)}^{*}(\widetilde{r}_{i})=r_i$.
\end{hypothesis}

\bigskip 

\noindent Now, provided the hypothesis above is satisfied we have the Leray-Hirsch isomorphism in the continuum:
\begin{gather}
\nonumber
    \Delta_{\mathrm{O}(d)}: H^{*}(B\mathrm{O}(d), \Z_{2})\;\otimes  H^{*}(BH, \Z_{2}) \rightarrow H^{*}(B\Gamma_{\mathrm{O}(d)}, \Z_{2}),\\
    \quad b\;\otimes r \;\;\;\longmapsto \;\;\;p_{\mathrm{O}(d)}^{*}(b)\smile \widetilde{r}.\label{eq:internal_continuum_mod2}
\end{gather}

\medskip

\noindent Before we move on we make an important assertion here. Recall that for any space $X$, the mod-2 cohomology $H^{*}(X,\Z_{2})$ is acted upon by a Steenrod algebra with operations called the Steenrod squares, denoted $\{\mathrm{Sq}^{i}\}$. We choose the non-canonical lift such that the subspace spanned by the Leray-Hirsch lifts $\mathrm{Span}(\widetilde{r}_{i})$ is stable under $\mathrm{Sq}^{1}$. For the classes of internal symmetries considered below, such a choice of the lift function $s_{\mathrm{O}(d)}$ exists. Indeed, in all of the examples we consider, the Leray-Hirsch generators may be chosen as characteristic classes of associated bundles induced by representations into $\mathrm{O}(m)$, and their $\mathrm{Sq}^{1}$-action is explicit and remains within the same fiber-generated subalgebra. This is the only compatibility with Steenrod operations needed in the following discussion.

\bigskip 

\noindent \noindent We call this compatibility of the lifts with $\mathrm{Sq}^{1}$ the Steenrod stability condition, represented concisely by
\begin{equation}
\label{eq:Steenrod_stability_cont}
    \mathrm{Sq}^{1}\left(\mathrm{Im}(s_{\mathrm{O}(d)})\right) \subseteq \mathrm{Im}(s_{\mathrm{O}(d)}).
\end{equation}
Equivalently, there is a degree-$1$ endomorphism
\begin{equation}
\label{eq:delta_cont}
    \delta_{R^*}:R^*\to R^{*+1},
\qquad R^*:=H^*(BH,\mathbb Z_2),
\end{equation}
defined by $\mathrm{Sq}^1(s_{\mathrm{O}(d)}(r))=s_{\mathrm{O}(d)}(\delta_{R^*}r)$ for all $r\in R^*$.

\bigskip 

Given that we have a Leray-Hirsch isomorphism on the continuum group $B\Gamma_{\mathrm{O}(d)}$, we now show that this implies we have Leray-Hirsch isomorphisms for all lattice subgroup objects $B\Gamma_{E}$ where $E\leq \mathrm{O}(d)$, and where the fibration $B\Gamma_{E}$ is induced through restriction from $B\Gamma_{\mathrm{O}(d)}$. 

\bigskip

\noindent Recall, we defined the fibration $BH\to B\Gamma_{\mathrm{O}(d)}\to B\mathrm{O}(d)$ using the action $\mu_{\mathrm{O}(d)}: \mathrm{O}(d)\to \mathrm{Aut}(H)$. This continuum group action induces actions 
\begin{equation*}
    \mu_{E}: E \xhookrightarrow[]{i_{E}} \mathrm{O}(d)\xrightarrow[]{\mu_{\mathrm{O}(d)}} \mathrm{Aut}(H),\quad \forall\;\; E\leq \mathrm{O}(d).
\end{equation*}
And therefore, we have the induced fibrations $BH \to B\Gamma_{E} \to BE$ where $\Gamma_{E}:= H\rtimes_{\mu_{E}}E$. More precisely, we have the following naturality square induced by the inclusion $i_E: E\hookrightarrow \mathrm{O}(d)$:
\begin{equation*}
    \begin{tikzcd}
        B\Gamma_{E} \arrow[r, "B\iota_{E}"] \arrow[d, "p_{E}"] & B\Gamma_{\mathrm{O}(d)} \arrow[d, "p_{\mathrm{O}(d)}"]\\
        BE \arrow[r, "Bi_{E}"] & B\mathrm{O}(d)
    \end{tikzcd}
\end{equation*}
where we have the induced map on fibrations $B\iota_{E}: B\Gamma_{E}\to B\Gamma_{\mathrm{O}(d)}$. This square is naturally a homotopy pullback-square: recall that we started with just the continuum fibration(right vertical arrow) $p_{\mathrm{O}(d)}$ and the induced inclusion(bottom arrow) of classifying spaces $Bi_{E}$, and we constructed the pullback object for the square\footnote{the homootpy pullback is defined as $BE\times^{h}_{B\mathrm{O}(d)}B\Gamma_{\mathrm{O}(d)}=\{(b_E, x_{\mathrm{O}(d)})\in BE \times B\Gamma_{\mathrm{O}(d)}\;|\; \exists\;\gamma: [0,1] \to B\mathrm{O}(d)\;\;\text{st}\;\; \gamma(0)=Bi_E(b_E),\; \gamma(1)=p_{\mathrm{O}(d)}(x_{\mathrm{O}(d)})\}$. This follows more generally from the fact that $\Gamma_{E}=E\times_{\mathrm{O}(d)}\Gamma_{\mathrm{O}(d)}:=\{(e,x)\in E\times H\;|\; i_{E}(e)=p_{\mathrm{O}(d)}(x)\}$, i.e. $\Gamma_{E}$ is the ordinary pullback of groups.}: 
\begin{equation*}
    B\Gamma_{E}\simeq BE\times^{h}_{B\mathrm{O}(d)}B\Gamma_{\mathrm{O}(d)}.
\end{equation*}
We denote the induced fiber inclusion on $B\Gamma_{E}$ as 
\begin{equation*}
    f_{E}: BH \hookrightarrow B\Gamma_{E}, \quad (f_{E})^{*}: H^{*}(B\Gamma_{E}, \Z_{2}) \to H^{*}(BH, \Z_{2}).
\end{equation*}
Now, recall the Leray-Hirsch isomorphism on the fibration $BH\to B\Gamma_{\mathrm{O}(d)}\to B\mathrm{O}(d)$ gives us the global lifts $\{\widetilde{r}_{i}\}\in H^{*}(B\Gamma_{\mathrm{O}(d)},\Z_{2})$ with restrictions $\{f_{\mathrm{O}(d)}^{*}(\widetilde{r}_{i})=r_{i}\}\in H^{*}(BH, \Z_{2})$ which form a $\Z_{2}$-basis. Then, using the pullback square, we can define the global classes on $B\Gamma_{E}$:
\begin{equation*}
    \widetilde{r}_{j}^{(E)}:=(B\iota_{E})^{*}(\widetilde{r}_{j})\in H^{*}(B\Gamma_{E}, \Z_{2}).
\end{equation*}
Now, we can use the fiber inclusion $f_{E}: BH\hookrightarrow B\Gamma_{E}$  followed by $B\iota_{E}$ to show that the restriction of the above classes to $BH$ satisfy:
\begin{equation*}
    (f_{E})^{*}\left(\widetilde{r}^{(E)}_{i}\right) = f_{\mathrm{O}(d)}^{*}(\widetilde{r}_{i})= r_{i}.
\end{equation*}
using the relation $f_{\mathrm O(d)}\simeq B\iota_E\circ f_E$,
which follows from the corresponding equality of group inclusions
$H\hookrightarrow\Gamma_E\hookrightarrow\Gamma_{\mathrm O(d)}$. Hence, we have the specification of global lifts on $B\Gamma_{E}$, given by $\{\widetilde{r}_{i}^{(E)}\}\in H^{*}(B\Gamma_{E},\Z_{2})$ which restrict under $f_{E}^{*}$ to the \textit{same} $\Z_{2}$-basis $\{r_{i}\}\in H^{*}(BH)$ of the fiber cohomology that we had assigned when formulating the Leray-Hirsch isomorphism for the continuum fibration $B\Gamma_{\mathrm{O}(d)}$. Therefore, proving the Leray-Hirsch isomorphism for the continuum fibration $BH \to B\Gamma_{\mathrm{O}(d)} \to B\mathrm{O}(d)$ proves the Leray-Hirsch isomorphism for every induced fibration $BH \to B\Gamma_{E} \to BE$ with $E\leq \mathrm{O}(d)$. Not only does the Leray-Hirsch isomorphism on $B\Gamma_{\mathrm{O}(d)}$ induce Leray-Hirsch isomorphisms on $B\Gamma_{E}$ for $E\leq \mathrm{O}(d)$, but it does so such that the system of isomorphisms
\begin{gather}
    \label{eq:LH_BE}
    \Delta_{E}: H^{*}(BE)\otimes H^{*}(BH) \to H^{*}(B\Gamma_{E}),\quad E\leq \mathrm{O}(d)\\
    s_{E}:  H^{*}(BH,\Z_{2}) \to H^{*}(B\Gamma_{E}, \Z_{2})\nonumber
\end{gather}
are compatible or natural with respect to each other. Here, we note that importantly the procedure above produces given $s_{\mathrm{O}(d)}$ all other lifts $s_{E}$. 

\bigskip 

More explicitly, by naturality we mean the following: for any morphism $\lambda:E\to E'$ in $\mathcal A$, represented by $\lambda\in \mathrm{O}(d)$ with $\lambda E\lambda^{-1}\le E'$, there is an induced homomorphism\footnote{Indeed, this respects the semidirect-product multiplication because
\begin{equation*}
\mu_{\mathrm O(d)}(\lambda e\lambda^{-1}) \circ \mu_{\mathrm O(d)}(\lambda)
= \mu_{\mathrm O(d)}(\lambda) \circ \mu_{\mathrm O(d)}(e).
\end{equation*}}
\begin{equation*}
\iota_\lambda:\Gamma_E\to \Gamma_{E'}, \quad (h,e)\mapsto (\mu_{\mathrm{O}(d)}(\lambda)(h),\; \lambda e \lambda^{-1}), 
\end{equation*}
and hence a pullback map
\begin{equation*}
     (B\iota_\lambda)^*:H^*(B\Gamma_{E'};\mathbb Z_2)\to H^*(B\Gamma_E;\mathbb Z_2).
\end{equation*}
By construction of the Leray–Hirsch lifts from the continuum classes, we have
\begin{equation*}
    s_E=(B\iota_E)^*\circ s_{\mathrm{O}(d)},\quad s_{E'}=(B\iota_{E'})^*\circ s_{\mathrm{O}(d)}
\end{equation*}
and therefore the lifts are compatible with restriction:
\begin{equation*}
    (B\iota_\lambda)^*\circ s_{E'} = (B\iota_{\lambda})^{*}\circ (B\iota_{E'})^{*}\circ s_{\mathrm{O}(d)} = (B(\iota_{E'}\circ \iota_{\lambda}))^{*}\circ s_{\mathrm{O}(d)}
\end{equation*}
The two homomorphisms into $\Gamma_{\mathrm O(d)}$ satisfy $\iota_{E'}\circ\iota_\lambda = c_{(1,\lambda)}\circ\iota_E$, where $c_{(1,\lambda)}$ is conjugation by $(1,\lambda)\in\Gamma_{\mathrm O(d)}$. Since inner automorphisms induce the identity on cohomology, it follows that
\begin{equation*}
    (B\iota_\lambda)^*\circ s_{E'} = (B\iota_E)^*\circ s_{\mathrm{O}(d)} =s_{E}.
\end{equation*}
Equivalently, for each $r_i\in H^*(BH;\mathbb Z_2)$,
\begin{equation*}
    (B\iota_\lambda)^*(\widetilde r_i^{(E')})=\widetilde r_i^{(E)}.
\end{equation*}
It follows that the corresponding Leray–Hirsch maps
\begin{gather*}
    \Delta_E:H^{*}(BE;\mathbb Z_2)\otimes H^{*}(BH;\mathbb Z_2)\to H^{* }(B\Gamma_E;\mathbb Z_2),\\
    \Delta_E: b\otimes r \longmapsto p_{E}^{*}(b)\smile s_E(r),
\end{gather*}
form a natural system with the relation:
\begin{equation*}
    (B\iota_\lambda)^*\circ \Delta_{E'} = \Delta_{E}\circ ((Bc_\lambda)^{*}\otimes \mathrm{id}_{H^{*}(BH)}). 
\end{equation*}

Since the system $(\Delta_{E}, s_{E})_{E\in \mathcal{A}}$ is natural with respect to the morphisms of $\mathcal{A}$, it induces an isomorphism on inverse limits:
\begin{equation}
    \label{eq:internal_lim_mod2}
    \Delta_{\lim}: \left( \varprojlim_{E\in \mathcal{A}}\; H^{*}(BE, \Z_{2})\right)\otimes H^{*}(BH, \Z_{2}) \to \varprojlim_{E\in \mathcal{A}}\; H^{*}(B\Gamma_{E}, \Z_{2}),
\end{equation}
\noindent We make one final remark about the Steenrod compatibility of the natural Leray-Hirsch system. By Eqs.~\ref{eq:Steenrod_stability_cont},\ref{eq:delta_cont}, the continuum lift $s_{\mathrm{O}(d)}$ is chosen so that $\operatorname{Im}(s_{\mathrm{O}(d)})$ is stable under $\mathrm{Sq}^{1}$. Therefore there is a degree-$1$ operator
\begin{equation}
\delta_{R^{*}}:R^{*}\to R^{*+1},
\qquad R^{*}:=H^{*}(BH,\mathbb Z_{2}),
\end{equation}
defined by the identity
\begin{equation}
\mathrm{Sq}^{1}(s_{\mathrm{O}(d)}(r))=s_{\mathrm{O}(d)}(\delta_{R^{*}}r).
\end{equation}
Since every induced lift is obtained by pullback, $s_E=(B\iota_E)^*s_{\mathrm{O}(d)}$, and Steenrod squares are natural, the same operator satisfies
\begin{equation}
\label{eq:indep_E}
\mathrm{Sq}^{1}(s_E(r))=s_E(\delta_{R^{*}}r)
\end{equation}
for every $E\le \mathrm{O}(d)$. Moreover, the compatibility of the lifts with all morphisms in $\mathcal A$ implies that this operator is independent of $E$ and is well defined after passing to the inverse limit.

\subsection{Classes of internal symmetries}
\label{app:class_internal}

Since, there are infinite possible internal symmetries that we could construct and which act on various different types of degrees of freedom, verifying the properties above for all cases is infeasible. We do, however, verify the properties for a few important classes of internal symmetries which we list below:
\bigskip 

\textit{1. Flavor symmetries:} if the group $\mathrm{O}(d)$ acts trivially on the internal symmetry $H$, and we have a direct product group $\Gamma_{\mathrm{O}(d)}= H\times \mathrm{O}(d)$. Here 
\begin{equation*}
B\Gamma_{\mathrm{O}(d)}\cong BH\times B\mathrm{O}(d),
\end{equation*} 
and the fiber inclusion $f_{\mathrm{O}(d)}:BH\hookrightarrow BH\times B\mathrm{O}(d)$ is $x\mapsto (x,*)$ for some chosen base-point $*\in B\mathrm{O}(d)$. Hence 
\begin{equation*}
f_{\mathrm{O}(d)}^{*}:H^{*}(BH\times B\mathrm{O}(d);\mathbb Z_2)\to H^{*}(BH;\mathbb Z_2)  
\end{equation*}
is the projection on the $BH$-factor and is surjective. In particular, for any chosen $\mathbb Z_2$-basis ${r_j}$ of $R^{*}$, the classes $\widetilde{r}_j:=r_j\otimes 1\in H^{*}(BH\times B\mathrm{O}(d),\mathbb Z_2)$ restrict to $r_j$. Thus Leray–Hirsch isomorphism holds trivially.

\bigskip
 
\textit{2. Unitary symmetries:} if we have unitary groups $H=\U(1)\;\text{or}\;\mathrm{SU}(N)$, then orientation-reversing spatial symmetries act on $H$ by complex conjugation. Fix an orientation(twist) character $\det:\mathrm{O}(d)\to\{\pm1\}$ recording which spatial symmetries are orientation reversing. Define
\begin{equation*}
    \mu_{g}=\begin{cases}
    \mathrm{id}_H,& \det(g)=+1,\\[2pt]
    \text{complex conjugation on }H,& \det(g)=-1.
    \end{cases}
\end{equation*}
We discuss the $H=\mathrm{SU}(N)$ case here, which can be readily used to treat $H=\U(1)$. Consider the real $2N$-dimensional orthogonal representation  
\begin{equation*}
\Gamma_{\mathrm{O}(d)}=\mathrm{SU}(N)\rtimes_{\mu_{\mathrm{O}(d)}} \mathrm{O}(d)\longrightarrow O(2N)  
\end{equation*}
given by the defining complex $N$-dimensional representation of $\mathrm{SU}(N)$ on $\mathbb C^N$, and letting $\det(g)=-1$ act by complex conjugation on $\mathbb C^N$ (viewed as $\mathbb R^{2N}$). This yields a map $B\Gamma_{\mathrm{O}(d)}\to BO(2N)$ and hence a real rank-$2N$ bundle 
\begin{equation*}
\xi\to B\Gamma_{\mathrm{O}(d)}.
\end{equation*}
Restricting $\xi$ along the fiber inclusion $f_{\mathrm{O}(d)}:B\mathrm{SU}(N)\hookrightarrow B\Gamma_{\mathrm{O}(d)}$ recovers the underlying real bundle of the universal complex $N$-plane bundle over $B\mathrm{SU}(N)$. Consequently,  
\begin{equation*}
f_{\mathrm{O}(d)}^{*}\big(w_{2k}(\xi)\big)=\overline{c_k}\in H^{2k}(B\mathrm{SU}(N);\mathbb Z_2),\quad  f_{\mathrm{O}(d)}^{*}\big(w_{2k+1}(\xi)\big)=0,  
\end{equation*}
where $\overline{c_k}$ is the mod-2 reduction of the Chern class. Since 
\begin{equation*}
H^{*}(B\mathrm{SU}(N);\mathbb Z_2)\cong \mathbb Z_2[\overline c_2,\dots,\overline c_N],
\end{equation*}
the global classes $\widetilde{r}_k:=w_{2k}(\xi)\in H^{2k}(B\Gamma_{\mathrm{O}(d)};\mathbb Z_2)$ for $k=2,\dots,N$ restrict to the polynomial generators of $R^{*}=H^{*}(B\mathrm{SU}(N),\Z_{2})$. Thus Leray–Hirsch holds.

\bigskip 
    
\textit{3. Real internal symmetry $H=\mathrm{SO}(N)$.} This case is relevant for systems with an $N$-component real order parameter carrying an $\SO(N)$ internal symmetry. The choice of $\mu_{\mathrm{O}(d)}$ below is the natural $\det$-coupling in which orientation-reversing spatial symmetries act by conjugation implemented by a reflection. For example, this would be an appropriate description for a system with an $N$-component order parameter $\phi$ with a discrete spatial reflection acting as $\phi \mapsto r\phi$. 
Fix a reflection $r\in O(N)\setminus \SO(N)$, so that $r^{2}=1$, and define $\mu_{\mathrm{O}(d)}:\mathrm{O}(d)\to \mathrm{Aut}(\SO(N))$ by  
\begin{equation*}
\mu_g(A)=\begin{cases}  
A,& \det(g)=+1,\\  
rAr^{-1},& \det(g)=-1.  
\end{cases}  
\end{equation*}
Define $z:\mathrm{O}(d)\to O(N)$ by $z(g)=1$ if $\det(g)=+1$ and $z(g)=r$ if $\det(g)=-1$. Then  
\begin{equation*}
\phi:\Gamma_{\mathrm{O}(d)}=\SO(N)\rtimes_{\mu_{\mathrm{O}(d)}} \mathrm{O}(d)\to O(N),\qquad \phi(A,g)=A\cdot z(g)  
\end{equation*}
is a homomorphism (the semidirect product identity is exactly $z(g)A'z(g)^{-1}=\mu_g(A')$). This yields a map $B\Gamma_{\mathrm{O}(d)}\to BO(N)$ and a rank-$N$ bundle $\eta\to B\Gamma_{\mathrm{O}(d)}$. Restricting to the fiber $f_{\mathrm{O}(d)}:B\SO(N)\hookrightarrow B\Gamma_{\mathrm{O}(d)}$, the bundle $\eta$ becomes the universal oriented $N$-plane bundle. Hence the Stiefel–Whitney classes satisfy  
\begin{equation*}
f_{\mathrm{O}(d)}^{*}(w_j(\eta))=w_j \in H^j(B\SO(N);\mathbb Z_2),\qquad j=2,\dots,N,  
\end{equation*}
and these classes generate the mod-2 cohomology ring  
\begin{equation*}
H^{*}(B\SO(N);\mathbb Z_2)\cong \mathbb Z_2[w_2,\dots,w_N].
\end{equation*}
Therefore the global classes $\widetilde{r}_j:=w_j(\eta)\in H^{*}(B\Gamma_{\mathrm{O}(d)};\mathbb Z_2)$ restrict to generators of $R^{*}= H^{*}(B\SO(N), \Z_{2})$. Thus Leray–Hirsch holds.

\bigskip 

\noindent In all of the cases above, the key point is the same: one realizes the total group $\Gamma_{\mathrm{O}(d)}$ inside an orthogonal group $O(m)$ (or in the trivial product case, as a direct product), and then uses characteristic classes of the associated bundle over $B\Gamma_{\mathrm{O}(d)}$ to produce global classes whose restrictions generate $H^{*}(BH, \Z_{2})$. This is exactly the input needed for the Leray–Hirsch hypothesis. In the nontrivial cases, these lifts are characteristic classes of associated bundles, so their $\mathrm{Sq}^{1}$-behavior is explicit and remains inside the same fiber-generated sub-algebra; in the flavor case, the same stability is immediate from the product decomposition. We do not attempt a general classification here, but these examples suggest a general template: whenever the coupling $\mu_{\mathrm{O}(d)}$
admits a sufficiently well-behaved representation-theoretic realization that produces such global characteristic classes, the same method should apply.

\subsection{Formal proof of $H$-enriched stratification theorems}
\label{app:proof_internal}

Now we prove the $H$-enriched analogs of the stratification theorems. This subsection has two parts. First, we prove the mod-$2$ Quillen restriction theorem for the $H$-enriched spaces $B\Gamma_E$. This result follows from the natural Leray-Hirsch system constructed above together with the usual Quillen isomorphism for $\mathrm{O}(d)$. Second, we prove the $H$-enriched main theorem in the group cohomology model. The latter proof requires several preparatory claims concerning the lifted twisted integral theory, which are collected in App.~\ref{app:prep_res_H} before the final proof in App.~\ref{app:final_proof_H}.

\bigskip 

\noindent Let us consider the continuum fibration:
\begin{equation*}
    BH \xhookrightarrow[]{f_{\mathrm{O}(d)}} B\Gamma_{\mathrm{O}(d)} \xrightarrow[]{p_{\mathrm{O}(d)}} B\mathrm{O}(d),
\end{equation*}
and the induced fibrations on classifying spaces of subgroups $E\leq \mathrm{O}(d)$:
\begin{equation*}
    BH \xhookrightarrow[]{f_{E}} B\Gamma_{E} \xrightarrow[]{p_{E}} BE.
\end{equation*}
Let us denote the mod-$2$ cohomology of the fiber as $R^{*}=H^{*}(BH,\Z_{2})$. 

\bigskip 

\begin{proposition}[$H$-enriched Quillen isomorphism in mod-$2$ cohomology]
\label{prop:quillen_isom_H}
    Assume Hypothesis~\ref{hyp:1}. Then the $H$-enriched analog of the Quillen isomorphism in mod-$2$ cohomology
    \begin{equation}
    \label{eq:internal_mod2_isom}
    \res_{H}: H^{*}(B\Gamma_{\mathrm{O}(d)}, \Z_{2}) \longrightarrow \varprojlim\limits_{E\in \mathcal{A}}\; H^{*}(B\Gamma_{E}, \Z_{2})
    \end{equation}
    is an isomorphism.
\end{proposition}

\begin{proof}[\textnormal{\textbf{Proof of Prop.~\ref{prop:quillen_isom_H}}}]  
Recall that provided Hypothesis~\ref{hyp:1} is satisfied, we have the Leray-Hirsch isomorphism Eq.~\ref{eq:internal_continuum_mod2}, which is given by the two maps $(\Delta_{\mathrm{O}(d)}, s_{\mathrm{O}(d)})$ where we recall that $s_{\mathrm{O}(d)}$ is the non-canonical section which defined the global lifts of the $\Z_{2}$-basis of generators on the fiber cohomology. Furthermore, this isomorphism induces a system of Leray-Hirsch isomorphisms $(\Delta_{E},s_{E})$ for all $E\in \mathcal{A}$ in a compatible or natural manner. This system of isomorphisms can be used to construct a "limit" isomorphism given in Eq.~\ref{eq:internal_lim_mod2}. We can then construct the following square:
\begin{equation}
\label{eq:app_diagram}
    \begin{tikzcd}[column sep = large]
        H^{*}(B\mathrm{O}(d), \Z_{2})\;\otimes  R^{*} \arrow[d, "\Delta_{\mathrm{O}(d)}", swap, "\cong"'] \arrow[r, "\res\otimes \mathrm{id}_{R^{*}}", "\cong"'] &  \left( \varprojlim\limits_{E\in \mathcal{A}}\; H^{*}(BE, \Z_{2})\right)\otimes R^{*} \arrow[d, "\Delta_{\lim}", "\cong"']\\
        H^{*}(B\Gamma_{\mathrm{O}(d)}, \Z_{2}) \arrow[r, "\res_{H}"] &\varprojlim\limits_{E\in \mathcal{A}}\; H^{*}(B\Gamma_{E}, \Z_{2})
    \end{tikzcd}
\end{equation}
Here, the two vertical maps and the top map are all isomorphisms, therefore, we obtain the mod-$2$ result which is the the bottom arrow is an isomorphism:
\begin{equation}
    \res_{H}: H^{*}(B\Gamma_{\mathrm{O}(d)}, \Z_{2}) \longrightarrow \varprojlim\limits_{E\in \mathcal{A}}\; H^{*}(B\Gamma_{E}, \Z_{2}).
\end{equation}
This is the analog of Eq.~\ref{eq:quillen_mod2}(and Eq.~\ref{eq:qisom_mod2} in App.~\ref{app:refinement}) derived for the no-internal-symmetry case.
\end{proof}

The rest of the subsection is organized as follows. We first state the $H$-enriched main theorem. Before trying to prove the theorem we impose another hypothesis needed for the proof of the main theorem. After that, App.~\ref{app:prep_res_H} collects the preparatory claims. Finally, we use these claims to prove the main theorem in App.~\ref{app:final_proof_H}.

\bigskip 

\begin{theorem}[$H$-enriched main theorem in the group cohomology model]
\label{appG:main_theorem_H}
Assume Hypotheses~\ref{hyp:1} and~\ref{hyp:2}. Then, for every spatial dimension $d\geq 1$, the stratification map
\begin{equation}
F_{2,H}:
\tor_{2}\bigl( \mathrm{cSPT}^{d+1}_{\mathrm{univ}, H} \bigr)
\;\twoheadrightarrow\;
\tor_{2}\left(\Theta^{d+1} \bigl(B(H\rtimes_{\rho} \mathrm{O}(d))\bigr)\right).
\end{equation}
is surjective. Here 
\begin{equation*}
\mathrm{cSPT}^{d+1}_{\mathrm{univ},H}=\varprojlim\limits_{G\in \ofin(\mathrm{O}(d))} \; \tor_{2} \left(H^{d+2}(B(H\rtimes_{\rho}G),\Z_{w_1})\right).
\end{equation*}
\end{theorem}

\bigskip

\noindent We now impose the additional hypothesis needed for the proof of the main theorem. Its role is to ensure that the primary mod-$2$ Bockstein contains all of the $2$-primary torsion information required in the lifted argument. Equivalently, it guarantees that mod-$2$ reduction is injective on the order-$2$ subgroup of the relevant twisted integral cohomology groups. This property is automatic in the no-internal-symmetry case, but in the presence of internal symmetry it must be assumed explicitly.

\bigskip 

\begin{hypothesis}[No higher torsion from $H$]
    \label{hyp:2}
    We assume the following identities hold:
    \begin{gather*}
        \mathrm{Tor}_{2}(H^{n}(B\Gamma_{\mathrm{O}(d)}, \mathbb{Z}_{w_{1}}))= H^{n}(B\Gamma_{\mathrm{O}(d)}, \mathbb{Z}_{w_{1}})[2]\\ 
        \mathrm{Tor}_{2}(H^{n}(B\Gamma_E, \mathbb{Z}_{w_{1}}))= H^{n}(B\Gamma_E, \mathbb{Z}_{w_{1}})[2]\quad \forall\;\; E\in \mathcal{A}_{2}(\mathrm{O}(d)).
    \end{gather*}
    Recall that $\tor_{2}(A)$ denotes the $2$-primary torsion subgroup of abelian group $A$, whereas $A[2]$ denotes order-$2$ torsion. 
\end{hypothesis}

\bigskip 

\noindent Under Hypothesis~\ref{hyp:2}, each objectwise reduction map
\begin{equation*}
    \rho_{E}^{H}:H^{n}(B\Gamma_E,\mathbb Z_{w_1})[2]\to H^{n}(B\Gamma_E,\mathbb Z_2)
\end{equation*}
is injective. Indeed, if an order-$2$ class lies in the kernel of reduction, exactness of the coefficient long exact sequence implies that it is twice an integral class, which would require nontrivial torsion of order at least $4$. Hypothesis~\ref{hyp:2} excludes this possibility. The hypothesis enters the lifted argument in two related ways. First, the injectivity above is needed when passing from mod-$2$ information back to order-$2$ integral torsion. Second, it ensures that the order-$2$ Bockstein complex already contains all of the relevant $2$-primary torsion information, so that no higher Bockstein differentials are required in the computation of $H^{*}(R^{*},\delta_{R^{*}})$.

\bigskip

\noindent In the no-internal-symmetry case, the corresponding statement follows from the fact that the relevant $2$-primary torsion groups for $B\mathrm{O}(d)$ and for elementary abelian subgroups are already order-$2$. With internal symmetry, higher $2$-torsion can in principle appear, for example when the integral cohomology of $BH$ contains classes of order $4$ or higher. We therefore record the assumption explicitly. We note that this hypothesis holds for all classes of internal symmetry groups considered in App.~\ref{app:class_internal}, and more importantly will hold for general $H$ unless they are made up of groups like $H=\Z_{4}$ which could give rise to higher order-$2$ torsion. We claim that the hypothesis holds for the great majority of internal symmetry groups which are considered in physical theories.

\bigskip 

\noindent If Hypothesis~\ref{hyp:2} fails, the underlying strategy can be generalized using the full Bockstein spectral sequence associated to multiplication by $2$ on integral coefficients.\footnote{The first differential of this spectral sequence is the primary Bockstein
\begin{equation*}
d_{1}=\rho\circ\beta=\mathrm{Sq}^{1},
\end{equation*}
whereas the higher differentials detect successively deeper $2$-primary torsion. See Ch. 10 in Ref.~\cite{McCleary2000} or Ch.7 in Ref.~\cite{Mosher2008}.}
In that setting, the mod-$2$ Leray-Hirsch isomorphism must lift compatibly to the corresponding Bockstein exact couples, so that the spatial-internal decomposition persists through the higher Bockstein differentials. We do not impose this stronger structure in the general theorem, since it is unnecessary whenever Hypothesis~\ref{hyp:2} holds. For the higher-torsion internal symmetry used in Sec.~\ref{sec:example_2_highertorsion}, the required compatibility and the resulting secondary-Bockstein reconstruction are established separately in App.~\ref{app:higher_order_example}.

\bigskip

\subsubsection{Preparatory results}
\label{app:prep_res_H}

\begin{claim}
The continuum Bockstein map
    \begin{equation}
    \label{eq:internal_beta_Od}
        \beta_{\mathrm{O}(d)}^{H}: H^{d+1}(B\Gamma_{\mathrm{O}(d)}, \mathbb{Z}_{2}) \to H^{d+2}(B\Gamma_{\mathrm{O}(d)}, \mathbb{Z}_{w_{1}})[2]
    \end{equation}
is a surjection.
\end{claim}
\begin{proof}[\textnormal{\textbf{Proof}}]
The proof follows from the exactness of the LES in cohomology induced by SES in coefficients $0\to \Z_{w_{1}}\to \Z_{w_{1}} \to \Z_{2}\to 0$:
\begin{equation*}
    \begin{tikzcd}
    H^{d+1}(X,\Z_{w_{1}}) \rar[]{\times 2} & H^{d+1}(X,\Z_{w_{1}}) \rar[]{\rho^{H}_{\mathrm{O}(d)}}
             \ar[draw=none]{d}[name=X, anchor=center]{}
    & H^{d+1}(X,\Z_{2}) \ar[rounded corners,
            to path={ -- ([xshift=2ex]\tikztostart.east)
                      |- (X.center) \tikztonodes
                      -| ([xshift=-2ex]\tikztotarget.west)
                      -- (\tikztotarget)}]{dll}[at end]{\beta_{\mathrm{O}(d)}^{H}} \\      
  H^{d+2}(X,\Z_{w_{1}}) \rar[]{\times 2} & H^{d+2}(X,\Z_{w_{1}}) \rar[]{\rho^{H}_{\mathrm{O}(d)}} & H^{d+2}(X,\Z_{2})
\end{tikzcd}
\end{equation*}
where $X=B\Gamma_{\mathrm{O}(d)}$. Exactness then implies:
\begin{equation*}
    \mathrm{Im}\left(\beta^{H}_{\mathrm{O}(d)}\right) = \mathrm{ker}\left(\times 2\right) = H^{d+2}(B\Gamma_{\mathrm{O}(d)},\Z_{w_{1}})[2]
\end{equation*}
where the RHS is the order-2 subgroup of the twisted integral cohomology of the space $B\Gamma_{\mathrm{O}(d)}$. And therefore, indeed the map Eq.~\ref{eq:internal_beta_Od} is surjective. 
\end{proof}

\bigskip 

Unlike the proof of the main theorem in App.~\ref{app:lift}, in this section it is necessary to begin with the generalized construction of the inverse limit complex, and its cohomology in arbitrary degree. In the no internal symmetry case, one only needs to understand the cohomology of spatial classes in total degree $(d+2)$. In the presence of internal symmetries, however, the total degree can split between internal symmetry classes and spatial classes in many ways. Thus making it necessary to have the general construction from the outset.

Similar to the proof of the main theorem without internal symmetry, the analysis depends on the construction of the analog of the complex $(\Delta_{\lim}, D_{\lim})$ (see Eq.~\ref{eq:Lambda_complex}) to include internal symmetries. We first define the underlying mod-$2$ inverse-limit group
\begin{equation}
{\bf \Lambda}_{H} := \varprojlim_{E\in\mathcal A} H^{*}(B\Gamma_E,\Z_{2}).
\label{eq:Lambda_H_def}
\end{equation}
For each $E\in\mathcal A$, the short exact sequence of local coefficient systems $0\to \mathbb Z_{w_1} \xrightarrow{\times 2} \mathbb Z_{w_1}
\xrightarrow{\rho} \mathbb Z_2 \to 0$ on $B\Gamma_E$ gives a twisted Bockstein map
\begin{equation}
\beta_{E}^{H}: H^{*}(B\Gamma_E,\Z_{2}) \to H^{*+1}(B\Gamma_E,\Z_{w_1})
\end{equation}
and a reduction map
\begin{equation}
\rho_{E}^{H}: H^{*}(B\Gamma_E,\Z_{w_1}) \to H^{*}(B\Gamma_E,\Z_{2}).
\end{equation}
The maps $\beta_{E}^{H},\rho_{E}$ are natural with respect to morphisms in $\mathcal A$, and therefore assemble into inverse-limit operators:
\begin{gather}
    \label{eq:beta_lim_H}
    \beta_{\lim}^{H}: \lim_{E\in \mathcal{A}}H^{*}(B\Gamma_E,\Z_{2}) \to \lim_{E\in \mathcal{A}}H^{*+1}(B\Gamma_E,\Z_{w_1}),\\
    \rho_{\lim}^{H}: \lim_{E\in \mathcal{A}} H^{*}(B\Gamma_E,\Z_{w_1}) \to \lim_{E\in \mathcal{A}} H^{*}(B\Gamma_E,\Z_{2}).
    \label{eq:rho_lim_H}
\end{gather}

\bigskip 
\noindent We define the operator
\begin{equation}
D_{\Gamma_E}^{H}:= \rho_{\Gamma_E}^{H}\circ\beta_{\Gamma_E}^{H}.
\label{eq:D_Gamma_E_H_def}
\end{equation}
Equivalently, for $x\in H^{*}(B\Gamma_E,\Z_2)$,
\begin{equation}
D_{\Gamma_E}^{H}(x) = p_E^{*}w_1\cup x+\mathrm{Sq}^{1}(x),\quad (w_{1})_{B\Gamma_{E}}=p_{E}^{*}(w_{1})_{BE}.
\label{eq:D_Gamma_E_H_formula}
\end{equation}
where the local system $(w_{1})_{B\Gamma_{E}}$ is induced from the local system on the base space $BE$, see Eq.~\ref{eq:coeff_pullback}, using the projection
\begin{equation*}
    p_{E}: B\Gamma_{E} \to BE, \quad p_{E}^{*}: H^{*}(BE, \Z_{2}) \to H^{*}(B\Gamma_E, \Z_{2}).
\end{equation*}
The maps $D_{\Gamma_E}^{H}$ are natural with respect to morphisms in $\mathcal A$, and therefore assemble into an inverse-limit operator
\begin{equation}
D_{\lim}^{H}: {\bf \Lambda}_{H}^{*} \to {\bf \Lambda}_{H}^{*+1}.
\label{eq:D_lim_H_def}
\end{equation}

\begin{definition}[$H$-enriched inverse-limit Bockstein complex]
\label{def:LambdaH_complex}
The $H$-enriched inverse-limit Bockstein complex is the pair
\begin{equation}
({\bf \Lambda}_{H},D_{\lim}^{H}),
\end{equation}
where ${\bf \Lambda}_{H}$ is defined in Eq.~\eqref{eq:Lambda_H_def} and $D_{\lim}^{H}$ is the inverse-limit operator defined in Eq.~\eqref{eq:D_lim_H_def}.
\end{definition}

\bigskip 

We now identify this complex using the natural Leray-Hirsch isomorphism. Recall from Eq.~\eqref{eq:internal_lim_mod2} that
\begin{equation}
\Delta_{\lim}: \left( \varprojlim_{E\in\mathcal A} H^{*}(BE,\Z_{2}) \right)\otimes R^{*}
\xrightarrow{\cong} \varprojlim_{E\in\mathcal A} H^{*}(B\Gamma_E,\Z_{2}) = {\bf \Lambda}_{H},
\label{eq:Delta_lim_again}
\end{equation}
where $R^{*}:=H^{*}(BH,\Z_{2})$. 

\bigskip 

\begin{claim}[Decomposition of the $H$-enriched inverse-limit differential]
\label{claim:DlimH_decomposition}
Under the isomorphism Eq.~\ref{eq:Delta_lim_again}, and using the Steenrod-compatible Leray-Hirsch system from Eq.~\eqref{eq:indep_E}, the differential decomposes as
\begin{equation}
\Delta_{\lim}^{-1}\circ D_{\lim}^{H}\circ \Delta_{\lim} =
D_{\lim}\otimes \mathrm{id}_{R^{*}} + \mathrm{id}_{{\bf \Lambda}}\otimes \delta_{R^{*}},
\label{eq:DlimH_decomposition}
\end{equation}
where ${\bf \Lambda}:=\varprojlim_{E\in\mathcal A}H^{*}(BE,\Z_{2})$ (see Eq.~\ref{eq:Lambda_complex}).
\end{claim}

\begin{proof}[\textnormal{\textbf{Proof}}]
It is enough to prove the corresponding statement objectwise for each $E\in\mathcal A$, and then pass to the inverse limit. Let $b\in H^{*}(BE,\Z_{2})$ and $r\in R^{*}=H^{*}(BH,\Z_{2})$. Under the Leray-Hirsch map $\Delta_E$, the element $b\otimes r$ is sent to
\begin{equation}
\Delta_E(b\otimes r)=p_E^{*}(b)\smile s_E(r).
\end{equation}
Using Eq.~\eqref{eq:D_Gamma_E_H_formula}, the Cartan formula, and naturality of Steenrod squares, we obtain
\begin{align*}
    D_{E}^{H}(p_{E}^{*}(b)\smile \widetilde{r}^{(E)}) &= (w_{1})_{B\Gamma_E} \cup (p_{E}^{*}(b)\smile \widetilde{r}^{(E)}) \\
    &\qquad +\;\; \mathrm{Sq}^{1}_{B\Gamma_{E}}(p_{E}^{*}(b)\smile \widetilde{r}^{(E)})\\
    &= p_{E}^{*} (w_{1})_{BE} \cup (p_{E}^{*}(b)\smile \widetilde{r}^{(E)}) \\
    &\;\;+ p_{E}^{*}\mathrm{Sq}^{1}_{BE}(b)\smile \widetilde{r}^{(E)} + p_{E}^{*}(b)\smile \mathrm{Sq}^{1}_{B\Gamma_{E}}(\widetilde{r}^{(E)})\\
    &= p_{E}^{*}D_{E}(b)\smile \widetilde{r}^{(E)} + p_{E}^{*}(b)\smile \mathrm{Sq}^{1}_{B\Gamma_{E}}(\widetilde{r}^{(E)}).
\end{align*}
By the definition of $D_E$ on the base and by Eq.~\eqref{eq:indep_E}, this becomes
\begin{equation}
D_{\Gamma_E}^{H}\bigl(p_E^{*}(b)\smile s_E(r)\bigr)
= p_E^{*}D_E(b)\smile s_E(r) + p_E^{*}(b)\smile s_E(\delta_{R^{*}}r).
\end{equation}
Applying $\Delta_E^{-1}$ gives
\begin{equation}
\Delta_E^{-1}\circ D_{\Gamma_E}^{H}\circ \Delta_E(b\otimes r)
= D_E(b)\otimes r + b\otimes \delta_{R^{*}}(r).
\end{equation}
Hence
\begin{equation}
\Delta_E^{-1}\circ D_{\Gamma_E}^{H}\circ \Delta_E
= D_E\otimes \mathrm{id}_{R^{*}} + \mathrm{id}_{H^{*}(BE)}\otimes \delta_{R^{*}}.
\end{equation}
The maps $\Delta_E$, $D_E$, $D_{\Gamma_E}^{H}$, and $\delta_{R^{*}}$ are compatible with restriction morphisms in $\mathcal A$. Therefore these objectwise identities assemble into the inverse-limit identity
\begin{equation}
\Delta_{\lim}^{-1}\circ D_{\lim}^{H}\circ \Delta_{\lim}
= D_{\lim}\otimes \mathrm{id}_{R^{*}} + \mathrm{id}_{{\bf \Lambda}}\otimes \delta_{R^{*}}.
\end{equation}
This proves the claim.
\end{proof}

\bigskip 

\begin{claim}[Cohomology of the $H$-enriched inverse-limit complex]
\label{claim:LambdaH_cohomology}
If $d=2m$ is even, then
\begin{equation}
H^{*}({\bf \Lambda}_{H},D_{\lim}^{H})
\cong
y_m\mathbb F_2[y_1^2,\dots,y_m^2]\otimes R_{\mathrm{free},2}^{*}.
\label{eq:LambdaH_cohomology_even}
\end{equation}
If $d$ is odd, then
\begin{equation}
H^{*}({\bf \Lambda}_{H},D_{\lim}^{H})=0.
\label{eq:LambdaH_cohomology_odd}
\end{equation}
\end{claim}

\begin{proof}[\textnormal{\textbf{Proof}}]
By Claim~\ref{claim:DlimH_decomposition}, the complex $({\bf \Lambda}_{H},D_{\lim}^{H})$ is identified with the tensor-product complex
\begin{equation}
\left( {\bf \Lambda}\otimes R^{*}, D_{\lim}\otimes \mathrm{id}_{R^{*}} + \mathrm{id}_{{\bf \Lambda}}\otimes \delta_{R^{*}} \right).
\end{equation}
Since all groups involved are $\mathbb F_2$-vector spaces, the K\"unneth theorem gives
\begin{equation}
H^{*}({\bf \Lambda}_{H},D_{\lim}^{H})
\cong H^{*}({\bf \Lambda},D_{\lim}) \otimes H^{*}(R^{*},\delta_{R^{*}}).
\label{eq:LambdaH_Kunneth}
\end{equation}

We first identify the internal factor. Since the trivial group belongs to $\mathcal A$, Hypothesis~\ref{hyp:2} applied to $E=\{1\}$ implies that the $2$-primary torsion in $H^{*}(BH,\Z)$ is all order-$2$. Consider the untwisted coefficient sequence $0\to \Z \xrightarrow{\times 2} \Z \xrightarrow{\rho} \Z_2 \to 0$
on $BH$. The associated Bockstein $\beta_R$ satisfies
\begin{equation}
\delta_{R^{*}}=\rho_{R^{*}}\circ\beta_{R^{*}}.
\end{equation}
Because reduction is injective on order-$2$ torsion under Hypothesis~\ref{hyp:2}, we have
\begin{equation}
\ker(\delta_{R^{*}}) = \ker(\beta_{R^{*}}) =
\operatorname{Im}\left(H^{*}(BH,\Z)\xrightarrow{\rho_{R^{*}}}H^{*}(BH,\Z_2)\right).
\end{equation}
Moreover,
\begin{equation}
\operatorname{Im}(\delta_{R^{*}}) = \rho_{R^{*}}\left(\operatorname{Im}(\beta_{R^{*}})\right)
= \rho_{R^{*}}\left(H^{*}(BH,\Z)[2]\right).
\end{equation}
Therefore
\begin{equation}
H^{*}(R^{*},\delta_{R^{*}})
\cong \frac{\rho_{R^{*}}\left(H^{*}(BH,\Z)\right)}{\rho_{R^{*}}\left(H^{*}(BH,\Z)[2]\right)}.
\end{equation}
After quotienting by the reduction of the order-$2$ torsion subgroup, only the mod-$2$ reductions of the free integral classes remain. Hence
\begin{equation}
H^{*}(R^{*},\delta_{R^{*}}) = R_{\mathrm{free},2}^{*}.
\label{eq:Rfree2_internal}
\end{equation}

\bigskip 

\noindent Now we use the spatial computation from App.~\ref{app:lift}. For $d=2m$ even,
\begin{equation}
H^{*}({\bf \Lambda},D_{\lim})
\cong
y_m\mathbb F_2[y_1^2,\dots,y_m^2].
\end{equation}
Combining this with Eqs.~\eqref{eq:LambdaH_Kunneth} and~\eqref{eq:Rfree2_internal} gives
\begin{equation}
H^{*}({\bf \Lambda}_{H},D_{\lim}^{H})
\cong
y_m\mathbb F_2[y_1^2,\dots,y_m^2]\otimes R_{\mathrm{free},2}^{*}.
\end{equation}
If $d$ is odd, then the spatial cohomology $H^{*}({\bf \Lambda},D_{\lim})$ vanishes. Hence Eq.~\eqref{eq:LambdaH_Kunneth} implies
\begin{equation}
H^{*}({\bf \Lambda}_{H},D_{\lim}^{H})=0.
\end{equation}
This proves the claim.
\end{proof}

\bigskip 

We now define the free-shadow subgroup for the $H$-enriched case. Consider the map:
\begin{equation}
\label{eq:res_full_H}
    \res_{H,\mathcal{A}}^{\mathrm{full}}: H^{n}(B\Gamma_{\mathrm{\mathrm{O}(d)}}, \Z_{w_{1}}) \to \varprojlim_{E\in \mathcal{A}} H^{n}(B\Gamma_{E}, \Z_{w_{1}}),
\end{equation}
where in contrast to Eq.~\ref{eq:internal_resA} the map has domain the full group including the free part. 

\begin{definition}[$H$-enriched free-shadow subgroup]
\label{def:Sfree_H}
Fix a degree $n$. The $H$-enriched free-shadow subgroup is defined by
\begin{equation}
S_{\mathrm{free},H}^{n}
:=
\operatorname{Im}
\left(
\res^{\mathrm{full}}_{H,\mathcal A}
\big\vert_{\mathrm{Free}\left(H^{n}(B\Gamma_{\mathrm{O}(d)},\mathbb Z_{w_1})\right)}
\right)
\subseteq
\varprojlim_{E\in\mathcal A}
H^{n}(B\Gamma_E,\mathbb Z_{w_1})[2].
\label{eq:def_Sfree_H}
\end{equation}
Thus $S_{\mathrm{free},H}^{n}$ consists of the elementary order-$2$ compatible families obtained by restricting twisted free continuum classes.
\end{definition}

\begin{claim}[Exact sequence measuring the failure of $H$-enriched Bockstein surjectivity]
\label{claim:bockstein_failure_sequence_H}
There is a short exact sequence
\begin{equation}
\label{eq:ses_Bockstein_failure_H}
0
\rightarrow
\operatorname{Im}\left(\left(\beta_{\lim}^{H}\right)^{n-1}\right)
\rightarrow
\mathrm{cSPT}_{\mathrm{elem},H}^{n}
\xrightarrow{\left(\overline{\rho}_{\lim}^{H}\right)^{n}}
\dfrac{
\operatorname{Im}\left(\left(\rho_{\lim}^{H}\right)^{n}\right)
}{
\operatorname{Im}\left(\left(D_{\lim}^{H}\right)^{n-1}\right)
}
\rightarrow
0.
\end{equation}
Here
\begin{equation}
\left(\overline{\rho}_{\lim}^{H}\right)^{n}(x)
:=
\left[
\left(\rho_{\lim}^{H}\right)^n(x)
\right],
\end{equation}
where $[\;\;]$ denotes the quotient map
\begin{equation}
\operatorname{Im}\left(\left(\rho_{\lim}^{H}\right)^n\right)
\longrightarrow
\dfrac{
\operatorname{Im}\left(\left(\rho_{\lim}^{H}\right)^n\right)
}{
\operatorname{Im}\left(\left(D_{\lim}^{H}\right)^{n-1}\right)
}.
\end{equation}
\end{claim}

\begin{proof}[\textnormal{\textbf{Proof}}]
The proof is identical to the proof of Claim~\ref{claim:bockstein_failure_sequence}, with the maps $\beta_{\lim},\rho_{\lim},D_{\lim}$
replaced by their $H$-enriched analogs $\beta_{\lim}^{H},\rho_{\lim}^{H},D_{\lim}^{H}$. The only point to check is the injectivity of $(\rho_{\lim}^{H})^{n}$ on $\mathrm{cSPT}_{\mathrm{elem},H}^{n}$. By Hypothesis~\ref{hyp:2}, each group $H^{n}(B\Gamma_E,\mathbb Z_{w_1})$ has no higher $2$-primary torsion. Therefore the objectwise reduction map
\begin{equation*}
\rho_E^{H}: H^{n}(B\Gamma_E,\mathbb Z_{w_1})[2] \longrightarrow H^{n}(B\Gamma_E,\mathbb Z_2)
\end{equation*}
is injective. Since inverse limits in $\mathbf{Ab}$ are left exact, the induced inverse-limit map
\begin{equation*}
\left(\rho_{\lim}^{H}\right)^{n}:
\mathrm{cSPT}_{\mathrm{elem},H}^{n}
\longrightarrow
{\bf \Lambda}_{H}^{n}
\end{equation*}
is injective.

\noindent With this injectivity established, the argument of Claim~\ref{claim:bockstein_failure_sequence} applies word-for-word. Namely,
\begin{equation*}
\ker\left((\overline{\rho}_{\lim}^{H})^{n}\right)
= \operatorname{Im}\left((\beta_{\lim}^{H})^{n-1}\right),
\end{equation*}
and $\left(\overline{\rho}_{\lim}^{H}\right)^{n}$ is surjective by definition of its target. Hence the sequence in Eq.~\eqref{eq:ses_Bockstein_failure_H} is exact.
\end{proof}

\begin{claim}[Free shadows identify the quotient obstruction]
\label{claim:Sfree_H_quotient_iso}
The restriction of $\left(\overline{\rho}_{\lim}^{H}\right)^{n}$ to $S_{\mathrm{free},H}^{n}$ is an isomorphism:
\begin{equation}
\left(\overline{\rho}_{\lim}^{H}\right)^{n}\big\vert_{S_{\mathrm{free},H}^{n}}:
S_{\mathrm{free},H}^{n}
\xrightarrow{\cong}
\frac{\operatorname{Im}\left(\left(\rho_{\lim}^{H}\right)^{n}\right)}
{\operatorname{Im}\left(\left(D_{\lim}^{H}\right)^{n-1}\right)}.
\end{equation}
\end{claim}

\begin{proof}[\textnormal{\textbf{Proof}}]
We prove the even-dimensional case $d=2m$; the odd-dimensional case follows immediately because $H^{*}({\bf \Lambda}_{H},D_{\lim}^{H})=0$. By Claim~\ref{claim:LambdaH_cohomology}, the cohomology of the complex $({\bf \Lambda}_{H},D_{\lim}^{H})$ is represented by
\begin{equation}
H^{*}({\bf \Lambda}_{H},D_{\lim}^{H})
\cong
y_m\mathbb F_2[y_1^2,\dots,y_m^2]\otimes R_{\mathrm{free},2}^{*}.
\end{equation}
Moreover, by the definition of $S_{\mathrm{free},H}^{n}$ as the elementary restriction of twisted free continuum classes (see Def.~\ref{def:Sfree_H}), its mod-$2$ reduction is precisely the degree-$n$ part of this representative subspace:
\begin{equation}
\left(\rho_{\lim}^{H}\right)^{n}
\left(S_{\mathrm{free},H}^{n}\right)
=
\left(
y_m\mathbb F_2[y_1^2,\dots,y_m^2]\otimes R_{\mathrm{free},2}^{*}
\right)^{n}.
\label{eq:rho_Sfree_H_image}
\end{equation}
Indeed, the twisted free continuum classes are generated by twisted spatial free classes multiplied by untwisted free internal classes.\footnote{In the even-dimensional spatial case, the basic twisted spatial free generator is the Euler class $E_d\in H^d(B\mathrm{O}(d),\mathbb Z_{w_1})$. Thus a typical twisted free class has the form $E_d\cdot P(p_1,\dots,p_m)\cdot \widetilde r$, where $P(p_1,\dots,p_m)$ is an untwisted free spatial class and $\widetilde r$ is an untwisted free internal class. After mod-$2$ reduction and restriction to elementary subgroups, this gives $y_m\cdot P(y_1^2,\dots,y_m^2)\otimes r$.}
Under mod-$2$ reduction and restriction to elementary subgroups, the twisted spatial free part gives $y_m\mathbb F_2[y_1^2,\dots,y_m^2]$, while the untwisted free internal classes give $R_{\mathrm{free},2}^{*}$.

\bigskip 

\noindent We first show surjectivity. Let
\begin{equation}
[z]\in
\frac{\operatorname{Im}\left(\left(\rho_{\lim}^{H}\right)^{n}\right)}
{\operatorname{Im}\left(\left(D_{\lim}^{H}\right)^{n-1}\right)}
\end{equation}
be arbitrary, and choose a representative $z\in \operatorname{Im}\left(\left(\rho_{\lim}^{H}\right)^{n}\right)$. Since $D_{\lim}^{H}=\rho_{\lim}^{H}\circ\beta_{\lim}^{H}$ and $\beta_{\lim}^{H}\circ\rho_{\lim}^{H}=0$ by exactness, we have
\begin{equation}
\left(D_{\lim}^{H}\right)^{n}(z)=0.
\end{equation}
Thus $z$ defines a cohomology class in $H^{n}({\bf \Lambda}_{H},D_{\lim}^{H})$. By Claim~\ref{claim:LambdaH_cohomology}, this class has a representative in
\begin{equation}
\left(
y_m\mathbb F_2[y_1^2,\dots,y_m^2]\otimes R_{\mathrm{free},2}^{*}
\right)^{n}.
\end{equation}
By Eq.~\eqref{eq:rho_Sfree_H_image}, this representative is equal to $\left(\rho_{\lim}^{H}\right)^{n}(s)$ for some $s\in S_{\mathrm{free},H}^{n}$. Therefore
\begin{equation}
[z]
=
\left[
\left(\rho_{\lim}^{H}\right)^{n}(s)
\right]
=
\left(\overline{\rho}_{\lim}^{H}\right)^{n}(s),
\end{equation}
which proves surjectivity.

\noindent We now prove injectivity. Suppose $s\in S_{\mathrm{free},H}^{n}$ and
\begin{equation}
\left(\overline{\rho}_{\lim}^{H}\right)^{n}(s)=0.
\end{equation}
Then
\begin{equation}
\left(\rho_{\lim}^{H}\right)^{n}(s)
\in
\operatorname{Im}\left(\left(D_{\lim}^{H}\right)^{n-1}\right).
\end{equation}
But by Eq.~\eqref{eq:rho_Sfree_H_image}, the same element lies in the chosen cohomology-representative subspace
\begin{equation}
\left(
y_m\mathbb F_2[y_1^2,\dots,y_m^2]\otimes R_{\mathrm{free},2}^{*}
\right)^{n}.
\end{equation}
This representative subspace intersects the boundary subspace $\operatorname{Im}\left(\left(D_{\lim}^{H}\right)^{n-1}\right)$ trivially. Hence
\begin{equation}
\left(\rho_{\lim}^{H}\right)^{n}(s)=0.
\end{equation}
By Hypothesis~\ref{hyp:2}, the map $\left(\rho_{\lim}^{H}\right)^{n}$ is injective on $\mathrm{cSPT}_{\mathrm{elem},H}^{n}$. Since $S_{\mathrm{free},H}^{n}\subseteq \mathrm{cSPT}_{\mathrm{elem},H}^{n}$, it follows that $s=0$.

\noindent Thus the restriction of $\left(\overline{\rho}_{\lim}^{H}\right)^{n}$ to $S_{\mathrm{free},H}^{n}$ is both injective and surjective, and is therefore an isomorphism.
\end{proof}

\begin{claim}[Free-shadow splitting]
\label{claim:Sfree_H_splitting}
There is a direct-sum decomposition
\begin{equation}
\mathrm{cSPT}_{\mathrm{elem},H}^{n} = \operatorname{Im}\left(\left(\beta_{\lim}^{H}\right)^{n-1}\right)
\oplus S_{\mathrm{free},H}^{n}.
\end{equation}
\end{claim}

\begin{proof}[\textnormal{\textbf{Proof}}]
The proof is the same as the proof of Claim~\ref{claim:Sfree_splitting}, with $\beta_{\lim}$, $\rho_{\lim}$, $D_{\lim}$, and $S_{\mathrm{free}}^{n}$ replaced by their $H$-enriched analogs.

\noindent Indeed, by Claim~\ref{claim:bockstein_failure_sequence_H}, we have the short exact sequence
\begin{equation}
0
\rightarrow
\operatorname{Im}\left(\left(\beta_{\lim}^{H}\right)^{n-1}\right)
\rightarrow
\mathrm{cSPT}_{\mathrm{elem},H}^{n}
\xrightarrow{\left(\overline{\rho}_{\lim}^{H}\right)^{n}}
\frac{
\operatorname{Im}\left(\left(\rho_{\lim}^{H}\right)^{n}\right)
}{ \operatorname{Im}\left(\left(D_{\lim}^{H}\right)^{n-1}\right) } \rightarrow 0.
\end{equation}
By Claim~\ref{claim:Sfree_H_quotient_iso}, the restriction of $\left(\overline{\rho}_{\lim}^{H}\right)^{n}$ to $S_{\mathrm{free},H}^{n}$ is an isomorphism onto the quotient term. Therefore $S_{\mathrm{free},H}^{n}$ gives a complement to $\operatorname{Im}\left(\left(\beta_{\lim}^{H}\right)^{n-1}\right)$, and hence
\begin{equation}
\mathrm{cSPT}_{\mathrm{elem},H}^{n} =
\operatorname{Im}\left(\left(\beta_{\lim}^{H}\right)^{n-1}\right)
\oplus S_{\mathrm{free},H}^{n}.
\end{equation}
\end{proof}

\bigskip 

\noindent Recall that the map $\res_{H,\mathcal A}^{n}[2]$ has been defined in Eq.~\ref{eq:internal_resA} and is the restriction of the full map $\left(\res_{H,\mathcal A}^{\mathrm{full}}\right)^{n}$, defined in Eq.~\ref{eq:res_full_H}, to the order-$2$ subgroup.

\begin{claim}[Identification of the $H$-enriched Bockstein-detected summand]
\label{claim:bockstein_summand_H}
The map $\res_{H,\mathcal A}^{n}[2]$ induces an isomorphism
\begin{equation}
\res_{H,\mathcal A}^{n}[2]: H^n(B\Gamma_{\mathrm{O}(d)},\mathbb Z_{w_1})[2]
\xrightarrow{\cong} \operatorname{Im}\left(\left(\beta_{\lim}^{H}\right)^{n-1}\right)
\subseteq \mathrm{cSPT}_{\mathrm{elem},H}^{n}.
\end{equation}
\end{claim}

\begin{proof}[\textnormal{\textbf{Proof}}]
The proof is identical to the proof of Claim~\ref{claim:bockstein_summand}, with $B\mathrm{O}(d)$, $\Lambda$, $\beta_{\lim}$, and $\rho_{\lim}$ replaced respectively by $B\Gamma_{\mathrm{O}(d)}$, ${\bf \Lambda}_{H}$, $\beta_{\lim}^{H}$, and $\rho_{\lim}^{H}$.

\end{proof}

\begin{claim}[Quotient by $H$-enriched free shadows]
\label{claim:free_shadow_quotient_H}
There is a canonical isomorphism
\begin{equation}
\frac{\mathrm{cSPT}_{\mathrm{elem},H}^{n}}
{S_{\mathrm{free},H}^{n}}
\cong
H^n(B\Gamma_{\mathrm{O}(d)},\mathbb Z_{w_1})[2].
\end{equation}
\end{claim}

\begin{proof}[\textnormal{\textbf{Proof}}]
By Claim~\ref{claim:Sfree_H_splitting},
\begin{equation}
\frac{\mathrm{cSPT}_{\mathrm{elem},H}^{n}}
{S_{\mathrm{free},H}^{n}}
\cong
\operatorname{Im}\left(\left(\beta_{\lim}^{H}\right)^{n-1}\right).
\end{equation}
The result follows from Claim~\ref{claim:bockstein_summand_H}.
\end{proof}

\subsubsection{Formal proof of $H$-enriched straftification theorem}
\label{app:final_proof_H}

\begin{proposition}[$H$-enriched general-degree elementary retraction]
\label{prop:general_degree_retraction_H}
There exists a canonical surjective homomorphism
\begin{equation}
\widetilde F_{2,H}^{n}:
\mathrm{cSPT}_{\mathrm{elem},H}^{n}
\twoheadrightarrow
H^{n}(B\Gamma_{\mathrm{O}(d)},\mathbb Z_{w_1})[2]
\end{equation}
with kernel $S_{\mathrm{free},H}^{n}$, satisfying
\begin{equation}
\widetilde F_{2,H}^{n}\circ \res_{H,\mathcal A}^{n}[2]
=
\mathrm{id}_{H^{n}(B\Gamma_{\mathrm{O}(d)},\mathbb Z_{w_1})[2]}.
\end{equation}
\end{proposition}

\begin{proof}[\textnormal{\textbf{Proof}}]
The proof is identical to the proof of Proposition~\ref{prop:general_degree_retraction}, with all objects replaced by their $H$-enriched analogs. By Claim~\ref{claim:Sfree_H_splitting}, every element $x\in \mathrm{cSPT}_{\mathrm{elem},H}^{n}$ has a unique decomposition
\begin{equation}
x=x_{\mathrm{Bock}}+x_{\mathrm{free}},
\end{equation}
where
\begin{equation}
x_{\mathrm{Bock}}\in \operatorname{Im}\left(\left(\beta_{\lim}^{H}\right)^{n-1}\right),
\qquad
x_{\mathrm{free}}\in S_{\mathrm{free},H}^{n}.
\end{equation}
By Claim~\ref{claim:bockstein_summand_H}, the restriction map
\begin{equation}
\res_{H,\mathcal A}^{n}[2]:
H^{n}(B\Gamma_{\mathrm{O}(d)},\mathbb Z_{w_1})[2]
\xrightarrow{\cong}
\operatorname{Im}\left(\left(\beta_{\lim}^{H}\right)^{n-1}\right)
\end{equation}
is an isomorphism. Define
\begin{equation}
\widetilde F_{2,H}^{n}(x)
:=
\left(\res_{H,\mathcal A}^{n}[2]\right)^{-1}(x_{\mathrm{Bock}}).
\end{equation}
This is well-defined by the uniqueness of the splitting.

Its kernel is exactly $S_{\mathrm{free},H}^{n}$, since $\widetilde F_{2,H}^{n}(x)=0$ if and only if $x_{\mathrm{Bock}}=0$. Finally, if $a\in H^{n}(B\Gamma_{\mathrm{O}(d)},\mathbb Z_{w_1})[2]$, then $\res_{H,\mathcal A}^{n}[2](a)$ lies entirely in the Bockstein-detected summand, and hence
\begin{equation}
\widetilde F_{2,H}^{n}\left(\res_{H,\mathcal A}^{n}[2](a)\right)=a.
\end{equation}
Thus $\widetilde F_{2,H}^{n}$ is a left inverse to $\res_{H,\mathcal A}^{n}[2]$, and in particular is surjective.
\end{proof}

\bigskip

\begin{proof}[\textnormal{\textbf{Proof of $H$-enriched main theorem~\ref{appG:main_theorem_H}}}]
In the group cohomology model, we identify
\begin{equation}
\Theta^{d+1}\bigl(B(H\rtimes_{\rho}\mathrm{O}(d))\bigr)
\cong
H^{d+2}(B\Gamma_{\mathrm{O}(d)},\mathbb Z_{w_1}).
\end{equation}
By Hypothesis~\ref{hyp:2}, the relevant $2$-primary torsion subgroup agrees with the order-$2$ subgroup. Thus it is enough to construct a surjection
\begin{equation}
F_{2,H}:
\tor_{2}\bigl( \mathrm{cSPT}^{d+1}_{\mathrm{univ},H} \bigr)
\twoheadrightarrow
H^{d+2}(B\Gamma_{\mathrm{O}(d)},\mathbb Z_{w_1})[2].
\end{equation}

\noindent Let
\begin{equation}
\Pi_{H}^{d+2}:
\tor_{2}\bigl( \mathrm{cSPT}^{d+1}_{\mathrm{univ},H} \bigr)
\longrightarrow
\mathrm{cSPT}_{\mathrm{elem},H}^{d+2}
\end{equation}
be the canonical projection obtained by restricting a compatible family over $\mathcal O_{\mathrm{fin}}(\mathrm{O}(d))$ to the elementary subcategory $\mathcal A_2(\mathrm{O}(d))$. By Proposition~\ref{prop:general_degree_retraction_H}, there is a canonical surjection
\begin{equation}
\widetilde F_{2,H}^{d+2}:
\mathrm{cSPT}_{\mathrm{elem},H}^{d+2}
\twoheadrightarrow
H^{d+2}(B\Gamma_{\mathrm{O}(d)},\mathbb Z_{w_1})[2],
\end{equation}
satisfying
\begin{equation}
\widetilde F_{2,H}^{d+2}\circ \res_{H,\mathcal A}^{d+2}[2]
=
\mathrm{id}_{H^{d+2}(B\Gamma_{\mathrm{O}(d)},\mathbb Z_{w_1})[2]}.
\end{equation}
Define
\begin{equation}
F_{2,H}
:=
\widetilde F_{2,H}^{d+2}\circ \Pi_{H}^{d+2}.
\end{equation}

It remains to show that this map is surjective. The restriction from the continuum to the elementary category factors through the full finite-subgroup category:
\begin{equation}
\res_{H,\mathcal A}^{d+2}[2]
=
\Pi_{H}^{d+2}\circ \res_{H,\mathcal O}^{d+2}[2].
\end{equation}
Therefore
\begin{align}
F_{2,H}\circ \res_{H,\mathcal O}^{d+2}[2]
&=
\widetilde F_{2,H}^{d+2}\circ \Pi_{H}^{d+2}\circ \res_{H,\mathcal O}^{d+2}[2] \\
&=
\widetilde F_{2,H}^{d+2}\circ \res_{H,\mathcal A}^{d+2}[2] \\
&=
\mathrm{id}_{H^{d+2}(B\Gamma_{\mathrm{O}(d)},\mathbb Z_{w_1})[2]}.
\end{align}
Thus $F_{2,H}$ admits $\res_{H,\mathcal O}^{d+2}[2]$ as a right section, and hence $F_{2,H}$ is surjective. This proves the theorem.
\end{proof}

\subsubsection{Higher-Bockstein extension for $H=\mathbb Z_4\times\mathbb Z_4$}
\label{app:higher_order_example}

We now establish the specialized extension used in
Sec.~\ref{sec:example_2_highertorsion}. Let
\begin{equation*}
H=\mathbb Z_4^{(1)}\times\mathbb Z_4^{(2)},
\qquad
\Gamma_K=H\times K,
\end{equation*}
where $K\in\{O(2)\}\cup\mathcal A_2(O(2))$. The orientation local system on $B\Gamma_K$ is pulled back from $BK$. Hypothesis~\ref{hyp:2} does not hold in this example because
$H^{*}(BH,\mathbb Z)$ contains order-$4$ torsion. Consequently, the primary Bockstein does not by itself reconstruct the full
$2$-primary torsion. The mod-$2$ Leray-Hirsch and Quillen arguments, however, remain unchanged. The only modification is that the inverse-limit Bockstein construction must be continued through its secondary differential.

\bigskip

\begin{proposition}[Internal secondary-Bockstein calculation]
\label{prop:Z4Z4_internal_Bockstein}
Let $R^{*}:=H^{*}(BH,\mathbb Z_2)$. There is an isomorphism of graded algebras
\begin{equation}
R^{*} \cong
\Lambda_{\mathbb Z_2}(u_1,u_2)
\otimes \mathbb Z_2[v_1,v_2],
\qquad
|u_i|=1, \qquad |v_i|=2.
\label{eq:Z4Z4_mod2_cohomology}
\end{equation}
The primary internal Bockstein vanishes,
\begin{equation}
d_1^{H}=\mathrm{Sq}^{1}=0,
\label{eq:Z4Z4_d1_zero}
\end{equation}
whereas the secondary Bockstein is the derivation determined by
\begin{equation}
d_2^{H}(u_i)=v_i,
\qquad
d_2^{H}(v_i)=0.
\label{eq:Z4Z4_d2_generators}
\end{equation}
Consequently,
\begin{equation}
H^{*}(R^{*},d_2^{H})
\cong
\mathbb Z_2,
\label{eq:Z4Z4_d2_cohomology}
\end{equation}
is concentrated in degree zero.
\end{proposition}

\begin{proof}
For a single factor $\Z_{4}=\langle g\mid g^4=1\rangle$, let
\begin{equation*}
N:=1+g+g^2+g^3\in\mathbb Z[\Z_{4}].
\end{equation*}
The trivial $\mathbb Z[\Z_{4}]$-module $\mathbb Z$ admits the standard
periodic projective resolution
\begin{equation}
\cdots \xrightarrow{N} \mathbb Z[\Z_{4}]
\xrightarrow{g-1} \mathbb Z[\Z_{4}] \xrightarrow{N}
\mathbb Z[\Z_{4}] \xrightarrow{g-1} \mathbb Z[\Z_{4}]
\xrightarrow{\varepsilon} \mathbb Z
\longrightarrow0;
\label{eq:projres}
\end{equation}
see, for example, Ch.6 Ref.~\cite{Weibel1994}. Applying
$\operatorname{Hom}_{\mathbb Z[\Z_{4}]}(-,\mathbb Z)$ to Eq.~\ref{eq:projres}, with $\mathbb Z$ regarded as the trivial $\Z_{4}$-module, gives the cochain complex
\begin{equation*}
0
\longrightarrow
\mathbb Z
\xrightarrow{0}
\mathbb Z
\xrightarrow{\times4}
\mathbb Z
\xrightarrow{0}
\mathbb Z
\xrightarrow{\times4}
\mathbb Z
\longrightarrow\cdots.
\end{equation*}
It follows that
\begin{equation*}
H^*(B\Z_{4},\mathbb Z)
\cong
\mathbb Z[t]/(4t),
\qquad
|t|=2.
\end{equation*}

\noindent For $\mathbb Z_2$ coefficients, applying $\operatorname{Hom}_{\mathbb Z[\Z_{4}]}(-,\mathbb Z_2)$ to Eq.~\ref{eq:projres}, one can show that both cochain differentials vanish in the resulting cochain complex, thus proving $H^{n}(B\Z_{4}, \Z_{2})\cong \Z_{2}$ for all $n\geq 0$. Choosing generators
\begin{equation*}
u\in H^1(B\Z_{4},\mathbb Z_2),
\qquad
v:=\rho_2(t)\in H^2(B\Z_{4},\mathbb Z_2),
\end{equation*}
one obtains\footnote{for the degree-1 class we have $u^{2}=\mathrm{Sq}^{1}(u)= d_{1}^{H}(u)=0$ which leads to the anti-symmetric ring for the degree-1 generator.}
\begin{equation*}
H^*(B\Z_{4},\mathbb Z_2)
\cong \Lambda_{\mathbb Z_2}(u)\otimes\mathbb Z_2[v],
\qquad |u|=1, \qquad |v|=2.
\end{equation*}

We next compute the first two Bockstein differentials. Let
$\widetilde u$ be an integral cochain lifting a cocycle representative
of $u$. In the integral cochain complex,
\begin{equation*}
\delta\widetilde u=4c,
\end{equation*}
where $c$ represents the integral class $t$. The primary integral
Bockstein associated to
\begin{equation*}
0\longrightarrow\mathbb Z \xrightarrow{\times2} \mathbb Z \xrightarrow{\rho_2} \mathbb Z_2 \longrightarrow 0
\end{equation*}
therefore satisfies
\begin{equation*}
\beta_2(u) = \left[ \frac{\delta\widetilde u}{2} \right] = 2t.
\end{equation*}
Consequently, the first differential in the Bockstein spectral
sequence is
\begin{equation*}
d_1^H(u) = \rho_2\circ \beta_2(u) = \rho_2(2t) = 0.
\end{equation*}
Since $u$ survives to the second page, its secondary Bockstein is represented by
\begin{equation*}
d_2^H(u) = \left[ \frac{\delta\widetilde u}{4}
\bmod2 \right]
= \rho_2(t) = v.
\end{equation*}
Finally, since $v=\rho_2(t)$ is the reduction of an integral class, all of its Bockstein differentials vanish
\begin{equation*}
d_{1}^{H}(v)=0, \quad d_2^H(v)=0.
\end{equation*}

\noindent For each $i$, the complex
\begin{equation*}
\left(
\Lambda_{\mathbb Z_2}(u_i)\otimes\mathbb Z_2[v_i],
d_2^{H}
\right)
\end{equation*}
has cohomology $\mathbb Z_2$ in degree zero: every positive power of
$v_i$ is a boundary because
\begin{equation*}
d_2^{H}(u_i v_i^{k})=v_i^{k+1},
\end{equation*}
and no class containing $u_i$ is closed. The K\"unneth theorem over
$\mathbb Z_2$ therefore gives
\begin{equation*}
H^{*}(R^{*},d_2^{H})
\cong
\mathbb Z_2.
\end{equation*}
\end{proof}

We next continue the inverse-limit Bockstein calculation of App.~\ref{app:prep_res_H} through the secondary differential. The first page of the elementary Bockstein spectral sequence is the mod-$2$ inverse-limit group 
\begin{equation} 
\mathcal B_{1,\mathrm{elem}}^{*} := {\bf\Lambda}_{H} = \varprojlim_{E\in\mathcal A_2(O(2))} H^{*}(B\Gamma_E,\mathbb Z_2), 
\label{eq:Z4Z4_elementary_B1} 
\end{equation} 
equipped with the primary inverse-limit Bockstein differential 
\begin{equation} 
d_{1,\lim} := D_{\lim}^{H} = \rho_{\lim}^{H}\circ\beta_{\lim}^{H}. 
\label{eq:Z4Z4_elementary_d1_def} 
\end{equation} 
Recall that the natural direct-product Leray-Hirsch isomorphism identifies this first page with 
\begin{equation} 
\Delta_{\lim}: {\bf\Lambda}\otimes R^{*} \xrightarrow{\cong} {\bf\Lambda}_{H}, \qquad {\bf\Lambda} := \varprojlim_{E\in\mathcal A_2(O(2))} H^{*}(BE,\mathbb Z_2), 
\label{eq:Z4Z4_limit_Kunneth} 
\end{equation} 
where $R^{*}=H^{*}(BH,\mathbb Z_2)$. Under this identification, Claim~\ref{claim:DlimH_decomposition} gives 
\begin{equation} 
\Delta_{\lim}^{-1} \circ d_{1,\lim}\circ \Delta_{\lim} = D_{\lim}\otimes\mathrm{id}_{R^{*}} + \mathrm{id}_{{\bf\Lambda}}\otimes d_1^{H}. 
\label{eq:Z4Z4_limit_d1_decomposition} 
\end{equation} 
By Proposition~\ref{prop:Z4Z4_internal_Bockstein}, $d_1^{H}=\delta_{R^{*}}=0$, and therefore the primary differential acts only on the spatial factor: 
\begin{equation} 
\Delta_{\lim}^{-1} \circ d_{1,\lim}\circ \Delta_{\lim} = D_{\lim}\otimes\mathrm{id}_{R^{*}}. 
\label{eq:Z4Z4_product_d1} 
\end{equation}

The second page is obtained by taking the cohomology of this first-page complex: 
\begin{equation*} 
\mathcal B_{2,\mathrm{elem}}^{*} = H^{*}({\bf\Lambda}_{H},d_{1,\lim}) \cong H^{*}({\bf\Lambda},D_{\lim})\otimes R^{*}. 
\label{eq:Z4Z4_elementary_B2_general} 
\end{equation*} 
For $d=2$, the spatial computation of App.~\ref{app:lift} gives 
\begin{equation} 
H^{*}({\bf\Lambda},D_{\lim}) \cong y_1\mathbb Z_2[y_1^2], \qquad |y_1|=2. 
\label{eq:O2_primary_spatial_cohomology} 
\end{equation} 
Hence 
\begin{equation} 
\mathcal B_{2,\mathrm{elem}}^{*} \cong y_1\mathbb Z_2[y_1^2]\otimes R^{*}. 
\label{eq:Z4Z4_elementary_B2} 
\end{equation}
The spatial classes in $y_1\mathbb Z_2[y_1^2]$ are reductions of twisted free integral classes and hence support no higher Bockstein differentials. Therefore the secondary differential acts only on the internal factor: 
\begin{equation} 
d_{2,\lim} = \mathrm{id}\otimes d_2^{H}. 
\label{eq:Z4Z4_elementary_d2} 
\end{equation} 
Proposition~\ref{prop:Z4Z4_internal_Bockstein} now gives 
\begin{equation} 
\mathcal B_{3,\mathrm{elem}}^{*} = H^{*}(\mathcal B_{2,\mathrm{elem}}^{*}, d_{2,\lim}) \cong y_1\mathbb Z_2[y_1^2], 
\label{eq:Z4Z4_elementary_B3} 
\end{equation} 
by Proposition~\ref{prop:Z4Z4_internal_Bockstein}, since $H^{*}(R^{*},d_{2}^{H})\cong\Z_{2}$ is concentrated in degree zero. Hence, only the degree-zero internal class survives, there are no further differentials.

\noindent The same two-stage calculation applies to the continuum group 
\begin{equation*} 
\mathcal B_{1,\mathrm{cont}}^{*} := H^{*}(B(H\times O(2)),\mathbb Z_2), 
\end{equation*} 
equipped with the primary continuum Bockstein differential $d_{1,\mathrm{cont}}=D_{O(2)}^{H}$. Indeed, the direct-product K\"unneth isomorphism gives 
\begin{equation*} 
\mathcal B_{1,\mathrm{cont}}^{*} \cong H^{*}(BO(2),\mathbb Z_2)\otimes R^{*}, 
\end{equation*} 
under which $d_{1,\mathrm{cont}} = D_{O(2)}\otimes\mathrm{id}_{R^{*}}$, because $d_1^{H}=0$. The restriction map 
\begin{equation} 
\res_{H,\mathcal A}: \mathcal B_{1,\mathrm{cont}}^{*} \longrightarrow \mathcal B_{1,\mathrm{elem}}^{*} 
\end{equation} 
is an isomorphism by Proposition~\ref{prop:quillen_isom_H}. Naturality of the reduction and Bockstein maps implies \begin{equation*} \res_{H,\mathcal A}\circ d_{1,\mathrm{cont}} = d_{1,\lim}\circ\res_{H,\mathcal A}. \end{equation*} Hence restriction is an isomorphism of the first-page complexes and therefore induces an isomorphism \begin{equation*} \mathcal B_{2,\mathrm{cont}}^{*} \xrightarrow{\cong} \mathcal B_{2,\mathrm{elem}}^{*}. \end{equation*} Moreover, restriction defines a morphism of the underlying Bockstein exact couples. It therefore intertwines the secondary differentials: \begin{equation*} \res_{2}\circ d_{2,\mathrm{cont}} = d_{2,\mathrm{elem}}\circ\res_{2}. \end{equation*} Consequently, the continuum and elementary third pages are also canonically isomorphic: \begin{equation*} \mathcal B_{3,\mathrm{cont}}^{*} \xrightarrow{\cong} \mathcal B_{3,\mathrm{elem}}^{*} \cong y_1\mathbb Z_2[y_1^2]. \end{equation*} Thus the continuum and elementary Bockstein spectral sequences agree through the secondary stage.

\bigskip 

\begin{lemma}[Order-two elementary sector with spatial support]
\label{lem:Z4Z4_spatial_kernel_order2}
Let $H=\mathbb Z_4^{(1)}\times\mathbb Z_4^{(2)}$ and $\Gamma_{E}=H \times E$ where
$E\in\mathcal A_2(O(2))$. Then
\begin{equation}
2\ker\left[
\res^{H\times E}_{H}:
\tor_2 H^4\left(B\Gamma_E,\mathbb Z_{w_1}\right)
\to
\tor_2 H^4(BH,\mathbb Z)
\right]
=0.
\label{eq:Z4Z4_spatial_kernel_order2}
\end{equation}
\end{lemma}

\begin{proof}
Under the K\"unneth decomposition for
$B(H\times E)\simeq BH\times BE$, restriction to $BH$ retains the
purely internal contribution and kills all terms of positive spatial
degree. Hence the kernel in
Eq.~\eqref{eq:Z4Z4_spatial_kernel_order2} consists precisely of the
purely spatial and mixed spatial-internal contributions.

\medskip 

\noindent Since $E$ is elementary abelian, its positive-degree integral
cohomology, with either the trivial or orientation-twisted
coefficients appearing here, is annihilated by $2$. The tensor and
$\operatorname{Tor}$ terms involving these positive-degree spatial
groups are therefore also annihilated by $2$. It follows that every
class in the kernel has order at most $2$.
\end{proof}

\bigskip

\begin{proposition}[Specialized degree-$4$ elementary reconstruction]
\label{prop:Z4Z4_degree4_reconstruction}
Define
\begin{equation}
\mathrm{cSPT}_{\mathrm{elem},H}^{4,(2)}
:=
\varprojlim_{E\in\mathcal A_2(O(2))}
\tor_2
\left(
H^{4}(B(H\times E),\mathbb Z_{w_1})
\right).
\label{eq:Z4Z4_full_elementary_group}
\end{equation}
Then restriction induces a canonical isomorphism
\begin{equation}
\res_{H,\mathcal A}^{4,(2)}:
\tor_2
\left(
H^{4}(B(H\times O(2)),\mathbb Z_{w_1})
\right)
\xrightarrow{\;\cong\;}
\mathrm{cSPT}_{\mathrm{elem},H}^{4,(2)}.
\label{eq:Z4Z4_degree4_restriction_iso}
\end{equation}
\end{proposition}

\begin{proof}
Set
\begin{equation*}
T_{\mathrm{cont}}
:=
\tor_2
\left(
H^4(B(H\times O(2)),\mathbb Z_{w_1})
\right),
\;\;
T_{\mathrm{elem}}
:=
\mathrm{cSPT}_{\mathrm{elem},H}^{4,(2)}.
\end{equation*}
By Proposition~\ref{prop:quillen_isom_H}, restriction is an
isomorphism on the first pages of the continuum and elementary
Bockstein spectral sequences. Naturality of the Bockstein
construction implies that this isomorphism commutes with both
$d_1$ and $d_2$. Hence the continuum and elementary spectral
sequences agree through the third page. By Eq.~\eqref{eq:Z4Z4_elementary_B3}, the terminal elementary page is
\begin{equation*}
\mathcal B_{\infty,\mathrm{elem}}^{*}
\cong
y_1\mathbb Z_2[y_1^2].
\end{equation*}
Its nonzero homogeneous elements have degrees $2,6,10,\ldots$, and
therefore
\begin{equation}
\mathcal B_{\infty,\mathrm{elem}}^{4}
=
\mathcal B_{\infty,\mathrm{cont}}^{4}
=
0.
\label{eq:Z4Z4_terminal_degree4_zero}
\end{equation}
The vanishing of the terminal degree-$4$ group means that there is no
free integral contribution in degree $4$. Moreover, since the
spectral sequences collapse after the secondary differential, there
is no degree-$4$ $2$-primary torsion of order $8$ or higher.
Consequently,
\begin{equation*}
4T_{\mathrm{cont}}=0,
\qquad
4T_{\mathrm{elem}}=0.
\end{equation*}

The Bockstein construction equips each of these groups with the
descending filtration by images of multiplication by powers of $2$:
\begin{equation*}
T \supseteq 2T \supseteq
4T=0,\quad \text{where}\quad
2^rT
:=
\{2^r x\mid x\in T\}.
\end{equation*}
The two successive layers of this filtration are
\begin{equation*}
T/2T
\qquad\text{and}\qquad
2T/4T=2T.
\end{equation*}
The preceding comparison of the continuum and elementary Bockstein
spectral sequences therefore implies that restriction induces
isomorphisms
\begin{equation}
T_{\mathrm{cont}}/2T_{\mathrm{cont}}
\xrightarrow{\cong}
T_{\mathrm{elem}}/2T_{\mathrm{elem}}
\label{eq:Z4Z4_quotient_layer_iso}
\end{equation}
and
\begin{equation}
2T_{\mathrm{cont}}
\xrightarrow{\cong}
2T_{\mathrm{elem}}.
\label{eq:Z4Z4_double_layer_iso}
\end{equation}

It remains to pass from these two layers to the full groups. Suppose
$x\in T_{\mathrm{cont}}$ restricts to zero. Its image in
$T_{\mathrm{cont}}/2T_{\mathrm{cont}}$ maps to zero, and
Eq.~\eqref{eq:Z4Z4_quotient_layer_iso} is injective. Hence
\begin{equation*}
x\in 2T_{\mathrm{cont}}.
\end{equation*}
Equation~\eqref{eq:Z4Z4_double_layer_iso} is also injective, so the
vanishing of the restriction of $x$ implies $x=0$. Thus restriction
is injective.

Now let $y\in T_{\mathrm{elem}}$. By the surjectivity of
Eq.~\eqref{eq:Z4Z4_quotient_layer_iso}, there exists
$x\in T_{\mathrm{cont}}$ such that
\begin{equation*}
y-\res_{H,\mathcal A}^{4,(2)}(x)
\in
2T_{\mathrm{elem}}.
\end{equation*}
By the surjectivity of Eq.~\eqref{eq:Z4Z4_double_layer_iso}, there
exists $z\in 2T_{\mathrm{cont}}$ satisfying
\begin{equation*}
\res_{H,\mathcal A}^{4,(2)}(z)
=
y-\res_{H,\mathcal A}^{4,(2)}(x).
\end{equation*}
It follows that
\begin{equation*}
\res_{H,\mathcal A}^{4,(2)}(x+z)=y.
\end{equation*}
Thus restriction is surjective. Therefore
\begin{equation*}
\res_{H,\mathcal A}^{4,(2)}:
T_{\mathrm{cont}}
\xrightarrow{\cong}
T_{\mathrm{elem}}
\end{equation*}
is an isomorphism.
\end{proof}

\bigskip

Because the degree-$4$ terminal page vanishes, there is no
free-shadow quotient in this case. We may therefore define the
specialized elementary retraction directly by
\begin{equation}
\widetilde F_{2,H}^{4,(2)}
:=
\left(
\res_{H,\mathcal A}^{4,(2)}
\right)^{-1}.
\label{eq:Z4Z4_specialized_retraction}
\end{equation}

\noindent Let
\begin{equation}
\Pi_H^{4,(2)}:
\tor_2\left(
\mathrm{cSPT}_{\mathrm{univ},H}^{3}
\right)
\longrightarrow
\mathrm{cSPT}_{\mathrm{elem},H}^{4,(2)}
\label{eq:Z4Z4_specialized_projection}
\end{equation}
denote restriction from the full finite-subgroup category to the
elementary subcategory. Define
\begin{equation}
F_{2,H}
:=
\widetilde F_{2,H}^{4,(2)}
\circ
\Pi_H^{4,(2)}.
\label{eq:Z4Z4_specialized_F2}
\end{equation}

\noindent The remainder of the argument is identical to the proof of
Theorem~\ref{appG:main_theorem_H}. Indeed,
\begin{equation*}
\res_{H,\mathcal A}^{4,(2)}
=
\Pi_H^{4,(2)}
\circ
\res_{H,\mathcal O}^{4,(2)},
\end{equation*}
and hence
\begin{equation}
F_{2,H}
\circ
\res_{H,\mathcal O}^{4,(2)}
=
\mathrm{id}.
\label{eq:Z4Z4_specialized_section}
\end{equation}
Thus $F_{2,H}$ is surjective and, by construction,
\begin{equation}
\ker\left(\Pi_H^{4,(2)}\right)
\subseteq
\ker(F_{2,H}).
\label{eq:Z4Z4_specialized_kernel_inclusion}
\end{equation}
This proves the specialized higher-Bockstein extension required in Sec.~\ref{sec:example_2_highertorsion}.

\section{Cobordism models}
\label{app:cob_proof}

\noindent \textbf{Note}: we follow notation~\ref{notation:mod2_coefficients_cobordism}, and suppress the coefficients in cohomology $H^{*}(X)=H^{*}(X,\Z_{2})$.

\bigskip

\noindent In this appendix, we use the unoriented bordism classification model
\begin{equation}
    \Theta_{\mathrm{cob}}^{d+1}(BG)
    :=
    \widehat{\Omega}_{\mathrm O}^{\,d+1}(BG)
    =
    \operatorname{Hom}
    \bigl(
        \Omega_{d+1}^{\mathrm O}(BG),
        \mathrm U(1)
    \bigr),
    \label{eq:cobordism_classification_model}
\end{equation}
where $\Omega_{d+1}^{\mathrm O}(BG)$ is the unoriented bordism group
of closed $(d+1)$-manifolds equipped with maps to $BG$, and the
right-hand side is its Pontryagin dual. This agrees with the
classification proposed by Kapustin~\cite{Kapustin2014},
\begin{equation*}
    \Omega_{\zeta}^{d+1}(BG)
    =
    \operatorname{Hom}
    \bigl(
        (\Omega_{d+1}^{\zeta}(BG))_{\mathrm{tor}},
        \mathrm U(1)
    \bigr),
\end{equation*}
specialized to $\zeta=\mathrm O$, since the unoriented bordism groups
appearing here are entirely $2$-torsion. In
App.~\ref{app:cob}, we further showed that this Pontryagin-dual
classification agrees with the stable-homotopy-theoretic
classification of Freed and Hopkins. Thus
Eq.~\eqref{eq:cobordism_classification_model} gives the common
unoriented torsion classification model used below.

\bigskip  

\begin{proposition}[Cobordism restriction isomorphism]
\label{prop:cobordism_restriction_isomorphism}
For every degree $n$, restriction along the subgroup inclusions $E\hookrightarrow\mathrm O(d)$ induces an isomorphism
\begin{equation}
    \res^{\mathcal A}_{\mathrm{cob}}:
    \widehat{\Omega}_{\mathrm O}^{\,n}
    \bigl(B\mathrm O(d)\bigr)
    \xrightarrow{\;\cong\;}
    \varprojlim_{E\in\mathcal A_2(\mathrm O(d))}
    \widehat{\Omega}_{\mathrm O}^{\,n}(BE).
    \label{eq:quillen_cob}
\end{equation}
The map $\res^{\mathcal A}_{\mathrm{cob}}$ is canonical.
\end{proposition}

\begin{proof}
Fix the stable equivalence of spectra from
Theorem~\ref{thm:MO_stable_splitting},
\begin{equation*}
    \omega:
    \mathrm{MO}
    \xrightarrow{\;\simeq\;}
    \bigvee_{\alpha\in\mathcal B}
    \Sigma^{|\alpha|}H\mathbb Z_2.
\end{equation*}
Evaluating the homology theories represented by the two sides of this
stable equivalence on a space $X$ means smashing with
$\Sigma^\infty X_+$ and taking homotopy groups. Thus,
\begin{align}
    \mathrm{MO}_n(X)
    &:=
    \pi_n\bigl(
        \mathrm{MO}\wedge\Sigma^\infty X_+
    \bigr)
    \nonumber\\
    &\cong
    \pi_n\left(\bigvee_{\alpha\in\mathcal B}
        \Sigma^{|\alpha|}H\mathbb Z_2
        \wedge\Sigma^\infty X_+ \right)
    \nonumber\\
    &\cong \bigoplus_{\alpha\in\mathcal B} \pi_{n-|\alpha|}
    \bigl( H\mathbb Z_2\wedge\Sigma^\infty X_+ \bigr)
    \nonumber\\
    &= \bigoplus_{\alpha\in\mathcal B} H_{n-|\alpha|}(X,\mathbb Z_2).
    \label{eq:MO_homology_from_splitting}
\end{align}
For each $q$, let
\begin{equation*}
    \mathcal B_q := \{\alpha\in\mathcal B\mid |\alpha|=q\}.
\end{equation*}
Since $\mathcal B$ is a homogeneous basis of the coefficient ring $\Omega_*^{\mathrm O}$, there is an identification
\begin{equation*}
    \Omega_q^{\mathrm O} \cong
    \operatorname{Span}_{\mathbb Z_2}(\mathcal B_q).
\end{equation*}
Grouping the summands in Eq.~\eqref{eq:MO_homology_from_splitting} by the degree $q=|\alpha|$, and writing $p=n-q$, we get:
\begin{equation*}
    \bigoplus_{\alpha\in\mathcal B_q}
    H_{p}(X,\mathbb Z_2)\cong H_{p}(X, \Z_{2})\otimes_{\Z_{2}}\Omega^{\mathrm O}_{q}.
\end{equation*}
Therefore gives the natural additive decomposition
\begin{equation}
    \Omega_n^{\mathrm O}(X)
    \cong
    \mathrm{MO}_n(X)
    \cong
    \bigoplus_{p+q=n}
    H_p(X;\mathbb F_2)
    \otimes_{\mathbb F_2}
    \Omega_q^{\mathrm O}.
    \label{eq:MO_homology_additive_decomposition}
\end{equation}

Since the groups appearing in each fixed degree are finite-dimensional, dualizing gives a natural additive isomorphism
\footnote{The dual decomposition uses four standard facts. First, for fixed $n$ the sum over $p+q=n$ is finite, so its linear dual is again the corresponding direct sum. Second, since each $\Omega_q^{\mathrm O}$ is finite-dimensional, $\mathrm{Hom}_{\Z_{2}}(H_p(X)\otimes\Omega_q^{\mathrm O}, \Z_{2}) \cong \mathrm{Hom}(H_p(X),\Z_{2})\otimes\mathrm{Hom}_{\Z_{2}}(\Omega_q^{\mathrm O},\Z_{2})$. Third, the universal coefficient theorem over $\mathbb Z_2$ gives $\mathrm{Hom}_{\Z_{2}}(H_p(X),\Z_{2})\cong H^p(X)$. Finally, because $\Omega_q^{\mathrm O}$ has exponent $2$, its $\mathbb Z_2$-linear dual agrees with its $\mathrm U(1)$-valued Pontryagin dual.}
\begin{equation}
    \widehat{\Omega}_{\mathrm O}^{\,n}(X)
    \cong
    \bigoplus_{p+q=n}
    H^p(X)\otimes_{\mathbb Z_2}
    \widehat{\Omega}_{\mathrm O}^{\,q}(\{\mathrm{pt}\}).
    \label{eq:dual_MO_additive_decomposition}
\end{equation}

Under these identifications, the restriction map
$\res^{\mathcal A}_{\mathrm{cob}}$ fits into the commutative diagram
\begin{equation}
\begin{tikzcd}
\widehat{\Omega}_{\mathrm O}^{\,n}
\bigl(B\mathrm O(d)\bigr)
\arrow[r,"\res^{\mathcal A}_{\mathrm{cob}}"]
\arrow[d,"\cong"']
&
\displaystyle
\varprojlim_{E\in\mathcal A_2(\mathrm O(d))}
\widehat{\Omega}_{\mathrm O}^{\,n}(BE)
\arrow[d,"\cong"]
\\
\displaystyle
\bigoplus_{p+q=n}
H^p\bigl(B\mathrm O(d)\bigr)
\otimes
\widehat{\Omega}_{\mathrm O}^{\,q}
\arrow[r,"\res_{\mathcal A}\otimes\mathrm{id}"']
&
\displaystyle
\varprojlim_{E\in\mathcal A_2(\mathrm O(d))}
\bigoplus_{p+q=n}
H^p(BE)
\otimes
\widehat{\Omega}_{\mathrm O}^{\,q}.
\end{tikzcd}
\label{eq:cobordism_restriction_diagram}
\end{equation}
where we use $\widehat{\Omega}_{\mathrm O}^{\,q}=\widehat{\Omega}_{\mathrm O}^{\,q}(\{\mathrm{pt}\})$. The bottom map is an isomorphism because the mod-$2$ Quillen restriction map
\begin{equation*}
    \res_{\mathcal A}:
    H^p\bigl(B\mathrm O(d)\bigr)
    \xrightarrow{\;\cong\;}
    \varprojlim_{E\in\mathcal A_2(\mathrm O(d))}
    H^p(BE)
\end{equation*}
is an isomorphism in every degree. Hence
$\res^{\mathcal A}_{\mathrm{cob}}$ is an isomorphism.

\noindent The vertical identifications depend on the stable splitting chosen above, but the upper restriction map is defined directly by the subgroup inclusions and is independent of this choice. The splitting is used only to prove that this canonical map is an isomorphism.
\end{proof}

\bigskip 

\noindent Proposition~\ref{prop:cobordism_restriction_isomorphism} proves the elementary detection theorem, Theorem~\ref{theorem:intermediate}, for the unoriented bordism classification model. As explained in Sec.~\ref{sec:rigor}, once the elementary restriction map is known to be an isomorphism, the main theorem in the corresponding classification model follows. We state this conclusion explicitly below.

\bigskip

\begin{theorem}[Main theorem in the unoriented bordism model]
\label{appJ:main_theorem_cobordism}
For every spatial dimension $d\geq 1$, the stratification map
\begin{equation}
    F_{\mathrm{cob}}:
    \mathrm{cSPT}_{\mathrm{univ},\mathrm{cob}}^{d+1}
    \twoheadrightarrow
    \widehat{\Omega}_{\mathrm O}^{\,d+1}
    \bigl(B\mathrm O(d)\bigr)
\end{equation}
is surjective, where
\begin{equation*}
    \mathrm{cSPT}_{\mathrm{univ},\mathrm{cob}}^{d+1}
    :=
    \varprojlim_{G\in\mathcal O_{\mathrm{fin}}(\mathrm O(d))}
    \widehat{\Omega}_{\mathrm O}^{\,d+1}(BG).
\end{equation*}
\end{theorem}

\begin{proof}[
\textnormal{\textbf{Proof of
Theorem~\ref{appJ:main_theorem_cobordism}}}]
Let
\begin{equation*}
    \mathcal O
    :=
    \mathcal O_{\mathrm{fin}}(\mathrm O(d)),
    \qquad
    \mathcal A
    :=
    \mathcal A_2(\mathrm O(d)),
\end{equation*}
and define the universal and elementary inverse limits
\begin{equation*}
    \mathrm{cSPT}_{\mathrm{univ},\mathrm{cob}}^{d+1}
    :=
    \varprojlim_{G\in\mathcal O}
    \widehat{\Omega}_{\mathrm O}^{\,d+1}(BG),
\end{equation*}
and
\begin{equation*}
    \mathrm{cSPT}_{\mathrm{elem},\mathrm{cob}}^{d+1}
    :=
    \varprojlim_{E\in\mathcal A}
    \widehat{\Omega}_{\mathrm O}^{\,d+1}(BE).
\end{equation*}
By Proposition~\ref{prop:cobordism_restriction_isomorphism}, the
elementary restriction map is an isomorphism:
\begin{equation*}
    \res_{\mathrm{cob}}^{\mathcal A}:
    \widehat{\Omega}_{\mathrm O}^{\,d+1}
    \bigl(B\mathrm O(d)\bigr)
    \xrightarrow{\;\cong\;}
    \mathrm{cSPT}_{\mathrm{elem},\mathrm{cob}}^{d+1}.
\end{equation*}
We may therefore define the elementary continuum-limit map by
\begin{equation*}
    \widetilde F_{\mathrm{cob}}
    :=
    \bigl(\res_{\mathrm{cob}}^{\mathcal A}\bigr)^{-1}:
    \mathrm{cSPT}_{\mathrm{elem},\mathrm{cob}}^{d+1}
    \longrightarrow
    \widehat{\Omega}_{\mathrm O}^{\,d+1}
    \bigl(B\mathrm O(d)\bigr).
\end{equation*}
The inclusion
$\mathcal A\hookrightarrow\mathcal O$ induces the canonical projection
\begin{equation*}
    \Pi_{\mathrm{cob}}:
    \mathrm{cSPT}_{\mathrm{univ},\mathrm{cob}}^{d+1}
    \longrightarrow
    \mathrm{cSPT}_{\mathrm{elem},\mathrm{cob}}^{d+1}.
\end{equation*}
By Proposition~\ref{prop:Pi_surjective}, this projection is
surjective. Hence the stratification map
\begin{equation*}
    F_{\mathrm{cob}}
    :=
    \widetilde F_{\mathrm{cob}}
    \circ
    \Pi_{\mathrm{cob}}
\end{equation*}
is surjective, since it is the composition of a surjective map with
an isomorphism.
\end{proof}

\subsection{Cobordism classification: adding internal symmetries}
\label{app:cob_int}

We now extend the preceding cobordism argument to include a decoupled internal symmetry $H$ (see Def.~\ref{def:decoupled_internal} for definition). Let
\begin{equation*}
    \rho:
    \mathrm O(d)\longrightarrow\operatorname{Aut}(H)
\end{equation*}
describe the action of the spatial symmetry on $H$, and define
\begin{equation*}
    \Gamma_{\mathrm O(d)}
    := H\rtimes_{\rho}\mathrm O(d),
    \quad
    \Gamma_E
    := H\rtimes_{\rho|_E}E
\end{equation*}
for every finite subgroup $E\leq\mathrm O(d)$. We assume
Hypothesis~\ref{hyp:1} of Sec.~\ref{sec:internal_sym}. As shown in
App.~\ref{app:canon_factor_H}, this gives compatible Leray-Hirsch
isomorphisms
\begin{equation*}
    \alpha_E:
    H^*(B\Gamma_E;\mathbb Z_2)
    \xrightarrow{\;\cong\;}
    H^*(BE;\mathbb Z_2)\otimes_{\mathbb Z_2}R^*,
    \quad
    R^*:=H^*(BH;\mathbb Z_2),
\end{equation*}
including the case $E=\mathrm O(d)$.

\bigskip 

\noindent By Proposition~\ref{prop:quillen_isom_H}, Hypothesis~\ref{hyp:1}
implies that the $H$-enriched restriction map
\begin{equation}
    \res_{\mathcal A}^{H}:
    H^n\bigl(B\Gamma_{\mathrm O(d)};\mathbb Z_2\bigr)
    \xrightarrow{\;\cong\;}
    \varprojlim_{E\in\mathcal A_2(\mathrm O(d))}
    H^n(B\Gamma_E;\mathbb Z_2)
    \label{eq:internal_cohomology_restriction}
\end{equation}
is an isomorphism in every degree $n$.

\bigskip 

\begin{proposition}[Cobordism restriction with internal symmetry]
\label{prop:cobordism_restriction_internal}
For every degree $n$, restriction along the subgroup inclusions
$\Gamma_E\hookrightarrow\Gamma_{\mathrm{O}(d)}$ induces a canonical isomorphism
\begin{equation}
    \res_{\mathrm{cob}}^{\mathcal A,H}:
    \widehat{\Omega}_{\mathrm O}^{\,n}(B\Gamma_{\mathrm{O}(d)})
    \xrightarrow{\;\cong\;}
    \varprojlim_{E\in\mathcal A_2(\mathrm O(d))}
    \widehat{\Omega}_{\mathrm O}^{\,n}(B\Gamma_E).
    \label{eq:cobordism_restriction_internal}
\end{equation}
\end{proposition}

\begin{proof}
Fix the stable splitting of $\mathrm{MO}$ from
Theorem~\ref{thm:MO_stable_splitting}. By
Eq.~\eqref{eq:dual_MO_additive_decomposition}, it induces natural
additive isomorphisms
\begin{equation*}
    \widehat{\Omega}_{\mathrm O}^{\,n}(X)
    \cong
    \bigoplus_{p+q=n}
    H^p(X)\otimes_{\mathbb F_2}
    \widehat{\Omega}_{\mathrm O}^{\,q}(\{\mathrm{pt}\}).
\end{equation*}
Under these identifications, the restriction map fits into the
commutative diagram
\begin{equation}
\begin{tikzcd}
\widehat{\Omega}_{\mathrm O}^{\,n}(B\Gamma_{\mathrm{O}(d)})
\arrow[r,"\res_{\mathrm{cob}}^{\mathcal A,H}"]
\arrow[d,"\cong"']
&
\displaystyle
\varprojlim_{E\in\mathcal A_2(\mathrm O(d))}
\widehat{\Omega}_{\mathrm O}^{\,n}(B\Gamma_E)
\arrow[d,"\cong"]
\\
\displaystyle
\bigoplus_{p+q=n}
H^p(B\Gamma_{\mathrm{O}(d)})\otimes
\widehat{\Omega}_{\mathrm O}^{\,q}
\arrow[r,"\res_{\mathcal A}^{H}\otimes\mathrm{id}"']
&
\displaystyle
\varprojlim_{E\in\mathcal A_2(\mathrm O(d))}
\bigoplus_{p+q=n}
H^p(B\Gamma_E)\otimes
\widehat{\Omega}_{\mathrm O}^{\,q}.
\end{tikzcd}
\label{eq:cobordism_restriction_internal_diagram}
\end{equation}
The bottom map is an isomorphism by
Eq.~\eqref{eq:internal_cohomology_restriction}. Hence
$\res_{\mathrm{cob}}^{\mathcal A,H}$ is an isomorphism.

The vertical identifications depend on the fixed stable splitting, but
the upper restriction map is defined directly by the subgroup
inclusions and is independent of this choice. The splitting is used
only to prove that this canonical map is an isomorphism.
\end{proof}

\bigskip 

\noindent The inverse of Eq.~\eqref{eq:cobordism_restriction_internal} defines the partial stratification map
\begin{equation}
    \widetilde F_{\mathrm{cob}}^{H}
    :=
    \left(
        \res_{\mathrm{cob}}^{\mathcal A,H}
    \right)^{-1}:
    \varprojlim_{E\in\mathcal A_2(\mathrm O(d))}
    \widehat{\Omega}_{\mathrm O}^{\,n}(B\Gamma_E)
    \xrightarrow{\;\cong\;}
    \widehat{\Omega}_{\mathrm O}^{\,n}(B\Gamma_{\mathrm{O}(d)}).
\end{equation}
Let
\begin{equation*}
    \Pi_H:
    \varprojlim_{P\in\mathcal O_{\mathrm{fin}}(\mathrm O(d))}
    \widehat{\Omega}_{\mathrm O}^{\,n}(B\Gamma_P)
    \longrightarrow
    \varprojlim_{E\in\mathcal A_2(\mathrm O(d))}
    \widehat{\Omega}_{\mathrm O}^{\,n}(B\Gamma_E)
\end{equation*}
be the canonical restriction map induced by the inclusion of indexing
categories. The full stratification map is
\begin{equation}
    F_{\mathrm{cob}}^{H}
    :=
    \widetilde F_{\mathrm{cob}}^{H}\circ\Pi_H.
    \label{eq:full_cobordism_stratification_internal}
\end{equation}
Proposition~\ref{prop:cobordism_restriction_internal} proves the
$H$-enriched elementary detection theorem for the unoriented bordism
model. The general argument of Sec.~\ref{sec:rigor} therefore implies
the corresponding main theorem.

\bigskip

\begin{theorem}[Main theorem in the unoriented bordism model with internal symmetry]
\label{appJ:main_theorem_cobordism_internal}
For every spatial dimension $d\geq1$, the stratification map
\begin{equation}
    F_{\mathrm{cob}}^{H}:
    \varprojlim_{P\in\mathcal O_{\mathrm{fin}}(\mathrm O(d))}
    \widehat{\Omega}_{\mathrm O}^{\,d+1}(B\Gamma_P)
    \twoheadrightarrow
    \widehat{\Omega}_{\mathrm O}^{\,d+1}
    \bigl(B\Gamma_{\mathrm O(d)}\bigr)
\end{equation}
is surjective.
\end{theorem}

\begin{proof}
By Proposition~\ref{prop:cobordism_restriction_internal},
$\widetilde F_{\mathrm{cob}}^{H}$ is an isomorphism. By
Proposition~\ref{prop:Pi_surjective}, the canonical projection
$\Pi_H$ from the full finite-subgroup inverse limit to the elementary
inverse limit is surjective. Therefore
\begin{equation*}
    F_{\mathrm{cob}}^{H}
    =
    \widetilde F_{\mathrm{cob}}^{H}\circ\Pi_H
\end{equation*}
is surjective.
\end{proof}

\section{Universal continuum limit of invertible bosonic TQFTs with $\SO$-structure}
\label{app:so_str}

In the main text we prove surjectivity of the universal continuum-limit map for unoriented ($\mathrm O$-structured) invertible bosonic theories. In this appendix we explain how the corresponding statement for oriented ($\mathrm{SO}$-structured) theories follows formally, using functoriality of the classification model together with naturality of the Freed-Quinn construction.

\subsection{Setup}

We use the full group cohomology classification model~\eqref{eq:grpcohtormodel} defined in Sec.~\ref{sec:classification_models}. As explained there, we regard $\Theta^{d+1}$ as a contravariant classification functor on the classifying spaces of the symmetry groups under consideration, with the relevant coefficient system understood from the spatial representation; see also Remark~\ref{rmk:convention}. Thus,
\begin{equation*}
    \Theta^{d+1}(BG)
    :=
    \tor_2\bigl(
        H^{d+2}(BG,\mathbb Z_{w_1})
    \bigr).
\end{equation*}
For $G=\mathrm O(d)$, $w_1$ is the universal orientation character, whereas for $G=\SO(d)$ its pullback is trivial, so that
\begin{equation*}
    \Theta^{d+1}(B\SO(d))
    =
    \tor_2\bigl(
        H^{d+2}(B\SO(d),\mathbb Z)
    \bigr).
\end{equation*}
Elements of these groups are interpreted as monoidal natural isomorphism classes of invertible response theories with the corresponding tangential structure.

\bigskip 

Let $\mathcal O_{\mathrm{fin}}(G)$ denote the orbit/category of finite subgroups of a model Lie group $G$,
and define the corresponding universal ``lattice'' inverse limits
\begin{gather*}
\mathrm{cSPT}^{d+1}_{\mathrm O}
:= \varprojlim_{P\in \mathcal O_{\mathrm{fin}}(\mathrm{O}(d))}\Theta^{d+1}(BP),\\
\mathrm{cSPT}^{d+1}_{\mathrm{SO}}
:= \varprojlim_{P\in \mathcal O_{\mathrm{fin}}(\SO(d))}\Theta^{d+1}(BP).
\end{gather*}
The global universal continuum-limit map in the unoriented model,
\begin{equation*}
    F_{\mathrm O}: \mathrm{cSPT}^{d+1}_{\mathrm O}
    \longrightarrow \Theta^{d+1}(B\mathrm O(d)),
\end{equation*}
is constructed and shown to be surjective in Sec.~\ref{sec:rigor}. This map should be distinguished from the universal continuum-limit theory developed in Sec.~\ref{sec:continuum_limit}. The latter is the natural transformation $\kappa^{\mathrm{univ}}:\Theta_{\mathrm{univ}}^{d+1}\Longrightarrow\Phi$, whose component at a finite lattice symmetry group $G$ is the local universal continuum-limit map
\begin{equation*}
    \kappa_G^{\mathrm{univ}}:
    \operatorname{Im}(\pi_G)
    \longrightarrow
    \Theta^{d+1}(B\mathrm O(d))_G.
\end{equation*}
As explained in Sec.~\ref{sec:inverse_limit_construction} and proved
in App.~\ref{app:continuum_limit_component}, these local maps are
obtained canonically from $F_{\mathrm O}$ by the componentwise
factorization through the projections $\pi_G$. Whenever a local
universal continuum-limit prescription exists, the corresponding
commutative square therefore determines it uniquely from the global
map.

The same framework motivates considering a global oriented
continuum-limit map
\begin{equation*}
    F_{\mathrm{SO}}:
    \mathrm{cSPT}^{d+1}_{\mathrm{SO}}
    \longrightarrow
    \Theta^{d+1}(B\SO(d)).
\end{equation*}
We do not construct $F_{\mathrm{SO}}$ independently in this appendix.
Instead, we assume that the oriented universal continuum-limit
prescription exists and is natural under the inclusion
$\SO(d)\hookrightarrow\mathrm O(d)$. We then show that this naturality,
together with the explicitly constructed surjective map
$F_{\mathrm O}$, forces $F_{\mathrm{SO}}$ to be surjective. The
required compatibility is expressed by the comparison square
in Eq.~\eqref{eq:O_to_SO_univ_square}.

\subsection{From $\mathrm O$ to $\mathrm{SO}$: the comparison diagram}

Let $\iota:B\SO(d)\hookrightarrow B\mathrm{O}(d)$ be the canonical inclusion.
Functoriality of $\Theta^{d+1}$ gives a homomorphism
\begin{equation*}
\iota^{*}:\ \Theta^{d+1}(B\mathrm{O}(d))\longrightarrow \Theta^{d+1}(B\SO(d)).
\end{equation*}
On the lattice side, the inclusion of indexing categories
$\mathcal O_{\mathrm{fin}}(\SO(d))\hookrightarrow \mathcal O_{\mathrm{fin}}(\mathrm{O}(d))$ induces a canonical projection between inverse limits
\begin{equation*}
\pi^{\mathrm O}_{\mathrm{SO}}:\ \mathrm{cSPT}^{d+1}_{\mathrm O}\longrightarrow \mathrm{cSPT}^{d+1}_{\mathrm{SO}}.
\end{equation*}
Naturality of the universal continuum-limit construction with respect to the inclusion $\mathrm{SO}(d)\hookrightarrow\mathrm O(d)$ implies that these maps fit into the commutative square
\begin{equation}
\label{eq:O_to_SO_univ_square}
\begin{tikzcd}[column sep=large,row sep=large]
\mathrm{cSPT}^{d+1}_{\mathrm O} \arrow[r, "\kappa^{\mathrm{univ}}_{\mathrm O}"] \arrow[d, "\pi^{\mathrm O}_{\mathrm{SO}}"'] &
\Theta^{d+1}(B\mathrm{O}(d)) \arrow[d, "\iota^{*}"]\\
\mathrm{cSPT}^{d+1}_{\mathrm{SO}} \arrow[r, "\kappa^{\mathrm{univ}}_{\mathrm{SO}}"'] &
\Theta^{d+1}(B\SO(d)).
\end{tikzcd}
\end{equation}
By the unoriented main theorem proved in the main text, $\kappa^{\mathrm{univ}}_{\mathrm O}$ is surjective.

\bigskip

\begin{lemma}
\label{lem:iota_surj_theta}
In the group cohomology classification model, restriction along
$\iota:B\SO(d)\to B\mathrm O(d)$ induces a surjection
\begin{equation}
\iota^*: \tor_2\left(H^{d+2}(B\mathrm O(d),\Z_{w_1}) \right)
\twoheadrightarrow \tor_2\left(H^{d+2}(B\SO(d),\Z)
\right).
\label{eq:iota_surj_assumption}
\end{equation}
Equivalently,
\begin{equation*}
\iota^*: \Theta^{d+1}(B\mathrm O(d))
\twoheadrightarrow \Theta^{d+1}(B\SO(d))
\end{equation*}
is surjective.
\end{lemma}

\begin{proof}
Consider the short exact sequence of local coefficient systems $
0 \to \Z_{w_1} \xrightarrow{\times 2} \Z_{w_1} \to \Z_2 \to 0$ on $B\mathrm O(d)$. Its pullback to $B\SO(d)$ is the untwisted sequence $0 \to \Z \xrightarrow{\times 2} \Z \to \Z_2 \to 0,
$ since $\iota^*w_1=0$. Naturality of the associated Bockstein homomorphisms gives the commutative square
\begin{equation}
\label{eq:bock_sq}
\begin{tikzcd}[column sep=large,row sep=large]
H^{d+1}(B\mathrm O(d),\Z_2)
\arrow[r,"\iota^*"]
\arrow[d,"\beta_{\mathrm O}"']
&
H^{d+1}(B\SO(d),\Z_2)
\arrow[d,"\beta_{\mathrm{SO}}"]
\\
H^{d+2}(B\mathrm O(d),\Z_{w_1})
\arrow[r,"\iota^*"']
&
H^{d+2}(B\SO(d),\Z).
\end{tikzcd}
\end{equation}

\noindent The upper horizontal map is surjective. Indeed,
\begin{gather*}
H^*(B\mathrm O(d),\Z_2)
\cong \Z_2[w_1,\ldots,w_d],\\
H^*(B\SO(d),\Z_2)
\cong \Z_2[w_2,\ldots,w_d],
\end{gather*}
and $\iota^*$ sends $w_1$ to zero and $w_j$ to $w_j$ for $j\geq2$.

\bigskip 

\noindent By exactness of the two Bockstein sequences,
\begin{gather*}
\operatorname{Im}(\beta_{\mathrm O})
= H^{d+2}(B\mathrm O(d),\Z_{w_1})[2],\\
\operatorname{Im}(\beta_{\mathrm{SO}})
= H^{d+2}(B\SO(d),\Z)[2].
\end{gather*}
By Theorems~\ref{thm:BH_no_odd_torsion} and~\ref{thm:Greenblatt_twisted_no_odd_torsion}, all torsion in these two integral cohomology groups has order $2$. Consequently,
\begin{gather*}
H^{d+2}(B\mathrm O(d),\Z_{w_1})[2]
= \tor_2\left( H^{d+2}(B\mathrm O(d),\Z_{w_1}) \right),\\
H^{d+2}(B\SO(d),\Z)[2] =
\tor_2\left( H^{d+2}(B\SO(d),\Z) \right).
\end{gather*}

\bigskip 

\noindent Now let $y\in \tor_2\left( H^{d+2}(B\SO(d),\Z) \right)$. Since $\beta_{\mathrm{SO}}$ is surjective onto this subgroup, choose $\overline y\in H^{d+1}(B\SO(d),\Z_2)$ such that $\beta_{\mathrm{SO}}(\overline y)=y$. Surjectivity of the upper horizontal map in Eq.~\eqref{eq:bock_sq} gives $\overline x\in H^{d+1}(B\mathrm O(d),\Z_2)$ such that $\iota^*(\overline x)=\overline y$. Set
$x:=\beta_{\mathrm O}(\overline x)$. Then $x$ lies in
$\tor_2(H^{d+2}(B\mathrm O(d),\Z_{w_1}))$, and commutativity of
Eq.~\eqref{eq:bock_sq} gives
\begin{equation*}
\iota^*(x)
= \iota^*\bigl(\beta_{\mathrm O}(\overline x)\bigr)
= \beta_{\mathrm{SO}}\bigl(\iota^*(\overline x)\bigr)
= \beta_{\mathrm{SO}}(\overline y) = y.
\end{equation*}
Therefore $\iota^*$ is surjective.
\end{proof}

\bigskip 

\begin{theorem}[Main theorem for oriented systems in group cohomology model]
\label{theorem:main_SO}
For every spatial dimension $d\geq1$, assume that the global oriented continuum-limit map
\begin{equation*}
    F_{\mathrm{SO}}:
    \mathrm{cSPT}^{d+1}_{\mathrm{SO}}
    \longrightarrow
    \Theta^{d+1}(B\SO(d))
\end{equation*}
exists and is natural with respect to the inclusion
$\SO(d)\hookrightarrow\mathrm O(d)$, as expressed by
Eq.~\eqref{eq:O_to_SO_univ_square}. Then
\begin{equation}
    F_{\mathrm{SO}}:
    \mathrm{cSPT}^{d+1}_{\mathrm{SO}}
    \twoheadrightarrow
    \Theta^{d+1}(B\SO(d))
    \label{eq:maintheorem_SO}
\end{equation}
is surjective.
\end{theorem}

\begin{proof}
Let $\in\Theta^{d+1}(B\SO(d))$. By Lemma~\ref{lem:iota_surj_theta}, there exists $x\in\Theta^{d+1}(B\mathrm O(d))$ such that $\iota^*(x)=y$. Since the global unoriented continuum-limit map $F_{\mathrm O}$ is surjective, there exists
\begin{equation*}
    \widetilde x\in\mathrm{cSPT}^{d+1}_{\mathrm O}\quad \text{such that}\quad 
    F_{\mathrm O}(\widetilde x)=x.
\end{equation*}
Commutativity of Eq.~\eqref{eq:O_to_SO_univ_square} then gives
\begin{equation*}
    F_{\mathrm{SO}} \bigl(
        \pi^{\mathrm O}_{\mathrm{SO}}(\widetilde x)\bigr)
    = \iota^*\bigl(F_{\mathrm O}(\widetilde x)\bigr) = \iota^*(x) = y.
\end{equation*}
Therefore $F_{\mathrm{SO}}$ is surjective. The argument does not require the projection $\pi^{\mathrm O}_{\mathrm{SO}}$ itself to be surjective.
\end{proof}

\begin{theorem}[Main theorem for oriented systems with internal symmetry]
\label{theorem:main_SO_internal}
Let $H$ be a decoupled internal symmetry equipped with an action
\begin{equation*}
    \rho:
    \mathrm O(d)\longrightarrow\operatorname{Aut}(H),
\end{equation*}
and define
\begin{equation*}
    \Gamma_{\mathrm O(d)}
    :=
    H\rtimes_{\rho}\mathrm O(d),
    \qquad
    \Gamma_{\mathrm{SO}(d)}
    :=
    H\rtimes_{\rho|_{\mathrm{SO}(d)}}\mathrm{SO}(d).
\end{equation*}
Assume Hypotheses~\ref{hyp:1} and~\ref{hyp:2}. Suppose that the global
oriented continuum-limit map with internal symmetry exists and is
natural under the inclusion
$\Gamma_{\mathrm{SO}(d)}\hookrightarrow\Gamma_{\mathrm O(d)}$.
Then
\begin{equation}
    F_{\mathrm{SO}}^{H}:
    \mathrm{cSPT}^{d+1}_{\mathrm{SO},H}
    \twoheadrightarrow
    \Theta^{d+1}
    \bigl(B\Gamma_{\mathrm{SO}(d)}\bigr)
    \label{eq:maintheorem_SO_internal}
\end{equation}
is surjective.
\end{theorem}

\begin{proof}
By Hypothesis~\ref{hyp:1}, the compatible Leray-Hirsch
isomorphisms identify restriction along
\begin{equation*}
    \iota_H:
    B\Gamma_{\mathrm{SO}(d)}
    \longrightarrow
    B\Gamma_{\mathrm O(d)}\quad \text{with}\quad 
    \iota^*\otimes\mathrm{id}_{H^*(BH;\mathbb Z_2)}.
\end{equation*}
Since
\begin{equation*}
    \iota^*:
    H^*(B\mathrm O(d);\mathbb Z_2)
    \twoheadrightarrow
    H^*(B\mathrm{SO}(d);\mathbb Z_2)
\end{equation*}
is surjective, it follows that
\begin{equation}
    \iota_H^*:
    H^*(B\Gamma_{\mathrm O(d)};\mathbb Z_2)
    \twoheadrightarrow
    H^*(B\Gamma_{\mathrm{SO}(d)};\mathbb Z_2)
    \label{eq:internal_O_SO_mod2_surjection}
\end{equation}
is surjective.

\noindent Naturality of the Bockstein homomorphisms gives the commutative
square
\begin{equation}
\begin{tikzcd}[column sep=large,row sep=large]
H^{d+1}(B\Gamma_{\mathrm O(d)},\mathbb Z_2)
\arrow[r,"\iota_H^*"]
\arrow[d,"\beta_{\mathrm O(d)}^H"']
&
H^{d+1}(B\Gamma_{\mathrm{SO}(d)},\mathbb Z_2)
\arrow[d,"\beta_{\mathrm{SO}(d)}^H"]
\\
H^{d+2}(B\Gamma_{\mathrm O(d)},\mathbb Z_{w_1})
\arrow[r,"\iota_H^*"']
&
H^{d+2}(B\Gamma_{\mathrm{SO}(d)},\mathbb Z).
\end{tikzcd}
\label{eq:internal_O_SO_Bockstein_square}
\end{equation}
Exactness of the corresponding Bockstein sequences gives
\begin{gather*}
    \operatorname{Im}\bigl(\beta_{\mathrm O(d)}^H\bigr)
    =
    H^{d+2}
    \bigl(B\Gamma_{\mathrm O(d)},\mathbb Z_{w_1}\bigr)[2]\\
    \operatorname{Im}\bigl(\beta_{\mathrm{SO}(d)}^H\bigr)
    =
    H^{d+2}
    \bigl(B\Gamma_{\mathrm{SO}(d)},\mathbb Z\bigr)[2].
\end{gather*}
By Hypothesis~\ref{hyp:2}, these order-$2$ subgroups agree with the
corresponding $2$-primary classification groups. The same lifting
argument used in Lemma~\ref{lem:iota_surj_theta}, together with
Eq.~\eqref{eq:internal_O_SO_mod2_surjection}, therefore proves that
\begin{equation}
    \iota_H^*:
    \Theta^{d+1}
    \bigl(B\Gamma_{\mathrm O(d)}\bigr)
    \twoheadrightarrow
    \Theta^{d+1}
    \bigl(B\Gamma_{\mathrm{SO}(d)}\bigr)
    \label{eq:internal_O_SO_classification_surjection}
\end{equation}
is surjective.

\noindent Finally, apply the same element chase as in the proof of
Theorem~\ref{theorem:main_SO} to the $H$-enriched comparison square.
Surjectivity of the global unoriented map $F_{\mathrm O}^{H}$,
surjectivity of Eq.~\eqref{eq:internal_O_SO_classification_surjection},
and commutativity of that square imply that
$F_{\mathrm{SO}}^{H}$ is surjective.
\end{proof}

\subsection{Why \eqref{eq:iota_surj_assumption} is a statement about theories, not just abstract classification groups}
\label{app:O_SO_theory_naturality}

The proof in the preceding subsection was formulated entirely in terms
of classification groups and the homomorphisms between them. We now
explain why these maps represent corresponding operations on the
invertible TQFTs classified by those groups. The essential input is
the naturality of the Freed-Quinn construction established in
Secs.~\ref{app:naturality_bg_freedquinn} and~\ref{app:naturality_coeff_freedquinn}. A cocycle
\begin{equation*}
    \alpha
    \in
    Z^{d+1}\bigl(B\mathrm O(d),\mathbb R/\mathbb Z_{w_1}\bigr)
\end{equation*}
determines, through the Freed-Quinn construction, an invertible
unoriented TQFT
\begin{equation*}
    \mathcal Z_\alpha:
    \bordd^{\mathrm O}\bigl(B\mathrm O(d)\bigr)
    \longrightarrow
    \Line.
\end{equation*}
The inclusion
\begin{equation*}
    \iota:
    B\SO(d)
    \longrightarrow
    B\mathrm O(d)
\end{equation*}
pulls back both the background map and the coefficient system. Since $\iota^*w_1=0$, the pulled-back sign local system becomes the ordinary constant coefficient system on $B\SO(d)$. The resulting cocycle
\begin{equation*}
    \iota^*\alpha
    \in
    Z^{d+1}\bigl(B\SO(d),\mathbb R/\mathbb Z\bigr)
\end{equation*}
therefore defines an oriented invertible TQFT
\begin{equation*}
    \mathcal Z_{\iota^*\alpha}:
    \bordd^{\mathrm{SO}}\bigl(B\SO(d)\bigr)
    \longrightarrow
    \Line.
\end{equation*}
At the level of bordism categories, the inclusion $\iota$ induces a
symmetric-monoidal functor
\begin{equation*}
    \iota_*:
    \bordd^{\mathrm{SO}}\bigl(B\SO(d)\bigr)
    \longrightarrow
    \bordd^{\mathrm O}\bigl(B\mathrm O(d)\bigr),
\end{equation*}
which forgets the chosen orientation and postcomposes the background
map with $\iota$. Naturality of the Freed-Quinn construction gives the canonical
monoidal natural isomorphism
\begin{equation}
\begin{tikzcd}
\bordd^{\mathrm{SO}}\bigl(B\SO(d)\bigr)
\arrow[r,"\mathcal Z_{\iota^*\alpha}"]
\arrow[d,"\iota_*"']
&
\Line
\arrow[d,equal]
\\
\bordd^{\mathrm O}\bigl(B\mathrm O(d)\bigr)
\arrow[r,"\mathcal Z_\alpha"']
&
\Line
\end{tikzcd}
\Longleftrightarrow
\mathcal Z_{\iota^*\alpha}
\cong
\mathcal Z_\alpha\circ\iota_*.
\label{eq:FQ_SO_pullback}
\end{equation}
Thus the restriction homomorphism
\begin{equation*}
    \iota^*:
    \Theta^{d+1}\bigl(B\mathrm O(d)\bigr)
    \longrightarrow
    \Theta^{d+1}\bigl(B\SO(d)\bigr)
\end{equation*}
is not merely an abstract map between cohomology groups. Under the Freed-Quinn interpretation, it sends an unoriented TQFT to an oriented TQFT whose response agrees with the original theory on oriented backgrounds, where both theories are defined. The relevant compatibility may be summarized
by the commutative square
\begin{equation}
\begin{tikzcd}[column sep=huge,row sep=large]
\Theta^{d+1}\bigl(B\mathrm O(d)\bigr)
\arrow[r,"\text{[}\alpha\text{]}\mapsto\text{[}\mathcal Z_\alpha\text{]}"]
\arrow[d,"\iota^*"']
&
\pi_0\!\left(
\tqftd^{\mathrm O,\times}\bigl(B\mathrm O(d)\bigr)
\right)
\arrow[d,"(-)\circ\iota_*"]
\\
\Theta^{d+1}\bigl(B\SO(d)\bigr)
\arrow[r,"\text{[}\alpha\text{]}\mapsto\text{[}\mathcal Z_\alpha\text{]}"']
&
\pi_0\!\left(
\tqftd^{\mathrm{SO},\times}\bigl(B\SO(d)\bigr)
\right).
\end{tikzcd}
\label{eq:FQ_O_SO_classification_square}
\end{equation}
Here $\pi_0$ denotes monoidal natural isomorphism classes of invertible TQFT functors.

\bigskip 

Lemma~\ref{lem:iota_surj_theta} implies that every oriented response theory represented in the full group-cohomology model is, up to canonical monoidal natural isomorphism, obtained by restricting an unoriented response theory to oriented backgrounds. Thus, if
\begin{equation*}
    y=\iota^*(x),
\end{equation*}
then the corresponding Freed-Quinn theories satisfy
\begin{equation*}
    \mathcal Z_y
    \cong
    \mathcal Z_x\circ\iota_*.
\end{equation*}
The comparison square~\eqref{eq:O_to_SO_univ_square} states that this
restriction is compatible with the continuum limit:
\begin{equation}
    \iota^*\bigl(F_{\mathrm O}(\widetilde x)\bigr)
    =
    F_{\mathrm{SO}}
    \bigl(
        \pi^{\mathrm O}_{\mathrm{SO}}(\widetilde x)
    \bigr).
    \label{eq:physical_O_SO_continuum_compatibility}
\end{equation}
Accordingly, the proof of Theorem~\ref{theorem:main_SO} has the
following interpretation. Given an oriented continuum response theory,
one first lifts it to an unoriented response theory, then uses the
surjectivity of $F_{\mathrm O}$ to realize that theory by compatible
unoriented lattice data, and finally restricts those data to
$\SO(d)$-preserving systems. Equation
\eqref{eq:physical_O_SO_continuum_compatibility} guarantees that the
resulting oriented continuum limit is the original theory.

\noindent The Freed-Quinn construction therefore identifies the restriction
maps between cohomological classification groups with the corresponding
pullbacks of invertible TQFT functors. In this sense, the maps used in
Theorem~\ref{theorem:main_SO} relate the classified response theories
themselves, rather than only their abstract classification groups.

\begin{remark}[Scope of the interpretation]
The discussion above concerns the $2$-primary torsion sector of the
group-cohomology classification. Unlike the unoriented case, the full
classification of oriented invertible bosonic theories can contain
free, integer-quantized phases. Such phases, including the
$(2+1)$-dimensional $E_8$ phase, are not represented in the
classification model used in this appendix and are therefore not
included in Theorems~\ref{theorem:main_SO}
and~\ref{theorem:main_SO_internal}.
\end{remark}

\bigskip 

\subsubsection{Dependence on the structure group and the $C_2$ example}
\label{app:O_SO_admissible_domain_example}

We conclude this appendix by recording how the continuum-admitting domain depends on whether the target continuum theory is required to have $\mathrm O(d)$- or $\mathrm{SO}(d)$-structure, and by applying this distinction to the example of Sec.~\ref{sec:compatible_families}.

\begin{remark}[Dependence of the admissible domain on the structure group]
\label{rmk:O_SO_admissible_domains}
For every finite subgroup $G\leq\mathrm{SO}(d)$, let
\begin{equation*}
\pi_G^{\mathrm O}:
\mathrm{cSPT}^{d+1}_{\mathrm O}
\longrightarrow
\Theta^{d+1}(BG),
\quad
\pi_G^{\mathrm{SO}}:
\mathrm{cSPT}^{d+1}_{\mathrm{SO}}
\longrightarrow
\Theta^{d+1}(BG)
\end{equation*}
denote the corresponding component projections. Restricting an
$\mathrm O(d)$-compatible family to the orientation-preserving
indexing category gives
\begin{equation}
\pi_G^{\mathrm O}
=
\pi_G^{\mathrm{SO}}
\circ
\pi_{\mathrm{SO}}^{\mathrm O} \implies 
\operatorname{Im}(\pi_G^{\mathrm O})
\subseteq
\operatorname{Im}(\pi_G^{\mathrm{SO}}).
\label{eq:O_SO_domain_inclusion}
\end{equation}
The inclusion need not be an equality. A class in
\begin{equation*}
\operatorname{Im}(\pi_G^{\mathrm{SO}})
\setminus
\operatorname{Im}(\pi_G^{\mathrm O})
\end{equation*}
extends to a compatible family over the finite subgroups of
$\mathrm{SO}(d)$, but not after orientation-reversing symmetry sectors
are included. Thus, assuming the oriented universal continuum-limit
prescription exists, such a class may admit an
$\mathrm{SO}(d)$-structured continuum limit while being obstructed
from admitting an $\mathrm O(d)$-structured one. This does not
conflict with Theorem~\ref{theorem:main_SO}, which concerns
surjectivity onto the oriented continuum classification and does not
require $\pi_{\mathrm{SO}}^{\mathrm O}$ to be surjective.
\end{remark}

The class considered in Sec.~\ref{sec:compatible_families} gives a
concrete illustration in $d=2$. In the pre-TQFT model,
\begin{equation*}
[c^3]\in H^3(BC_2,\mathbb Z_2)
\end{equation*}
does not extend to a compatible family over the relevant
$\mathrm O(2)$ indexing category. It is therefore obstructed from
belonging to the $\mathrm O(2)$-structured universal domain $\mathrm{cSPT}^{2+1}_{\mathrm{univ}, \mathrm{O}}$. In the full group-cohomology model, the corresponding lattice TQFT class is
the Bockstein lift
\begin{equation*}
\beta_{C_2}(c^3)
\in
H^4(BC_2,\mathbb Z).
\end{equation*}
For the oriented structure group, $\mathrm{SO}(2)\cong\mathrm U(1)$, let
\begin{equation*}
u=c_1
\in
H^2(B\mathrm{SO}(2),\mathbb Z).
\end{equation*}
Then the free class $u^2 \in H^4(B\mathrm{SO}(2),\mathbb Z)$ restricts to compatible torsion classes on the finite cyclic subgroups of $\mathrm{SO}(2)$, and its $C_2$-component is
\begin{equation}
\operatorname{res}^{\mathrm{SO}(2)}_{C_2}(u^2)
=
\beta_{C_2}(c^3).
\label{eq:C2_free_shadow_restriction}
\end{equation}
Consequently,
\begin{equation*}
\beta_{C_2}(c^3)
\in
\operatorname{Im}(\pi_{C_2}^{\mathrm{SO}})
\setminus
\operatorname{Im}(\pi_{C_2}^{\mathrm O}).
\end{equation*}
The class therefore admits an oriented compatible extension, even
though it does not admit an unoriented one.

Finally, there is an additional subtlety. In the strict torsion model used in
this appendix,
\begin{equation*}
\Theta^3(B\mathrm{SO}(2))
=
\tor_2
H^4(B\mathrm{SO}(2),\mathbb Z)
=
0.
\end{equation*}
Hence the oriented continuum image of this compatible family is
trivial. This is precisely consistent with its interpretation as a
free-shadow family: its finite-subgroup components are torsion
classes, but they arise by restricting the free continuum class
$u^2$. The lattice class $\beta_{C_2}(c^3)$ itself does not require
differential data, but retaining its nontrivial free continuum parent
does. A global response associated with $u^2$ requires a differential
refinement that includes a connection and its curvature data, as
explained in App.~\ref{app:diff_refinement}. Since such refinements are
not part of the strict torsion classification used here, the free
continuum response is projected out.

\bigskip 

This example also shows that the tangential structure and classification model must be chosen according to the continuum structure and response data one wishes to retain or characterize. Using an $\mathrm O(d)$-structured model for a problem requiring only $\mathrm{SO}(d)$-structure may impose irrelevant orientation-reversing extension conditions, while restricting to the torsion model when free responses are physically relevant may discard nontrivial differentially refined continuum data.

\bibliography{draft}

\begin{thebibliography}{134}
\bibitem [{not({\natexlab{b}})}]{note_free}%
  \BibitemOpen
  \href@noop {} {}\bibinfo {howpublished} {These integer responses are often called “Schwarz-type” or “secondary” TQFTs. We refer to them as the free/secondary sector to highlight their dependence on extra tangential structure; with that extra structure they define good framed/structured invertible TQFTs, but they are not the diffeomorphism-invariant objects studied here.} ({\natexlab{b}})\BibitemShut {NoStop}%
\bibitem [{not({\natexlab{d}})}]{note_universalbundle}%
  \BibitemOpen
  \href@noop {} {}\bibinfo {howpublished} {Explicit constructions of universal bundles were given by Milnor via the Milnor join construction~\cite{MilnorJoin}, and later by Segal~\cite{SegalClassifyingSpace}, whose approach connects naturally with modern homotopy-theoretic and categorical formulations.} ({\natexlab{d}})\BibitemShut {NoStop}%
\bibitem [{not({\natexlab{f}})}]{note_GCT}%
  \BibitemOpen
  \href@noop {} {}\bibinfo {howpublished} {\textcolor{red}{Note additional conditions that define GCT...}} ({\natexlab{f}})\BibitemShut {NoStop}%
\bibitem [{not({\natexlab{g}})}]{note_conjugation}%
  \BibitemOpen
  \href@noop {} {}\bibinfo {howpublished} {Let $G\le O(d)$ be finite and let $\lambda\in O(d)$. Conjugation defines an isomorphism $c_\lambda :\ G \to \lambda G\lambda^{-1}$ and hence induces an isomorphism on classifying spaces $Bc_\lambda: BG\to B(\lambda G\lambda^{-1})$ and on classifications \[\Theta(Bc_\lambda):\ \Theta\bigl(B(\lambda G\lambda^{-1})\bigr)\xrightarrow{\ \cong\ }\Theta(BG).\] Naturality of restriction along the inclusion $G\hookrightarrow O(d)$ gives the commutative relation $\mathrm{res}^{O(d)}_{G} \;=\; \Theta(Bc_\lambda )\circ \mathrm{res}^{O(d)}_{\lambda G\lambda ^{-1}}.$ Since $\Theta(Bc_\lambda )$ is an isomorphism, we have \[\ker\bigl(\mathrm{res}^{O(d)}_{G}\bigr) = \ker\bigl(\mathrm{res}^{O(d)}_{\lambda G\lambda ^{-1}}\bigr) \subseteq \Theta(BO(d)).\] The same argument applies to any subgroup $P\le G$ (noting $\lambda P\lambda ^{-1}\le \lambda G\lambda ^{-1}$), yielding $ \ker\bigl(\mathrm{res}^{O(d)}_{P}\bigr) = \ker\bigl(\mathrm{res}^{O(d)}_{\lambda P\lambda ^{-1}}\bigr).$
  Therefore the $G$-invisible subgroup $K_{\mathcal O(G)}$ is conjugation-invariant, and we have $K_{\mathcal O(G)} = K_{\mathcal O(\lambda G\lambda ^{-1})}.$} ({\natexlab{g}})\BibitemShut {NoStop}%
\bibitem [{not({\natexlab{h}})}]{note_fit}%
  \BibitemOpen
  \href@noop {} {}\bibinfo {howpublished} {We can define the map $\mathrm{res}^{O(d)}_{H}: \Theta(BO(d))\to \mathrm{im}(\mathrm{res}^{O(d)}_{H})$ which is by definition a surjection. This map has the kernel which we identified as $\mathrm{ker}(\mathrm{res}^{O(d)}_{H})=K_{\mathcal{O}(H)}$, and the First isomoprhism theorem states that given a surjective map $f:A\to B$ with kernel $K$ there is an isomorphism $\overline{f}:\dfrac{A}{K}\to B$.} ({\natexlab{h}})\BibitemShut {NoStop}%
\bibitem [{not({\natexlab{i}})}]{note_kill}%
  \BibitemOpen
  \href@noop {} {}\bibinfo {howpublished} {Recall that one can define a reduction map $\rho: H^{d+2}(BO(2), \mathbb{Z}_{w_{1}})[2]\to H^{d+2}(BO(2), \mathbb{Z}_{2})$ and the Bockstein map $\beta_{w_{1}}: H^{d+2}(BO(2), \mathbb{Z}_{2}) \to H^{d+2}(BO(2), \mathbb{Z}_{w_{1}})[2]$ where $H^{d+2}(BO(2), \mathbb{Z}_{w_{1}})[2]$ represents the order-$2$ subgroup of $H^{d+2}(BO(2), \mathbb{Z}_{w_{1}})$. Then, one can show that $\rho\circ \beta_{w_{1}}(w_{1}^{3})= \mathrm{Sq}^{1}(w_{1}^{3})+ w_{1}\cup w_{1}^{3}=0$, and since $\rho$ is injective(follows from the LES in cohomology) we have $\beta_{w_{1}}(w_{1}^{3})=0$. This consequently means the restriction to all elementary groups of this class vanish when lifted to twisted integral cohomology too.} ({\natexlab{i}})\BibitemShut {NoStop}%
\end{thebibliography}
\end{document}